%% file: PhD_Thesis_TS.tex
\documentclass[10pt]{thesis_mw2}
\usepackage[utf8]{inputenc}   

\usepackage[usenames,dvipsnames,svgnames,table]{xcolor} 
\usepackage[obeyspaces,hyphens,spaces]{url}
\usepackage{thesisfrontmatter_TS}
\usepackage{pdfpages}

\definecolor{Blue}{rgb}{0.25, 0.41, 0.88}
\definecolor{Red}{rgb}{0.92,0.,0.}
\definecolor{darkorange}{rgb}{1.0,0.549,0.}
\definecolor{cobalt}{RGB}{44, 98, 120}
\definecolor{Mathematica1}{rgb}{0.368417, 0.506779, 0.709798}
\definecolor{Mathematica2}{rgb}{0.880722, 0.611041, 0.142051}
\definecolor{Mathematica3}{rgb}{0.560181, 0.691569, 0.194885}
\definecolor{Mathematica3Dark}{rgb}{0.320, 0.450, 0.110}
\definecolor{Mathematica4}{rgb}{0.922526, 0.385626, 0.209179}
\definecolor{Mathematica4Dark}{rgb}{0.5, 0.2, 0.1}
\definecolor{Mathematica4Bright}{rgb}{1.0, 0.5, 0.3}
\definecolor{Mathematica5}{rgb}{0.528488, 0.470624, 0.701351}
\definecolor{Mathematica5Light}{rgb}{0.750, 0.650, 0.850}
\definecolor{Mathematica6}{rgb}{0.772079, 0.431554, 0.102387}
\definecolor{Mathematica7}{rgb}{0.363898, 0.618501, 0.782349}
\definecolor{Mathematica8}{rgb}{1, 0.75, 0}
\definecolor{Mathematica9}{rgb}{0.647624, 0.37816, 0.614037}
\definecolor{MathematicaBrown}{rgb}{0.6, 0.4, 0.2}
\definecolor{plotBlue}{RGB}{94, 130, 181}
\definecolor{plotRed}{RGB}{233, 85, 54}
\definecolor{plotGreen}{RGB}{142, 176, 50}
\definecolor{plotPurple}{RGB}{135, 120, 178}
\definecolor{stormyocean}{RGB}{50,100,130}
\definecolor{cornellRed}{HTML}{B31B1B}
\definecolor{cornellRed2}{RGB}{152, 22, 22}
\definecolor{cornellBlue}{HTML}{0068AC}
\definecolor{cornellGreen}{HTML}{6EB43F}
\definecolor{dullpurple}{rgb}{0.431,0.188,0.534}
\definecolor{darkgreen}{rgb}{0.075,0.302,0.047}
\definecolor{darkergreen}{rgb}{0,0.196,0.125}
\definecolor{darkergreen2}{rgb}{0,0.294,0.188}
\definecolor{dullred}{rgb}{0.706,0.208,0.192}
\definecolor{darkred}{rgb}{0.545,0,0}
\definecolor{antiquefuchsia}{rgb}{0.57, 0.36, 0.51}
\definecolor{MaroonC}{rgb}{0,0.502,0.502}
\definecolor{dullblue}{rgb}{0,0.298,0.49}
\definecolor{blue3}{RGB}{31, 119, 180}
\definecolor{red3}{RGB}{	214, 39, 40}
\definecolor{orange3}{RGB}{255, 127, 14}
\definecolor{green3}{RGB}{44, 160, 44}
\definecolor{BetterYellow}{rgb}{0.85, 0.6, 0}
\definecolor{deeppurple}{RGB}{128, 0, 64}
\definecolor{redpurple}{RGB}{200, 10, 30}
\definecolor{brightblue}{RGB}{0, 0, 255}
\definecolor{charcoal}{RGB}{51,51,51}
\definecolor{peachy}{RGB}{238, 180, 143}
\definecolor{mintgreen}{RGB}{90, 170, 130}
\definecolor{softred}{RGB}{220, 100, 110}
\definecolor{forestgreen}{RGB}{50, 120, 50}
\definecolor{mutedblue}{rgb}{0.227, 0.396, 0.549}
\definecolor{mutedred}{rgb}{0.549, 0.227, 0.227}
\definecolor{darkmutedblue}{RGB}{75, 93, 120}
\definecolor{darkblue1}{RGB}{33, 31, 78}
\definecolor{mediumblue1}{RGB}{80, 137, 190}
\definecolor{tealgreen1}{RGB}{82, 173, 157}
\definecolor{plotblue1}{RGB}{128, 128, 250}

\usepackage{ifthen}
\usepackage{fancyhdr}
\usepackage{enumitem}
\usepackage[colorlinks=true
,urlcolor=mutedblue 
,anchorcolor=blue3
,citecolor=mutedblue
,filecolor=blue3
,linkcolor=mutedred
,menucolor=blue3
,pagecolor=blue3
,linktocpage=true
,pdfproducer=medialab
]{hyperref}
\usepackage{booktabs}
\usepackage[english, american, dutch, greek, swedish]{babel}
\usepackage{amsmath, amssymb, amsbsy, amstext, amsthm, simplewick}
\usepackage{cleveref}
\usepackage{graphicx}
\usepackage{amsfonts}
\usepackage{enumitem}
\usepackage{upgreek}
\usepackage{framed}
\usepackage{tensor}
\usepackage{pifont}
\usepackage{latexsym, mathrsfs}
\usepackage{array}
\usepackage{xspace}
\usepackage{longtable}
\usepackage{multirow}
\usepackage{cases}
\usepackage{empheq}

\usepackage{bm}
\usepackage{amsmath}
\usepackage{bbm}
\usepackage{appendix}
\usepackage{mathrsfs}

\renewcommand{\vec}[1]{\boldsymbol{\mathbf{#1}}}

\newcommand{\dd}{\mathop{\mathrm{d}\!}{}}
\newcommand{\floq}[1]{{\scalebox{0.65}{$(#1)$}}}
\newcommand{\ped}[1]{\textormath{\textsubscript{#1}}{_{\mathrm{#1}}}}
\newcommand{\ap}[1]{\textormath{\textsuperscript{#1}}{^{\mathrm{#1}}}}

\DeclarePairedDelimiter{\abs}{\lvert}{\rvert}
\newcommand{\slab}[1]{{\textsc{#1}}}
\usepackage{nicematrix}
\newcommand*{\mline}[1]{%
\begingroup
    \renewcommand*{\arraystretch}{1.5}%
   \begin{tabular}[c]{@{}>{\raggedright\arraybackslash}p{2.5cm}@{}}#1\end{tabular}%
  \endgroup
}

\usepackage{braket}
\usepackage{yfonts}

\usepackage[load-configurations=astronomy, range-units=brackets, range-phrase=-, per-mode=reciprocal, mode=math]{siunitx}
\usepackage{threeparttable}
\usepackage{subfig}
\usepackage{ragged2e}
\usepackage{hhline}
\usepackage{array}
\newcolumntype{P}[1]{>{\centering\arraybackslash}p{#1}}
\usepackage{booktabs}
\usepackage{tikz}
\usetikzlibrary{calc,shadings,patterns,tikzmark,fadings}

\newcolumntype{C}[1]{>{\centering\let\newline\\\arraybackslash\hspace{0pt}}m{#1}}

\numberwithin{equation}{section}

\def\beq{\begin{equation}}
\def\eeq{\end{equation}}

\def\bea{\begin{eqnarray}}
\def\eea{\end{eqnarray}}

\newcommand{\minus}{{\scalebox{0.75}[1.0]{$-$}}}

\newcommand{\blackasteriskfootnote}[1]{%
  \begingroup
  \renewcommand{\thefootnote}{\textcolor{black}{\dag}}%
  \footnote{#1}%
  \endgroup
}

\DeclareRobustCommand{\SkipTocEntry}[4]{}

\usepackage{colortbl}

\usepackage[T1]{fontenc}
\usepackage{lmodern}
\definecolor{blue2}{cmyk}{1, 0.1, 0.1, 0}

\definecolor{pyBlue}{RGB}{31, 119, 180}
\definecolor{pyRed}{RGB}{214, 39, 40}
\definecolor{pyGreen}{RGB}{44, 160, 44}
\definecolor{pyBlue2}{RGB}{0, 111, 237}
\definecolor{pyRed2}{RGB}{224, 52, 36}

\makeatletter
\def\Ddots{\mathinner{\mkern1mu\raise\p@
\vbox{\kern7\p@\hbox{.}}\mkern2mu
\raise4\p@\hbox{.}\mkern2mu\raise7\p@\hbox{.}\mkern1mu}}
\makeatother

\usepackage{chngcntr}
\usepackage{etoolbox}

\AtBeginEnvironment{subappendices}{%
}

\thesisfont{cmr} 
\thesissectionstyle{\rmfamily\bfseries}
\thesissectionsizes{\Large}{\large}{}

\usepackage[explicit]{titlesec}
\usepackage{lmodern}
\newlength\chapnumb
\newlength\chapnumbless
\titleformat{\chapter}[block]
{\normalfont\sffamily\scshape}{}{0pt}
{\parbox[b]{\chapnumb}{
\fontsize{60}{0}\selectfont\thechapter}
\parbox[b]{\dimexpr\textwidth-\chapnumb\relax}{
\raggedleft
{\fontsize{20}{24}\selectfont#1 \\[4pt]}
\rule{\dimexpr\textwidth-\chapnumb\relax}{0.4pt} \vskip -9pt
\rule{\dimexpr\textwidth-\chapnumb\relax}{0.4pt}}}    
                     
\titleformat{name=\chapter,numberless}[block]
{\normalfont\sffamily\scshape}{}{0pt}
{\parbox[b]{\chapnumbless}{
\mbox{}}
\parbox[b]{\dimexpr\textwidth-\chapnumbless\relax}{
\raggedleft
\hfill{\Huge#1}\\
\rule{\dimexpr\textwidth-\chapnumbless\relax}{0.4pt} \vskip -10pt
\rule{\dimexpr\textwidth-\chapnumbless\relax}{0.4pt}}}
              
\usepackage{fix-cm}
\newcommand\HUGE{\@setfontsize\Huge{38}{47}} 

\title{{\Huge \sffamily Exploring Black Hole Environments\\[10pt]}}   
\thesiscolofon{\input{Colofon}}
\author{Thomas F.~M.~Spieksma}
\placeofbirth{Maastricht}
\theday{Thursday}
\date{4th September 2025}
\thetime{14.00}

\promotor{
\committeeentry{prof.~dr.~V.~Cardoso}{Niels Bohr Institutet, K{\o}benhavns Universitet},
\committeeentry{}{Instituto Superior T\'{e}cnico da Universidade de Lisboa}
}
\copromotor{\committeeentry{}{}}
\othercomitteemembers{ 
\committeeentry{dr.~K.~Clough}{Queen Mary University of London},
\committeeentry{dr.~T.~Harmark}{Niels Bohr Institutet, K{\o}benhavns Universitet},
\committeeentry{prof.~dr.~P.~Pani}{Universit\`{a} degli Studi di Roma ``La Sapienza''}
}
\department{\small Niels Bohr Institute}

\usepackage{pifont}
\usepackage[nottoc]{tocbibind}

\newcommand{\committeeentry}[2]{\begin{tabular*}{0.78\textwidth}{p{0.28\textwidth}l}#1 & #2\end{tabular*}}

\begin{document}
\selectlanguage{english}
{
\includepdf[fitpaper]{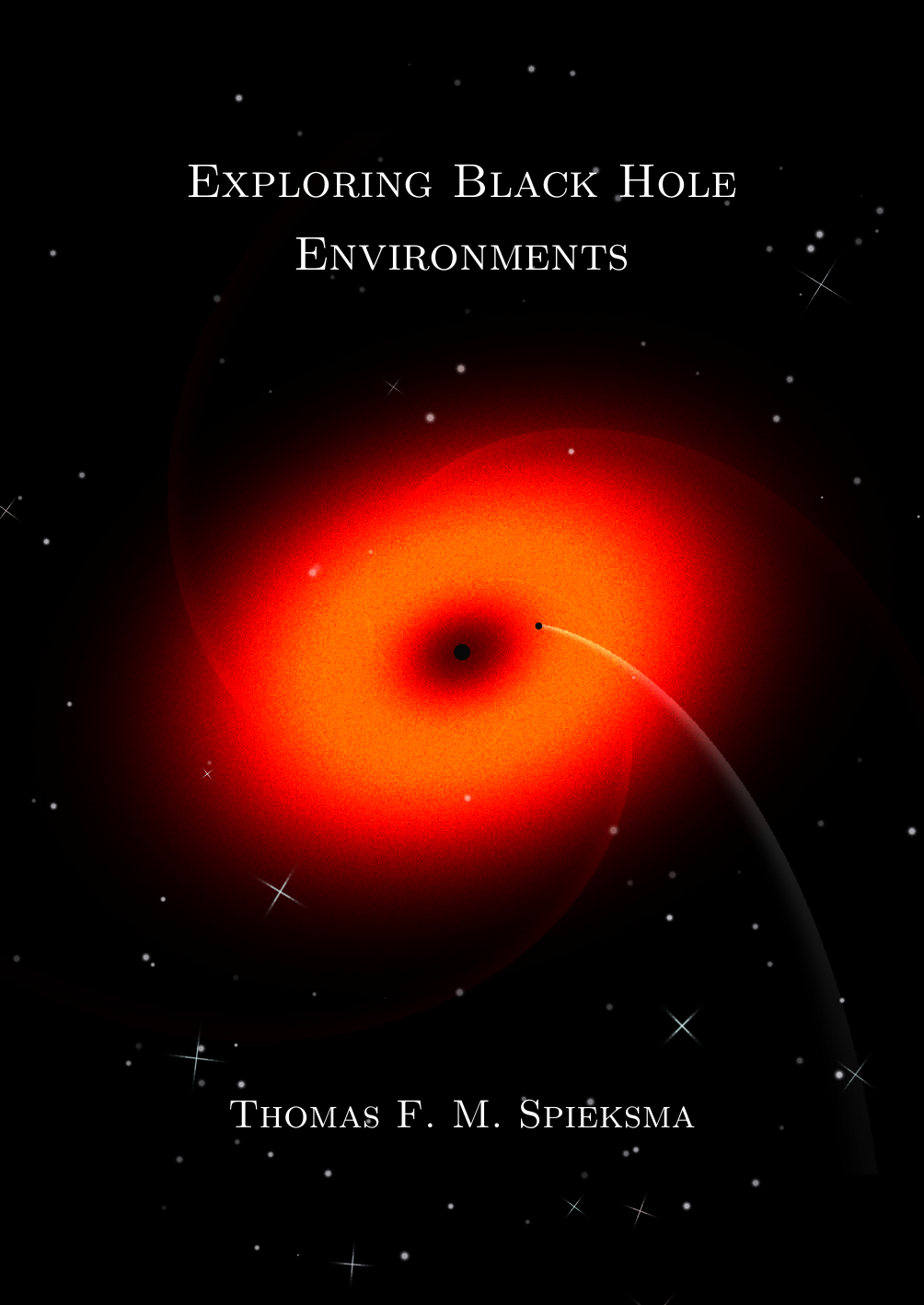}
\let\cleardoublepage\clearpage
\frontmatter
	\maketitle
	}
\poetry{}{\input{Poem}}
\summary{Abstract / Abstrakt}{\input{Sammenfatning2}}
\basedon{Publications}{\input{Basedon}}
\acknowledgements{Acknowledgements}{\input{Acknowledgements}}
\setcounter{tocdepth}{2}
\tableofcontents
\prologue{Prologue}{\input{Prologue}}
\mainmatter
\chapter{Introduction}
\vspace{-0.8cm}
\hfill \emph{Mine is the migrating bird}

\hfill \emph{Winging over perilous regions of the ocean}

\hfill \emph{Ever tracing out the age-old path of the}

\hfill \emph{Wandering waves}
\vskip 5pt

\hfill A Tuamotuan \emph{fangu}
\vskip 35pt
\noindent The power of theoretical physics lies in its ability to explore a world beyond our reach. The quintessential example is the black hole. By definition, it marks a region of spacetime that is causally disconnected from us:~no one will ever venture inside and return to tell the tale. When black holes were first predicted over a century ago, they seemed little more than a mathematical curiosity, mere artefacts of General Relativity. That scepticism, in retrospect, was misplaced. Not only do we now have compelling evidence for their existence from multiple observational channels~\cite{Remillard:2006fc,Ghez:2008ms,2009ApJ...692.1075G,LIGOScientific:2016aoc,EventHorizonTelescope:2019dse,EventHorizonTelescope:2022wkp}, but it is also becoming increasingly clear that they drive some of the most powerful phenomena in the Universe and play a crucial role in its evolution~\cite{Blandford:1977ds,King:2003ix,Volonteri:2010wz,Fabian:2012xr,Kormendy:2013dxa}.
\vskip 2pt
The observational triumph of black holes was crowned by the first detection of gravitational waves in 2015~\cite{LIGOScientific:2016aoc} -- an achievement whose significance is hard to overstate. These waves were produced by two black holes merging 1.4 billion years ago, sending ripples through spacetime that, upon reaching Earth, displaced two mirrors by just one hundredth the width of a proton. From this minuscule shift, their coalescence could be inferred -- an extraordinary feat of experimental precision and technological ingenuity. Though the physical disturbance has long passed, its scientific reverberations persist. Until then, our understanding of the Universe had relied entirely on light:~photons travelling across space into our eyes and telescopes. This detection fundamentally changed that paradigm:~these waves are not light but ripples of spacetime itself, offering a way to explore the \emph{dark Universe}.
\clearpage
The last decade marked the dawn of a new era in gravitational physics, an \emph{era of observations}. During this time, LIGO and Virgo have detected gravitational waves from over a hundred mergers of binary black holes and neutron stars~\cite{LIGOScientific:2018mvr,LIGOScientific:2020ibl,KAGRA:2021vkt}. These detections have not only confirmed our theory of gravity, General Relativity, in the strong-field, dynamical regime~\cite{LIGOScientific:2019fpa,LIGOScientific:2020tif,LIGOScientific:2021sio}, but also serve as astrophysical messengers, revealing events from across the Universe. A landmark event was the first detection of a binary neutron star merger~\cite{LIGOScientific:2017vwq}, accompanied by electromagnetic counterparts~\cite{LIGOScientific:2017ync,Cowperthwaite:2017dyu,Troja:2017nqp}, which marked a pivotal moment for the field of \emph{multi-messenger astronomy}. It provided compelling evidence that neutron star mergers forge heavy elements~\cite{Kasen:2017sxr,Pian:2017gtc}, enabled precise measurements of the speed of gravitational waves~\cite{LIGOScientific:2017zic} and introduced a new method for measuring the expansion rate of the Universe~\cite{LIGOScientific:2017adf}.
\vskip 2pt
Building on this exciting potential, the next generation of gravitational-wave detectors is being developed to extend the frequency range and enhance sensitivity. While current detectors primarily observe compact-object mergers of nearly equal mass, these are not the only sources of gravitational waves in our Universe. Of particular interest are \emph{extreme mass ratio inspirals} (EMRIs), in which a small object orbits a much larger one. Due to the disparity in masses, the smaller object moves through spacetime like a rubber duck drifting along a river, slowly spiralling inwards as it radiates gravitational waves. Whereas existing detectors catch only the final moments of a binary coalescence -- the crescendo of its ``song’' -- EMRIs could remain within a detector’s sensitivity band for years, allowing us to hear their entire melody. Moreover, unlike the violent, waterfall-like mergers of equal-mass binaries, EMRIs are more sensitive to disturbances along the way. Precisely tracking their motion could reveal subtle effects from surrounding matter, turning them into powerful tools for mapping their \emph{environment}. Such matter distributions are ubiquitous around black holes, especially in galactic centres~\cite{Sadeghian:2013laa}, making EMRIs natural \emph{(astro)particle detectors}.
\vskip 2pt
The \emph{Laser Interferometer Space Antenna} (LISA) is one of the key detectors planned to observe EMRIs. With its 2.5 million-kilometre-long laser arms, LISA will operate in the millihertz regime~\cite{Colpi:2024xhw}. Much like removing earplugs, LISA will reveal a cacophony of cosmic signals, demanding major advances in theoretical modelling and statistical analysis to disentangle them~\cite{Cornish:2005qw,Littenberg:2023xpl}. Environments complicate this task by modifying the waveform relative to vacuum, introducing additional parameters and potential degeneracies into the analysis. On top of that, environments themselves are dynamical:~binary systems can disrupt or deplete them, even before entering a detector’s sensitivity band. Addressing these challenges calls for careful modelling of both the environment and the binary's evolution, from formation to final merger. Without this, gravitational-wave signals might be missed, parameters misinterpreted, and opportunities lost to probe dark matter or new fundamental fields~\cite{Barausse:2014tra,Cole:2022yzw}.
\vskip 2pt
Most environments can be classified as either gaseous or composed of dark matter. Gaseous environments, such as accretion disks and plasma, are extensively studied through electromagnetic observations and offer exciting prospects for multi-messenger astronomy. Accretion disks in active galactic nuclei, for example, may act as ``nurseries'' for black hole binaries~\cite{Bartos:2016dgn,Stone:2016wzz,Tagawa:2019osr,Pan:2021ksp,Derdzinski:2022ltb}, capturing them and \emph{driving} their inspiral by aligning the orbit with the disk plane and shaping their eccentricity evolution~\cite{Spieksma:2025wex}. Additionally, gaseous environments can influence fundamental fields near black holes, particularly when those fields are coupled to the electromagnetic sector. Specifically, while putative ultralight bosonic fields could give rise to powerful electromagnetic radiation, surrounding plasmas can suppress the conversion of bosons to photons, a phenomenon known as \emph{in-medium suppression}~\cite{Cannizzaro:2024hdg}. As a result, only very strong couplings can lead to detectable electromagnetic signatures~\cite{Spieksma:2023vwl}. Plasmas also play a crucial role for charged black holes, especially in the final stages of binary coalescence, where they may alter the characteristic ``ringing'' frequencies of the remnant black hole or even generate \emph{echoes} in the gravitational-wave signal~\cite{Cannizzaro:2024yee}.
\begin{figure}[t!]
\centering
\includegraphics[scale=1.10, trim = 23 0 0 12]{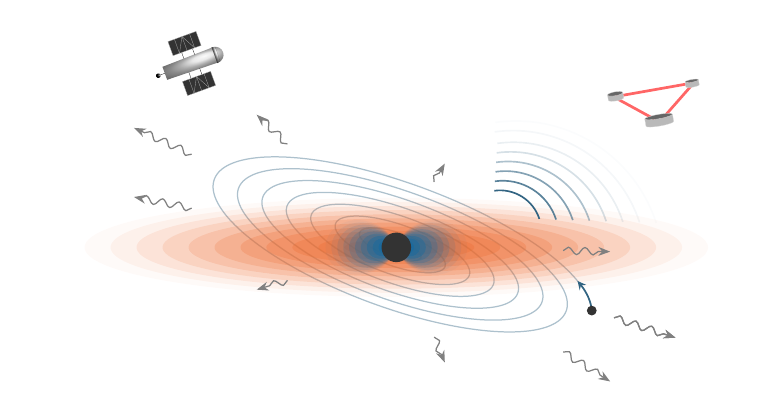}
\caption{Illustration of the central theme of this thesis:~exploring black hole environments. Black holes may drive some of the most energetic processes in our Universe and offer a unique window into a regime of physics that is largely uncharted.}
\label{fig:introfigure}
\end{figure}
\vskip 2pt
Dark matter structures around black holes represent another important avenue for exploration~\cite{Bertone:2004pz,Cirelli:2024ssz}. The nature and properties of dark matter are among the most important open issues in science, and black holes are key in probing it. On large scales, simulations have provided a solid understanding of the behaviour of dark matter~\cite{Navarro:1995iw}, however, its impact on gravitational-wave astronomy is not well understood. For relatively low-density environments, such as on galactic scales, the only detectable effect, at least in the ringdown, is a redshift of the gravitational waves as they climb the gravitational potential~\cite{Spieksma:2024voy}. However, in regions with higher dark matter densities, their impact may be more pronounced. 
\vskip 2pt
There are several ways in which such densities can build up around black holes. If dark matter consists of cold, collisionless particles, adiabatic black hole growth can lead to overdensities, or \emph{spikes}~\cite{Gondolo:1999ef,Ferrer:2017xwm}. Alternatively, dark matter could be composed of light bosonic fields such as \emph{axions}~\cite{Hu:2000ke,Marsh:2015xka,Hui:2016ltb}, which may simultaneously offer solutions to longstanding problems in the Standard Model~\cite{Peccei:1977hh,Weinberg:1977ma,Wilczek:1977pj}. Large quantities of such particles can be generated around black holes through \emph{black hole superradiance}~\cite{ZelDovich1971,ZelDovich1972,Starobinsky:1973aij,Arvanitaki:2009fg,Brito:2015oca}, a process which allows the boson to extract energy and angular momentum from a rotating black hole, forming a dense cloud around it. For superradiance to be effective, the Compton wavelength of the boson needs to match the black hole size, requiring the putative field to be ultralight. Unlike high-energy collider experiments, which are limited to particles with strong interactions, superradiance thus provides a natural mechanism to probe particles at the \emph{weak-coupling frontier}, relying solely on gravitational interactions.
\vskip 2pt
The presence of such clouds in binary systems leads to a rich phenomenology~\cite{Baumann:2018vus,Baumann:2019ztm,Baumann:2021fkf,Tomaselli:2023ysb,Tomaselli:2024bdd}. To fully characterise their observational signatures, it is crucial to study the entire evolution of these systems, even before they enter the sensitivity band of gravitational-wave detectors. A systematic exploration of this ``history'' shows that, in many cases, the cloud is destroyed early in the inspiral, leaving strong imprints on the binary's configuration and providing an indirect observational channel~\cite{Tomaselli:2024bdd,Tomaselli:2024dbw}. Should the cloud survive until close to merger, it becomes directly detectable through effects like accretion or dynamical friction~\cite{Baumann:2021fkf,Tomaselli:2023ysb}. In this regime, a fully relativistic, self-consistent and generic approach to EMRIs and environments in Kerr spacetime is essential. Such a framework, when applied to the boson cloud scenario, reveals large deviations from Newtonian and Schwarzschild predictions~\cite{Dyson:2025dlj}.
\vskip 2pt
The advent of next-generation detectors~\cite{ET:2019dnz,Colpi:2024xhw} highlights the importance and urgency of systematically studying the environments surrounding black holes. These systems act as cosmic laboratories, generating unique physical phenomena and probing regimes inaccessible on Earth. This thesis embraces that opportunity, leveraging all available observational tools to explore these systems. Figure~\ref{fig:introfigure} illustrates this perspective. In doing so, the synergy between theoretical developments and search strategies is essential, not only for detecting signals but also for guiding our focus~\cite{Cole:2022yzw,Khalvati:2024tzz}. The future of gravitational-wave and black hole physics is bright! From formation to ringdown, black hole binaries shout out their presence to the Universe. Only by tracing their evolution from the first note to the final fading chord, can we begin to understand the cosmic symphony they compose. I hope this thesis contributes to that effort.
\clearpage
\subsubsection{Outline of this thesis}
This thesis is structured as follows. In Chapter~\ref{chap:GravAstro}, I introduce key aspects of gravitational-wave astrophysics, covering black holes, gravitational waves, and current and future gravitational-wave observatories. In Chapter~\ref{chap:BHenv}, I discuss the various environments that may be present around black holes, including dark matter halos, boson clouds, plasmas, and accretion disks. After a general introduction, each case is examined in detail, outlining the relevant theoretical background. The chapter concludes by exploring how black holes can serve as probes of these environments, both in isolation and in binary systems.
\vskip 2pt
The remaining chapters present the main results of this thesis. The first two focus on black holes in isolation. In Chapter~\ref{chap:SR_Axionic}, I investigate the evolution of superradiant boson clouds coupled to the electromagnetic sector, showing that these systems can lead to a stationary emission of light. I describe the observational prospects and consider the impact of plasma on this emission. Chapter~\ref{chap:in_medium_supp} further examines the interactions between plasma and new fundamental fields. I demonstrate how plasma can induce in-medium suppression, a mechanism that strongly inhibits the mixing between ultralight bosons and the electromagnetic sector, with implications for both superradiance and observations.
\vskip 2pt
The following five chapters are focused on binary systems. In Chapter~\ref{chap:plasma_ringdown}, I study the ringdown of charged black holes in the presence of plasma. I show that plasma can modify the fundamental quasi-normal mode in the black hole ringdown and, when localised away from the black hole, may induce gravitational-wave echoes. In Chapter~\ref{chap:BHspec}, I investigate whether galactic dark matter halos can influence black hole ringdown and, consequently, the black hole spectroscopy programme. From a data-analysis perspective, I assess if future gravitational-wave detectors will be sensitive to the presence of the halo and explore the role of spectral instabilities. In Chapters~\ref{chap:legacy} and~\ref{chap:inspirals_selfforce}, I shift to the inspiral phase of black hole binaries, focusing on intermediate to extreme mass ratio inspirals, where the larger black hole hosts a superradiant boson cloud. In Chapter~\ref{chap:legacy}, I examine the inspiral in the Newtonian regime, far from merger, exploring resonances between bound states of the cloud for eccentric and inclined orbits. I analyse the sequence of resonances encountered during the evolution of the system and characterise their observational signatures, showing that the boson cloud can imprint detectable effects on the binary. In Chapter~\ref{chap:inspirals_selfforce}, I develop the first self-consistent, fully relativistic framework to study perturbations induced by an inspiralling secondary in the Kerr geometry. Applying this to a superradiant boson cloud reveals a rich wake structure, and computing scalar fluxes in Kerr shows significant deviations from the Schwarzschild case. In Chapter~\ref{chap:capture}, I explore the evolution of binaries captured by accretion disks in active galactic nuclei. I show that binaries rapidly align with the disk plane, whereas the evolution of their eccentricity is more complex and strongly dependent on other orbital parameters. Finally, in Chapter~\ref{chap:Conclusions}, I summarise the main findings of this thesis and discuss future directions.
\vskip 2pt
Several appendices contain technical details. In Appendix~\ref{app:NR}, I detail the numerical relativity simulations used in Chapter~\ref{chap:SR_Axionic}. In Appendix~\ref{app:Mathieu}, I expand on the Mathieu equation, which explains some of the results from Chapter~\ref{chap:SR_Axionic} analytically. In Appendix~\ref{app:BHPT}, I provide an overview of black hole perturbation theory and describe the numerical framework for evolving plunging particles. In Appendix~\ref{app:flatinstab}, I discuss axionic instabilities in the presence of electric fields in flat spacetime. In Appendix~\ref{app:DPbasis}, I elaborate on the choice of basis for the dark photon. In Appendix~\ref{app:GA}, I give additional details on resonances in gravitational atoms. Finally, in Appendix~\ref{app:BHPT_Kerr}, I expand on the relativistic perturbative framework introduced in Chapter~\ref{chap:inspirals_selfforce}.
\subsubsection*{Notation and conventions} 
Throughout this thesis, I adopt the  ``mostly plus'' metric signature $(-,+, +, +)$. The black hole mass is denoted by $M$ and, in the case of a binary system, refers to the more massive object. The mass ratio of the binary is given by $q$, with $q\leq1$. Dimensionless quantities, normalised by the black hole mass, are indicated with tildes, e.g., the spin of the black hole is given by $\tilde{a} \equiv a/M$. The covariant derivative associated with the metric is denoted by $\nabla$. Spacetime indices are represented by Greek letters ($\mu,\nu, \ldots$), while Latin letters ($i, j, \ldots$) denote spatial indices. The shorthand $\sum_{\ell m} \equiv \sum_{\ell = 0}^{\infty} \sum_{m = -\ell}^{\ell}$ is adopted when summing over harmonics. Unless stated otherwise, I use natural units where $G=\hbar = c = 1$.
\chapter{Gravity and Astrophysics}\label{chap:GravAstro}
\vspace{-0.8cm}
\hfill \emph{E eu que era triste}

\hfill \emph{Descrente deste mundo}

\hfill \emph{Ao encontrar voc\^{e} eu conheci}

\hfill \emph{O que \'{e} felicidade meu amor}
\vskip 5pt

\hfill Ant\^{o}nio Carlos Jobim

\vskip 35pt
\noindent In this thesis, I explore the interactions between black holes and their environments. While Chapter~\ref{chap:BHenv} provides a broader discussion of the latter, this chapter lays the theoretical groundwork necessary to understand key concepts related to black holes and gravitational waves. Rather than taking a purely mathematical or theoretical approach, I focus on these topics from an astrophysical and observational perspective.
\vskip 2pt
I begin with an overview of General Relativity in Section~\ref{GravAstro_sec:GR}, highlighting two of its most significant predictions:~black holes (Section~\ref{GravAstro_subsec:BH}) and gravitational waves (Section~\ref{GravAstro_subsec:GW}). I then examine the dynamics of compact binaries in Section~\ref{GravAstro_sec:binary_cbc}, with a focus on two crucial stages of binary coalescence, which play a recurring role in this thesis:~the inspiral (Section~\ref{GravAstro_subsec:binary_insp}) and the ringdown (Section~\ref{GravAstro_subsec:ringdown}). Finally, in Section~\ref{GravAstro_sec:observ}, I review the current state of gravitational-wave detectors and outline proposals for next-generation observatories.
\section{General Relativity}\label{GravAstro_sec:GR}
General Relativity (GR) stands as one of the most elegant and powerful theories in all of physics. Formulated by Albert Einstein in 1915~\cite{Einstein:1915ca}, it extends the principles of Special Relativity to incorporate gravity, describing it not as a conventional force but as a manifestation of spacetime curvature.
\clearpage
A key insight that set the stage for GR stems from a simple yet profound observation, known since before Newton:~objects of different masses fall at the same rate. In other words, an object's ``inertial'' mass (its resistance to acceleration) and ``gravitational'' mass (its response to gravity) are equal. While this equivalence is an unexplained coincidence in Newtonian gravity, GR promotes it to a guiding concept:~the \emph{weak equivalence principle} (WEP). It implies that gravity acts in a \emph{universal} manner:~any two particles, with the same initial position and velocity will follow the same trajectory in a gravitational field, regardless of their mass or composition. This universality stands in contrast to forces like electromagnetism, where a test-particle with a different charge moves on a different trajectory.
\vskip 2pt
The question then arises:~if all objects fall the same way under gravity, how can one tell whether a gravitational field is present at all? This question led Einstein to propose one of his most famous \emph{gedankenexperiments}. Imagine someone inside a sealed elevator in free fall. From their perspective, everything inside appears weightless. Crucially, the same experience would occur if the elevator were in empty space, being accelerated at a constant rate. This leads to an important insight:~a uniform gravitational field is \emph{indistinguishable} from uniform acceleration. It forms the basis of the \emph{Einstein equivalence principle}, which extends the WEP by stating that, within any sufficiently small region of spacetime, it is always possible to choose a coordinate system in which the laws of physics reduce to those of Special Relativity. In such a local inertial frame, the effects of gravity vanish and spacetime appears flat. This suggests that gravity should be understood as a property of spacetime itself -- specifically, its \emph{curvature}.
\vskip 2pt
This geometric perspective naturally leads to the transition from the flat Minkowski spacetime of Special Relativity to the curved spacetime of GR. In a curved geometry, a fundamental quantity is the proper length between two distinct points, which is expressed by the \emph{line element},
\begin{equation} \label{eqn:lineelement}
ds^2 = g_{\mu \nu} \dd x^\mu \dd x^\nu\,,
\end{equation}
where $x^\mu$ represents a choice of coordinate and $g_{\mu \nu}$ is the \emph{spacetime metric}, encoding the geometric structure of spacetime. In a local inertial frame, the metric reduces to the flat Minkowski form, $g_{\mu \nu} = \eta_{\mu \nu} = \text{diag} (-1, 1, 1, 1)$.
\vskip 2pt
The motion of free-falling test particles in curved spacetime follows the \emph{geodesic equation}:
\begin{equation} \label{eqn:geodesic}
\frac{\dd v^\mu}{\dd \tau} + \Gamma^{\mu}_{\alpha \beta} v^\alpha v^\beta =0\,.
\end{equation}
Here, $\tau$ represents the proper time of the particle, $v^\mu \equiv \dd x^\mu / \dd \tau$ is its four-velocity, and $\Gamma^{\mu}_{\alpha \beta} = g^{\mu \rho} (\partial_\alpha g_{\rho \beta} + \partial_\beta g_{\rho \alpha} - \partial_\rho g_{\alpha \beta} )/2$ are the Christoffel symbols. While this equation governs the motion in a given gravitational field, the field itself is determined by the \emph{Einstein field equations}~\cite{Einstein:1915ca,Einstein:1916vd}:
\begin{equation}\label{eq:EFE}
\boxed{R_{\mu \nu} - \frac{1}{2} g_{\mu \nu} R = 8 \pi T_{\mu \nu}\,,}
\end{equation}
where $R_{\mu\nu}$ refers to the Ricci tensor, $R=R^\mu{}_\mu$ is the Ricci scalar, and $T_{\mu\nu}$ represents the energy-momentum tensor. The left-hand side of~\eqref{eq:EFE} is known as the \emph{Einstein tensor} $G_{\mu \nu}$ and encodes the geometry of the spacetime, while the right-hand side describes the energy and matter content. Thus, this equation shows how matter shapes the curvature of spacetime, which, in turn, dictates the evolution of matter itself.
\vskip 2pt
GR has superseded Newtonian gravity as our most accurate description of gravity, successfully passing every experimental and observational test to date. Instead of attempting a comprehensive review (see~\cite{Misner:1973prb,Wald:1984rg,Schutz:1985jx} for excellent textbooks), the remainder of this chapter will focus on aspects of GR most relevant for this thesis:~black holes and gravitational waves.
\subsection{Black Holes}\label{GravAstro_subsec:BH}
\begin{figure}[t!]
\centering
\includegraphics[scale=1.4]{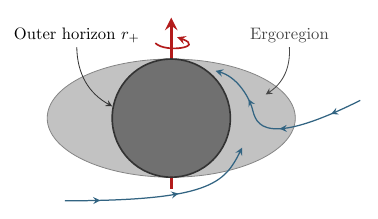}
\caption{Schematic illustration of a rotating (Kerr) BH, showing the event horizon and the surrounding ergoregion, where spacetime is dragged by the BH's rotation. Within the ergoregion, all particles [{\color{stormyocean}{blue}}] -- regardless of their initial motion -- are forced to co-rotate with the BH. Here, the timelike Killing vector $K^{\mu}_{(t)}$ becomes spacelike, making it possible to extract energy and angular momentum from the BH.}
\label{fig:ergoregion}
\end{figure}

One of the most striking predictions of GR is \emph{black holes} (BHs). They correspond to vacuum ($T_{\mu\nu} = 0$) solutions of the Einstein field equations~\eqref{eq:EFE}. Initially regarded as mere mathematical curiosities, their existence is now firmly established through observational evidence~\cite{Remillard:2006fc,Ghez:2008ms,2009ApJ...692.1075G,LIGOScientific:2016aoc,EventHorizonTelescope:2019dse,EventHorizonTelescope:2022wkp}. It is only in recent years that we have begun to fully appreciate their astrophysical and cosmological significance -- BHs may be the engines that drive some of the most energetic phenomena in the Universe~\cite{Blandford:1977ds,Fabian:2012xr}.
\vskip 2pt
The first exact BH solution was discovered in 1916 by Karl Schwarzschild, shortly after Einstein formulated his field equations. It describes a static, spherically symmetric BH and, in Schwarzschild-Droste coordinates $(t,r,\theta,\varphi)$~\cite{Schwarzschild1916,Droste1917}, the metric takes the form:\footnote{Historical footnote:~Schwarzschild originally derived his solution using a different radial coordinate. This form of the metric~\eqref{eq:metric_schwarzschild} was first written down in 1917 by Johannes Droste, a student of Hendrik Lorentz, who independently found the solution.}
\begin{equation}\label{eq:metric_schwarzschild}
ds^2=-\left(1-\frac{2M}r\right)\dd t^2+\left(1-\frac{2M}r\right)^{-1}\dd r^2+r^2\left(\dd\theta^2+\sin^2\theta\dd\varphi^2\right)\,,
\end{equation}
where $M$ is the BH mass. Examining this solution~\eqref{eq:metric_schwarzschild}, two special locations immediately stand out:~the metric appears singular at $r=0$ and $r=2M$. The singularity at $r=2M$ is a \emph{coordinate singularity} which can be removed by transforming to a different coordinate system, such as Eddington-Finkelstein coordinates~\cite{Finkelstein:1958zz}. Still, $r=2M$ marks a physically important surface, namely the \emph{event horizon}. This is a three-dimensional null hypersurface that defines the boundary within which all trajectories, even those of light, are irrevocably trapped. In other words, anything inside the event horizon $r<2M$ can never escape to the exterior $r>2M$. The singularity at $r=0$ instead, is a true singularity, where the spacetime curvature diverges, signalling a breakdown in our current understanding of physics.
\vskip 2pt
Besides its elegance and simplicity, the Schwarzschild metric~\eqref{eq:metric_schwarzschild} has far-reaching implications. It is the \emph{unique} spherically symmetric vacuum solution to Einstein’s field equations, as stated by \emph{Birkhoff’s theorem}~\cite{Birkhoff,Jebsen}, which asserts that any such solution must be \emph{static} and \emph{asymptotically flat}. This makes the Schwarzschild metric~\eqref{eq:metric_schwarzschild} not only relevant to our understanding of BHs, but also applicable to describing the outer regions of many astrophysical objects, where spherical symmetry is often a good approximation. Indeed, it played a key role in the early observational evidence of GR:~the Schwarzschild metric~\eqref{eq:metric_schwarzschild} correctly accounts for the anomalous precession of Mercury’s perihelion~\cite{Einstein1916mercury} and successfully predicts the bending of starlight passing near the Sun, as famously confirmed during the solar eclipse of 1919~\cite{Dyson:1920cwa}.
\vskip 2pt
Astrophysically, BHs are thought to form through the collapse of massive stars, a process that would require them to possess angular momentum. Although the Schwarzschild solution was derived soon after the formulation of Einstein's equations, its rotating counterpart was not discovered until nearly half a century later, when Roy Kerr~\cite{Kerr1963} derived it using the Newman-Penrose formalism~\cite{Newman:1961qr}. In Boyer-Lindquist coordinates~\cite{BL1967}, which we also denote by $(t,r,\theta,\varphi)$, the metric is:
\begin{equation}\label{eq:metric_kerr}
ds^2=-\frac\Delta{\rho^2}(\dd t-a\sin^2\theta\dd\varphi)^2 + \frac{\rho^2}\Delta\dd r^2+\rho^2\dd\theta^2+\frac{\sin^2\theta}{\rho^2}(a\dd t-(r^2+a^2)\dd\varphi)^2\,,
\end{equation}
which describes a rotating BH of mass $M$ and angular momentum $J$. Here, $a = J/M$ is the spin parameter, bounded by $0 \leq a \leq M$, and we define $\Delta \equiv r^2-2Mr+a^2$ and $\rho^2 = r^2 + a^2\cos^2{\theta}$. In the non-spinning limit $a\to0$, the Kerr solution correctly reduces to the Schwarzschild metric~\eqref{eq:metric_schwarzschild}. Unlike the Schwarzschild case, the Kerr metric has two horizons:~an \emph{inner} and an \emph{outer} horizon, given by the roots of $\Delta$ and located at $r_\pm = M \pm \sqrt{M^2 -a^2}$. The inner (or ``Cauchy'') horizon is not accessible for external observers, and thus not of interest for us.
\vskip 2pt
Many other vacuum solutions to Einstein's field equations exist (see, e.g.,~\cite{Stephani_Kramer_MacCallum_Hoenselaers_Herlt_2003}). However, the significance of the Schwarzschild and Kerr solutions is underscored by the \emph{no-hair theorem}~\cite{Israel:1967wq,Carter:1971zc,Robinson:1975bv}, which states that all BHs in GR are fully characterised by just three parameters:~mass, spin, and charge.\footnote{With ``charge'' we refer to an \emph{electric charge}. In principle, if magnetic monopoles exist in the Universe~\cite{Preskill:1984gd}, BHs may also possess a \emph{magnetic charge}, adding a fourth independent parameter for BHs.} While we have not explicitly introduced the charged extensions of the Schwarzschild and Kerr solutions\footnote{Astrophysical BHs are expected to be electrically neutral due to neutralisation by surrounding plasmas, quantum discharge effects~\cite{Gibbons:1975kk} or electron-positron pair production~\cite{1969ApJ...157..869G,Eardley:1975kp,Blandford:1977ds}. There may still be ways to charge BHs by invoking beyond the Standard Model physics, which will be relevant in Chapters~\ref{chap:in_medium_supp} and~\ref{chap:plasma_ringdown}.} -- the Reissner-Nordstr\"{o}m~\cite{Reissner:1916cle,1918KNAB...20.1238N} and Kerr-Newman metrics~\cite{Newman:1965my}, respectively -- these four solutions encompass the entire set of relevant BH spacetimes. This is a remarkable result:~it means that if a star collapses into a BH, almost all of the parameters that described it vanish, leaving only its mass, spin and charge. This makes BHs the simplest macroscopic objects in the Universe and ideal candidates for high-precision observational studies. As Chandrasekhar put it~\cite{MTB}:
\vskip 2pt
\begin{quote} ``The black holes of nature are the most perfect macroscopic objects there are in the Universe:~the only elements in their construction are our concepts of space and time.''
\end{quote}
\vskip 2pt
Let us now examine the Kerr solution~\eqref{eq:metric_kerr} in more detail. Different from the Schwarzschild case, the Kerr metric contains an off-diagonal component, $g_{t \varphi}$, which gives rise to the \emph{Lense-Thirring effect}~\cite{Lens-Thirring1918}. It causes free-falling test particles on purely radial trajectories to co-rotate with the BH, as seen by an observer at infinity. When a particle approaches the event horizon ($r\to r_+$), it asymptotically acquires the angular velocity of the BH:
\begin{equation}\label{eq:OmegaH}
\Omega\ped{H} = \frac{a}{2 M r_+}\,.
\end{equation}
While it could, in principle, be given a boost in the counter-rotating direction to prevent it from co-rotating with the BH, there exists a region called the \emph{ergoregion}, where this is no longer possible:~all observers -- even light -- \emph{must} co-rotate with the BH. This can be understood by examining the norm of the Killing vector associated to the time translation invariance of the Kerr metric, $K^{\mu}_{(t)}$, which at the horizon satisfies $K^{\mu}_{(t)}K^{\nu}_{(t)}g_{\mu\nu}|_{r = r_+} = a^2\sin^2{\theta}/\rho^2 \geq 0$. In other words, outside the outer horizon, $K^{\mu}_{(t)}$ transitions from being timelike at infinity to spacelike. This transition point defines the \emph{ergosphere}, which is located at:
\begin{equation}
r\ped{erg}=M+\sqrt{M^2-a^2\cos^2\theta}\,.
\end{equation}
Within the ergosphere ($r_+<r<r\ped{erg}$), objects can still escape the BH's gravity but are forced to co-rotate with it. A schematic illustration of the ergoregion is shown in Figure~\ref{fig:ergoregion}.
\vskip 2pt
A crucial feature of the ergosphere is that a particle's energy $E=-K_{(t)}^\mu p_\mu$ (where $p^\mu$ is its four-momentum), can be \emph{negative} inside this region, as measured by an observer at infinity. This observation, first made by Penrose~\cite{Penrose:1971uk}, forms the basis of the \emph{Penrose process}, a mechanism by which a rotating BH can lose energy and angular momentum. We will examine this phenomenon in Section~\ref{BHenv_subsec:superrad} in more detail. However, such energy extraction does not happen arbitrarily, and is limited by the second law of BH thermodynamics~\cite{Bardeen:1973gs}. In particular, the event horizon of the BH $A\ped{BH}=8\pi Mr_+$ can never decrease. Consequently, the mass of a BH cannot fall below the so-called \emph{irreducible mass}~\cite{Christodoulou:1970}:
\begin{equation}\label{eq:irreducible}
M^2\ped{irr} = \frac{1}{2}\left(M^2+\sqrt{M^4-J^2}\right)\,.
\end{equation}
The extractable energy is thus $M-M\ped{irr}$, which may be interpreted as the rotational energy of the BH. This fraction can be substantial -- up to $29$\% of the total mass for an extremal BH with $J=M^2$~\eqref{eq:irreducible}. Remarkably, as we shall see in Section~\ref{BHenv_subsec:superrad}, energy extraction from rotating BHs is also possible through bosonic fields, via a process known as \emph{black hole superradiance}.
\vskip 2pt
Since the discovery of the Schwarzschild solution in 1916, physicists have debated whether BHs are real astrophysical objects or theoretical constructs. The first compelling observational evidence came from X-ray binaries~\cite{Giacconi:1962zz,1967ApJ...148L.119G,1972Natur.235...37W}, which revealed extremely compact objects with stellar masses. A major breakthrough followed from long-term monitoring of stellar orbits near the centre of our galaxy. Tracking the so-called ``S-stars'' led to the firm conclusion that an invisible, compact object with a mass of $4.3\times10^6\,M_\odot$~\cite{2016ApJ...830...17B} resides at the galactic centre, widely accepted to be a supermassive BH. Today, there is strong evidence for the existence of such supermassive BHs ($M > 10^5\,M_\odot$) at the centres of most galaxies~\cite{2021NatRP...3..732V}. Notably, the Event Horizon Telescope recently produced the first direct images of the luminous matter surrounding the central BHs in the M87 galaxy and our own Milky Way (Sgr\,$\mathrm{A}^*$), using very long-baseline interferometry~\cite{EventHorizonTelescope:2019dse,EventHorizonTelescope:2022wkp}. Yet, the most direct probe of BHs and strong-field gravity comes from gravitational waves -- a topic we now turn to. Together, these discoveries have ushered in a true \emph{observational era} in BH physics.
\subsection{Gravitational Waves}\label{GravAstro_subsec:GW}
In the previous section, we examined exact solutions of the Einstein field equations~\eqref{eq:EFE} assuming vacuum and certain symmetries. Solving the full set of ten coupled partial differential equations for a generic dynamical spacetime, by contrast, is notoriously difficult. Fortunately, approximations can be made in certain regimes. One particularly important case arises when the gravitational fields are weak, leading to one of GR’s major observational predictions:~gravitational waves (GWs). This section establishes the theoretical foundation of GWs (see~\cite{Maggiore:2007ulw,Maggiore:2018sht} for comprehensive textbooks), while later sections will discuss their role in compact binary systems and their detection by current and future observatories. 
\vskip 2pt
To study the generation and properties of GWs, it is useful to consider the \emph{weak-field approximation}, where the metric is treated as a small perturbation around flat spacetime, writing
\begin{equation}
g_{\mu\nu}=\eta_{\mu\nu}+h_{\mu\nu}\,,
\end{equation}
where $\eta_{\mu\nu}$ is the Minkowski metric and $\abs{h_{\mu\nu}}\ll1$ represents the \emph{metric perturbation}. By consistently expanding the Einstein tensor (which involves derivatives of the Christoffel symbols, which themselves involve derivatives of the metric), and keeping only leading-order terms in $h_{\mu\nu}$, the full Einstein equations~\eqref{eq:EFE} reduce to their linearised form:
\begin{equation}\label{eq:linearisedh}
-\Box h_{\mu\nu}+\partial^\alpha\partial_\mu h_{\nu\alpha}+\partial^\alpha\partial_\nu h_{\mu\alpha}-\partial_\mu\partial_\nu h - (\partial^\alpha\partial^\beta h_{\alpha\beta}-\Box h)\eta_{\mu\nu}=16\pi\,T_{\mu\nu}\,,
\end{equation}
where $\Box=\partial^\alpha\partial_\alpha$ is the flat-space d'Alembertian operator and $h\equiv h^\alpha{}_\alpha$ is the trace of $h_{\mu\nu}$. Although eq.~\eqref{eq:linearisedh} is still complicated, it also holds gauge dependencies which can be exploited. By imposing the \emph{harmonic gauge} (or \emph{de Donder gauge}), $\partial^\mu h_{\mu\nu}-\partial_\nu h/2=0$, the linearised field equations~\eqref{eq:linearisedh} reduce to
\begin{equation}\label{eq:linsource}
\Box\bar h_{\mu\nu}=-16\pi\,T_{\mu\nu}\,,
\end{equation}
where $\bar h_{\mu\nu}=h_{\mu\nu}-\eta_{\mu\nu}h/2$ is the trace-reversed metric perturbation. Equation~\eqref{eq:linsource} immediately shows how a matter distribution $T_{\mu\nu}$ sources the metric perturbation $h_{\mu\nu}$ in a wave-like manner.
\vskip 2pt
In vacuum ($T_{\mu\nu} = 0$), eq.~\eqref{eq:linsource} reduces to a homogeneous wave equation. A natural class of solutions is that of plane waves, which take the form: 
\begin{equation}\label{eq:plane_w}
\bar h_{\mu\nu}=\mathrm{Re}(H_{\mu\nu}e^{ik_\sigma x^\sigma}) = \mathrm{Re}(H_{\mu\nu}e^{ik_i x^i}e^{-i\omega t})\,,
\end{equation}
where the polarisation tensor $H_{\mu\nu}$ is constant and symmetric and $k^{\sigma}$ is the wave vector with $\omega = -k_0$. Applying the d'Alembertian operator to~\eqref{eq:plane_w} shows that $ k^{\sigma}k_{\sigma} = 0$, which means that $k^{\sigma}$ is a null vector and GWs in vacuum propagate at the speed of light. Enforcing the harmonic gauge condition further constrains the system, as it requires the polarisation tensor to be transverse to the direction of propagation, i.e., $k^{\mu}H_{\mu\nu}=0$. This reduces the independent components of $H_{\mu\nu}$ from ten to six. However, the harmonic gauge condition alone does not completely fix the gauge freedom in $\bar{h}_{\mu\nu}$. The residual freedom allows us to impose the \emph{transverse traceless} (TT) gauge, where we set $H_{0\mu}=0$ and $H^\mu{}_\mu=0$. This choice reduces the number of independent components to two. For a wave propagating in the $\hat{z}$--direction, the polarisation tensor takes the following form:
\begin{equation}
H_{\mu\nu}=\begin{pmatrix}
0 & 0 & 0 & 0\\
0 & H_+ & H_\times & 0\\
0 & H_\times & -H_+ & 0\\
0 & 0 & 0 & 0
\end{pmatrix}\,,
\end{equation}
where the components $H_+$ and $H_\times$ correspond to the two independent polarisation modes of the GWs, namely the \emph{plus} and \emph{cross} polarisations.
\vskip 2pt
We now turn to the emission of GWs from sources ($T_{\mu\nu} \neq 0$). In the weak-field, slow-motion regime, where the Newtonian potential $\Phi \ll 1$ and the source's velocity $v \ll 1$, eq.~\eqref{eq:linsource} can be solved using a retarded Green’s function, analogous to electromagnetism. For a source located at a large distance $r$, the solution takes the form~\cite{Einstein1918}:
\begin{equation}\label{eq:metric_sourced_sol}
\bar h^{\rm TT}_{ij}(t,x^k)=\frac{2}{r}\frac{\dd^2}{\dd t^2}\,\mathcal{I}_{ij}(t-r)\,,
\end{equation}
where
\begin{equation}\label{eq:red_qm}
\mathcal{I}_{ij}=\int \dd^3x  \left(x_i x_j-\frac{1}{3}r^2\delta_{ij}\right)T_{00}\,,
\end{equation}
is the \emph{reduced quadrupole moment}. This expression, referred to as the \emph{quadrupole approximation}, shows that GWs are proportional to the second derivative of the \emph{quadrupole moment} of the energy density. This is in contrast with electromagnetism, where radiation is sourced by changes in the \emph{dipole moment} of the charge distribution. 
\vskip 2pt
The energy carried away by GWs can be determined by integrating the energy flux across a sphere at large distances from the source
\begin{equation}\label{eq:GW_power}
P\ped{GW} \equiv -\frac{\dd E}{\dd t} = -\lim _{r \rightarrow \infty} \int T_{t r}^{\mathrm{GW}} r^2 \dd \Omega\,,
\end{equation}
where $\dd \Omega$ is the solid angle element and $T^{\rm GW}_{\alpha\beta} = 1/(32\pi)\langle \partial_{\alpha} h^{\rm TT}_{ij} \partial_{\beta} h^{\mathrm{TT}, ij}\rangle$ is the effective stress-energy tensor for GWs. Substituting eq.~\eqref{eq:metric_sourced_sol} into~\eqref{eq:GW_power} yields the total power emitted in GWs, known as the \emph{quadrupole formula}~\cite{Einstein1918}:
\begin{equation}\label{eq:quadrupole_formula}
P\ped{GW}=\frac15\Braket{\frac{\dd^3 \mathcal{I}_{ij}}{\dd t^3}\frac{\dd^3\mathcal{I}_{ij}}{\dd t^3}}\,.
\end{equation}
Although derived over a century ago, the validity of this formula remained uncertain until the late 20th century.\footnote{More generally, the existence of GWs was debated for decades due to gauge ambiguities and the linear approximation, even by Einstein himself~\cite{EinsteinRosen1937}. Moreover, there were many concerns on whether GWs could carry any energy as their energy-momentum tensor is not gauge-invariant. This is a consequence of the fact that locally, one can always choose an inertial frame where the metric and its first derivative vanish. The resolution lies in averaging over a sufficiently large region of spacetime, which does yield a gauge-invariant quantity.} The breakthrough came with the discovery of the Hulse-Taylor binary pulsar (PSR 1913+16) in 1974~\cite{Hulse:1974eb}. By the early 1980s, long-term radio measurements showed that its orbital decay matched the energy loss predicted by eq.~\eqref{eq:quadrupole_formula}~\cite{Taylor1982}, providing strong support for GW emission.
\section{Compact Binary Coalescences}\label{GravAstro_sec:binary_cbc}
The quadrupole formula~\eqref{eq:quadrupole_formula} reveals something remarkable:~\emph{any object} with a time-varying quadrupole moment emits GWs. In practice, however, these waves are extraordinarily weak, which is why you do not perceive their effects while reading this thesis. Gravity is far weaker than any other fundamental force, making the detection of GWs extremely challenging. To do so, we must focus on the ``loudest'' sources in the Universe:~compact binary coalescences. These systems consist of two compact objects, typically neutron stars or BHs, that orbit each other, radiating energy and angular momentum through GW emission. As a result, their orbit gradually shrinks, and the objects spiral towards one another. To quantify the strength of GWs, it is useful to introduce the \emph{strain}, defined as the fractional change in distance between two test masses, $h \equiv \Delta L / L$. For a binary consisting of a central BH of mass $M$, and a smaller companion with mass ratio $q < 1$, located at a distance $r$, the strain can be estimated as
\begin{equation}\label{eq:strain}
h \simeq \frac{(GM)^{5/3}q f^{2/3}}{c^4 r}\simeq \SI{5e-23}{} \left(\frac{M}{10^6 M_{\odot}}\right)^{5/3}\left(\frac{q}{10^{-5}}\right)\left(\frac{1\,\mathrm{Gpc}}{r}\right) \left(\frac{f}{10^{-3}\,\mathrm{Hz}}\right)^{2/3}\,,
\end{equation}
where $f$ is the \emph{gravitational-wave frequency} and we have temporarily restored units for clarity. When two bodies spiral inwards, their motion becomes increasingly relativistic, yielding rapidly evolving quadrupole moments and stronger GW signals. As astrophysical BHs are characterised by only a few parameters (see Section~\ref{GravAstro_subsec:BH}) -- mass and spin -- their dynamics can be modelled accurately. Therefore, the waveforms produced by compact binaries can be predicted to high precision, making them powerful tools for probing both the binary’s evolution and the underlying physics. Crucially for this thesis, any deviation from the expected (vacuum) waveform could reveal new astrophysical effects influencing the binary.
\vskip 2pt
A compact binary coalescence progresses through three stages:~inspiral, merger, and ringdown. During the inspiral, the two objects are relatively far apart, gradually spiralling inwards as they lose energy and angular momentum. Meanwhile, the GW amplitude and frequency increase over time in a characteristic \emph{chirp}. Though this phase can last millions of years, current detectors can only capture its final moments. The inspiral takes place in the weak-field regime, and is well-described by the post-Newtonian formalism (Section~\ref{GravAstro_subsec:binary_insp}). This phase, central to Chapters~\ref{chap:legacy},~\ref{chap:inspirals_selfforce}, and~\ref{chap:capture}, is particularly relevant for studying environments, as subtle effects from the surrounding matter can accumulate over time, affecting the binary’s motion. As the inspiral progresses, strong-gravity effects will dominate, leading to a brief \emph{merger} phase (Section~\ref{GravAstro_subsec:merger}), requiring numerical relativity to accurately describe the dynamics and emitted GWs. While this thesis does not focus on the merger phase, Chapter~\ref{chap:SR_Axionic} employs numerical relativity to explore strong-gravity effects. Finally, after the merger, the remnant object settles into a stationary state during the \emph{ringdown} phase (Section~\ref{GravAstro_subsec:ringdown}). This occurs through a characteristic set of frequencies called the \emph{quasi-normal modes}. The ringdown phase is typically modelled using BH perturbation theory and will be relevant in Chapters~\ref{chap:plasma_ringdown} and~\ref{chap:BHspec}.
\vskip 2pt
One final remark is in order. Constructing gravitational waveforms always involves approximations, whether from truncating perturbative expansions or numerical errors. The key question from an observational standpoint is:~to what \emph{precision} must waveforms be computed? The answer depends on the detector for which they are designed. Ideally, a modelled waveform should be indistinguishable from the true signal within the measurement accuracy of a given detector. This requirement is typically quantified by the \emph{faithfulness}, $\mathcal{F}$, between two waveforms:
\begin{equation}
\mathcal{F} < \frac{\mathcal{O}(1)}{\mathrm{SNR}^2}\,,
\end{equation}
where $\mathrm{SNR}$ is the \emph{signal-to-noise ratio}, set by detector's sensitivity. This highlights the challenge ahead:~next-generation detectors will reach $\mathrm{SNR}$ values orders of magnitude higher than current ones, demanding significantly greater accuracy in waveform modelling~\cite{Purrer:2019jcp,LISAConsortiumWaveformWorkingGroup:2023arg}.
\subsection{Binary Inspirals}\label{GravAstro_subsec:binary_insp}
The inspiral phase is the longest epoch in the evolution of a binary system. Consider a binary system with component masses $M$ and $qM$ (with $q < 1$ denoting the mass ratio), and orbital separation $R \gg M$. As the binary emits GWs, it loses energy and angular momentum according to the quadrupole formula~\eqref{eq:quadrupole_formula}, causing the orbit to shrink. In the inspiral phase, the orbital frequency is well described by Kepler's law:
\begin{equation}
\Omega=\sqrt{\frac{M(1+q)}{R^3}}\,.
\end{equation}
For circular orbits, the rate of energy loss due to GW emission is
\begin{equation}\label{eq:Pgw_f}
\frac{\dd E\ped{orb}}{\dd t}= - P\ped{GW} = -\frac{32}{5}\frac{M^5q^2(1+q)}{R^5}\,,
\end{equation}
where $E\ped{orb}= M^2q/(2R)$ is the total orbital energy, and $P\ped{GW}$ is found by substituting the reduced quadrupole moment~\eqref{eq:red_qm} into the quadrupole formula~\eqref{eq:quadrupole_formula}. Thus, the orbital separation evolves as
\begin{equation}
R(t)=R_0\left(1-\frac{t}{t_0}\right)^{1/4}\,,
\end{equation}
where the inspiral timescale is
\begin{equation}\label{eq:t_inspirals}
t_0=\frac{5R_0^4}{256M^3 q(1+q)} \simeq \SI{2e5}{yrs}\left(\frac{10^6M_{\odot}}{M}\right)^3\left(\frac{10^{-5}}{q}\right)\left(\frac{R_0}{10^{-3}\,\mathrm{pc}}\right)^4\,.
\end{equation}
\vskip 2pt
This description can be extended to eccentric binaries, in which the semi-major axis $a$ and the eccentricity $\varepsilon$ evolve according to Peters' equations~\cite{Peters:1963ux,Peters:1964zz}:
\begin{equation}\label{eq:Peters}
\begin{aligned}
\left\langle \frac{\dd a}{\dd t}\right\rangle &= - \frac{64}{5}\frac{M^3 q(1+q)}{a^3(1-\varepsilon^2)^{7/2}}\left(1+\frac{73}{24}\varepsilon^2+\frac{37}{96}\varepsilon^4\right)\,,\\
\left\langle \frac{\dd \varepsilon}{\dd t}\right\rangle &= - \frac{304}{15}\varepsilon\frac{M^3 q(1+q)}{a^4(1-\varepsilon^2)^{5/2}}\left(1+\frac{121}{304}\varepsilon^2\right)\,,
\end{aligned}
\end{equation}
which are time-averaged over one full orbit. The semi-major axis thus decreases faster for larger eccentricity, meaning that highly eccentric binaries merge more quickly than their circular counterparts.  Moreover, GW emission gradually reduces the eccentricity, resulting in nearly circular orbits by the time the binary enters the sensitivity band of current detectors. However, next-generation detectors -- capable of detecting binaries earlier in the inspiral -- may observe systems before circularisation is complete, making the evolution of eccentricity an important factor to consider and one we will return to later in this thesis.
\vskip 2pt
In the early stages of the inspiral, the gravitational field is weak, and the components move slowly. As the binary spirals inwards, its orbital frequency increases, and GW emission intensifies~\eqref{eq:Pgw_f}. Eventually, the assumption of weak-field and slow-velocity dynamics breaks down, and more sophisticated methods are required to describe the system's evolution. A natural approach is to treat the dynamics of GR as perturbations of the Newtonian limit. Two main frameworks are used for this purpose, differing in their choice of expansion parameter. The \emph{post-Newtonian} (PN) expansion, which orders terms in powers of the orbital velocity $v$, and the \emph{post-Minkowskian} (PM) expansion, which uses Newton’s constant $G$. Specifically, in the PN framework, the metric and energy-momentum tensor are expanded in terms of
\begin{equation}
\epsilon \sim \frac{v}{c}\sim\sqrt{\frac{GM}{c^2r}}\,,
\end{equation}
where we restored units for clarity. For bound systems, the virial theorem implies that $v^2 \sim GM/r$, justifying $v/c$ as a perturbative parameter. Terms of order $\epsilon^n$ are referred to as $(n/2)$--PN corrections, with $0$PN corresponding to the Newtonian limit. The metric and orbital motion are then solved order by order. For non-spinning binaries on circular orbits, the evolution is currently known up to 4.5PN order~\cite{Blanchet:2023bwj}, while eccentric, spinning, and precessing systems are known to lower orders.
\vskip 2pt
When one of the binary components is much more massive than the other, the gravitational effects of the heavier object dominate at leading order, and the background spacetime can no longer be approximated as Minkowski. In this regime, the PN and PM expansions break down, requiring a different approach. Such systems, known as \emph{extreme mass ratio inspirals} (EMRIs), typically have mass ratios $q < 10^{-4}$, making $q$ a natural perturbative parameter. This forms the basis of the \emph{self-force programme}.\footnote{Even for intermediate mass ratio inspirals, $10^{-4}<q<10^{-2}$, this framework remains effective~\cite{Wardell:2021fyy}.} In this approach, the spacetime metric is expanded as
\begin{equation}
g_{\mu\nu} = g^{(0)}_{\mu\nu} + q h^{(1)}_{\mu\nu}+ q^2 h^{(2)}_{\mu\nu}+\cdots\,,
\end{equation}
where $g^{(0)}_{\mu\nu}$ is the metric of the heavier object, typically a Kerr BH. In the test particle limit ($q \to 0$), the trajectory of the smaller object traces a geodesic in the background spacetime~\cite{Bardeen1972}. However, for finite mass ratios, the smaller body perturbs the background, generating a gravitational self-force that alters its trajectory. Thus, the trajectory of the smaller object is influenced both by the gravitational field of the larger object and by the gravitational radiation it emits, which is called the \emph{radiation-reaction}~\cite{PhysRevD.50.3816,Hughes:1999bq,Mino:2003yg,Barack:2018yvs,Pound:2021qin}. Schematically, the geodesic equation~\eqref{eqn:geodesic} is modified as
\begin{equation}
v^\alpha \nabla_\alpha v^\mu=f^\mu\left[h_{\alpha \beta}^{(1)}, h_{\alpha \beta}^{(2)}, \cdots ; g_{\alpha \beta}^{(0)}, v^\mu\right]\,,
\end{equation}
where $f^\mu$ is the self-force, which depends on the background metric, metric perturbations and the small body's motion. Extending this scheme to higher orders in $q$ is challenging. Estimates suggest that second-order ($q^2 h^{(2)}_{\mu\nu}$) self-force corrections are necessary for matched filtering searches of EMRIs~\cite{Hinderer:2008dm,Hughes:2016xwf}. Such waveforms have recently been computed for a subset of the relevant parameter space~\cite{Wardell:2021fyy}.
\vskip 2pt
An alternative, highly efficient approach for modelling the inspiral exploits a key result from Newtonian dynamics:~the two-body problem can be mapped to a one-body problem. This idea extends to GR via the \emph{effective one-body} (EOB) approach, which maps the conservative PN dynamics of a binary to an effective system describing a test particle moving in a deformed Kerr metric~\cite{Buonanno:1998gg, Buonanno:2000ef}. The EOB formalism introduces free parameters, which must be calibrated against numerical relativity simulations~\cite{Taracchini:2013rva}. By doing so, the EOB approach provides an accurate description of the binary in regimes where PN alone is insufficient. 
\vskip 2pt
The Newtonian, PN, PM, self-force, and EOB approaches each have their own range of validity, depending on the stage of the inspiral and the mass ratio of the binary. However, all of these methods break down as the system approaches merger, at which point gravitational fields become extremely strong and nonlinear effects dominate.
\subsection{Binary Mergers}\label{GravAstro_subsec:merger}
The final moments of a binary coalescence mark the onset of the \emph{merger phase}, where the two compact objects plunge together. This regime lies beyond the reach of perturbative techniques, and \emph{numerical relativity} becomes essential to evolve the system accurately. Numerical relativity aims to solve the field equations~\eqref{eq:EFE} directly through computational methods and has become the most powerful tool for probing highly dynamical and strong-gravity spacetimes (see, e.g.,~\cite{Gourgoulhon:2007ue,Alcubierre,Baumgarte:2010ndz}).
\vskip 2pt
Solving Einstein’s field equations~\eqref{eq:EFE} -- a set of ten coupled, nonlinear partial differential equations -- presents several unique challenges. One difficulty arises from the choice of coordinates. In GR, coordinates are arbitrary labels with no intrinsic physical meaning, but numerically, a poor choice can lead to instabilities. This issue becomes particularly pronounced near singularities, which may be either physical (such as those inside BHs) or coordinate-dependent [such as the $r=2M$ singularity in eq.~\eqref{eq:metric_schwarzschild}]. Physical singularities are problematic as curvature and energy densities diverge, which requires special techniques such as excising the singularity region from the numerical domain. Additionally, computing gravitational waveforms has its own set of complexities. Since GWs waves must be measured at large distances from the binary, simulations require large computational domains. However, evolving the system for long enough to allow the waves to reach the far-field region is computationally expensive. Numerical schemes must therefore be capable of handling vastly different length and time scales. Finally, poorly chosen gauge conditions can also introduce instabilities, leading to numerical errors that grow uncontrollably over time. These are just a few examples of the many difficulties encountered in numerical relativity simulations. Overcoming them requires a careful consideration of several factors, including initial data construction, horizon tracking, and coordinate choices. 
\vskip 2pt
The evolution of the system is governed by Einstein’s equations~\eqref{eq:EFE}, which not only describe the dynamics but also impose constraints through the Bianchi identity, $\nabla_\mu G^{\mu \nu} = 0$.  In the $3+1$ ADM formalism~\cite{Arnowitt:1959ah}, these manifest as the Hamiltonian and momentum constraints, which must be enforced throughout the evolution and serve as a consistency check on the simulation. Additionally, convergence tests are essential to verify that the numerical solution approaches the continuum limit.
\vskip 2pt
Despite all the challenges, numerical relativity offers a unique window into the strong-field dynamics of BH mergers. This has proven indispensable for GW detections. The first stable simulations of binary BH mergers were achieved in 2005~\cite{Pretorius:2005gq,Baker:2005vv,Campanelli:2005dd}, and since then, various methods have been employed to perform these simulations~\cite{Kidder:2000yq,Loffler:2011ay,Moesta:2013dna,Brugmann:2008zz,Clough:2015sqa}. Due to their high computational cost, simulations typically cover only the final $\sim100$ orbits before merger (see~\cite{Jani:2016wkt,Healy:2019jyf,Boyle:2019kee} for different catalogues).
\subsection{Black Hole Ringdown}\label{GravAstro_subsec:ringdown}
Following the merger of two compact objects, the newly formed remnant BH is in a highly perturbed state. It relaxes to a stationary configuration by emitting GWs in what is known as the \emph{ringdown} phase. These GWs have characteristic frequencies and damping times that encode information about the spacetime geometry and the fundamental nature of gravity~\cite{Vishveshwara:1970zz,Press:1971wr,Chandrasekhar:1975zza}. As such, the ringdown provides a remarkably clean observational probe of GR.
\vskip 2pt
The ringdown phase is best described within the framework of BH perturbation theory, which treats the spacetime as a background solution perturbed by small fluctuations (see Appendix~\ref{app:BHPT}). The theoretical foundations of this approach were laid down by Regge and Wheeler~\cite{ReggeWheeler} and Zerilli~\cite{Zerilli:1970se,Zerilli:1970wzz} for the Schwarzschild case, and later extended to Kerr BHs by Press and Teukolsky~\cite{Teukolsky:1972my,Teukolsky:1973ha,Press:1973zz}. Building on these foundations, Vishveshwara demonstrated that a perturbed BH responds to an incoming pulse of radiation with a superposition of damped exponentials, each with a discrete frequency and damping time, similar to the fading tones of a vibrating bell~\cite{Vishveshwara:1970zz}. This damping arises because BHs absorb gravitational radiation at the horizon and emit it towards spatial infinity, making the problem inherently dissipative. These oscillation modes are therefore known as \emph{quasi-normal modes} (QNMs), in contrast with the normal modes of conservative systems~\cite{Kokkotas:1999bd,Nollert:1999ji,Berti:2009kk,Konoplya:2011qq}.
\vskip 2pt
In linearised BH perturbation theory, each multipolar component of the waveform can be accurately expressed as a sum of QNMs~\cite{Leaver:1986gd}. The GW strain takes the form\footnote{Strictly speaking, the damped exponentials only describe the signal well at intermediate times, after an initial transient phase, known as ``the prompt response'' and before a power-law decay, known as ``the tail''. These tails result from radiation backscattering off the spacetime curvature~\cite{Price:1971fb,Leaver:1986gd,Gundlach:1993tp,Hintz:2020roc}.}
\begin{equation}
h(t) \simeq \mathrm{Re} \sum_{\ell m n} A_{\ell m n} (r) e^{-i(\omega_{\ell m n} t + \phi_{\ell m n}) - t/\tau_{\ell m n}}\,,
\end{equation}
where $A_{\ell m n}$ is the mode amplitude at a distance $r$, $\phi_{\ell m n}$ is the phase, $\omega_{\ell m n}$ is the characteristic oscillation frequency and $\tau_{\ell m n}$ is the damping time. The time coordinate $t$ refers to the post-merger evolution when the system has entered the linear regime. The indices ($\ell$, $m$) define the angular distribution of the radiation, with $|m|\leq \ell$, while $n$ labels the overtones. For a given $(\ell,m)$, the QNMs form a ``tower of modes'' ordered by their damping times, with the fundamental mode ($n=0$) being the longest-lived. Higher overtones ($n>0$) decay more rapidly but can still play a role in the early ringdown phase. The frequencies and damping times of the QNMs are known to high precision even for large $n$~\cite{GRITJHU}. In contrast, the amplitudes and phases depend on the astrophysical processes that excite the perturbation as well as on the observer's orientation~\cite{Lim:2019xrb}. A detailed computation of QNMs in the frequency domain is provided in Appendix~\ref{app:BHPT}.
\vskip 2pt
The study of QNMs, known as \emph{black hole spectroscopy}~\cite{Dreyer:2003bv,Berti:2005ys,Baibhav:2023clw}, offers a powerful tool for probing the nature of the remnant BH. Crucially, the fundamental QNM frequency of a Kerr BH depends only on its mass and spin~\cite{Detweiler:1977gy}. Measuring it thus provides a direct test of GR~\cite{LIGOScientific:2016lio,Berti:2015itd,Yunes:2016jcc}. Moreover, the detection of multiple QNMs enables tests of the \emph{no-hair theorem} (see Section~\ref{GravAstro_subsec:BH}) and the BH paradigm~\cite{Will:2014kxa,Cardoso:2019rvt}.
\section{Current and Future Observatories}\label{GravAstro_sec:observ}
In the previous sections, we discussed two remarkable predictions of GR:~black holes and gravitational waves. Until 2015, evidence for both was only \emph{indirect}, such as through X-ray binaries~\cite{Remillard:2006fc} or the Hulse-Taylor pulsar~\cite{Hulse:1974eb,Taylor1982}. This changed in September 2015, when the LIGO observatories made the first \emph{direct detection} of GWs from a binary BH merger~\cite{LIGOScientific:2016aoc}. Since then, over a hundred GW events have been detected, primarily from BH binaries, but also from neutron star binaries and BH-neutron star systems~\cite{LIGOScientific:2018mvr,LIGOScientific:2020ibl,KAGRA:2021vkt}. In this section, we review the status of current detectors, highlight key discoveries, and explore the prospects of future observatories.
\vskip 2pt
\begin{figure}[t!]
\centering
\includegraphics[scale=1]{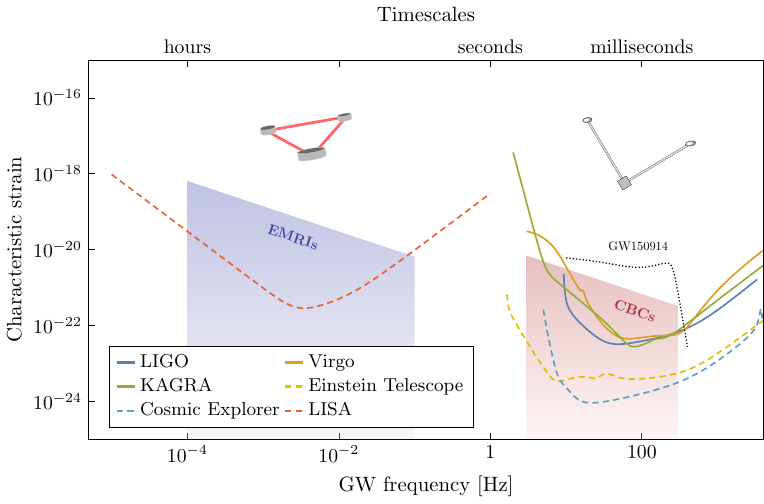}
\caption{The sensitivity curves of current (solid) and planned (dashed) GW detectors. The characteristic strain for two types of sources is shown:~extreme mass ratio inspirals (EMRIs) and compact binary coalescences (CBCs). For comparison, the first detected binary BH event, GW150914, is also displayed. The area between the strain of the source and the sensitivity curve of the detector indicates the \emph{strength} of the signal, i.e., the signal-to-noise ratio. Data for these curves was taken from~\cite{GWPlotter}.}
\label{fig:detectorcurves}
\end{figure} 
Currently, four GW detectors are in operation:~the two Advanced LIGO detectors in the United States~\cite{TheLIGOScientific:2014jea}, the Advanced Virgo detector in Italy~\cite{TheVirgo:2014hva}, and the KAGRA detector in Japan~\cite{KAGRA:2020tym}. These are L-shaped Michelson interferometers with kilometre-scale arms, capable of detecting tiny variations in arm length caused by passing GWs~\eqref{eq:strain}. Their peak sensitivity lies in the $10-1000\,\mathrm{Hz}$ range, making them ideal for observing mergers of compact objects with nearly equal masses between $1-100\,M_{\odot}$. Such signals typically last from milliseconds to seconds. Figure~\ref{fig:detectorcurves} shows the sensitivity curves of these detectors at their design sensitivity, along with the strain of typical compact binary coalescences (CBCs) and the first detected signal, GW150914.
\vskip 2pt
Ground-based detectors have already enabled stringent tests of GR, including measurements of BH QNM frequencies~\cite{LIGOScientific:2016lio,LIGOScientific:2019fpa}, constraints on the speed of GWs~\cite{LIGOScientific:2017zic}, and insights into the mass distribution of BHs~\cite{LIGOScientific:2020kqk}. A landmark event was the detection of a binary neutron star merger, GW170817~\cite{LIGOScientific:2017vwq}, which was accompanied by an electromagnetic counterpart. This event was precisely localised, leading to the identification of a coincident short gamma-ray burst by the Fermi GBM telescope and a multi-wavelength follow-up that revealed a kilonova~\cite{LIGOScientific:2017ync}. These observations not only provided new insights into the internal structure of neutron stars~\cite{LIGOScientific:2018cki} but also confirmed that neutron star mergers are key sites for heavy-element nucleosynthesis~\cite{Kasen:2017sxr,Pian:2017gtc} and enabled an independent measurement of the Hubble constant~\cite{LIGOScientific:2017adf}, demonstrating the potential of GWs as a cosmological probe.
\vskip 2pt
The detections from LIGO and Virgo mark only the beginning of GW astronomy. Plans are underway to expand the global detector network with LIGO-India~\cite{Unnikrishnan:2013qwa,Saleem:2021iwi,Unnikrishnan:2023uou}, which will improve sky localisation and overall sensitivity. There are also proposals to push into higher frequencies such as the Neutron star Extreme Matter Observatory (NEMO)~\cite{Ackley:2020atn}, designed to reach the kilohertz regime. Beyond these, a new generation of GW observatories is on the horizon, promising a dramatic leap in sensitivity. The Einstein Telescope~\cite{Punturo:2010zz,Sathyaprakash:2012jk,ET:2019dnz}, an underground interferometer featuring a triangular configuration with 10 km arms, and Cosmic Explorer~\cite{Reitze:2019iox,Evans:2021gyd}, a L-shaped detector with 40 km arms, are designed to broaden the accessible frequency range and increase detection rates by orders of magnitude (see Figure~\ref{fig:detectorcurves}). These advancements will enable the observation of new sources and improve precision tests of gravity.
\vskip 2pt
Space-based detectors will further revolutionise GW astronomy, with the recently adopted Laser Interferometer Space Antenna (LISA)~\cite{LISA:2017pwj,Colpi:2024xhw} leading the way. LISA’s 2.5 million-kilometre-long laser beams will open up the millihertz frequency regime $10^{-4}-10^{-1}\,\mathrm{Hz}$, granting access to an entirely new class of astrophysical sources such as EMRIs, massive BH mergers at high redshifts, stochastic GW backgrounds and compact galactic binaries~\cite{Colpi:2024xhw}. Similar proposed space-based missions include TianQin~\cite{TianQin:2015yph,TianQin:2020hid} and Taiji~\cite{Gong:2021any}, while proposals such as the Advanced Laser Interferometer Antenna (ALIA) or the Big Bang Observer (BBO)~\cite{Crowder:2005nr} are even more ambitious.
\vskip 2pt
To bridge the gap between the millihertz band of space-based detectors and the hertz band of ground-based detectors, a new class of decihertz detectors has also been proposed. These include the Decihertz Interferometer Gravitational Wave Observatory (DECIGO)~\cite{Kawamura:2011zz} and TianGO~\cite{Kuns:2019upi}, aimed at detecting, e.g.,~primordial GWs or measuring the Hubble constant. Finally, at very low frequencies, the global network of pulsar timing arrays is sensitive to GWs in the nanohertz regime $10^{-10} - 10^{-6}\,\mathrm{Hz}$, targeting sources such as the stochastic GW background and supermassive BH binaries. In 2023, the NANOGrav collaboration announced the first evidence of a stochastic GW background~\cite{NANOGrav:2023gor}.
\vskip 2pt
Gravitational-wave astronomy is still in its infancy, yet it has already transformed our understanding of the Universe. Although not all proposed detectors may come to fruition, they highlight the immense potential for discovery. With greater sensitivity and access to new frequency bands, the coming decades promise to uncover entirely new astrophysical phenomena. Importantly, next-generation detectors will offer unprecedented opportunities to study matter configurations around BHs -- the focus of the next chapter.
\chapter{Black Hole Environments}\label{chap:BHenv}
\vspace{-0.8cm}
\hfill \emph{The wonder is that we can see these trees and not wonder more}
\vskip 5pt

\hfill Ralph Waldo Emerson
\vskip 35pt
\noindent Most of the discussion so far has focused on BHs and GW emission in vacuum. In reality, however, GW sources are embedded in astrophysical environments, which take many different forms and range from dilute to extremely dense. Understanding how these environments influence BHs and how, in turn, BHs affect their environment is the central theme of this thesis.
\vskip 2pt
This chapter provides an overview of the astrophysical environments relevant to GW astrophysics (Section~\ref{BHenv_sec:taxonomy}), followed by a closer examination of those most pertinent to this thesis. I begin with dark matter (Section~\ref{BHenv_sec:DM}) followed by ultralight bosons (Section~\ref{BHenv_sec:boson_clouds}), which not only serve as a promising dark matter candidate but may also help resolve open questions in fundamental physics. Next, I turn to astrophysical plasmas (Section~\ref{BHenv_sec:Plasma}) and their behaviour in curved spacetime under electromagnetic perturbations. This is followed by a discussion of accretion disks and active galactic nuclei (Section~\ref{BHenv_sec:accretion_disks_AGN}), some of the few BH environments with direct observational data. Finally, I explore how BHs can aid in probing these environments (Section~\ref{BHenv_sec:BH_lab}).
\section{Taxonomy}\label{BHenv_sec:taxonomy}
Compact objects, such as BHs, are rarely isolated in the Universe. They interact with their surroundings, which can include stars, gas or electromagnetic fields. In fact, it was observations of these \emph{environments} that provided the first evidence for BHs. For instance, the motion of stars orbiting a weak radio source at the centre of our galaxy offered strong evidence for a supermassive BH (Sgr\,$\mathrm{A}^*$) residing there~\cite{Ghez:2008ms,2009ApJ...692.1075G}. More recently, this was confirmed by the image of the Event Horizon Telescope of the luminous matter surrounding it~\cite{EventHorizonTelescope:2022wkp}. The term ``environment'' can thus be taken very broadly, but for the purposes of this thesis, a more precise working definition is useful.
\begin{figure}[t!]
\centering
\includegraphics[scale=1]{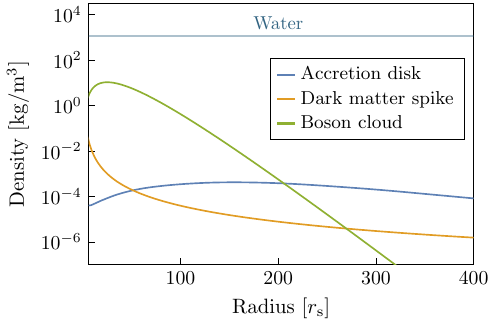}
\caption{Density profiles of various astrophysical environments surrounding a supermassive BH with a mass of $10^6 M_{\odot}$. The accretion disk (Section~\ref{BHenv_sec:accretion_disks_AGN}) follows the Sirko-Goodman model~\cite{Sirko:2002ex}, with its inner region described by the $\alpha$-disk prescription of Shakura \& Sunyaev~\cite{1973A&A....24..337S}, assuming a viscosity parameter $\alpha\ped{visc} = 0.01$. The dark matter spike (Section~\ref{BHenv_sec:DM}) is modelled as $\rho = \rho_6(r_6/r)^{\gamma\ped{sp}}$ with $r_6 = 10^{-6}\,\mathrm{pc}$, $\rho_6 \sim 10^{17}M_{\odot}/\mathrm{pc}^3$~\cite{Cole:2022yzw} and $\gamma\ped{sp} = 7/3$~\cite{Gondolo:1999ef}.  The boson cloud (Section~\ref{BHenv_sec:boson_clouds}) is assumed to occupy its dominant growing mode, with a total mass equal to 10\% of the BH mass and a boson mass parameter of $\mu M = 0.2$. For reference, the density of water is also shown.}
\label{fig:taxonomy}
\end{figure} 
\vskip 2pt
I will define ``astrophysical environments'' as \emph{matter distributions around BHs, either gaseous or dark in nature}. This definition is more restrictive than some found in the literature, where any non-vacuum or beyond-GR effect that influences the system's dynamics is considered an ``environmental effect''. Examples include third-body interactions in BH triplets~\cite{Bonetti:2018tpf}, tidal forces from a nearby companion~\cite{Yang:2017aht}, and dynamics in dense stellar systems where multi-body encounters can play a significant role~\cite{Wen:2002km,Blaes:2002cs,Miller:2002pg,Amaro-Seoane:2011rdr,2013MNRAS.431.2155N}, such as through Kozai-Lidov resonances~\cite{Kozai,LIDOV1962719}. Other phenomena sometimes labelled as ``environmental'' are Doppler shifts, gravitational lensing~\cite{Ezquiaga:2020gdt} or spin-orbit couplings in triples~\cite{Fang:2019mui}. 
\vskip 2pt
While this thesis does not attempt to cover all possible environmental effects, a comprehensive understanding of waveform modifications is essential for successful GW astrophysics. Extracting signals from detectors like LISA requires accounting for \emph{all} potential sources of dephasing, rather than selectively modelling only certain effects. Failing to do so could introduce biases in parameter estimation, misinterpreting data or even lead to missed detections~\cite{Roy:2024rhe}. Additionally, understanding degeneracies between different environments, as well as those arising from beyond-GR effects~\cite{Barausse:2014tra}, is crucial. Nonetheless, the first priority is to understand the interplay between BHs, the surrounding matter, and the resulting GWs, which forms the central focus of this thesis.
\vskip 2pt
Among astrophysical environments, gaseous structures such as accretion disks and plasmas are particularly well studied through electromagnetic observations. Accretion disks consist mainly of baryonic matter -- gas and dust -- accreted onto the BH from a companion star or the interstellar medium. These disks can influence binary BH dynamics in multiple ways:~accretion onto the BHs alters their mass and spin, while dynamical friction and planetary migration drive gradual inward motion. Furthermore, the disk’s self-gravity can modify the binary’s orbital evolution. We will return to accretion disks, particularly in active galactic nuclei, in Section~\ref{BHenv_sec:accretion_disks_AGN}.
\vskip 2pt
Dark matter structures are another important class of astrophysical environments. High dark matter densities can form around BHs in various ways. For example, if dark matter consists of cold, collisionless particles, the adiabatic growth of BHs leads to the formation of overdensities, known as ``spikes''~\cite{Gondolo:1999ef}, discussed further in Section~\ref{BHenv_sec:DM}. Alternatively, if dark matter is composed of light bosonic fields such as axions -- motivated by both particle physics~\cite{Peccei:1977hh, Weinberg:1977ma, Wilczek:1977pj} and astrophysics~\cite{Hu:2000ke, Hui:2016ltb} -- it can form a macroscopic Bose-Einstein condensate around the BH through a process called ``superradiance''. This structure, known as a boson cloud or ``gravitational atom''~\cite{Brito:2015oca}, can contain up to 10\% of the BH's mass~\cite{Brito:2014wla,East:2017ovw,Herdeiro:2021znw}. We will explore these systems in more detail in Section~\ref{BHenv_sec:boson_clouds}. To give a quantitative sense of scale, Figure~\ref{fig:taxonomy} shows their density profiles around a supermassive BH with a mass of $10^6 M_{\odot}$.
\section{Dark Matter}\label{BHenv_sec:DM}
One of the greatest unsolved mysteries in modern physics is the nature of dark matter. Astrophysical and cosmological observations have revealed that the visible matter in our Universe accounts for only about 15\% of the total matter content. The remaining 85\% consists of an unknown, invisible component which is called \emph{dark matter} (DM) (see~\cite{Bertone:2004pz,Cirelli:2024ssz} for reviews). Unlike ordinary matter, DM does not seem to interact with the strong, weak, or electromagnetic force. However, by virtue of the equivalence principle, it must interact gravitationally and indeed its gravitational influence has been shaping galaxies, clusters, and the large-scale structure of the Universe.
\vskip 2pt
The story of DM dates back to 1933, when Swiss astronomer Fritz Zwicky studied a large collection of galaxies called the Coma Cluster~\cite{Zwicky:1933gu}. He found that the velocity dispersion of galaxies within that cluster was so high that, in order for the galaxies in the cluster to remain gravitationally bound, there had to be significantly more mass than what was visible. Similar conclusions were drawn soon after by Holmberg and Smith, who studied the galaxies and mass of the Virgo Cluster~\cite{Smith:1936mlg,1937AnLun...6....1H}. By now, the existence of DM is supported by a wide range of observations. One of the most striking pieces of evidence comes from galaxy rotation curves. Observations of spiral galaxies, including the Milky Way, show that their outer regions rotate at nearly constant speeds rather than slowing down as expected from Keplerian motion $v = \sqrt{M/r}$~\cite{Rubin:1970zza,Freeman:1970mx,Ostriker:1973uit,Roberts1975,Rubin:1980zd,Persic:1995ru,Iocco:2015xga}. This suggests the presence of an extended, invisible mass distribution, now understood as a \emph{dark matter halo}. Dark matter also plays a crucial role on cosmological scales. The detailed structure of the cosmic microwave background (CMB) and its temperature anisotropies provide strong constraints on the amount and nature of DM~\cite{2013ApJS..208...19H,Planck:2018vyg}. Further support comes from large-scale galaxy clustering, gravitational lensing observations~\cite{2004IAUS..220..439H,2006ApJ...648L.109C}, and baryon acoustic oscillations~\cite{2010MNRAS.404...60R,2011ApJS..192...18K}. Taken together, these diverse lines of evidence paint a consistent picture:~DM is an important component of the Universe’s mass-energy budget and is essential to our understanding of its evolution and large-scale structure formation.
\vskip 2pt
Yet despite overwhelming indirect evidence, its fundamental nature remains unknown. Two key arguments suggest that DM cannot be baryonic. First, Big Bang nucleosynthesis constrains the baryon density, as the observed abundances of light elements such as helium and deuterium would be drastically different if DM were composed of baryons~\cite{Cyburt:2004cq,Steigman:2007xt,Iocco:2008va,Fields:2011zzb,Cyburt:2015mya}. Second, the temperature fluctuations in the CMB encode information about the density of baryons~\cite{Peebles:1970ag,Bond:1980ha,Hu:2001bc}. If DM were baryonic, it would have influenced the acoustic peaks in the CMB spectrum in a way that is inconsistent with observations~\cite{Planck:2018vyg}. These considerations support the view that DM is cold and interacts only through gravity, a premise that underpins the $\Lambda$CDM model~\cite{Carroll:2000fy,Copeland:2006wr,Cirelli:2024ssz}. This model successfully explains the formation of galaxies and galaxy clusters, as well as the large-scale structure of the Universe and its accelerated expansion driven by dark energy~\cite{Planck:2018vyg}. The name ``$\Lambda$CDM'' reflects these two primary components:~a cosmological constant $\Lambda$ accounting for dark energy~\cite{Carroll:2000fy,Copeland:2006wr} and a pressureless cold DM component~\cite{Bertone:2004pz,Cirelli:2024ssz}. Together with General Relativity, they form a consistent theoretical framework that describes the large-scale behaviour of the Universe~\cite{Planck:2018vyg}.
\vskip 2pt
Despite decades of effort, no particle with the required properties has been found, suggesting the need for new physics. Given that most galaxies are thought to harbour supermassive BHs at their centres~\cite{2021NatRP...3..732V}, DM inevitably interacts with these strong-gravity environments, making GW astronomy a unique tool for probing its nature. In Section~\ref{BHenv_subsec:DMStruct}, I will first discuss how DM clusters on scales relevant for GW physics and then, in Section~\ref{BHenv_subsec:candidates}, I will outline some of the most compelling candidates for DM.
\subsection{Structures}\label{BHenv_subsec:DMStruct}
While the distribution of DM on cosmological scales is well understood, this thesis focuses on its detection through BHs and GWs. Accordingly, the emphasis is on smaller scales -- galactic and below -- where DM may accumulate around BHs or form compact objects, referred to as ``dark compact objects''. In this section, we will describe the distribution of DM on such scales, focusing progressively on smaller length scales.
\vskip 2pt
Our understanding of the DM distribution on large scales is guided by numerical N-body simulations, which suggest a \emph{universal} density profile for DM halos, exhibiting similar characteristics across different halo masses, cosmic epochs, and initial density fluctuations in the early Universe (the ``input power spectra'')~\cite{Navarro:1995iw}. The density of DM halos can be parameterised as
\begin{equation}\label{eq:DM_halo_gener}
\rho(r)=\frac{\rho\ped{H}}{(r / a\ped{H})^\gamma\left[1+(r / a\ped{H})^\alpha\right]^{(\beta-\gamma) / \alpha}}\,, 
\end{equation}
where $a\ped{H}$ is a scale length, $\rho\ped{H}$ the central halo density and the parameters ($\alpha, \beta, \gamma$) depend on the choice profile, which remains a subject of debate. One of the most popular models is the Navarro-Frenk-White (NFW) profile~\cite{Navarro:1995iw}, which follows an $r^{-1}$ power-law behaviour at small radii and transitions to an $r^{-3}$ decay at large radii. The observed density profiles of galactic bulges and elliptical galaxies are well-described by the Hernquist profile~\cite{1990ApJ...356..359H}, corresponding to $(\alpha, \beta, \gamma) = (1,4,1)$. Its density is given by
\begin{equation}\label{eq:Hernquist}
\rho(r) = \frac{M\ped{H}a\ped{H}}{2\pi r (r+a\ped{H})^3}\,,
\end{equation}
where $M\ped{H}$ is the total mass of the halo. Within a radius $r$, the mass of the halo is
\begin{equation}\label{eq:Hernquist_mass}
M\ped{H}(r) = \int_{0}^{r} 4\pi \rho(r')\,(r')^2\dd r' = \frac{2\pi r^2 \rho\ped{H}a\ped{H}^3}{(a\ped{H}+r)^2}\,.
\end{equation}
Taking the limit $r\rightarrow \infty$, this expression yields the total mass:~$M\ped{H} = 2\pi \rho\ped{H}a\ped{H}^3$. We will encounter this profile again in Chapter~\ref{chap:BHspec}. Other commonly used density profiles are those of King~\cite{King:1962wi}, Einasto~\cite{1965TrAlm...5...87E}, Jaffe~\cite{Jaffe:1983iv}, Kravtsov et al.~\cite{Kravtsov:1997dp}, and Moore et al.~\cite{Moore:1999gc}.
\vskip 2pt
All of these profiles predict an increasing DM density towards the inner regions of the halo. However, both Newtonian and relativistic analyses show that when a BH resides at the centre, the DM density vanishes at the horizon~\cite{Gondolo:1999ef,Sadeghian:2013laa}. This creates a steep inner cusp whose length scale is set by the BH mass. The total mass contained in this cusp is negligibly small, and at larger radii, the profile smoothly transitions to one of the aforementioned halo models.
\subsubsection{The Einstein cluster}
For the purposes of GW physics, we now turn to a self-consistent model of a BH surrounded by a DM halo, following the approach outlined in~\cite{Cardoso:2021wlq,Cardoso:2022whc}, and making use of the so-called ``Einstein Cluster''.
\vskip 2pt
We assume a spherically symmetric spacetime with the line element 
\begin{equation}\label{eqn:line_element_theory}
ds^2= -A(r)\dd t^2 + \frac{\dd r^2}{1-2m(r)/r} + r^2 \dd \Omega^2\,,
\end{equation}
where $\dd \Omega^2 = \dd \theta ^2 + \sin \theta^2 \dd \varphi^2$, and $m(r)$ is the mass function. The aim is to find a geometry that describes a BH on small scales and a matter distribution following a DM profile, such as the Hernquist model~\eqref{eq:Hernquist}, on large scales. To achieve this, we extend Einstein’s construction of a stationary system of gravitating masses, known as an \emph{Einstein Cluster}~\cite{Einstein_cluster,Einstein_cluster_2}, to include a central BH. In this framework, particles move on circular geodesics, and the system is described using an averaged stress-energy tensor given by $\langle T^{\mu\nu}\rangle = n/m\ped{p}\langle p^{\mu}p^{\nu}\rangle$, where $n$ is the number density of particles with mass $m\ped{p}$, and $p^{\mu}$ is the four-momentum satisfying the geodesic equation. This configuration is mathematically equivalent to modelling an anisotropic fluid with a tangential pressure $P\ped{t}$ and no radial pressure ($P\ped{r} = 0$), which leads to the stress-energy tensor:
\begin{equation}\label{eq:stress_energy_appendix}
    T_{\mu\nu}=(\rho+P\ped{t}) v_\mu v_\nu+P\ped{t}(g_{\mu\nu}-r_\mu r_\nu)\,,
\end{equation}
where $\rho$ and $v^\mu$ are the density and four-velocity of the fluid, respectively, $g_{\mu \nu}$ the metric and $r^\mu$ is a unit radial vector. 
\vskip 2pt
With this setup in place, we can assign a mass function to the system. A choice inspired by the Hernquist profile~\eqref{eq:Hernquist} is given by
\begin{equation}\label{eq:BHenv_HernquistMass}
m(r)=M+\frac{M\ped{H} r^2}{\left(a\ped{H}+r\right)^2}\left(1-\frac{2 M}{r}\right)^2\,.
\end{equation}
At small radii $r \ll a\ped{H}$, this profile is dominated by the BH gravity, while at large radii, the mass profile aligns with the Hernquist profile~\eqref{eq:Hernquist}, consistent with the astrophysical scenario we have in mind. Although the Hernquist model serves as a useful example, similar constructions can be applied to other DM profiles~\cite{Konoplya:2022hbl,Figueiredo:2023gas,Speeney:2024mas}, all of which share the same qualitative behaviour. In realistic astrophysical settings, DM halos are significantly more massive than the central BH they host, with characteristic scales satisfying $a\ped{H} \gg M\ped{H} \gg M$~\cite{Navarro:1995iw,Kim:2004tc}.
\vskip 2pt
Given the mass function~\eqref{eq:BHenv_HernquistMass}, the metric functions can be solved to obtain an analytical solution for the background spacetime. The full expressions are provided in~\cite{Cardoso:2021wlq}, but the key properties are summarised here. The spacetime features an horizon at $r = 2M$ and a singularity at $r = 0$. At large distances, the Newtonian potential matches that of the Hernquist profile, and the spacetime is asymptotically flat with an ADM mass $M_{\scalebox{0.7}{$\mathrm{ADM}$}} = M+ M\ped{H}$. Finally, in the vicinity of the BH, the redshift factor goes as $\sim 1 - 2M\ped{H}/a\ped{H}$. 
\vskip 2pt
Although idealised, the Einstein cluster offers a useful proxy for exploring the dynamics of BHs in generic (dark) matter distributions and can serve as a stepping stone toward more realistic models. In Chapter~\ref{chap:BHspec}, we will explore how such a DM halo affects BH ringdown signals and assess its potential implications for GW observations.
\subsubsection{Dark matter around black holes}
On smaller scales, DM can accumulate around BHs, forming overdensities that peak close to the BH horizon~\cite{Gondolo:1999ef, Sadeghian:2013laa}. In the absence of any disruptive astrophysical processes, such structures would evolve from an initial DM halo~\eqref{eq:DM_halo_gener} into a profile that is orders of magnitude denser. This process has been studied extensively for supermassive BHs at the centres of galaxies~\cite{Gondolo:1999ef, Gondolo:2000pn, Ullio:2001fb, Bertone:2001jv, Merritt:2002vj, Bertone:2005hw, Merritt:2006mt}, as well as for intermediate-mass BHs~\cite{Bertone:2005xz, Zhao:2005zr, 2009PhRvL.103p1301B}.
\vskip 2pt
In particular, for BHs with masses $\leq 10^5 M_{\odot}$ residing at the centre of a DM halo, the BH gradually accretes matter onto a seed, which could originate from the direct collapse of a supermassive star. The DM distribution evolves in response to the BH's adiabatic growth by forming a steep \emph{dark matter spike}, with a density profile following a power law~\cite{Gondolo:1999ef}:
\begin{equation}
\rho \sim r^{-\gamma\ped{sp}}\,,\quad \text{where}\quad \gamma\ped{sp} = \frac{9-2\alpha}{4-\alpha}\,.
\end{equation}
Here, $\gamma\ped{sp}$ denotes the spike slope, which depends on the initial inner slope $\alpha$ of the DM halo~\eqref{eq:DM_halo_gener}. For example, an initial NFW profile with $\alpha = 1$ gives $\gamma\ped{sp} = 7/3$. A representative example of this profile using fiducial parameters is shown in Figure~\ref{fig:taxonomy}.
\vskip 2pt
However, as the growth of such spikes occurs over long astrophysical timescales, various processes may disturb them. Baryonic feedback and DM self-annihilation can significantly flatten the central profile~\cite{Merritt:2002vj, Sadeghian:2013laa, Ferrer:2017xwm}, while their survival through galactic mergers -- especially those involving BHs of comparable masses -- remains uncertain~\cite{Merritt:2002vj, Kavanagh:2018ggo}. Several refinements and extensions to the spike model have been proposed, including relativistic corrections~\cite{Sadeghian:2013laa, Ferrer:2017xwm} and alternative structures such as DM mounds~\cite{Bertone:2024wbn}, crests~\cite{Merritt:2006mt}, and mini-spikes~\cite{Bertone:2005xz, Zhao:2005zr}. In any case, the DM density around BHs could still be much larger than the average DM density in our Universe, making them promising targets for GW physics. A qualitatively different form of DM overdensities arises when the surrounding matter consists of ultralight bosonic fields rather than classical particles. This case will be examined in Section~\ref{BHenv_sec:boson_clouds}.
\clearpage
\subsubsection{Dark compact objects}
If DM coalesces into objects of astrophysical size, it could form \emph{dark compact objects} (DCOs). These objects are of interest both as a potential form of DM and as a means to test the BH paradigm.
\vskip 2pt
One class of DCOs consists of \emph{exotic stars}, which can arise from massive bosonic fields -- both real and complex -- that are well-motivated DM candidates (as we will discuss in Section~\ref{BHenv_subsec:u_bosons}). These fields can form ultra-compact, coherent solitonic configurations bound by gravity or self-interactions. If the fields are complex, they form \emph{boson stars}~\cite{Kaup1968,Ruffini1969,Breit:1983nr, Colpi1986,Schunck:2003kk,Liebling:2012fv}, whereas real fields give rise to \emph{oscillatons}~\cite{Seidel1991,Copeland:1995fq,Braaten:2015eeu,Lee1987-2,Lee1987-3,Lee1987-4,Lynn:1988rb}. Unlike boson stars, oscillatons are inherently time dependent and gradually dissipate through GW emission. The compactness $\mathcal{C}$ of these objects depends on the specific properties of the boson, but can reach values comparable to those of  neutron stars, with $\mathcal{C} \sim 0.1$~\cite{Ozel:2016oaf,LIGOScientific:2018cki}. They can form via gravitational collapse and evolve into ultra-compact configurations through a process known as \emph{gravitational cooling}~\cite{Bianchi:1990mha,Seidel:1993zk,Liebling:2012fv,Brito:2015yga,DiGiovanni:2018bvo}, with self-interactions playing a crucial role in efficient formation. Additionally, fermionic DM can give rise to \emph{fermion stars}~\cite{Narain:2006kx,Kouvaris:2015rea,Gresham:2018rqo,Chang:2018bgx}, where Fermi degeneracy pressure counteracts gravitational collapse. However, without strong attractive interactions, fermion stars tend to be less compact than boson stars. 
\vskip 2pt
A second category of DCOs is \emph{primordial black holes} (PBHs)~\cite{HawkingPBH,Chapline:1975ojl,Carr:2016drx}, which form from highly overdense regions in the early Universe~\cite{Zeldovich:1967lct,Carr:1974nx,1975ApJ...201....1C}. Unlike stellar-mass BHs, PBHs are not subject to the Tolman--Oppenheimer--Volkoff lower mass bound of $\gtrsim 3 M_\odot$~\cite{Tolman1939}, as they do not originate from stellar collapse. Since the impact of non-baryonic matter is already inferred from Big Bang Nucleosynthesis~\cite{Cyburt:2015mya}, PBHs must have formed within the first second after the Big Bang to be viable DM candidates. Furthermore, their initial mass must have been high enough to avoid complete evaporation through Hawking radiation. These two factors together limit their available mass range. Constraints from astrophysical and cosmological observations, including extragalactic gamma-ray backgrounds~\cite{Carr:2009jm}, gravitational microlensing~\cite{EROS-2:2006ryy,2013MNRAS.435.1582C,Niikura:2017zjd}, supernova lensing~\cite{Zumalacarregui:2017qqd}, and imprints on the CMB~\cite{Ali-Haimoud:2016mbv,Poulin:2017bwe}, significantly limit the fraction of DM that PBHs can constitute.
\vskip 2pt
A final category of DCOs includes \emph{black hole mimickers}, which are horizonless compact objects that closely resemble BHs at large distances and near the photon ring (see~\cite{Cardoso:2019rvt} for a review). Many possibilities exist, such as \emph{gravastars}~\cite{Giddings:1992hh,Posada:2018agb}, \emph{wormholes}~\cite{Einstein:1935tc}, \emph{anisotropic stars} and compact objects composed of entirely different fields or particles~\cite{Raidal:2018eoo,Deliyergiyev:2019vti}.
\clearpage
\subsection{Candidates}\label{BHenv_subsec:candidates}
Numerous candidates have been proposed for DM, sometimes collectively referred to as the ``DM zoo''~\cite{Bertone:2004pz}. A key contender is the \emph{weakly interacting massive particle} (WIMP), which interacts via the weak nuclear force (see~\cite{Arcadi:2017kky} for a review). WIMPs gained prominence through the so-called ``WIMP miracle'', which refers to the fact that their predicted thermal relic abundance matches the observed DM density when their cross section is characteristic of the weak interaction~\cite{Srednicki:1988ce,Gondolo:1990dk}. Efforts to detect WIMPs typically involve two strategies. One focuses on directly observing nuclear recoils from WIMP interactions in ultra-sensitive underground experiments~\cite{Akerib:2016vxi,Cui:2017nnn}, for example the XENONnT experiment~\cite{Aprile:2018dbl,XENON:2024wpa}. However, despite decades of refinement, no conclusive signal has been found, and searches are approaching the ``neutrino floor'', where backgrounds from solar and atmospheric neutrinos become dominant~\cite{Schumann:2019eaa}. The other strategy looks for indirect signals from WIMP annihilation or decay products, which should generate Standard Model particles such as photons or neutrinos in regions with high DM density, e.g., the Galactic centre. In particular, gamma-ray emissions from WIMP annihilation could appear as an excess in observational data. The lack of such a signal in Fermi satellite data imposes strong constraints on the self-annihilating cross-section of WIMPs~\cite{FermiLAT2009,Adriani:2008zr,Gaskins:2016cha}. This had led to an increased interest in alternative DM candidates.
\vskip 2pt
Among these, \emph{axions} and \emph{axion-like particles} emerge as compelling candidates. These are ultralight, feebly interacting particles which may solve various problems in high-energy physics. Their motivation and potential will be further discussed in Section~\ref{BHenv_subsec:u_bosons}. Another proposal is the \emph{sterile neutrino}~\cite{Shi:1998km}, which interacts with standard neutrinos via the weak nuclear force and can remain stable over cosmological timescales. Multiple mechanisms can produce them in the early Universe with the appropriate abundance~\cite{Shi:1998km,Laine:2008pg,Boyarsky:2009ix,Drewes:2016upu}. Although most sterile neutrinos must be long-lived to serve as DM, some inevitably decay, emitting mono-energetic photons. An observed $3.55\,\mathrm{keV}$ X-ray line in galaxy clusters has been proposed as a possible signature of this decay~\cite{Bulbul:2014sua,Boyarsky:2014jta}, though this remains a subject of ongoing debate~\cite{2015MNRAS.450.2143J,Dessert:2018qih,Silich:2021sra}.
\vskip 2pt
A particularly intriguing class of DM candidates -- relevant to this thesis -- are particles with a small electric charge, known as \emph{millicharged dark matter}~\cite{Goldberg:1986nk,DeRujula:1989fe,Cheung:2007ut, Feldman:2007wj}. Such particles arise naturally in extensions of the Standard Model featuring a hidden $U(1)$-gauge symmetry. Their charge is typically much smaller than the electron charge and their coupling to the Maxwell sector is suppressed. This reduces the otherwise large charge-to-mass ratio of the electron, allowing them to evade standard discharge mechanisms and potentially charge BHs. This leads to an interesting phenomenology, which is further explored in Chapters~\ref{chap:in_medium_supp} and~\ref{chap:plasma_ringdown}. Astrophysical searches and collider experiments have constrained some of their parameter space~\cite{Dimopoulos:1989hk,Davidson:2000hf,Essig:2013lka,Berlin:2018bsc}. A related proposal concerns \emph{dark photons}, gauge bosons that kinetically mix with ordinary photons~\cite{Holdom:1986eq,Okun:1982xi,Galison:1983pa}. Denoting the dark photon by $A^\prime_\mu$, the relevant terms in the Lagrangian are
\begin{equation}\label{eqn:darkphotonL} 
\mathcal{L}_{A^\prime} \supset -\frac{1}{4} F^\prime_{\mu \nu} F^{\prime \mu \nu} - \frac{\epsilon}{2} F^\prime_{\mu \nu} F^{\mu \nu}  - \frac{1}{2} \mu_{\gamma'}^2 A^\prime_\mu A^{\prime \mu}\,,  
\end{equation}
where $F^\prime_{\mu \nu}$ is the dark photon field strength, $\mu_{\gamma'}$ its mass, and $\epsilon \ll 1$ is the kinetic mixing parameter.\footnote{While $\epsilon$ is used here for the kinetic mixing parameter, a different but equivalent notation is adopted in Chapter~\ref{chap:in_medium_supp}. Additionally, Appendix~\ref{app:DPbasis} further explores different basis choices for the dark photon.} Current constraints apply mainly to heavy dark photons, leaving the ultralight regime largely unexplored~\cite{Caputo:2021eaa} (see Figure~\ref{fig:constraints}).
\vskip 2pt
The proposed candidates for DM span an enormous mass range, from ultralight bosons at $10^{-22}$ eV to PBHs of $\mathcal{O}(10M_{\odot})$, reflecting the uncertainty and elusiveness about its nature. Part of the challenge arises from its weak interactions with ordinary matter. Gravity, however, acts as a universal messenger, which has prompted a growing effort in recent years to explore how gravitational physics, especially GWs, can be used to uncover the properties of DM.
\section{Boson Clouds}\label{BHenv_sec:boson_clouds}
While DM is one of the most important unresolved puzzles in physics and astronomy, it is far from the only one. In what follows, I highlight several other open challenges (Section~\ref{BHenv_subsec:BSM}) and introduce a class of particles that appear in some proposed solutions (Section~\ref{BHenv_subsec:u_bosons}). I then present a mechanism capable of producing large numbers of these particles around BHs (Section~\ref{BHenv_subsec:superrad}), before describing the resulting structures (Section~\ref{BHenv_subsec:gatoms}).
\subsection{Physics Beyond the Standard Model}\label{BHenv_subsec:BSM}
Whereas Chapter~\ref{chap:GravAstro} focused on the gravitational sector of our Universe as described by General Relativity, all other forces are governed by the Standard Model (SM) of particle physics. Though it has achieved remarkable experimental success~\cite{Aad:2012tfa,Chatrchyan:2012xdj,Tanabashi:2018oca}, it is still widely regarded as incomplete, with observational evidence pointing strongly towards ``physics beyond the SM''. For example, even when accounting for DM, ordinary and dark matter together comprise only about 30\% of the total energy budget of the Universe. The remaining 70\% is attributed to dark energy, which is responsible for the observed acceleration of the Universe's expansion~\cite{Riess:1998cb, Perlmutter:1998np}. In the $\Lambda$CDM model~\cite{Planck:2018vyg}, dark energy is modelled as a cosmological constant -- a constant energy density that fills all of spacetime~\cite{Carroll:2000fy,Copeland:2006wr}. A natural explanation involves the vacuum energy predicted by quantum field theory. However, if quantum fluctuations up to the Planck scale contributed to the cosmological constant, its expected value would be about 120 orders of magnitude larger than the observed one. This extreme discrepancy, known as the \emph{cosmological constant problem}~\cite{Weinberg:1988cp, Padmanabhan:2002ji}, suggests an extraordinary degree of \emph{fine tuning}.
\vskip 2pt
Similar fine-tuning problems appear elsewhere in the SM. A striking example is the \emph{hierarchy problem}, which stems from the huge difference between the electroweak scale and gravity. The characteristic energy scale of gravity is set by the Planck mass, $M\ped{Pl}\sim\SI{e+19}{GeV}$, which is much larger than the electroweak scale, $\Lambda\ped{EW}\sim\SI{e+2}{GeV}$, the highest energy scale associated with the SM. As a result, a scalar field like the Higgs receives quantum corrections from all energy scales up to $M\ped{Pl}$, so without a stabilising mechanism, it would naturally be expected to be of order $M\ped{Pl}$, rather than the observed $m\ped{H} = \SI{125}{GeV}$. This again represents a problem of fine-tuning:~unless a symmetry or other stabilising mechanism exists, the Higgs mass is extremely sensitive to high-energy physics. In principle, ultraviolet physics could introduce counter-terms that give the Higgs its ``small mass'', yet this requires a fine-tuning on the level $\sim (m\ped{H}/M\ped{Pl})^2 \sim 10^{-34}$.
\vskip 2pt
A final example of fine-tuning lies in the strong sector. One of the parameters of the SM is the vacuum angle of quantum chromodynamics (QCD) $\theta_{\scalebox{0.7}{$\mathrm{QCD}$}}$, which appears in the CP-violating term of the Lagrangian:
\begin{equation}
\mathcal L\ped{SM}\supset\theta_{\scalebox{0.7}{$\mathrm{QCD}$}}\frac{g_s^2}{32\pi^2}G_{\mu \nu}\,{}^{*}\!G^{\mu\nu}\,,
\end{equation}
where $g_s$ is the strong coupling constant, $G_{\mu\nu}$ is the gluon field strength and ${}^{*}\!G^{\mu\nu}$ represents its dual. \emph{A priori}, $\theta_{\scalebox{0.7}{$\mathrm{QCD}$}}$ can take any value between $0$ and $2\pi$. Given that most dimensionless parameters in the SM are naturally of order unity, one might expect $\theta_{\scalebox{0.7}{$\mathrm{QCD}$}}$ to be of a comparable size. However, measurements of the neutron electric dipole moment indicate that $\theta_{\scalebox{0.7}{$\mathrm{QCD}$}}$ is extremely small, i.e., $\abs{\theta_{\scalebox{0.7}{$\mathrm{QCD}$}}}<10^{-10}$~\cite{Baker:2006ts, Afach:2015sja}. This absence of CP violation in the strong interaction is referred to as the \emph{strong CP problem}. In the next section, we will discuss a compelling resolution:~promoting $\theta_{\scalebox{0.7}{$\mathrm{QCD}$}}$ to a dynamical field.
\subsection{Ultralight Bosons}\label{BHenv_subsec:u_bosons}
\begin{figure}[t!]
\centering
\includegraphics[scale=0.85]{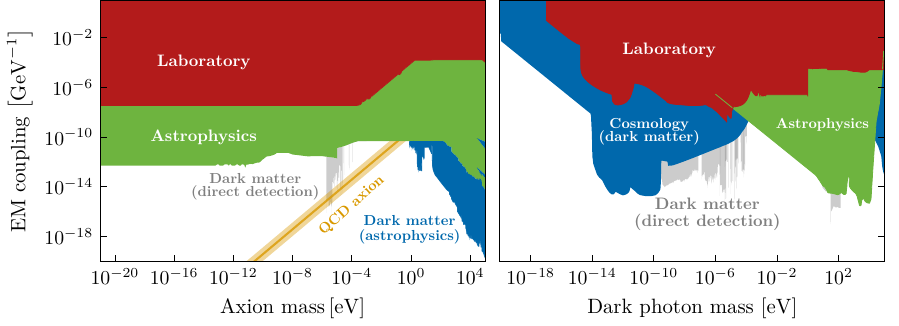}
\caption{Schematic overview of the constraints on the axion (\emph{left panel}) and dark photon (\emph{right panel}) from their coupling to the electromagnetic sector, realised through the axion-photon coupling $g_{a\gamma\gamma}$ and the kinetic dark photon-photon coupling $\epsilon$, respectively. The experiments contributing to these constraints are listed in~\cite{AxionLimits}. The yellow diagonal band indicates the predicted parameter space for the QCD axion~\eqref{eqn:QCDaxionmass}. It is evident that the weak-coupling and small-mass regime is largely unconstrained.}
\label{fig:constraints}
\end{figure}

Despite a wealth of evidence pointing to physics beyond the SM, the search for new degrees of freedom is challenging. For decades, this search has largely focused on the \emph{high-energy frontier}, where particle collisions at accelerators create heavy particles from lighter ones. However, many extensions to the SM predict the existence of particles that interact only weakly with ordinary matter, placing them beyond the reach of collider experiments. This is where gravity comes into play. Unlike other forces, gravity is universal;~all particles, regardless of their interaction strength with the SM, must couple to it, making it a powerful probe of new physics. In particular, \emph{ultralight bosons} stand out, as they arise naturally in extensions of the SM and could help resolve other outstanding problems, such as the nature of dark matter.
\vskip 2pt
The lack of CP violation in the strong interaction requires an extreme level of fine-tuning. The most compelling resolution to this problem is the Peccei-Quinn (PQ) mechanism~\cite{Peccei:1977hh,Weinberg:1977ma,Wilczek:1977pj}, which proposes to eliminate the CP-violating parameter $\theta_{\scalebox{0.7}{$\mathrm{QCD}$}}$ in a dynamical manner. Specifically, $\theta_{\scalebox{0.7}{$\mathrm{QCD}$}}$ is promoted to a dynamical field, the axion field $a$, which is associated with a new global $U(1)$-symmetry that spontaneously breaks at an energy scale $f_{a}$, known as the \emph{axion decay constant}. The resulting Nambu-Goldstone boson, \emph{the axion}, drives $\theta_{\scalebox{0.7}{$\mathrm{QCD}$}}$ towards zero as it evolves to the minimum of its potential, naturally explaining the smallness of $\theta_{\scalebox{0.7}{$\mathrm{QCD}$}}$ and resolving the strong CP problem. At low energies, the effective Lagrangian of the axion takes the form
\begin{equation}\label{eqn:axionLagrangian}
\mathcal{L}_{a} = - \frac{1}{2} \partial_\mu a \, \partial^\mu a + \frac{a}{f_a} \frac{g_s^2}{32\pi^2} G_{\mu \nu}\,{}^{*}\!G^{\mu\nu}+ \frac{\partial_\mu a}{f_a} j^\mu_a + \frac{1}{4}  g_{a \gamma \gamma} \, a \,F_{\mu \nu}\,{}^{*}\!F^{\mu\nu}\,,
\end{equation}
where $j^\mu_a$ represents the axial current generated by the quarks~\cite{Srednicki:1985xd}, $F_{\mu \nu}$ is the electromagnetic field strength and $g_{a \gamma\gamma} \propto f_{a}^{-1}$ is the axion-photon coupling constant. The properties of the axion are governed by the symmetry-breaking scale $f_a$. Below the PQ scale, the axion acquires a potential due to non-perturbative QCD effects, causing it to oscillate around its minimum with an effective mass~\cite{Weinberg:1977ma,diCortona:2015ldu}:\footnote{Higher order terms in the axion potential give rise to self-interactions. They appear as powers of the axion field, such as $a^n$, where $n \geq 3$, and can affect the dynamics of axion oscillations, especially at high field values.}
\begin{equation}\label{eqn:QCDaxionmass}
\mu\sim\SI{6e-10}{eV}\left(\frac{\SI{e+16}{GeV}}{f_a}\right)\,.
\end{equation}
A larger decay constant thus results in a lighter axion. If $f_a$ approaches the Grand Unified Theory scales $\sim 10^{16}$ GeV, the axion mass can be as low as $\lesssim 10^{-10}$ eV.
\vskip 2pt
Currently, most experimental searches for the QCD axion target its coupling to photons. Figure~\ref{fig:constraints} summarises the relevant constraints for generic ultralight scalar and vector fields, with the yellow band indicating the predicted mass-coupling relation for the QCD axion. While eq.~\eqref{eqn:QCDaxionmass} is well-defined, the precise width of the mass-coupling relation is model-dependent~\cite{Kim:1979if,Shifman:1979if,Dine:1981rt,Zhitnitsky:1980tq}. Even if the interaction is weak, sufficiently large quantities of axions could still produce detectable signals -- a prospect we will consider in Chapters~\ref{chap:SR_Axionic} and~\ref{chap:in_medium_supp}.
\vskip 2pt
Beyond the QCD axion, ultralight (pseudo-)scalar fields also naturally arise in string theory and are referred to as \emph{axion-like particles} (ALPs)~\cite{Arvanitaki:2009fg,Acharya:2010zx,Cicoli:2012sz}. When the extra dimensions predicted by string theory are compactified, they give rise to Kaluza-Klein zero modes~\cite{Green:1987mn,Svrcek:2006yi,Dine:2007zp}, which can acquire masses through non-perturbative effects, much like the QCD axion. Instead of a single field, string theory typically predicts a whole spectrum of ALPs, a scenario known as the \emph{string axiverse}~\cite{Arvanitaki:2009fg}. Their masses span from sub-eV scales down to the Hubble scale. Although their low-energy effective dynamics are still governed by parameters like $\mu$ and $f_a$, they do not follow the strict relation~\eqref{eqn:QCDaxionmass}. 
\vskip 2pt
Axion-like particles also present a well-motivated candidate for DM~\cite{Hui:2021tkt}. They fall into the category of \emph{wave dark matter} or \emph{fuzzy dark matter}~\cite{Hu:2000ke}. Fuzzy DM specifically refers to ultralight particles with masses in the range $\mu \sim [10^{-22}-10^{-20}]\,\mathrm{eV}$, while wave DM encompasses any form of DM that exhibits wave-like behaviour, generally with masses up to $\mu \lesssim 30\,\mathrm{eV}$. Anything above that is generally considered particle-like, such as the WIMP (see  Section~\ref{BHenv_subsec:candidates}). The relevant length scale for wave DM is set by the \emph{de Broglie wavelength}: 
\begin{equation}\label{eqn:DBwavelength}
\lambda\ped{dB} \equiv \frac{2\pi}{\mu v}\simeq 10\,\mathrm{kpc} \left(\frac{10^{-21}\mathrm{eV}}{\mu}\right)\left(\frac{120\,\mathrm{km}/\mathrm{s}}{v} \right)\,, 
\end{equation}
where $v$ is the velocity dispersion in a galactic DM halo. As can be seen in eq.~\eqref{eqn:DBwavelength}, for ultralight masses, the field exhbitis a characteristic length scale on the order of typical galaxies ($\lesssim 10\,\mathrm{kpc}$). On larger scales, the field behaves like a pressureless fluid -- just like cold DM -- preserving the success of the $\Lambda$CDM model in describing the formation of large-scale structure~\cite{Hui:2021tkt}. On smaller scales instead, where standard cold DM models show tensions with observations~\cite{Weinberg:2013aya}, the wave-like nature of the field leads to distinct behaviour. In particular, fuzzy DM stabilises against collapse on scales smaller than $\lambda\ped{dB}$, smoothing out inhomogeneities and suppressing structure formation~\cite{Hu:2000ke}. This helps resolve typical small-scale issues faced by particle DM models, such as the missing satellite~\cite{Klypin:1999uc}, ``too big to fail''~\cite{Boylan-Kolchin:2011qkt}, or cusp-core problems~\cite{Flores:1994gz}. Although Lyman-$\alpha$ forest constraints have limited the viability of the lower mass window $\lesssim 10^{-22}\,\mathrm{eV}$~\cite{Irsic:2017yje, Rogers:2020ltq}, fuzzy or wave DM remains an attractive DM candidate due to its unique phenomenological features, such as the formation of solitons, as discussed in Section~\ref{BHenv_subsec:DMStruct}.
\vskip 2pt
While the focus so far has been on ultralight scalar fields, several extensions of the SM predict bosons with different spins. In Section~\ref{BHenv_subsec:candidates}, we introduced \emph{dark photons}, an example of vector bosons. Current constraints on dark photons primarily arise from their mixing with the electromagnetic sector, and are shown in Figure~\ref{fig:constraints}. Like with scalar bosons, the low-mass, weak-coupling regime is difficult to constrain.
\vskip 2pt
Throughout this thesis, we will explore scenarios in which these ultralight particles either couple to the electromagnetic sector or interact purely gravitationally. In both cases, large quantities of these particles are necessary for any observable effect to manifest itself. The next section will discuss how spinning BHs facilitate that process.
\subsection{Black Hole Superradiance}\label{BHenv_subsec:superrad}
Ultralight bosons are of particular interest in BH physics due to \emph{superradiance}, a kinematic phenomenon that arises across various areas of physics (see~\cite{PhysRevD.58.064014,Brito:2015oca} for reviews). The concept was first introduced by Dicke in 1954 in the context of coherent emission in quantum optics~\cite{PhysRev.93.99}. Two decades later, in 1971, Zel'dovich showed that rotating bodies can amplify incident waves through a dissipative process. Specifically, when radiation scatters off a rotating cylinder with a conductive surface, the incoming wave can extract energy and angular momentum from the cylinder -- provided it is spinning fast enough -- resulting in a reflected wave with a larger amplitude~\cite{ZelDovich1971,ZelDovich1972}. This phenomenon is what we refer to as ``rotational superradiance''. More recently, rotational superradiance has even been demonstrated experimentally by scattering water waves off vortex flows in draining water tanks~\cite{2017NatPh..13..833T}.
\vskip 2pt
Remarkably, rotational superradiance is also triggered by spinning BHs~\cite{Starobinsky:1973aij,Bekenstein:1973mi,Starobinskil:1974nkd}. In this case, the dissipative mechanism is provided by the event horizon, which facilitates the transfer of energy and angular momentum. One of the powerful aspects of BH superradiance is its purely gravitational nature:~any bosonic field satisfying the appropriate conditions can undergo superradiant amplification. Even quantum vacuum fluctuations can spontaneously trigger superradiance~\cite{Unruh:1974bw}. 
\vskip 2pt
Energy extraction from BHs relies on the presence of an \emph{ergosphere} (see Figure~\ref{fig:ergoregion}). Within this region, objects are forced to co-rotate with the BH, yet still retain the ability to escape to infinity. Crucially, the time Killing vector $K^{\mu}_{(t)}$ becomes spacelike inside the ergosphere, allowing particles to attain negative energy states relative to an observer at infinity, i.e., $E = -K^{\mu}_{(t)}p_{\mu} < 0$, where $p^{\mu}$ is the four-momentum of the particle. This observation led Penrose to propose the following \emph{gedankenexperiment}~\cite{Penrose:1971uk}:~consider a particle falling in from infinity with energy $E$, which then splits into two particles, $A$ and $B$, within the ergosphere, carrying energies $E_{A}$ and $E_{B}$, respectively. If particle $A$ acquires negative energy ($E_{A}<0$), it must fall into the BH, while particle $B$ can still escape. By energy conservation, the escaping particle must satisfy $E_B = E-E_A>E$, meaning it emerges with more energy than the original infalling particle. In this way, the particle extracts energy from the BH itself! While the Penrose process is unlikely to play a major role in astrophysics, it illustrates the remarkable fact that energy extraction from rotating BHs is indeed possible. This principle has been instrumental in explaining relativistic jets powered by BHs through e.g., the Blandford-Znajek process~\cite{Blandford:1977ds}, magnetohydrodynamic Penrose process~\cite{1992ApJ...386..455H}, or modified Hawking radiation~\cite{1999Sci...284..115V}. 
\vskip 2pt
The wave analogue of the Penrose process is known as \emph{black hole superradiance}.\footnote{Black hole superradiance is an \emph{analogue} rather than an exact counterpart of the Penrose process. While the Penrose process demonstrates the possibility of energy extraction, superradiance involves a field whose wavelength is larger than the size of the BH, meaning that it does not directly ``see'' the ergoregion.} It naturally emerges in the context of scalar fields in a Kerr background~\eqref{eq:metric_kerr}. To investigate this in more detail, we consider a massive, real scalar field $\Phi$ with mass $\mu$, minimally coupled to gravity. It satisfies the Klein-Gordon equation:
\begin{equation}\label{eq:env_KG_eqn}
\left(g^{\alpha\beta}\nabla_{\alpha}\nabla_{\beta} - \mu^2\right)\Phi = 0\,.
\end{equation}
In Boyer-Lindquist coordinates, this equation is separable~\cite{Carter:1968,Brill:1972xj} using the \emph{ansatz}
\begin{equation}\label{eq:scala_ansatz}
\Phi = R(r)S(\theta)e^{im\varphi}e^{-i\omega t}\,,
\end{equation}
which yields decoupled ordinary differential equations for the radial $R$ and angular $S$ parts~\cite{Teukolsky:1973ha,Teukolsky:1974yv}. Transforming to the tortoise coordinate $r_*$ (which maps the event horizon at $r=r_+$ to $r_*\to-\infty$), the radial equation takes on a Schr\"{o}dinger-like form:
\begin{equation}\label{eq:env_Schro_eqn}
\frac{\dd^2R}{\dd r_*^2}+V\ped{eff}(r_*)R=0\,,
\end{equation}
where $V\ped{eff}$ is an effective potential that depends on the specific scenario under consideration.
\vskip 2pt
To find the condition under which superradiance occurs, we consider a scattering experiment involving monochromatic waves with frequency $\omega$ and azimuthal angular momentum $m$~\cite{Brito:2015oca}. The asymptotic solutions to eq.~\eqref{eq:env_Schro_eqn} take the form:
\begin{equation}\label{eqn:scattering_exp}
R(r_*)\sim\begin{dcases}
\mathcal Ie^{-ik_\infty r_*}+\mathcal Re^{ik_\infty r_*} & r_*\to\infty\,,\\
\mathcal Te^{-ik\ped{H}r_*}+\mathcal Oe^{ik\ped{H}r_*} & r_*\to-\infty\,,
\end{dcases}
\end{equation}
where $k^2\ped{H} = V\ped{eff}(r_*\to-\infty)$ and $k^2_{\infty} = V\ped{eff}(r_*\to\infty)$. Here, $\mathcal{I}$ represents an incoming wave from infinity, which is partially reflected ($\mathcal{R}$) and partially transmitted ($\mathcal{T}$). The outgoing flux ($\mathcal{O}$) vanishes due to purely ingoing boundary conditions at the BH horizon. Using that the Wronskian $(\dd R/\!\dd r_*)R^*-(\dd R^*/\!\dd r_*)R$ is independent of $r_*$, eq.~\eqref{eqn:scattering_exp} gives:
\begin{equation}\label{eq:condition_ampl}
\abs{\mathcal R}^2=\abs{\mathcal I}^2-\frac{k\ped{H}}{k_\infty}\abs{\mathcal T}^2\,.
\end{equation}
Consequently, when $k\ped{H}/k_\infty <0$, it follows that $\abs{\mathcal R}^2 > \abs{\mathcal I}^2$, meaning that the reflected wave has a larger amplitude than the incoming one. In other words, these waves have undergone \emph{superradiant amplification}.
\vskip 2pt
To identify the conditions under which $k\ped{H}/k_\infty <0$, it is useful to evaluate the radial Teukolsky equation~\eqref{eq:env_Schro_eqn} near the horizon $r=r_+$, where it simplifies to 
\begin{equation}\label{eq:radial_T_hor}
\frac{\dd^2R}{\dd r_*^2}+(\omega-m\Omega\ped{H})^2R=0\,.
\end{equation}
Here, $\Omega\ped{H}$ is the angular velocity of the horizon~\eqref{eq:OmegaH}. On the other hand, since $k_\infty=\sqrt{\omega^2-\mu^2}>0$ as $r\rightarrow \infty$, it follows from~\eqref{eq:condition_ampl} and~\eqref{eq:radial_T_hor} that superradiance occurs whenever 
\begin{equation}\label{eq:env_SR_condition}
\frac{\omega}{m} < \Omega\ped{H}\,.
\end{equation}
That is, Kerr BHs amplify waves whose angular phase velocity is smaller than the angular velocity of the horizon. While the above reasoning applies to all bosonic fields, it does not extend to fermionic fields. This is because superradiant amplification relies on increasing the occupation number of a single quantum state, which is forbidden for fermions due to Pauli's exclusion principle.
\vskip 2pt
While the superradiance process is intriguing, its direct astrophysical impact seems modest at first glance. This is evident from the amplification factor, $\mathcal{Z} = |\mathcal{R}|^{2}/|\mathcal{I}|^2-1$. For a massless scalar field scattering off a near-extremal Kerr BH, $\mathcal{Z}$ peaks at only 0.4\% (see, e.g., Figure~14 of~\cite{Brito:2015oca}), and is even smaller for massive fields. This suggests that superradiance, by itself, is not particularly efficient. However, Press and Teukolsky~\cite{Press:1972zz} first pointed out that if the amplified field were reflected back, the process could repeat, enabling sustained energy extraction from the BH. While they envisioned a futuristic giant mirror, Nature itself provides a mechanism for such reflection:~if the scalar field has a mass, it becomes gravitationally bound to the BH, forming a natural confinement. This leads to the requirement that the length scale of the field, its Compton wavelength $\lambda\ped{c}$, should be comparable to the length scale of the BH, its gravitational radius $r\ped{g}$. This is usually expressed in terms of $\alpha$, the \emph{gravitational fine-structure constant}:
\begin{equation}\label{eq:gravitational_fineS}
\alpha \equiv \frac{r\ped{g}}{\lambda\ped{c}} = M\mu\simeq 0.2\left(\frac{M}{10^6M_{\odot}}\right)\left(\frac{\mu}{3\times10^{-17}\,\mathrm{eV}}\right)\,.
\end{equation}
When $\alpha$ is of order unity, superradiance operates at peak efficiency and the boson field continuously extracts energy and angular momentum from the BH, allowing its occupation number to grow exponentially, with all particles accumulating in the same quantum state. The result is a macroscopic Bose-Einstein condensate around the BH. We will refer to this structure as a ``boson cloud'', and to the combined system of the BH and its surrounding cloud as a ``gravitational atom''. Its unique properties will be the focus of the next section.
\subsection{Gravitational Atoms}\label{BHenv_subsec:gatoms}
Superradiance shuts off when the bosonic field has spun down the BH sufficiently, such that the condition~\eqref{eq:env_SR_condition} is no longer satisfied. At that point, a quasi-stationary cloud of ultralight bosons surrounding the BH has formed. This BH-cloud system resembles a hydrogen atom and is therefore often called a \emph{gravitational atom}. Since these systems appear frequently throughout this thesis, we introduce their properties in detail here. Most of their qualitative features can be obtained by working in the non-relativistic regime, which we will adopt in this section. However, in some cases, such as those discussed in Chapters~\ref{chap:SR_Axionic} and~\ref{chap:inspirals_selfforce}, a relativistic treatment will be necessary. The relevant details are provided in Appendix~\ref{appNR_sec:freescalars}.
\subsubsection{Schr\"{o}dinger equation}
To study gravitational atoms, we return to the Klein-Gordon equation~\eqref{eq:env_KG_eqn} in a Kerr background~\eqref{eq:metric_kerr}. When $\alpha \ll 1$, the boson cloud is localised far away from the BH, allowing us to work in the non-relativistic regime. In this case, it is convenient to adopt the \emph{ansatz}~\cite{Arvanitaki:2010sy},
\begin{equation}\label{eqn:ansatzwf}
\Phi(t, \vec{r})=\frac{1}{\sqrt{2 \mu}}\left[\psi(t, \vec{r}) e^{-i \mu t}+\psi^*(t, \vec{r}) e^{i \mu t}\right]\,,
\end{equation}
where $\psi$ is a complex scalar field that evolves on timescales much longer than $\mu^{-1}$, referred to as the \emph{non-relativistic field}. As a result, we can neglect the second derivative with respect to time, i.e., $|\partial^2_t \psi| \ll \mu |\partial_t \psi|$, and extract the slowly varying component by substituting~\eqref{eqn:ansatzwf} into the Klein-Gordon equation~\eqref{eq:env_KG_eqn}. Expanding in powers of $\alpha$, the field $\psi$ satisfies a Schr\"{o}dinger equation:
\begin{equation}\label{eq:env_Schrodinger}
i\frac{\partial \psi}{\partial t} =\left(-\frac{1}{2\mu}\nabla^2-\frac{\alpha}{r} +\cdots\right)\psi\,.
\end{equation}
The dominant term appearing in~\eqref{eq:env_Schrodinger} is a Coulomb-like potential, while subleading terms feature higher powers of $\alpha$ and $1/r$.
\subsubsection{Bound states}
To leading order, eq.~\eqref{eq:env_Schrodinger} is mathematically identical to the Schr\"{o}dinger equation for the hydrogen atom. Therefore, its bound state solutions are the familiar hydrogenic eigenfunctions, characterised by three integers:~the principal quantum number $n$, orbital angular momentum number $\ell$, and azimuthal angular momentum number $m$, with $n > \ell $, $\ell \geq 0$, and $\ell \geq |m|$. These solutions are given by
\begin{equation}\label{eqn:BHenv_eigenstates}
\psi_{n \ell m}(t, \vec{r})=R_{n \ell}(r) Y_{\ell m}(\theta, \varphi) e^{-i\left(\omega_{n \ell m}-\mu\right) t}\,,
\end{equation}
where $Y_{\ell m}$ denotes the spherical harmonics and $R_{n \ell}$ are the hydrogenic radial functions:
\begin{equation}\label{eqn:HydrogenicRadial}
R_{n \ell}(r)=\sqrt{\left(\frac{2 \mu \alpha}{n}\right)^3 \frac{(n-\ell-1) !}{2 n(n+\ell) !}}\,\left(\frac{2 \alpha \mu r}{n}\right)^{\ell} \exp \left(-\frac{\mu \alpha r}{n}\right) L_{n-\ell-1}^{2 \ell+1}\left(\frac{2 \mu \alpha r}{n}\right)\,,
\end{equation}
where $L_{n-\ell-1}^{2 \ell+1}(x)$ is the associated Laguerre polynomial. The radial profile peaks around $r \sim n^{2}r\ped{c}$, where $r\ped{c} \equiv(\mu \alpha)^{-1}$ is the Bohr radius, and decays exponentially as $r \rightarrow\infty$.
\vskip 2pt
Using a quantum mechanics analogy, the wavefunction~\eqref{eqn:BHenv_eigenstates} can be written in bra-ket notation, i.e., $\ket{n\ell m}$, with the normalisation $\braket{n\ell m|n'\ell'm'} = \delta_{nn'}\delta_{\ell \ell'}\delta_{m m'}$. The mass density of the cloud with wavefunction $\psi(t, \vec{r})$ is given by $\rho(t,\vec{r}) = 2 M\ped{c}|\mathrm{Re}[\psi(t,\vec{r})]|^2$,\footnote{For a complex field, the mass density is $\rho(t,\vec{r}) = M\ped{c}|\psi(t,\vec{r})|^2$.} where $M\ped{c}$ is the total mass of the cloud. The resulting cloud densities can far exceed those typically found in other astrophysical environments, as shown in Figure~\ref{fig:taxonomy}. An illustration of a gravitational atom in its dominant $\ket{n\ell m}=\ket{211}$ state is depicted in Figure~\ref{fig:spectrum}. 
\vskip 2pt
The Schr\"{o}dinger equation~\eqref{eq:env_Schrodinger} also admits unbound (continuum) state solutions, characterised by a positive, real-valued wavenumber $k$. While these solutions are important for calculating effects like dynamical friction (see Appendix~\ref{app_BHPT_Kerr:scalar_Newt} and~\cite{Baumann:2021fkf}), they are not relevant for the current discussion.
\subsubsection{Spectrum}
\begin{figure}[t!]
\centering
\includegraphics[scale=0.85]{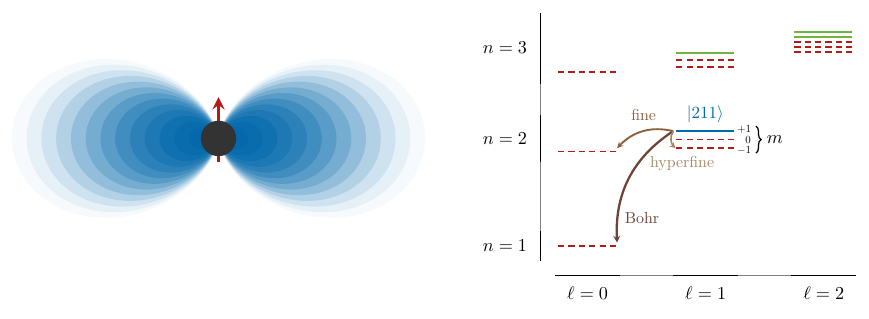}
\caption{Schematic illustration of the gravitational atom in its dipolar $\ket{211}$ state (\emph{left}) and the eigenfrequency spectrum of the cloud for first few states (\emph{right}). Dashed lines [{\color{cornellRed}red}] indicate decaying modes, while solid lines [{\color{cornellGreen}green}] correspond to the growing modes that are naturally populated via superradiance. The $\ket{211}$ state is highlighted in [{\color{cornellBlue}blue}] to match the visual representation on the left. For each combination of $n$ and $\ell$, there exists a multiplet of states spanning $-\ell \leq m \leq \ell$, with energies determined by eq.~\eqref{eq:BHenv_eigenenergy}. The energy splittings between different modes are indicated by brown lines in a few example cases.}
\label{fig:spectrum}
\end{figure}
Although the hydrogen and gravitational atom share qualitative similarities, an important difference is that gravitational atoms exist in the presence of an event horizon, which modifies the boundary conditions. Specifically, rather than being regular at the origin as in the hydrogen case, the bosonic field must satisfy purely ingoing behaviour at the horizon and decay exponentially at infinity. These conditions can only be satisfied by a discrete and complex set of eigenfrequencies for the eigenstates of the boson cloud,
\begin{equation}\label{eq:boson_eigenfreq}
    \omega_{n \ell m} = (\omega_{n\ell m})\ped{R}+i(\omega_{n\ell m})\ped{I}\,,
\end{equation}
which is why they are referred to as \emph{quasi-bound} states. Here, the subscripts $\mathrm{R}$ and $\mathrm{I}$ denote the real and imaginary parts of $\omega_{n\ell m}$, respectively, and we assume $(\omega_{n\ell m})\ped{R}>0$ without loss of generality. The real part determines the energy levels of the bound states, i.e., $\epsilon_{n\ell m}\equiv(\omega_{n\ell m})\ped{R}$, which to leading order follow the hydrogenic spectrum~\cite{Baumann:2019eav}:
\begin{equation}\label{eq:BHenv_eigenenergy}
\epsilon_{n\ell m}=\mu\left(1-\frac{\alpha^{2}}{2 n^{2}}-\frac{\alpha^{4}}{8 n^{4}}-\frac{(3 n-2 \ell-1) \alpha^{4}}{n^{4}(\ell+1 / 2)}+\frac{2 \tilde{a} m \alpha^{5}}{n^{3} \ell(\ell+1 / 2)(\ell+1)}+\mathcal{O}(\alpha^{6})\right)\,.
\end{equation}
Different types of energy splittings between two states are commonly referred to as \emph{Bohr} ($\Delta n \neq 0$), \emph{fine} ($\Delta n = 0$, $\Delta \ell \neq 0$), or \emph{hyperfine} ($\Delta n = 0$, $\Delta \ell = 0$, $\Delta m \neq 0$). The imaginary part in~\eqref{eq:boson_eigenfreq} determines whether a certain mode grows or decays. Using a matched asymptotic expansion, the growth (or decay) rate, $\Gamma_{n\ell m}\equiv(\omega_{n\ell m})\ped{I}$, is given by~\cite{Detweiler:1980uk,Baumann:2019eav}
\begin{equation}\label{eqn:Gamma_nlm}
\Gamma_{n\ell m}=\frac{2r_+}MC_{n\ell m}(m\Omega\ped{H}-\omega_{n\ell m})\alpha^{4\ell+5}\,,
\end{equation}
where
\begin{align}
C_{n\ell m}&=\frac{2^{4\ell+1}(n+\ell)!}{n^{2\ell+4}(n-\ell-1)!}\left(\frac{\ell!}{(2\ell)!(2\ell+1)!}\right)^2\prod_{j=1}^\ell\left[j^2(1-\tilde a^2)+(\tilde{a} m-2 r_+ \omega_{n\ell m})^2\right]\,.
\end{align}
Equation~\eqref{eqn:Gamma_nlm} clearly illustrates the superradiance condition~\eqref{eq:env_SR_condition}:~when it is satisfied, eq.~\eqref{eqn:Gamma_nlm} is positive, causing the scalar mode to grow;~otherwise, it is negative and no amplification occurs. An illustration of the bound state spectrum is shown in Figure~\ref{fig:spectrum}.
\subsubsection{Formation and saturation}
In the relevant superradiance regime ($\alpha < 1$), the growth timescale~\eqref{eqn:Gamma_nlm} dramatically increases with higher $\ell$ modes. Therefore, the fastest-growing superradiant modes generally satisfy $m=\ell=n-1$. In a multiplet of states $\ket{n\ell m'}$ with $m'\le m$, this mode occupies the highest-energy state, as evident from~\eqref{eq:BHenv_eigenenergy}. The ``dominant'' growing mode is thus the dipolar one $\ket{n\ell m} = \ket{211}$ (see Figure~\ref{fig:spectrum}), which we will assume for the remainder of this section.\footnote{The extension to higher modes such as $\ket{322}$ is trivial. However, due to the strong dependence of $\Gamma_{n\ell m}$ on $\ell$~\eqref{eqn:Gamma_nlm}, the growth timescales quickly exceeds astrophysical timescales.} For $\alpha \ll 1$, its growth timescale is given by
\begin{equation}\label{eqn:211growthtime}
\Gamma^{-1}\ped{c} \simeq \frac{10\,\text{yrs}}{\tilde{a}} \left( \frac{M}{10^6 M_\odot} \right) \left( \frac{0.2}{\alpha} \right)^{9}\,.
\end{equation}
Imposing the growth to take place on astrophysically relevant timescales $\sim \mathcal{O}(\mathrm{Gyr})$ sets a lower bound on $\alpha$, while the superradiance condition $m\Omega\ped{H} > \omega$ places an upper bound. This yields
\begin{equation}\label{eqn:211bounds}
0.03 \left(\frac{M}{10^6M_{\odot}}\right)^{1/9} \lesssim \alpha < 0.5\,.
\end{equation}
The viable range for $\alpha$ covers just one order of magnitude, making superradiance astrophysically relevant only for specific ``mass pairs'' of the boson and the BH. Since astrophysical BHs range in mass from $M \sim 10$ to $10^{10} M_{\odot}$, the corresponding boson masses fall within $\mu \sim 10^{-20}$ to $10^{-10}\,\mathrm{eV}$. In this way, superradiance turns BHs into astrophysical probes of \emph{ultralight bosons}.
\vskip 2pt
As superradiance is ongoing, energy and angular momentum is transferred from the BH to the cloud until the superradiance condition~\eqref{eq:env_SR_condition} is met. The BH spin at that point is
\begin{equation}\label{eq:spinatsat}
\tilde{a}\ped{s} = \frac{4 m (M\ped{s}\omega\ped{s})}{m^2 + 4(M\ped{s}\omega\ped{s})^2}\,,
\end{equation}
where quantities \emph{at saturation} are indicated with a subscript $\mathrm{s}$. Note that, at this point, the frequency $\omega\ped{s}$ is purely real. Conservation of mass and angular momentum determines the cloud mass at saturation. Taking $\omega\ped{s} \approx\mu$ in~\eqref{eq:env_SR_condition}, and a BH with initial mass $M\ped{i}$ and spin $\tilde{a}\ped{i}$, this yields~\cite{East:2017ovw,Herdeiro:2021znw,Hui:2022sri}:
\begin{equation}\label{eq:massatsat}
\frac{M\ped{c,s}}{M\ped{i}-M\ped{c,s}}= \frac{2 \tilde{a}\ped{i} \alpha\ped{i}-m\left(1-\sqrt{1-\frac{16\alpha\ped{i}^2(m-\tilde{a}\ped{i}\alpha\ped{i})^2}{m^4}}\right)}{2(m-\tilde{a}\ped{i}\alpha\ped{i})}\,.
\end{equation}
The maximal value occurs at $\alpha\approx0.24$ and $\tilde{a}\ped{i}=1$, yielding $M\ped{c,s}/(M\ped{i}-M\ped{c,s}) \approx 10.78\%$ and a final spin $\tilde{a}\ped{s} = 0.78$ (for $m = 1$).
\subsubsection{Gravitational-wave emission}
The description above applies to both real and complex scalar fields. However, their stability properties differ depending on the nature of the field. Specifically, when the scalar field $\Phi$ is real, it is time-dependent and non-axisymmetric, inducing a time-varying quadrupole moment.\footnote{When the scalar field is complex instead, the time dependence in the stress-energy cancels out, i.e., $T_{00}\sim\Phi^*\Phi\sim e^{i\mu t}e^{-i\mu t}$. Hence, they do not emit GWs and are a form of ``BH hair''.} As a consequence, the cloud loses mass by emitting GWs as~\cite{Arvanitaki:2010sy,Yoshino:2013ofa,Brito:2017zvb}
\begin{equation}\label{eqn:GWemission}
M\ped{c}(t) = \frac{M\ped{c,i}}{1 + t/\tau\ped{c}}\,, 
\end{equation}
where $M\ped{c,i}$ is the initial mass of the cloud and $\tau\ped{c}$ its \emph{lifetime}. The inverse of $\tau\ped{c}$ can be interpreted as the ``GW power'', given by
\begin{equation}\label{eqn:GWlifetime}
\tau\ped{c}^{-1} = G_{n \ell} \frac{M\ped{c,i}}{M^2} \alpha^{4\ell+10}\,,
\end{equation}
where the coefficient $G_{n \ell}$ can be found in~\cite{Yoshino:2013ofa}. For $\alpha\gtrsim 0.1$, nonlinear effects suppress the emission power, making eq.~\eqref{eqn:GWlifetime} an upper bound~\cite{Yoshino:2013ofa,Brito:2017zvb}. Comparing the $\alpha$-dependence in~\eqref{eqn:GWlifetime} with~\eqref{eqn:Gamma_nlm}, it is clear that the cloud always forms faster than it depletes through GW emission. Moreover, the timescale $\tau\ped{c}$ of a given mode scales similarly to the instability rate of the next superradiant state, i.e., $(n,\ell,m) \to (n+1,\ell+1,m+1)$, ensuring that by the time the cloud depletes its initial state (say, $\ket{211}$), a new state has had sufficient time to grow (say, $\ket{322}$). For all practical purposes, it is thus a good approximation to assume the cloud occupies a single state. For the dominant growing mode $\ket{211}$, eq.~\eqref{eqn:GWlifetime} gives
\begin{equation}\label{eqn:211lifetime}
\tau\ped{c} \simeq \SI{6e+6}{yrs}\,\left(\frac{M}{10^6M_\odot}\right)\left(\frac{0.1}{M\ped{c}/M}\right)\left(\frac{0.2}{\alpha}\right)^{14}\,,
\end{equation}
which is indeed much longer than its growth timescale~\eqref{eqn:211growthtime} and demonstrates that the cloud can persist over astrophysical timescales. While self-interactions could affect this conclusion~\cite{Baryakhtar:2020gao}, they will be neglected throughout this thesis. The GW emission from the cloud is nearly monochromatic and continuous, with a frequency set by the scalar field mass,
\begin{equation}\label{eq:mono_freq}
    f\ped{c} \simeq \frac{\mu}{\pi} = 483\, \mathrm{Hz} \left(\frac{\mu}{10^{-12}\, \mathrm{eV}}\right)\,.
\end{equation}
These provide a direct observational probe of the boson cloud, a point we revisit in Section~\ref{BHenv_subsec:isol}.
\vskip 2pt
As has hopefully become clear, gravitational atoms are unique objects with a rich spectrum and high densities. Their distinctive properties have motivated various searches, through the GWs they emit or the impact on the spin of the host BH. Such \emph{isolated} configurations are explored in Section~\ref{BHenv_subsec:isol} and form the focus of Chapter~\ref{chap:SR_Axionic}. A promising new direction involves gravitational atoms in binary systems, where they can leave detectable imprints on the orbital dynamics and resulting waveforms~\cite{Chia:2020dye,Tomaselli:2024faa}. This scenario is examined in Section~\ref{BHenv_subsec:binary} and is a key theme of this thesis (Chapters~\ref{chap:legacy} and~\ref{chap:inspirals_selfforce}).
\section{Plasma}\label{BHenv_sec:Plasma}
While ultralight bosons are an intriguing area of study, their existence is speculative. Plasma, on the other hand, is a well-established component of the Universe, making up roughly 99\% of all visible matter. Stars, the interstellar medium, and astrophysical jets are just a few examples of plasma-rich environments. Despite their prevalence in astrophysics, they have received comparatively little attention in GW astronomy and BH physics, especially regarding their interactions with fundamental fields. This section introduces key properties of plasmas (Section~\ref{BHenv_subsec:plasmafeatures}) before exploring the propagation of electromagnetic waves through plasma (Section~\ref{BHenv_subsec:EM_waves}).
\vskip 2pt
Plasma consists of ionised gas, where electrons are stripped from atoms, leaving behind positively charged nuclei, or \emph{ions}. However, not all ionised gas qualifies as plasma;~this is only the case if the gas satisfies the following conditions~\cite{Chen_plasma,Fitz_plasma}
\begin{enumerate}[label=(\roman*)]
\item The number of charged particles is large enough for long-range Coulomb interactions ($\propto 1/r^2$) to dominate their dynamics, while remaining low enough for short-range interactions between individual particles to be negligible.
\item The gas is quasi-neutral:~electrons and ions are balanced in such a way that the overall charge density is approximately zero, although local variations may generate electric and magnetic fields.
\end{enumerate}
Because of these properties, plasma exhibits \emph{collective behaviour}, meaning that the charged particles interact over large scales under the influence of long-range forces.
\subsection{Fundamental Properties}\label{BHenv_subsec:plasmafeatures}
To better understand this collective behaviour, we outline some key plasma properties, beginning with Debye shielding.
\subsubsection{Debye shielding}
Consider a plasma in thermal equilibrium, where the number densities of electrons and ions are approximately balanced, $n\ped{e} \simeq Zn\ped{ion}$ with $Z=1$ for a hydrogen plasma. In this state, the electrostatic potential $\phi$ vanishes. Introducing a test charge disturbs the equilibrium and induces a nonzero electrostatic potential governed by Poisson’s equation:
\begin{equation}\label{eq:gradient_plasma}
\nabla^2\phi = -\frac{e}{\varepsilon_0} (n\ped{ion}-n\ped{e})\,,
\end{equation}
where $e$ is the elementary charge and $\varepsilon_0$ the vacuum permittivity. Assuming the electrons follow a Maxwell-Boltzmann distribution, their density takes the form
\begin{equation}
n\ped{e} = n\ped{ion}e^{e\phi/(k\ped{B}T\ped{e})}\,,
\end{equation}
where $k\ped{B}$ is the Boltzmann constant, and $T\ped{e}$ the electron temperature. Since ions are much heavier than electrons, their thermal velocities are smaller by a factor $v\ped{ion}/v\ped{e} =\sqrt{m\ped{e}/m\ped{ion}} \ll 1$, making them nearly stationary on electron timescales. Under these assumptions, solving eq.~\eqref{eq:gradient_plasma} yields:
\begin{equation}
\phi\ped{D} = \frac{1}{4\pi\varepsilon_0} \frac{e^{-r/\lambda\ped{D}}}{r}\,,
\end{equation}
where $\lambda\ped{D}$, the \emph{Debye length},\footnote{Historical footnote:~during the writing of this thesis, I discovered that Peter Debye was born in my hometown of Maastricht, making him our only Nobel Prize winner to date.} is defined as
\begin{equation}\label{eq:DebyeLength}
\lambda\ped{D} = \sqrt{\frac{\varepsilon_0 k\ped{B}T\ped{e}}{n\ped{e}e^2}} = 74\,\sqrt{\frac{T\ped{e}}{0.1\,\mathrm{eV}}\,\frac{10^{-3}\,\mathrm{cm}^{-3}}{n\ped{e}}}\,\text{m}\,,
\end{equation}
where we normalised to typical values for the temperature and density of the interstellar medium~\cite{Saintonge:2022tfq}. The Debye length thus characterises the scale over which a test charge influences the plasma. At distances larger than $\lambda\ped{D}$, a point charge is said to be \emph{shielded}, leaving the bulk plasma free from any significant electric fields as $n\ped{e} \simeq n\ped{ion}$~\eqref{eq:gradient_plasma} [satisfying condition (ii) from earlier]. At this scale, the collective plasma behaviour dominates over individual particle interactions [condition (i)]. It is important to emphasise that this shielding is not a static condition but rather a \emph{dynamic response} of the plasma. On smaller scales (sub-Debye), the small imbalance of charges may still generate potentials on the order of $k\ped{B}T\ped{e}/e$, allowing electromagnetic forces to play a role while maintaining quasi-neutrality. 
\subsubsection{Plasma frequency}
It is insightful to explore this active response from the plasma in more detail. When electrons in a plasma are displaced from equilibrium, the resulting charge separation generates an electric field that pulls them back towards their original position. Due to their inertia, the electrons overshoot, leading to oscillations at a characteristic frequency known as the \emph{plasma frequency}. As these oscillations occur on timescales far shorter than the ion response time, the ions can be treated as a stationary background. For a small displacement $\delta x$ from the equilibrium position $x_0$, the induced electric field is $E = -e n\ped{e} \delta x/\varepsilon_0$, and applying Newton’s second law gives
\begin{equation}
m\ped{e}\frac{\dd^2\delta x}{\dd t^2} = -eE = \frac{e^2n\ped{e}\delta x}{\varepsilon_0}\,.
\end{equation}
This describes harmonic motion with a characteristic frequency:
\begin{equation}\label{eq:BHenv_plasmafreq}
\begin{aligned}
&\frac{\mathrm{d}^2\delta x}{\mathrm{d}t^2} + \omega\ped{p}^2\,\delta x= 0\,, \: \text{where}\\
&\omega\ped{p} = \sqrt{\frac{n\ped{e} e^2}{\varepsilon_0m\ped{e}}} \simeq 1.8\times 10^{-5}\,\text{s}^{-1}\sqrt{\frac{n\ped{e}}{10^{-3}\text{cm}^{-3}}}\simeq 10^{-12}\,\text{eV}\sqrt{\frac{n\ped{e}}{10^{-3}\text{cm}^{-3}}}\,,
\end{aligned}
\end{equation}
is the plasma frequency. Notably, as we will see later, these oscillations have zero group velocity, meaning that they do not propagate.
\subsubsection{Modelling choices}
Plasmas are some of the most complex astrophysical systems, and there are a number of possible modelling approaches depending on the length and time scales of interest. Moreover, systems can be described with different degrees of approximation based on the specific situation and the available computational power.
\vskip 2pt
In principle, a plasma is a many-particle system whose full description requires tracking the position and velocity of every particle. However, this is computationally intractable for most practical applications. A more feasible approach is to adopt a statistical framework, which marginalises over microscopic degrees of freedom. This leads to \emph{kinetic theory}, where the Boltzmann equation governs plasma dynamics. Kinetic theory is particularly powerful for capturing short-timescale phenomena, such as plasma oscillations at a frequency $\omega\ped{p}$~\eqref{eq:BHenv_plasmafreq}. However, solving the Boltzmann equation numerically often requires particle-in-cell methods~\cite{BirdsallLangdon,Arber_2015}, which become computationally prohibitive for large-scale or long-timescale problems.
\vskip 2pt
A widely used alternative in astrophysics is \emph{magnetohydrodynamics} (MHD)~\cite{2012ASSL..391.....S,osti_5612158}, which treats the plasma as a single fluid under the assumption of quasi-neutrality, $n\ped{e} = Zn\ped{ion}$.  This restricts MHD to length scales much larger than the Debye length, $L\ped{MHD} \gg \lambda\ped{D}$, while the single-fluid treatment limits its validity to timescales much longer than the plasma oscillation period $\tau_{\scalebox{0.55}{MHD}} \gg \omega\ped{p}^{-1}$. As a result, MHD sacrifices important information about electron dynamics, particularly the plasma frequency, which is important for the scenarios considered in this thesis.
\vskip 2pt
To incorporate short-timescale effects while maintaining a computationally feasible model, we adopt the \emph{two-fluid formalism}, where electrons and ions are treated as separate fluids. Each obeys its own continuity and momentum equations while interacting via the electromagnetic field. The governing equations can be derived from kinetic theory~\cite{1991JPlPh..45..135K,1995ppai.book.....D} and will be shown in the next section [see eqs.~\eqref{eq:total_plasma_eq}].
\subsection{Electromagnetic Wave Propagation}\label{BHenv_subsec:EM_waves}
Understanding the propagation of electromagnetic waves in plasma is essential for studying plasma-rich environments near BHs. Many key features can be captured by examining a simple plane-wave scenario in a cold, collisionless plasma using the two-fluid formalism. Denoting equilibrium quantities with a subscript $\mathrm{b}$ and assuming the plasma is initially field-free ($E\ped{b} = B\ped{b} = 0$), we can write the electric field ($E$), magnetic field ($B$), number density ($n$) and velocity ($v$) as
\begin{equation}
E = \bar{E} e^{-i\omega t}\,, \quad B = \bar{B} e^{-i\omega t}\,, \quad  n = n\ped{b} + \bar{n} e^{-i\omega t}\,, \quad v = \bar{v} e^{-i\omega t}\,,
\end{equation}
where we adopt a harmonic time dependence and label perturbations with an overhead bar. Given their much larger inertia, ions are assumed to remain stationary compared to electrons. Under these assumptions, the first-order Maxwell equations in Lorentz-Heaviside units reduce to:
\begin{equation}
\begin{aligned}
\nabla \times \bar{B} &= -i\omega \bar{E} + en\ped{b}\bar{v}\,,\\
\nabla \times \bar{E} &= i\omega \bar{B}\,,\\
i\omega \bar{v} &= -\frac{e}{m\ped{e}}\bar{E}\,.
\end{aligned}
\end{equation}
Assuming wave propagation along the $\hat{z}$--direction, these equations simplify in the frequency domain to the form:
\begin{equation}\label{eq:dispersion_mass}
\begin{pmatrix}
\omega^2-\omega\ped{p}^2 - k^2& 0 & 0 \\
0 & \omega^2-\omega\ped{p}^2 - k^2& 0  \\
0 & 0 & \omega^2-\omega\ped{p}^2 
\end{pmatrix} 
\begin{pmatrix}
\bar{E}_x\\
\bar{E}_y\\
\bar{E}_z
\end{pmatrix} = 0\,.
\end{equation}
The three independent solutions correspond to different wave behaviours. The first two are transverse electromagnetic waves, with oscillations confined to the $x-y$ plane. Their dispersion relation follows $\omega^2 = k^2 +\omega\ped{p}^2$, introducing a frequency cutoff:~waves with $\omega < \omega\ped{p}$ are unable to propagate and are reflected, making the plasma \emph{overdense}. In contrast, waves with $\omega > \omega\ped{p}$ propagate freely. This cutoff behaves similarly to a massive field, with the plasma frequency $\omega\ped{p}$ acting as an effective mass. The third solution corresponds to a longitudinal electrostatic mode, characterised by $\omega^2 = \omega\ped{p}^2$. As advertised, this mode has zero group velocity $v\ped{gr} = \partial\omega / \partial k = 0$, meaning it does not propagate. This contrasts Proca theory, where the longitudinal mode shares the same dispersion relation as the transverse ones, allowing it to propagate. While Proca-like theories often serve as convenient approximations, they may thus not always be reliable in realistic plasma scenarios.
\subsubsection{Curved space}
To account for plasma effects in curved spacetime without resorting to full numerical simulations, we adopt the model from~\cite{Cannizzaro:2020uap,Cannizzaro:2021zbp}, based on~\cite{1981AA....96..293B}. This framework describes a cold, non-relativistic, collisionless electron-ion plasma. Extensions to include collisions or thermal effects~\cite{Cannizzaro:2021zbp} are generally unnecessary for the astrophysical scenarios explored in this thesis. For instance, accretion disks can be treated as cold plasmas~\cite{1973A&A....24..337S,Novikov:1973kta}, and electron thermal velocities are typically much smaller than the phase velocity of electromagnetic waves, $v\ped{e} \equiv \sqrt{2k\ped{B}T\ped{e}/m\ped{e}}\ll \omega/k$.
\vskip 2pt
We model the plasma as a two-species system composed of electrons and ions. The electrons are characterised by a number density $n\ped{e}$ and four-velocity $u^{\mu}\ped{e}$, while the ions contribute a current $j^{\mu}\ped{ion}$. The evolution of this system in curved spacetime is governed by the following set of equations:
\begin{equation}\label{eq:total_plasma_eq}
\begin{aligned}
\nabla_\nu F^{\mu \nu} &= en\ped{e}u\ped{e}^{\mu} + j^{\mu}\ped{ion}\,,\\
u\ped{e}^{\nu}\nabla_{\nu}u\ped{e}^{\mu} &= \frac{e}{m\ped{e}}F^{\mu}_{\ \nu}u^{\nu}\ped{e}\,,\\ 
\nabla_\mu (n\ped{e} u\ped{e}^\mu)&=0\,,\\
u^{\mu}\ped{e}u_{\mathrm{e}, \mu} &= -1\,.
\end{aligned}
\end{equation}
These represent Maxwell’s equations in presence of sources, the electron momentum equation, the continuity equation, and the normalisation of the electron’s four-velocity, in covariant form. Similar to before, we can study the linearised dynamics by introducing small perturbations $\bar{n}$, $\bar{u}^{\mu}$, $\bar{F}^{\mu\nu}$ (and likewise for other quantities) and neglect second-order terms, metric perturbations (as gravitational backreaction is negligible), and ion perturbations, which are suppressed by a factor $\propto m\ped{e}/m\ped{ion} \ll 1$. The effective metric tensor can then be introduced as
\begin{equation}
\gamma_{\mu\nu} = g_{\mu\nu}+u_{\mu}u_{\nu}\,,
\end{equation}
which projects vectors and tensors onto hypersurfaces orthogonal to the electron four-velocity. From the momentum and Maxwell’s equations, it is possible to derive an evolution equation for the perturbed electromagnetic tensor $\bar{F}_{\mu\nu}$, incorporating the effects of gravity and plasma motion~\cite{1981AA....96..293B}:
\begin{equation}\label{eq:full_equation_Breuer}
\gamma^{\alpha}_{\beta}u^{\delta}\nabla_{\delta}\nabla_{\gamma}\bar{F}^{\beta\gamma}-\omega\ped{p}^2\bar{F}^{\alpha\beta}u_{\beta}+(\omega^{\alpha}_{\beta} +\omega_{\mathrm{L}, \beta}^{\alpha}+\theta^{\alpha}_{\beta}+\theta^{\mu}_{\mu} \gamma^{\alpha}_{\beta}+\frac{e}{m\ped{e}}E^{\alpha}u_{\beta})\nabla_{\gamma}\bar{F}^{\beta\gamma} = 0\,.
\end{equation}
Here, $\omega_{\mu\nu}$ and $\theta_{\mu\nu}$ are the vorticity and deformation tensors, respectively, $\omega_{\mathrm{L}, \mu\nu}$ is the Larmor tensor and $E^{\mu} = F^{\mu}_{\nu}u^{\nu}$ is the electric field in the fluid frame.
\vskip 2pt
Equation~\eqref{eq:full_equation_Breuer} is valid for any stationary background geometry, including a Kerr BH surrounded by a cold, collisionless plasma. As shown in~\cite{Cannizzaro:2020uap}, in flat spacetime it reproduces the expected dispersion relations:~longitudinal modes with $\omega = \omega\ped{p}$ and transverse modes with $\omega^2 = |k|^2+\omega\ped{p}^2$. The framework outlined above thus provides a robust and tractable approach for studying plasmas in curved spacetime, which we will use in Chapters~\ref{chap:SR_Axionic},~\ref{chap:in_medium_supp} and~\ref{chap:plasma_ringdown}.
\section{Accretion Disks and Active Galactic Nuclei}\label{BHenv_sec:accretion_disks_AGN}
Accretion disks and active galactic nuclei (AGNs) are arguably the most well-known astrophysical environments associated with BHs. In fact, they represent one of the few BH environments for which there is \emph{direct} observational evidence. These systems are complex, and highly relevant to GW astrophysics, especially for future detectors (see, e.g.,~\cite{Kocsis:2011dr,Cole:2022yzw,Speri:2022upm}). While a full treatment lies beyond the scope of this thesis, this section introduces the essential aspects of AGNs and accretion disks. We will revisit this topic in Chapter~\ref{chap:capture}.
\vskip 2pt
An AGN refers to the entire region surrounding a supermassive BH ($\geq 10^6 M_{\odot}$) at the centre of an active galaxy, where large amounts of gas are accreted and radiated away (see~\cite{1990agn..conf.....B,2013peag.book.....N} for reviews). The inner AGN consists of an \emph{accretion disk}, where viscosity, primarily driven by the magnetorotational instability~\cite{1991ApJ...376..214B}, transports angular momentum outwards. This allows gas to spiral deeper into the gravitational well, converting potential energy into heat (with temperatures reaching $\sim 10^5\,\mathrm{K}$) and radiation. This powers the AGN’s intense luminosity -- up to $\sim 10^{48}\,\mathrm{erg/s}$ -- producing emission across the entire electromagnetic spectrum and fuelling additional structures such as relativistic jets and outflows~\cite{2013peag.book.....N}. Foundational work on this accretion process was laid out by Zel’dovich~\cite{1964SPhD....9..195Z} and Salpeter~\cite{1964ApJ...140..796S}. 
\vskip 2pt
The disk itself can reach sub-parsec scales and is often surrounded by optically thick material that further interacts with the disk. In some systems, the outer disk may extend up to $1-10\,\mathrm{pc}$~\cite{2015ARAA..53..365N}, coupling with the surrounding interstellar medium. However, not all accretion disks form AGNs;~only those with sufficiently high accretion rates can sustain the intense activity that defines them.
\vskip 2pt
AGN disks are of interest in GW astronomy, not only due to dynamical effects in a binary inspiral~\cite{Kocsis:2011dr,Cole:2022yzw,Speri:2022upm}, but also because they are proposed as sites for the formation and merger of compact binaries~\cite{2012MNRAS.425..460M,2011MNRAS.417L.103M,Yang:2019cbr,Secunda:2018kar,Fabj:2020qqc,Tagawa:2019osr}. This is partly due to high escape velocities near supermassive BHs, which can trap stellar-mass BHs in the disk. Once embedded, these can interact with the dense gas, accelerating inspirals or even leading to hierarchical mergers~\cite{Gerosa:2021mno,Santini:2023ukl,Whitehead:2023hmh,Vaccaro:2023cwr}. Understanding these processes is crucial in modelling GW sources for current and future detectors. While the broader AGN structure is relevant for interpreting electromagnetic observations, in the context of GW astrophysics, the sub-parsec accretion disk is typically the main component of interest. Early theoretical models of AGN accretion disks relied on simplified, steady-state solutions, but computational advances have enabled more detailed studies incorporating radiative transfer, magnetic fields, and GR (e.g.,~\cite{1972ApJ...173..431W,1991MNRAS.250..581F,Schartmann:2005pe,Schartmann:2008qb,2009ApJ...702...63W,Husko:2022uwx}). Despite these developments, accretion disk physics remains highly uncertain, often requiring expensive numerical simulations for realistic modelling (see, e.g.,~\cite{Jiang:2019bxn}). For our purposes, a fully numerical approach is unnecessary, and we will rely on analytical models. An example of the density profile of an accretion disk for fiducial parameters is shown in Figure~\ref{fig:taxonomy}.
\subsubsection{Shakura-Sunyaev}
\begin{figure}[t!]
    \centering
    \includegraphics[scale = 1]{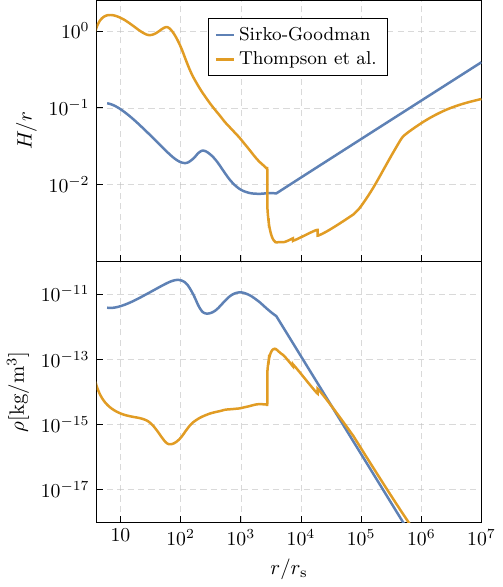}
    \caption{Aspect ratio (\emph{top panel}) and density profile (\emph{bottom panel}) for AGN disk models from Sirko-Goodman~\cite{Sirko:2002ex} and Thompson et al.~\cite{Thompson:2005mf}, assuming a central BH mass of $M = 10^7 M_{\odot}$, based on~\cite{Gangardt:2024bic}. The inner region of the disk is modelled using the standard $\alpha$-prescription of the Shakura-Sunyaev model. For the Sirko-Goodman case, we adopt $\alpha\ped{visc} = 0.01$, an Eddington ratio $f_{\rm Edd} = 0.5$, radiative efficiency $\eta = 0.1$ and hydrogen fraction $X = 0.7$. The Thompson et al.~model introduces additional parameters, including a supernova feedback parameter $\chi = 1$, an angular momentum efficiency $m = 2$ and a star formation radiative efficiency $\eta_{\rm star} = 0.001$.}
    \label{fig:densities_AGN}
\end{figure}

The Shakura-Sunyaev model provides a simplified yet powerful framework for describing accretion disks around BHs. It assumes a geometrically thin, optically thick disk that is radiatively efficient and in steady-state flow~\cite{1973A&A....24..337S}. Although initially formulated in the Newtonian regime, it was later extended to incorporate GR~\cite{Novikov:1973kta,Page:1974he,1995ApJ...450..508R,2012MNRAS.420..684P}. The model comes in the form of the well-known $\alpha$ and $\beta$-viscosity prescriptions, which continue to be widely used for understanding the inner regions of accretion disks where radiation pressure dominates~\cite{Abramowicz2013}.\footnote{The stability of these disks remains an open question~\cite{Jiang:2019bxn,1978ApJ...221..652P,1974ApJ...187L...1L,1976MNRAS.175..613S,1977A&A....59..111B}. In particular, $\beta$-disks have been proposed to avoid thermal instabilities that appear in the analytical solutions of $\alpha$-disks. Despite these instabilities, the $\alpha$-disk model is still the most widely used analytical approximation for turbulent accretion flows in the radiation-dominated regime.} 
\vskip 2pt
The key distinction between these viscosity prescriptions is how they parameterise the turbulent stresses that are responsible for angular momentum transport. The $\alpha$-disk model relates the viscous stress to the total pressure, incorporating both gas and radiation pressure $\propto \alpha\ped{visc}(p\ped{gas} + p\ped{rad})$~\cite{1973A&A....24..337S}. In contrast, the $\beta$-disk model assumes the stress depends only on the gas pressure $\propto \alpha\ped{visc} p\ped{gas}$~\cite{1981ApJ...247...19S}. Here, $\alpha\ped{visc}$ is the dimensionless viscosity parameter, representing the efficiency of angular momentum transport and thus encapsulates most of the complex magnetohydrodynamic processes. For AGN disks, observational and theoretical studies indicate $\alpha\ped{visc}$ lies in the range $0.001$ to $0.1$~\cite{2010ApJ...713...52D,2012A&A...545A.115K,Jiang:2019bxn,Nouri:2023nss}.
\vskip 2pt
The power of the Shakura-Sunyaev approach lies in its simplicity. By making a few physically motivated assumptions -- such as steady-state flow, radiative efficiency, and the phenomenological viscosity prescription -- it reduces the complex problem of accretion onto BHs to a tractable set of algebraic equations, while still providing a robust first-order approximation for a wide range of astrophysical accretion disks. Even if it does not capture all the nuances of MHD simulations~\cite{Jiang:2019bxn}, it provides valuable insights into the disk's structure and behaviour. For a comprehensive derivation and discussion, see~\cite{Abramowicz2013}.
\vskip 2pt
The disk's properties can be characterised through its \emph{surface density} $\Sigma$ and \emph{scale height} $H$, which depend on the accretion rate relative to the Eddington rate ($f\ped{Edd}$) and the radiative efficiency ($\eta$). For $\alpha$ and $\beta$-disks, the surface density is given by
\begin{equation}\label{eq:surface_density}
\begin{aligned}
\Sigma_\alpha \approx & \,5.4 \times 10^3\,\frac{\mathrm{kg}}{\mathrm{~m}^2}\left(\frac{0.1}{\alpha\ped{visc}}\right)\left(\frac{0.1}{f_{\mathrm{Edd}}}\frac{\eta}{0.1}\right)\left(\frac{r}{10 M}\right)^{3/2}\,, \\
\Sigma_\beta \approx & \,2.1 \times 10^7\,\frac{\mathrm{kg}}{\mathrm{~m}^2}\left(\frac{0.1}{\alpha\ped{visc}}\right)^{4 / 5}\left(\frac{0.1}{f_{\mathrm{Edd}}} \frac{\eta}{0.1}\right)^{-3 / 5} \left(\frac{M}{10^6 M_{\odot}}\right)^{1 / 5}
    \left(\frac{r}{10M}\right)^{-3/5}\,.
\end{aligned}
\end{equation}
The characteristic scale height of the disk, which sets its vertical extent, is expressed as
\begin{equation}\label{eq:scale_height}
H \approx 0.78\,r\ped{s}\left(\frac{f_{\mathrm{Edd}}}{0.1} \frac{0.1}{\eta}\right)\approx 2.3 \times 10^9\,\mathrm{m}\left(\frac{f_{\mathrm{Edd}}}{0.1} \frac{0.1}{\eta}\right)\left(\frac{M}{10^6 M_{\odot}}\right)\,.
\end{equation}
The thin-disk assumption requires that $H \ll r$, and using the scale height, the mid-plane density can be estimated as $\rho = \Sigma/(2H)$. To describe the vertical structure, a simple piecewise profile is usually adopted:
\begin{equation}\label{eq:verticalpiecewise}
     \rho(r,z) = 
     \begin{dcases}
        \rho(r) &  -H(r)/2\leq z\leq H(r)/2\,, \\
        0 &   {\rm otherwise}\,.\\
    \end{dcases}
\end{equation}
Alternatively, a smoother and more physically motivated profile used in the literature is a Gaussian distribution in the vertical direction:
\begin{equation}\label{eq:verticalgaussian}
    \rho(r,z) = \rho(r)\exp{\left(-\frac{z^2}{2H^2(r)}\right)}\,.
\end{equation}
\vskip 2pt
To account for the high luminosities observed in AGNs~\cite{1981ARA&A..19..137P}, several models have been developed that extend the Shakura-Sunyaev framework to larger radii. Notable examples include the Sirko-Goodman~\cite{Sirko:2002ex} and Thompson et al.~\cite{Thompson:2005mf} models, which offer more comprehensive treatments of AGN disks while maintaining analytical tractability. Figure~\ref{fig:densities_AGN} shows the disk height and density profiles for these two models, which we now explore in more detail.
\subsubsection{Sirko-Goodman}
At larger radii, the gravitational pull in the vertical direction can cause the disk to become unstable, leading to fragmentation and star formation. This process depletes the gas supply necessary to sustain accretion onto the BH. The Sirko-Goodman model resolves this by introducing an external heating mechanism that reduces the gas density in the outer regions, counteracting gravitational collapse~\cite{Sirko:2002ex}. While the exact source of this heating is not explicitly defined, it is likely linked to stellar feedback -- such as supernovae or radiation from young stars -- that injects energy into the disk and helps maintain thermal support.
\vskip 2pt
The model relies on the \emph{Toomre parameter}~\cite{1964ApJ...139.1217T} to characterise the stability of the disk:
\begin{equation}\label{eq:Toomre_p}
Q\ped{T} =\frac{\Omega\ped{vel}^2}{2\pi \rho}\,, 
\end{equation}
where $\Omega\ped{vel}$ is its angular velocity. When $Q\ped{T} < 1$, the disk is gravitationally unstable and prone to fragmentation. The Sirko-Goodman model enforces a marginally stable configuration in the outer regions by maintaining $Q\ped{T} \sim 1$, thus suppressing fragmentation without requiring detailed knowledge of the exact heating mechanism. The disk is therefore divided into two regimes:~(i) an inner region where $Q\ped{T} \gg 1$ due to high temperatures and angular frequencies, resembling the standard Shakura-Sunyaev disk;~(ii) an outer region, where $Q\ped{T} \sim 1$ and the disk is maintained near marginal stability. In this outer zone, the density follows $\rho \propto \Omega\ped{vel}^2 \propto r^{-3}$.
\subsubsection{Thompson et al.}
The Thompson et al.~model builds on the Sirko-Goodman framework by specifying the physical mechanism behind disk stabilisation and refining the treatment of angular momentum transport~\cite{Thompson:2005mf}. Rather than relying on unspecified external heating, vertical support in the outer disk is explicitly attributed to radiation pressure from newly formed stars. Additionally, it replaces local viscous stresses with large-scale gravitational torques as the primary driver of mass transport, and incorporates mass loss due to star formation, leading to a radially varying accretion rate, $\dot{M}(r)$.
\vskip 2pt
In this framework, the central BH is fed by material -- such as interstellar gas -- flowing inwards from beyond an outer radius at a constant rate. As star formation progresses, it gradually depletes this inflow, reducing $\dot{M}$ with decreasing radius. At the same time, radiation pressure from these young stars provides the thermal support needed to maintain the stability of the outer disk. Large-scale torques also modify the rotation profile of the disk, so that the angular velocity deviates from a purely Keplerian form:
\begin{equation}\label{eq:vel_non-kep}
\Omega\ped{vel}= \sqrt{\frac{M}{r^3} + \frac{2\sigma^2}{r^2}}\,,
\end{equation}
where $\sigma$ denotes the velocity dispersion. Disk stability is still described by the Toomre parameter~\eqref{eq:Toomre_p}. Similar to the Sirko-Goodman case, in the inner regions, where $Q\ped{T} \gg 1$, star formation is effectively suppressed. In contrast, the outer regions remain near marginal stability with $Q\ped{T} \sim 1$, where radiation pressure from stellar feedback plays a crucial role in preventing gravitational collapse.
\section{Black Holes as Cosmic Laboratories}\label{BHenv_sec:BH_lab}
Black holes are fascinating objects even in \emph{vacuum}, but, as we have seen throughout this chapter, they also host a wide range of astrophysical environments. This section provides an overview of the various methods and tools available to study such systems and highlights how BHs can act as natural laboratories for exploring new frontiers in physics, offering opportunities to test theories of gravity, quantum mechanics, and high-energy physics. Unlike traditional experiments, which are limited by technological and energy constraints, BHs naturally reside in regimes that are difficult to reproduce in particle accelerators or labs. Notably, the interaction between BHs and their environment is a two-way street:~while this thesis focuses on using BHs to probe their environments, these same environments have long been central to uncovering BH properties. Indeed, before the advent of GWs detectors, such indirect methods were the only way to explore BHs~\cite{Remillard:2006fc,Ghez:2008ms,2009ApJ...692.1075G}.
\vskip 2pt
We now consider how different environments may affect BH physics and give rise to observational signatures. Section~\ref{BHenv_subsec:isol} examines isolated systems, while Section~\ref{BHenv_subsec:binary} focuses on binary systems. 
\subsection{Isolated Systems}\label{BHenv_subsec:isol}
Since the first detection of GWs in 2015, much attention has centred on BH pairs in the dynamical regime. Nonetheless, even in isolation, BHs are powerful tools for probing strong gravity and testing fundamental physics.
\vskip 2pt
A recent breakthrough in observational techniques has been the high-resolution imaging of BHs by the Event Horizon Telescope~\cite{EventHorizonTelescope:2019dse,EventHorizonTelescope:2022wkp}, which opens a new channel for testing gravity in the strong-field regime~\cite{Ayzenberg:2023hfw}. The shape and polarisation structure of BH shadows can reveal deviations from the Kerr metric~\cite{EventHorizonTelescope:2021bee,EventHorizonTelescope:2022xqj}, potentially induced by DM distributions~\cite{Nampalliwar:2021tyz,Ezquiaga:2020dao,2013A&A...554A..36L,Profumo:2006im,Yang:2023tip,Liu:2023oab} or exotic forms of matter~\cite{Davoudiasl:2019nlo,Saha:2022hcd}. Similarly, precision measurements of stellar orbits around BHs -- such as those performed by the GRAVITY collaboration -- have been used to constrain DM profiles and search for ultralight fields~\cite{GRAVITY:2020gka,Zakharov:2007fj,Nampalliwar:2021tyz,Lacroix:2018zmg,Shen:2023kkm,2022NatSR..1215258C,GRAVITY:2023cjt,Zuriaga-Puig:2023imf,John:2023knt,Lechien:2023psa}.
\vskip 2pt
A particularly promising direction involves the interaction between BHs and new fundamental fields. As discussed in Section~\ref{BHenv_sec:boson_clouds}, ultralight bosonic fields can extract large amounts of energy from BHs, forming dense clouds around them. These clouds give rise to a variety of observational signatures. For instance, as shown in eqs.~\eqref{eqn:GWlifetime} and~\eqref{eq:mono_freq}, real scalar fields can emit monochromatic GWs that could be detectable through continuous wave searches~\cite{Arvanitaki:2010sy,Arvanitaki:2014wva,Yoshino:2014wwa,Arvanitaki:2016qwi,Baryakhtar:2017ngi,Brito:2017zvb,Brito:2017wnc,Ng:2020jqd}. Such searches have been performed during the first three observation runs of Advanced LIGO and Virgo~\cite{Tsukada:2018mbp,Palomba:2019vxe,LIGOScientific:2021rnv,Yuan:2022bem}, placing constraints on the scalar boson mass around $10^{-13}\,\mathrm{eV}$, depending on the assumed BH spin and source distance. Future GW detectors, which can access different frequency ranges, are expected to extend these constraints~\cite{Brito:2017zvb}. Additionally, spontaneous transitions between different bound states of the clouds could produce GW bursts~\cite{Arvanitaki:2014wva}, offering yet another observational channel. Furthermore, the formation of the cloud leads to BH spin-down, implying that the absence of rapidly rotating BHs in certain mass ranges can serve as indirect evidence for ultralight bosons. Although measuring BH spin is generally challenging, constraints on the scalar boson mass have been placed around $10^{-13}-10^{-12}\,\mathrm{eV}$ and $10^{-19}-10^{-18}\,\mathrm{eV}$~\cite{Arvanitaki:2014wva,Stott:2018opm,Fernandez:2019qbj,Ng:2019jsx,Stott:2020gjj,Ng:2020ruv,Baryakhtar:2020gao,Mehta:2021pwf,Wen:2021yhz,Hoof:2024quk}. A similar approach has been used to place limits on the masses of the dark photon~\cite{Pani:2012bp,Pani:2012vp,Cardoso:2018tly} and the graviton~\cite{Brito:2013wya}.
\vskip 2pt
The discussion above ignores potential interactions between ultralight bosonic fields and the Standard Model. While current bounds indicate that these couplings are weak (see Figure~\ref{fig:constraints}), the large particle number in superradiant clouds may still lead to observable effects. Particularly intriguing are interactions with the electromagnetic sector. For the QCD axion and its axionic couplings [see eq.~\eqref{eqn:axionLagrangian}], stimulated decay into photon pairs becomes relevant for axion masses $\geq 10^{-8}\,\mathrm{eV}$, corresponding to primordial BHs with masses $\leq 0.01M_{\odot}$. Such systems could produce powerful electromagnetic radiation in the radio band~\cite{Rosa:2017ury,Boskovic:2018lkj,Ikeda:2018nhb,Spieksma:2023vwl}. This presents a compelling signature in which the environment acts as a \emph{portal} channelling rotational energy from the BH into electromagnetic radiation. We will revisit this phenomenon in detail in Chapter~\ref{chap:SR_Axionic}. Similarly, spin-1 fields such as the dark photon naturally couple to the electromagnetic sector~\eqref{eqn:darkphotonL}. Their behaviour, particularly in the presence of plasma, will be explored in Chapter~\ref{chap:in_medium_supp}.
\vskip 2pt
There are many other observational channels that go beyond the scope of this thesis, particularly those based on electromagnetic signals (see~\cite{Miller:2025yyx} for a recent review). For example, tidal disruption events -- where stars are torn apart by BHs -- offer a way of studying accretion physics and may serve as promising sources for multi-messenger astronomy~\cite{Olejak:2025hcx}. More broadly, the dynamics of accretion flows and plasma around BHs can provide crucial insights into jet production and high-energy electromagnetic emission. An exciting challenge for the future is to leverage the centuries of experience with electromagnetic observations to better understand regions of strong gravity, particularly by finding ways to correlate electromagnetic signatures with GW detections. All in all, isolated BHs are far more than remnants of stellar collapse;~they are powerful testbeds for uncovering new physics.
\subsection{Binary Systems}\label{BHenv_subsec:binary}
\begin{figure}[t!]
\centering
\includegraphics[scale=0.9, trim = 0 5 0 10]{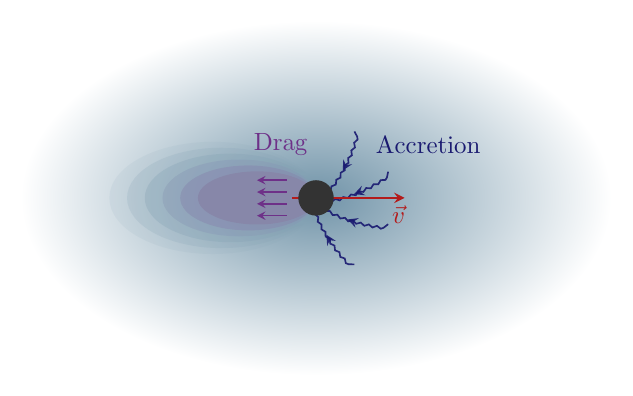}
\caption{Schematic illustration of a BH moving through a medium. As the BH moves, it creates an overdensity of matter behind it, which pulls it back -- a process known as dynamical friction. At the same time, some of the matter is accreted by the BH. Both of these effects influence the motion of the BH  and, in turn, modify the GWs it emits when part of a binary system.}
\label{fig:AccretionFriction}
\end{figure}
Irrespective of the nature of the environment, a binary system evolving within it will be subject to various effects. These can subtly alter the orbital evolution of the binary, imprinting a shift on the GW phase compared to vacuum. Since this shift depends on the properties of the environment, it provides an opportunity to extract information about them. Detecting such effects requires both identifying \emph{modifications} to the gravitational waveform and \emph{distinguishing} them from ``standard'' GW signals. Two of the most prominent mechanisms affecting a BH moving through a medium are \emph{accretion} and \emph{dynamical friction}, schematically shown in Figure~\ref{fig:AccretionFriction}.\footnote{In accretion disks, gas torques are in fact the dominant effect, driving \emph{planetary migration}~\cite{1980ApJ...241..425G,Kocsis:2011dr,Yunes:2011ws,Derdzinski:2020wlw}.} Accretion is straightforward:~as a BH moves through the medium, it captures some of the surrounding material. At the same time, it exerts a gravitational pull on the medium, leading to a friction effect, which was first studied by Chandrasekhar~\cite{Chandrasekhar:1943ys,1943ApJ....97..263C,1943ApJ....98...54C}. The precise impact of these processes depends sensitively on the nature of the environment. For instance, the drag generated by collisionless DM~\cite{Chandrasekhar:1943ys,1943ApJ....97..263C,1943ApJ....98...54C} is different from that of ultralight scalar fields~\cite{Macedo:2013jja,Macedo:2013qea,Hui:2016ltb,Traykova:2021dua,Vicente:2022ivh,Tomaselli:2023ysb}. To study these processes without solving the full system of equations governing both the binary and its environment, effects such as accretion, dynamical friction, gravitational radiation, and the self-gravity of the medium are typically treated individually, and then incorporated sequentially according to their PN order. However, many existing approaches rely on approximations or restrictive assumptions, such as neglecting the backreaction on the environment. A systematic, fully relativistic method for studying environments will be presented in Chapter~\ref{chap:inspirals_selfforce}.
\vskip 2pt
Environments can also modify the structure of the compact objects themselves, altering the waveform through so-called \emph{finite-size effects}. A general astrophysical object can be described by its multipolar structure~\cite{Geroch:1970cd,Hansen:1974zz,Thorne:1980ru}. For example, a spinning object acquires a quadrupole moment, whose strength is quantified by the dimensionless parameter $\kappa$, representing the degree of rotational deformation. Kerr BHs have $\kappa = 1$~\cite{Hansen:1974zz,Thorne:1980ru}, whereas superradiant boson clouds or boson stars can exhibit values as large as $\kappa \sim 10^3$~\cite{Chia:2020dye,DeLuca:2021ite}. These effects contribute at 2PN order. Moreover, tidal interactions lead to a tidally-induced quadrupole moment, characterised by the \emph{Love number} $\lambda$. Famously, the Love numbers of BHs vanish~\cite{Binnington:2009bb,Damour:2009vw,Gurlebeck:2015xpa, Landry:2015zfa,Pani:2015hfa,Chia:2020yla}, yet this is not case when environments are present~\cite{Baumann:2018vus,Cardoso:2019upw,DeLuca:2021ite,Cardoso:2021wlq,Brito:2023pyl,Cannizzaro:2024fpz}. Such effects enter at 5PN order.
\vskip 2pt
So far, all GW detections and most analyses have been conducted on the premise that binary coalescences occur in vacuum. This is generally well-justified for the current LIGO-Virgo-KAGRA (LVK) data, which primarily involve stellar-mass BHs in comparable mass ratio binaries.\footnote{Although efforts have been made to study environments using the available LVK data~\cite{CanevaSantoro:2023aol,Roy:2024rhe}.} Several reasons support this assumption;~(i) the types of sources that LVK detectors are sensitive to are generally not ideal candidates for detecting environmental effects. Apart from superradiance, which can occur for BHs of any mass (given a suitable boson mass), environments like accretion disks or DM spikes are more commonly associated with intermediate to supermassive BHs;~(ii) even when an environment is present, it may be swept away in an equal-mass binary coalescence before the GW signal enters the detector's sensitivity band;\footnote{However, when the environment itself carries angular momentum, this is not necessarily true. It is currently unclear how much of the environment is depleted in such cases (for scalar fields, see e.g.,~\cite{Ikeda:2020xvt,Bamber:2022pbs,Aurrekoetxea:2023jwk,Aurrekoetxea:2024cqd,Tomaselli:2024ojz}).} (iii) most observed GW signals last only a few cycles, leaving little time for subtle environmental effects to accumulate while the system is in the detectors' sensitivity band;~(iv) finally, LVK detectors are most sensitive to the final moments of coalescence, when the gravitational field is strongest, making it more likely that gravitational effects dominate over those from the environment.
\vskip 2pt
This situation is expected to change with the advent of future GW observatories, capable of tracking extreme mass ratio inspirals over long timescales. In particular, millihertz detectors such as LISA will track these binaries over many years, enabling precise measurements of their motion and interactions with surrounding matter. Moreover, these systems typically involve intermediate-mass or supermassive BHs, which -- unlike stellar-mass BHs -- are typically found at galactic centres~\cite{2021NatRP...3..732V}, where dense gas, accretion disks, and DM structures are common (see Section~\ref{BHenv_sec:taxonomy}). Given this potential, substantial efforts are being directed towards modelling environmental effects. Studies of accretion disks~\cite{1973A&A....24..337S,Novikov:1973kta,Tanaka_2002,Barausse:2007dy,Abramowicz2013,Yunes:2011ws,Kocsis:2011dr,Barausse:2014tra,Derdzinski:2018qzv,Duffell:2019uuk,Derdzinski:2020wlw,Pan:2021oob,Zwick:2021dlg,Cole:2022yzw,Speri:2022upm,Garg:2022nko,Tiede:2023cje,Zwick:2024yzh,Garg:2024oeu,Garg:2024yrs,Garg:2024zku,Khalvati:2024tzz,Duque:2024mfw,Copparoni:2025jhq}, active galactic nuclei~\cite{Tagawa:2019osr,Tagawa:2020qll,Pan:2021ksp,Derdzinski:2022ltb,Morton:2023wxg}, and DM structures~\cite{Macedo:2013qea,Eda:2013gg,Eda:2014kra,Barausse:2014tra,Yue:2018vtk,Hannuksela:2019vip,Kavanagh:2020cfn,Coogan:2021uqv,Dai:2021olt,Cardoso:2021wlq,Coogan:2021uqv,Cole:2022yzw,Cardoso:2022whc,Speeney:2022ryg,Becker:2022wlo,Berezhiani:2023vlo,Karydas:2024fcn,Kavanagh:2024lgq,Bertone:2024wbn,Gliorio:2025cbh} -- potentially consisting of ultralight particles~\cite{Ferreira:2017pth,Traykova:2021dua,Kim:2022mdj,Vicente:2022ivh,Bamber:2022pbs,Buehler:2022tmr,Aurrekoetxea:2023jwk,Traykova:2023qyv,Aurrekoetxea:2023jwk,Wang:2024cej,Aurrekoetxea:2024cqd,Dyson:2024qrq}, bosonic clouds~\cite{Zhang:2018kib,Baumann:2018vus,Zhang:2019eid,Baumann:2019ztm,Baumann:2021fkf,Takahashi:2021eso,Cole:2022yzw,Baumann:2022pkl,Takahashi:2023flk,Tomaselli:2023ysb,Brito:2023pyl,Duque:2023seg,Tomaselli:2024bdd,Tomaselli:2024dbw,Boskovic:2024fga,Khalvati:2024tzz} or merging exotic compact objects~\cite{Palenzuela:2006wp,Palenzuela:2017kcg,Helfer:2018vtq,Bezares:2018qwa,Clough:2018exo} -- suggest that GW observations can provide unique insights into these environments, and with that help probe the nature of DM or reveal new fundamental fields.
\vskip 2pt
Environmental effects, however, present not only opportunities but also significant challenges. If not properly accounted for, they can bias inferred binary parameters or degrade the signal-to-noise ratio, potentially leading to missed detections~\cite{Cole:2022yzw,Zwick:2022dih,Roy:2024rhe,Garg:2024qxq,Khalvati:2024tzz}. Furthermore, distinguishing genuine environmental effects from modifications to GR will be a key challenge for future GW astrophysics.
\vskip 2pt
This thesis explores a range of environments in the presence of binary systems. In Chapters~\ref{chap:plasma_ringdown} and~\ref{chap:BHspec}, we examine the ringdown in presence of plasma and galactic DM halos, respectively, while in Chapters~\ref{chap:legacy} and~\ref{chap:inspirals_selfforce}, we study the inspiral of scalar clouds around BHs in the Newtonian and relativistic regime. In Chapter~\ref{chap:capture}, we focus on the evolution of binaries in AGN disks.
\chapter{Superradiance:~Axionic Couplings and Plasma Effects} \label{chap:SR_Axionic}
\vspace{-0.8cm}
\hfill \emph{Don't you know that I hold the sea and its ways in my hand,}

\hfill \emph{and the heavens are my chart?}
\vskip 5pt

\begin{figure}[t!]
  \centering
    \includegraphics[width = \textwidth]{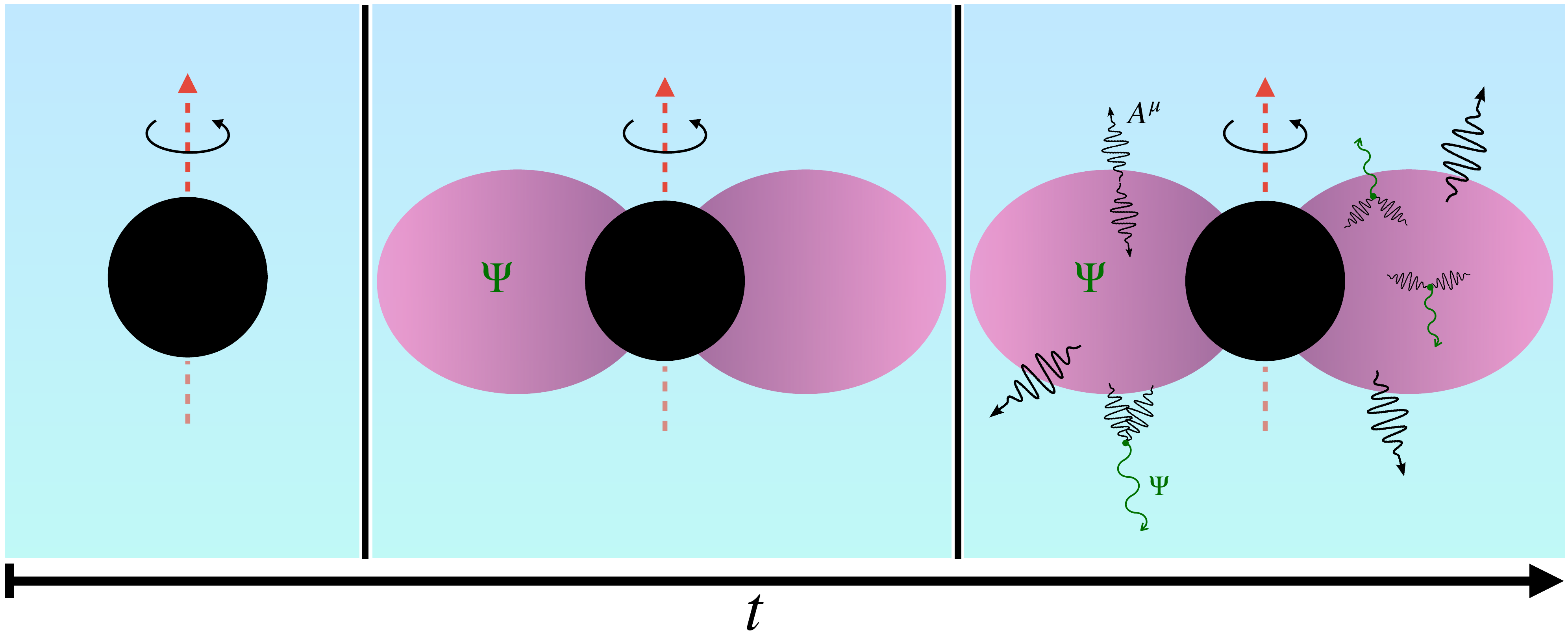}
    \caption{Schematic illustration of our setup. Starting from a spinning BH of mass $M$ (\emph{left panel}), a superradiant cloud of mass $M\ped{c}$ is formed in the dominant (dipolar) growing mode (\emph{centre panel}). For sufficiently large couplings to the Maxwell sector, the configuration is unstable:~any small EM fluctuation will trigger emission of EM radiation [\textbf{black}] (\emph{right panel}). Some of these waves recombine to create axion waves [{\color{forestgreen}{green}}]. The blue background indicates the presence of a plasma.}
    \label{fig:Evolution}
\end{figure}

\hfill Ru, people of Tupua'i
\vskip 35pt
\noindent
In Section~\ref{BHenv_subsec:superrad}, I discussed how spinning BHs can transfer a significant fraction of their energy to ultralight bosonic fields via superradiance, condensing them in a co-rotating structure or ``cloud''. The precise development of this instability is well understood in vacuum and in absence of couplings to the Standard Model. However, BHs are surrounded by interstellar matter or accretion disks and couplings between bosonic fields and the Standard Model may be non-vanishing (see Figure~\ref{fig:constraints}).
\vskip 2pt
It was argued analytically and with numerical simulations, that axionic couplings to the Maxwell sector might trigger \emph{parametric instabilities}, whereby the scalar cloud transfers energy to electromagnetic (EM) radiation~\cite{Rosa:2017ury,SenPlasma,Boskovic:2018lkj,Ikeda:2018nhb}. Additionally, the presence of a surrounding plasma may quench the parametric instability due to the high energy (large ``effective mass'') of typical astrophysical environments~\cite{Cardoso:2020nst,Cannizzaro:2021zbp,Blas:2020kaa}. The previous works left important gaps:~(i) the parametric instability was shown to give rise to periodic bursts of light, but its period and amplitude were not studied. In fact, the effect of a superradiantly growing cloud was also not understood properly.~(ii)~The role of plasmas in the development of EM instabilities is known poorly, but could have a drastic effect (see e.g.,~recent works on dark photon superradiance~\cite{Caputo:2021efm,Siemonsen:2022ivj}), since the plasma frequency is rather large in most astrophysical circumstances.
\clearpage
In this chapter, I present the work done in~\cite{Spieksma:2023vwl}, where these couplings were studied using numerical relativity. By evolving the coupled axion-Maxwell system on a BH background, taking into account the axionic coupling concurrently with the growth of the cloud, a new stage emerges:~that of a \emph{stationary state} where a constant flux of electromagnetic waves is fed by superradiance, for which accurate analytical estimates are found. Moreover, I show how the existence of electromagnetic instabilities in the presence of plasma is entirely controlled by the axionic coupling;~even for dense plasmas, an instability is triggered for high enough couplings.
\vskip 2pt
The outline of this chapter is as follows. In Section~\ref{sec_SR_Axionic:setup}, I set up the relevant equations of motion and discuss the modelling of the cloud as well as the plasmic environment. In Section~\ref{sec_SR_Axionic:withoutSR}, I study the evolution of the axion-Maxwell system in the absence of superradiance, while in Section~\ref{sec_SR_Axionic:withSR}, I carry out a similar analysis, now including superradiance. In Section~\ref{sec_SR_Axionic:surroundingplasma}, I describe the influence of a plasma on the EM instability, and in Section~\ref{sec_SR_Axionic:observational}, I explore possible observational signatures. Finally, in Section~\ref{sec_SR_Axionic:conclusions}, I provide a summary and outlook. Additional details are contained in Appendices~\ref{app:NR} and~\ref{app:Mathieu}. For the purposes of this chapter, I retain the explicit dependence on $\hbar$, and thus work in geometric units where $G=c=1$, while using rationalised Heaviside-Lorentz units for the Maxwell equations. Finally, in this chapter, I use the tilde to denote the Fourier transform, rather than its usual role as a marker for dimensionless quantities elsewhere in the thesis. An illustration of the setup can be seen in Figure~\ref{fig:Evolution}.
\clearpage
\section{Setup}\label{sec_SR_Axionic:setup}
\subsection{The Theory}
We consider a real, massive (pseudo)scalar field $\Psi$ with axionic couplings to the EM field. In addition, the EM field is coupled to a cold, collisionless electron-ion plasma. In this setup, the Lagrangian takes the following form:
\begin{equation}\label{eq:SRax_lagrangian}
\mathcal{L}=\frac{R}{16\pi}-\frac{1}{4}F_{\mu\nu}F^{\mu\nu}-\frac{1}{2}\nabla_{\mu}\Psi\nabla^{\mu}\Psi-\frac{\mu^{2}}{2}\Psi^{2}-\frac{k\ped{a}}{2}\Psi\,{}^{*}\!F^{\mu\nu}F_{\mu\nu}+A^\mu j_\mu + \mathcal{L}\ped{m}\,.
\end{equation}
The mass of the scalar field $\Psi$ is given by $m\ped{a} = \mu \hbar$, $A_{\mu}$ is the vector potential, $F_{\mu\nu} \equiv \nabla_{\mu}A_{\nu}-\nabla_{\nu}A_{\mu}$ is the Maxwell tensor and ${}^{*}\!F^{\mu\nu} \equiv \frac{1}{2}\epsilon^{\mu\nu\rho\sigma}F_{\rho\sigma}$ is its dual. We use the definition $\epsilon^{\mu\nu\rho\sigma}\equiv (1/\sqrt{-g})E^{\mu\nu\rho\sigma}$, where $E^{\mu\nu\rho\sigma}$ is the totally anti-symmetric Levi-Civita symbol with $E^{0123}=1$. We define the Lagrangian for the plasma as $\mathcal{L}\ped{m}$, while $k\ped{a}$ quantifies the axionic coupling which we take to be constant.\footnote{This quantity corresponds to $g_{a\gamma\gamma}$ in eq.~\eqref{eqn:axionLagrangian}, and the notation is chosen for consistency with previous works~\cite{Boskovic:2018lkj,Ikeda:2018nhb}.} There exists a wide variety of theories predicting axions and axion-like particles, and generically $k\ped{a}$ is independent of the boson mass [unlike the case of the QCD axion~\eqref{eqn:QCDaxionmass}]. Therefore, we take $k\ped{a}$ to be an additional free parameter of the theory. Notice that we do not consider self-interactions, which could appear as an expansion of the axion's periodic potential. This corresponds to a region $k\ped{a} f\ped{a} \geq \mathcal{O}(1)$ predicted in models such as clockwork axions~\cite{Kaplan:2015fuy,Farina:2016tgd} and magnetic monopoles in the anomaly loop~\cite{Sokolov:2021ydn,Sokolov:2022fvs}, where $f_a$ is the decay constant of the axion.\footnote{The strong self-interaction regime was discussed in e.g.,~\cite{Yoshino:2012kn, Baryakhtar:2020gao,Omiya:2022gwu,Chia:2022udn}, where transitions to various cloud modes and distortion of bound state wave functions are expected.} In principle, a similar analysis could be performed for scalar couplings ($\mathcal{L} \supset k\ped{s} \Psi F^{\mu \nu} F_{\mu \nu}$), at least when the coupling strength is weak~\cite{Boskovic:2018lkj}.
\vskip 2pt
Finally, $j_\mu$ is the plasma current, and captures both the contributions of the electrons and the much heavier ions. We adopt a two-fluid formalism model for the plasma, where electrons and ions are treated as two different fluids, coupled through the Maxwell equations (see Section~\ref{BHenv_sec:Plasma} for details). Hence, the plasma current is given by $j^\mu= \sum\ped{s} q\ped{s} n\ped{s} u\ped{s}^\mu$, where the index $\mathrm{s}$ represents the sum over the two different species, electrons and ions, and $q\ped{s}$, $n\ped{s}$ and $u\ped{s}^\mu$ are the charge, number density and four velocity of the fluids, respectively.
\vskip 2pt
An axion cloud produced from superradiance can grow to be $\lesssim 10\%$ of the BH mass~\eqref{eq:massatsat}. We will consider the cloud's backreaction on the geometry to be small and thus evolve the system on a fixed background. The gravitational coupling $\mu M$ determines the strength of the interaction between the BH and the axion and is a crucial quantity. In order for superradiance to be efficient on astrophysical timescales, the gravitational coupling must be $\mathcal{O} (1)$~\eqref{eqn:211bounds}. For $\mu M \ll 0.1$, the exponential growth is too slow~\eqref{eqn:Gamma_nlm}, while the instability is exponentially suppressed for $\mu M \gg 1$~\cite{Zouros:1979iw,Berti:2009kk}. Consequently, we will perform simulations in the range $\mu M \sim 0.1 -0.3$.
\vskip 2pt
From the Lagrangian of our theory~\eqref{eq:SRax_lagrangian}, we obtain the equations of motion for the scalar and EM field. In order to close the system, we also need to consider the continuity and momentum equation of the fluids, which come from the conservation of the energy-momentum tensor for the Maxwell-plasma sector. Ignoring the backreaction of the fields in the spacetime, we obtain:
\begin{equation}\label{eq:SRAxion_evoleqns}
\begin{aligned}
\left(\nabla^\mu \nabla_\mu-\mu^2\right) \Psi &= \frac{k\ped{a}}{2}\,{ }^*\!F^{\mu \nu} F_{\mu \nu}\,,\\
\nabla_\nu F^{\mu \nu} &=j^{\mu} -2 k\ped{a}{ }^*\!F^{\mu \nu} \nabla_\nu \Psi\,,\\
u\ped{s}^{\nu}\nabla_{\nu}u\ped{s}^{\mu} &= \frac{q\ped{s}}{m\ped{s}}F^{\mu}_{\ \nu}u^{\nu}\ped{s}\,,\\ 
\nabla_\mu (n\ped{s} u\ped{s}^\mu)&=0\,.
\end{aligned}
\end{equation}
Finally, we impose the Lorenz condition on the vector field
\begin{equation}
    \nabla_{\mu}A^{\mu} = 0\,,
\end{equation}
thereby fixing our gauge freedom.
\subsection{Modelling Superradiance}
Even though we are interested in an axion cloud that grows through superradiance, and thus requires a spinning BH described by the Kerr metric~\eqref{eq:metric_kerr}, we will instead mimic superradiant growth without the need of a spinning BH. The reason is of a practical nature:~timescales to superradiantly grow an axion cloud are larger than $\sim 10^6M$~\cite{Cardoso:2005vk,Dolan:2007mj,Berti:2009kk}, a prohibitively large timescale for numerical purposes. Therefore, we mimic superradiant growth following Zel'dovich~\cite{ZelDovich1971,ZelDovich1972,Cardoso:2015zqa}, by adding a simple Lorentz-invariance-violating term to the Klein-Gordon equation,
\begin{equation}\label{eq:ASR}
    \left(\nabla^\mu \nabla_\mu-\mu^2\right) \Psi = C \frac{\partial \Psi}{\partial t}+\frac{k\ped{a}}{2}\,{ }^*\!F^{\mu \nu} F_{\mu \nu}\,.
\end{equation}
Here, $C$ is a constant, which in the absence of the axionic coupling gives rise to a linear instability on a timescale of the order $1/C$, where we can tune $C$ to be within our numerical limits. For further details, we refer to Appendix~\ref{appNR_subsec:artificial_super}.
\subsection{Modelling Plasma}\label{sec_SR_Axionic:SRax_Plasma}
One of the most important characteristics of plasmas is their peculiar response to external perturbations. As we discussed in Section~\ref{BHenv_sec:Plasma}, when plasma is perturbed by an EM wave, electrons are displaced and start oscillating around their equilibrium position at the so-called \emph{plasma frequency}: 
\begin{equation}\label{eq:SR_Axionic_plasmafreq}
     \omega\ped{p}= \sqrt{\frac{n\ped{e} q\ped{e}^2}{m\ped{e}}}\approx \frac{10^{-12}}{\hbar}\sqrt{\frac{n\ped{e}}{10^{-3}\,\mathrm{cm}^{-3}}}\,\text{eV}\,.
\end{equation}
Remarkably, the dispersion relation of the transverse modes of a photon propagating in a plasma are modified by a gap which corresponds to the plasma frequency, i.e.,~$\omega^2=k^2 + \omega\ped{p}^2$. For this reason, the plasma frequency acts as an \emph{effective mass} for the transverse polarisations of the photons. This effect is crucial to take into account when studying parametric instabilities as the axion decay into photons could be suppressed in a dense plasma, i.e.,~when $\omega\ped{p}\gg\mu$. Throughout this chapter, we work under the following assumptions regarding the plasma. 
\vskip 2pt
\begin{enumerate}[label=(\roman*)]
\item We neglect nonlinear terms in the axion-photon-plasma system. That is to say, as long as the EM field remains small, it suffices to consider only the linear response of the medium. Consequently, the backreaction of the EM field on the axion field is omitted.
\item We ignore ion oscillations induced by the EM field. For non-relativistic plasmas, this assumption is justified because the ions have much larger inertia than the electrons, allowing them to be treated as a neutralising background. In the relativistic regime, however, this approximation breaks down when $\Gamma\ped{e} \gg m\ped{ion}/m\ped{e}$ where $\Gamma\ped{e}$ is the Lorentz factor of the electrons~\cite{KrallTrivelpiece1973}. 
\item We assume the plasma to be locally quasi-neutral, i.e.,~$Zn\ped{ion}\approx n\ped{e}$, where $Z$ is the atomic number. This results in a vanishing charge density. The global neutrality of the plasma ensures the validity of this assumption at sufficiently large length scales, particularly in systems where the length scales are much larger than the Debye length~\eqref{eq:DebyeLength}.
\item We model the plasma as cold and collisionless. In principle, our formalism can be extended to include thermal and collisional effects by adding appropriate terms to the momentum equation~\cite{Cannizzaro:2021zbp}. Thermal effects can be incorporated by introducing a non-negligible pressure and an equation of state, while electron-ion collisions can be modelled by adding a term proportional to the relative velocity of the two fluid species.
\item We neglect the evolution of the fluid's four-velocity relative to an \emph{Eulerian observer}\footnote{An Eulerian observer follows a worldline orthogonal to the spacelike hypersurface.\label{ft:Eulerian}} due to gravity. The 3+1 decomposition of the momentum equation~\eqref{eq:momeqnlinear} shows that, after linearisation, the evolution is governed by a gravitational term $a_i$ and an EM term $E_i$. In a Schwarzschild background, the only nonzero gravitational component is $a^r=M/r^2$, representing the familiar gravitational acceleration near a spherical object.\footnote{At the horizon, this corresponds to the surface gravity.} Since the axion cloud is localised around the Bohr radius, gravitational effects remain small on the relevant timescales. For instance, for a cloud with $\mu M = 0.1$ at $r=200 M$, the acceleration is $a^r=2.5 \times 10^{-5} M^{-1}$, which would only induce significant velocity changes on a timescale $t \sim 10^5 M$, far exceeding the growth timescales of the EM field. Additionally, the gravitational term is suppressed relative to the EM term by a factor of $m\ped{e}/q\ped{e} \sim 10^{-22}$. Neglecting gravity simplifies the system in two ways. First, while the electron momentum equation retains its EM term, the ionic momentum equation becomes trivial under assumption (ii), as ions with initially zero velocity remain stationary. Thus, they serve as a neutralising background without further evolution. Second, with the gravitational term absent, the electron momentum constraint~\eqref{eq:constrainMomeqn} reduces to $u^\nu \partial_\nu \Gamma=0$. Since $\Gamma = 1$ to linear order, this constraint is automatically satisfied. As a consequence, dropping the gravitational term simplifies our evolution scheme considerably.
\end{enumerate}
\subsection{Numerical Procedure}
\begin{table}[t!]
\centering
\renewcommand*{\arraystretch}{1.03}
\resizebox{0.45\linewidth}{!}{%
\begin{tabular}{ | c || c | c | c | c | } 
  \hline
  Run & $k\ped{a}\Psi_{0}$ & $\mu M$ &  $10^{3}CM$ & $ 10^{4} E_{0}M/\Psi_{0}$\\ 
  \hline \hline
  $\mathcal{I}_{1}$ & $0.0$ & $0.3$ & $0.0$ &  $8.1$ \\ 
  \hline
  $\mathcal{I}_{2}$ & $0.0295$ & $0.3$ & $0.0$ &  $8.1$ \\ 
  \hline
  $\mathcal{I}_{3}$ & $0.147$ & $0.3$ & $0.0$ & $8.1$ \\ 
  \hline \hline
  $\mathcal{J}_{1}$ & $0.0737$ & $0.2$ & $4.0$ &  $100.0$ \\ 
  \hline
  $\mathcal{J}_{2}$ & $0.0737$ & $0.2$ & $4.0$ &  $1.0$ \\ 
  \hline
  $\mathcal{J}_{3}$ & $0.0737$ & $0.2$ & $4.0$ &  $0.01$ \\ 
  \hline
  $\mathcal{J}_{4}$ & $0.0737$ & $0.2$ & $4.0$ &  $0.0001$ \\ 
  \hline \hline
  $\mathcal{J}_{5}$ & $0.0737$ & $0.2$ & $4.0$ &  $8.1$ \\ 
  \hline
  $\mathcal{J}_{6}$ & $0.0563$ & $0.2$ & $4.0$ &  $8.1$ \\ 
  \hline
  $\mathcal{J}_{7}$ & $0.0328$ & $0.2$ & $4.0$ &  $8.1$ \\ 
  \hline
  $\mathcal{J}_{8}$ & $0.00737$ & $0.2$ & $4.0$ & $8.1$ \\ 
  \hline \hline
  $\mathcal{J}_{9}$ & $0.0737$ & $0.2$ & $0.08$ &  $8.1$ \\ 
  \hline
  $\mathcal{J}_{10}$ & $0.0737$ & $0.2$ & $0.2$ &  $8.1$ \\ 
  \hline
  $\mathcal{J}_{11}$ & $0.0737$ & $0.2$ & $0.8$ &  $8.1$ \\ 
  \hline
  $\mathcal{J}_{12}$ & $0.0737$ & $0.2$ & $1.0$ &  $8.1$ \\ 
  \hline
  $\mathcal{J}_{13}$ & $0.0737$ & $0.2$ & $2.0$ &  $8.1$ \\ 
  \hline
  $\mathcal{J}_{14}$ & $0.0737$ & $0.2$ & $8.0$ &  $8.1$ \\ 
  \hline \hline
   &  &  &  &  $\omega\ped{p} M$ \\ 
  \hline
  $\mathcal{K}_{1}$ & $0.147$ & $0.3$ & $0.0$ &  $0.01$ \\ 
  \hline
  $\mathcal{K}_{2}$ & $0.147$ & $0.3$ & $0.0$ &  $0.1$ \\ 
  \hline
  $\mathcal{K}_{3}$ & $0.147$ & $0.3$ & $0.0$ &  $0.15$ \\ 
  \hline
  $\mathcal{K}_{4}$ & $0.147$ & $0.3$ & $0.0$ &  $0.2$ \\ 
  \hline
  $\mathcal{K}_{5}$ & $0.147$ & $0.3$ & $0.0$ &  $0.3$ \\ 
  \hline
  $\mathcal{K}_{6}$ & $0.147$ & $0.3$ & $0.0$ &  $0.4$ \\ 
  \hline
  $\mathcal{K}_{7}$ & $0.295$ & $0.3$ & $0.0$ &  $0.2$ \\ 
    \hline
  $\mathcal{K}_{8}$ & $0.590$ & $0.3$ & $0.0$ &  $0.2$ \\ 
  \hline
  $\mathcal{K}_{9}$ & $0.0737$ & $0.1$ & $2.0$ &  $0.02$ \\ 
  \hline
  $\mathcal{K}_{10}$ & $0.0737$ & $0.1$ & $2.0$ &  $0.07$ \\ 
  \hline
\end{tabular}}
\caption{A summary of the simulations discussed in this chapter.  Simulations labelled $\mathcal{I}_{i}$ do not include plasma or superradiant growth. Simulations labelled $\mathcal{J}_{i}$ include superradiant growth but still exclude plasma, while simulations labelled $\mathcal{K}_{i}$ include the plasma. The parameters we consider include the axionic coupling $k\ped{a}\Psi_{0}$, the mass coupling $\mu M$, the artificial superradiance parameter $C$, and the ratio between the initial amplitude of the electric field $E_{0}$~\eqref{eq:InitialElectric} and the scalar field $\Psi_{0}$~\eqref{eq:normalisation}. For simulations that include plasma ($\mathcal{K}_{i}$), the initial EM amplitude is $10^{4}E_{0}M/\Psi_0 = 8.1$ and we also report the plasma frequency $\omega\ped{p}$. In all our simulations, the EM pulse is initialised at $r_{0} = 40M$ with $\sigma = 5M$.}
\label{tb:simulations}
\end{table}
\vskip 2pt
To evolve the system, we numerically solve the equations of motion~\eqref{eq:SRAxion_evoleqns} around a Schwarzschild BH of mass $M$~\eqref{eq:metric_schwarzschild}. We introduce the spatial components of the Maxwell field denoted by $\mathcal{A}_{i}$, the electric field $E^{i}$, the magnetic field $B^{i}$, an auxiliary field $\mathcal{Z}$, and the conjugate momentum $\Pi$ of the scalar field. By applying the 3+1 decomposition to the equations of motion, we derive the evolution equations for the scalar field, the EM field, and the plasma. A detailed account of the formulation of our system as a Cauchy problem can be found in Appendix~\ref{appNR_sec:Cauchy}.
\vskip 2pt
The 3+1 decomposition also yields a set of constraint equations, which are shown explicitly in~\eqref{eq:constrainteqn}. The initial data we construct should satisfy these equations. Following~\cite{Boskovic:2018lkj,Ikeda:2018nhb}, we assume the following profile for the electric field,:
\begin{equation}
\begin{aligned}
\label{eq:InitialElectric}
E^r &=E^\theta=\mathcal{A}_i=0\,,\\
E^{\varphi} &=E_{0}e^{-\left(\frac{r-r_{0}}{\sigma}\right)^{2}} M\,,
\end{aligned}
\end{equation}
where $E^{i}=F^{i\mu}n_{\mu}\,(i=r,\theta,\varphi)$, with $n_{\mu}$ defined as the normal vector of the spacetime foliation. Here, $E^{\varphi}$ can be an arbitrary function of $r$ and $\theta$. In the profile~\eqref{eq:InitialElectric}, $E_{0}$, $r_{0}$, and $\sigma$ represent the typical amplitude, radius and width of the Gaussian, respectively. The EM pulse is initialised at $r_{0} = 40M$ with $\sigma = 5M$ in \emph{all} our simulations. We have verified that our results do not depend on these parameters, validating the generality of our findings. For the initial data of the scalar field, we use a quasi-bound state constructed via Leaver's method (see Appendix~\ref{appNR_subsec:boundstates}). We assume the cloud to occupy the dominant (dipolar) growing mode with an amplitude $\Psi_{0}$, whose normalisation is given by eq.~\eqref{eq:normalisation}. Finally, the constraint equation for the plasma is trivially satisfied, as detailed in Appendix~\ref{appNR_subsec:evolPlasma}, and for simplicity, we take a constant density plasma as initial data.
\vskip 2pt
To keep track of the scalar and EM field during the time evolution, we perform a multipolar decomposition (see Appendix~\ref{appNR_sec:waveextraction}). In the scalar case, we directly project the field $\Psi$ onto spheres of constant coordinate radius using the spherical harmonics with spin weight $s\ped{w} = 0$ to obtain $\Psi_{\ell m}$~\eqref{eq:scalarextract}. In the EM case, we track the evolution of the field using the Newman-Penrose scalar $\Phi_{2}$, which captures the outgoing EM radiation at spatial infinity. Analogous to the scalar case, we project these using spherical harmonics, yet now using spin weight $s\ped{w} = -1$ to obtain $(\Phi_{2})_{\ell m}$~\eqref{eq:NPextract}. In most figures, we will show $|(\Phi_{2})_{\ell m}| = \sqrt{(\Phi_{2})^{*}_{\ell m}(\Phi_{2})_{\ell m}}$. Since these are massless EM waves, $|(\Phi_{2})_{\ell m}| \propto 1/r$ at large spatial distances.
\vskip 2pt
Throughout this chapter, we will discuss various simulations. In Table~\ref{tb:simulations}, the specific parameters of these simulations are listed. Furthermore, a schematic illustration of our setup can be seen in Figure~\ref{fig:Evolution}.
\section{Superradiance Turned Off}\label{sec_SR_Axionic:withoutSR}
An axion cloud coupled to the Maxwell sector can give rise to a burst of EM radiation. The first exploration of this phenomenon was presented in~\cite{Rosa:2017ury}, while the full numerical exercise followed in~\cite{Boskovic:2018lkj, Ikeda:2018nhb}. In this section, our goal is to carefully perform a further analysis and, as we will see, find some new features of the system. Throughout this section, we assume superradiant growth to be absent, as in~\cite{Boskovic:2018lkj, Ikeda:2018nhb}. Even though this is clearly an artificial assumption, as it means that the cloud was allowed to grow without being coupled to the Maxwell sector, it allows us to isolate and understand better some of the phenomena. The full case will be dealt with afterwards.
\vskip 2pt
As shown analytically in flat spacetime, but also numerically in a Kerr background~\cite{Boskovic:2018lkj, Ikeda:2018nhb}, upon growing the cloud to some predetermined value, an EM instability is triggered depending on the quantity $k\ped{a}\Psi_{0}$. In particular, there exist two regimes, a \emph{subcritical} regime and a \emph{supercritical} regime.  In the former, no instability is triggered and some initial EM fluctuation does not experience exponential growth. Conversely, in the supercritical regime, an instability is triggered and the axion field ``feeds'' the EM field, which grows exponentially, resulting in a burst of radiation. The threshold between these regimes is set by two competing effects:~the parametric production rate of the photon, $\propto \mu k\ped{a} \Psi_0$, and their escape rate from the cloud, $\propto \mu^{2} M$. The latter is approximated by the inverse of the cloud size. Similarly to previous works, we find this threshold to be on the order $k\ped{a} \Psi_0 \sim 0.1-0.4$, for $\mu M \sim 0.2-0.3$.
\subsection{The Process at Large}
\begin{figure}[t!]
  \centering
    \includegraphics[scale=0.45]{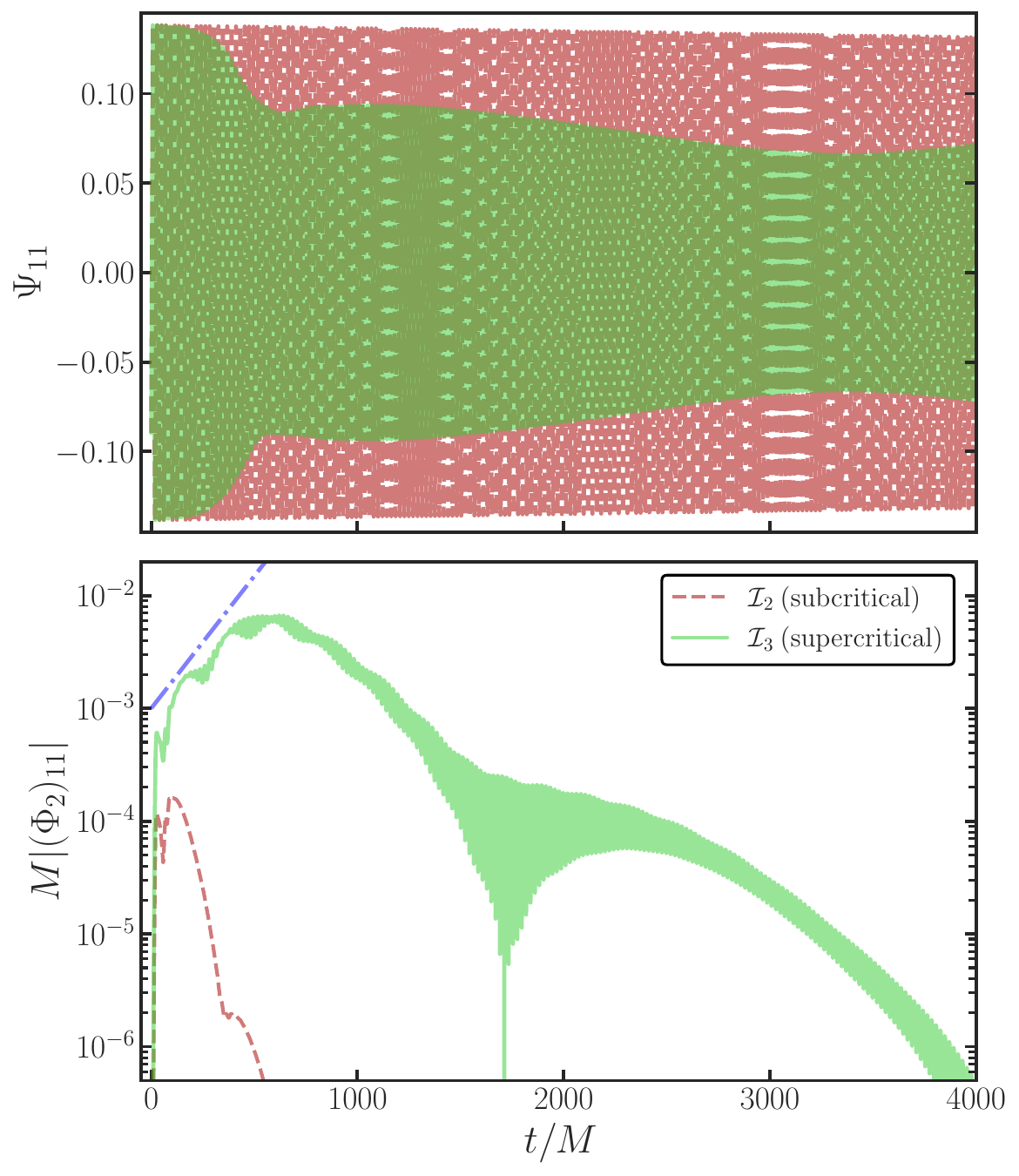}
    \caption{\emph{Top panel}:~The time evolution of the (real part of the) dipolar, $\ell = m = 1$, bound state component of an axion cloud around a Schwarzschild BH in two scenarios:~coupling is subcritical [{\color{cornellRed}{red}} dashed] (simulation $\mathcal{I}_{2}$) or supercritical [{\color{cornellGreen}{green}}] (simulation $\mathcal{I}_{3}$). As a consistency check, we also evolve with a vanishing coupling to the Maxwell sector (simulation $\mathcal{I}_{1}$), which we find to be almost indistinguishable from the subcritical case.
    \emph{Bottom panel}:~The time evolution of the absolute value of the $\ell = m = 1$ component of the Newman-Penrose scalar $\Phi_{2}$ for a subcritical and supercritical coupling. Dash-dotted line [{\color{plotblue1}{blue}}] shows the growth rate, $\lambda = 0.0054/M$, as estimated from eq.~$(77)$ of~\cite{Boskovic:2018lkj}. In both panels, the field is extracted at $r\ped{ex} = 20M$ and $\mu M = 0.3$.}
    \label{fig:BurstmuM03r20}
\end{figure}

In the following, we explore these two regimes by evolving the coupled system describing a superradiant cloud of axions coupled to the Maxwell sector, while initialising it with a small vector fluctuation $E_{0}$. 
\vskip 2pt
Figure~\ref{fig:BurstmuM03r20} summarises well the possible outcomes, which depend on the strength of the coupling $k\ped{a} \Psi_0$. For sufficiently small couplings, the axion field is left unaffected, and remains in a bound state of (near) constant amplitude around the BH. For large couplings however, in what we term the \emph{supercritical} regime, the amplitude of the axion field decreases. This transition signals a parametric instability whereby axions are quickly converted into photons. 
\vskip 2pt
The \emph{bottom panel} of Figure~\ref{fig:BurstmuM03r20} shows the behaviour of the EM radiation during this process (we show only the dipolar component $\ell = m = 1$ of the Newman-Penrose scalar, but we find that higher modes are also excited to important amplitudes, see Appendix~\ref{appNR_sec:higherorder}). In the subcritical regime, any initial EM fluctuation decays on short timescales. In contrast, in the supercritical regime a burst is initiated:~these are the photons that are created by the axion cloud. We find that the growth rate of this instability follows earlier estimates~\cite{Boskovic:2018lkj} and can be approximated as the difference between the photon production and escape rates: $\lambda \sim \lambda_{*} - \lambda\ped{esc}$, where $\lambda_{*} \sim \frac{1}{2}\mu k\ped{a} \Psi_{0}$ and $\lambda\ped{esc} \sim 1/d$, with $d$ the size of the cloud. This estimate is indicated by the blue dash-dotted line in Figure~\ref{fig:BurstmuM03r20}.
\vskip 2pt
At late times, the system settles to a final, stationary state. In the subcritical regime, this final state is almost the same as its initial state since the axion cloud is barely affected by the EM perturbation. Conversely, in the supercritical regime, the parametric instability has driven the axion field to decrease to a subcritical value. Therefore, in the absence of superradiant growth, no further instability can be triggered and the axion cloud settles on a final state with a lower amplitude than its initial value, while the created photons travel outwards.
\begin{figure}[t!]
  \centering
    \includegraphics[scale=0.45, trim = 0 0 0 0]{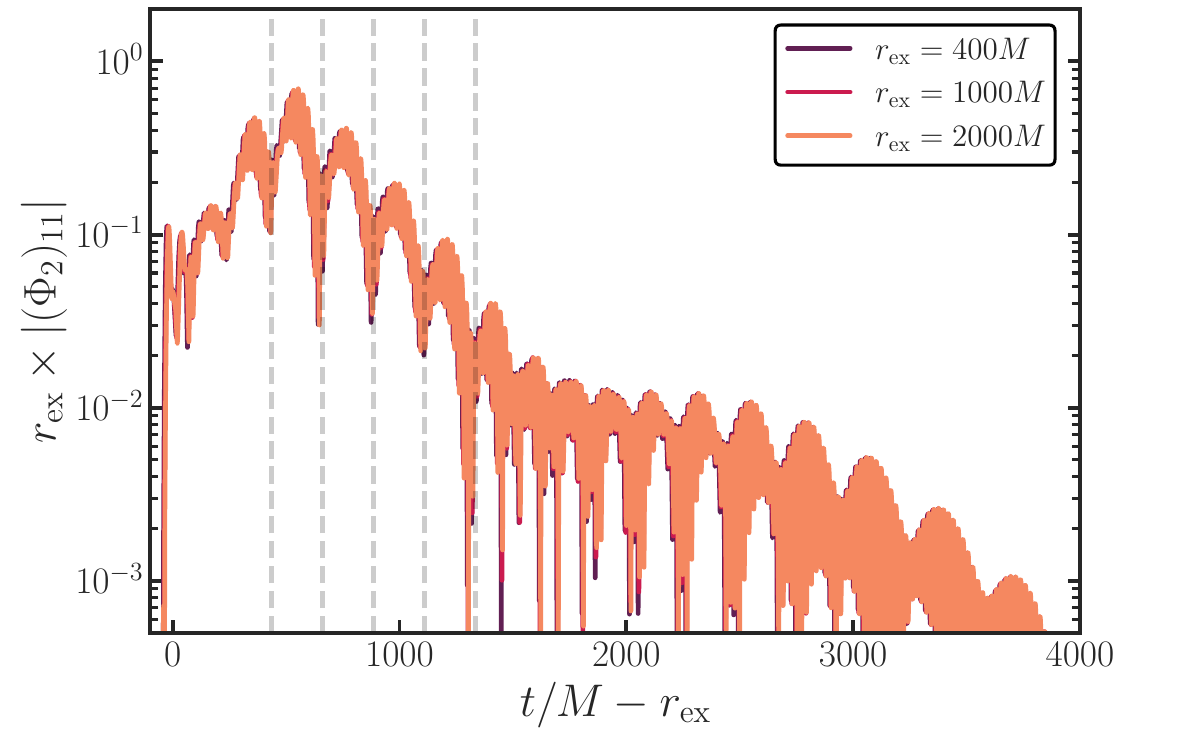}	
    \includegraphics[scale=0.45, trim = 0 0 0 0]{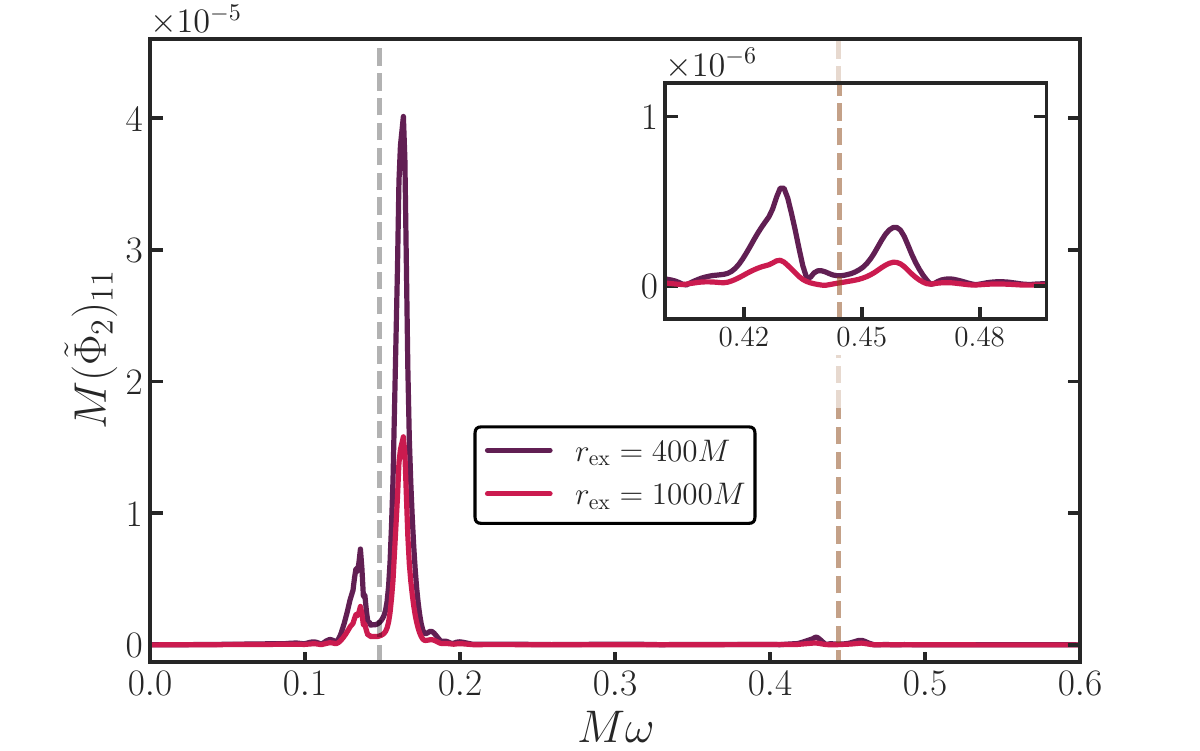}
    \caption{\emph{Top panel}:~The dipolar component $|(\Phi_{2})_{11}|$ of the EM field in the supercritical regime with $\mu M = 0.3$, for simulation $\mathcal{I}_{3}$. The alignment of the waveforms shows that we are dealing with EM radiation. High-frequency oscillations are set by the mass scale $\mu$, whereas ``beatings'' (indicated by vertical dashed lines) are controlled by $1/\mu^2$.
    \emph{Bottom panel}:~Fourier transform of the signal (denoted by a tilde), taken on the entire time domain, showing the dominant frequencies in the problem. Dashed lines show $N (\omega_0 / 2)$ for $N = 1$ [{\color{gray}{grey}}], or $N = 3$ [{\color{MathematicaBrown}{brown}}], where $\omega_0 \approx \mu$ is the frequency of the fundamental mode.}    \label{fig:EMfieldBurstmuM03r481020}
\end{figure}
\subsection{Axion and Photon Emission}
\begin{figure}[t!]
  \centering
    \includegraphics[scale=0.45, trim = 0 0 0 10]{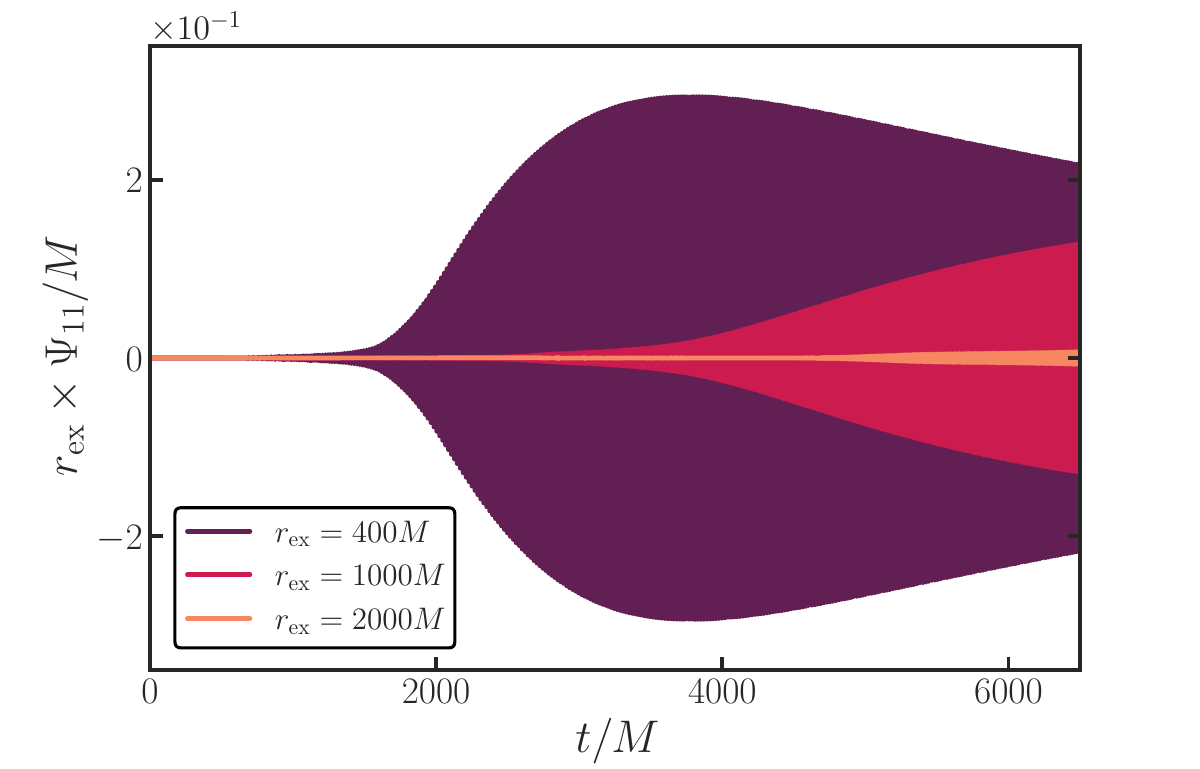}
    \includegraphics[scale=0.45, trim = 0 0 0 0]{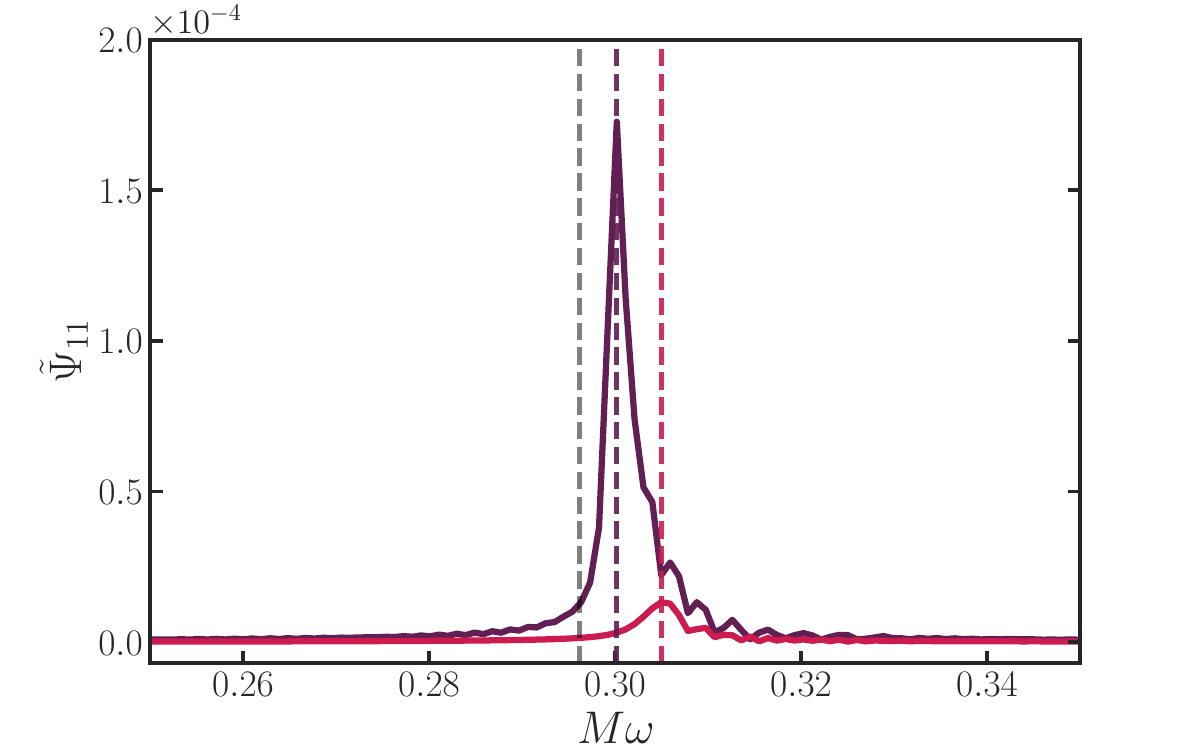}
    \caption{\emph{Top panel}:~The dipolar component of the axion field in the supercritical regime (simulation $\mathcal{I}_{3}$). When extracted at large radii, the nonzero value of $\Psi_{11}$ is only present in the supercritical regime and explained by the production of axion waves. Due to dispersion, the morphology of the wave changes while travelling outwards. 
    \emph{Bottom panel}:~The Fourier transform of the dipolar component, taken on the entire time domain shown in the \emph{top panel}. Dashed lines show the frequency of the fundamental mode of the bound state [{\color{gray}{grey}}], and the frequencies of the signal at $r\ped{ex} = 400M$ [{\color{deeppurple}{purple}}] and $r\ped{ex} = 1000M$ [{\color{redpurple}{red}}], whose peaks are situated at $\omega > \mu$, confirming these are \emph{axion waves}.}
    \label{fig:BurstAxionmuM03r40100}
\end{figure}

Although the axion is massive, EM waves are massless and allowed to travel freely once outside the cloud. To study EM wave propagation, we monitor the system at large radii. The \emph{top panel} of Figure~\ref{fig:EMfieldBurstmuM03r481020} summarises our findings for EM radiation, where we align waveforms in time. Some features are worth noting:~(i) we find that $\Phi_{2}$ decays like the inverse of the distance to the BH, as might be expected for EM waves;~(ii) the pattern of the waveform is not changing as it propagates, typical of massless fields. The radiation travels at the speed of light, as it should.
\vskip 2pt
Additionally, we observe an interesting morphology in the EM burst. It has a high-frequency component slowly modulated by a beating pattern. While the high-frequency component is set by the boson mass $\mu$, with oscillation period $\sim 2\pi/(\mu/2)$, the beating frequency scales with $1/\mu^{2}$ as its origin lies with the presence of the cloud. Specifically, when the photons are produced inside the cloud, they have to travel through it, allowing for further interactions. These photon \emph{echoes} exhibit a symmetric frequency distribution with respect to the primary photons, lying around $\omega \sim \mu/2$. This can be seen in the Fourier transform in the \emph{bottom panel} of Figure~\ref{fig:EMfieldBurstmuM03r481020}. The frequency difference between these peaks, $\Delta \omega$, corresponds to the observed beating timescale $\sim 2\pi/\Delta\omega$, indicated by the dashed lines in the \emph{top panel} of Figure~\ref{fig:EMfieldBurstmuM03r481020}. In addition to the bulk of photon frequencies near half the axion mass, there are other peaks in the frequency domain, namely two around $M\omega \sim 0.45$. We believe these higher order peaks do not originate from a parametric resonance, as one would expect peaks for each integer $N$ at $N\mu/2$, while we find the peaks at \emph{even} $N$ to be absent. Furthermore, the bandwidth of higher order parametric resonances is extremely small, making it hard to trigger those. We instead attribute these additional peaks to result from photon echoes as well, generated at later times, i.e.,~photons produced by the parametric mechanism that are up-scattered by the axion cloud. These results are different from a homogeneous axion background, where only echoes with the same frequency are produced. The discrepancy is due to the large momentum tail of the axion cloud.
\vskip 2pt
Besides the expected EM radiation, the reverse process -- two photons combining to create an axion -- may also provide a non-negligible contribution. In fact, this process has been explored in the context of axion clusters~\cite{Kephart1995}, where energetic axions are created that can not be stimulated anymore, hence they escape the cluster (so-called \emph{sterile axions}). Applying this scenario in the context of superradiant instabilities leads to the creation of unbound axion states, with frequencies $\omega>\mu$, that are thus able to escape to infinity. Such axion waves are indeed produced in our setup, as can be seen in Figure~\ref{fig:BurstAxionmuM03r40100}. The large scalar field contribution far away from the cloud is \emph{only} present in the supercritical regime. Since these are massive waves, the dependence with time and distance from the source is less simple due to dispersion. As components with different frequencies travel with different velocities, the wave changes morphology when travelling to infinity, which is apparent in the \emph{top panel} of Figure~\ref{fig:BurstAxionmuM03r40100}. 
\vskip 2pt
The Fourier transform of the axion waves is shown in the \emph{bottom panel} of Figure~\ref{fig:BurstAxionmuM03r40100}. It indeed contains components with frequency $\omega > \mu$, showing that the field is energetic enough to travel away from the source. The frequencies of the peaks correspond to a group velocity $v=\sqrt{1-\mu^2/\omega^2} = 0.036$ and $v = 0.18$ for $r\ped{ex} = 400M$ and $r\ped{ex} = 1000M$, respectively. Note that this Fourier transform is taken over the full time domain and thus dominated by the late signal of the axion waves, consisting of larger amplitude, non-relativistic waves. This also explains why the peak for the $r\ped{ex} = 1000M$ curve is at higher frequency:~the slower waves did not have time to arrive yet at this larger radius. If we instead calculate the Fourier transform on only the first part of the signal, we capture the (more) relativistic components. These emitted axion waves can in principle be detected by terrestrial axion detectors if the BH is close enough to Earth.
\vskip 2pt
Besides the dipole component, higher order scalar multipoles are also created by the photons. In fact, from our initial data, only scalar multipoles with odd $\ell$ can be produced. This selection rule is detailed in Appendix~\ref{appNR_sec:selectionrules}. The higher multipoles for both the axion and photon radiation are shown in Appendix~\ref{appNR_sec:higherorder}, where it can also be seen that excited photons can recombine to create axion waves with twice the axion mass. 
\section{Superradiance Turned On}\label{sec_SR_Axionic:withSR}
The formation of an EM burst depends on whether the photon production via the parametric instability exceeds the escape rate from the cloud, or vice versa. The initialisation of the system in a supercritical state however, is artificial. Instead, it starts in the subcritical regime and potentially grows supercritical through superradiance. Previous works claimed that this process developed through a ``burst-quiet sequence'':~a burst of EM and axion waves would deplete the cloud, which would then grow on a superradiant timescale before another burst occurred~\cite{Ikeda:2018nhb}. We argue that in fact bursts do not occur, and that the process is smoother than thought. As we will show, the presence of superradiance introduces two important differences:~(i) the growth rate of the EM field is modified and~(ii) the system is forced into a stationary phase.
\subsection{Numerical Results}\label{sec_SR_Axionic:sim}
We numerically evolve the coupled axion-photon system under the influence of a superradiantly growing cloud. In these simulations, we start the system in the  subcritical regime, and let it evolve to supercriticality via (artificial) superradiance, since now $C\neq 0$. 
\begin{figure}[t!]
  \centering
    \includegraphics[scale=0.45, trim = 0 0 0 0]{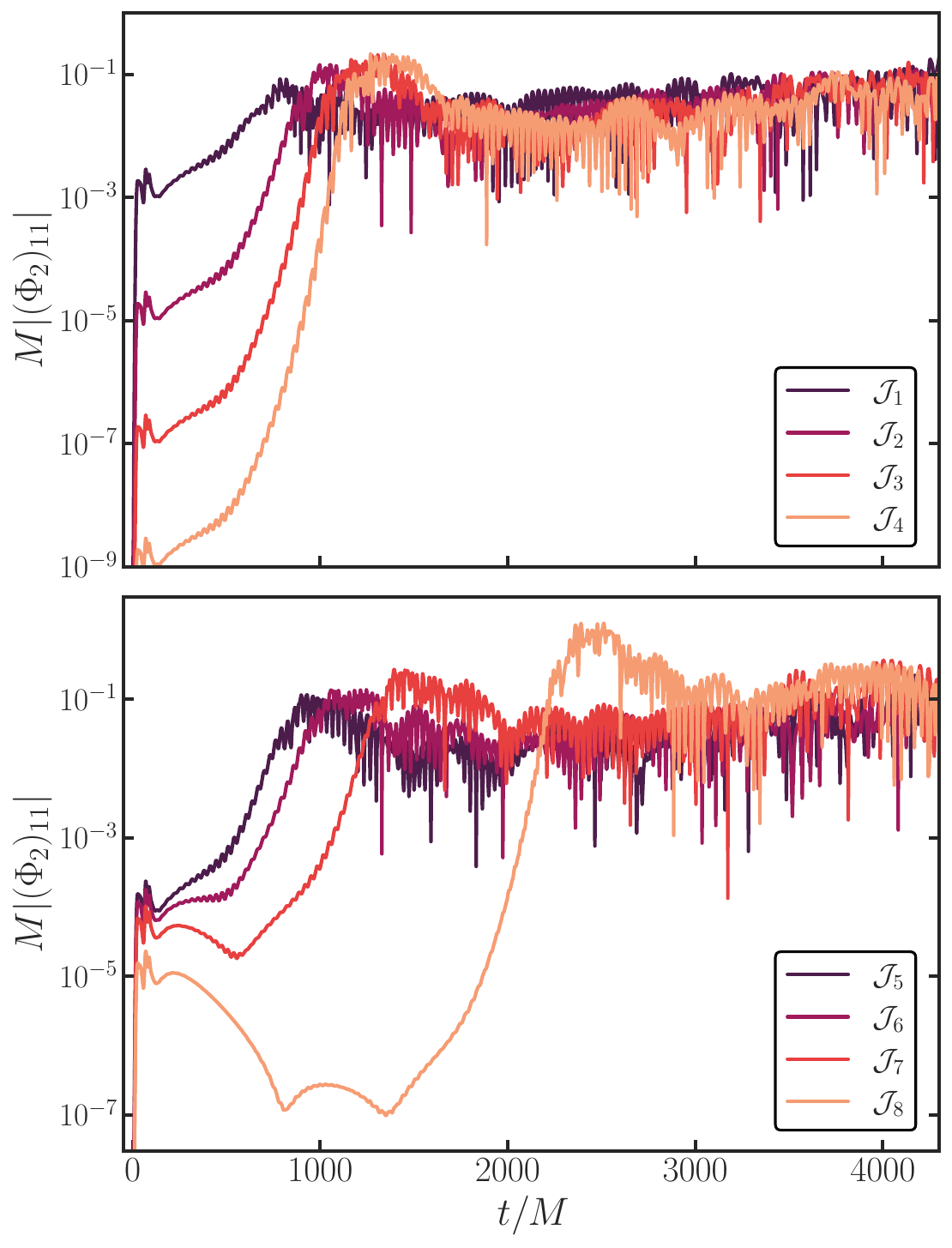}
    \caption{Time evolution of the dipolar component of the EM field including superradiant growth for different strengths of the initial EM pulse (\emph{top panel}) or the initial coupling strength $k\ped{a}\Psi_{0}$ (\emph{bottom panel}) (see Table~\ref{tb:simulations}). The field is extracted at $r\ped{ex} = 20M$ and $\mu M = 0.2$. Notice how a stationary state is reached instead of a burst of EM waves. Moreover, the final EM value is independent of initial conditions, even though the timescale required to reach saturation does depend on how the fields were initialised.}   
\label{fig:Phi2VaryingEMpulse_Phi2VaryingInitialScalar}
\end{figure}
\vskip 2pt
Figure~\ref{fig:Phi2VaryingEMpulse_Phi2VaryingInitialScalar} illustrates the behaviour of the system. We evolve different initial conditions, corresponding to different seed EM fields $E_{0}/\Psi_{0}$ and different couplings $k\ped{a}\Psi_{0}$, and we see a \emph{saturation} of the EM field, to a value which is independent of the initial conditions. This stability is simply achieved by turning on superradiant growth like in eq.~\eqref{eq:ASR}. In contrast to the previous section, where the bound state solely loses energy, the supplement to the axion cloud is dominant at first, resulting in exponential growth. As the cloud approaches the critical value, parametric decay to the EM field begins to compensate for the energy gain from the BH, ultimately reaching a phase where energy gain and loss are balanced. As a result, the entire system consisting of the axion cloud and the EM field is constantly pumped by superradiant growth, with a steady emission of EM waves travelling outwards.
\begin{figure}[t!]
  \centering
\includegraphics[scale=0.45, trim = 0 0 0 0]{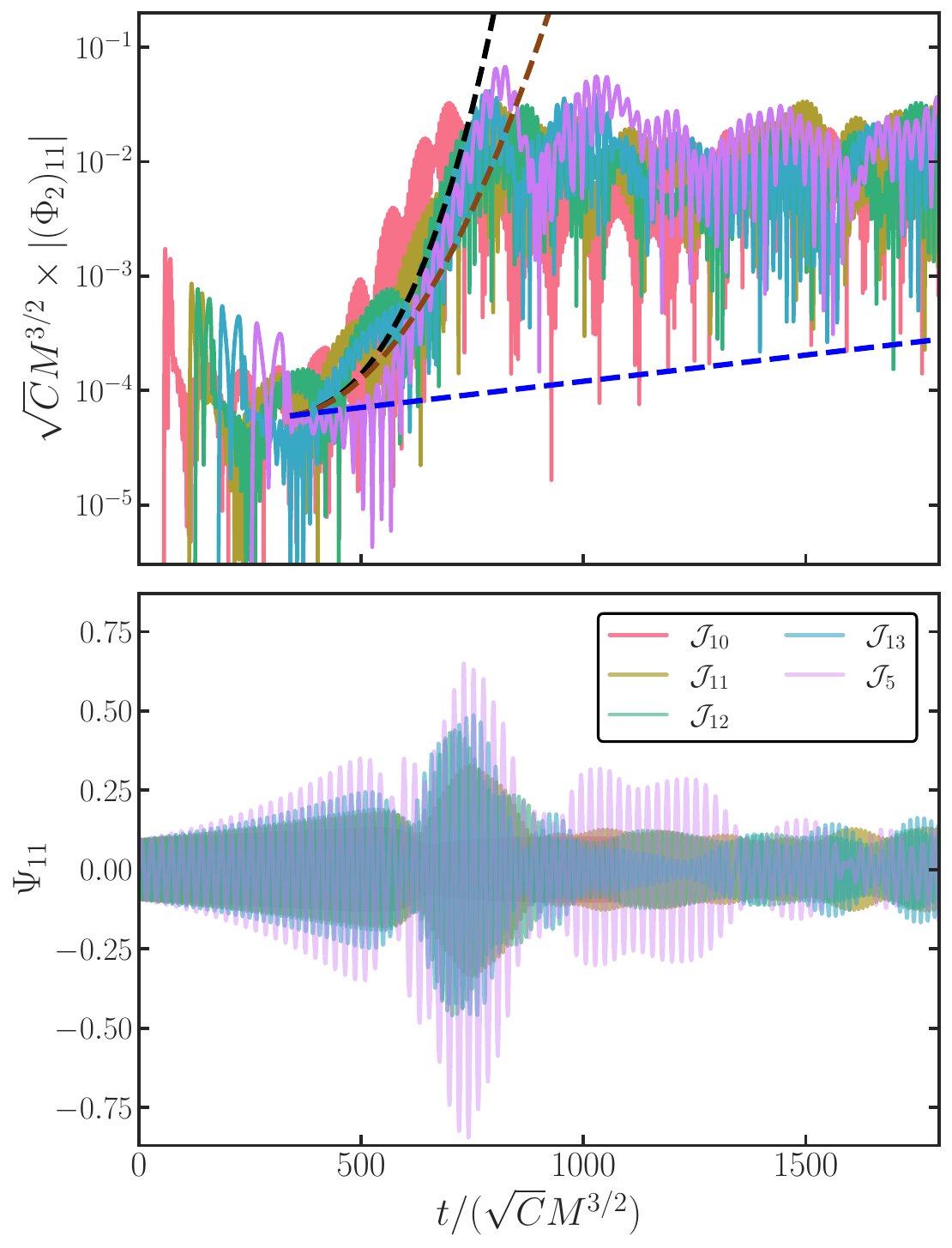}
\caption{Dipolar component of the EM field (\emph{top panel}, extracted at $r\ped{ex} = 400M$) or the scalar field (\emph{bottom panel}, extracted at $r\ped{ex} = 20M$) with superradiance turned on and $\mu M = 0.2$. Dashed lines show the growth rate predicted by the standard Mathieu equation [{\color{brightblue}{blue}}], the superradiant Mathieu equation~\eqref{eq:modifiedgrowthrate} [\textbf{black}] and its first-order expansion~\eqref{eq:modifiedgrowthrateFirst} [{\color{MathematicaBrown}{brown}}]. The value of $C$ is varied for simulations $\mathcal{J}_{i}$, see Table~\ref{tb:simulations}. Both the horizontal and vertical axis are rescaled by $\sqrt{C}$, aligning all simulations with $\mathcal{J}_{14}$. This reveals a clear and simple dependence on the superradiance rate, which we investigate analytically in Sections~\ref{sec_SR_Axionic:analyticalgrow} and~\ref{sec_SR_Axionic:analytical}.}
    \label{fig:Phi2VaryingC}
\end{figure}
\vskip 2pt
The saturation value of the EM field \emph{does} depend on the superradiance parameter $C$. This is simply due to the fact that the more axions that are created by superradiance, the more photons that can be produced through the parametric mechanism. We find the saturation value to be proportional to $\sqrt{C}$, shown in the \emph{top panel} of Figure~\ref{fig:Phi2VaryingC}. This result is also supported by analytical estimates in Section~\ref{sec_SR_Axionic:analytical}, in particular eq.~\eqref{eq:saturationvalue}. Additionally, our results demonstrate that the timescale required to reach saturation (at fixed initial field values), scales with $\sqrt{C}$ as well. This behaviour is explained in Section~\ref{sec_SR_Axionic:analyticalgrow}.
\vskip 2pt
These simulations provide us with robust evidence that the saturation phase is~(i) independent of the initial data, and~(ii) occurring for all tested values of $C$ that span two orders of magnitude, allowing for universal predictions. In the following, we discuss various features related to the saturation phase. 
\subsubsection{Evolution of the cloud's morphology}
\begin{figure}[t!]
    \includegraphics[width = \linewidth]{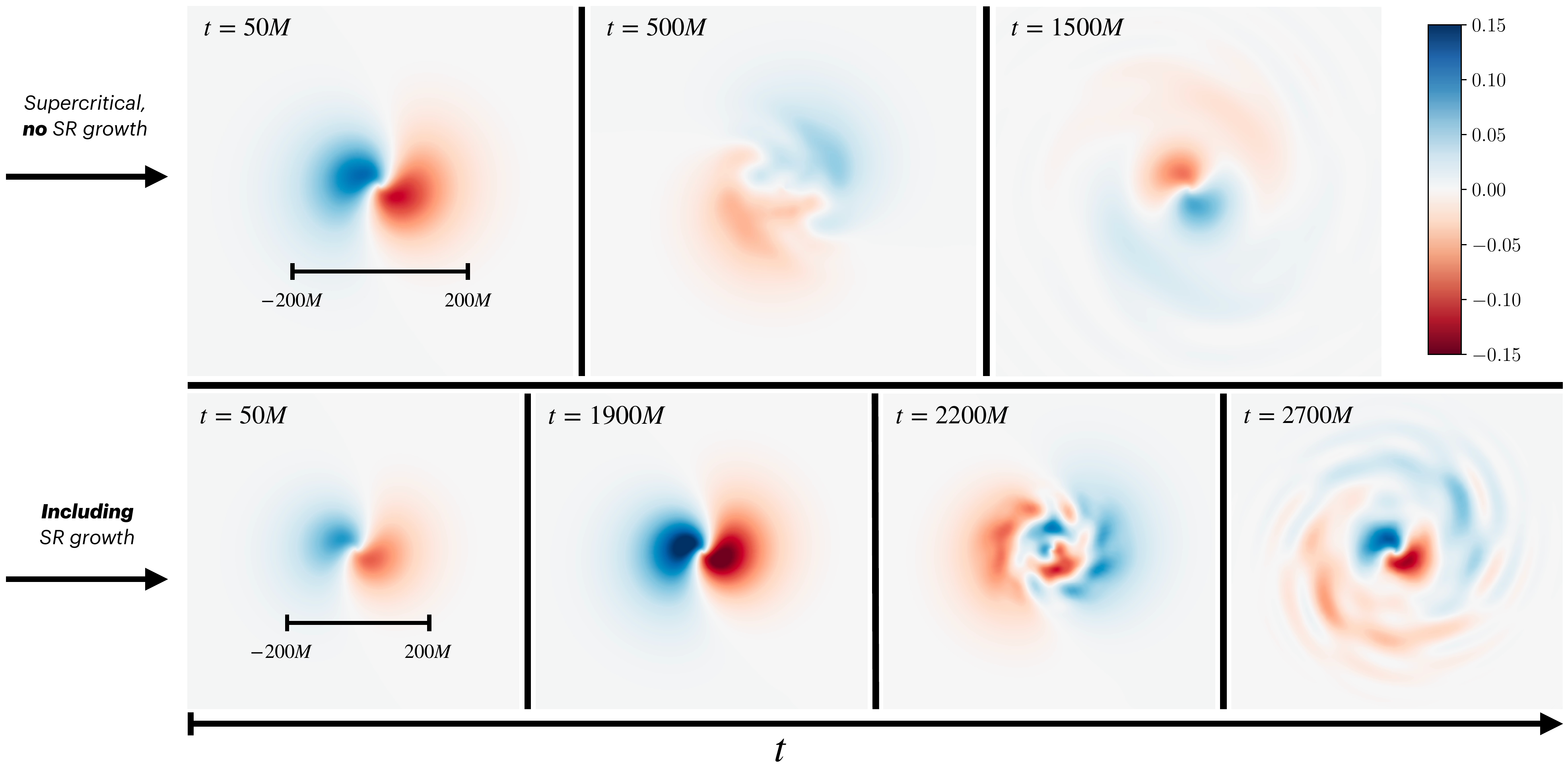}
    \caption{Snapshots of the axion profile during the evolution of the axion-photon system. \emph{Top row} shows the system in the supercritical regime (simulation $\mathcal{I}_{3}$), but \emph{without} superradiant growth. The cloud starts in its initial dipole state, and gets disrupted by the parametric instability. Afterwards the configuration settles down while axion waves propagate to infinity. On the \emph{bottom row}, we show an initially subcritical cloud, yet \emph{with} superradiant growth turned on, $C = 10^{-3} M^{-1}$ (simulation $\mathcal{J}_{12}$). Again, the cloud starts in its dipole mode, yet now it grows in amplitude due to superradiance. Afterwards, the cloud is disrupted due to the EM instability, eventually settling down to a saturation phase during which axion waves are continuously produced.}
    \label{fig:Snapshots}
\end{figure}
When the system has just reached the critical threshold, the EM field starts growing (super-)exponentially until it reaches the saturation value. When this happens, the nonlinear backreaction in the Klein-Gordon equation becomes important~\eqref{eq:SRAxion_evoleqns}. In absence of superradiance, the EM field quickly decays in time after reaching its maximum and with that its backreaction onto the axion field, allowing the cloud to settle back to a stable configuration at late times (see Figure~\ref{fig:BurstmuM03r20}). Conversely, in presence of superradiance, the EM field settles to a large and constant value that continuously backreacts onto the axion field. Consequently, it exhibits strong deviations from the initial pure bound state configuration as overtones are triggered, i.e.,~it acquires a beating-like pattern. This can be seen in Figure~\ref{fig:Phi2VaryingC} (\emph{bottom panel}), where around $t\sim 700 \sqrt{C}M^{3/2}$ the saturation phase ensues and there is no relaxation to the pure quasi-bound state. We show a series of snapshots from the cloud's evolution in the two distinct scenarios in Figure~\ref{fig:Snapshots}.
\subsubsection{Angular structure of outgoing EM waves}
\begin{figure}[t!]
  \centering
    \includegraphics[scale=0.45, trim = 0 0 0 0]{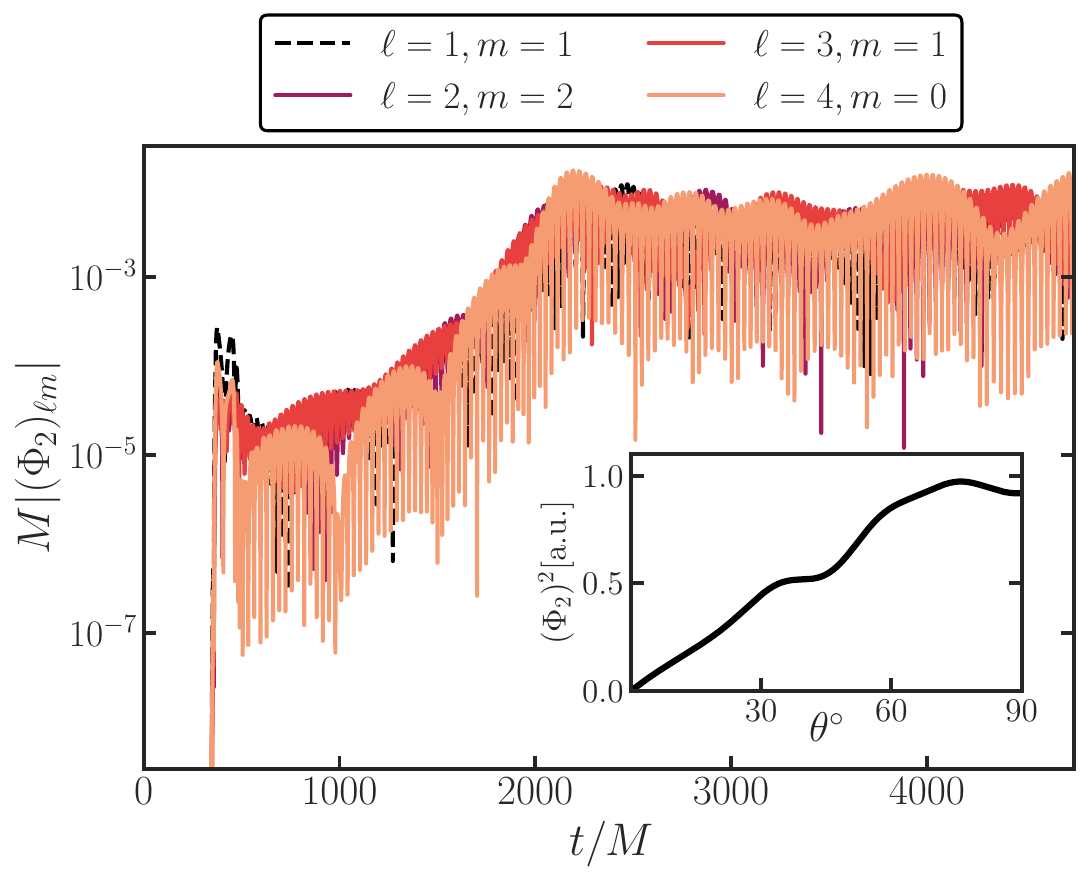}
    \caption{A subset of the multipoles of the EM field for simulation $\mathcal{J}_{12}$, extracted at $r\ped{ex} = 400M$. The inset shows the angular structure of the outgoing EM waves in the saturation phase (where all multipoles up to $\ell \leq 8$ are taken into account). The vertical axis is reported in arbitrary units [a.u.]. Photons are predominantly produced on the equatorial plane where the cloud's density is highest.}
    \label{fig:Phi2MultipolesLargeradii}
\end{figure}
During the saturation phase, there is a constant emission of EM waves. For observational purposes, we study the angular structure of the outgoing radiation. We do this through the multipole components of $\Phi_{2}$, while up to now only the dipole was considered. A subset of these multipoles is shown Figure~\ref{fig:Phi2MultipolesLargeradii}. From the differences in amplitude between different modes, we conclude that the radiation is not isotropic. In fact, we find that the dominant radiation is on the equatorial plane (see inset of Figure~\ref{fig:Phi2MultipolesLargeradii}), where the density of the axion cloud is highest. To strengthen this result, we also compute analytically the excitation coefficients of different multipoles given our initial data. We report them in Appendix~\ref{appNR_sec:selectionrules}. 
\subsection{Growth Rate}\label{sec_SR_Axionic:analyticalgrow}
In previous work~\cite{Boskovic:2018lkj}, it was shown that in absence of superradiance, the growth rate of the EM field can be approximated by a simple, analytical expression. This is a consequence of the fact that when the background spacetime is Minkowski and the background axion field is a coherently oscillating, homogeneous condensate, the Maxwell equations can be rearranged in the form of a Mathieu equation~\cite{Boskovic:2018lkj, hertzberg:2018zte}. The growth rate is then found by taking the production rate of the photons ($\sim$~the Floquet exponent of the dominant, unstable mode of the Mathieu equation) and subtracting the escape rate of the photons from the cloud ($\sim$~inverse of the cloud size). While this approach yields accurate predictions in absence of superradiance (see Figure~\ref{fig:BurstmuM03r20}), it breaks down in presence of superradiance (blue dashed line in Figure~\ref{fig:Phi2VaryingC}). Remarkably, as we will show, a simple adjustment to the Minkowski toy model restores its validity. Details are provided in Appendix~\ref{app_Mathieu_sec:SR}.
\vskip 2pt
Let us consider the Maxwell equations in flat spacetime. We adopt Cartesian coordinates and assume the following \emph{ansatz} for the EM field,
\begin{equation}
\label{eq:MathieuEM}
    A_\mu (t,\vec{x})=\alpha_\mu(t, \vec{p}) e^{i (\vec{p} \cdot \vec{x} - \omega t)}\,,
\end{equation}
where $\vec{p}$ is the wave vector which we assume to be aligned in the $\hat{z}$--direction without loss of generality, i.e.,~$\vec{p}=(0,0,p_z)$. To mimic the amplification of the axion field via superradiance, we consider a homogeneous condensate that exponentially grows in time:\footnote{We adopt a different notation to distinguish the amplitude of the homogeneous axion field in this toy model, $\psi_0$, with the one of the axion cloud around the BH, $\Psi_0$.}
\begin{equation}
    \Psi=\frac{1}{2}(\psi_0 e^{-i \mu t}+ \psi_0^* e^{i \mu t}) e^{C t}\,.
\end{equation}
Adopting the field redefinition $y_k =e^{i \omega t}\alpha_k$, rescaling the time as $T=\mu t$ and projecting along a circular polarisation basis $e_\pm$ such that $y=y_\omega e_\pm$, we obtain a ``superradiant'' Mathieu-like (SRM) equation:
\begin{equation}\label{eq:modifiedMathieu}
    \partial_T^2 y_\omega\!+\!\frac{1}{\mu^{2}}\Big(p_z^2+ 2  p_z e^{\frac{CT}{\mu}}\psi_0 k\ped{a}(C \cos T\!-\!\mu \sin T)\Big)y_\omega=0\,.
\end{equation}
Unsurprisingly, for $C=0$, this equation reduces to the original Mathieu equation.\footnote{Our result includes a sine instead of the cosine found in~\cite{Boskovic:2018lkj}, which originates from a small sign mistake in their derivation, see eq.~(19). This has no consequences for the physics, as it only induces a $\pi/2$ phase shift.} From~\eqref{eq:modifiedMathieu}, two new features emerge:~an extra oscillating term $\sim C \cos T$, and, most importantly, an exponentially growing factor $\sim e^{CT/\mu}$. By solving~\eqref{eq:modifiedMathieu} numerically, we find that, similar to the standard Mathieu equation, this equation admits instability bands, albeit with a larger growth rate. Fitting the exponent of the numerical solution, we conclude that the solution to the superradiant Mathieu equation is well-described by a super-exponential expression $y_\omega \sim e^{\lambda_{\scalebox{0.55}{$\mathrm{SRM}$}} t}$, with
\begin{equation}
\label{eq:modifiedgrowthrate}
    \lambda_{\scalebox{0.65}{$\mathrm{SRM}$}}=\frac{\mu}{2}k\ped{a }\psi_0  e^{C t/2}\,.
\end{equation}
In Appendix~\ref{app_Mathieu_sec:SR}, we compare this expression against the numerical solutions, and we derive~\eqref{eq:modifiedgrowthrate} analytically using a multiple-scale method. 
\vskip 2pt
We compare the growth rate of the EM field in the full axion-photon system on a Schwarzschild background with the prediction from our toy model~\eqref{eq:modifiedgrowthrate}, as shown in Figure~\ref{fig:Phi2VaryingC}. Here, the standard Mathieu growth rate [{\color{brightblue}{blue}}], the full solution [\textbf{black}] and first-order expansion in $C$ [{\color{MathematicaBrown}{brown}}] can be seen, where the latter is defined by
\begin{equation}
\label{eq:modifiedgrowthrateFirst}
    \lambda_{\scalebox{0.65}{$\mathrm{SRM}$}} t \approx \frac{\mu}{2}k\ped{a}\psi_0 \left(t+\frac{C t^2}{2}\right)\,.
\end{equation}
In all curves, the time it takes for photons to escape the cloud has been taken into account. Furthermore, we use the critical value for the coupling $k\ped{a}\Psi_{0}$. As can be seen, the SRM growth rate closely matches the numerical results. In addition, the $C t^2$ term from~\eqref{eq:modifiedgrowthrateFirst} that appears at first-order, naturally explains why the timescale to reach saturation scales as $\sqrt{C}$. 
\vskip 2pt
In summary, when considering the axion-photon system under the influence of superradiance, a simple extension to Mathieu equation allows for elegant, analytic predictions of our numerical results. Note that the true value for superradiance is many orders of magnitude lower than the one considered in our simulations and thus we expect the correction to be subdominant in realistic astrophysical scenarios. Finally, our prediction neglects the backreaction onto the axion cloud. Hence, it naturally breaks down when the rate of energy loss due to conversion to photons becomes comparable to superradiant growth, i.e.,~when the saturation phase ensues.
\subsection{Saturation Phase}\label{sec_SR_Axionic:analytical}
As can be seen in Figures~\ref{fig:Phi2VaryingEMpulse_Phi2VaryingInitialScalar} and~\ref{fig:Phi2VaryingC}, turning on superradiant growth forces the system into a stationary configuration. Here, the energy loss of the cloud due to the parametric instability balances the superradiant pump sourced by the rotational energy of the BH. A description of this phase is remarkably simple as we will show below. A similar conclusion was found in~\cite{Rosa:2017ury}. 
\vskip 2pt
For the photons to reach an equilibrium phase, it is required that the parametric decay rate, $\lambda\ped{pd}$, equals the escape rate of the photon, $\lambda\ped{esc}$. Assuming that the former can be approximated by the decay rate in the homogeneous condensate case~\cite{Boskovic:2018lkj,hertzberg:2018zte}, we have
\begin{equation}\label{eq:saturation}
    \lambda\ped{pd} \approx \frac{k\ped{a} \Psi\ped{sat} \mu}{2} \approx \frac{1}{d} \approx \lambda\ped{esc}\,,
\end{equation}
where $\Psi\ped{sat}$ is the average amplitude of the scalar field \emph{within the cloud} at saturation. In the non-relativistic regime, the size of the cloud is well-approximated by the standard deviation of the radius $d = \langle r \rangle \approx 2 \sqrt{3} r\ped{c} = 4 \sqrt{3} / (\mu^{2}M)$. This yields a relation for which the cloud reaches saturation, namely
\begin{equation}\label{eq:scalarsaturation}
    k\ped{a} \Psi\ped{sat}  \approx \frac{\mu M}{2\sqrt{3}}\,.
\end{equation}
This is consistent with what we observe in the \emph{bottom panel} of Figure~\ref{fig:Phi2VaryingC}, where $k\ped{a}\Psi\ped{sat} \approx 0.2/(2\sqrt{3}) \approx 0.06$.
\vskip 2pt
Additionally, we consider the equilibrium condition of the axion cloud. It is sourced by the superradiance rate $\Gamma\ped{SR}$, yet loses energy due to the parametric instability, $\Gamma\ped{PI}$. In our setup, these two rates are
\begin{equation}\label{eq:rates}
    \Gamma_{\scalebox{0.65}{$\mathrm{SR}$}} = \frac{C}{2} \quad \mathrm{and} \quad \Gamma_{\scalebox{0.75}{$\mathrm{PI}$}} \approx 2\Gamma_{\scalebox{0.65}{$\Psi \rightarrow \gamma \gamma$}} f_{\gamma}\,,
\end{equation}
where $\Gamma_{\scalebox{0.65}{$\Psi \rightarrow \gamma \gamma$}} = \hbar k\ped{a}^2 \mu^3/(16 \pi)$ is the perturbative decay width of the axion-to-photon conversion and $f_\gamma$ is the photon occupation number. When the dominant production is in a narrow band around $p_{\gamma} = \mu/2$, we have~\cite{Carenza:2019vzg}
\begin{equation}
    f_{\gamma} (p_\gamma=\mu/2) = \frac{8\pi^3 n_{\gamma}}{4\pi p^{2}_{\gamma} \Delta p_{\gamma}} \approx \frac{2\pi^2 A^{2}_{\gamma}}{\hbar k\ped{a} \Psi\ped{sat} \mu^2}\,,
\end{equation}
where $\Delta p_{\gamma} \approx 2 k\ped{a} \Psi\ped{sat} \mu$ is the photon dispersion bandwidth, approximated to be the resonant bandwidth at $p_{\gamma} = \mu/2$. Moreover, $A_{\gamma}$ is the photon amplitude, which relates to our measure of the EM field $\Phi_{2}$ via $A_{\gamma} \sim 2\Phi_{2}/\omega$, where $\omega\sim \mu$ is the frequency of the axion field. Finally, $n_\gamma = 2\rho/(\hbar \omega)$ is the photon number density, with $\rho =  A_{\gamma}^{2}\omega^{2}/2$ the energy density. Substituting these relations into~\eqref{eq:rates}, we find that inside the cloud
\begin{equation}\label{eq:saturationvalue}
    \frac{A^{2}_{\gamma}}{\Psi\ped{sat}^2} \approx \frac{2C}{\pi \mu k\ped{a}\Psi\ped{sat}} \approx \frac{C}{\pi \lambda\ped{esc}}\,,
\end{equation}
where we used again~\eqref{eq:saturation}. Hence, this simple analytical estimate shows that the EM field stabilises to a value proportional to $\sqrt{C}$. This result is in excellent agreement with our simulations (see Figure~\ref{fig:Phi2VaryingC}) and it allows us to consider a case in which $C$ coincides with the superradiant growth timescale.
\vskip 2pt
The total energy flux of the photons with frequency $\mu/2$ is defined as~\cite{Rosa:2017ury} 
\begin{equation}
    \frac{\mathrm{d}E}{\mathrm{d}t}=\frac{\hbar\mu}{2}n_{\gamma}\lambda\ped{esc}\,\chi r\ped{c}^{3}\,,
\end{equation}
where $\chi r^{3}\ped{c}$ is the volume of the cloud. Here, $\chi$ is a numerical factor which in the non-relativistic regime yields $\chi \approx \mathcal{O}(10^{2})$.\footnote{
To obtain $\chi$, we introduce a threshold value $\epsilon$ for the absolute value of the scalar field, and define
\begin{equation}
    \chi =\int_{0}^{\infty}\mathrm{d}r\,r^{2}\int\mathrm{d}\Omega\,\Theta \left(|\Psi|-\epsilon \right)\nonumber\,,
\end{equation}
where $\dd \Omega$ is the solid angle element, $\Theta$ is the Heaviside step function and $\epsilon\sim 0.5  \max \left(\Psi\right)$. We checked that the order of $\chi$ does not strongly depend on $\epsilon$.}
Assuming that the superradiance rate $\Gamma_{\scalebox{0.65}{$\mathrm{SR}$}}$ equals the decay rate $\Gamma_{\scalebox{0.65}{$\mathrm{PI}$}}$ in the saturation phase~\eqref{eq:rates}, we obtain
\begin{equation}\label{eq:analyticL}
    \frac{\mathrm{d}E}{\mathrm{d}t}\approx 7.6\times 10^{45}\left(\frac{\chi}{100}\right)\left(\frac{2.5\times 10^{2}M}{\tau\ped{s}}\right)\left(\frac{0.2}{\mu M}\right)^{2}\left(\frac{10^{-13}{\rm GeV}^{-1}}{k\ped{a}}\right)^{2}{\rm erg/s}\,,
\end{equation}
where $\tau\ped{s}=C^{-1}$. To probe the superradiance regime, we tune $C$ to match the well-known superradiant growth rate~\eqref{eqn:Gamma_nlm} in the dominant growing mode~\cite{Detweiler:1980uk}
\begin{equation}\label{eq:SRrate}
    \Gamma_{\scalebox{0.65}{$\mathrm{SR}$}} \approx \frac{a_{\scalebox{0.65}{$\mathrm{J}$}}\,(\mu M)^9}{24 M}\,, \quad \text{when} \quad \mu M \ll 1\,,
\end{equation}
where $a_{\scalebox{0.65}{$\mathrm{J}$}}$ is the spin of the BH. Substituting this into~\eqref{eq:analyticL} yields
\begin{equation}\label{eq:analyticSRLuminosity}
\frac{\mathrm{d}E}{\mathrm{d}t}\approx 8.1\times 10^{40}\left(\frac{\chi}{100}\right)\left(\frac{a_{\scalebox{0.65}{$\mathrm{J}$}}}{M}\right) \left(\frac{\mu M}{0.2}\right)^{7}\left(\frac{10^{-13}{\rm GeV}^{-1}}{k\ped{a}}\right)^{2}{\rm erg/s}\,.
\end{equation}
Counterintuitively, lower couplings result in higher fluxes. This occurs because, for lower couplings, the axion field reaches saturation at higher values. As the EM flux is proportional to the axion field value, this leads to a higher flux.\footnote{A similar behaviour was found in the context of dark photons with kinetic mixings to the Standard Model photon~\cite{Caputo:2021efm}.} As a consequence, the saturation phase opens a channel to constrain axionic couplings ``from below''. Finally, the divergence of~\eqref{eq:analyticL} and~\eqref{eq:analyticSRLuminosity} at small couplings indicates when our model breaks down. In particular, the equilibrium condition~\eqref{eq:scalarsaturation} suggests a minimum value for the coupling for which the cloud's mass becomes larger than its maximum of $M\ped{c}  = 0.1M$. For example, for $\mu M = 0.2$, the minimum coupling for which our description holds is $k\ped{a} > 3.8\times 10^{-18}~{\rm GeV}^{-1}$. Lower couplings then this yield an unphysical situation and thus a breakdown of our description.
\subsection{Implications for Superradiance}\label{sec_SR_Axionic:realSR}
\begin{figure}[t!]
  \centering
    \includegraphics[scale=0.45, trim = 0 0 0 20]{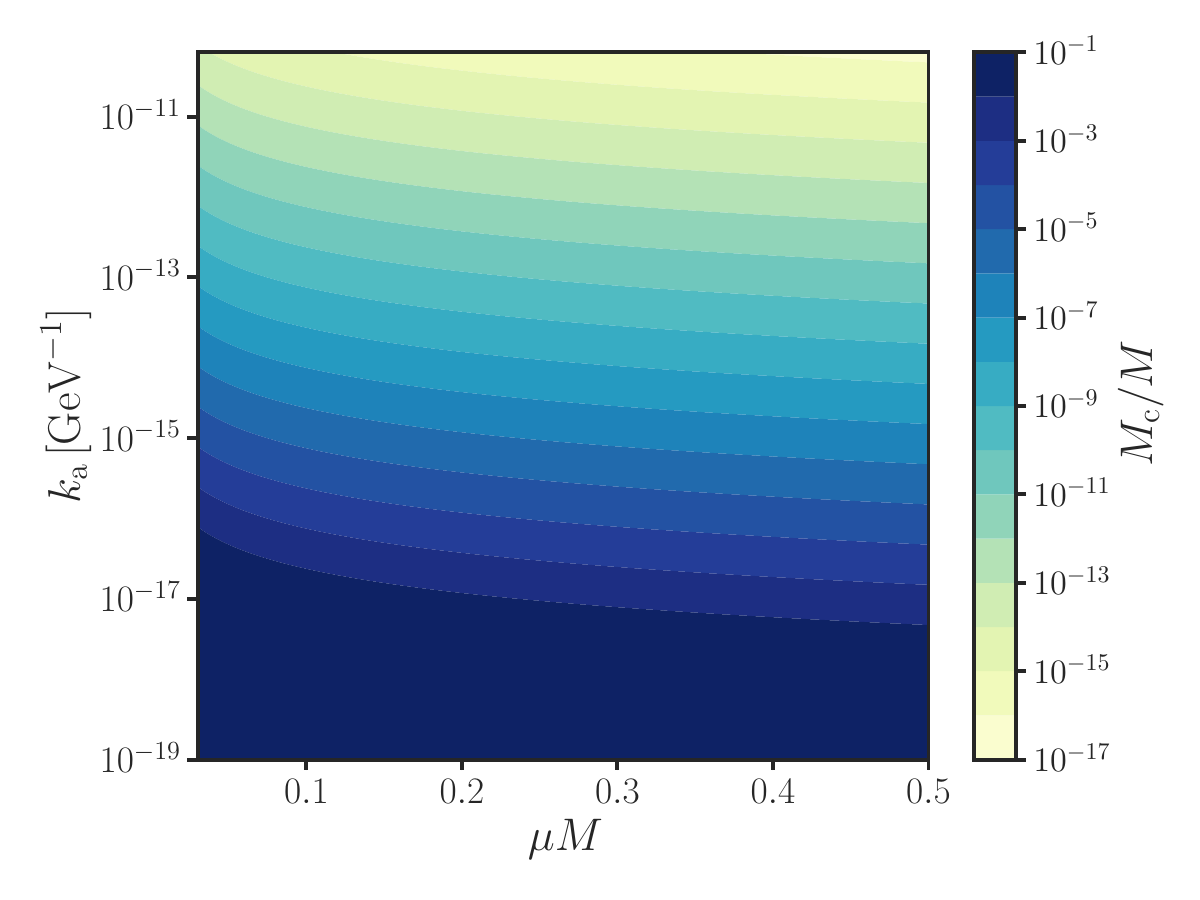}
    \caption{Contour plot of the mass of the cloud ($M\ped{c}$) \emph{at saturation} depending on the axionic and mass coupling, $k\ped{a}\Psi\ped{sat}$ and $\mu M$, respectively. The dark blue area at the bottom denotes the maximum mass of $10\%$ that the cloud can achieve in the purely gravitational case.}
    \label{fig:ContouronmuMKaMs}
\end{figure}

Using the scaling relation~\eqref{eq:saturationvalue}, we can explore the system in the superradiant regime and thereby test the validity of~\eqref{eq:analyticSRLuminosity} against our numerical simulations. Before doing so, however, we must justify our ability to extrapolate beyond the parameter space directly probed for $C$.
\vskip 2pt
Both ends of the explored $C$-range present numerical challenges. At high $C$, the system exhibits large growth on short timescales, leading to numerical instabilities. Conversely, at low $C$, the evolution timescales become prohibitively long, making simulations computationally expensive. In addition to the simulations shown in Figure~\ref{fig:Phi2VaryingC}, we carried out two more simulations -- $\mathcal{J}_{9}$ (low $C$) and $\mathcal{J}_{14}$ (high $C$) -- to better characterise the behaviour at the boundaries of our parameter space. For $\mathcal{J}_{14}$, corresponding to high $C$, we observe signs that the saturation phase breaks down:~the EM field resumes growth after a transient saturation period. Physically, this behaviour is expected. In the regime of extreme superradiant growth, the balance between photon production and escape is disrupted:~the photons simply do not have time to escape the cloud while plenty of axions are produced. This imbalance could potentially lead to a burst-like radiation pattern.\footnote{Such behaviour might occur in a Bosenova-like collapse~\cite{Yoshino:2012kn}, where the axion density sharply rises on short timescales.} Since our focus here is on the superradiance regime, we do not investigate this scenario further.
\vskip 2pt
The regime of low $C$ is of interest as the superradiance rate is at significantly lower values than what we can probe numerically~\eqref{eq:SRrate}. From the lowest $C$ we probe, $\mathcal{J}_{9}$, we find that even though there is an apparent decrease after the super-exponential growth, the $\sqrt{C}$ scaling is respected at late times. Physically, this behaviour is consistent with the persistence of a saturation phase. When the growth rate is small, the system becomes adiabatic; as the axions are slowly produced, the system steadily approaches the critical value at which the system is in equilibrium and a saturation phase ensues.
\vskip 2pt
To extract the energy flux from our simulations, we exploit the properties of the Newman-Penrose scalar. In particular, we can define 
\begin{equation}\label{eq:NPscalarEnergy}
    \frac{\mathrm{d}^{2}E}{\mathrm{d}t \mathrm{d}\Omega} = \lim_{r\rightarrow \infty}\frac{r^{2}}{2\pi}|\Phi_{2}|^{2}\,,
\end{equation}
where $\mathrm{d}\Omega \equiv \sin{\theta}\mathrm{d}\theta\mathrm{d}\varphi$. Decomposing~\eqref{eq:NPscalarEnergy} in terms of spin-weighted spherical harmonics, we obtain
\begin{equation}\label{eq:energyflux}
    \frac{\mathrm{d}E}{\mathrm{d}t} = \sum_{\ell m} \int \mathrm{d}\Omega\,\frac{1}{2\pi} | (\Phi_{2}^{\circ})_{\ell m} \> {}_{{\scalebox{0.65}{$-$}}1}\mkern-2mu Y_{\ell m}|^{2}\,,
\end{equation}
where $\Phi_{2} = \Phi_{2}^{\circ}/r$. From our simulations, we extract $(\Phi_{2})_{\ell m, \rm sim}$ for a certain $C\ped{sim}$ at large radii ($r = r\ped{ex}$) by averaging over a sufficient period in the saturation phase. Then, we scale these multipoles to match their saturation value in the case of superradiance according to 
\begin{equation}\label{eq:Phi2RealSR}
\begin{aligned}
    (\Phi_{2})_{\ell m, \scalebox{0.65}{$\mathrm{SR}$}} &\approx \frac{(\Phi_{2})_{\ell m, \rm sim}}{\sqrt{C\ped{sim}/(2\Gamma_{\scalebox{0.65}{$\mathrm{SR}$}})} }\,,
\end{aligned}
\end{equation}
where $\Gamma_{\scalebox{0.65}{$\mathrm{SR}$}}$ is defined in~\eqref{eq:SRrate}. We do this for each multipole and sum them according to~\eqref{eq:energyflux} to obtain the total flux. As the contribution to the energy flux becomes smaller for higher multipoles, we sum each multipole until the increment is less than 5\%. In practice, this means summing over the first $\sim 8$ values of $\ell$. Following this procedure, we find an estimate from our simulations for the total, nearly constant, energy flux in the saturation phase:
\begin{equation}
    \frac{\mathrm{d}E}{\mathrm{d}t}  = 9.10 \times 10^{40}\;\left(\frac{10^{-13}~\mathrm{GeV}^{-1}}{k\ped{a}}\right)^{2}\; \mathrm{erg/s}\,,
    \label{eq:dEdt}
\end{equation}
where we assumed $\mu M = 0.2$ and the BH to be maximally spinning. This matches closely the theoretical prediction from~\eqref{eq:analyticSRLuminosity}.
\vskip 2pt
Besides the photon production, the parametric instability also affects the axion cloud. As we showed in eq.~\eqref{eq:scalarsaturation}, the axion amplitude at saturation is independent of $C$. By translating the amplitude of the axion field to the mass of the cloud, the impact of coupling axions to photons becomes much more apparent~\cite{Brito:2014wla}. In the purely gravitational case, the cloud is able to obtain a maximum mass of $M\ped{c} \lesssim 0.1 M$~\eqref{eq:massatsat}. As can be seen in Figure~\ref{fig:ContouronmuMKaMs}, through the coupling to the Maxwell sector, the cloud's mass can saturate significantly below this maximum. Note that this estimate assumes a hydrogen-like profile for the cloud, which is strictly valid only in the no-coupling case. As a result, when the cloud is disrupted by the strong backreaction on the axion field in the saturation phase, this approximation breaks down. However, for the (much) lower superradiant growth rate, the associated EM flux is weaker, leading to less disruption of the cloud. 
\vskip 2pt
Figure~\ref{fig:ContouronmuMKaMs} has implications for existing constraints on the mass of ultralight bosons derived from either GW searches~\cite{LIGOScientific:2021rnv,Tsukada:2018mbp, Palomba:2019vxe, Yuan:2022bem,Ng:2020jqd} or BH spin measurements~\cite{Arvanitaki:2010sy,Arvanitaki:2014wva,Brito:2014wla,Brito:2017zvb,Cardoso:2018tly,Stott:2020gjj,Ng:2020ruv,Ng:2019jsx,Davoudiasl:2019nlo,Wen:2021yhz,Fernandez:2019qbj}. Due to the reduced cloud mass, the backreaction on the BH spin-down may become negligible, rendering current constraints -- which typically assume no couplings -- potentially inapplicable. Additionally, the environmental impact of the superradiant cloud on gravitational waveforms in BH binaries is expected to be suppressed~\cite{Zhang:2018kib,Baumann:2018vus,Zhang:2019eid,Baumann:2019ztm,  Baumann:2021fkf,Baumann:2022pkl,Takahashi:2021eso,Cole:2022yzw,Baumann:2022pkl,Takahashi:2023flk,Tomaselli:2023ysb,Brito:2023pyl,Duque:2023seg,Tomaselli:2024bdd,Tomaselli:2024dbw,Boskovic:2024fga,Khalvati:2024tzz}.
\section{Surrounding Plasma}\label{sec_SR_Axionic:surroundingplasma}
The presence of plasma affects the axion-to-photon conversion in the parametric instability mechanism, as the transverse polarisations of the photon are dressed with an effective mass~\eqref{eq:dispersion_mass}, i.e., the plasma frequency $\omega\ped{p}$~\eqref{eq:SR_Axionic_plasmafreq}. Therefore, when $\omega\ped{p}> \mu/2$, the process $\Psi \rightarrow \gamma + \gamma$ is kinematically forbidden. Even though it is common lore to approximate the photon-plasma system with a Proca toy model, the full physics is more involved:~already in the simplest case of a cold, collisionless plasma, the longitudinal degrees of freedom are electrostatic, unlike the Proca case (see Section~\ref{BHenv_subsec:EM_waves}). In curved space, these transverse and longitudinal modes are coupled and thus the Proca model cannot assumed to be correct \emph{a priori}. Moreover, nonlinearities provide additional couplings between the modes, and also the inclusion of collisions or thermal corrections create strong deviations from a Proca theory. Hence, a consistent approach from first principles is imperative. In this section, we take a first step in that direction by studying a linearised axion-photon-plasma system. The underlying physical assumptions of our plasma model were discussed before in Sections~\ref{BHenv_sec:Plasma} and~\ref{sec_SR_Axionic:SRax_Plasma}, while the numerical implementation is detailed in Appendices~\ref{appNR_subsec:evolPlasma}--\ref{appNR_subsec:DecompMom}.
\subsection{Without Superradiance}
We start by studying the axion-photon-plasma system in absence of superradiance, and initialise the axion cloud in a supercritical state with $k\ped{a} \Psi_0 \ll 1$. We evolve the system on a BH background for different values of the plasma frequency (see Table~\ref{tb:simulations}). Note that there is no backreaction onto the axion field in our linearised setup.
\begin{figure}[t!]
  \centering
    \includegraphics[scale=0.45, trim = 0 0 0 0]{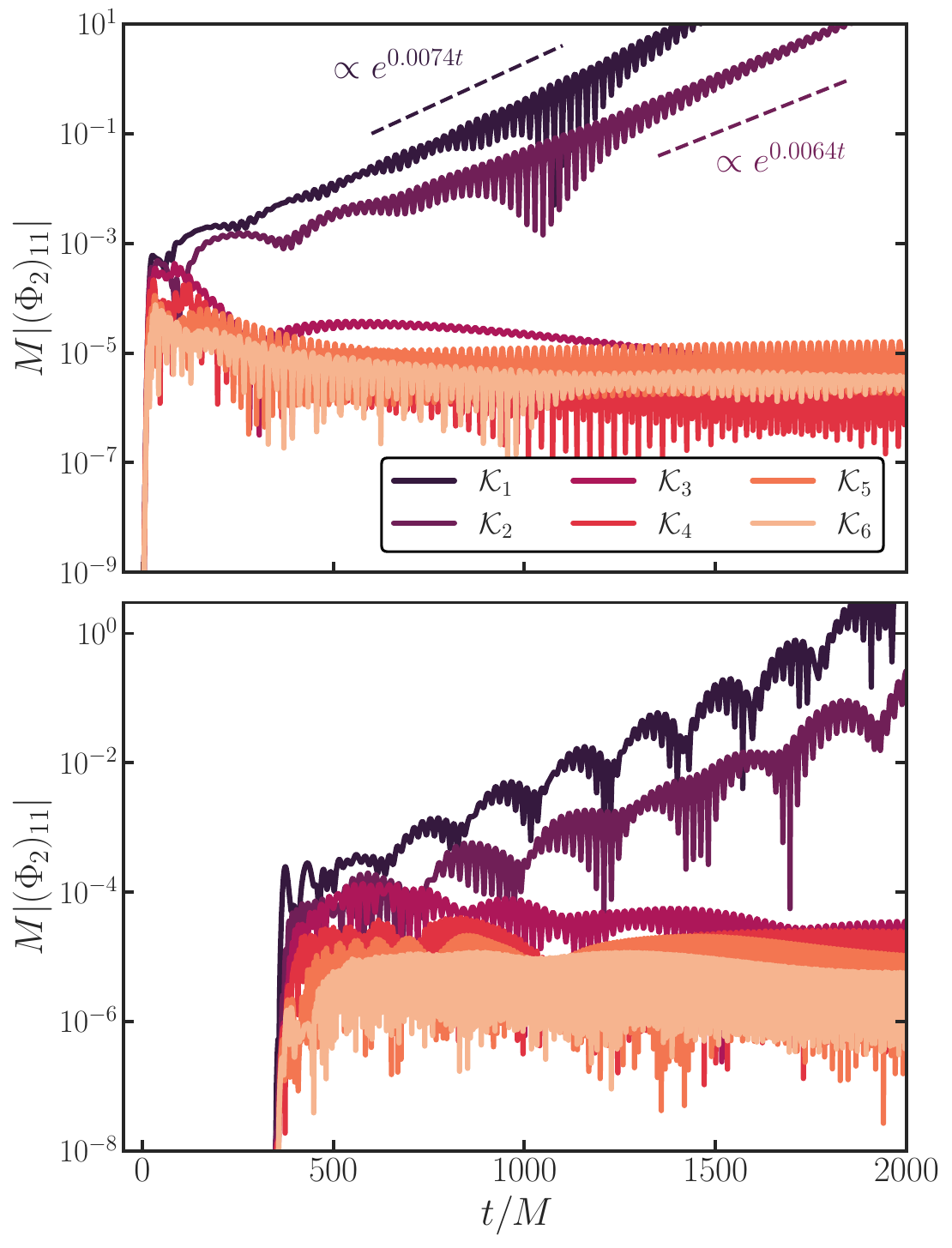}
    \caption{Time evolution of the dipolar component $|(\Phi_{2})_{11}|$ of the EM field in the presence of plasma for $\mu M = 0.3$, extracted at $r\ped{ex} = 20M$ (\emph{top panel}) or $r\ped{ex} = 400M$ (\emph{bottom panel}). The plasma frequency $\omega\ped{p}$ is progressively larger for simulations $\mathcal{K}_{1}\!-\!\mathcal{K}_{6}$, see Table~\ref{tb:simulations}. The exponential growth rate (dashed lines) is determined from~\eqref{eq:sengrowth}. The modulations at large radii arise from scattering of photons with the axion cloud, similar to Figure~\ref{fig:EMfieldBurstmuM03r481020}.}
    \label{fig:BurstPlasmamuM03r20}
\end{figure}
\vskip 2pt
Figure~\ref{fig:BurstPlasmamuM03r20} summarises the main results. When $\omega\ped{p} < \mu/2$, the plasma has little impact on the system and the parametric instability ensues. When $\omega\ped{p} \geq \mu/2$ instead, a suppression of the photon production is seen. We find the growth rate estimated in eq.~(19) of~\cite{SenPlasma} to fit our simulations well, when taking into account the finite-size effect of the cloud as $\lambda\ped{esc} \sim 1/d$, i.e.,
\begin{equation}\label{eq:sengrowth}
    \lambda \approx \frac{\mu^{2}\sqrt{\mu^{2}-4\omega\ped{p}^{2}}}{2\mu^{2}-4\omega\ped{p}^{2}}k\ped{a} \Psi_0 - \lambda\ped{esc}\,.
\end{equation}
Moreover, the beating pattern in the EM radiation at larger radii (see \emph{bottom panel} of Figure~\ref{fig:BurstPlasmamuM03r20}) originates from the photons having to travel through the cloud, thereby scattering of the axions.
\vskip 2pt
Additionally, we show the Fourier decomposition of the signal at $r\ped{ex} = 20M$ in Figure~\ref{fig:FourierPlasma}. As we concluded from Figure~\ref{fig:BurstPlasmamuM03r20}, for low $\omega\ped{p}$, the parametric instability is barely hindered and a clear peak arises at half the boson mass. However, when $\omega\ped{p}>\mu/2$, we observe the presence of modes with a frequency very close to $\omega\ped{p}$. We find good agreement between these peaks and the plasma-driven quasi-bound states computed in a similar setup~\cite{Cannizzaro:2020uap}. Note however, that these bound states are extremely fragile and geometry dependent, and may disappear if more realistic plasma models are considered~\cite{Dima:2020rzg}. We conjecture the origin of the two additional peaks at $\omega\ped{p}\pm \mu$ to be up and down-scattering from the quasi-bound state photons with the axion cloud. Due to the fact that modes with frequency $\omega\ped{p}- \mu$ are decaying, their amplitude is highly suppressed compared to the up-converted ones. 
\begin{figure}[t!]
    \centering
    \includegraphics[scale=0.45, trim = 0 0 0 0]{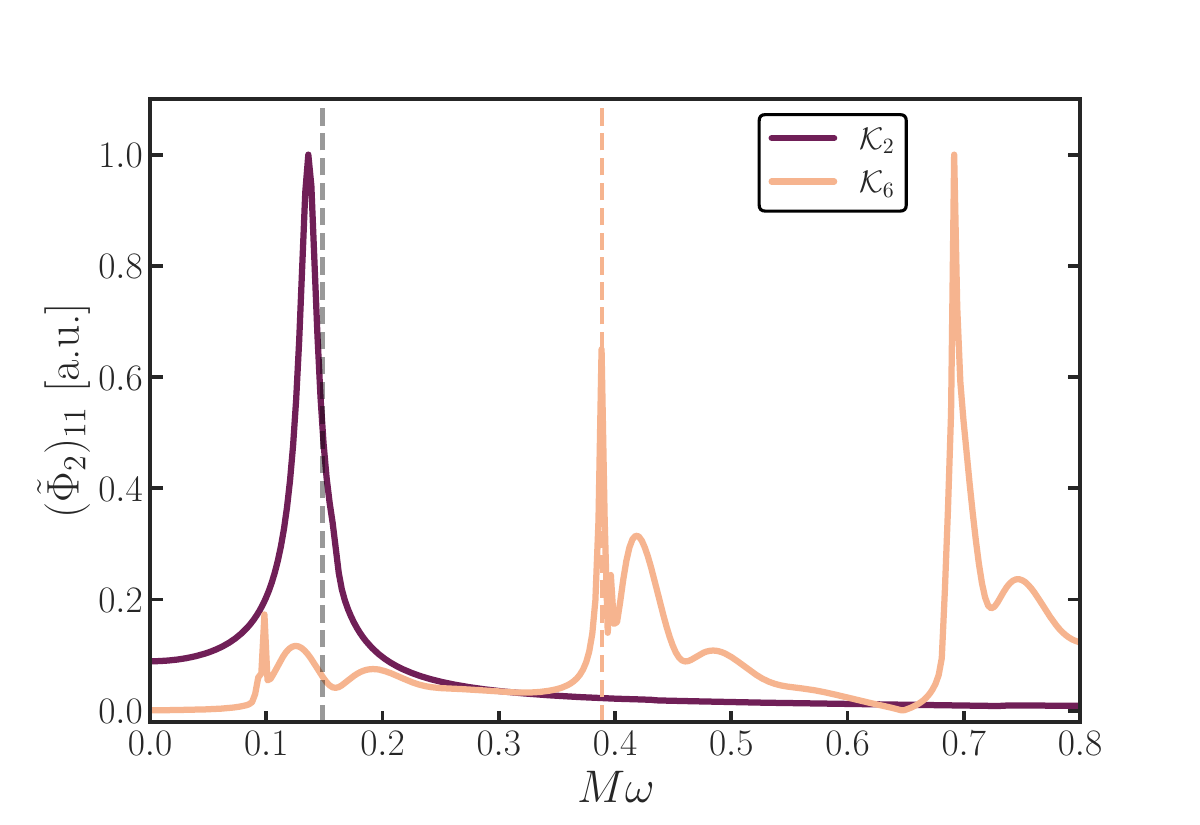}
    \caption{Fourier transform of $(\Phi_{2})_{11}$ extracted at $r\ped{ex} = 20M$ for simulations $\mathcal{K}_{2}$ ($\omega\ped{p} < \mu/2$) and $\mathcal{K}_{6}$ ($\omega\ped{p} > \mu/2$) with $\mu M = 0.3$. The vertical axis is shown in arbitrary units such that both curves are visible on the same figure. Dashed lines denote half the boson mass [{\color{gray}{grey}}] and the plasma-driven quasi-bound state frequency for $\omega\ped{p} = 0.4$, given by $M\omega=0.3887-0.0016i$ [{\color{peachy}{orange}}].}
    \label{fig:FourierPlasma}
\end{figure}
\vskip 2pt
We now focus on the high axionic coupling regime. In the toy model considered in~\cite{SenPlasma}, it was shown that even when $\omega\ped{p} \geq \mu/2$, an EM instability could be triggered for high enough $k\ped{a}\Psi_0$. In Figure~\ref{fig:FourierPlasmaSen}, we confirm this prediction numerically and show, for the first time, the presence of an instability in dense plasmas. This might seem in tension with the kinematic argument that for $\omega\ped{p}>\mu/2$ the axion decay into two photons is forbidden. However, as we show in the inset of Figure~\ref{fig:FourierPlasmaSen}, the frequency centres at $\omega=\mu$ instead of the usual $\omega=\mu/2$. This suggests the photon production to be dominated by a different process, namely $\Psi + \Psi \rightarrow \gamma + \gamma$.
\vskip 2pt
To support this hypothesis, we revisit the connection to the Mathieu equation. As detailed in Appendix~\ref{app_Mathieu_sec:SR}, in flat spacetime, the Maxwell equations in the presence of a plasma can be recast into the form of a Mathieu equation, which exhibits instability bands whenever $\omega^2 = p_z^2+\omega\ped{p}^2 = n^2\,\mu^2/4$, with $n \in \mathbb{N}$. Consequently, when $\omega\ped{p} > \mu/2$, the first instability band ($n=1$), corresponding to $\omega = \mu/2$, is no longer accessible. However, the second instability band ($n=2$), located at $\omega = \mu$, remains viable. This matches exactly the phenomenology observed in Figure~\ref{fig:FourierPlasmaSen}, leading us to identify the EM instability as arising from the second instability band of the Mathieu equation. This band indeed corresponds to the process $\Psi + \Psi \rightarrow \gamma + \gamma$, and remains kinematically viable for $\omega\ped{p} > \mu/2$~\cite{hertzberg:2018zte}. This analysis can, in principle, be extended to higher-order bands. However, these bands become increasingly narrow, and thus exciting them would likely require (extremely) high values of the axionic coupling.
\begin{figure}[t!]
    \centering    
    \includegraphics[scale=0.45, trim = 0 0 0 0]{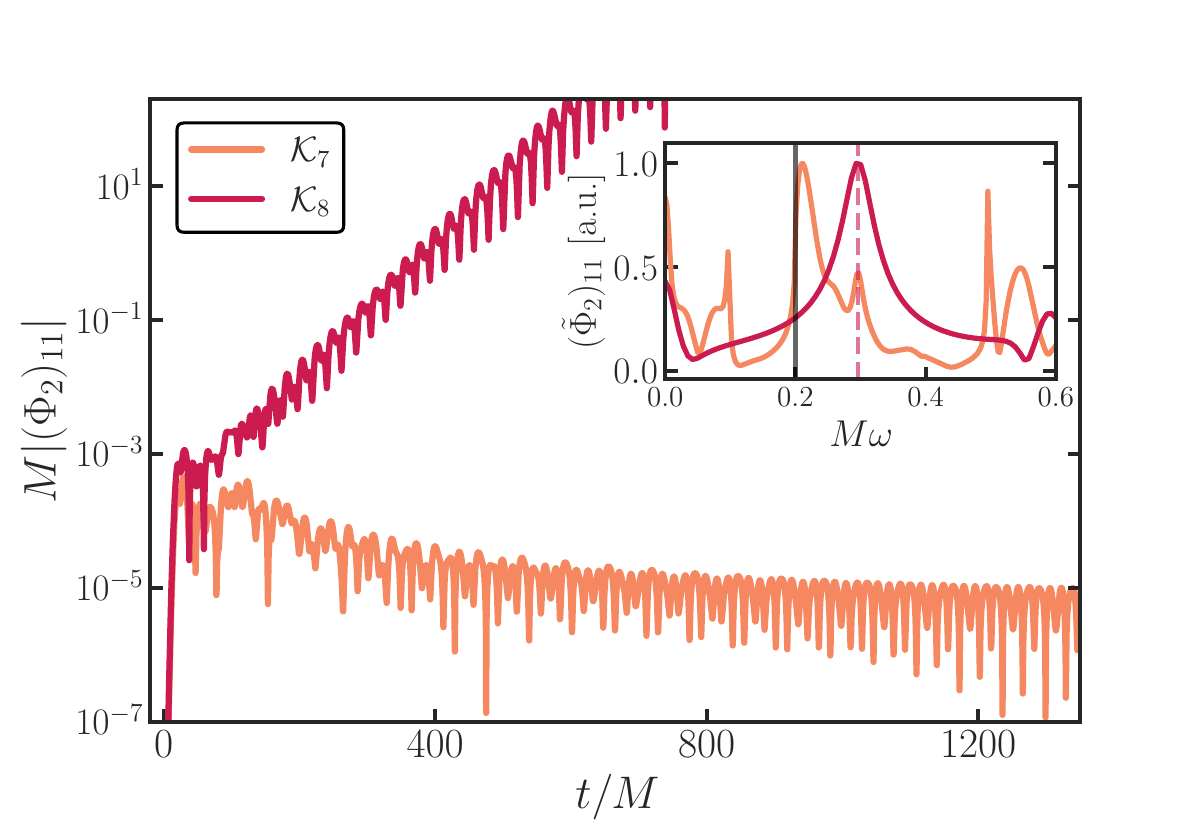}
    \caption{Time evolution of $|(\Phi_{2})_{11}|$ extracted at $r\ped{ex} = 20M$ for simulations $\mathcal{K}_{7}$ and $\mathcal{K}_{8}$ with $\mu M = 0.3$. Even though in both simulations $\omega\ped{p} = 0.2 > 0.15 = \mu/2$, the instability can be restored when $k\ped{a}\Psi_0$ is high enough. The inset shows the Fourier transform of both curves, with the solid line [{\color{gray}{grey}}] indicating the plasma frequency. The dashed line [{\color{cornellRed}{red}}] shows the frequency of the peak for $\mathcal{K}_8$, which is at $M\omega = 0.3$, indicating that the second instability band of the Mathieu solution is triggered.}
    \label{fig:FourierPlasmaSen}
\end{figure}
\subsection{With Superradiance}
\begin{figure}[t!]
    \centering
    \includegraphics[scale=0.45, trim = 0 0 0 0]{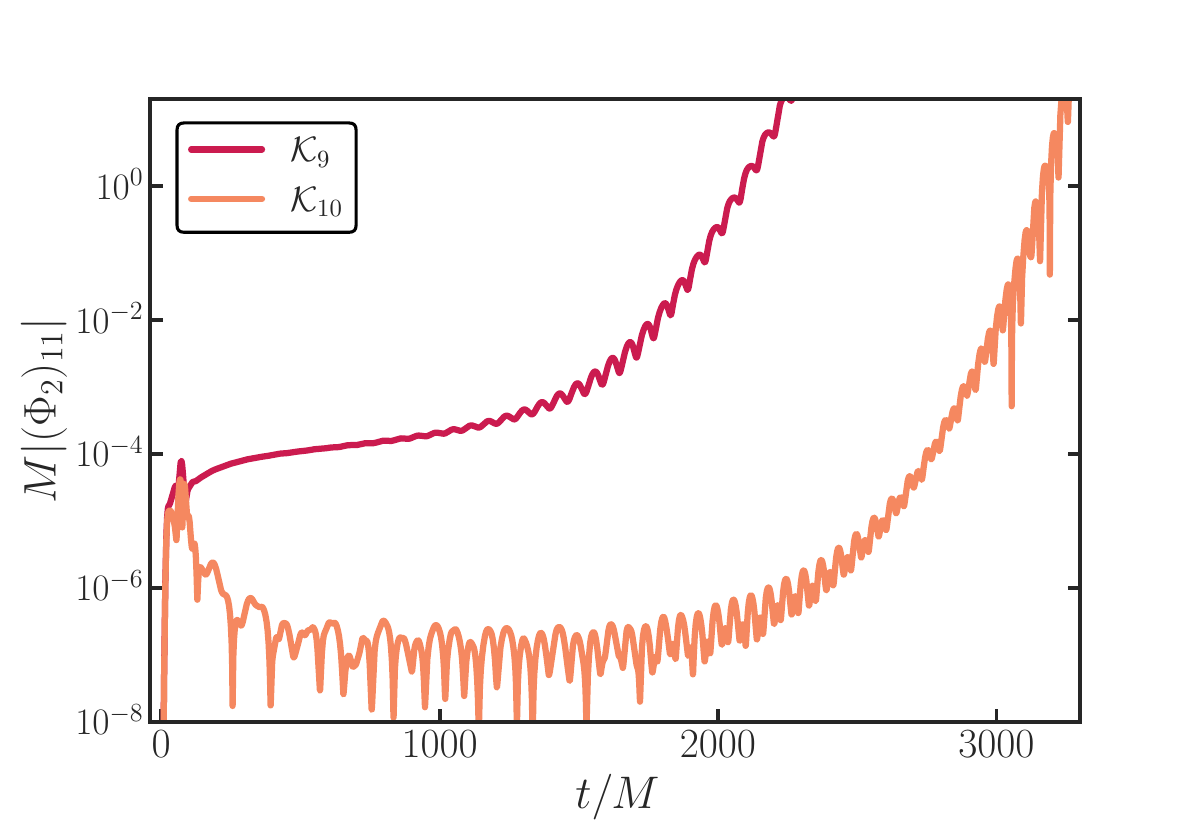}
    \caption{The dipolar component of EM radiation extracted at $r\ped{ex} = 20M$ for simulations $\mathcal{K}_{9}$ ($\omega\ped{p} = 0.02$) and $\mathcal{K}_{10}$ ($\omega\ped{p} = 0.07$) with $\mu M = 0.1$. Although initially subcritical, superradiance drives the axionic coupling high enough such that an instability can occur, even when $\omega\ped{p} = 0.07 > 0.05 = \mu/2$.}
    \label{fig:FourierPlasmawithC}
\end{figure}

We now probe the axion-photon-plasma system starting from a subcritical regime, yet letting it evolve to supercritical values via superradiance. Based on the previous section, we expect the system to turn unstable at some point, as the axionic coupling $k\ped{a} \Psi_0$ grows indefinitely. Due the longer timescales associated with this process, we anticipate assumption~(v) of our plasma model, namely neglecting the gravitational term, to be violated for the same parameter choices as before. Therefore, we evolve the system with $\mu M = 0.1$, such that all the assumptions are still justified. 
\vskip 2pt
In Figure~\ref{fig:FourierPlasmawithC}, we show two simulations, $\mathcal{K}_{9}$ ($\omega\ped{p} < \mu /2$) and $\mathcal{K}_{10}$ ($\omega\ped{p} > \mu /2$), which capture well the two distinct outcomes. In the former, the usual instability with $\omega = \mu /2$ ensues when the system has reached the supercritical threshold, yet in the latter, the time to reach this threshold is longer as the axionic coupling must grow sufficiently to trigger the second instability band. Note that, similar to the previous section, there is no backreaction onto the axion field, which therefore merely acts as a big reservoir for the EM field. This naturally explains the absence of a saturation phase. Should the backreaction be included however, there is no physical reason to expect that the saturation phase is ruined by the presence of plasma as it does not interfere with the balance between the energy inflow from the BH and energy outflow from the emitted photons. We therefore expect that the general outcome from the analysis in Section~\ref{sec_SR_Axionic:analytical} still holds, aside from minor modifications.\footnote{For example, the presence of a plasma affects the escape rate, as photons will travel more slowly through it, i.e.,~$v_{\gamma} < c$.}
\vskip 2pt
By allowing the axionic coupling to take on arbitrarily high values, an instability is thus always triggered, regardless of the plasma frequency. In practice however, it is bounded by constraints on the coupling constant $k\ped{a}$ and the mass of the cloud (which relates to $\Psi_0$) when superradiant growth is saturated. We can estimate the maximum axionic coupling, and therefore the maximum plasma frequency (= electron density) for which an instability occurs. We do this using the flat space toy model detailed in Appendix~\ref{app_Mathieu_sec:plasmaMathieu}. From eq.~\eqref{eq:PlasmaMathieu}, it is immediately clear that the critical value to trigger an instability in the presence of an overdense plasma (when $\omega\ped{p} \gtrsim p_z$) is given by
\begin{equation}
    k\ped{a} \psi_0 \gtrsim \frac{\omega\ped{p}^2+ p_z^2}{2\mu p_z}\approx \frac{\omega\ped{p}^2}{\mu^2}\,.
\end{equation}
This condition corresponds to the requirement that the harmonic term in the Mathieu-like equation dominates over the non-oscillatory one [cf.~eq.~\eqref{eq:PlasmaMathieu}]. We have confirmed that this flat spacetime model closely matches the simulations in curved spacetime. Therefore, we can safely use the (flat spacetime) relation between the axion amplitude and the mass of the cloud~\cite{Brito:2015oca} to obtain
\begin{equation}
    k\ped{a} \gtrsim 8 \times 10^{2} \left(\frac{\omega\ped{p}^2}{\mu^2}\right)\left(\frac{0.1}{M\ped{c}/M}\right)^{1/2}\left(\frac{0.2}{\mu M}\right)^{2}\,.
\end{equation}
Note that when $\omega\ped{p} \approx \mu$, this condition reduces to the one derived in~\cite{Ikeda:2018nhb}, while in the case $\omega\ped{p}\gg\mu$, stronger constrains on the coupling are imposed. We can translate this into the following condition for when an instability is triggered
\begin{equation}\label{eq:criticalcoupling}
\frac{10^{-13}~\mathrm{GeV}^{-1}}{k\ped{a}} \lesssim 8 \times 10^{5} \left(\frac{10^{-3}~\mathrm{cm}^{-3}}{n\ped{e}}\right)
\left(\frac{M\ped{c}/M}{0.1}\right)^{1/2} \left(\frac{1M_{\odot}}{M}\right)^{2}\left(\frac{\mu M}{0.2}\right)^{4}\,.
\end{equation}
Since current constraints on the coupling constant are around $10^{-13}~\mathrm{GeV}^{-1}$ (see Figure~\ref{fig:constraints}), this means a plasmic environment can at least be a few orders of magnitude higher than the interstellar medium ($n\ped{e} \sim [10^{-3}-1]~\mathrm{cm}^{-3}$)~\cite{Saintonge:2022tfq} and still an EM instability would be triggered. 
\section{Observational Prospects}\label{sec_SR_Axionic:observational}
\begin{figure}[t!]
    \centering
    \includegraphics[scale=0.45, trim = 0 0 0 0]{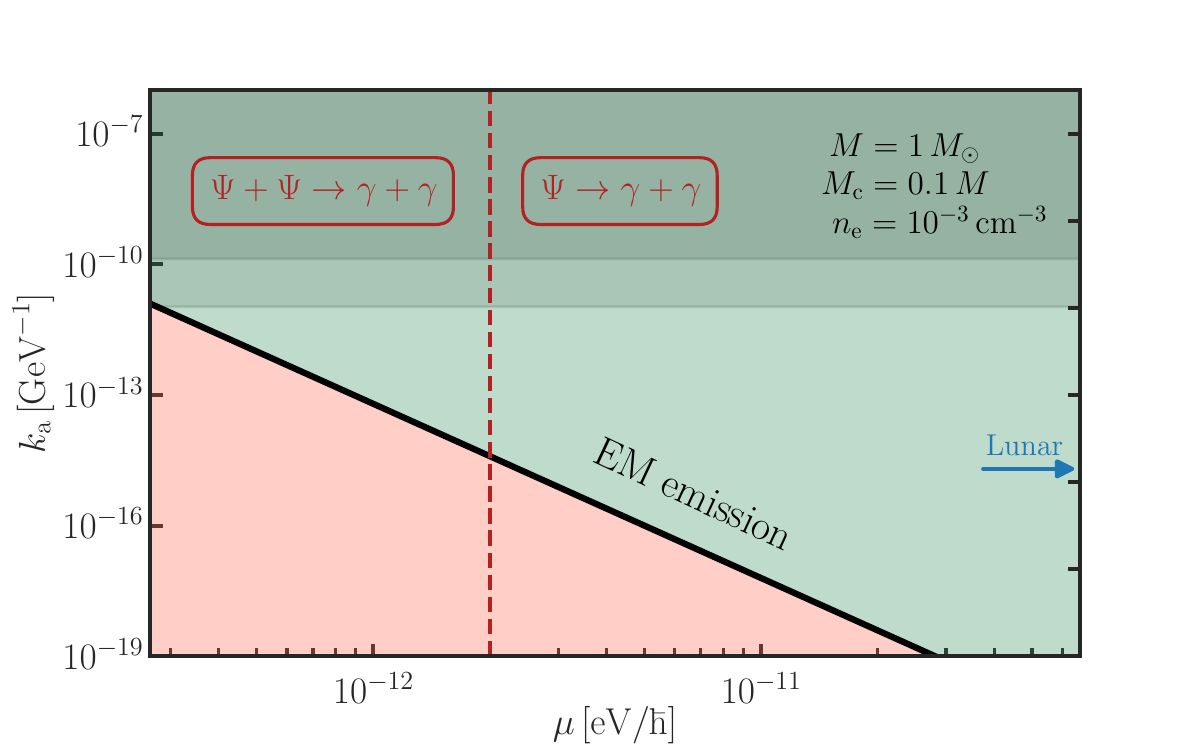}
    \caption{Possible outcomes of the axion-photon-plasma system depending on the boson’s mass and the axion-photon coupling. Above the threshold given by eq.~\eqref{eq:criticalcoupling}, EM emission is triggered [{\color{mintgreen}{green}}], while below it, the system never turns supercritical [{\color{softred}{red}}]. When the boson frequency is low, i.e.,~$\mu \lesssim 2\omega\ped{p}$, the EM instability can still be triggered in the presence of plasma, given a high enough axionic coupling. This happens when the system enters the second instability band, indicated by the dashed line. In the future, Lunar-based radio observatories could probe a part of this parameter space, denoted by the arrow [{\color{cornellBlue}{blue}}]. Finally, we show two of the most robust constraints on the axion-photon coupling, namely from the solar axion experiment CAST~\cite{CAST:2017uph} (darker region) and from measurements on supernova 1987A (lighter region)~\cite{Hoof:2022xbe}. This data was collected from~\cite{AxionLimits}.}
    \label{fig:SummaryBurst}
\end{figure}

Based on the previous section, there are two distinct outcomes for parametric photon production in presence of plasma; (i) the dominant instability for $\omega\ped{p} < \mu/2$, and (ii) higher band instabilities in the regime of large axionic couplings, for $\omega\ped{p} > \mu/2$. In situation (i), the plasma frequency establishes a threshold for the frequency of the emitted photons. In the case of the interstellar medium, characterised by an electron density of approximately $1~\mathrm{cm}^{-3}$~\cite{Ferriere:2001rg,Saintonge:2022tfq}, the value of $\omega\ped{p}$ is estimated to be around $10^{-11}~\mathrm{eV}/\hbar$~\eqref{eq:SR_Axionic_plasmafreq}, corresponding to a frequency of $7.6~\mathrm{kHz}$. This should be compared to e.g., a BH with mass $5 M_\odot$, which can effectively ($\mu M = 0.4)$ accumulate an axion cloud with the same frequency, i.e.,~$\mu \approx 10^{-11}~\mathrm{eV}$. In this case, the axion cloud would decay into pairs of photons with a frequency of approximately $3.8~\mathrm{kHz}$, which is close to the threshold value required for observation. For higher $\mu$, the plasma-induced effective mass can be considered negligible, and we anticipate that the primary photon flux will exhibit a nearly monochromatic energy of $\mu/2$ within the radio-frequency band. Note however, that for higher $\mu$, we need to invoke subsolar-mass BHs to grow the cloud on astrophysically relevant timescales. Besides the total photon flux derived analytically~\eqref{eq:analyticSRLuminosity} and extracted from our simulations~\eqref{eq:dEdt}, we also demonstrate, for the first time, the anisotropic emission morphology in the frame of the BH, see Figure~\ref{fig:Phi2MultipolesLargeradii}. Consequently, one expects varying observer inclination angles to result in quantitatively distinct signals.
\vskip 2pt
Situation (ii) presents an opportunity to observe photons produced by axion clouds beyond stellar-mass BHs. Still, the typical frequency of these photons fall below the $\mathrm{MHz}$ band, and thus poses a challenge for current Earth-based radio observations. However, the forthcoming moon-based radio observatories can potentially detect these signals~\cite{burns2019farside}. Moreover, in the case of a rapidly spinning BHs resulting from binary mergers, one can anticipate that the radio signals will follow strong GW emissions with a delay determined by the superradiance timescale. Consequently, by employing multi-messenger observations between GW detectors and lunar radio telescopes, constraints can be imposed on the axion-photon coupling. 
\vskip 2pt
Finally, the projected saturated value of $k\ped{a} \Psi\ped{sat}$ can induce a rotation in the linear polarisation emitted in the vicinity of BHs~\cite{Carroll:1989vb,Harari:1992ea}. This phenomenon has been investigated in the context of supermassive BHs~\cite{Chen:2019fsq,Chen:2021lvo,Chen:2022oad}. It should be noted however that due to the significant hierarchy between the ultra-low superradiance mass window and the plasma mass generated by a dense environment, higher-order instabilities are not expected to occur in supermassive BHs. Consequently, the axion cloud outside a supermassive BH remains robust against axion-photon couplings. A summary of our findings in the presence of plasma can be found in Figure~\ref{fig:SummaryBurst}.
\clearpage
\section{Summary and Outlook}\label{sec_SR_Axionic:conclusions}
The presence of ultralight bosons is ubiquitous in theories beyond the Standard Model. Originally proposed as solutions to the dark matter puzzle or the strong CP problem, they are of interest across various areas of physics (see Section~\ref{BHenv_subsec:u_bosons}). Detecting them however, is notoriously hard, especially when their coupling to the Standard Model is weak. Through superradiance, a new channel for detection opens up, turning BHs into powerful \emph{particle detectors} in the cosmos. 
\vskip 2pt
In this chapter, we presented a detailed numerical study of the dynamics of boson clouds around BHs, focusing on axionic couplings to the Maxwell sector. We confirm the existence of an EM instability, consistent with previous work. Crucially, whereas earlier studies typically assumed that the boson cloud forms \emph{before} turning on the coupling, we relax this assumption. Instead, we consider the axionic coupling \emph{simultaneously} with the growth of the cloud, conform to the superradiance mechanism. In this more realistic setup, we find that the system reaches a \emph{stationary state} in which every axion produced by superradiance is immediately converted into photons that steadily escape the cloud. This leads to strong observational signatures:~the nearly monochromatic and constant EM signal could have a luminosity comparable with some of the brightest sources in our Universe. Moreover, the depletion of the axion cloud impacts current constraints on the boson mass. 
\vskip 2pt
We also investigated the impact of a surrounding plasma on the EM instability. In the regime of small axionic couplings, we find the expected suppression of the instability when the plasma frequency exceeds half the boson mass. Surprisingly, however, the instability can be restored for sufficiently strong couplings. By demonstrating that the Maxwell equations in a plasma background reduce to a Mathieu equation, we offer a natural explanation for this behaviour:~higher-order instability bands become accessible, allowing the instability to persist. This interpretation suggests that even very dense plasmas do not necessarily quench the parametric instability;~it depends critically on the value of the axionic coupling. Since higher bands correspond to higher-frequency photon emission, this mechanism may extend beyond the stellar-mass BH regime, opening up new avenues for detection.
\vskip 2pt
To make concrete observational predictions, two key challenges must be addressed:~(i) the geometry of realistic plasmic environments and (ii) the nonlinear dynamics of the axion-photon-plasma system. In this chapter, we assumed a constant plasma density throughout space. However, astrophysical environments can be nearly planar, for example in the case of thin disks (see e.g.,~\cite{Abramowicz2013}), which will impact the resulting EM flux. Moreover, while our linearised framework accurately captures the initial impact of plasma on the instability, understanding the long-term evolution requires accounting for backreaction on the axion field. As we have shown, large axionic couplings may be required to trigger instabilities in dense plasmas. Therefore, a natural extension of this study is to investigate the fully nonlinear dynamics of the coupled axion-photon-plasma system. Indeed, the nonlinear propagation of large-amplitude EM waves in dense plasma is known to exhibit rich and complex behaviour (see e.g.,~\cite{KawDawson1970,1971PhRvL..27.1342M,Cardoso:2020nst,Cannizzaro:2023ltu}).
\chapter{Ultralight Bosons near Compact Objects:~In-Medium Suppression}\label{chap:in_medium_supp}
\vspace{-0.8cm}
\hfill \emph{Mirum est} 
\vskip -5pt
\hfill \scalebox{0.75}{\emph{It is remarkable}}

\hfill \emph{ut animus agitatione motuque corporis excitetur} 
\vskip -5pt
\hfill \scalebox{0.75}{\emph{how the mind is stimulated by the exercise and movement of the body}}
\vskip 5pt

\hfill Pliny the Younger, \emph{Epistulae}, Book I, Letter VI
\vskip 35pt
\noindent In the previous chapter, I studied the interactions between ultralight scalars and the electromagnetic sector, and how plasma affects their conversion rate. This chapter extends those investigations by considering different and intriguing astrophysical scenarios in which spin-$0$ or spin-$1$ bosons couple to Standard Model photons. While such couplings are typically weak (see Figure~\ref{fig:constraints}), large densities of the field -- achievable, for instance, through superradiance -- can give rise to striking observational signatures. In addition to the scenario explored in the previous chapter, other astrophysical environments offer promising opportunities to detect axion-like particles. For instance, axion-photon mixing is enhanced in the presence of strong electromagnetic fields, which are naturally found near neutron stars, making them compelling \emph{axion laboratories} (see e.g.,~\cite{Battye:2023oac,Huang:2018lxq,Hook:2018iia, Leroy:2019ghm,Witte:2020rvb,Witte:2021arp}). A striking phenomenology also occurs for charged BHs, which are unstable under axion-photon couplings~\cite{Boskovic:2018lkj}. This results in the formation of \emph{axionic hair} on very short timescales, a process now well-understood in both the perturbative and fully nonlinear regime~\cite{Boskovic:2018lkj,Burrage:2023zvk}.
\vskip 2pt
Crucially, many studies to date neglect the impact of environmental plasmas, despite their widespread presence around compact objects~\cite{Abramowicz2013,Barausse:2014tra}. Plasmas can strongly affect axion-photon conversion by endowing the transverse photon modes with an effective mass, determined by the plasma frequency $
\omega\ped{p}$~\eqref{eq:BHenv_plasmafreq}. Whenever $\omega\ped{p}\gg \mu$ (with $\mu$ the mass of the boson), the conversion is heavily suppressed, and stronger couplings are necessary to achieve any observable signal~\cite{PhysRevD.37.1237, Raffelt:1996wa, Mirizzi:2006zy, Redondo:2008aa, An:2013yfc}. This phenomenon, known as \emph{in-medium suppression}, effectively weakens the interaction between bosons and photons in dense plasma environments.
\vskip 2pt
In this chapter, I discuss the work of~\cite{Cannizzaro:2024hdg} and investigate the mixing between ultralight bosons and Standard Model photons in the presence of plasma on BH spacetimes. I focus on two relevant cases:~(i) \emph{axionic instabilities} in Reissner-Nordstr\"{o}m spacetimes and (ii)~\emph{photon production} from superradiant dark photon clouds around Schwarzschild BHs. While astrophysical BHs are not expected to be charged due to various discharge mechanisms~\cite{1969ApJ...157..869G,Eardley:1975kp,Gibbons:1975kk,Blandford:1977ds}, case (i) serves as a useful proxy to study axion dynamics around magnetised neutron stars. Moreover, the possibility that even a small charge, achievable via various mechanisms~\cite{Wald:1974np,Komissarov:2021vks,Cardoso:2016olt}, could lead to the formation of axionic hair is appealing. Case (ii), concerning dark photon clouds, was partially addressed in~\cite{Caputo:2021efm} using a plane-wave approximation. However, a complete analysis in curved spacetime is still lacking. As highlighted in~\cite{Caputo:2021efm}, plasmas could play an important role in the evolution of superradiant instabilities, and prevent significant interaction of the dark photon cloud with the Standard Model, \emph{especially} for small dark photon masses or weak couplings. 
\vskip 2pt
The rest of this chapter is organised as follows:~in Section~\ref{sec_IMS:theory}, I set up of the theoretical framework. Then, in Section~\ref{sec_IMS:flat}, I discuss the impact of plasma on the boson-to-photon conversion in flat spacetime. In Section~\ref{sec_IMS:curved}, I extend the discussion to BH spacetimes, and finally, in Section~\ref{sec_IMS:summary}, I present the conclusions. Additional details are contained in Appendices~\ref{app:BHPT},~\ref{app:flatinstab} and~\ref{app:DPbasis}. The reader is reminded that quantities with a tilde are dimensionless, e.g., the BH charge~$\tilde{Q} = Q/M$.
\section{The Theory}\label{sec_IMS:theory}
We consider a general Lagrangian describing a massive axion field $\Psi$ and a dark photon field $A'_{\mu}$, both coupled to the Standard Model electromagnetic field $A_{\mu}$, which is sourced by a cold, collisionless plasma. The dark photon sector is formulated in the so-called \emph{interaction basis}~\cite{Jaeckel:2012mjv, Siemonsen:2022yyf}.\footnote{When redefining the fields in the interaction basis, we neglect terms of the order $\left(\sin{\chi_0}\right)^{2}$~\cite{Jaeckel:2012mjv}. Note that typical values of the coupling are in the range $\sin{\chi_0} \lesssim \mathcal{O}(10^{-4})$ (see Figure~\ref{fig:constraints}).} Other possible choices of basis are further discussed in Appendix~\ref{app:DPbasis}. The Lagrangian then takes the form:
\begin{equation}\label{eq:lagrangian}
    \begin{aligned}
        &\mathcal{L}=\frac{R}{16\pi}-\frac{1}{2}\nabla_{\mu}\Psi\nabla^{\mu}\Psi
-\frac{\mu\ped{a}^{2}}{2}\Psi^{2}-\frac{k\ped{a}}{2}\Psi\,{}^{*}\!F^{\mu\nu}F_{\mu\nu}+ j_\mu A^\mu+\mathcal{L}\ped{m}\\&-\frac{1}{4}\left(F_{\mu\nu}F^{\mu\nu}+F'_{\mu\nu}F'^{ \mu\nu}\right)-\frac{\mu_{\gamma '}^2}{2} A'^\mu A'_\mu- \mu_{\gamma'}^2\sin\chi_0 A'_\mu A^\mu \,,
\end{aligned}
\end{equation}
where $\mu\ped{a}$ and $\mu_{\gamma'}$ are the axion and dark photon masses, $F_{\mu\nu}=\partial_\mu A_\nu-\partial_\nu A_\mu$ is the Maxwell tensor with an equivalent definition for $F'_{\mu\nu}$. Furthermore, ${}^{*}\!F^{\mu\nu}\equiv (1/2)\epsilon^{\mu\nu\rho\sigma}F_{\rho\sigma}$ is the dual Maxwell tensor and $R$ the Ricci scalar. The plasma Lagrangian and current are given by $\mathcal{L}\ped{m}$ and $j_\mu= e n\ped{e} v_\mu$, respectively, where $e$ is the electron charge, $n\ped{e}$ the number density and $v^\mu$ the four velocity. The couplings between the bosons and photons are realised through the axionic coupling $k\ped{a}$ and the kinetic mixing term $\sin \chi_0$. 
\vskip 2pt
We model the plasma using the Einstein cluster~\cite{Einstein_cluster,Einstein_cluster_2,Cardoso:2021wlq,Cardoso:2022whc,Feng:2022evy}, in which plasma particles are assumed to be in circular orbits in all possible orientations around the BH. We refer to Section~\ref{BHenv_subsec:DMStruct} for details on this setup, while here we simply note that the cluster is equivalent to an anisotropic fluid with only a tangential pressure $P\ped{t}$, described by the following stress-energy tensor: 
\begin{equation}
\label{eq:plasmaSET}
    T^{\rm p}_{\mu\nu}=(\rho+P\ped{t}) v_\mu v_\nu+P\ped{t}(g_{\mu\nu}-r_\mu r_\nu)\,,
\end{equation}
where $\rho=n\ped{e} m\ped{e}$ is the energy density of the fluid, $g_{\mu \nu}$ the metric of the underlying spacetime and $r^\mu$ a unit vector in the radial direction.
\vskip 2pt
From the Lagrangian~\eqref{eq:lagrangian}, we can infer the equations of motion for the scalar, electromagnetic, dark photon and gravitational fields: 
\begin{equation}\label{eq:IMS_evoleqns}
\begin{aligned}
\left(\nabla^\mu \nabla_\mu-\mu\ped{a}^2\right) \Psi &= \frac{k\ped{a}}{2}\,{ }^*\!F^{\mu \nu} F_{\mu \nu}\,,\\
\nabla_\nu F^{\mu \nu} &=j^{\mu} -2 k\ped{a}{ }^*\!F^{\mu \nu} \nabla_\nu \Psi-\sin\chi_0 \mu_{\gamma'}^2 A' {}^{\mu}\,,\\
\nabla_\nu F'^{\mu \nu} &=-\mu_{\gamma'}^2 A'^\mu-\sin\chi_0 \mu_{\gamma'}^2 A^{\mu}\,,\\
R_{\mu\nu}-\frac{1}{2}g_{\mu\nu}R&= 8 \pi \Big(T^{\rm \Psi}_{\mu\nu}+T^{\rm EM}_{\mu\nu}+T^{\rm DP}_{\mu\nu}+T^{\rm p}_{\mu\nu} \Big)\,,
\end{aligned}
\end{equation}
where in the last line, we introduced the Ricci tensor $R_{\mu\nu}$, along with the stress-energy tensors for the axionic, electromagnetic and dark photon sectors. 
\vskip 2pt
To close the system, we need the continuity and momentum equations, which we infer from the conservation of the stress-energy tensor and the current:
\begin{equation}\label{eq:momentum_continuity}
\nabla^\nu T^{\rm p}_{\mu\nu}=e n\ped{e} F_{\mu\nu}v^\nu\,, \quad \nabla_\mu (n\ped{e} v^\mu)=0\,.
\end{equation}
As we model the dark photon in the interaction basis, the hidden field $A'_\mu$ is \emph{sterile}, meaning that it does not couple to the plasma directly.
\section{Plasma Suppression in Flat Spacetime}\label{sec_IMS:flat}
Before studying the mixing in BH spacetimes, we first consider the impact of plasma in flat spacetime. This simpler setting helps isolate and clarify key aspects of the dynamics. In flat space, the Einstein cluster model reduces to a plasma at rest, with a vanishing tangential pressure. We can thus treat the plasma as pressureless dust, with a momentum equation given by
\begin{equation}\label{eq:momentumflat}
    v^\nu \partial_\nu v^\mu=\frac{e}{m\ped{e}}F^{\mu\nu}v_\nu\,,
\end{equation}
where $m\ped{e}$ is the electron mass. In the presence of an external magnetic field, there is mutual conversion between the axion and the propagating modes of the photon~\cite{PhysRevLett.51.1415,PhysRevD.37.1237, Raffelt:1996wa, Mirizzi:2006zy}, which parallels the mixing of neutrinos. In contrast, dark photon-photon mixing can arise even in the absence of background electromagnetic fields.
\vskip 2pt
First, we consider the axionic case, setting $\sin\chi_0=0$. As a background configuration, we take a static, homogeneous electron-ion plasma in a constant magnetic field along the $\hat{y}$--direction. We then adopt linear perturbation theory to study the propagation of plane waves. As detailed in~\cite{PhysRevD.37.1237}, the conversion between axions and photons requires a change in the azimuthal angular momentum. A longitudinal magnetic field would preserve azimuthal symmetry, thus stopping any conversion. Hence, we consider the propagation of waves along the $\hat{z}$--axis without loss of generality. 
\vskip 2pt
We denote the components of the electromagnetic potential as parallel ($A_y=A_\parallel$) and perpendicular ($A_x=A_\perp$) to the background magnetic field. As for the axionic sector, we consider a vanishing axion background $\Psi = 0$, and denote its perturbation by $\psi$. From parity considerations, one can readily see that the perpendicular component of the electromagnetic field decouples from the axion. Indeed, the two photon states $A_\perp$ and $A_\parallel$ are even and odd, respectively, under parity in the $y-z$ plane, while the axion plane wave state is odd. Thus, only the parallel component of the electromagnetic field mixes with the axion~\cite{PhysRevD.37.1237}. Considering only the dynamics along this direction, solving the momentum equation~\eqref{eq:momentumflat} yields
\begin{equation}\label{eq:mom_parallel}
    v_\parallel=-\frac{e}{m\ped{e}}A_\parallel\,.
\end{equation}
Therefore, the current can be expressed in terms of the electromagnetic field. One is then left with a coupled system involving the parallel component of the Maxwell equations and the Klein-Gordon equation. In the frequency domain and assuming for simplicity relativistic axions ($\omega\gg \mu\ped{a}$), it can be expressed as\footnote{In the following, we neglect the effects of Faraday rotation. Although it can be important for polarisation effects, it does not impact the conversion probability between axions and photons~\cite{Mirizzi:2006zy}, which is the relevant subject here.}
\begin{equation}\label{eq:axion-photon-mixing-flat}
    (\omega- i\partial_z+\mathcal{M}\ped{a\gamma})\begin{bmatrix}
    A_\parallel \\
    \psi
    \end{bmatrix} =0\,,
\end{equation}
where we introduced the \emph{axion-photon mixing matrix}:
\begin{equation}
\label{eq:Edispersion}  
\mathcal{M}\ped{a\gamma }=
    \begin{bmatrix}
    -\omega\ped{p}^2/2 \omega & B_y k\ped{a}  \\
    B_y k\ped{a} & -\mu\ped{a}^2/2\omega 
    \end{bmatrix}\,.
\end{equation}
Here, the off-diagonal terms couple the two fields, and are thus responsible for the mixing. Clearly, in the case $k\ped{a}=0$ or $B_y=0$, i.e., when the matrix is diagonal, the two fields are decoupled, and eq.~\eqref{eq:axion-photon-mixing-flat} simply returns the dispersion relation of the photon and the axion. To simplify the dynamics, we can perform a field redefinition that diagonalises the matrix~\eqref{eq:Edispersion} through a rotation in the field basis, with the rotation angle given by\footnote{This resembles redefining the fields in Reissner-Nordstr\"{o}m spacetimes to decouple the gravito-electromagnetic perturbations in terms of two master functions (see e.g.,~\cite{MTB,Pani:2013wsa,Berti:2005eb}).}
\begin{equation}\label{eq:angle-axion-photon}
    \theta= \frac{1}{2}\arctan{\left(\frac{4 \omega B_y k\ped{a}}{-\omega\ped{p}^2+\mu\ped{a}^2} \right)}\,.
\end{equation}
This angle is proportional to the off-diagonal terms in~\eqref{eq:Edispersion} and quantifies the coupling between the modes. Specifically, the axion-photon conversion rate is proportional to $P(a \rightarrow \gamma)\propto \rm{sin}^2 (2 \theta)$~\cite{Mirizzi:2006zy, PhysRevD.37.1237}. In the limits $B_y \rightarrow 0$ or $k\ped{a} \rightarrow 0$, the probability naturally goes to zero and no conversion is possible. However, even in the presence of magnetic fields and couplings, a large plasma frequency, i.e., $\omega\ped{p}\gg \mu\ped{a}, k\ped{a} B_y$, kinematically disfavours the conversion from axions to photons, as the mixing angle drops to zero. This quenching is referred to as \emph{in-medium suppression}. Moreover, when $\mu=\omega\ped{p}$, the conversion probability is maximised, which is termed a \emph{resonant conversion} between the two states.
\vskip 2pt
In the dark photon case ($k\ped{a}=0$), a similar procedure yields the dark photon-photon mixing matrix~\cite{An:2023mvf}:
\begin{equation}
\label{eq:DPEdispersion}  
\mathcal{M}_{\gamma \gamma' }=\frac{1}{2\omega}
    \begin{bmatrix}
    -\omega\ped{p}^2 & \sin{\chi_0}\, \mu_{\gamma'}^2  \\
    \sin{\chi_0}\,\mu_{\gamma'}^2 & -\mu_{\gamma'}^2
    \end{bmatrix}\,.
\end{equation}
The same conclusions apply as in the axionic case:~in the presence of a sufficiently dense plasma, i.e., $\omega\ped{p}\gg \mu_{\gamma'}$, the in-medium conversion angle goes to zero, suppressing the conversion even for large couplings. When $\mu_{\gamma'}=\omega\ped{p}$ instead, a resonant conversion is triggered.
\section{Plasma Suppression in Curved Spacetime}\label{sec_IMS:curved}
We now turn to a fully relativistic setup, studying the impact of plasma on the mixing using BH perturbation theory (see Appendix~\ref{app:BHPT}). In this context, we neglect the backreaction of the plasma on the Einstein and Maxwell background equations~\cite{Cannizzaro:2024yee}. This is justified because the plasma's energy density is typically low in astrophysical environments, and its source terms are further suppressed by the large charge-to-mass ratio of the electron. In addition, we consider the presence of an oppositely charged component in the plasma -- ions -- inducing a current with the opposite sign in the Maxwell equations, neutralising the plasma background~\cite{Cannizzaro:2020uap,Cannizzaro:2021zbp,Cannizzaro:2023ltu,Spieksma:2023vwl}.
\subsection{Charged Black Holes:~Axionic Instabilities}\label{sec_IMS:chargedBHs}
We start by analysing axionic instabilities around charged compact objects. Assuming a spherically symmetric spacetime, the background geometry is described by the standard Reissner-Nordstr\"{o}m solution:
\begin{equation}\label{eq:RN_Sol}
ds^2 =-f\mathrm{d} t^2+f^{-1} \mathrm{d} r^2+r^2\mathrm{d}\Omega^{2}\,, \ \text{with}\quad f= 1-\frac{2 M}{r}+\frac{Q^2}{r^2}\,, 
\end{equation}
where $M$ and $Q$ are BH mass and charge respectively, and the background electromagnetic field is given by $A_\mu=(Q/r,0,0,0)$. 
\vskip 2pt
We model the plasma as non-relativistic, characterised by a macroscopic plasma frequency satisfying $M\omega\ped{p}\gtrsim 1$. However, as we will show in the next chapter [see eq.~\eqref{eq:plasmafreqKaw}], significant BH charge induces relativistic plasma motion, which in turn reduces the photon's effective mass via \emph{relativistic plasma transparency}~\cite{KawDawson1970, Cardoso:2020nst, Cannizzaro:2024yee}. This effect arises through the Lorentz factor $\gamma\ped{e}$, which modifies the electron's relativistic mass $m\ped{e}\rightarrow \gamma\ped{e} m\ped{e}$, thereby lowering the plasma frequency~\eqref{eq:BHenv_plasmafreq}. Nevertheless, several mechanisms -- such as BH charge screening~\cite{Feng:2022evy} and magnetic pressure -- can oppose this relativistic motion. Furthermore, this (non-relativistic) model is intended as a proxy to study compact objects embedded in electromagnetic fields in more realistic scenarios. A prime example is magnetised neutron stars, where surrounding plasma is expected to dress the photon with a large effective mass (see e.g.,~\cite{Leroy:2019ghm, Witte:2020rvb, McDonald:2023shx}). Although more work is needed to model realistic astrophysical conditions, our approach represents the first consistent study of axion-photon conversion in curved spacetime with plasma effects, using BH perturbation theory.
\vskip 2pt
In electro-vacuum, background electromagnetic fields of charged BHs can trigger axionic instabilities, leading to new ``hairy'' BH solutions~\cite{Ikeda:2018nhb}. Here, we investigate whether such axion-photon dynamics persist in more astrophysically motivated setups where BHs are surrounded by plasma. For intuition, Appendix~\ref{app:flatinstab} provides a simplified flat spacetime analysis that, while analytically tractable, captures essential aspects of the phenomenon.
\vskip 2pt
To study the system in a Reissner-Nordstr\"{o}m background, we linearise the field equations~\eqref{eq:IMS_evoleqns} and perform a multipolar decomposition of the fields. As the geometry is spherically symmetric, perturbations can be recasted in two decoupled sectors, axial and polar, depending on their behaviour under parity transformations (see Appendix~\ref{app:BHPT})~\cite{ReggeWheeler,Zerilli:1970wzz,Zerilli:1970se,Zerilli:1974ai}. Owing to its pseudo-scalar nature, the axion couples \emph{only} to the axial sector, which thus becomes the focus of our analysis. The polar sector instead, is described by the standard gravito-electromagnetic perturbations (in the presence of plasma). Details of the perturbation scheme and derivation of the evolution equations are provided in Appendix~\ref{appBHPT_subsec:plasma}. In the axial sector, plasma perturbations can be solved analytically. As the pressure and density perturbation are scalar quantities with a polar symmetry, they vanish identically. Meanwhile, the axial fluid velocity -- $v_4$ -- can be related to the axial electromagnetic mode -- $u_4$ -- via the linear relation 
\begin{equation}
\label{eq:IMS_momentumcons}
v_{4}=-\frac{e}{m\ped{e}}u_{4}\,,
\end{equation}
similar to the flat spacetime case~\eqref{eq:mom_parallel}. The system is then described by three variables:~the gravitational Moncrief-like master variable $\Psi$, the electromagnetic axial degree of freedom $u_{4}$ and the axion multipole field $\psi$. These three functions obey a set of coupled, second-order partial differential equations:
\begin{equation}
\label{eq:IMS_wavelike-eqn}
\begin{aligned}
\hat{\mathcal{L}}\Psi &=\Bigg(\frac{4 Q^4}{r^6}+ \frac{Q^2(-14 M + r (4+\lambda))}{r^5} + \left(1-\frac{2M}{r}\right)\left[\frac{\lambda}{r^{2}}-\frac{6M}{r^{3}}
\right]\!\Bigg)\Psi-\frac{8 Q f}{r^3 \lambda} u_{4}\,, \\
\hat{\mathcal{L}}u_{4}&=f\left(\omega\ped{p}^2 + \frac{\lambda}{r^{2}} + \frac{4 Q^2}{r^4}\right) u_{4}-\frac{(\ell-1)\lambda(\ell+2) Q f}{2 r^3}\Psi 
+\frac{2 \lambda Q f k\ped{a}}{r^3}\psi\,,
 \\
\hat{\mathcal{L}}\psi&=f\Bigg(\frac{-2 Q^2+r(2M+r \lambda +r^3 \mu\ped{a}^2)}{r^4}\Bigg) \psi +\frac{2 Q  f k\ped{a} }{r^2} u_{4}\,,
\end{aligned}
\end{equation}
where $\lambda=\ell(\ell+1)$, the wave operator $\hat{\mathcal{L}} = \partial^{2}/\partial r_*^{2} - \partial^{2}/\partial t^{2}$ and the tortoise coordinate is defined as $\mathrm{d}r_*/\mathrm{d}r = f^{-1}$. In the limit $Q \rightarrow 0$, the gravitational sector is described by the Regge--Wheeler equation~\eqref{eq:potentials_RW_Z} and the axion completely decouples from the system. In other words, in the absence of background electromagnetic fields, no mixing with the photon is possible.
\vskip 2pt
We evolve eqs.~\eqref{eq:IMS_wavelike-eqn} in time with a two-step Lax-Wendroff algorithm that uses second-order finite differences~\cite{Zenginoglu:2011zz}, following earlier work~\cite{Krivan:1997hc,Pazos_valos_2005,Zenginoglu:2011zz,Zenginoglu:2012us,Cardoso:2021vjq}. Our grid is uniformly spaced in tortoise coordinates $r_*$, with the boundaries placed sufficiently far away such that boundary effects cannot have an impact on the evolution of the system at the extraction radius. Further details on our numerical scheme are reported in Appendix~\ref{appBHPT_sec:num_frame}, including convergence tests. Analogous to~\cite{Cannizzaro:2024yee}, we initialise each sector with a Gaussian, choosing its amplitude, frequency, width and location to be $A = 1$, $M\Omega_0 = 0.4$, $\sigma = 4.0M$, $r_0 = 20M$. 
\begin{figure}[t!]
    \centering
    \includegraphics[scale=1]{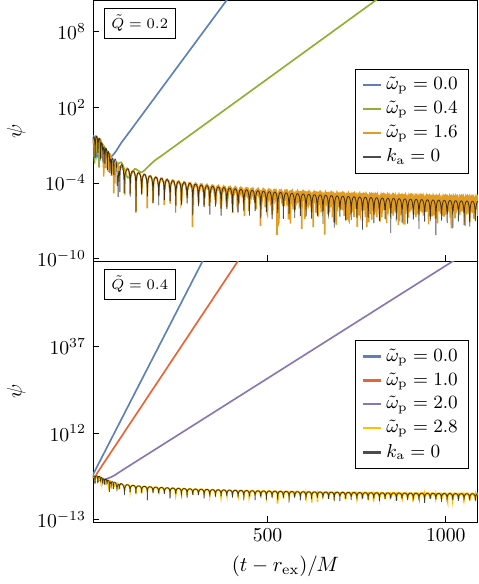}
    \caption{Evolution of the axion sector $\psi$ for $\tilde{Q} = 0.2$ (\emph{top panel}) and $\tilde{Q} = 0.4$ (\emph{bottom panel}), for various choices of the plasma frequency. We initialise with a Gaussian in all sectors (gravitational, electromagnetic and axion) and choose $k\ped{a} = 20$, $\ell = 2$, $\tilde{\mu}\ped{a} = 0.2$, while we extract at $r\ped{ex} = 30M$. In absence of plasma [{\color{Mathematica1}{blue}}] ($\tilde{\omega}\ped{p} = 0$), the instability rates follow eq.~(47) in~\cite{Boskovic:2018lkj}. For fixed couplings, the critical plasma frequency necessary to quench the instability scales as $\tilde{\omega}^{\rm crit}\ped{p} \propto \tilde{Q}^{3/2}$. When the photon production is heavily suppressed (large $\omega\ped{p}$), the behaviour of the axionic sector limits towards that of a free scalar [{\color{black}{\textbf{black}}}].}
    \label{fig:axion_hair}
\end{figure}
\vskip 2pt
Figure~\ref{fig:axion_hair} shows the evolution of the axion field for different values of the plasma frequency. The two panels corresponds to different values of the BH charge $\tilde{Q}=0.2, 0.4$, where we denote dimensionless quantities with a tilde. As evident from the figure, the system is unstable to axionic perturbations for $\tilde{\omega}\ped{p}=0$, provided that the BH charge or the axionic coupling is sufficiently high. We find a good agreement with the instability condition found in~\cite{Ikeda:2018nhb} in this limit. Crucially, increasing $\omega\ped{p}$ progressively weakens the instability, quenching it altogether after a critical value. This is similar to the flat spacetime analysis reported in Appendix~\ref{app:flatinstab}: large values of $\omega\ped{p}$ tend to stabilise the system, and higher values of $Q$ or $k\ped{a}$ are needed to restore the instability. Indeed, as shown in the \emph{bottom panel}, for a higher value of $Q$, a larger $\omega\ped{p}$ is needed to quench the growth. Specifically, at fixed couplings $k\ped{a}$ the critical plasma frequency necessary to quench the instability scales as $\tilde{\omega}^{\rm crit}\ped{p} \propto \tilde{Q}^{3/2}$.
\vskip 2pt
Finally, whenever the instability is quenched, the axion behaves as a free scalar field in a Reissner-Nordstr\"{o}m geometry:~this is because plasma effectively decouples it from the photon. This is also shown in Figure~\ref{fig:axion_hair}, where the axion field in the regime $k\ped{a} \neq 0$, $\omega\ped{p} \gg \mu\ped{a}$ matches that of an axion field with $k\ped{a}=0$, as indicated by the black line. This is in agreement with the discussion in the previous section and Appendix~\ref{app:flatinstab} (see Figure~\ref{fig:flatmodes}).
\subsection{Dark Photon Clouds:~Photon Production}\label{sec_IMS:darkphotons}
Consider then dark photon-photon mixing in a BH background. Since no background electromagnetic fields are necessary for the mixing to occur, we stick to a Schwarzschild spacetime~\eqref{eq:metric_schwarzschild}. Similar to before, we consider a non-relativistic background plasma. 
\vskip 2pt
Setting $k\ped{a} = 0$, we linearise the field equations~\eqref{eq:IMS_evoleqns} and perform a multipolar expansion using the \emph{ansatz} reported in Appendix~\ref{appBHPT_sec:perturbations}. In contrast to the axion case, dark photon perturbations appear in both the axial and polar sectors, and they couple directly to the electromagnetic perturbations. For simplicity, we still focus on the axial sector, as fluid perturbations can be treated analytically here. In the absence of a background charge, gravitational perturbations decouple from the system. As a consequence, the dynamics reduce to two coupled master functions: the axial electromagnetic mode $u_{4}$ and the dark photon mode $u'_{4}$. They satisfy the following system of coupled partial differential equations:
\begin{equation}
\label{eq:wavelike-eqn_DP}
\begin{aligned}
\hat{\mathcal{L}}u_{4}&=f\left(\omega\ped{p}^2 + \frac{\lambda}{r^{2}}\right) u_{4}+f \mu_{\gamma'}^2 \sin{\chi_0}\,u'_{4}\,,
 \\
\hat{\mathcal{L}}u'_{4}&=f\left(\mu_{\gamma'}^2 + \frac{\lambda}{r^{2}}\right) u'_{4}+f \mu_{\gamma'}^2 \sin{\chi_0}\,u_4\,,
\end{aligned}
\end{equation}
where $f = 1-2M/r$ in the Schwarzschild case. As evident from~\eqref{eq:wavelike-eqn_DP}, the mixing between modes is proportional to both the vacuum mixing angle $\sin{\chi_0}$ and the dark photon mass, in agreement with the flat spacetime result~\eqref{eq:DPEdispersion}. Note that the present analysis is carried out in the \emph{interaction basis}. For completeness, the corresponding form of~\eqref{eq:wavelike-eqn_DP} in the mass basis is provided in Appendix~\ref{appDPbasis:BHPT}. We now proceed to evolve the equations of motion in time, initialising the system with perturbations purely in the dark photon sector. 
\begin{figure}[t!]
    \centering
    \includegraphics[scale=1]{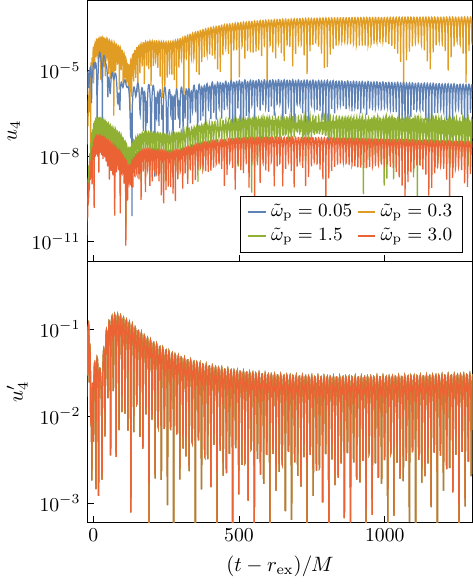}
    \caption{Evolution of the electromagnetic (\emph{top panel}) and dark photon (\emph{bottom panel}) field for various choices of the plasma frequency. We initialise purely in the dark photon sector, and take $\sin{\chi_0} = 0.0001$, $\tilde{\mu}_{\gamma'} = 0.3$ and $\ell = 1$. In the \emph{top} (\emph{bottom}) \emph{panel}, we extract at $r\ped{ex} = 400\,(50)\,M$. The dark photon field settles on a QBS, whose frequency we have checked explicitly: $\tilde{\omega}\ped{R} = 0.29564$ (time domain) vs. $\tilde{\omega}\ped{R} = 0.29598$ (frequency domain).}
    \label{fig:DarkPhotonPhoton}
\end{figure}
\vskip 2pt
The results are presented in Figure~\ref{fig:DarkPhotonPhoton}. In the \emph{bottom panel}, we show the evolution of the dark photon field, which remains largely unaffected by the mixing due to the small coupling. The dynamics are dominated by outwards-propagating waves and the formation of a quasi-bound state (QBS) near the BH. The figure specifically highlights the latter, as the field is extracted at $r\ped{ex} = 50M$, just beyond the peak density of the bound state. We have explicitly verified that its frequency matches that of the QBS (see the caption for details). In the \emph{top panel} instead, we show the electromagnetic field, which is clearly affected by the plasma. Its production is peaked at $\tilde{\omega}\ped{p}=\tilde{\mu}_{\gamma'}=0.3$, where the conversion probability is expected to be resonantly enhanced. When increasing the plasma density, i.e., $\omega\ped{p}\gg \mu_{\gamma'}$, the in-medium suppression acts and the electromagnetic field decays. In this regime, the two fields decouple more and more, resulting in less photons being produced. The rate at which this happens has been studied analytically in flat spacetime~\cite{Dubovsky:2015cca,Caputo:2021efm}. It was found that for large plasma frequencies, the suppression factor should be $\propto \mu_{\gamma'}^2/\omega\ped{p}^{2}$.
\begin{figure}[t!]
    \centering
    \includegraphics[scale=1]{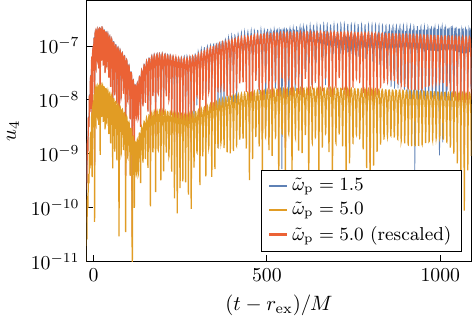}
    \caption{Similar setup as Figure~\ref{fig:DarkPhotonPhoton} for $\tilde{\omega}\ped{p} = 1.5, 5.0$. The suppression goes as $\propto \omega\ped{p}^{-2}$, which is the rescaling applied to the red curve.}
    \label{fig:DP_scaling}
\end{figure}
\vskip 2pt
We verify this scaling numerically in a BH background in Figure~\ref{fig:DP_scaling}, which shows the electromagnetic field evolution for a fixed boson mass ($\tilde{\mu}_{\gamma'}=0.3$) and $\tilde{\omega}\ped{p}=1.5, 5$. Both cases correspond to a regime where $\omega\ped{p}\gg \mu_{\gamma'}$, such that the scaling in~\cite{Dubovsky:2015cca, Caputo:2021efm} should hold. The two curves [{\color{Mathematica1}{blue}} and {\color{Mathematica4}{red}}] show remarkable agreement, particularly in the early part of the signal. During this stage, the dynamics are dominated by relativistic modes with $\omega \gg \omega\ped{p}$, such that dispersion effects are negligible. At later times instead, the signal is ``contaminated'' by non-relativistic modes with $\omega \gtrsim \omega\ped{p}$, leading to mild deviations from the expected scaling due to dispersive corrections.
\vskip 2pt
By validating the predicted scaling from~\cite{Dubovsky:2015cca, Caputo:2021efm}, we can estimate the strength of the electric field generated by a superradiant dark photon cloud. In particular, we find the ratio between the electromagnetic and dark photon field amplitudes to be
\begin{equation}
\frac{A_{\gamma}}{A_{\gamma'}} = 10^{-17}\left(\frac{\sin{\chi_0}}{10^{-7}}\right)\left(\frac{\mu_{\gamma'}}{10^{-15}\,\mathrm{eV}}\right)^{2}\left(\frac{10^{-10}\,\mathrm{eV}}{\omega\ped{p}}\right)^{2}\,.
\end{equation}
Considering the scenario of a fully grown dark photon cloud~\cite{East:2017ovw,East:2017mrj} (such that the mass of the cloud is $10$\% of the BH mass), the strength of the observable electric field, which will depend on the ``in-medium suppression factor'' ($\mu_{\gamma'}/\omega\ped{p}$), is estimated to be
\begin{equation}\label{eq:electricfield_DPgenerated}
E_0 \simeq 6.3 \left(\frac{\sin{\chi_0}}{10^{-7}}\right) \left(\frac{\mu_{\gamma'}/\omega\ped{p}}{10^{-7}}\right) \left(\frac{\mu_{\gamma'} M}{0.2}\right)^3 \left(\frac{10^6 M_{\odot}}{M}\right)\,\frac{\mathrm{V}}{\mathrm{m}}\,.
\end{equation}
Note that, fixing $\mu_{\gamma'} M$ and $M$ implies a value for the boson mass~\eqref{eq:gravitational_fineS}, e.g.,~for $\mu_{\gamma'} M = 0.2$ and $M = 10^{6}M_{\odot}$, the mass of the boson is $\mu_{\gamma'} = 2.6 \times 10^{-17}\,\mathrm{eV}$. To convert from an electronic density to the plasma frequency, one can then make use of eq.~\eqref{eq:BHenv_plasmafreq}.
\clearpage
\section{Summary and Outlook}\label{sec_IMS:summary}
Hypothetical new ultralight degrees of freedom are expected to couple weakly to the Standard Model, making their detection inherently challenging. However, in regions of strong gravity, such as those surrounding compact objects, ultralight fields may accumulate or be amplified, allowing them to play an important role. This prospect has driven considerable efforts to characterise possible observational signatures, with many scenarios involving electromagnetic radiation.
\vskip 2pt
Despite this, the role of plasma is often either entirely neglected or treated using oversimplified flat spacetime approximations. As a large fraction of astrophysical compact objects is expected to be surrounded by plasma, e.g.,~in the form of accretion disks, a more detailed analysis is imperative. In this chapter, we introduced a fully relativistic and self-consistent framework to study how plasma affects the dynamics of ultralight bosons near compact objects. Two cases of interest are studied explicitly:~axion-photon mixing around charged BHs and dark photon-photon mixing around neutral BHs. Both scenarios yield similar conclusions:~the presence of plasma strongly impacts the conversion rate and suppresses it in realistic regimes. This effect, termed \emph{in-medium suppression}, could have an impact on the observational signatures from such systems and thus on constraints that have been set on the coupling constants. Moreover, in-medium suppression can stabilise superradiant systems by inhibiting the depletion of bosonic clouds into photons (see Figure~\ref{fig:ContouronmuMKaMs}). This strengthens the prospects of detecting superradiant clouds via alternative channels, e.g.,~through binary systems~\cite{Zhang:2018kib,Baumann:2018vus, Zhang:2019eid,Baumann:2019ztm,  Baumann:2021fkf,Baumann:2022pkl,Takahashi:2021eso,Cole:2022yzw,Baumann:2022pkl,Takahashi:2023flk,Tomaselli:2023ysb,Brito:2023pyl,Duque:2023seg,Tomaselli:2024bdd,Tomaselli:2024dbw,Boskovic:2024fga,Khalvati:2024tzz}.
\vskip 2pt
As an example, consider typical electron densities in the interstellar medium $n\ped{e}\!\sim\! [10^{-3}-10]\,\mathrm{cm}^{-3}$~\cite{Saintonge:2022tfq}, corresponding to plasma frequencies $\omega\ped{p}\!\sim\! [10^{-12}\!-\!10^{-10}]\, \textrm{eV}$~\eqref{eq:BHenv_plasmafreq}. These densities are already sufficient to suppress mixing across much of the superradiant mass range $\mu\sim [10^{-20}-10^{-10}]\,\textrm{eV}$. In denser environments -- like accretion disks -- the suppression becomes even more severe~\cite{Abramowicz2013, Barausse:2014tra}
\vskip 2pt
A natural extension of this study would be to generalise the framework to magnetised neutron stars. Recent studies have shown that such systems can exhibit a rich axion phenomenology, particularly in small, localised regions of the magnetosphere known as \emph{polar caps}, where strong electromagnetic fields may efficiently produce axions~\cite{Prabhu:2021zve, Noordhuis:2022ljw}. These can then stream away and resonantly convert into photons, generating broadband radio fluxes. Alternatively, they could be bounded gravitationally to the star, forming axion clouds with extreme densities~\cite{Noordhuis:2023wid}. Such clouds can generate an axion-induced electric field, leading to striking electromagnetic observables like periodic nulling in the pulsar emission or complementary radio emissions~\cite{Caputo:2023cpv}. Notably, the strength of this electric field scales with the \emph{in-medium suppression factor} $(\mu\ped{a}/\omega\ped{p})$, similar to the dark photon case~\eqref{eq:electricfield_DPgenerated}. Applying this framework to such systems requires a careful modelling of the background system, in particular the magnetic field geometry and the magnetosphere phenomenology. In addition, there will be mixing between $\ell$--modes in the presence of magnetic fields (see~\cite{Brito:2014nja, Day:2019bbh}).
\begin{figure}[t!]
    \centering
    \includegraphics[scale=1]{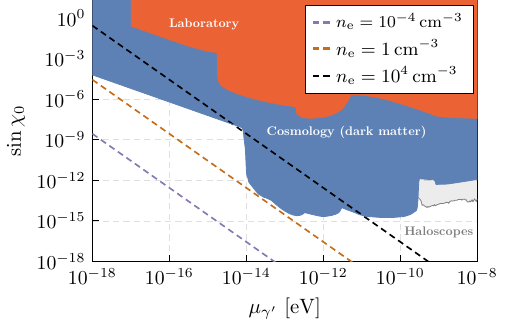}
    \caption{Available mass-coupling parameter space for the dark photon. Below the dashed lines, in-medium suppression is efficient and cannot be removed by nonlinear effects. Typical densities of the interstellar medium $\left(n\ped{e}\sim [10^{-3}\!-\!10]\,\mathrm{cm}^{-3}\right)$~\cite{Saintonge:2022tfq} are sufficient to suppress the mixing in most of the unconstrained regions of the parameter space. Shaded regions show constraints from cosmological nature [{\color{Mathematica1}{blue}}], experiments on Earth [{\color{Mathematica4}{red}}] or haloscopes [{\color{gray}{grey}}], see~\cite{ Caputo:2021eaa,AxionLimits} and references therein.}
    \label{fig:DP_bounds}
\end{figure}
\vskip 2pt
Finally, another interesting direction is nonlinear photon-plasma interactions in the context of dark photon superradiance. It has been argued~\cite{Caputo:2021efm} that in-medium suppression may be alleviated in a promising region of parameter space, specifically for $\mu_{\gamma'}\gtrsim 10^{-16}\,\rm{eV}$ and $\sin{\chi_0} \gtrsim 10^{-8}\!-\!10^{-7}$, due to relativistic transparency effects induced by the dark photon cloud. A similar assumption was made in~\cite{Siemonsen:2022ivj}, which argues that the dark photon field may generate a pair plasma via the Schwinger mechanism. In that scenario, the plasma frequency is assumed to vanish shortly after the plasma forms, due to the large ambient electric field. In particular, using eq.~\eqref{eq:electricfield_DPgenerated}, the critical field amplitude required to suppress the plasma frequency can be estimated as $E\ped{NL} = m\ped{e}\omega\ped{p}/e$, at which point nonlinear plasma dynamics are triggered~\cite{Cardoso:2020nst,Cannizzaro:2023ltu}. Figure~\ref{fig:DP_bounds} shows a lower bound on the mixing value above which nonlinear effects can remove the in-medium suppression. As seen there, typical densities of the interstellar medium are sufficient to suppress the mixing in most of the unconstrained regions of the parameter space. In denser environments, such as accretion disks, the required boson mass for nonlinear effects may lie entirely outside the allowed superradiant range. Accounting for nonlinearities could therefore refine our understanding of relativistic transparency and open the door to exploring a broader range of potentially important plasma effects (see e.g.,~\cite{Cannizzaro:2023ltu}).
\chapter{Impact of Plasma on the Relaxation of Black Holes} \label{chap:plasma_ringdown}
\vspace{-0.8cm}
\hfill \emph{Nothingness is being and being nothingness...}

\hfill \emph{Our limited mind can not grasp or fathom this,}

\hfill \emph{for it joins infinity}
\vskip 5pt

\begin{figure}[t!]
    \centering
    \includegraphics[scale=1]{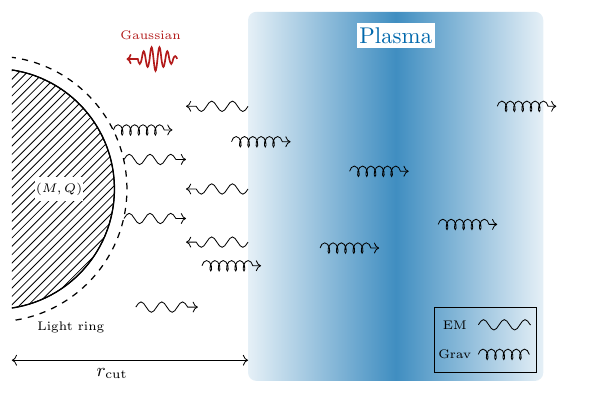}
    \caption{Schematic illustration of our setup:~a charged BH surrounded by plasma is stimulated by external processes (an initial Gaussian wavepacket), emitting electromagnetic and gravitational radiation. While GWs are able to travel through the plasma to distant observers, low frequency electromagnetic waves are not. Instead they excite further GWs, echoes of the original burst.}  
    \label{fig:schematicillustration}
\end{figure}

\hfill Azrael of Gerona
\vskip 35pt
\noindent In the previous chapters, I explored the interplay between ultralight bosons, photons, and plasma. I now turn to a different subject:~over the next two chapters, I focus on the impact of environments on the final stage of a binary coalescence -- the ringdown. 
\vskip 2pt
Plasma may also play a role here, particularly if black holes carry electric charge. While significant amounts of electromagnetic charge are not expected to survive long for accreting systems (due to selective accretion, Hawking radiation or pair production~\cite{Eardley:1975kp,Gibbons:1975kk,Cardoso:2016olt}), exceptions exist. For instance, a subset of primordial BHs formed in the early Universe can carry a large amount of charge, suppressing Hawking radiation and potentially allowing for electric or colour-charged BHs to survive to our days~\cite{deFreitasPacheco:2023hpb,Alonso-Monsalve:2023brx}. Additionally, BH mergers might be accompanied by strong magnetic fields pushing surrounding plasma to large radii, and preventing neutralisation processes. 
\vskip 2pt
Beyond the realm of Standard Model physics, BHs could be charged in a variety of models, by circumventing in different ways discharge mechanisms. These models include millicharged dark matter or hidden vector fields (see Section~\ref{BHenv_subsec:candidates}), constructed to be viable cold dark matter candidates~\cite{Bai:2019zcd,Kritos:2021nsf,DeRujula:1989fe,Holdom:1985ag,Davidson:2000hf,Dubovsky:2003yn,Sigurdson:2004zp,Gies:2006ca,Gies:2006hv,Burrage:2009yz,Ahlers:2009kh,McDermott:2010pa,Dolgov:2013una,Haas:2014dda,Cardoso:2016olt,Khalil:2018aaj,Caputo:2019tms,Gupta:2021rod,Carullo:2021oxn,Fiorillo:2024upk}. Finally, some BH mimickers are globally neutral while possessing a non-vanishing dipole moment, thus emitting electromagnetic radiation. Examples include topological solitons in string-theory fuzzballs scenarios~\cite{Bah:2020ogh, Bah:2021owp}. 
\vskip 2pt
Charge constraints via GW dephasing in the inspiral phase of two compact objects or via BH spectroscopy, assume implicitly that photons propagate freely from source to observer~\cite{Cardoso:2016olt,Khalil:2018aaj,Gupta:2021rod,Carullo:2021oxn}. But the Universe is filled with matter. Even if dilute, the interstellar plasma prevents the propagation of electromagnetic waves with frequencies smaller than the plasma frequency, which effectively behaves as an effective mass~\eqref{eq:BHenv_plasmafreq}.
\vskip 2pt
In this chapter, I discuss the work of~\cite{Cannizzaro:2024yee} and investigate the ringdown of charged BHs in the presence of plasma. The emission of GWs and electromagnetic waves during mergers of compact, charged objects is a coupled phenomenon. Hence, the BH \emph{gravitational} spectrum contains electromagnetic-driven modes~\cite{Leaver:1990zz,Berti:2009kk}. However, if electromagnetic modes are unable to propagate, their impact on GW generation and propagation could be important, affecting spectroscopy tests to an unknown degree. In this chapter, I will show from first principles that~(i) electromagnetic waves are indeed screened by plasma, which filters out electromagnetic-led modes from GWs and~(ii) in certain plasma-depleted environments, GW echoes are triggered, serving as a clear observational signature of plasmas surrounding charged BHs. A schematic illustration of our setup is shown in Figure~\ref{fig:schematicillustration}.
\vskip 2pt
The structure of this chapter is as follows. In Section~\ref{sec_QNMplasma:setup}, I lay out the theoretical framework and governing equations. In Section~\ref{sec_QNMplasma:results}, I present results for plasmas localised near the light ring, while Section~\ref{sec_QNMplasma:echoes} considers plasmas positioned farther away. Finally, I summarise the findings in Section~\ref{sec_QNMplasma:summary}. Additional details on perturbation theory and the numerical methods are provided in Appendix~\ref{app:BHPT}. Throughout this chapter, I adopt Gaussian units for Maxwell's equations, while dimensionless quantities will be denoted with a tilde as usual, e.g., the charge $\tilde{Q} = Q/M$.
\section{Setup}\label{sec_QNMplasma:setup} 
As in the previous chapter, we consider an ``Einstein cluster'' -- equivalent to an anisotropic fluid -- surrounding a charged BH~\cite{Einstein_cluster,Einstein_cluster_2,Cardoso:2021wlq,Cardoso:2022whc} (see Section~\ref{BHenv_subsec:DMStruct}). Focusing on a fluid consisting of electrons, the stress-energy tensor is given by eq.~\eqref{eq:plasmaSET}.\footnote{The following discussion also applies to millicharged DM. For clarity purposes, we focus on electrons.} We then consider Einstein-Maxwell theory in the presence of this fluid. The relevant field equations are~\eqref{eq:EFE}
\begin{equation}
\label{eq:einstein_maxwell}
G_{\mu\nu}= 8 \pi \left(T_{\mu\nu}^{\rm EM}+T_{\mu\nu}^{\rm p}\right)\,, \quad \nabla_\nu F^{\mu\nu}=j^\mu\,,
\end{equation}
where $G_{\mu\nu}$ and $F_{\mu\nu}$ are the Einstein and Maxwell tensor, respectively, $j^\mu=e n\ped{e} v^\mu$ the plasma current and $T_{\mu\nu}^{\rm EM}$ the stress-energy tensor for the electromagnetic sector:
\begin{equation}
\begin{aligned}
T_{\mu\nu}^{\rm EM}&=\frac{1}{4 \pi}\left(g^{\rho \sigma}F_{\rho \mu}F_{\sigma \nu}-\frac{1}{4}g_{\mu\nu}F_{\rho \sigma}F^{\rho \sigma}\right)\,.
\end{aligned}
\end{equation}
Finally, to close the system, we use the momentum and continuity equation of the charged fluid, as in eq.~\eqref{eq:momentum_continuity}. Following the approach of Section~\ref{sec_IMS:curved}, we neglect the backreaction of the fluid in the Einstein and Maxwell equations. We also assume the presence of ions in the plasma, which can be considered as a stationary, neutralising background~\cite{Cannizzaro:2020uap,Cannizzaro:2021zbp,Cannizzaro:2023ltu,Spieksma:2023vwl}. Under these assumptions, eqs.~\eqref{eq:einstein_maxwell} yield the Reissner-Nordstr\"{o}m solution~\eqref{eq:RN_Sol}, with the event horizon located at $r_+=M+\sqrt{M^2-Q^2}$ and the light ring at $r\ped{LR}=3M/2+\sqrt{9M^{2}-8Q^{2}}/2$. 
\vskip 2pt
For a non-relativistic fluid, i.e., $P\ped{t}\ll\rho$, eq.~\eqref{eq:plasmaSET} reduces to $T^{\rm p}_{\mu\nu}\approx \rho v_\mu v_\nu+P\ped{t}(g_{\mu\nu}-r_\mu r_\nu)$, and the left hand side of the momentum equation~\eqref{eq:momentum_continuity} resembles the non-relativistic Euler equation. Solving the momentum equation~\eqref{eq:momentum_continuity} then yields the tangential pressure
\begin{equation}
\label{eq:tangentialpressure}
    P\ped{t}=-\frac{n\ped{e}(e Q r\sqrt{f}+(Q^2-Mr)m\ped{e})}{Q^2-3Mr+2r^2}\,.
\end{equation}
For the non-relativistic assumption $P\ped{t} \ll \rho = n\ped{e} m\ped{e}$ to hold, we must have either $Q/M<m\ped{e}/e$ or $r\gg M$. Given the large charge-to-mass ratio of electrons ($e/m\ped{e}\approx 10^{22}$), the former condition is only satisfied for extremely weakly charged BHs. Nevertheless, a number of effects can affect this outcome, such as magnetic fields, the formation of a cavity in the plasma due to mergers~\cite{Armitage:2005xq,Artymowicz:1994bw,Armitage:2005xq,Grobner:2020drr,Farris:2014zjo,Ishibashi:2020zzy,2013MNRAS.436.2997D,2017AA...598A..43C} or the partial screening of the BH charge by plasma over a Debye length~\eqref{eq:DebyeLength}~\cite{Feng:2022evy, Alonso-Monsalve:2023jfq}. In the following, we consider high values of $Q$ as a proxy to model these scenarios, which are too complicated to be included in a self-consistent way. Moreover, for millicharged DM, the charge-to-mass ratio of the particles can be arbitrarily small. 
\subsubsection{Relativistic regime}
In principle, one can extend this analysis to the relativistic regime by considering the full momentum equation.  Following the same approach as in the non-relativistic case, we solve the background momentum equation, yielding:
\begin{equation}
    P\ped{t}=-\frac{n\ped{e}\left(e Q r\sqrt{f}+(Q^2-Mr)m\ped{e}\right)}{2fr^{2}}\,.
\end{equation}
The axial component of the perturbed momentum equation gives the relationship between the perturbed four-velocity and the electromagnetic axial mode:
\begin{equation}
    v_4=\frac{2 e r^2 f}{e Q r\sqrt{f}-\left(Q^2-3Mr+2r^{2}\right)m\ped{e}}u_4\,.
\end{equation}
In the limit of large $r$, i.e., in the non-relativistic regime, this expression correctly reduces to eq.~\eqref{eq:momentumcons}, recovering $e n^{(0)}\ped{e} v_4=-\omega\ped{p}^2 u_4$. In contrast, in the strong field regime where $e\gg m\ped{e}$, the current entering Maxwell's equations is:
\begin{equation}
    e n^{(0)}\ped{e}v_4 \approx \frac{2 e n^{(0)}\ped{e} r\sqrt{f}}{Q} u_4\,.
\end{equation}
This leads to an \emph{effective plasma frequency}:
\begin{equation}
\label{eq:effectiveplasmafreq}
\omega\ped{p}^2= \frac{2 e n^{(0)}\ped{e} \sqrt{f}}{A_0}\,,
\end{equation}
where $A_{0}=Q/r$ is the electromagnetic potential of the BH.
\vskip 2pt
Compare this to a well-known result from plasma physics:~in the presence of a strong electromagnetic potential $\vec{A}$, electrons are unable to oscillate due to their large relativistic mass. In this scenario, the \emph{relativistic plasma frequency} is given by~\cite{KawDawson1970}
\begin{equation}\label{eq:plasmafreqKaw}
    \omega\ped{p}^2= \frac{\overline{n}\ped{e} e^2}{\overline{m}\ped{e} \gamma\ped{e}}\,,
\end{equation}
where $\gamma\ped{e}$ is the Lorentz factor of the electrons under the influence of an electromagnetic field, while $\overline{m}\ped{e}$ and $\overline{n}\ped{e}$ are the rest mass and rest-mass density, respectively. In our model, the Lorentz factor can be estimated by equating the (relativistic) centrifugal force of the circular orbits of the electrons to the Coulomb attraction between the electrons and the charged BH. This yields $\gamma\ped{e}\approx \sqrt{1+ A_0 e/ \overline{m}\ped{e}}$. 
\vskip 2pt
Crucially, if one uses $e A_0/\overline{m}\ped{e}\gg 1$ and $\overline{n}\ped{e}=n\ped{e}/\gamma\ped{e}$, eqs.~\eqref{eq:plasmafreqKaw} and~\eqref{eq:effectiveplasmafreq} coincide (modulo a redshift factor). This reveals a \emph{relativistic transparency effect}:~the background charge $Q$ suppresses the plasma frequency in the vicinity of the BH. This is the first time, to the best of our knowledge, that a consistent model of plasmas around charged BHs leads to this effect. In the relativistic regime, we thus expect the same phenomenology to hold as in the non-relativistic case, albeit with a largely suppressed effective mass. To simplify the discussion, we will proceed in the non-relativistic regime.
\subsubsection{Evolution equations}
Consider then the linearisation of the field equations~\eqref{eq:einstein_maxwell} around the Reissner-Nordstr\"{o}m geometry~\eqref{eq:RN_Sol}, the background fields and fluid variables. Perturbations can then be decomposed in two sectors -- \emph{axial} (or odd) and \emph{polar} (or even) -- depending on their behaviour under parity transformations. These two sectors decouple in spherically symmetric geometries (see Appendix~\ref{app:BHPT})~\cite{ReggeWheeler,Zerilli:1970wzz,Zerilli:1970se,Zerilli:1974ai}.
\vskip 2pt
The axial sector is completely determined by two functions, a Moncrief-like ``master gravitational'' variable $\Psi$~\cite{Moncrief:1974ng,PhysRevD.9.2707,Moncrief:1975sb} and a ``master electromagnetic'' variable $u_4$, which obey a coupled set of second-order, partial differential wavelike equations,
\begin{equation}
\label{eq:QNM_wavelike-eqn}
\begin{aligned}
\hat{\mathcal{L}}\Psi &=\Bigg(\frac{4 Q^4}{r^6}+ \frac{Q^2(-14 M + r (4+\lambda))}{r^5} + \left(1-\frac{2M}{r}\right)\left[\frac{\lambda}{r^{2}}-\frac{6M}{r^{3}}
\right]\Bigg)\Psi-\frac{8 Q f}{r^3 \lambda} u_{4}\,, \\
\hat{\mathcal{L}}u_{4}&=f\left(\omega\ped{p}^2 + \frac{\lambda}{r^{2}} + \frac{4 Q^2}{r^4}\right) u_{4}-\frac{(\ell-1)\lambda(\ell+2) Q f}{2 r^3}\Psi\,,
\end{aligned}
\end{equation}
where $\lambda=\ell(\ell+1)$, $\hat{\mathcal{L}} = \partial^{2}/\partial r_*^{2} - \partial^{2}/\partial t^{2}$ and the tortoise coordinate is defined as $\mathrm{d}r_*/\mathrm{d}r = f^{-1}$~\eqref{eq:RN_Sol}. In the limit $Q\rightarrow 0$, the equations decouple:~the first one reduces to the Regge--Wheeler equation~\eqref{eq:potentials_RW_Z} while the second one coincides with the axial mode of an electromagnetic field in Schwarzschild in the presence of plasma~\cite{Cannizzaro:2020uap}. 
\vskip 2pt
The polar sector is more intricate, with electromagnetic and fluid perturbations being coupled. As detailed in the Appendix~\ref{appBHPT_subsec:plasma}, at large radii and neglecting metric fluctuations, we recover the dispersion relation $(\omega^2-k^2-\omega\ped{p}^2)\,\delta F_{12}=0$, where $k$ is the wave vector in Fourier space and $\delta F_{12}$ the perturbed Maxwell tensor. Even in the polar sector, the plasma frequency thus acts as an effective mass for the propagating degree of freedom. As the dynamics emerging in the axial sector are precisely contingent upon this fact, we expect the phenomenology to be similar~\cite{Cannizzaro:2020uap, Cannizzaro:2021zbp} and we hereafter focus only on the axial sector.
\subsubsection{Initial conditions}
The wavelike equations~\eqref{eq:QNM_wavelike-eqn} are evolved in time using the same numerical scheme as in the previous chapter (see Appendix~\ref{appBHPT_sec:num_frame} for details). We consider a plasma profile truncated at a radius $r\ped{cut}$, smoothened by a sigmoid-like function:
\begin{equation}
\label{eq:plasma-profile2}
\omega\ped{p} = \omega^{\rm (c)}\ped{p} \frac{1}{1+e^{- (r-r\ped{cut})/d}}\,.
\end{equation}
Here, $\omega^{\rm (c)}\ped{p}$ is the (constant) amplitude of the plasma barrier and $d$ determines how ``sharp'' the cut is. We choose $d = M$, but we verified that the results are not sensitive to this parameter.\footnote{As we will see, the outcome depends on a critical value for $\omega\ped{p}$ (the fundamental electromagnetic quasi-normal mode), making the density distribution after the barrier ($r > r\ped{cut}$) or a tenuous plasma before the barrier ($r < r\ped{cut}$) unimportant for the phenomenology.} Profile~\eqref{eq:plasma-profile2} allows us to consider two distinct scenarios;~(i) plasmas that ``permeate'' the light ring $(r\ped{cut} < r_{\scalebox{0.60}{$\mathrm{LR}$}})$, hence possibly affecting the \emph{generation} of quasi-normal modes (QNMs) and (ii)~plasmas localised away from the BH $(r\ped{cut} \gg r_{\scalebox{0.60}{$\mathrm{LR}$}})$, affecting at most the \emph{propagation} of the signal. We consider the initial conditions $\Psi(0,r) =\Psi_0$, $u_{4}(0,r)=u_{40}$ with~\cite{Cardoso:2020nst}
\begin{equation}
\label{eq:ID}
\begin{aligned}
\mkern-22mu(\Psi_0,u_{40}) &= (A\ped{g},A\ped{EM})\exp{\left[-\frac{(r_*-r_0)^2}{2 \sigma^2}-i\Omega_{0} r_*\right]} \,, \\
\partial_t \Psi_0 &= -i\Omega_0 \Psi_0\,, \quad \partial_t u_{40} = -i\Omega_0 u_{40}\,,
\end{aligned}
\end{equation}
where $(A\ped{g}, A\ped{EM})=(1,0), (0,1), (1,1)$ for $\mathrm{ID}\ped{g},\mathrm{ID}\ped{EM}$ and $\mathrm{ID}\ped{2}$, respectively. Throughout this chapter, we initialise at $r_0 = 20M$ and we extract the signal at $r\ped{ex} = 300M$. We pick $\sigma = 4.0M$ and wavepacket frequency $\Omega_{0} = 0.1$, yet tested extensively that our results are independent of these factors.
\section{Impact of Plasma on Quasi-Normal Modes}\label{sec_QNMplasma:results} 
\begin{figure}[t!]
    \centering
    \includegraphics[scale=1]{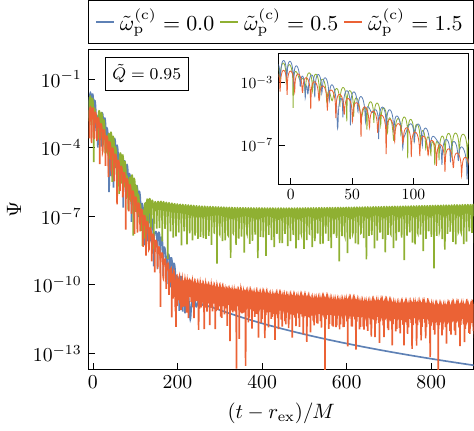}
    \caption{Gravitational waves for $\mathrm{ID}\ped{EM}$ and $\tilde{Q} = 0.95$. For sufficiently dense plasmas (when $\tilde{\omega}^{\rm (c)}\ped{p}$ exceeds the electromagnetic QNM frequency), the electromagnetic mode is screened. This is apparent in the inset:~only one -- gravitational-led mode -- is present at large plasma frequency [{\color{Mathematica4}{red}}]. Associated QNM frequencies can be found in Table~\ref{tab:frequencies}. At late times, long-lived modes are excited, in contrast to the usual power-law tail in vacuum. These originate from the quasi-bound states formed in the electromagnetic sector and are subsequently imprinted onto the GW signal.} \label{fig:contaminationID_EM}
\end{figure}
\begin{table}[b!]
\centering
\renewcommand*{\arraystretch}{1.4}
\resizebox{0.7\linewidth}{!}{
\tabcolsep=0.17cm
\begin{tabular}{|c||c|c|} 
\hline
$\tilde{Q} = 0.95$ & $\tilde{\omega}\ped{QNM}$ (time domain) & $\tilde{\omega}\ped{QNM}$ (frequency domain) \\
\hline \hline
$\tilde{\omega}^{\rm (c)}\ped{p} = 0.0$ 
& \mline{\begin{tabular}[t]{@{}c@{}}$0.42170~(0.086647)$\\$0.65475~(0.094609)$\end{tabular}} & \mline{\begin{tabular}[t]{@{}c@{}}$0.42169~(0.086659)$\\$0.65476~(0.094605)$\end{tabular}} \\ 
\hline
$\tilde{\omega}^{\rm (c)}\ped{p} = 1.5$  &\mline{$0.45902~(0.090143)$} & \mline{$0.45902~(0.090146)$} \\
\hline
\end{tabular}}
\caption{Real (imaginary) part of the fundamental QNM frequencies as calculated from a time- and frequency-domain approach. We consider: (i) no plasma (\emph{top row}), and we find two modes contributing to the signal and (ii) plasma (\emph{bottom row}), for which the electromagnetic mode is screened and only the gravitational one remains, with a shifted frequency. We obtain similar results in the time domain, regardless of the chosen $\mathrm{ID}$.}
\label{tab:frequencies}
\end{table}
When plasma permeates the light ring, BH relaxation is expected to change \emph{in the electromagnetic channel}. We indeed find a total suppression of the electromagnetic signal at large distances for large $\omega\ped{p}$. However, we find something more significant, summarised in Figure~\ref{fig:contaminationID_EM}, which shows the \emph{gravitational} waveform for $\tilde{Q}=0.95$ and different plasma frequencies $\omega^{\rm (c)}\ped{p}$. In absence of plasma, the signal [{\color{Mathematica1}{blue}}] is described by a superposition of gravitational- and electromagnetic-led modes, clearly visible (see inset) due to the high coupling $Q$. A best-fit to the signal shows the presence of two dominant modes, with (complex) frequencies reported in Table~\ref{tab:frequencies}. The plasma suppresses propagation of electromagnetic modes when $\omega\ped{p}$ exceeds the fundamental electromagnetic QNM frequency. More importantly, our results show that the coupling to GWs also affects the gravitational signal to an important degree. In fact, as apparent in Figure~\ref{fig:contaminationID_EM}, GWs now carry mostly a single gravitational-led mode [{\color{Mathematica4}{red}}], but with a shifted frequency, see Table~\ref{tab:frequencies}. This shift is surprising, and it originates from the coupling between gravity and electromagnetism. The presence of plasma thus affects the QNM frequencies of the \emph{gravitational signal}.    
\vskip 2pt
We confirm these results by frequency-domain calculations (where QNMs are obtained by direct integration with a shooting method) in Table~\ref{tab:frequencies}. Note that we impose purely outgoing boundary conditions at infinity in vacuum, while in the presence of plasma, we consider exponentially decaying electromagnetic modes at large distances, to account for quasi-bound states (QBS). Clearly, the results from time and frequency domain are in good agreement.
\vskip 2pt
As can be seen in Figure~\ref{fig:contaminationID_EM}, on longer timescales, the gravitational signal is ``polluted''. This is due to electromagnetic QBSs that are formed in the presence of plasma~\cite{Lingetti:2022psy,Dima:2020rzg}. These are long-lived states which are prevented from leaking to infinity due to the plasma effective mass, and are thus similar to QBS of massive fundamental fields~\cite{Brito:2015oca}. At late times, we indeed observe a signal ringing at a frequency comparable (yet slightly smaller) than the plasma frequency $\omega\ped{R} \lesssim \omega\ped{p}$. As the plasma frequency is increased, the QBSs form at progressively late times, and thus at lower amplitudes, unreachable for observations. This phenomenology is similar to the toy model considered in~\cite{Cardoso:2020nst}, but here explored from first principles.
\vskip 2pt
To study these late-time features in more detail, we consider the uncharged case $\tilde{Q} = 0$, such that the gravitational and electromagnetic sector decouple and the plasma can only affect the latter. Initialising the system with $\mathrm{ID}_{2}$, while placing the plasma \emph{away} from the BH, we vary the plasma frequency to understand its impact on the electromagnetic sector. 
\begin{figure}[t!]
    \centering
    \includegraphics[scale=1]{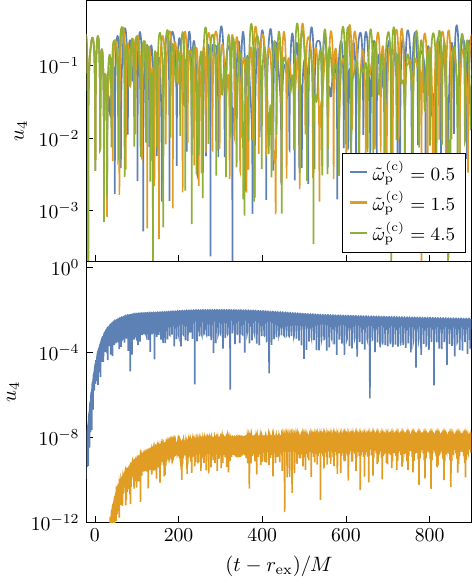}
    \caption{We show the electromagnetic sector $u_{4}$ with a plasma at $r\ped{cut} = 50M$, while initialising in both sectors ($\mathrm{ID}_{2}$) and varying the amplitude of the plasma barrier $\omega^{\rm (c)}\ped{p}$. In the \emph{top panel}, $u_{4}$ is extracted at $r\ped{ex} = 30M$ and we see trapped waves in the cavity. In the \emph{bottom panel}, the electromagnetic component is extracted at $r\ped{ex} = 300M$. As expected, while increasing the plasma barrier, less radiation is able to travel through. In fact, we do not show $\tilde{\omega}^{\rm (c)}\ped{p} = 4.5$, as it reaches the noise level.}
    \label{fig:unchargedplasma}
\end{figure}
In Figure~\ref{fig:unchargedplasma}, we show said system both at small (\emph{top panel}) and large radii (\emph{bottom panel}). In both cases, the formation of long-lived modes can be seen, yet these have a different origin. At small radii, there are modes that do not decay in time and have frequency \emph{smaller} than the plasma frequency. These are electromagnetic waves that are trapped between the gravitational potential and the plasma barrier~\cite{Lingetti:2022psy,Dima:2020rzg}. At large radii, we find modes with a frequency \emph{higher} than the plasma frequency, which confirms these are travelling waves. They originate from the initial Gaussian, which has a tail with frequencies $\omega > \omega\ped{p}^{\rm (c)}$. Indeed, by increasing the plasma frequency, these modes become less and less prominent, as shown in the \emph{bottom panel} of Figure~\ref{fig:unchargedplasma}. This behaviour highlights the ``filtering effect'' of plasma, which suppresses modes lying below a threshold determined by the plasma frequency.
\section{Propagation Effects:~Echoes in Waveforms}\label{sec_QNMplasma:echoes} 
\begin{figure}[t!]
    \centering
    \includegraphics[scale=1]{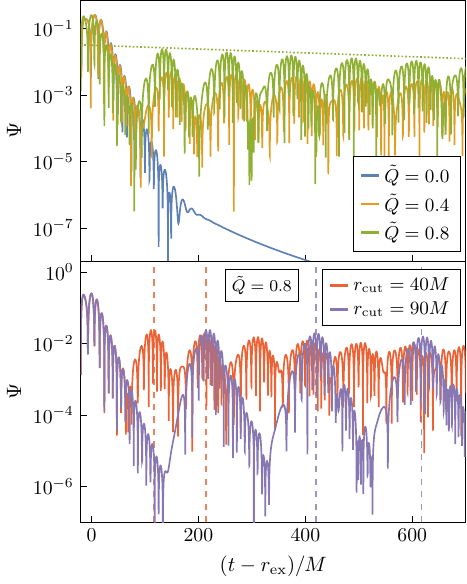}
    \caption{GW signal generated in the presence of a plasma localised away from the BH. In the \emph{top panel} $r\ped{cut} = 40M$; the amplitude of the echoes increases concurrently with the coupling $Q$. Dotted line indicates the decay rate of the signal, $M\Gamma = -0.00133$, as predicted from~\eqref{eq:amplitude-fall}. \emph{Bottom panel} illustrates how the echo timescale depends on the position of the plasma barrier. The estimates from~\eqref{eqn:timescale-barrier} are indicated by vertical dashed lines. Both panels are initialised with $\mathrm{\mathrm{ID}\ped{g}}$ and $\tilde{\omega}^{\rm (c)}\ped{p} = 1.5$.}
    \label{fig:echoes}
\end{figure}
When the plasma is localised away from the BH, a new phenomenology emerges. BH ringdown is associated mostly with light ring physics, hence prompt ringdown is no longer affected~\cite{Cardoso:2016rao,Cardoso:2019rvt}. However, upon exciting the BH, both electromagnetic and GWs travel outwards. While GWs travel through the plasma, electromagnetic waves are reflected, interacting with the BH again and exciting one more stage of ringdown and corresponding GW \emph{echoes}. Such echoes have been found before in the context of (near-) horizon quantum structures~\cite{Cardoso:2016rao,Cardoso:2016oxy,Oshita:2018fqu,Wang:2019rcf}, exotic states of matter in ultracompact or neutron stars~\cite{Ferrari:2000sr,Pani:2018flj,Buoninfante:2019swn} and modified theories of gravity~\cite{Buoninfante:2019teo,Delhom:2019btt,Zhang:2017jze} (see~\cite{Cardoso:2017cqb,Cardoso:2019rvt} for reviews). We find them in a General Relativity setting.
\vskip 2pt
In the \emph{top panel} of Figure~\ref{fig:echoes}, we show the GW signal for different values of the BH charge. In contrast to vacuum (where exponential ringdown gives way to a power-law tail), in the presence of plasma prompt ringdown is followed by echoes of the original burst. For higher BH charge, the reflected electromagnetic signal is more strongly coupled, increasing the amplitude of the GW echoes. The main features of the echoing signal are simple to understand. For $r\ped{cut} \gg r_{\scalebox{0.60}{$\mathrm{LR}$}}$, the time between consecutive echoes can be estimated as
\begin{equation}
\label{eqn:timescale-barrier}
\Delta t = 2\int_{r_{\scalebox{0.60}{$\mathrm{LR}$}}}^{r\ped{cut}} \frac{\mathrm{d}r}{f(r)} \approx 2\,r\ped{cut}\,.
\end{equation}
This interval is shown by the vertical dashed lines in Figure~\ref{fig:echoes} (\emph{bottom panel}) and clearly in good agreement with the numerics.
\vskip 2pt
The echo amplitude decays in time, since the BH absorbs part of the reflected waves, and part of the energy is carried to infinity by GWs. The amplitude $A(t)$ of trapped modes in a cavity of length $\sim r\ped{cut}$ is expected to fall off as  
\begin{equation}
\label{eq:amplitude-fall}
A(t) = A_0 e^{-\Gamma t}\,, \quad \text{with} \quad \Gamma \sim \frac{{\cal A}^2}{r\ped{cut}}\,,
\end{equation}
where $A_0$ is the initial amplitude and ${\cal A}^2$ the absorption coefficient of the BH (neglecting losses to GWs at infinity). For simplicity, we take the absorption coefficient of low-frequency monochromatic waves for neutral BHs, given by ${\cal A}^2\sim 256(M\omega\ped{R})^6/225$~\cite{Starobinskil:1974nkd,Brito:2015oca}, where $\omega\ped{R}$ is the frequency of the trapped electromagnetic waves. As the BH absorbs high-frequency modes first, the decay rate decreases over time, asymptoting to a QBS, while the trapped wavepacket broadens. Taking $\omega\ped{R}$ as the highest-frequency peak in the spectrum, we obtain a decay rate~\eqref{eq:amplitude-fall} in agreement of $\mathcal{O}(1)$ for the first few echoes in Figure~\ref{fig:echoes}. We confirmed that at later times, the high frequency components of the electromagnetic field are indeed lost and the decay rate is decreased accordingly. A similar phenomenology can be found for any mechanism that places a BHs in a confining box, e.g.,~AdS BHs where the AdS radius is much larger than the horizon radius, or Ernst BHs immersed in a magnetic field $B \ll 1/M$~\cite{Brito:2014nja}.
\section{Summary and Outlook}\label{sec_QNMplasma:summary}  
Plasmas are ubiquitous in the Universe, but their impact on our ability to do precision GW physics is poorly understood. We have studied plasma physics in curved spacetime from first principles, capturing their impact on the ringdown of charged BHs. Our results are surprising at first sight. We find an important impact of plasma physics on the \emph{gravitational waves} generated by charged BHs, changing BH spectroscopy to a measurable extend. We see a ringing frequency going up, and the lifetime of the ringdown going down, a behaviour that would be important to dissect. We also find that plasmas may trigger measurable echoes in GWs. As the amplitude of these echoes decays slowly, they could be in reach of current or future detectors. A dedicated data-analysis study to uncover whether these effects can be detected in a realistic setup would be valuable for future research.
\vskip 2pt
In our studies, we focused on values of the plasma frequency $\omega\ped{p}\sim \mathcal{O}(1/M)$. Note however, that larger values yield similar outcomes. Specifically, a denser plasma present at the light ring causes a greater shift in the gravitational QNM frequency, while a denser plasma localised outside the light ring increases the height of the plasma barrier, making the reflection of photons and thus the echoes, more prominent. Most of our results would also apply to magnetic BHs, which share many similarities with charged BHs in the ringdown phase~\cite{Pereniguez:2023wxf,Dyson:2023ujk}.
\chapter{Black Hole Spectroscopy in Environments:~Detectability Prospects} \label{chap:BHspec}
\vspace{-0.8cm}
\hfill \emph{Als ik zou willen dat je het begreep, had ik het wel beter uitgelegd}
\vskip 5pt

\hfill Johan Cruijff
\vskip 35pt
\noindent The presence of matter around charged BHs can alter the fundamental QNM frequency, an effect observed even in time-domain evolutions, as shown in the previous chapter. A comprehensive set of works have established that QNMs are in general \emph{spectrally unstable} under small perturbations in the underlying spacetime~\cite{Nollert:1996rf,Nollert:1998ys,Leung:1997was,Barausse:2014tra,Jaramillo:2021tmt,Cardoso:2019rvt,Jaramillo:2020tuu,Cheung:2022rbm,Konoplya:2022pbc,Konoplya:2022hll,Cardoso:2022whc,Cardoso:2024mrw,Yang:2024vor,Ianniccari:2024ysv}, which might correspond to the one caused by astrophysical environments. However, time-domain analyses in these same geometries suggest that the prompt, early-time ringdown signal is not affected by spectral instabilities, questioning its relevance for GW astronomy~\cite{Berti:2022xfj,Cardoso:2024mrw,Yang:2024vor}. Despite these theoretical insights, no quantification of the impact of realistic environments in BH spectroscopy from a data-analysis point of view has ever been attempted.\footnote{This statement concerns the dominant mode and especially higher overtones, which are generically afflicted by stronger instabilities~\cite{Cardoso:2024mrw}. While data-analysis oriented studies exist~\cite{Leong:2023nuk,Redondo-Yuste:2023ipg,DeLuca:2024uju}, they focus on ad hoc matter profiles, or accreting Vaidya spacetimes, less relevant from an astrophysical viewpoint.}
\vskip 2pt
In this chapter, I present the work of~\cite{Spieksma:2024voy}, and examine the ringdown signal in an astrophysically motivated environment:~a DM halo surrounding a supermassive BH at the centre of a galaxy. The focus is on understanding how astrophysical environments influence BH spectroscopy. Do they affect the \emph{detectability} of a signal? Can they be \emph{distinguished} from the pure-vacuum case? I specifically consider highly asymmetric binaries, for which the environment is expected to survive the inspiral phase. This is a conservative assumption, since environments of comparable mass binaries are significantly more depleted~\cite{Aurrekoetxea:2024cqd}. The taken approach intends to be agnostic regarding the nature of a possible instability, using only the well-understood vacuum waveform as a ruler to measure how well environmental effects in the ringdown can be differentiated.
\vskip 2pt
This chapter is organised as follows. In Section~\ref{sec_BHspec:DBHs}, I introduce the necessary definitions and equations. In Section~\ref{sec_BHspec:methods}, I describe the methods for the time-domain evolution and signal analysis. Then, in Section~\ref{sec_BHspec:GWsignal}, I present the GW signal in the time domain. In Section~\ref{sec_BHspec:faith}, I analyse its distinguishability from the vacuum case. In Section~\ref{sec_BHspec:spect}, I address spectral instabilities from an observational viewpoint. Finally, in Section~\ref{sec_BHspec:summary}, I summarise the findings. Additional details on perturbation theory and the numerical evolution are provided in Appendix~\ref{app:BHPT}.
\section{Dirty Black Holes}\label{sec_BHspec:DBHs} 
As a proxy for a galactic environment, we consider a BH at the centre of some halo matter distribution, following Section~\ref{BHenv_subsec:DMStruct}. The spacetime is taken to be spherically symmetric spacetime, with line element 
\begin{equation}\label{eqn:line_element}
ds^2= -A(r)\mathrm{d}t^2 + \frac{\mathrm{d}r^2}{1-2m(r)/r} + r^2 \mathrm{d}\Omega^2\,,
\end{equation}
where $A(r)$ is given explicitly in~\cite{Cardoso:2021wlq}, $\mathrm{d}\Omega^2 = \mathrm{d}\theta ^2 + \sin \theta^2 \mathrm{d}\varphi^2 $ and $m(r)$ is the mass function of the system. The ``Einstein cluster'' approach provides, yet again, a consistent way to construct stationary solutions~\cite{Einstein_cluster,Einstein_cluster_2}, and is characterised by the energy-momentum tensor of an anisotropic fluid with vanishing radial pressure and a tangential pressure $P\ped{t}$, which reads
\begin{equation}
T^{\mu}_{\nu} = \mathrm{diag}(-\rho, 0, P\ped{t}, P\ped{t})\,.
\end{equation}
The chosen mass function is inspired by the Hernquist halo profile~\eqref{eq:Hernquist}, which is commonly used to model elliptical galaxies and galactic bulges~\cite{1990ApJ...356..359H,Quinlan:1994ed}:
\begin{equation}
m(r)=M+\frac{M\ped{H} r^2}{\left(a\ped{H}+r\right)^2}\left(1-\frac{2 M}{r}\right)^2\,, \label{eq:HernquistMass}
\end{equation}
where $M\ped{H}$ and $a\ped{H}$ are, respectively, the mass and characteristic length of the halo. We stress that $M$ refers to the BH mass and not the total ADM mass of the system. This profile is dominated by the BH gravity at $r \ll a\ped{H}$ and at large distances recovers the mass profile of the Hernquist model~\eqref{eq:Hernquist}. We also define the compactness and density of the halo as
\begin{equation}\label{eq:compactness}
\mathcal{C} = \frac{M\ped{H}}{a\ped{H}}\quad \text{and} \quad \rho\sim \frac{{\cal C}^3}{M\ped{H}^2}\,,
\end{equation}
respectively. Both quantities affect the GW-response of the system. Halos are expected to be much more massive than the central BH they host, i.e., $M \ll M\ped{H}$, and have low compactness ($\mathcal{C}\lesssim 10^{-4}$). Other astrophysical environments, such as boson clouds composed by ultralight fields, can have much higher compactness (and density, see Figure~\ref{fig:taxonomy}). The Hernquist profile is one of many possible choices, and it is straightforward to repeat the same procedure to find generic stationary spacetimes describing BHs dressed by matter~\cite{Figueiredo:2023gas,Speeney:2024mas}. Yet, all of these exhibit the same qualitative behaviour and therefore we focus on the mass function in eq.~\eqref{eq:HernquistMass}, treat both $M\ped{H}$ and $a\ped{H}$ as free parameters, and take it as a proxy for generic distributions of matter around a BH. 
\vskip 2pt
We consider a barotropic equation of state, for which changes in pressure, $\delta P\ped{t, r}$ (tangential and radial), and density $\delta \rho$, are related by the speed of sound:
\begin{equation}\label{eqn:speedofsound1}
\delta P\ped{t, r}=c\ped{s_{t, r}}^2 \delta \rho\,, \quad \text{with} \quad c\ped{s,r} = \left(\frac{2M+a\ped{H} }{r+a\ped{H}}\right)^{4}\,.
\end{equation}
Small sound speeds lead to problems regarding the well-posedness of the system~\cite{Schoepe:2017cvt}, making it challenging to solve numerically. Following~\cite{Cardoso:2022whc,Datta:2023zmd,Speeney:2024mas}, we choose the \emph{ad hoc} profile in eq.~\eqref{eqn:speedofsound1}, and from previous works we do not expect major qualitative changes for other profiles.
\section{Methods}\label{sec_BHspec:methods}  
We perturb the spacetime~\eqref{eqn:line_element} by plunging a particle into the BH~\cite{ReggeWheeler,Zerilli:1970wzz,Zerilli:1970se, Barack:2018yvs, Pound:2021qin, Cardoso:2022whc, Duque:2023seg,Cardoso:2021wlq, Zenginoglu:2011zz}. Without loss of generality, the plunge is along the radial $\hat{z}$--direction, exciting only axially symmetric polar modes. The corresponding waveform is computed using recently developed techniques of BH perturbation theory in non-vacuum (spherically symmetric) spacetimes~\cite{Barack:2018yvs, Pound:2021qin,Cardoso:2021wlq,Cardoso:2022whc,Duque:2023seg}. Further details are found in Appendix~\ref{appBHPT_subsec:galactic}. After the plunge, the GW signal is well-described by
a superposition of exponentially damped sinusoids (see Section~\ref{GravAstro_subsec:ringdown}):\footnote{As the polarisation axes are oriented along the $\hat{\theta}$ and $\hat{\varphi}$--direction, the cross-polarisation ($h_{\times}$) is zero and the GW radiation is purely captured by $h_+$.}
\begin{equation}
h_+=\frac{M}{r}\,\mathrm{Re}\left[\,\sum_{n = 0}^{\infty}\sum_{\ell = 2}^{\infty} A_{\ell 0 n}\,e^{-i (\omega_{\ell 0 n} t-\phi_{\ell 0 n})}\!{}_{{\scalebox{0.65}{$-$}}2}\mkern-2mu Y_{\ell 0}(\theta, \varphi)\right]\,,
\label{eq:hplus_ringdown}
\end{equation}
where $h_+=h_+(t,r,\theta,\varphi)$, ${}_{{\scalebox{0.65}{$-$}}2}\mkern-2mu Y_{\ell 0}(\theta,\varphi)$ are spin-weighted spherical harmonics and $\omega_{\ell 0 n} = \omega_{\mathrm{R},\ell 0 n}+i\omega_{\mathrm{I},\ell 0 n}$ are the QNM frequencies. For radial plunges and our choice of axis, $m=0$ in eq.~\eqref{eq:hplus_ringdown}. This signal corresponds to light-ring relaxation~\cite{Cardoso:2019rvt}. In vacuum, it eventually gives way, at late times, to power-law tails from curvature backscattering, either when considering vacuum perturbations~\cite{Price:1971fb,Leaver:1986gd,Gundlach:1993tp,Hintz:2020roc} or infalling particles~\cite{Albanesi:2023bgi,Cardoso:2024jme,DeAmicis:2024not,Islam:2024vro}, but the structure in the presence of surrounding matter is richer, as we will see below.
\vskip 2pt
To understand if the ringdown of a BH surrounded by an astrophysical environment can be distinguished from its vacuum counterpart, we compute the \emph{faithfulness} between two waveforms $h_{1}$ and $h_{2}$, defined as
\begin{equation}\label{eq:faithfulness}
\mathcal{F} \equiv \mathop{\text{max}}_{t\ped{c},\phi\ped{c}} \frac{(h_1|h_2)}{\sqrt{(h_1|h_1)(h_2|h_2)}}\,,
\end{equation}
where $t\ped{c}$ and $\phi\ped{c}$ are, respectively, time and phase of the signal at the coalescence. The inner product $(h_1|h_2)$ is
\begin{equation}\label{eq:inner_product}
(h_1|h_2) = 4\,\mathrm{Re} \int_{0}^{\infty} \frac{\bar{h}_{1}(f)\bar{h}^{*}_{2}(f)}{S_{n}(f)} \mathrm{d}f\,,
\end{equation}
where the overhead bar indicates a Fourier transform and $S_{n}(f)$ is the one-sided power spectral density, which depends on the specific detector. For reference, we consider the LISA sensitivity curve~\cite{Robson:2018ifk}. 
We maximise over time and phase, with phase maximisation done by taking the absolute value instead of the real part in eq.~\eqref{eq:faithfulness}~\cite{Owen:1995tm}. We always take $h_{2}$ to be the vacuum waveform, which serves as the fiducial signal to compare against. Finally, we define the signal-to-noise ratio (SNR) as 
\begin{equation}
\mathrm{SNR}^{2} = 4\int_{0}^{\infty}\frac{\bar{h}^{*}(f)\bar{h}(f)}{S_{n}(f)}\mathrm{d}f\,.
\end{equation}
If two waveforms fulfil the criterion:
\begin{equation}\label{eq:criterion}
    1-\mathcal{F} < \frac{D}{2\,\mathrm{SNR}^{2}}\,,
\end{equation}
for a certain choice of $S_{n}(f)$ and respective SNR, they are classified as \emph{indistinguishable}~\cite{Flanagan:1997kp,Lindblom:2008cm,McWilliams:2010eq,Chatziioannou:2017tdw,Purrer:2019jcp} in the sense that the deviation between two waveforms $\delta h$ satisfies $\braket{\delta h|\delta h} < 1$ (see eqs.~(8.1)--(8.2)--(8.13) in~\cite{Flanagan:1997kp}). Here, $D$ denotes the dimension of the parameter space one considers, amounting to e.g., $D \sim 10$ for a two damped sinusoids analysis. Note that this criterion formally holds in the limit of high SNR which, with increased sensitivity of future detectors, is a well-justified assumption. 
\vskip 2pt
We expect astrophysical setups with a hierarchy of scales $M \ll M\ped{H} \ll a\ped{H}$. These scales and the need to extract the signal far away from the system due to the slow polynomial decay of the halo mass [see eq.~\eqref{eq:HernquistMass}], pose a numerical challenge and restrict the range of halo mass and size that we can study. We thus focus on relatively large halo compactnesses. As we will show, this choice leads to an \emph{overestimate} of the impact of the environment, which strengthens our conclusions. Specifically, we take $M\ped{H} = \left[0.1,0.3,0.5,1,10\right]M$, while varying $a\ped{H}$, ensuring the GW signal is always extracted \emph{outside} the halo, $r\ped{ex} \gg a\ped{H}$. In all our simulations, the particle starts the plunge at $r\ped{p}(t=0) = 100M$, and we extract the signal at $r\ped{ex} = \left[20,40,80,100\right]a\ped{H}$. 
At each radius, we locate the peak of the strain $h_{+}$ and we truncate the waveform roughly $\sim\!10M$ before the peak until the tail sets in.\footnote{As the trajectory of the particle depends on the ``halo density'' it encounters, i.e., $\mathrm{d}t/\mathrm{d}\tau \propto 1/\sqrt{A(r)}$ (see eq.~\eqref{eqn:line_element}, where $\tau$ is the proper time), the strain amplitude and instant at which the waveform peaks do depend on the choice of compactness.} Subsequently, the waveforms are ``rescaled'' relative to the waveform extracted at the largest radius. In particular, labelling the latter as $h_{1}$, we allow the (to-be-shifted) waveform $h_{2}$ to undergo a stretching $\alpha$ and a time shift $t_0$ according to
\begin{equation}\label{eq:shifted_waveform}
    \hat{h}_{2} = \int\!\mathrm{d}t\, e^{i\omega (t-t_0)}h_2(\alpha t) =e^{-i\hat{\omega}\alpha t_0}\bar{h}_{2}(\hat{\omega})\,,
\end{equation}
where an overhead bar denotes the Fourier transform and $t = \hat{t}/\alpha$ and $\hat{\omega} = \omega\alpha$. The parameters $\alpha$ and $t_0$ are then determined by maximising the faithfulness $\mathcal{F}(\hat{h}_{2}, h_{1})$~\eqref{eq:faithfulness}, i.e., $\mathop{\text{max}}_{\alpha, t_0} \mathcal{F}$. Since the waveforms are extracted at varying radii but correspond to the same halo configuration, the deviations in $\alpha$ and $t_0$ are minimal:~typically $\alpha \sim 1$ and $t_0 \sim 0$. To avoid errors associated with finite extraction radii, the waveform is then extrapolated to infinity by fitting a Chebyshev polynomial $rh(t) = h_{\infty} + a_{1}(1/r) + a_{2}(1/r^2) + \cdots$, where $h_{\infty}$ represents the waveform at future null infinity. In our fit, we include the first two orders, yet an error estimate coming from the first and third order will also be included.
\vskip 2pt
With the waveforms extrapolated to infinity, we can now compare the ringdown signal from different halo configurations. Conform to realistic GW searches, we again apply the transformation from eq.~\eqref{eq:shifted_waveform}, taking the reference waveform to be the vacuum one. In this context, $\alpha$ acquires a clear physical meaning:~it represents the overall redshift induced by the presence of the halo. 
\section{Gravitational-Wave Signals}\label{sec_BHspec:GWsignal}
\begin{figure}[t!]
    \centering
    \includegraphics[scale=1]{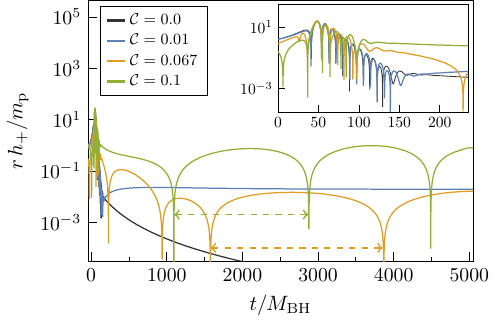}	
    \caption{Gravitational-wave signal $h_+$ when a point-like particle collides with a BH ``dressed'' by a halo of varying compactness and mass $M\ped{H} = 10M$. Particle begins from rest at $r\ped{p}= 100M$ and the signal is extracted at $r\ped{ex} = 3000M$. Vacuum signal is shown in black. All waveforms are aligned in time and amplitude. The oscillations at late times [{\color{Mathematica2}{orange}} and {\color{Mathematica3}{green}}] indicate the presence of a \emph{fluid-driven} mode, imprinted on the GW signal. Its period is well-approximated by $T\sim a\ped{H}/c\ped{s,r}$ (using eq.~\eqref{eqn:speedofsound1} with $r\sim a\ped{H}$). Inset shows zoom-in of the prompt ringdown.}
    \label{fig:Plunge_fluid}
\end{figure}
Consider a binary of mass ratio $q = m\ped{p}/M$, where $m\ped{p}$ is the mass of the smaller body. We excite the ringdown with a simple head-on collision. Figure~\ref{fig:Plunge_fluid} shows the time evolution of the GW strain when a particle collides with a BH in the presence of a halo with different compactnesses. The main features are:~(i) the early-time, dominant component is a prompt ringdown stage corresponding to light-ring excitation and trapping. This stage is very similar for different halo compactnesses, but \emph{is} affected by the environment (for ${\cal C}=0.1$ the signal is noticeably different). By lowering compactness, the signal tends towards the vacuum one (black line);~(ii) after the prompt ringdown, halo modes set in (on scales set by $a\ped{H}$) and dominate the late-time signal. This contribution is \emph{fluid-driven} (seen before in~\cite{Cardoso:2022whc}) and originates from the coupling between the matter and gravity sector. The amplitude and frequency of halo modes depend on the compactness. We find that they oscillate with period $T \sim a\ped{H}/c\ped{s,r}$~\cite{Cardoso:2022whc} for all setups studied. For example, for $M\ped{H}=10M$ and $a\ped{H} = 100M$ ($\mathcal{C} = 0.1$) [{\color{Mathematica3}{green}}], it corresponds to $T\sim 1500 M$, even when evolving the system for longer timescales than shown in Figure~\ref{fig:Plunge_fluid}. We expect a power-law tail on yet larger timescales, which our current numerical infrastructure cannot probe;~(iii) keeping halo compactness fixed while varying halo mass and size, we find that compactness is the dominant factor determining changes in the ringdown signal with respect to vacuum, at least for the parameter space probed.
\section{Faithfulness of Vacuum Templates}\label{sec_BHspec:faith}
\begin{figure}[t!]
    \centering
    \includegraphics[scale=1]{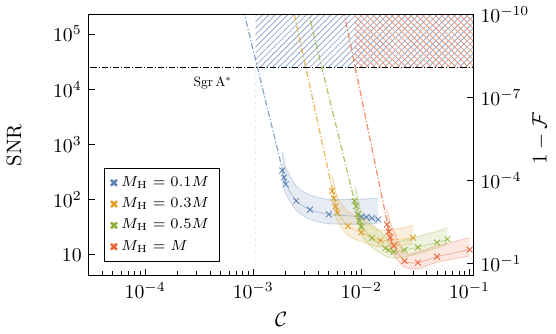}	
    \caption{$\mathrm{SNR}$ required to distinguish the ringdown signal from vacuum for different compactnesses and a given value of halo mass, using the criterion~\eqref{eq:criterion} and $D = 10$. 
    For a given SNR (determining an horizontal line) and halo mass curve, only the region of compactness right of the intersection between the horizontal line and the curve is distinguishable from vacuum.
    The black horizontal line represents a putative signal from Sgr\,$\mathrm{A}^*$ with $q = 10^{-5}$~\eqref{eq:SNR_analytic}. 
    Given its $\mathrm{SNR}$, the signal can be distinguished from vacuum for high compactness, as indicated by the diagonal-line region for $M\ped{H} = 0.1M$ [{\color{Mathematica1}{blue}}] and $M\ped{H} = M$ [{\color{Mathematica4}{red}}]. 
    The right axis shows the corresponding mismatch value (without assuming a value for $D$). 
    Coloured dash-dotted lines are a power-law fit through the last few points.
    The error in the shaded regions comes from extrapolating the waveform keeping $1/r$ and up to $1/r^3$ terms.}
    \label{fig:Detectability_fluid}
\end{figure}

Figure~\ref{fig:Detectability_fluid} shows the faithfulness~\eqref{eq:faithfulness} for varying halo mass and compactness. It approaches unity in the limit of zero compactness, even if the total halo mass is large:~the signal is simply redshifted to lower frequencies and a vacuum template will bias the BH mass, which is confirmed in a frequency-domain approach~\cite{Pezzella:2024tkf}. Indeed, for e.g.,~$M\ped{H}=M$, $a\ped{H}=58M$, our best match is achieved for a redshift $\alpha=0.978$, compared to the ``expected'' redshift $\sim 1-M\ped{H}/a\ped{H}=0.982$. For small $\mathcal{C}$, the mismatch $1-\mathcal{F}$ decreases as a power-law. While determining the precise scaling analytically would be insightful, waveform stretching leads to some loss of information. Nevertheless, it can be shown that when perturbations of the potential scale as $\mathcal{C}^n$, the faithfulness scales as $\mathcal{C}^{2n}$. Counterintuitively, lower halo masses allow to probe a larger range of compactness. This can be explained from eq.~\eqref{eq:compactness}:~for fixed compactness, lower halo masses actually increase the density of the halo close to the BH, making the impact on the ringdown more severe. However, for low enough halo masses the rise of the plateau (displayed by the curves at high compactness) will prevent the environment discrimination, as intuitively expected. For large compactnesses, the results showcase a complex behaviour:~as we allow the template to redshift, the result displays nontrivial features at large densities and compactnesses.
\vskip 2pt
To understand whether the ringdown signal can be distinguished from vacuum, the faithfulness should be compared against the expected $\mathrm{SNR}$ from astrophysical events~\eqref{eq:criterion}. Taking LISA as an example, using eq.~(3.12b) in~\cite{Berti:2005ys} and eq.~(4.12) in~\cite{Berti:2007fi}, we find
\begin{equation}\label{eq:SNR_analytic}
    \mathrm{SNR} = 2.5\times 10^{4} \left(\frac{q}{10^{-5}}\right)\left(\frac{M}{4.3\times 10^{6}M_{\odot}}\right)^{3/2}\left(\frac{8.3\,\mathrm{kpc}}{D\ped{L}}\right)
    \left(\frac{3.8
    \times 10^{-40}\,\mathrm{Hz}^{-1}}{S_{n}(f)}\right)^{1/2}\,,
\end{equation}
where we consider the fundamental mode $M\omega_{200} = 0.374$, typical values for the LISA sensitivity curve~\cite{Robson:2018ifk}, and we take Sgr\,$\mathrm{A}^*$ as reference, with mass $M_{\mathrm{Sgr}\,\mathrm{A}^{*}} = 4.3\times 10^{6}M_{\odot}$ and located at $D\ped{L} = 8.3\,\mathrm{kpc}$~\cite{Ghez:2008ms,2019A&A...625L..10G,GRAVITY:2021xju}.\footnote{Equation~\eqref{eq:SNR_analytic} agrees with eq.~(3.21) in~\cite{Berti:2005ys} for equal-mass binaries and their (outdated) LISA noise curve. Note however, that the dependence on mass ratio in~\eqref{eq:SNR_analytic} is $q/(1+q)^2$, which becomes relevant only for equal masses.} Using eq.~\eqref{eq:criterion} with $D = 10$, we find that when $1-\mathcal{F} < 8.2\times 10^{-9}$ the signal cannot be distinguished from vacuum for the benchmark parameters in~\eqref{eq:SNR_analytic}. From Figure~\ref{fig:Detectability_fluid}, we see that, for e.g.,~$M\ped{H} = 0.1M$, any halo with $\mathcal{C} \lesssim \mathcal{O}(10^{-3})$ is indistinguishable using ringdown, even in the overly optimistic scenario of a signal from Sgr\,$\mathrm{A}^*$. Note that mergers of supermassive BHs are predicted to be exceptionally loud but also more distant, resulting in similar or lower SNR~\cite{Bhagwat:2021kwv}. As part of the surrounding environment could be depleted in such mergers, the scenario we consider is expected to be conservative, suggesting our conclusions hold even in this case.
\vskip 2pt
There are two interesting applications of our results in the context of the Milky Way. The GRAVITY Collaboration constrained the mass within the orbit of the S2 star (highly eccentric, we take it to have radius $\sim 10^4M$) to be $\lesssim 10^{-3} M$, using a Plummer profile or one appropriate for bosonic bound states~\cite{GRAVITY:2021xju,GRAVITY:2023cjt}. From the $M\ped{H}$ scaling of our results, we find that for a collision at the centre of our galaxy to discriminate an environment via ringdown, then $a\ped{H}\lesssim 10^2 M$. In addition, at larger scales, the Milky Way bulge stellar mass is of order $(2.0 \pm 0.3)\times  10^{10}M_\odot$ with a redshift of order $10^{-6}$~\cite{2016A&A...587L...6V}. Figure~\ref{fig:Detectability_fluid} leaves little room for doubt:~BH spectroscopy will not be able to inform us on physics at these scales.
\vskip 2pt
Sources farther away will have a $\mathrm{SNR}$ too low to distinguish any value of the compactness from the vacuum waveform. Since realistic galactic halos have $\mathcal{C}\lesssim 10^{-4}$~\cite{Navarro:1995iw,Kim:2004tc}, we conclude that environments cannot be distinguished with ringdown, using currently planned detectors.
\section{Spectral Instabilities}\label{sec_BHspec:spect}
\begin{figure}[t!]
    \centering
    \includegraphics[scale=1]{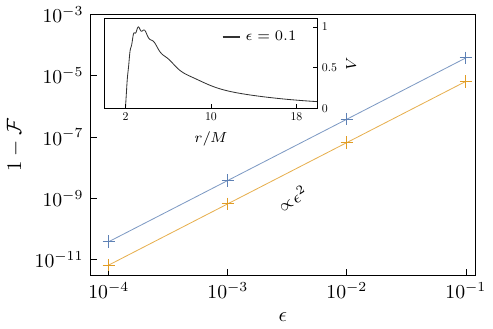}	
    \caption{Mismatch for various values of $\epsilon$, which quantifies the modification to the potential~\eqref{eq:perturbed_potential}. $1-\mathcal{F}$ goes down exactly as $\epsilon^{2}$, showcasing that mode instabilities do not affect observations to any significant degree. We use either ``polynomial initial data'' [{\color{Mathematica1}{blue}}] from~\cite{Jaramillo:2020tuu,Jaramillo:2021tmt,Ansorg:2016ztf}, or Gaussian initial data [{\color{Mathematica2}orange}] with a width $M$. The inset shows the effective potential experienced by GWs, with $\epsilon = 0.1$~\eqref{eq:perturbed_potential}.}
    \label{fig:Mismatch_epsilon}
\end{figure}
The QNM spectrum of BHs is unstable under small perturbations or couplings to matter~\cite{Nollert:1996rf,Cardoso:2024mrw}. Our case study contains instabilities for the fundamental mode, and we concluded for their nonobservability. We now want to show parenthetically that high-frequency, spectrally unstable perturbations are probably stable insofar as observations go. We stray from the galactic geometry~\eqref{eq:HernquistMass}, and consider instead the toy model of~\cite{Jaramillo:2020tuu,Jaramillo:2021tmt}, where the effective potential for GW propagation gets deformed by
\begin{equation}\label{eq:perturbed_potential}
\delta V=\left(1-\frac{2M}{r}\right)\frac{\epsilon}{r^2} \sin{\left(2\pi k \frac{2M}{r}\right)}\,,
\end{equation}
where we fix $k = 10$. Despite the presence of spectral instability (in high overtones)~\cite{Jaramillo:2020tuu,Jaramillo:2021tmt}, the faithfulness decreases with $\epsilon^{2}$ as shown in Figure~\ref{fig:Mismatch_epsilon}:~it is \emph{spectrally stable}. There will be no impact on observations.
\section{Summary and Outlook}\label{sec_BHspec:summary}
Black hole spectroscopy is an indispensable tool for studying astrophysical and fundamental properties of BHs. With the increased sensitivity of future GW detectors, it will become possible to probe ringdown more accurately, and in different frequency ranges. This opens the interesting possibility of using BH ringdown to probe environments, a prospect made even more exciting with the discovery that the BH spectrum is unstable~\cite{Nollert:1996rf,Nollert:1998ys,Leung:1997was,Barausse:2014tra,Jaramillo:2021tmt,Cardoso:2019rvt,Jaramillo:2020tuu,Cheung:2022rbm,Konoplya:2022pbc,Konoplya:2022hll,Cardoso:2022whc,Cardoso:2024mrw,Yang:2024vor,Ianniccari:2024ysv}. 
\vskip 2pt
In this chapter, we studied this possibility, using a model that resembles dark matter densities in typical galactic environments. For realistic values of environmental parameters, we find that the leading-order effect is simply a (gravitational) redshift of the fundamental frequency. This may be thought of as a \emph{propagation effect}, as the GW climbs out of the gravitational potential of matter. Our results are consistent with a high-frequency approximation to QNMs~\cite{Cardoso:2021wlq}, which can be used to argue that the compactness controls the QNMs of BHs in environments~\cite{Duque:2023seg}.
\vskip 2pt
We studied only nonspinning BHs, possibly a conservative approach, since spin may add one more degeneracy knob on the search. At leading order and large SNR, the error on the measurement of the BH mass yields~\cite{Berti:2005ys}
\begin{equation}
\frac{\sigma_{M}}{M}\approx \frac{2}{\mathrm{SNR}}\,,
\end{equation}
suggesting that, even if the BH mass is known \emph{a priori}, e.g.,~from an inspiral-only analysis or electromagnetic counterparts, $\mathrm{SNR}\sim a\ped{H}/M\ped{H}$ is required to distinguish an event from vacuum. For galactic environments, this requires unreasonably loud events~\cite{Berti:2016lat,Seoane:2021kkk}. The take-home message is that BH quality factors are too small for environments to significantly impact spectroscopy.
\vskip 2pt
Atomic and molecular spectroscopy in environments is well-understood. Among others, the Stark effect contributes to a distortion of spectral Balmer lines in plasmas~\cite{PhysRev.185.140,PhysRevA.6.1132,PhysRevA.11.1854}. Environments can, in principle, also affect BH spectroscopy, but our results suggest only those with extreme density or compactness could lead to detectable effects with planned detectors. Nevertheless, if used in conjunction with measurements of the inspiral, one may use this to our advantage to e.g.,~remove redshift degeneracies or obtain information on environmental properties through pre/post merger consistency tests. As seen in Figure~\ref{fig:Plunge_fluid}, environments do affect the very late time behaviour of coalescences, possibly amplifying tail amplitudes~\cite{Albanesi:2023bgi,Cardoso:2024jme,DeAmicis:2024not}, a topic which deserves further scrutiny.
\chapter{Resonant History of Boson Clouds in Black Hole Binaries} \label{chap:legacy}
\vspace{-0.8cm}
\hfill \emph{My theory stands as firm as a rock;}

\hfill \emph{every arrow directed against it will return quickly to its archer.}

\hfill \emph{How do I know this? I have studied it...}

\hfill \emph{I have followed its roots, so to speak,}

\hfill \emph{to the first infallible cause of all created things}
\vskip 5pt

\begin{figure}[t!] 
\centering
\includegraphics[width=\linewidth]{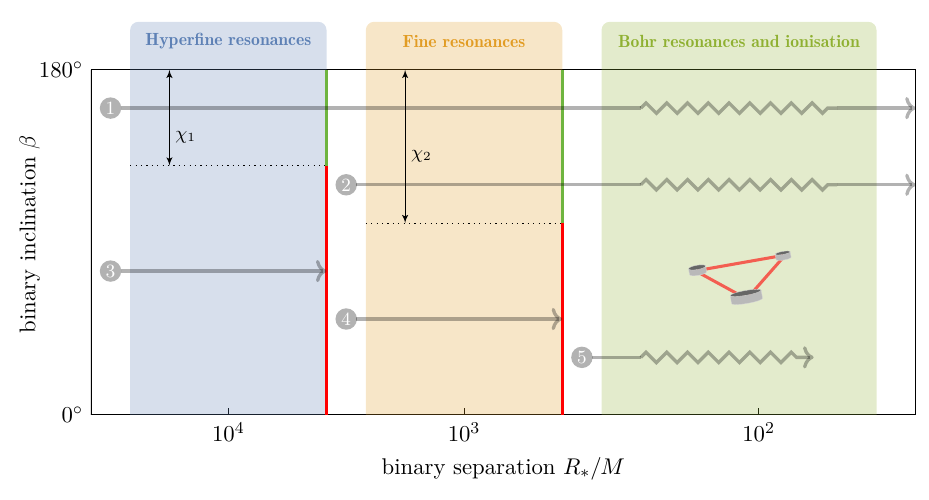}
\caption{Illustration of the possible outcomes of the resonant history of the cloud-binary system. The inspiral starts with the cloud in its initial state, either $\ket{211}$ or $\ket{322}$. Only systems {\normalsize \textcircled{\footnotesize 1}}--{\normalsize \textcircled{\footnotesize 2}} whose inclination angle is within intervals $\chi_1$ and $\chi_2$ from the counter-rotating ($\beta=\SI{180}{\degree}$) configuration can move past the hyperfine and fine resonances with the cloud still intact (green vertical lines). These later give rise to observational signatures in the form of ionisation and Bohr resonances. Others {\normalsize \textcircled{\footnotesize 3}}--{\normalsize \textcircled{\footnotesize 4}} are destroyed by the hyperfine or fine resonances (red vertical lines). Binary systems that form at small enough separations may be able to skip early resonances {\normalsize \textcircled{\footnotesize 5}}.}
\label{fig:history_211}
\end{figure}

\hfill Georg Cantor
\vskip 35pt
\noindent In the final three chapters of this thesis, I turn to the inspiral stage of a binary coalescence, focusing on intermediate and extreme mass ratio inspirals. These systems will spend thousands to millions of cycles in the millihertz regime, making them particularly sensitive to environmental effects that accumulate over time in the waveform~\cite{Barausse:2014tra,Cole:2022yzw}. Future GW detectors, such as LISA~\cite{LISA:2017pwj,Colpi:2024xhw}, will probe this frequency range, underscoring the need for precise modelling of the interactions between the binary and its surroundings. In this chapter, based on the work from~\cite{Tomaselli:2024bdd,Tomaselli:2024dbw}, I explore the early inspiral of a binary system where a boson cloud surrounds the heavier primary object.
\vskip 2pt
As detailed in Section~\ref{BHenv_sec:boson_clouds}, clouds of ultralight bosons may form around rotating BHs via superradiance. Although this mechanism applies to bosons of any spin, I focus here on scalars, both for their simplicity and stronger theoretical motivation. In contrast to Chapters~\ref{chap:SR_Axionic} and~\ref{chap:in_medium_supp}, the emphasis here is on bosons that interact purely gravitationally or have extremely weak couplings. Thanks to its universal nature, gravity -- and in particular GWs -- offers a rare probe of this largely unconstrained region of parameter space (see Figure~\ref{fig:constraints}). Furthermore, the rich phenomenology of boson clouds in binary systems~\cite{Zhang:2018kib,Baumann:2018vus,Zhang:2019eid,Baumann:2019ztm,Baumann:2021fkf,Baumann:2022pkl,Tomaselli:2023ysb,Brito:2023pyl,Duque:2023seg,Dyson:2025dlj}, makes them especially promising for precision tests of fundamental physics with GWs~\cite{LISA:2022kgy}.
\vskip 2pt
In particular, in previous works~\cite{Baumann:2018vus,Baumann:2019ztm} it was found that during a binary inspiral the cloud induces not only secular effects, such as dynamical friction or accretion, but also resonant behaviour. At certain \emph{resonance frequencies}, where the orbital frequency of the binary matches the energy difference between two eigenstates of the cloud~\eqref{eq:BHenv_eigenenergy}, the gravitational perturbation from the companion is resonantly enhanced and a full transfer from one state to another can be induced. The accompanied change in the angular momentum of the cloud must then be compensated for by the binary. Depending on the nature of the resonance, this leads to \emph{floating} or \emph{sinking} orbits, where the cloud releases or absorbs angular momentum from the binary, and as a result, the inspiral is either stalled or sped-up. Consequently, a potentially detectable dephasing is left in the GW signal. The orbital frequency at which these resonances happen can be predicted, and thus they serve as a direct and unique probe of the properties of the cloud. In addition to their floating or sinking nature, resonances can be divided into three different types, depending on the energy difference between the eigenstates involved. These are called hyperfine, fine, or Bohr resonances, which occur in this chronological order (see Figure~\ref{fig:spectrum}). The former two happen ``early'' in the inspiral (at radii far larger than the cloud's radius) and are all of the floating type. These will thus only affect the GW signal \emph{indirectly}:~the binary frequency is too low for GW detectors to pick up the signal, but the resonance can change the late-time evolution of the system. Conversely, Bohr resonances happen ``late'' in the inspiral (at radii comparable with the cloud's radius) and can be either floating or sinking. As these happen while the signal is in band, they can affect the GW signal \emph{directly}.
\vskip 2pt
Due to the exciting observational signatures, various works have studied these resonances. However, to date, they have all made simplifying assumptions, which turn out to crucially alter the behaviour of the system. This includes ignoring the backreaction from the resonance~\cite{Takahashi:2021eso}, assuming a quasi-circular and equatorial orbit, or including just the strongest multipole of the gravitational perturbation~\cite{Zhang:2018kib,Ding:2020bnl,Tong:2021whq,Du:2022trq,Takahashi:2023flk}. A nonzero eccentricity was only considered in~\cite{Berti:2019wnn}, yet at a time when the behaviour of the resonances had still not been fully understood. In this chapter, all of the aforementioned assumptions are relaxed to study resonances in gravitational atoms in full generality. This is then used to determine the evolution of the cloud-binary system throughout the inspiral. Crucially, by studying the nonlinear system, a critical threshold is found above which an \emph{adiabatic} floating resonance is initiated. Below the threshold, there is a negligible transfer, or a \emph{non-adiabatic} resonance. Furthermore, mechanisms are uncovered that induce a \emph{resonance breaking}, shutting down the transition before it is complete:~these are due to the decay of the state excited by the resonance, or to a slow time variation of the parameters. Additionally, by allowing for generically inclined and eccentric orbits, it is shown how resonances strongly impact the orbital parameters:~while the eccentricity is forced towards fixed points, the inclination angle is always tilted towards a co-rotating configuration. Finally, the impact of ionisation~\cite{Baumann:2021fkf,Baumann:2022pkl,Tomaselli:2023ysb} in the evolution of the system is taken into account as well as all relevant multipoles. 
\vskip 2pt
Using these results, I lay out, for the first time, a systematic study of resonances for realistic parameters, focusing on intermediate and extreme mass ratios. Starting from states commonly populated by superradiance, the binary is evolved taking into account energy losses from both GWs and ionisation. By doing so, the resonant evolution of the cloud and the impact of the resonances on the binary's orbit is determined. I find that, for generic orbital configurations, the cloud is often destroyed early in the inspiral. This is due to floating resonances that transfer the cloud to states that decay much quicker than the duration of the resonance. However, when the orbital inclination is within a certain interval centred around a counter-rotating configuration, all hyperfine and fine floating resonances are either non-adiabatic or break prematurely, allowing the cloud to survive until it enters the Bohr regime. Conversely, all sinking resonances for typical parameters are found to have a negligible impact on the cloud. A schematic illustration of these conclusions is given in Figure~\ref{fig:history_211}. This leads to (i) \emph{direct} observational signatures from the cloud, when it survives all resonances, and (ii) new, \emph{indirect} observational signatures, when the cloud is destroyed early on, thereby leaving a legacy on the binary through its eccentricity and inclination. 
\vskip 2pt
The outline of this chapter is as follows. In Section~\ref{sec_Legacy:setup}, I briefly review the setup and introduce necessary definitions and equations. Then, in Section~\ref{sec_Legacy:resonance-pheno}, I study resonances at the nonlinear level, determining when they are adiabatic and when they break, and extending the framework to eccentric and inclined orbits. In Section~\ref{sec_Legacy:types-of-resonances}, I discuss the different types of resonances. Then, in Section~\ref{sec_Legacy:history}, I turn to a realistic setting and unveil the full history of the cloud and binary. In Section~\ref{sec_Legacy:observational-signatures}, I discuss the observational signatures this leads to. I conclude in Section~\ref{sec_Legacy:conclusions}. Appendix~\ref{app:GA} contains technical details.
\vskip 2pt
Due to the length of this chapter, it is useful to establish some notation. The mass and radial distance of the smaller object are denoted by $M_*\equiv qM$ and $R_*$, where $q < 1$ is the mass ratio, while the orbital frequency is $\Omega$ and the mass of the cloud is $M\ped{c}$. The gravitational fine structure constant is $\alpha=\mu M$, where $\mu$ is the mass of the scalar field. Most of the results are written in an explicit scaling form, with respect to the following set of benchmark parameters:~$M=10^4M_\odot$, $M\ped{c}=0.01M$, $q=10^{-3}$ and $\alpha=0.2$. The eigenstate of the cloud before encountering a resonance will be denoted by $\ket{n_a\ell_am_a}$, while any other eigenstate involved in the resonance by $\ket{n_b\ell_bm_b}$. The cloud's wavefunction will then be expanded as a linear superposition of energy eigenstates:~$\ket{\psi}=c_a\ket{a}+\sum_bc_b\ket{b}$.
\section{Setup}\label{sec_Legacy:setup}
\begin{figure}[t!] 
\centering
\includegraphics[width=\textwidth]{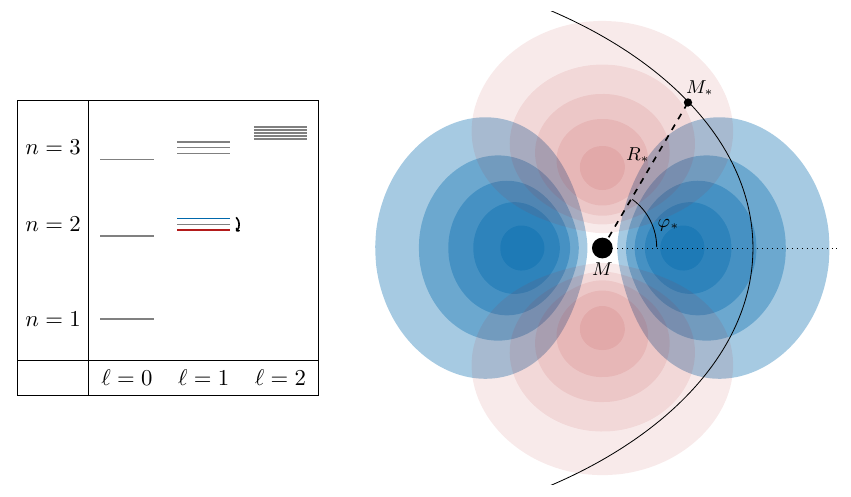}
\caption{Schematic illustration of the cloud-binary system. The primary object of mass $M$ is surrounded by a superradiantly grown scalar cloud of mass $M\ped{c}$. The secondary object of mass $M_*$ perturbs it through its gravitational potential, causing a mixing between different states of the cloud. The blue and red regions are a faithful representation of the mass densities of the states $\ket{211}$ and $\ket{21\,\minus1}$ on the equatorial plane, but the BH size is not to scale. In the box, we show the bound state spectrum of the gravitational atom for the first few values of $n$.}
\label{fig:SchematicIllustration}
\end{figure}
The goal of this section is to lay down our framework. We closely follow an earlier work~\cite{Tomaselli:2023ysb}, and thus we refer the reader there for more details.
\vskip 2pt
Via BH superradiance, bosonic fields can extract energy and angular momentum from rotating BHs. As we saw in Section~\ref{BHenv_subsec:superrad}, the key condition for this process to occur is that the boson's frequency $\omega_\slab{B}$ is smaller than the angular velocity of the event horizon $\Omega\ped{H}$, i.e., $\omega_\slab{B} < m\Omega\ped{H}$, where $m$ is the azimuthal quantum number in the BH frame~\eqref{eq:env_SR_condition}. If the boson is massive, then the superradiantly amplified waves can get trapped around the BH, allowing their number to grow exponentially. In this chapter, we will stay in the non-relativistic limit, and we thus refer to Section~\ref{BHenv_subsec:gatoms} for the description of the cloud-BH system, specifically the hydrogenic eigenfunctions~\eqref{eqn:BHenv_eigenstates} and the eigenenergy of different modes~\eqref{eq:BHenv_eigenenergy}. A schematic illustration of our setup is shown in Figure~\ref{fig:SchematicIllustration}. 
\vskip 2pt
We work in the reference frame of the central BH with mass $M$, where $\vec{r} = \{r, \theta, \phi\}$ and the coordinates of the companion with mass $M_{*}=qM$ are $\mathbf{R}_{*} = \{R_*, \theta_*, \varphi_*\}$. In our conventions, the gravitational perturbation from the companion in the Newtonian approximation is defined as
\begin{equation}
\label{eqn:V_star}
V_*(t,\vec r)=-\sum_{\ell_*=0}^\infty\sum_{m_*=-\ell_*}^{\ell_*}\frac{4\pi q\alpha}{2\ell_*+1}Y_{\ell_*m_*}(\theta_*,\varphi_*)Y_{\ell_*m_*}^*(\theta,\phi)\,F(r)\,,
\end{equation}
where
\begin{equation}
F(r)=
\begin{cases}
\dfrac{r^{\ell_*}}{R_*^{\ell_*+1}}\Theta(R_*-r)+\dfrac{R_*^{\ell_*}}{r^{\ell_*+1}}\Theta(r-R_*)&\text{for }\ell_*\ne1\,,\\[12pt]
\left(\dfrac{R_*}{r^2}-\dfrac{r}{R_*^2}\right)\Theta(r-R_*)&\text{for }\ell_*=1\,.
\end{cases}
\label{eqn:F(r)}
\end{equation}
On a generic orbit, the perturbation~\eqref{eqn:V_star} induces a mixing between the cloud's bound state $\ket{n_b \ell_b m_b}$ and another state $\ket{n_a\ell_am_a}$ through the matrix element~\cite{Baumann:2018vus,Baumann:2019ztm} 
\begin{equation}
\braket{n_a\ell_a m_a|V_*(t,\vec r)|n_b\ell_bm_b}=-\sum_{\ell_* m_*}\frac{4\pi\alpha q}{2\ell_*+1}Y_{\ell_*m_*}(\theta_*,\varphi_*)\, I_r(t) \, I_{\Omega}(t)\,,
\label{eqn:MatrixElement}
\end{equation}
where $I_r$ and $I_\Omega$ are integrals over radial and angular variables, respectively. The following \emph{selection rules} need to be satisfied in order for $I_{\Omega}$ to be nonzero
\begin{align}
(\text{S1})\qquad & m_*=m_b - m_a\,,\\
(\text{S2})\qquad & \ell_a+\ell_*+\ell_b=2 p\, \, \, \, \, \text{for}\, \, \, p \in \mathbb{Z}\,, \label{eqn:S2}\\
(\text{S3})\qquad & \abs{\ell_b-\ell_a} \leq \ell_* \leq \ell_b+\ell_a\,. \label{eqn:S3}
\end{align}
Furthermore, we will often expand the spherical harmonic $Y_{\ell_*m_*}$ appearing in~\eqref{eqn:MatrixElement} in terms of the form $Y_{\ell_* g}(\pi/2, 0)$ (where $g$ is a summation index), which is zero whenever $\ell_*$ and $g$ have opposite parity.
\section{Resonance Phenomenology}\label{sec_Legacy:resonance-pheno}
As first shown in~\cite{Baumann:2019ztm}, while the companion perturbs the cloud at a slowly increasing frequency, transitions between modes are induced, analogous to the ones described in quantum mechanics by Landau and Zener~\cite{zener1932non,landau1932theorie}. This process can exert a strong backreaction on the orbit, giving rise to ``floating'' and ``sinking'' orbits. In this section, we study these transitions for generic orbits and at the nonlinear level, by including the backreaction in the frequency evolution self-consistently. We do not, however, worry about the astrophysically relevant range of parameters just yet, nor about whether the phenomena we discover here can actually occur after well-motivated initial conditions. Such ``realistic'' cases, to which we often refer to, will only be defined and studied in Section~\ref{sec_Legacy:history}, where the general results found here will turn out to be crucial in determining the evolution of the cloud-binary system.
\vskip 2pt
In Section~\ref{sec:two-state-resonances} we review the setup and the well-known results, which we extend to inclined and eccentric orbits in Section~\ref{sec:eccentric-inclined-resonances}. Then, in Section~\ref{sec:backreaction} we include the backreaction, thus coupling the resonating states to the evolution of the binary parameters. The phenomenology of the resulting nonlinear system is explored in Section~\ref{sec:floating} and Section~\ref{sec:sinking}, for the floating and sinking cases, respectively.
\subsection{Two-State Resonances}\label{sec:two-state-resonances}
The matrix element~\eqref{eqn:MatrixElement} of the gravitational perturbation $V_*$ between two states $\ket a=\ket{n_a\ell_am_a}$ and $\ket b=\ket{n_b\ell_bm_b}$ is an oscillatory function of the true anomaly of the orbit $\varphi_*$,
\begin{equation}
\braket{a|V_{*}(t)|b} = \sum_{g\in \mathbb{Z}} \eta^\floq{g} e^{ig \varphi_*}\,.
\label{eqn:eta-circular}
\end{equation}
On equatorial co-rotating quasi-circular orbits, the only nonzero term is $g=m_b-m_a\equiv\Delta m$ (on counter-rotating orbits, $g$ has opposite sign), and the coefficients $\eta^\floq{g}$ only depend on time through $\Omega(t)$. Restricting our attention to the two-state system, the Hamiltonian is thus
\begin{equation}
\mathcal H=\begin{pmatrix}-\Delta\epsilon/2 & \eta^\floq{g}e^{ig\varphi_*}\\ \eta^\floq{g}e^{-ig\varphi_*} & \Delta\epsilon/2\end{pmatrix}\,,
\label{eqn:schrodinger-hamiltonian}
\end{equation}
where $\Delta\epsilon=\epsilon_b-\epsilon_a$ is the energy difference between $\ket{b}$ and $\ket{a}$. As in~\cite{Baumann:2019ztm}, it is useful to rewrite~\eqref{eqn:schrodinger-hamiltonian} in a \emph{dressed} frame, where the fast oscillatory terms $e^{ig\varphi_*}$ are traded for a slow evolution of the energies. This is done by means of a unitary transformation,
\begin{equation}
\begin{pmatrix}c_a\\ c_b\end{pmatrix}=\begin{pmatrix}e^{ig\varphi_*/2} & 0\\ 0 & e^{-ig\varphi_*/2}\end{pmatrix}\begin{pmatrix}\bar c_a\\ \bar c_b\end{pmatrix}\,,
\end{equation}
where $c_j=\braket{j|\psi}$ (with $j=a,b$) are the Schr\"{o}dinger frame coefficients, while $\bar c_a$ and $\bar c_b$ are the dressed frame coefficients. Because $\abs{c_i}^2=\abs{\bar c_i}^2$, we will drop the overhead bar in the following discussion. In the dressed frame, the Schr\"{o}dinger equation reads
\begin{equation}
\frac\dd{\dd t}\!\begin{pmatrix}c_a\\ c_b\end{pmatrix}=-i\mathcal H\ped{D}\begin{pmatrix}c_a\\ c_b\end{pmatrix},\qquad\mathcal H\ped{D}=\begin{pmatrix}-(\Delta\epsilon-g\Omega)/2 & \eta^\floq{g}\\ \eta^\floq{g} & (\Delta\epsilon-g\Omega)/2\end{pmatrix}\,.
\label{eqn:dressed-hamiltonian}
\end{equation}
When $\Omega(t)\equiv\dot\varphi_*$ is specified,~\eqref{eqn:dressed-hamiltonian} determines the evolution of the population of the two states.
\vskip 2pt
Without including the backreaction of the resonance on the orbit, $\Omega(t)$ is exclusively determined by external factors, such as the energy losses due to GW emission or cloud ionisation, which induce a frequency chirp. These effects typically have a nontrivial dependence on $\Omega$ itself, widely varying in strength at different points of the inspiral. However, the resonances described by~\eqref{eqn:dressed-hamiltonian} are restricted to a bandwidth $\Delta\Omega\sim\eta^\floq{g}$. This is typically narrow enough to allow us to approximate the external energy losses, as well as any other $\Omega$-dependent function, with their value at the resonance frequency,
\begin{equation}\label{eq:resonance_con}
\Omega_0=\frac{\Delta\epsilon}{g}\,.
\end{equation}
Around $\Omega_0$, we can linearise the frequency chirp and write $\Omega=\gamma t$. For concreteness, in this section we will assume that external energy losses are only due to GW emission, in which case
\begin{equation}
\gamma=\frac{96}5\frac{qM^{5/3}\Omega_0^{11/3}}{(1+q)^{1/3}}\,.
\label{eqn:gamma_gws}
\end{equation}
It is particularly convenient to rewrite the Schr\"{o}dinger equation in terms of dimensionless variables and parameters:
\begin{equation}
\label{eqn:dimensionless-dressed-schrodinger}
\frac\dd{\dd\tau}\!\begin{pmatrix}c_a\\ c_b\end{pmatrix}=-i\begin{pmatrix}\omega/2 & \sqrt{Z}\\ \sqrt{Z} & -\omega/2\end{pmatrix}\begin{pmatrix}c_a\\ c_b\end{pmatrix}\,,
\end{equation}
where the frequency chirp now reads $\omega=\tau$, and we defined
\begin{equation}
\label{eqn:coefficients}
\tau=\sqrt{\abs{g}\gamma}\,t\,,\qquad\omega=\frac{\Omega-\Omega_0}{\sqrt{\gamma/\abs{g}}}\,,\qquad Z=\frac{(\eta^\floq{g})^2}{\abs{g}\gamma}\,.
\end{equation}
The initial conditions at $\tau\to-\infty$ we are interested in are those where only one state is populated, say $c_a=1$ and $c_b=0$. The only dimensionless parameter of~\eqref{eqn:dimensionless-dressed-schrodinger} is the so-called ``Landau-Zener parameter'' $Z$, which determines uniquely the evolution of the system and its state at $\tau\to+\infty$. In fact, the populations at $\tau\to+\infty$ can be derived analytically and are given by the Landau-Zener formula:
\begin{equation}
\abs{c_a}^2=e^{-2\pi Z}\,,\qquad \abs{c_b}^2=1-e^{-2\pi Z}\,.
\label{eqn:lz}
\end{equation}
For $2\pi Z\gg1$ the transition can be classified as \emph{adiabatic}, meaning that the process is so slow that the cloud is entirely transferred from $\ket{a}$ to $\ket{b}$. Conversely, for $2\pi Z\ll1$, the transition is \emph{non-adiabatic}, with a partial or negligible transfer occurring.\footnote{The adiabaticity of a resonance is not related to the adiabaticity of the orbital evolution, which is always assumed to hold throughout this chapter.}
\vskip 2pt
There is one final remark to be made before we proceed. Dealing with two-state transitions is a good approximation as long as the frequency width of the resonance, $\Delta\Omega\sim\eta^\floq{g}$, is much narrower than the distance (in frequency) from the closest resonance. The latter becomes extremely small for hyperfine resonances, especially on generic orbits, where $g$ can take values different from $\Delta m$. In some cases formula~\eqref{eq:BHenv_eigenenergy} can indeed return an exact degeneracy of two resonances, up to $\mathcal O(\alpha^5)$. We have thoroughly checked, by numerical computation of the eigenfrequencies up to $\mathcal O(\alpha^6)$, that in all realistic cases the resonances are indeed narrow enough for the two-state approximation to hold.
\subsection{Resonances on Eccentric and Inclined Orbits}\label{sec:eccentric-inclined-resonances}
We now extend the treatment of Section~\ref{sec:two-state-resonances} to orbits with nonzero eccentricity or inclination, explaining what changes for the resonant frequencies and the overlap coefficients $\eta^\floq{g}$.
\vskip 2pt
Let us start with eccentric co-rotating orbits. In the quasi-circular case, eq.~\eqref{eqn:eta-circular} manifestly separates a \emph{fast} and a \emph{slow} motion:~the former originates from $\varphi_*$ varying over the course of an orbit, while the latter is due to the dependence of the coefficients $\eta^\floq{g}$ on $\Omega(t)$ (and can be safely neglected). It will be helpful to work with a variable that performs the same trick on eccentric orbits: the \emph{mean anomaly}
\begin{equation}
\hat{\varphi}_*(t)=\int^t\Omega(t')\dd t'\,.
\end{equation}
Because $\varphi_*$ itself is an oscillating function of $\hat\varphi_*$, we can write
\begin{equation}
\braket{a|V_{*}(t)|b} = \sum_{g\in \mathbb{Z}} \hat\eta^\floq{g} e^{ig\hat\varphi_*}\,,
\label{eqn:eta-tilde}
\end{equation}
where the coefficients $\hat\eta^\floq{g}$ only depend on time through $\Omega(t)$. For simplicity, in the following discussion we will drop the hats, with the different definition of $\eta^\floq{g}$ for nonzero eccentricity left understood.
\vskip 2pt
For a given eccentricity $\varepsilon\ne0$, multiple terms of~\eqref{eqn:eta-tilde}, each corresponding to a different value of $g$, can be nonzero. As a consequence, a resonance between two given states can be triggered at different points of the inspiral, at the frequencies $\Omega_0^\floq{g}=\Delta\epsilon/g$, for any integer $g$ (provided that it has the same sign as $\Delta\epsilon$). The numerical evaluation of the coefficients $\eta^\floq{g}$ requires to Fourier expand $V_*$ in the time domain, at the orbital frequency $\Omega=\Omega_0^\floq{g}$. This can be done with techniques similar to~\cite{Tomaselli:2023ysb}, where the same matrix element was evaluated between a bound and an unbound state. The coefficient $\eta^\floq{\Delta m}$ is special because it is the only one with a finite, nonzero limit for $\varepsilon\to0$, where it reduces to its circular-orbit counterpart. For all other values of $g$, instead, $\eta^\floq{g}$ vanishes for $\varepsilon\to0$. Even at moderately large $\varepsilon$, the coefficient $\eta^\floq{\Delta m}$ remains significantly larger than all the others
\vskip 2pt
Let us now look at circular but inclined orbits. Here, the Fourier coefficients $\eta^\floq{g}$ acquire a dependence on the inclination angle $\beta$, where $\beta=0$ and $\beta=\pi$ correspond to the co-rotating and counter-rotating scenarios. The functional dependence can be readily extracted by evaluating the perturbation~\eqref{eqn:V_star} using the identity~\cite{wigner}
\begin{equation}
Y_{\ell_*m_*}(\theta_*,\varphi_*)=\sum_{g=-\ell_*}^{\ell_*}d^\floq{\ell_*}_{m_*,g}(\beta)Y_{\ell_*g}\left(\frac\pi2,0\right)e^{ig\Omega t}\,.
\label{eqn:Y-decomposed}
\end{equation}
Here, $d^\floq{\ell_*}_{m_*,g}(\beta)$ is a Wigner small $d$-matrix and is responsible for the angular dependence of the coupling, $\eta^\floq{g}\propto d^\floq{\ell_*}_{m_*,g}(\beta)$. Its functional form takes on a simple expression in many of the physically interesting cases, as we will discuss explicitly in Section~\ref{sec_Legacy:types-of-resonances}. We thus see that inclined orbits also trigger resonances at $\Omega=\Omega_0^\floq{g}=\Delta\epsilon/g$, but this time $g$ can only assume a finite number of different values. Similar to the eccentric case, $g=\Delta m$ is special, because it is the only case where $d^\floq{\ell_*}_{m_*,g}(\beta)$ does not vanish for $\beta\to0$, as the resonance survives in the equatorial co-rotating limit. Similarly, in the counter-rotating case $\beta\to\pi$, the only surviving value is $g=-\Delta m$.
\vskip 2pt
Similar techniques can be applied in the eccentric \emph{and} inclined case, where the overlap can be expanded in two sums, each with its own index, say $g_\varepsilon$ and $g_\beta$. We do not explicitly compute $\eta^\floq{g}$ in the general case, as the understanding developed so far is sufficient to move forward and characterise the phenomenology in realistic cases.
\subsection{Backreaction on the Orbit}\label{sec:backreaction}
We now include the backreaction on the orbit, allowing for generic nonzero eccentricity \emph{and} inclination. During a resonance, the energy and angular momentum contained in the cloud change over time:~this variation must be compensated by an evolution of the binary parameters, the (dimensionless) frequency $\omega$, eccentricity $\varepsilon$ and inclination $\beta$. In turn, this backreaction impacts the Schr\"{o}dinger equation~\eqref{eqn:dimensionless-dressed-schrodinger}, which directly depends on $\omega$. The result is a coupled nonlinear system of ordinary differential equations, describing the co-evolution of the cloud and the binary, which we derive in this section.
\vskip 2pt
To describe the evolution of $\omega$, $\varepsilon$, and $\beta$ we need three equations. These are the conservation of energy and of two components of the angular momentum: the projection along the BH spin and the projection on the equatorial plane. The conservation of energy reads
\begin{equation}
\frac\dd{\dd t}\left(E+E\ped{c}\right)=-\gamma f(\varepsilon)\,\frac{qM^{5/3}}{3(1+q)^{1/3}\Omega_0^{1/3}}\,,
\label{eqn:E-balance}
\end{equation}
where $\gamma$ is defined in~\eqref{eqn:gamma_gws} and the binary's and cloud's energies are
\begin{equation}
E=-\frac{qM^{5/3}\Omega^{2/3}}{2(1+q)^{1/3}}\,,\qquad E\ped{c}=\frac{M\ped{c}}\mu(\epsilon_a\abs{c_a}^2+\epsilon_b\abs{c_b}^2)\,.
\end{equation}
The function
\begin{equation}
f(\varepsilon)=\frac{1+\frac{73}{24}\varepsilon^2+\frac{37}{96}\varepsilon^4}{(1-\varepsilon^2)^{7/2}}\,,
\end{equation}
quantifies the dependence of GW energy losses on the eccentricity~\cite{Peters:1963ux,Peters:1964zz}. Similarly, the conservation of the angular momentum components requires
\begin{align}
\label{eqn:Lz-balance}
\frac\dd{\dd t}\left(L\cos\beta+S\ped{c}\right)=-h(\varepsilon)\gamma\,\frac{qM^{5/3}}{3(1+q)^{1/3}\Omega_0^{4/3}}\cos\beta\,,\\
\label{eqn:Lx-balance}
\frac\dd{\dd t}\left(L\sin\beta\right)=-h(\varepsilon)\gamma\,\frac{qM^{5/3}}{3(1+q)^{1/3}\Omega_0^{4/3}}\sin\beta\,,
\end{align}
where
\begin{equation}
L=\frac{qM^{5/3}}{(1+q)^{1/3}}\frac{\sqrt{1-\varepsilon^2}}{\Omega^{1/3}}\,,\qquad S\ped{c}=\frac{M\ped{c}}\mu(m_a\abs{c_a}^2+m_b\abs{c_b}^2)\,,
\end{equation}
and
\begin{equation}
h(\varepsilon)=\frac{1+\frac78\varepsilon^2}{(1-\varepsilon^2)^2}\,.
\end{equation}
Before proceeding, there are two issues the reader might worry about. First, depending on the resonance, the spin of the cloud during the transition might also have equatorial components, and should thus appear in~\eqref{eqn:Lx-balance}. Second, the BH spin breaks spherical symmetry, therefore the equatorial projection of the angular momentum should not be conserved. Clearly, in the Newtonian limit this is not a problem, but one might still question whether it is consistent to treat within this framework hyperfine resonances, whose very existence is due to a nonzero BH spin in the first place. We address both these issues in Appendix~\ref{appGA_sec:hyperfine-angular-momentum}, where we justify our assumptions, and proceed here to study the dynamics of the previous equations.
\vskip 2pt
Equations~\eqref{eqn:E-balance},~\eqref{eqn:Lz-balance} and~\eqref{eqn:Lx-balance} can be put in a dimensionless form as follows:
\begin{align}
\label{eqn:dimensionless-omega-evolution}
\frac{\dd\omega}{\dd\tau}&=f(\varepsilon)-B\frac{\dd\abs{c_b}^2}{\dd\tau}\,,\\
\label{eqn:dimensionless-eccentricity-evolution}
C\frac\dd{\dd\tau}\sqrt{1-\varepsilon^2}&=\sqrt{1-\varepsilon^2}\left(f(\varepsilon)-B\frac{\dd\abs{c_b}^2}{\dd\tau}\right)+B\frac{\Delta m}g\frac{\dd\abs{c_b}^2}{\dd\tau}\cos\beta-h(\varepsilon)\,,\\
\label{eqn:dimensionless-inclination-evolution}
C\sqrt{1-\varepsilon^2}\frac{\dd\beta}{\dd\tau}&=-B\frac{\Delta m}g\frac{\dd\abs{c_b}^2}{\dd\tau}\sin\beta\,,
\end{align}
where we defined the dimensionless parameters
\begin{equation}
B=\frac{3M\ped{c}}M\frac{\Omega_0^{4/3}((1+q)M)^{1/3}}{q\alpha\sqrt{\gamma/\abs{g}}}(-g)\,,\qquad C=\frac{3\Omega_0}{\sqrt{\gamma/\abs{g}}}\,.
\label{eqn:BC}
\end{equation}
The Schr\"{o}dinger equation~\eqref{eqn:dimensionless-dressed-schrodinger} remains unchanged, but it should be kept in mind that $Z$ now depends on $\varepsilon$ and $\beta$ through $\eta^\floq{g}$ (instead, the dependence on $\omega$ can still be neglected if the resonance is narrow enough).
\vskip 2pt
The parameter $B$ controls the strength of the backreaction. As can be seen from~\eqref{eqn:dimensionless-omega-evolution}, a positive $B>0$ (i.e., $g<0$ and $\Delta\epsilon<0$) will slow down the frequency chirp, giving rise to a \emph{floating} orbit and generally making the resonance more adiabatic. Conversely, $B<0$ (i.e., $g>0$ and $\Delta\epsilon>0$) induces \emph{sinking} orbits and makes resonances less adiabatic. By extension, we will refer to floating resonances and sinking resonances to denote the type of backreaction they induce. A summary of the main variables used to describe the resonances and their backreaction is given in Appendix~\ref{appGA_sec:variables}.
\subsection{Floating Orbits}\label{sec:floating}
Backreaction of the floating type ($B>0$) turns out to be the most relevant case for realistic applications, so we make a detailed study of its phenomenology here. When the backreaction is strong, the evolution of the system exhibits a very well-defined phase of floating orbit. We are then concerned with three aspects.
\begin{enumerate}
\item Under what conditions is a floating resonance initiated? We answer this question with a simple analytical prescription, derive and discussed below.
\item How does the system evolve during the float? This is addressed later in the section, where we study the evolution of the eccentricity and inclination.
\item When does a floating resonance end? We show that several phenomena can \emph{break} (and end) the resonance before the transition from $\ket{a}$ to $\ket{b}$ is complete, and we compute accurately the conditions under which this phenomenon happens.
\end{enumerate}
\subsubsection{Adiabatic or non-adiabatic}\label{sec:floating-adiabatic}
\begin{figure}[t!] 
\centering
\includegraphics[width=\linewidth]{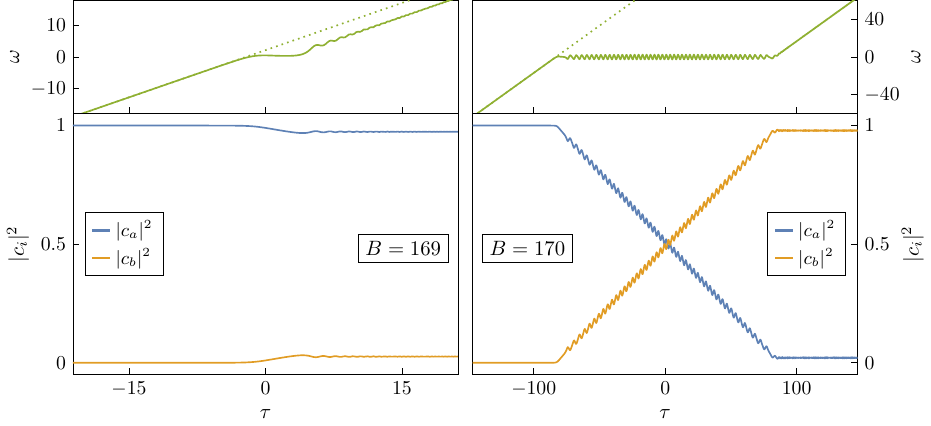}
\caption{Numerical solution of the nonlinear system~\eqref{eqn:dimensionless-dressed-schrodinger}--\eqref{eqn:dimensionless-omega-evolution}. In both panels we set $Z=0.001$, whereas we choose the values of $B$ to be $169$ (\emph{left panel}) and $170$ (\emph{right panel}), slightly below or above the adiabaticity threshold (the limit value of $ZB$ differs slightly from the one given in~\eqref{eqn:1/2piadiabatic}, due to finite-$Z$ corrections). In the \emph{left panel}, a non-adiabatic transition is observed. Conversely, in the \emph{right panel}, we find an adiabatic transition and the consequent formation of a floating orbit, whose duration matches the predicted $\Delta t\ped{float}=B/\sqrt{\abs{g}\gamma}$. The dotted lines represent the evolution of $\omega$ in absence of backreaction.}
\label{fig:floating-backreacted-resonance}
\end{figure}

From Section~\ref{sec:two-state-resonances}, we know that if $B=0$, then a fraction $1-e^{-2\pi Z}$ of the cloud is transferred during the resonances. For $2\pi Z\gg1$, this value is already very close to 1. Adding the backreaction does not change this conclusion:~the resonance stays adiabatic and a complete transfer from $\ket{a}$ to $\ket{b}$ is observed. Assuming, for simplicity, quasi-circular orbits ($\varepsilon=0$), the duration of the floating orbit can be easily read off~\eqref{eqn:dimensionless-omega-evolution}:
\begin{equation}
\Delta t\ped{float}=\frac{B}{\sqrt{\abs{g}\gamma}}=\frac{3M\ped{c}}M\frac{\Omega_0^{4/3}((1+q)M)^{1/3}}{q\alpha\gamma}(-g)\,.
\label{eqn:t-float}
\end{equation}
This is independent of the strength of the perturbation $\eta^\floq{g}$, and corresponds to the time it takes for the external energy losses to dissipate the energy of the two-state system. For nonzero eccentricity instead, one must integrate $f(\varepsilon)$ over time to determine the duration of the float.
\vskip 2pt
The situation for $2\pi Z\ll1$ is, in principle, much less clear:~with $B=0$ the resonance would be non-adiabatic, but backreaction tends to make it more adiabatic. Let us once again restrict to quasi-circular orbits for simplicity. By careful numerical study of eqs.~\eqref{eqn:dimensionless-dressed-schrodinger} and~\eqref{eqn:dimensionless-omega-evolution}, we find that the long-time behaviour of the system is predicted by the parameter $ZB$ alone. Depending on its value, two qualitatively different outcomes are possible:
\begin{equation}
\text{if}\quad2\pi Z\ll1\quad \mathrm{and} \quad
\begin{cases}
ZB<0.1686\ldots\quad\longrightarrow\quad\text{very non-adiabatic,}\\[12pt]
ZB>0.1686\ldots\quad\longrightarrow\quad\text{very adiabatic.}
\end{cases}
\label{eqn:1/2piadiabatic}
\end{equation}
In the upper case, a negligible fraction of the cloud is transferred and the time evolution of $\omega$ is almost exactly linear. Conversely, in the bottom case, the cloud is entirely transferred from $\ket{a}$ to $\ket{b}$ and $\omega$ is stalled for an amount of time $\Delta t\ped{float}$ given by~\eqref{eqn:t-float}, during which it oscillates around zero. Intermediate behaviours are not possible, unless the value of $ZB$ is extremely fine-tuned. Numerical solutions of~\eqref{eqn:dimensionless-dressed-schrodinger} and~\eqref{eqn:dimensionless-omega-evolution} are shown in Figure~\ref{fig:floating-backreacted-resonance}, choosing parameters in such a way to illustrate the two cases in~\eqref{eqn:1/2piadiabatic}.
\vskip 2pt
We can give an approximate derivation of the previous result as follows. As long as $\abs{c_b}^2$ is small enough, the backreaction term in~\eqref{eqn:dimensionless-omega-evolution} is negligible, hence $\omega$ evolves linearly and the final populations approximate the Landau-Zener result~\eqref{eqn:lz}, giving $\abs{c_b}^2\approx2\pi Z$. As the unbackreacted transition happens in the time window $\abs{\tau}\lesssim1$, we see from~\eqref{eqn:dimensionless-omega-evolution} that the backreaction becomes significant when $1\lesssim B\cdot2\pi Z\implies ZB\gtrsim1/(2\pi)\approx0.159\ldots$ Given the minimal numerical difference between this coefficient and the one given in~\eqref{eqn:1/2piadiabatic}, for simplicity we will often write the relevant condition for an adiabatic resonance simply as $2\pi ZB\gtrless1$. The slow-down effect on the evolution of $\omega(\tau)$ enjoys a positive-feedback mechanism:~the slower $\omega$ evolves, the more the transition is adiabatic, meaning that $\abs{c_b}^2$ is larger, which further slows down $\omega(\tau)$, and so on. This explains why no intermediate behaviours are observed:~once the backreaction goes over a certain critical threshold, the process becomes self-sustaining.
\vskip 2pt
The picture outlined so far changes slightly when the eccentricity is nonzero. First, if the binary had a constant eccentricity $\varepsilon_0$, we could simply replace $\gamma\to\gamma f(\varepsilon_0)$ to conclude that the critical threshold for adiabaticity becomes
\begin{equation}
2\pi ZB\gtrless f(\varepsilon_0)^{3/2}\,.
\label{eqn:2piZB-epsilon0}
\end{equation}
When the eccentricity is allowed to vary starting from the initial value $\varepsilon_0$, eq.~\eqref{eqn:2piZB-epsilon0} still correctly predicts whether the system enters a floating orbit phase. However, the transfer might no longer be complete, as the resonance might \emph{break}. This aspect will be discussed in Section~\ref{sec:resonance-breaking}.
\subsubsection{Evolution of eccentricity and inclination}\label{sec:floating-evolution-e-beta}
\begin{figure}[t!] 
\centering
\includegraphics[width=\linewidth]{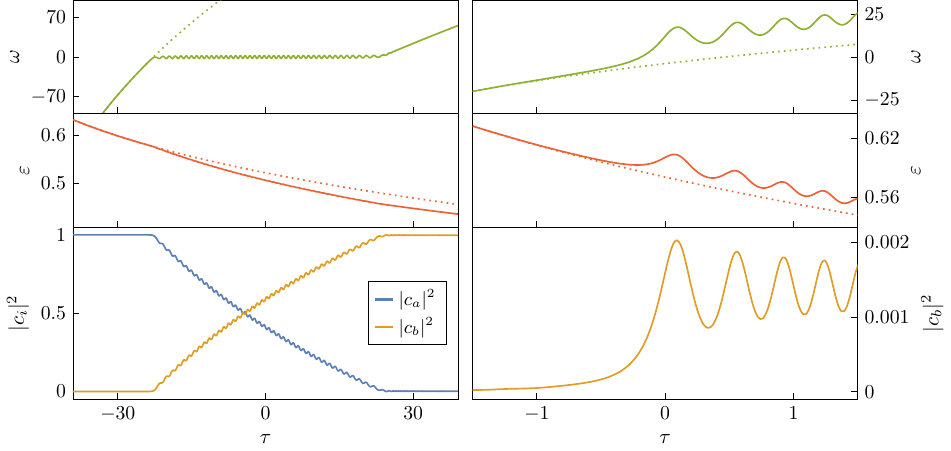}
\caption{Floating (\emph{left panel}) and sinking (\emph{right panel}) resonances on eccentric orbits, with $\Delta m/g=1$. We display the value of the frequency $\omega$, the eccentricity $\varepsilon$, and the populations $\abs{c_a}^2$ and $\abs{c_b}^2$ as function of $\tau$, obtained by solving eqs.~\eqref{eqn:dimensionless-dressed-schrodinger},~\eqref{eqn:dimensionless-omega-evolution}, and~\eqref{eqn:dimensionless-eccentricity-evolution} with $\beta=0$ numerically. The parameters used for the floating case are $Z=0.03$, $B=250$, $C=1000$, while for the sinking case we used $Z=0.01$, $B=-10000$, $C=100$. The dotted lines represent the evolution of $\omega$ and $\varepsilon$ in absence of backreaction. Even though the impact of the resonance on the eccentricity might look mild, the effect is actually dramatic when seen as a function of $\omega$, as shown in Figure~\ref{fig:omega-eccentricity}.}
\label{fig:eccentric-backreacted-resonance}
\end{figure}
\begin{figure}[t!] 
\centering
\includegraphics[width=\linewidth]{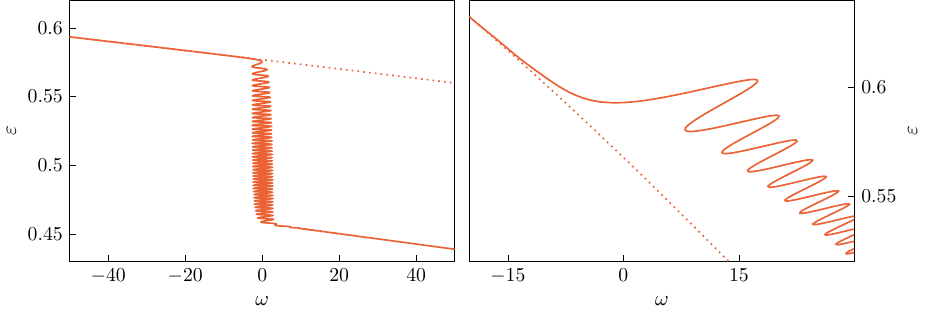}
\caption{Same resonances as in Figure~\ref{fig:eccentric-backreacted-resonance}, but now the evolution of eccentricity is shown as a function of the frequency, for floating (\emph{left panel}) and sinking (\emph{right panel}) orbits. The dashed lines represent the vacuum evolution.}
\label{fig:omega-eccentricity}
\end{figure}

The \emph{left panel} of Figure~\ref{fig:eccentric-backreacted-resonance} shows a numerical solution of the coupled nonlinear equations~\eqref{eqn:dimensionless-dressed-schrodinger},~\eqref{eqn:dimensionless-omega-evolution},~\eqref{eqn:dimensionless-eccentricity-evolution} and~\eqref{eqn:dimensionless-inclination-evolution}, for an equatorial co-rotating ($\beta=0$) but eccentric ($\varepsilon\ne0$) system, undergoing a floating orbit with $g=\Delta m$. The state dynamics is largely similar to what we described in Section~\ref{sec:floating-adiabatic}. The most interesting new effect concerns the evolution of the eccentricity, which can be seen to decrease during the float, at a rate faster than the circularisation provided by GW emission. The same numerical solution is shown as function of frequency in Figure~\ref{fig:omega-eccentricity}. 
\vskip 2pt
The evolution of the eccentricity during a float can be studied analytically by plugging $\dd\omega/\dd\tau\approx0$ into eqs.~\eqref{eqn:dimensionless-eccentricity-evolution} and~\eqref{eqn:dimensionless-inclination-evolution}, which become
\begin{align}
\label{eqn:eccentricity-evolution-floating}
C\frac\dd{\dd\tau}\sqrt{1-\varepsilon^2}&=\frac{\Delta m}gf(\varepsilon)\cos\beta-h(\varepsilon)\,,\\
\label{eqn:inclination-evolution-floating}
C\sqrt{1-\varepsilon^2}\frac{\dd\beta}{\dd\tau}&=-\frac{\Delta m}gf(\varepsilon)\sin\beta\,.
\end{align}
For resonances with $\beta=0$ and $g=\Delta m$, such as the one shown in Figures~\ref{fig:eccentric-backreacted-resonance} and~\ref{fig:omega-eccentricity}, a small-$\varepsilon$ expansion leads to the following solution:
\begin{equation}
\varepsilon(t)\approx\varepsilon_0\,e^{-\frac{22}{18}\gamma t/\Omega_0}\,.
\label{eqn:circulatization-floating}
\end{equation}
This result should be compared to the GW-induced circularisation in absence of backreaction,
\begin{equation}
\varepsilon(t)\approx\varepsilon_0\,e^{-\frac{19}{18}\gamma t/\Omega_0}\,.
\label{eqn:circulatization-GW}
\end{equation}
Therefore, not only is the orbit stalled at $\Omega(t)\approx\Omega_0$ for a potentially long time, given in~\eqref{eqn:t-float}, during which the eccentricity keeps reducing; but it also goes down at a faster rate than in the vacuum, as can be seen comparing~\eqref{eqn:circulatization-floating} with~\eqref{eqn:circulatization-GW}. The longer the resonance, the more the binary is circularised.
\vskip 2pt
This result holds for co-rotating resonances with $g=\Delta m$, which are the only ones surviving in the small-$\varepsilon$ limit and usually have the largest coupling $\eta^\floq{g}$ even at moderately large eccentricities. The dynamics are different in other cases. Remaining in the equatorial co-rotating case ($\beta=0$), eccentric binaries can also undergo (usually weaker) resonances where $g\ne\Delta m$. In this case,~\eqref{eqn:eccentricity-evolution-floating} has a different behaviour:~if $\abs{\Delta m/g}<1$, then there is a fixed point $\bar\varepsilon>0$ such that if $\varepsilon<\bar\varepsilon$ then $\varepsilon$ increases, while if $\varepsilon>\bar\varepsilon$ then $\varepsilon$ decreases. For example, for $\Delta m/g=1/2$, we have $\bar\varepsilon\approx0.46$ and the eccentricity approaches the fixed point according to
\begin{equation}
\varepsilon(t)\approx0.46+(\varepsilon_0-0.46)e^{-3.49\gamma t/\Omega_0}\,.
\label{eqn:eccentrification-floating-DeltamOverg=0.5}
\end{equation}
Floating resonances with $\abs{\Delta m/g}>1$ will instead circularise the binary even quicker than~\eqref{eqn:circulatization-floating}. As $\varepsilon$ decreases, however, so does $Z$:~eventually, the perturbation becomes too weak and the resonance stops, generically leaving the cloud in a mixed state as the inspiral resumes. This aspect will be discussed in Section~\ref{sec:resonance-breaking}.
\begin{figure}[t!] 
\centering
\includegraphics[width=\linewidth]{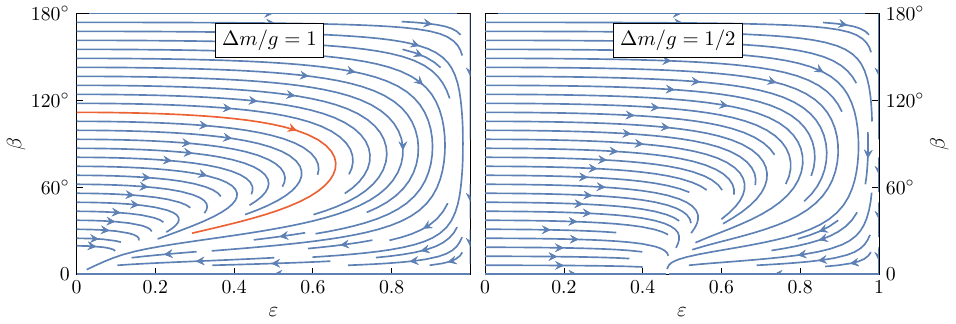}
\caption{Flow in the eccentricity-inclination plane $(\varepsilon,\beta)$ determined by eqs.~\eqref{eqn:dimensionless-eccentricity-evolution} and~\eqref{eqn:dimensionless-inclination-evolution} under the assumption that the system is on a floating orbit, i.e., $\dd\abs{c_b}^2/\dd\tau=f(\varepsilon)/B$, for two different values of $\Delta m/g$. The highlighted arrow [{\color{Mathematica4}red}] roughly depicts the trajectory followed by the system in Figure~\ref{fig:eccentric-inclined-backreacted-resonance}.}
\label{fig:streamplot_inclination-eccentricity}
\end{figure}
\vskip 2pt
The possibilities described so far are a particular case of the general dynamics, which includes the evolution of the inclination $\beta$. The flow induced by eqs.~\eqref{eqn:eccentricity-evolution-floating} and~\eqref{eqn:inclination-evolution-floating} in the $(\varepsilon,\beta)$-plane is shown in Figure~\ref{fig:streamplot_inclination-eccentricity}, where the dynamics on the horizontal axis are described by eqs.~\eqref{eqn:circulatization-floating} (\emph{left panel}) and~\eqref{eqn:eccentrification-floating-DeltamOverg=0.5} (\emph{right panel}). Perhaps the most striking feature of Figure~\ref{fig:streamplot_inclination-eccentricity} is the fact that the system is violently pulled away from inclined circular orbits (vertical axis). In fact, $\dd\varepsilon/\dd\tau$ diverges for $\varepsilon\to0$ and finite $\beta$, meaning that the validity of eqs.~\eqref{eqn:eccentricity-evolution-floating} and~\eqref{eqn:inclination-evolution-floating} must somehow break down in that limit. The explanation for this behaviour is that it is inconsistent to assume that the system undergoes an adiabatic floating resonance on inclined circular orbits:~eccentricity \emph{must} increase before the onset of the resonance. This is precisely the behaviour observed in Figure~\ref{fig:eccentric-inclined-backreacted-resonance}, where eqs.~\eqref{eqn:dimensionless-dressed-schrodinger},~\eqref{eqn:dimensionless-omega-evolution},~\eqref{eqn:dimensionless-eccentricity-evolution} and~\eqref{eqn:dimensionless-inclination-evolution} are solved numerically starting from $\varepsilon_0=0$ and $\beta_0\ne0$. If $2\pi ZB>f(\varepsilon_0)^{3/2}$, then the system enters the floating orbit and starts to follow the trajectories shown in Figure~\ref{fig:streamplot_inclination-eccentricity}. The total ``distance'' in the $(\beta,\varepsilon)$ plane travelled by the system by the time the transition completes depends on a single dimensionless ``distance parameter'',
\begin{equation}
D\equiv\frac{B}C=\frac{\gamma\Delta t\ped{float}}{\Omega_0}\,.
\label{eqn:D}
\end{equation}
However, it is also possible that the transition stops before fully completing, as shown in Figure~\ref{fig:eccentric-inclined-backreacted-resonance} (\emph{right panel}). This is the subject of Section~\ref{sec:resonance-breaking}.
\begin{figure}[t!] 
\centering
\includegraphics[width=\linewidth]{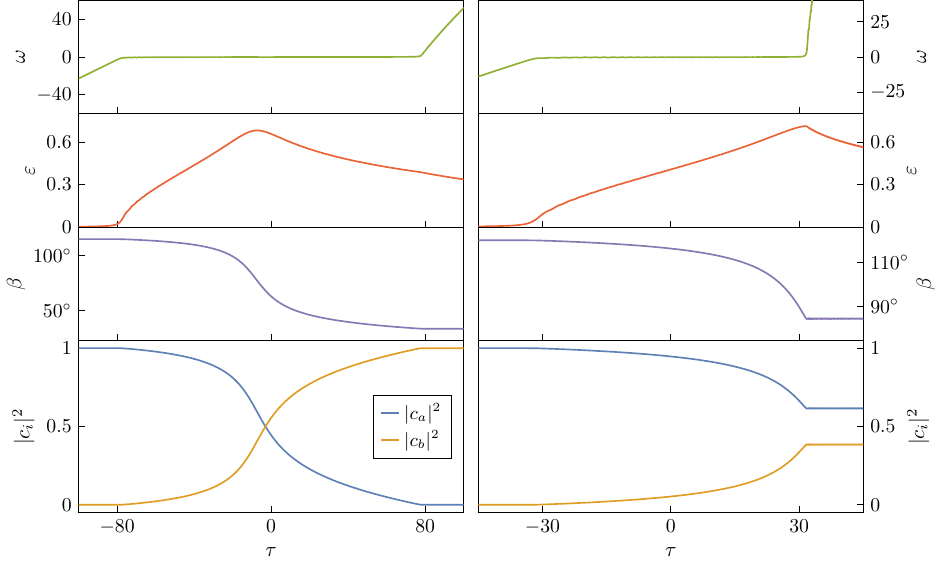}
\caption{Numerical solution of eqs.~\eqref{eqn:dimensionless-dressed-schrodinger},~\eqref{eqn:dimensionless-omega-evolution},~\eqref{eqn:dimensionless-eccentricity-evolution} and~\eqref{eqn:dimensionless-inclination-evolution} with parameters $Z=0.001$, $B=1000$, $D=4/3$ and $g=\Delta m$. For simplicity we ignore that in realistic cases $Z$ depends on the eccentricity, and we keep it constant instead. The system is initialised with eccentricity $\varepsilon_0=0$. A complete transition is achieved when the initial inclination is $\beta_0=\SI{115}{\degree}$ (\emph{left panel}), while a ``broken resonance'' is observed when $\beta_0=\SI{120}{\degree}$ (\emph{right panel}), with the float abruptly ending when~\eqref{eqn:epsilon-breaking} is satisfied. In both cases, the system follows the trajectories indicated in Figure~\ref{fig:streamplot_inclination-eccentricity} until the resonance ends or breaks.}
\label{fig:eccentric-inclined-backreacted-resonance}
\end{figure}
\subsubsection{Resonance breaking}\label{sec:resonance-breaking}
\begin{figure}[t!] 
\centering
\includegraphics[width=\linewidth]{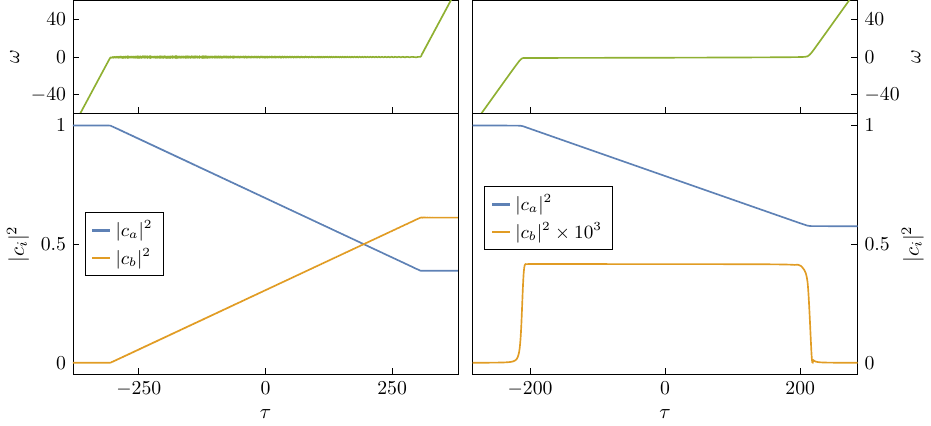}
\caption{Numerical solution of eqs.~\eqref{eqn:dimensionless-dressed-schrodinger} and~\eqref{eqn:dimensionless-omega-evolution} with (initial) parameters $Z=0.001$ and $B=1000$. A $Z$-breaking occurs when $Z$ is slowly reduced over time, with the resonance ending when~\eqref{eqn:Z-breaking} is satisfied (\emph{left panel}). A $\Gamma$-breaking is observed when $Z$ is kept fixed but state $\ket{b}$ is given a nonzero decay width $\Gamma=1.2$, with the resonance ending when~\eqref{eqn:gamma-breaking} is satisfied (\emph{right panel}).}
\label{fig:broken-resonance}
\end{figure}

When some parameters are allowed to vary with time, the floating orbit dynamics described in Section~\ref{sec:floating-adiabatic} and~\ref{sec:floating-evolution-e-beta} feature a new phenomenon, which we call \emph{resonance breaking}, and has been shown already in Figure~\ref{fig:eccentric-inclined-backreacted-resonance} (\emph{right panel}). The goal of this section is to determine analytically under which conditions a floating resonance breaks. Three different cases of parameter variation are encountered in realistic scenarios.
\begin{enumerate}
\item The binary eccentricity $\varepsilon$ changes with time, as seen in Figure~\ref{fig:eccentric-inclined-backreacted-resonance}. The eccentricity is the only binary parameter that appears explicitly in~\eqref{eqn:dimensionless-dressed-schrodinger} and~\eqref{eqn:dimensionless-omega-evolution}, while a change in $\beta$ only acts through a variation of $Z$.
\item As a consequence of changing $\varepsilon$ and $\beta$, the strength of the perturbation $\eta^\floq{g}$, and thus the Landau-Zener parameter $Z$, changes as well.
\item The total mass of the cloud changes with time if state $\ket{b}$ has $(\omega_{n\ell m})\ped{I}\ne0$: as a consequence, the Schr\"{o}dinger equation~\eqref{eqn:dimensionless-dressed-schrodinger} is modified to
\begin{equation}
\frac\dd{\dd\tau}\!\begin{pmatrix}c_a\\ c_b\end{pmatrix}=-i\begin{pmatrix}\omega/2 & \sqrt{Z}\\ \sqrt{Z} & -\omega/2-i\Gamma\end{pmatrix}\begin{pmatrix}c_a\\ c_b\end{pmatrix}\,,
\label{eqn:schrodinger-Gamma}
\end{equation}
where $\Gamma \equiv (\omega_{n\ell m})\ped{I}/\sqrt{\gamma\abs{g}}$, and care must be paid in the definition of $B$.
\end{enumerate}
All three effects come with two possible signs, one of which \emph{weakens} the resonance and potentially breaks it, while the other \emph{reinforces} it:~in the first category we have the increase of eccentricity, the decrease of $Z$ and the cloud decay ($\Gamma<0$).\footnote{The superradiant amplification of state $\ket{b}$, that is, $\Gamma>0$, is never encountered for floating resonances anyway.}
\vskip 2pt
To understand under what conditions a resonance breaks, it is insightful to study the evolution of $\omega$ during the float. To zeroth order, $\omega$ is identically zero, but Figures~\ref{fig:floating-backreacted-resonance},~\ref{fig:eccentric-backreacted-resonance}, and~\ref{fig:eccentric-inclined-backreacted-resonance} hint towards a nontrivial dynamics to higher order, with small oscillatory features of varying frequency. Let us try to find an equation of motion for the sole $\omega$, in the vanilla case with $\dd\varepsilon/\dd\tau=\dd Z/\dd\tau=\Gamma=0$, where no resonance break is expected. By taking the derivative of~\eqref{eqn:dimensionless-omega-evolution} and repeatedly using Schr\"{o}dinger's equation, we find
\begin{equation}
\frac{\dd^2\omega}{\dd\tau^2}=-B\frac{\dd\abs{c_b}^2}{\dd\tau^2}=-2ZB(1-2\abs{c_b}^2)+\sqrt ZB(c_a^*c_b+c_ac_b^*)\omega\,.
\label{eqn:d2omegadt2}
\end{equation}
Remarkably, the equation of motion obeyed by $\omega$ closely resembles a harmonic oscillator whose (squared) frequency is $-\sqrt ZB(c_a^*c_b+c_ac_b^*)$. It is thus natural to study this quantity:~by directly applying Schr\"{o}dinger's equation, we find
\begin{equation}
\sqrt Z\frac\dd{\dd\tau}(c_a^*c_b+c_ac_b^*)=\omega\frac{\dd\abs{c_b}^2}{\dd\tau}\,.
\label{eqn:dcacbcbcadt}
\end{equation}
We notice that eqs.~\eqref{eqn:d2omegadt2} and~\eqref{eqn:dcacbcbcadt} form a closed system of ordinary differential equations (because in the vanilla case $\abs{c_b}^2=(\tau-\omega)/B$), through which it is possible to prove mathematically a number of interesting properties of the system, such as the fact that at small $Z$ the evolution is entirely determined by $ZB$, as thoroughly described in Section~\ref{sec:floating-adiabatic}.
\vskip 2pt
For the scope of this section it is, however, sufficient to assume that the quantity $c_a^*c_b+c_ac_b^*$ evolves slowly during a float, with a timescale of $\Delta t\ped{float}$, similar to $\abs{c_b}^2$. Equation~\eqref{eqn:d2omegadt2} can then be solved in a WKB approximation as
\begin{equation}
\omega\approx\frac{2\sqrt Z(1-2\abs{c_b}^2)}{c_a^*c_b+c_ac_b^*}+\frac{A\,Z^{-1/8}B^{-1/4}}{(-c_a^*c_b-c_ac_b^*)^{1/4}}\cos\left(Z^{1/4}B^{1/2}\int_0^\tau\!\sqrt{-c_a^*c_b-c_ac_b^*}\dd\tau'+\delta\right)\,,
\label{eqn:omega-solution-oscillator}
\end{equation}
where $A$ is a constant and $\delta$ is a phase. As the fast oscillations average out, we can plug the first, non-oscillatory, term of~\eqref{eqn:omega-solution-oscillator} into~\eqref{eqn:dcacbcbcadt} and integrate to find $c_a^*c_b+c_ac_b^*\approx-\sqrt{1-(1-2\abs{c_b}^2)^2}$. The resulting solution for $\omega$,
\begin{equation}
\omega\approx-\frac{2\sqrt Z(1-2\abs{c_b}^2)}{\sqrt{1-(1-2\abs{c_b}^2)^2}}+\text{oscillatory terms}\,,
\end{equation}
is well-behaved for the entire duration of the float, only diverging before ($\abs{c_b}^2=0$) or after ($\abs{c_b}^2=1$) the resonance.
\vskip 2pt
The same analytical approach can be applied to the cases mentioned above, with varying $\varepsilon$ or $Z$, or $\Gamma\ne0$. A ``master equation'', where all three effects are turned on at the same time, is derived and shown in Appendix~\ref{appGA_sec:breaKING}. Here, we find it more illuminating to study them one at a time. The outcome in realistic cases may then be approximated by only retaining the strongest of the three effects.
\vskip 2pt
When the eccentricity is not a constant, the time derivative of~\eqref{eqn:dimensionless-omega-evolution} contains the additional term $\dd f(\varepsilon)/\dd\tau$. As a result, the equation of motion for $\omega$ and the expression of $c_a^*c_b+c_ac_b^*$ are both modified. The final result, which has been thoroughly checked against numerical solutions of the full system~\eqref{eqn:dimensionless-dressed-schrodinger}--\eqref{eqn:dimensionless-omega-evolution}--\eqref{eqn:dimensionless-eccentricity-evolution}--\eqref{eqn:dimensionless-inclination-evolution}, is
\begin{equation}
\omega\approx\frac{\frac{\dd f(\varepsilon)}{\dd\tau}-2ZB(1-2\abs{c_b}^2)}{\sqrt{ZB^2(1-(1-2\abs{c_b}^2)^2)-(f(\varepsilon)^2-f(\varepsilon_0)^2)}}+\text{oscillatory terms}\,.
\end{equation}
If $\varepsilon$ increases from its initial value $\varepsilon_0$, then the denominator can hit zero before the transition is complete, and the resonance breaks. The population remaining in state $\ket{a}$ and the binary eccentricity at resonance breaking satisfy
\begin{equation}
4ZB^2(\abs{c_a}^2-\abs{c_a}^4)=f(\varepsilon)^2-f(\varepsilon_0)^2\,,
\label{eqn:epsilon-breaking}
\end{equation}
which can be compared with the numerical solution in Figure~\ref{fig:eccentric-inclined-backreacted-resonance} (\emph{right panel}). Despite the simplicity of~\eqref{eqn:epsilon-breaking}, a numerical integration is still needed, in principle, to determine $\varepsilon$ as function of $\abs{c_a}^2$, and so whether a resonance will break. We can, however, make a simple conservative estimate by noting that the left-hand side can be at maximum $ZB^2$. If the system follows a trajectory in the $(\varepsilon,\beta)$-plane (cf.~Figure~\ref{fig:streamplot_inclination-eccentricity}) that significantly increases its eccentricity, such that
\begin{equation}
f(\varepsilon)>\sqrt ZB\,,
\label{eqn:epsilon-breaking-simple}
\end{equation}
the resonance must necessarily break.
\vskip 2pt
If, instead, $Z$ is allowed to vary while $\varepsilon$ is kept constant, then new terms appear when taking the time derivative of the Schr\"{o}dinger equation [used in the second equality of~\eqref{eqn:d2omegadt2}], and~\eqref{eqn:omega-solution-oscillator} becomes a damped harmonic oscillator. Similar to the previous case, the resonance breaks when $c_a^*c_b+c_ac_b^*=0$, which is equivalent to
\begin{equation}
4ZB^2(\abs{c_a}^2-\abs{c_a}^4)=f(\varepsilon)^2\left(1-\frac{Z}{Z_0}\right)\,.
\label{eqn:Z-breaking}
\end{equation}
We illustrate this phenomenon in Figure~\ref{fig:broken-resonance} (\emph{left panel}), by solving numerically~\eqref{eqn:dimensionless-dressed-schrodinger} and~\eqref{eqn:dimensionless-omega-evolution} while $Z$ slowly reduces over time. Analogous considerations as before can be applied to extract from~\eqref{eqn:Z-breaking} the approximate point of resonance breaking without performing a numerical integration.
\vskip 2pt
Taking into account a nonzero decay width $\Gamma$, while keeping $\varepsilon$ and $Z$ constant, requires more care. Because $\abs{c_a}^2+\abs{c_b}^2$ is no longer a constant, eq.~\eqref{eqn:dimensionless-omega-evolution} is now written as
\begin{equation}
\frac{\dd\omega}{\dd\tau}=f(\varepsilon)-\frac{B}{\Delta\epsilon}\left(\epsilon_a\frac{\dd\abs{c_a}^2}{\dd\tau}+\epsilon_b\frac{\dd\abs{c_b}^2}{\dd\tau}+2\Gamma\epsilon_b\abs{c_b}^2\right)=f(\varepsilon)+B\frac{\dd\abs{c_a}^2}{\dd\tau}\,,
\label{eqn:dimensionless-omega-evolution-Gamma}
\end{equation}
where the constant parameter $B$ is computed according to~\eqref{eqn:BC}, using the value of the mass of the cloud before the start of the resonance. Furthermore, due to the modified Schr\"{o}dinger equation, formula~\eqref{eqn:dcacbcbcadt} becomes
\begin{equation}
\sqrt Z\frac\dd{\dd\tau}(c_a^*c_b+c_ac_b^*)=-\omega\frac{\dd\abs{c_a}^2}{\dd\tau}-\Gamma\sqrt Z(c_a^*c_b+c_ac_b^*)\,.
\label{eqn:dcacbcbcadt_Gamma}
\end{equation}
As we will show later (cf.~Figure~\ref{fig:tfloat_tdecay}), in almost all realistic cases state $\ket{b}$ decays much faster than the duration of the resonance, i.e., $\tau\ped{decay}\equiv(2\Gamma)^{-1}\ll B$. As a consequence, its population $\abs{c_b}^2$ during a floating orbit stays approximately constant, at a value $\abs{c_b}^2=f(\varepsilon)/(2\Gamma B)$, where the state decay is balanced by the transitions from $\ket{a}$ to $\ket{b}$. As this saturation value is typically very small, we will neglect it. Under this assumption, we can solve~\eqref{eqn:dcacbcbcadt_Gamma} as
\begin{equation}
c_a^*c_b+c_ac_b^*\approx\sqrt{\frac{f(\varepsilon)(2ZB\abs{c_a}^2-f(\varepsilon)\Gamma)}{\Gamma ZB^2}}\,,
\end{equation}
and conclude that the resonance breaks when the remaining population in the initial state is
\begin{equation}
\abs{c_a}^2\approx\frac {f(\varepsilon)\Gamma}{2ZB}\,.
\label{eqn:gamma-breaking}
\end{equation}
This result is confirmed by a numerical solution of~\eqref{eqn:dimensionless-dressed-schrodinger} and~\eqref{eqn:dimensionless-omega-evolution} with nonzero $\Gamma$, as shown in Figure~\ref{fig:broken-resonance} (\emph{right panel}). Resonances where this quantity is larger than 1 do not exhibit a floating orbit at all, showing an ``immediate'' breaking.
\vskip 2pt
We refer to the three types of resonance breaking as $\varepsilon$-breaking, $Z$-breaking and $\Gamma$-breaking. A summary of the respective conditions is given below.
\begin{table}[h!]
\centering
\renewcommand*{\arraystretch}{1.1}
\begin{tabular}{c|c|c} 
$\varepsilon$-breaking & $Z$-breaking & $\Gamma$-breaking \\ 
\hline
$f(\varepsilon)\gtrsim\sqrt ZB$ & $Z/Z_0\lesssim1-ZB^2/f(\varepsilon)^2$ & $\abs{c_a}^2\lesssim  f(\varepsilon)\Gamma/(2ZB)$ \\ 
\end{tabular}
\end{table}
\subsection{Sinking Orbits}\label{sec:sinking}
Let us now turn our attention to sinking orbits, corresponding to $B<0$, where backreaction tends to make the resonance less adiabatic. This case turns out to not be as dramatically relevant as floating orbits for the resonant history of the system. However, it is important for direct GW signatures. For this reason, we will only study the aspects of it with observational consequences.
\vskip 2pt
All the observable sinking resonances have $2\pi Z\ll1$. In this case, the final population in state $\ket{b}$, as predicted by~\eqref{eqn:lz}, is very small, and this quantity is further reduced by the backreaction. In the regime where this correction is dominant, we can find a rough approximation for the total population transferred by only keeping the backreaction term in~\eqref{eqn:dimensionless-omega-evolution}. Further assuming $\abs{c_a}^2\approx1$ and $\dot c_b\approx0$, we can substitute in the second component of~\eqref{eqn:dimensionless-dressed-schrodinger} and obtain $\abs{c_b}^2\approx(Z/B^2)^{1/3}$, where we assumed, for simplicity, quasi-circular orbits.\footnote{The validity of the assumption will become clear in Section~\ref{sec_Legacy:history}.} This result is confirmed by numerical tests, modulo a multiplicative factor:
\begin{equation}
B\ll -\frac1Z\,,\qquad\abs{c_b}^2\approx3.7\,\left(\frac{Z}{B^2}\right)^{1/3}\,.
\label{eqn:sinkingpopulation}
\end{equation}
This formula is accurate for $2\pi Z\ll1$ and provides a slight underestimate of the final population for moderately large $Z$.
\vskip 2pt
Sinking orbits backreact on the orbit by increasing both the orbital frequency and the binary eccentricity, as shown in Figure~\ref{fig:eccentric-backreacted-resonance} and~\ref{fig:omega-eccentricity} (\emph{right panels}). At the same time, both $\Omega$ and $\varepsilon$ feature long-lived oscillations after the resonance. These oscillations slowly die out, so that a ``jump'' in the $\Omega$ and $\varepsilon$ is the only mark left after a long time. The non-monotonic behaviour of $\Omega$ was already observed in~\cite{Baumann:2019ztm}, where it was also speculated that sinking orbits could yield large eccentricities (becoming ``kicked orbits''). Our results confirm that the oscillations are not an artefact of having considered quasi-circular orbits and further show that the increase of the eccentricity is also not monotonic. However, for the realistic cases analysed in Section~\ref{sec_Legacy:history}, the increase in eccentricity due to sinking orbits turns out to be negligible.
\section{Three Types of Resonances}\label{sec_Legacy:types-of-resonances}
Resonances can be divided in three distinct categories, depending on the energy splitting between the two states, as computed from~\eqref{eq:BHenv_eigenenergy} and illustrated in Figure~\ref{fig:spectrum}. \emph{Hyperfine} resonances occur between states with same $n$ and $\ell$ but different $m$;~they have the smallest energy splitting and thus occur the earliest in the inspiral, as the corresponding resonant orbital frequency is smallest~\eqref{eq:resonance_con}. Then, \emph{fine} (same $n$, different $\ell$) and \emph{Bohr} resonances (different $n$) follow, the latter having the largest splittings. The tools developed in Section~\ref{sec_Legacy:resonance-pheno} apply to all of them:~the character of a resonance is only determined by the parameters $2Z$ and $B$;~its impact on eccentricity and inclination is quantified by $D$, and its duration (in case it is a floating resonance) is $\Delta t\ped{float}$. In principle, the recipe to determine the co-evolution of the binary and the cloud is clear:~(i) pick the earliest resonance, (ii) determine its character and backreaction by computing $Z$, $B$, $D$, and $\Delta t\ped{float}$, (iii) update the state of binary and cloud accordingly, and (iv) move to the next resonance and repeat. We will indeed execute this algorithm in Section~\ref{sec_Legacy:history}. To be as generic as possible and explore a wide parameter space, it will prove useful to find the scalings of the relevant quantities with $M$, $M\ped{c}$, $q$, $\alpha$, and $\tilde a$. Different types of resonances have different scalings, so we analyse them here systematically.
\subsection{Hyperfine Resonances}\label{sec:hyperfine}
Let us start with hyperfine resonances. From~\eqref{eq:BHenv_eigenenergy}, we see that the energy splitting (and thus the resonant frequency) scales as $\Omega_0\propto M^{-1}\alpha^6\tilde a$. The corresponding orbital separation is $R_0\propto M\alpha^{-4}\tilde a^{-2/3}$. This strong $\alpha$-dependence places hyperfine resonances at distances parametrically much larger than the cloud's size. At such large orbital separations, the cloud's ionisation is very inefficient, and thus the only significant mechanism for energy loss is the GW emission. As this too is a very weak effect, other phenomena might potentially be relevant, including astrophysical interactions connected to the binary formation mechanism. We will postpone the discussion of these complications to Section~\ref{sec_Legacy:history} and assume for now that formula~\eqref{eqn:gamma_gws} applies, giving a chirp rate of $\gamma\propto qM^{-2}\alpha^{22}\tilde a^{11/3}$. This information is already enough to determine the scaling of three key quantities:
\begin{equation}
B\propto\frac{M\ped{c}}Mq^{-3/2}\alpha^{-4}\tilde a^{-1/2}\,,\qquad \Delta t\ped{float}\propto M\ped{c}q^{-2}\alpha^{-15}\tilde a^{-7/3}\,,\qquad D\propto \frac{M\ped{c}}Mq^{-1}\alpha\tilde a^{1/3}\,.
\label{eqn:hyperfine-B-tfloat-exponent}
\end{equation}
\vskip 2pt
The scaling of the Landau-Zener parameter $Z$ depends instead on the overlap coefficient $\eta^\floq{g}$. Given the hierarchy of length scales, $R_0\gg r\ped{c}$, the ``inner'' term in~\eqref{eqn:F(r)} dominates the radial integral $I_r$. At fixed $\ell_*\ne1$, we thus have
\begin{equation}
\begin{split}
\eta^\floq{g}&\propto q\alpha\,d^\floq{\ell_*}_{\Delta m,g}(\beta)\, I_r=q\alpha\,d^\floq{\ell_*}_{\Delta m,g}(\beta)\int_0^\infty \frac{r^{\ell_*}}{R_0^{\ell_*+1}}R_{n\ell}(r)^2r^2\dd r\\
&\propto M^{-1}q\alpha^{2\ell_*+5}\tilde a^{2(\ell_*+1)/3}d^\floq{\ell_*}_{\Delta m,g}(\beta)\,,
\end{split}
\end{equation}
and so
\begin{equation}
Z\propto q\alpha^{4\ell_*-12}\tilde a^{(4\ell_*-7)/3}\left(d^\floq{\ell_*}_{\Delta m,g}(\beta)\right)^2\,.
\end{equation}
The dipole $\ell_*=1$ is an exception for two reasons: (a) its inner term in~\eqref{eqn:F(r)} vanishes, (b) its ``outer'' term is not simply $r^{\ell_*}/R_*^{\ell_*+1}$. However, hyperfine resonances connect states with same $\ell$:~from the selection rule~\eqref{eqn:S2}, only even values of $\ell_*$ contribute. We can thus safely ignore the dipole. The rest of the multipole expansion can be seen as a power series in the small parameter $r\ped{c}/R_0$, the smallest $\ell_*$ giving the strongest contribution. Because selection rules require $\ell_*\ge \abs{g}=-g$,\footnote{Strictly speaking, this constraint only applies on circular orbits. In general, the same inequality applies to $g_\beta$ instead.} a resonance with a given value of $g$ will be dominated by $\ell_*=-g$. The only two cases we will encounter in Section~\ref{sec_Legacy:history} are
\begin{align}
\label{eqn:hyperfine-g=-2-Z}
g=-2&\qquad Z\propto q\alpha^{-4}\tilde a^{1/3}\left(d^\floq{2}_{\Delta m,g}(\beta)\right)^2\,,\\
\label{eqn:hyperfine-g=-4-Z}
g=-4&\qquad Z\propto q\alpha^4\tilde a^{3}\left(d^\floq{4}_{\Delta m,g}(\beta)\right)^2\,.
\end{align}
Furthermore, the assumption $\ell_*=-g$ allows us to write the explicit expression for the angular dependence of $Z$ as
\begin{equation}
d^\floq{-g}_{\Delta m,g}(\beta)\propto\sin^{\Delta m-g}(\beta/2)\cos^{-\Delta m-g}(\beta/2)\,.
\label{eqn:dgg}
\end{equation}
\subsection{Fine Resonances}
Most of the assumptions made for hyperfine resonances work in the fine case too. The resonant frequency now scales as $\Omega_0\propto M^{-1}\alpha^5$ and, similar to before, we arrive to
\begin{equation}
B\propto\frac{M\ped{c}}Mq^{-3/2}\alpha^{-7/2}\,,\qquad \Delta t\ped{float}\propto M\ped{c}q^{-2}\alpha^{-38/3}\,,\qquad D\propto \frac{M\ped{c}}Mq^{-1}\alpha^{2/3}\,.
\end{equation}
The scaling of the overlap coefficient reads $\eta^\floq{g}\propto qM^{-1}\alpha^{(4\ell_*+13)/3}$, and we get
\begin{equation}
Z\propto q\alpha^{(8\ell_*-29)/3}\left(d^\floq{\ell_*}_{\Delta m,g}(\beta)\right)^2\,.
\end{equation}
The main difference with the previous case resides in the possible values of $\ell_*$. Fine resonances connect states with different values of $\ell$, and most of the cases we will study in Section~\ref{sec_Legacy:history} will have odd values of $\ell_*$. For $g=-3$, all the previous arguments apply and the octupole $\ell_*=3$ is the dominant contribution. For $g=-1$, the extreme weakness of the dipole at large distances again leaves the octupole as the most important term; because now $\ell_*\ne-g$, however, the angular dependence will have a form different from~\eqref{eqn:dgg}, which we will describe on a case-by-case basis in Section~\ref{sec_Legacy:history}. There is one further exception to this:~if $\ell_a+\ell_b=1$, then the selection rule~\eqref{eqn:S3} forbids all $\ell_*\ge2$. Only in this case (corresponding to the $\ket{211}\to\ket{200}$ resonance) the dipole is entirely responsible for the coupling between the two states. Its anomalous expression~\eqref{eqn:F(r)} endows $\eta^\floq{g}$ (and thus $Z$) with a non-power-law dependence on $\alpha$:~given the peculiarity of this case, we will treat it explicitly in Appendix~\ref{appGA_sec:211_200}.
\subsection{Bohr Resonances}\label{sec:Bohr_res}
\begin{figure}[t!] 
\centering
\includegraphics[width=\linewidth]{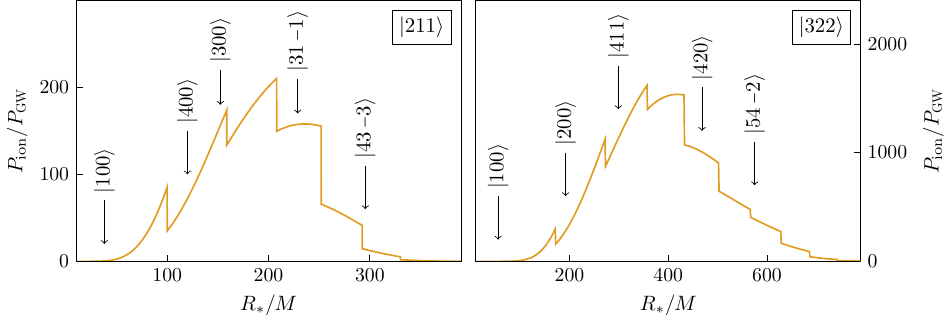}
\caption{Position of a few selected Bohr resonances, compared to $P\ped{ion}/P_\slab{gw}$, i.e., the ratio of the ionisation power to the power emitted in GWs, shown here for a circular counter-rotating orbit and for a cloud in the $\ket{211}$ (\emph{left panel}) or $\ket{322}$ (\emph{right panel}) state. We assumed $M\ped{c}/M=0.01$ and $\alpha=0.2$, but the relative position of the resonances and the shape of the curve do not depend on the parameters.}
\label{fig:ionisation-bohr-resonances}
\end{figure}

Bohr resonances are a different story. States with different principal quantum number $n$ have different energies to leading order, meaning that the resonant orbits are placed at distances comparable to the cloud's size. There is no parametric separation between the two, as now $R_0\propto M\alpha^{-2}\propto r\ped{c}$. At these orbital distances, the cloud's ionisation is generally a more effective mechanism for energy loss than GWs. We prove this point in Figure~\ref{fig:ionisation-bohr-resonances}, where the position of several Bohr resonances is shown on top of the ionisation-to-GWs power ratio, computed as in~\cite{Baumann:2021fkf,Baumann:2022pkl,Tomaselli:2023ysb} on circular orbits. This latter quantity scales as
\begin{equation}\label{eqn:P_ion}
\frac{P\ped{ion}}{P_\slab{gw}}\bigg|_{R_*=R_0}\propto\frac{M\ped{c}}M\alpha^{-5}\,.
\end{equation}
With the possible exception of transitions to $\ket{100}$, as they happen extremely late in the inspiral, Bohr resonances and ionisation thus happen \emph{at the same time}. This observation raises two points.
\begin{enumerate}
\item Formula~\eqref{eqn:gamma_gws} for the chirp rate $\gamma$ is no longer accurate, as ionisation must now be included.
\item The derivation of the expression for $P\ped{ion}$ laid down in~\cite{Baumann:2021fkf} assumes that the system is away from bound-to-bound state resonances.
\end{enumerate}
In Appendix~\ref{appGA_sec:ion-at-resonance} we extend the framework of~\cite{Baumann:2021fkf} to describe the ionisation of a system actively in resonance. Although this requires the addition of new terms, their effect is generally negligible for realistic parameters. It is thus a good approximation to simply adjust the value of $\gamma$ by a factor $1+P\ped{ion}/P_\slab{gw}\approx P\ped{ion}/P_\slab{gw}$, where $P\ped{ion}$ is computed as in~\cite{Baumann:2021fkf,Baumann:2022pkl,Tomaselli:2023ysb}. The last approximation holds whenever $P\ped{ion}\gg P_\slab{gw}$ and is always satisfied, unless the resonance involves $\ket{100}$ or the value of $\alpha$ is exceptionally large.
\vskip 2pt
Under these assumptions, we arrive to
\begin{equation}\label{eqn:B_bohr}
B\propto\sqrt{\frac{M\ped{c}}M}q^{-3/2}\,,\qquad \Delta t\ped{float}\propto Mq^{-2}\alpha^{-3}\,,\qquad D\propto \frac{M\ped{c}}Mq^{-1}\,.
\end{equation}
These quantities now also have a $\beta$-dependence, due to $P\ped{ion}$ having different values for different inclinations. However, we will see in Section~\ref{sec_Legacy:history} that this detail is not relevant, so we neglect it here. As for the overlap $\eta^\floq{g}$, there is now no clear hierarchy of multipoles. Luckily, $R_0$ has the same $\alpha$-scaling as the argument of the hydrogenic wavefunctions $R_{n\ell}$:~with an appropriate change of variable, we can show that
\begin{equation}
\eta^\floq{g}\propto M^{-1}q\alpha^3d^\floq{\ell_*}_{\Delta m,g}(\beta)\,.
\label{eqn:eta-bohr}
\end{equation}
The $\beta$-dependence in~\eqref{eqn:eta-bohr} can be written in terms of a Wigner small $d$-matrix only when there is a single value of $\ell_*$ that contributes. As this is the case for many of the Bohr resonances we will encounter in Section~\ref{sec_Legacy:history}, we keep that factor explicit here. Finally, the Landau-Zener parameter scales as
\begin{equation}\label{eqn:Z_bohr}
Z\propto \frac{M}{M\ped{c}}q\left(d^\floq{\ell_*}_{\Delta m,g}(\beta)\right)^2\,.
\end{equation}
One particularly interesting aspect of Bohr resonances is the disappearance of any $\alpha$-dependence from the Landau-Zener parameter $Z$ and from the backreaction $B$. This is in contrast with the steep power-laws found for hyperfine and fine resonances, and it means that the character of Bohr resonances is much more \emph{universal}.
\section{Resonant History of the Cloud}\label{sec_Legacy:history}
In this section we draw a consistent picture of the co-evolution of the cloud and the binary, using the tools developed in Sections~\ref{sec_Legacy:resonance-pheno} and~\ref{sec_Legacy:types-of-resonances}. Assuming a well-motivated initial state of the cloud (generally $\ket{211}$ or $\ket{322}$), an astrophysically relevant range for $\alpha$ [see eqs.~\eqref{eqn:alpha-211} and~\eqref{eqn:alpha-322}], and small $q$, the plethora of phenomena described in Section~\ref{sec_Legacy:resonance-pheno} only occur in recognisable and relatively simple patterns. These constitute the ``realistic'' cases, which we systematically explore in this section, with the goal of understanding the state of the system by the time it becomes observable:~for example, when it enters the LISA band. First, we discuss the generic behaviour of the different types of resonances in Section~\ref{sec:generalB}; then, in Section~\ref{sec:evolution-211} and~\ref{sec:evolution-322}, we study explicitly the history for a cloud initialised in the state $\ket{211}$ or $\ket{322}$.
\subsection{General behaviour}\label{sec:generalB}
The initial state $\ket{a}=\ket{n_a\ell_a m_a}$ of the cloud, populated by superradiance, generally has $m_a=\ell_a=n_a-1$. Within the multiplet of states $\ket{n_a\ell_am}$ with $m\le m_a$, this is the one with highest energy, as can be readily seen from~\eqref{eq:BHenv_eigenenergy}. Hyperfine resonances, which occur the earliest in the inspiral, thus necessarily have $\Delta\epsilon<0$ and are of the floating type. To understand their behaviour, it is important to keep in mind a few key points.
\begin{description}
\item[Adiabaticity.] The first question to answer is whether a given hyperfine resonance is adiabatic or not. We can apply the results of Section~\ref{sec:floating-adiabatic}. If $2\pi ZB>f(\varepsilon)^{3/2}$ then the resonance is adiabatic:~the binary starts evolving as described in Section~\ref{sec:floating-evolution-e-beta} until the transition completes after a time $\Delta t\ped{float}$, or the resonance breaks due to any of the conditions derived in Section~\ref{sec:resonance-breaking}. Almost all hyperfine resonances turn out to be adiabatic in the entire parameter space, except in a narrow interval of almost counter-rotating inclinations, say $\pi-\delta_1<\beta\le\pi$, where $\delta_1$ is the size of the interval. This is because, on floating orbits, the resonance condition $\Omega_0^\floq{g}=\Delta\epsilon/g$ forces $g$ to be negative; on the other hand, $\Delta m=m-m_a<0$, and from~\eqref{eqn:dgg} we see that for $\beta\to\pi$ the parameter $Z$ goes to zero as a (high) power of $\cos(\beta/2)$. The explicit determination of the angle $\delta_1$ as function of the parameters will be performed in Sections~\ref{sec:evolution-211} and~\ref{sec:evolution-322}.
\item[Cloud's decay and $\Gamma$-breaking.] After saturation of the dominant superradiant mode $\ket{n_a\ell_am_a}$, all states of the multiplet $\ket{n_a\ell_am}$ with $m\ne m_a$ have $\Im(\omega)<0$, meaning that they decay back in the BH with an $e$-folding time $t\ped{decay}\equiv\abs{2\Im(\omega_{n_a\ell_am})}^{-1}$. It is thus necessary to compare $\Delta t\ped{float}$ and $t\ped{decay}$. One of the most important results of this work is the following:~for intermediate or extreme mass ratios, and typical values of $M\ped{c}$ and $\alpha$, the decay timescale $t\ped{decay}$ is \emph{many orders of magnitude smaller} than floating timescale, $\Delta t\ped{float}$. It is not easy to prove this statement in full generality, due to the complicated dependence of $t\ped{decay}$ on the parameters. Nevertheless, for small $\alpha$ and $\tilde a$, using the Detweiler approximation~\eqref{eqn:Gamma_nlm} and the results from Section~\ref{sec:hyperfine}, we have
\begin{equation}
\frac{t\ped{decay}}{\Delta t\ped{float}}\propto\frac{M}{M\ped{c}}q^2\alpha^{10-4\ell_a}\tilde a^{4/3}\,,
\label{eqn:t-decay-t-float}
\end{equation}
where $\tilde a\propto\alpha$ at the superradiant threshold. For $\alpha\to0$ and small enough values of $\ell_a$, this ratio becomes very small. In fact, for small $q$, any possible value of $\alpha$ results in $t\ped{decay}\ll\Delta t\ped{float}$. A more detailed comparison is given in Figure~\ref{fig:tfloat_tdecay}, where $t\ped{decay}$ is computed numerically through Leaver's~\cite{Leaver:1985ax,Cardoso:2005vk,Dolan:2007mj,Berti:2009kk} (see also Appendix~\ref{appNR_subsec:boundstates}) and Chebyshev's~\cite{Baumann:2019eav} methods for various values of $\alpha$, and the spin $\tilde a$ is set to correspond to the boundary of the BH superradiant region.
This result has a dramatic consequence:~hyperfine transitions are never able to change the state of the cloud. Instead, the portion that is transferred to state $\ket{b}$ decays immediately back into the BH.\footnote{\label{fn:bh-parameters}As a consequence, the mass and spin of the BH change. Our framework is not able to capture this effect, which we accordingly ignore in this chapter.} The analysis of Section~\ref{sec:resonance-breaking} then applies, and the resonance $\Gamma$-breaks when the fraction of the cloud remaining in state $\ket{a}$ falls below the threshold determined in~\eqref{eqn:gamma-breaking}. In a relatively large portion of parameter space, generally around counter-rotating orbits, that formula returns $\abs{c_a}^2>1$, meaning that the resonance $\Gamma$-breaks immediately. The outcome is effectively similar to a non-adiabatic resonance, that never even starts the floating phase. Similar to before, we will define an angular interval $\pi-\chi_1<\beta\le\pi$, within which the resonance is not effective. The $\varepsilon$-breaking and $Z$-breaking are instead less relevant for realistic parameters.
\begin{figure}[t!] 
\centering
\includegraphics[width=\linewidth]{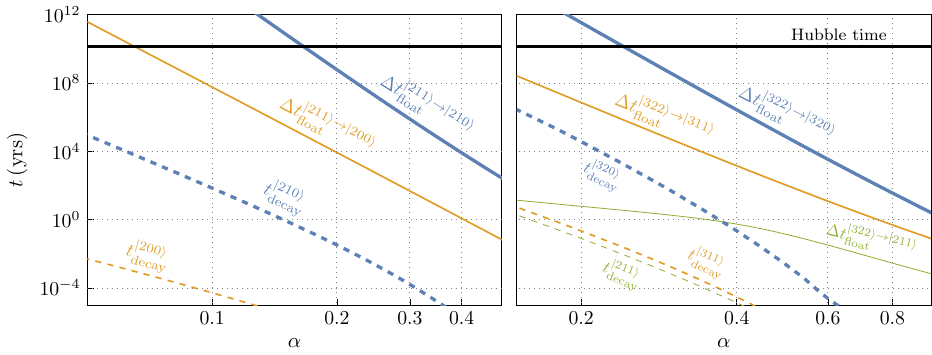}
\caption{Floating timescale $\Delta t\ped{float}$ (solid lines), compared to the decay timescale $t\ped{decay}$ (dashed lines) of the final state, for some selected resonances. We use benchmark parameters and determine the decay rate independently through Leaver's continued fraction method~\cite{Leaver:1985ax,Cardoso:2005vk,Dolan:2007mj,Berti:2009kk} and the Chebyshev method in~\cite{Baumann:2019eav}. Two resonances for a $\ket{211}$ initial state are shown (\emph{left panel}), namely $\ket{211}\to\ket{210}$ [{\color{Mathematica1}blue}] and $\ket{211}\to\ket{200}$ [{\color{Mathematica2}orange}]. Similarly, three resonances for a $\ket{322}$ initial state are shown (\emph{right panel}), namely $\ket{322}\to\ket{320}$ [{\color{Mathematica1}blue}], $\ket{322}\to\ket{311}$ [{\color{Mathematica2}orange}] and $\ket{322}\to\ket{211}$ [{\color{Mathematica3}green}]. The thick, normal and thin lines indicate hyperfine, fine or Bohr resonances respectively. Note the Bohr resonance falling outside of the ionisation regime for large $\alpha$, changing the scaling of $\Delta t\ped{float}$ from $\alpha^{-3}$ to $\alpha^{-8}$, as predicted by~\eqref{eqn:B_bohr} and~\eqref{eqn:P_ion}.}
\label{fig:tfloat_tdecay}
\end{figure}
\item[The strongest resonance.] As shown in Section~\ref{sec:eccentric-inclined-resonances}, on eccentric and inclined orbits a resonance between two given states is excited at many different orbital frequencies, depending on the value\footnote{As briefly mentioned in Section~\ref{sec:eccentric-inclined-resonances}, two separate indices, say $g_\varepsilon$ and $g_\beta$ are necessary when both eccentricity and inclination are not zero. However, this technicality is not crucial in understanding the history of the system.} of $\abs{g}=1,2,3,\ldots$ The strength of the coupling also depends on $\varepsilon$ and $\beta$. Keeping track of so many different resonances would be very complicated. However, the hierarchy $t\ped{decay}\ll\Delta t\ped{float}$ implies that as soon as an adiabatic floating resonance is encountered (and does not break early), the cloud is destroyed. This means that studying the ``strongest'' resonance (the one that destroys the cloud in the largest portion of parameter space) actually suffices to determine the fate of the cloud.
Up to moderate values of the eccentricity, the coupling $\eta^\floq{g}$ that remains nonzero in the limit of circular orbit is much larger than all the others. We can then approximate the ``strongest resonance'' by ignoring eccentricity altogether. Regarding inclined orbits instead, we observe that higher values of $g$ require contributions from higher values of the multipole index $\ell_*$:~at the separations of hyperfine resonances, the lowest value of $\ell_*$ (typically the quadrupole $\ell_*=2$) produces the strongest coupling. Given two states, we will then study the resonance with the smallest value of $\abs{g}$.
\end{description}
Applying the previous considerations to each possible hyperfine resonance, we are able to determine whether the cloud is destroyed in the process or survives to later stages of the inspiral. However, the binary might be able to ``skip'' hyperfine resonances for other reasons. This is because some of them are placed at extremely large binary separations:~typically $R_*/M\gtrsim\mathcal O(10^3)$ for a $\ket{211}$ initial state, and $R_*/M\gtrsim\mathcal O(10^4-10^5)$ for $\ket{322}$. These distances are large enough that not only other kinds of astrophysical interactions may play a role, but their presence is in some cases necessary, in order to bring the binary close enough for the merger to happen within a Hubble time. Rewriting eq.~\eqref{eq:t_inspirals}, the initial separation for a quasi-circular inspiral as function of the time-to-merger $t_0$ is given by
\begin{equation}
\frac{R_*}M=\SI{2.3e+4}{}\,\left(\frac{t_0}{\SI{e+10}{yrs}}\right)^{1/4}\left(\frac{10^4M_\odot}{M}\right)^{1/4}\left(\frac{q}{10^{-3}}\right)^{1/4}\,.
\end{equation}
In other words:~if we want the binary to merge within a Hubble time, we might be forced to assume that it ``starts'' its evolution too close for hyperfine resonances to be encountered, especially for a cloud initialised in the $\ket{322}$ state. This can be achieved by a variety of formation mechanisms, including dynamical capture~\cite{Amaro-Seoane:2012lgq,LISA:2022yao} and \emph{in-situ} formation~\cite{LISA:2022yao,Stone:2016wzz,Bartos:2016dgn,Mckernan:2017ssq,Levin:2006uc}.
\vskip 2pt
If the system is able to skip through hyperfine resonances because they are either all non-adiabatic, or they $\Gamma$-break early, or the binary is formed at small enough separations, then the cloud can be present when fine resonances are encountered. Their phenomenology is largely similar to hyperfine ones, as they too are all of the floating type. We defer the discussion of some state-dependent aspects to Sections~\ref{sec:evolution-211} and~\ref{sec:evolution-322}. For the purpose of the present general discussion, it suffices to say that, once again, the cloud can survive this stage if $\pi-\delta_2<\beta\le\pi$ (for some angle $\delta_2$ to be determined), if the resonance $\Gamma$-breaks early in an interval $\pi-\chi_2<\beta\le\pi$, or if the binary is formed $\emph{in situ}$ at very small radii.
\vskip 2pt
Finally, if the cloud makes it to this point, it becomes potentially observable:~the ``Bohr region'' can be in the LISA band and is rich of signatures of the cloud. These come in the form of ionisation and Bohr resonances, the vast majority of which are sinking and non-adiabatic. State-dependent details will be discussed in Sections~\ref{sec:evolution-211} and~\ref{sec:evolution-322} and a summary of the observational signatures will be given in Section~\ref{sec_Legacy:observational-signatures}. A diagrammatic representation of the three stages of the resonant history is shown in Figure~\ref{fig:history_211}.
\vskip 2pt
As a concluding remark, we note that the results derived here and in Section~\ref{sec_Legacy:resonance-pheno} are specific to resonances involving two states only. We have explicitly checked that this is the case for the resonances discussed in the next sections, so we apply the results of Section~\ref{sec_Legacy:resonance-pheno} without further modification.
\subsection{Evolution from a $\ket{211}$ Initial State}\label{sec:evolution-211}
The $\ket{211}$ state is the fastest-growing superradiant mode and represents therefore a natural assumption for the initial state of the cloud. As we saw in eq.~\eqref{eqn:211bounds}, the requirements that the superradiant amplification takes place, and does so on timescales no longer than a Gyr, set a constraint on $\alpha$:
\begin{equation}
\label{eqn:alpha-211}
0.02\left(\frac{M}{10^4M_\odot}\right)^{1/9}\lesssim\alpha<0.5\,.
\end{equation}
Once grown, the cloud will decay in GWs with a rate roughly proportional to $M\ped{c}^2\alpha^{14}$, assuming the scalar field is real [see eqs.~\eqref{eqn:GWemission}-\eqref{eqn:GWlifetime}]. The resulting decay of $M\ped{c}$ is polynomial, rather than exponential in time~\eqref{eqn:GWemission}; as such, we will not impose a further sharp bound on $\alpha$, and treat $M\ped{c}/M$ as an additional free parameter.
\vskip 2pt
There are two possible hyperfine resonances, with the states $\ket{210}$ and $\ket{21\,\minus1}$. Following the line of reasoning laid down in Section~\ref{sec:generalB}, we ignore the fact that the same resonances can be triggered at multiple points if the orbit is eccentric. Both resonances are then mediated by $g=-2$ and they are positioned at
\begin{align}
\ket{211}\overset{g=-2}{\longrightarrow}\ket{210} & \qquad \frac{R_0}M=\SI{8.3e+3}{}\,\left(\frac{0.2}{\alpha}\right)^{4}\left(\frac{0.5}{\tilde a}\right)^{2/3}\,,\\[8pt]
\ket{211}\overset{g=-2}{\longrightarrow}\ket{21\,\minus1} & \qquad \frac{R_0}M=\SI{5.2e+3}{}\,\left(\frac{0.2}{\alpha}\right)^{4}\left(\frac{0.5}{\tilde a}\right)^{2/3}\,,
\end{align}
where the value of the spin should be set equal to the threshold of superradiant instability of $\ket{211}$~\eqref{eq:spinatsat}. Both resonances become non-adiabatic in an interval $\pi-\delta_1<\beta\le\pi$, with the strongest constraint on $\delta_1$ given by $\ket{211}\to\ket{210}$. The value of $\delta_1$ is determined from~\eqref{eqn:2piZB-epsilon0}:~this means setting $2\pi ZB=f(\varepsilon_0)^{3/2}$, where $\varepsilon_0$ is the eccentricity at the onset of the resonance, and solving for $\beta$ as function of the parameters. Making use of the relations~\eqref{eqn:hyperfine-B-tfloat-exponent},~\eqref{eqn:hyperfine-g=-2-Z} and~\eqref{eqn:dgg}, and evaluating numerically the overlap $\eta^\floq{2}$ between the two states, we find
\begin{equation}
\delta_1=\SI{7.5}{\degree}\,\left(\frac{10^{-2}}{M\ped{c}/M}\right)^{1/6}\left(\frac{q}{10^{-3}}\right)^{1/12} \left(\frac{\alpha}{0.2}\right)^{4/3} \left(\frac{\tilde{a}}{0.5}\right)^{1/36}f(\varepsilon_0)^{1/4}\,.
\label{eqn:211-delta1}
\end{equation}
Although $\ket{211}\to\ket{210}$ is also non-adiabatic in a neighbourhood of $\beta=0$, such a co-rotating binary would still encounter the adiabatic floating resonance $\ket{211}\to\ket{21\,\minus1}$ later, so that the only ``safe'' inclinations are in the neighbourhood of counter-rotating determined in~\eqref{eqn:211-delta1}.
\begin{figure}[t!] 
\centering
\includegraphics[width=\linewidth]{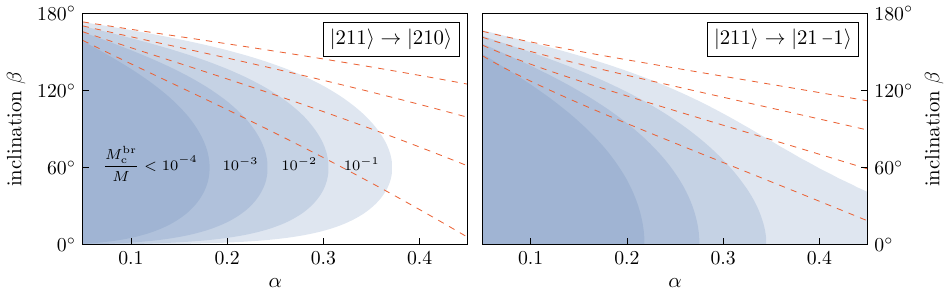}
\caption{Mass of the cloud $M\ped{c}\ap{br}$ at resonance $\Gamma$-breaking, as a function of $\alpha$ and $\beta$, for the two hyperfine resonances from the initial state $\ket{211}$. The mass of the cloud decreases during the resonance from its initial value $M\ped{c}$, and the resonance breaks when the value $M\ped{c}\ap{br}$ is reached. Values $M\ped{c}\ap{br}>M\ped{c}$ indicate that the resonance breaks immediately as it starts. The contours [{\color{Mathematica1}blue}] are calculated on circular orbits, as this gives a good approximation for the strongest constraint on $M\ped{c}\ap{br}$ even when overtones, due to orbital eccentricity, (i.e., higher values of $\abs{g}$ for the resonance between two given states) are taken into account. Due to the inaccuracy of the analytical approximations for the decay width $(\omega_{211})\ped{I}$, especially at large $\alpha$, we have determined the contours with Leaver's~\cite{Leaver:1985ax,Cardoso:2005vk,Dolan:2007mj,Berti:2009kk} and Chebyshev~\cite{Baumann:2019eav} methods. The dashed lines [{\color{Mathematica4}red}] are analytical approximations to the blue contours in the proximity of $\beta=\pi$, based on~\eqref{eqn:211_chi_1}.}
\label{fig:deltas211}
\end{figure}
\vskip 2pt
Having determined when hyperfine resonances can be adiabatic, we now calculate where they break, using the results of Section~\ref{sec:resonance-breaking}. As anticipated in Section~\ref{sec:generalB}, the $\Gamma$-breaking is the most relevant mechanism of resonance breaking. To assess its impact, we observe that, because $B\propto M\ped{c}$, eq.~\eqref{eqn:gamma-breaking} can be written as a relation for the final mass of the cloud at resonance breaking, $M\ped{c}\ap{br}=M\ped{c}\abs{c_a}^2$, which can be computed as a function of $\alpha$ and $\beta$. If $M\ped{c}\ap{br}>M\ped{c}$ is found, then the resonance breaks immediately as it starts, as if it was non-adiabatic. The value of $M\ped{c}\ap{br}$ as function of $\alpha$ and $\beta$ is shown in in Figure~\ref{fig:deltas211}. Note that, in principle, the resonance always $\Gamma$-breaks before the cloud is completely destroyed, but its observational impact becomes negligible when $M\ped{c}\ap{br}$ is too small.
\vskip 2pt
The combined constraints due to the $\Gamma$-breaking of $\ket{211}\to\ket{210}$ and $\ket{211}\to\ket{21\,\minus1}$ imply that the cloud survives in a neighbourhood of $\beta=\pi$, say $\pi-\chi_1<\beta<\pi$, similar to what we found for the adiabaticity of the resonances. An analytical approximation of $\chi_1$ for $\ket{211}\to\ket{210}$ based on Detweiler's formula~\eqref{eqn:Gamma_nlm} is
\begin{equation}
\chi_1\approx\SI{38}{\degree}\left(\frac{10^{-2}}{M\ped{c}\ap{br}/M}\right)^{1/6}\left(\frac\alpha{0.2}\right)^{7/6}\left(\frac{0.5}{\tilde a}\right)^{5/18}f(\varepsilon\ped{br})^{1/6}\,,
\label{eqn:211_chi_1}
\end{equation}
where $\varepsilon\ped{br}$ is the eccentricity at resonance breaking. Formula~\eqref{eqn:211_chi_1} significantly underestimates the result for large $\alpha$, as shown in Figure~\ref{fig:deltas211}. Because $\chi_1>\delta_1$, this angular interval overwrites~\eqref{eqn:211-delta1} as the portion of parameter space where the cloud survives hyperfine resonances.
\vskip 2pt
Finally, we check whether hyperfine resonances can $\varepsilon$-break or $Z$-break. Both $\varepsilon$ and $Z$ can vary significantly during the float, so we use the relation~\eqref{eqn:epsilon-breaking} as an order-of-magnitude estimate. For generic values of the inclination, both hyperfine resonances have
\begin{equation}
\sqrt ZB\sim10^6\left(\frac{M\ped{c}/M}{10^{-2}}\right)\left(\frac{10^{-3}}{q}\right)\left(\frac{0.2}{\alpha}\right)^{6}\left(\frac{0.5}{\tilde a}\right)^{1/3}\,.
\end{equation}
The resonances $\varepsilon$-breaks if $f(\varepsilon)=\sqrt ZB$, which is only satisfied at very high eccentricities, not smaller than $0.95$ for typical parameters. Such extreme eccentricities are only reachable if the initial inclination is very close to $\beta=\pi$, as can be seen from Figure~\ref{fig:streamplot_inclination-eccentricity}. But, as proved in~\eqref{eqn:211-delta1} and~\eqref{eqn:211_chi_1}, near-counter-rotating binaries do not undergo floating orbits at all, due to the resonances being either non-adiabatic or $\Gamma$-breaking immediately. As for the $Z$-breaking, one can conservatively ignore the term $Z/Z_0$ in~\eqref{eqn:Z-breaking}, falling back to the same relation as~\eqref{eqn:epsilon-breaking}.
\vskip 2pt
We conclude that the survival of the cloud to later stages of the inspiral is exclusively determined by the $\Gamma$-breaking. If the binary is outside the regions coloured in Figure~\ref{fig:deltas211}, and computed in~\eqref{eqn:211_chi_1}, it encounters the only possible fine resonance:
\begin{equation}
\ket{211}\overset{g=-1}{\longrightarrow}\ket{200}\qquad\frac{R_0}M=\SI{3.4e+2}{}\,\left(\frac{0.2}{\alpha}\right)^{10/3}\,,
\label{eqn:R_0-211-200}
\end{equation}
whose angular dependence is determined through~\eqref{eqn:dgg} as usual. This resonance, however, has anomalous behaviour for two reasons:
\begin{enumerate}
\item it is entirely mediated by the dipole $\ell_*=1$;
\item depending on the value of $\alpha$, it may fall inside the ionisation regime ($P\ped{ion}\gtrsim P_\slab{gw}$) despite not being a Bohr resonance.
\end{enumerate}
As a consequence, its Landau-Zener parameter $Z$ does not scale as a pure power-law in $\alpha$ (nor $M\ped{c}$), and must be computed numerically. The explicit result is reported in Appendix~\ref{appGA_sec:211_200}. Similar to hyperfine resonances, we can compute angular intervals $\delta_2$ and $\chi_2$ where the resonance is non-adiabatic and $\Gamma$-breaks, respectively. The extremely large decay width of $\ket{200}$ (as all states with $\ell=0$), however, makes $\chi_2$ as large as to correspond with the whole possible range of inclinations, from $\SI{0}{\degree}$ to $\SI{180}{\degree}$. Fine resonances are thus effectively never excited for a cloud in a $\ket{211}$ state.
\vskip 2pt
Finally, if the binary arrives to the Bohr region with the cloud still intact, then it encounters the Bohr resonances, all of which are of the sinking type and fall inside the ionisation regime (with the exception of $\ket{211}\to\ket{100}$). No extra circularisation is provided by the hyperfine resonances, if they do not significantly destroy the cloud. Nevertheless, by the time the binary arrives to the Bohr regime, not only has it presumably evolved for a long time under the circularising effect of GW radiation, but it also starts to ionise the cloud, further suppressing the eccentricity~\cite{Tomaselli:2023ysb}. We will therefore assume that quasi-circular orbits are a good approximation by this point. The final population after each sinking resonance can be found using the approximation~\eqref{eqn:sinkingpopulation}, which together with the scaling relations~\eqref{eqn:B_bohr} and~\eqref{eqn:Z_bohr}, implies
\begin{equation}
\abs{c_b}^2\approx3.7\,\left(\frac{Z}{B^2}\right)^{1/3}\propto\frac{M}{M\ped{c}}q^{4/3}\,.
\label{eqn:scaling-cb2}
\end{equation}
For the benchmark parameters, the values of $\abs{c_b}^2$ for the strongest sinking resonances (which are typically with states of the form $\ket{n00}$) are summarised in Figure~\ref{fig:sinking-from-211-and-322}, where we have assumed for simplicity a perfectly counter-rotating configuration ($\beta=\pi$). This is generally a good approximation, due to the relative smallness of the angle $\chi_1$. We see that all resonances are very non-adiabatic, in total transferring less than $1\%$ of the cloud to other states. Hence, ionisation of $\ket{211}$ happens with minimal disturbance from Bohr resonances.
\vskip 2pt
The only floating Bohr resonance is $\ket{211}\to\ket{100}$. It is worth noting that this is also the only Bohr resonance falling outside the ionisation regime (see Figure~\ref{fig:ionisation-bohr-resonances}) and that recent numerical studies~\cite{Brito:2023pyl,Dyson:2025dlj} have shown that it has a resonance width much larger than all other resonances (cf.~Figure~\ref{fig:Fluxes_Inf_Hor}). This last observation means that the resonance might partially evade the analysis of the present chapter, due to the nonlinear dependence of $P_\slab{gw}$ on $R_*$ playing an important role. In any case, we expect the extremely large decay width of $\ket{100}$ to $\Gamma$-break the resonance in most or all realistic cases, preventing the float from happening. 
\begin{figure}[t!] 
\centering
\includegraphics[width=\linewidth]{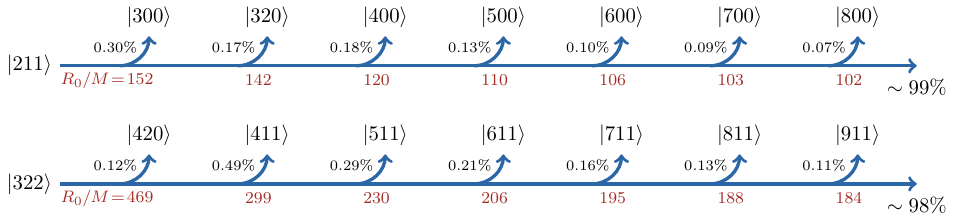}
\caption{Strongest sinking Bohr resonances on a counter-rotating orbit for a cloud in the $\ket{211}$ or $\ket{322}$ state. The percentages next to each resonance are the values of $\abs{c_b}^2$ for benchmark parameters, and they scale with $M\ped{c}$ and $q$ according to~\eqref{eqn:scaling-cb2}, while the red numbers below are the resonant orbital separations $R_0$, in units of $M$.}
\label{fig:sinking-from-211-and-322}
\end{figure}
\subsection{Evolution from a $\ket{322}$ Initial State}\label{sec:evolution-322}
\begin{figure}[t!] 
\centering
\includegraphics[width=\linewidth]{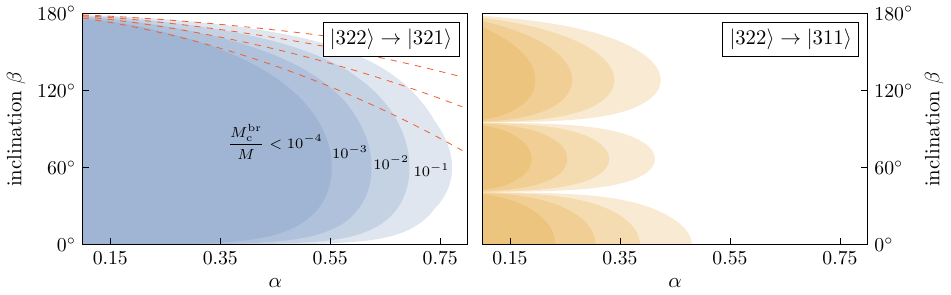}
\caption{Same as Figure~\ref{fig:deltas211}, for the strongest hyperfine (\emph{left panel}) and fine (\emph{right panel}) resonances from a $\ket{322}$ state. The analytical approximations of the contours are not shown in the latter case, as they quickly become inaccurate for moderate values of $\alpha$.}
\label{fig:deltas322}
\end{figure}

The second-fastest growing mode is $\ket{322}$. In this case, the constraint on $\alpha$ -- imposing that the superradiance timescale is shorter than a Gyr -- is
\begin{equation}
\label{eqn:alpha-322}
0.09\left(\frac{M}{10^4M_\odot}\right)^{1/13}\lesssim\alpha<1\,,
\end{equation}
while the rate of cloud decay in GWs is proportional to $M\ped{c}\alpha^{18}$.
\vskip 2pt
Compared to Section~\ref{sec:evolution-211}, a larger number of hyperfine resonances are possible, with any state of the form $\ket{32m_b}$, with $-2\le m_b\le1$. All of these can happen with $g=-4$, in which case the hexadecapole $\ell_*=4$ is entirely responsible for the mixing of the states. However, the cases $m_b=0$ and $m_b=1$ can also resonate, at different separations, with $g=-2$:~these are dominated by the quadrupole $\ell_*=2$ instead, which makes these resonances much stronger than the others. Their positions are
\begin{align}
\ket{322}\overset{g=-2}{\longrightarrow}\ket{321} & \qquad \frac{R_0}M=\SI{5.4e+4}{}\,\left(\frac{0.2}{\alpha}\right)^{4}\left(\frac{0.5}{\tilde a}\right)^{2/3}\,,\\[8pt]
\ket{322}\overset{g=-2}{\longrightarrow}\ket{320} & \qquad \frac{R_0}M=\SI{3.4e+4}{}\,\left(\frac{0.2}{\alpha}\right)^{4}\left(\frac{0.5}{\tilde a}\right)^{2/3}\,,
\end{align}
which should be evaluated at $\tilde a\approx2\alpha/(1+\alpha^2)$~\eqref{eq:spinatsat}. The most stringent constraint on $\delta_1$ is given by $\ket{322}\to\ket{321}$ and equals
\begin{equation}
\delta_1=\SI{5.4}{\degree}\,\left(\frac{10^{-2}}{M\ped{c}/M}\right)^{1/6}\left(\frac{q}{10^{-3}}\right)^{1/12} \left(\frac{\alpha}{0.2}\right)^{4/3} \left(\frac{\tilde{a}}{0.5}\right)^{1/36}f(\varepsilon_0)^{1/4}\,.
\label{eqn:322-delta1}
\end{equation}
The angle $\chi_1$, within which the same resonance $\Gamma$-breaks, is instead
\begin{equation}
\chi_1\approx\SI{4.8}{\degree}\left(\frac{10^{-2}}{M\ped{c}\ap{br}/M}\right)^{1/6}\left(\frac\alpha{0.2}\right)^{11/6}\left(\frac{0.5}{\tilde a}\right)^{5/18}f(\varepsilon\ped{br})^{1/6}\,,
\label{eqn:322_chi_1}
\end{equation}
also more accurately numerically computed and shown in Figure~\ref{fig:deltas322} (\emph{left panel}). Similar to the resonant history of $\ket{211}$, some resonances (such as $\ket{322}\to\ket{321}$) become weak around $\beta=0$, yet other resonances (such as $\ket{322}\to\ket{320}$) do not, thereby eliminating any possible ``safe interval'' around a co-rotating configuration.
\vskip 2pt
Differently from Section~\ref{sec:evolution-211}, there is no clear hierarchy between $\delta_1$ and $\chi_1$. Which one is largest depends not only on $\alpha$, but also on the chosen value of $M\ped{c}\ap{br}$. The angular interval that leads to the survival of the cloud in appreciable amounts is, however, generally dominated by the $\Gamma$-breaking, as even very light clouds, say $M\ped{c}\ap{br}/M<10^{-4}$, are able to give clear signatures in the Bohr region~\cite{Baumann:2021fkf}.
\vskip 2pt
As in the $\ket{211}$ case, the $\varepsilon$-breaking and $Z$-breaking prove to not be relevant for the resonant history:~the value
\begin{equation}
\sqrt ZB\sim10^7\left(\frac{M\ped{c}/M}{10^{-2}}\right)\left(\frac{10^{-3}}{q}\right)\left(\frac{0.2}{\alpha}\right)^{6}\left(\frac{0.5}{\tilde a}\right)^{1/3}\,,
\end{equation}
requires extremely high eccentricities ($\varepsilon\gtrsim0.98$) to give rise to a resonance breaking. The corresponding initial inclinations are extremely close to $\beta=\pi$ and would fall in the interval~\eqref{eqn:322-delta1}, where the resonance is not adiabatic.
\vskip 2pt
A cloud in the $\ket{322}$ state can experience fine resonances with states with $\ell\ne0$. Their decay width is smaller than those of the states with $\ell=0$:~as a consequence, fine resonances can destroy a significant portion of the cloud before they $\Gamma$-break. The fine resonance that gives the most stringent constraints on $\delta_2$ and $\chi_2$ is
\begin{equation}
\ket{322}\overset{g=-1}\longrightarrow\ket{311}\qquad\frac{R_0}M=\SI{2.3e+3}{}\,\left(\frac{0.2}{\alpha}\right)^{10/3}\,.
\end{equation}
Analytical approximations for $\beta\approx\pi$ give\footnote{For $\alpha\gtrsim0.5$, this resonance may marginally fall inside the ionisation regime. However, the value of $P\ped{ion}$ never becomes much larger than $P_\slab{gw}$. We therefore ignore this detail, which only slightly increases the value of $\delta_2$ compared to the one presented in~\eqref{eqn:322-delta2}.}
\begin{equation}
\delta_2=\SI{3.2}{\degree}\,\left(\frac{10^{-2}}{M\ped{c}/M}\right)^{1/4}\left(\frac{q}{10^{-3}}\right)^{1/8} \left(\frac{\alpha}{0.2}\right)^{31/24}f(\varepsilon_0)^{3/8}\,,
\label{eqn:322-delta2}
\end{equation}
and
\begin{equation}
\chi_2\approx\SI{9}{\degree}\left(\frac{10^{-2}}{M\ped{c}\ap{br}/M}\right)^{1/4}\left(\frac\alpha{0.2}\right)^{3/2}f(\varepsilon\ped{br})^{1/4}\,,
\label{eqn:322_chi_2}
\end{equation}
while a more accurate numerical determination of the mass of the cloud at resonance breaking is given in Figure~\ref{fig:deltas322} (\emph{right panel}). It is worth noting that the strength of the $\ket{322}\to\ket{311}$ resonance has a complicated $\beta$-dependence, due to the octupole $\ell_*=3\ne-g$ being the dominant term. Consequently, this resonance becomes weak not only around $\beta=\SI{180}{\degree}$, but also around $\beta=\SI{41}{\degree}$ and $\SI{95}{\degree}$ (as visible from Figure~\ref{fig:deltas322}). However, other fine resonances remain strong at these intermediate inclinations and so, once again, the cloud can only reach the Bohr region if the inclination is in a narrow interval around the counter-rotating configuration.
\vskip 2pt
In the Bohr region, the system encounters several sinking resonances, the strongest of which are with states of the form $\ket{n11}$. The final populations $\abs{c_b}^2$ are displayed in Figure~\ref{fig:sinking-from-211-and-322}. For benchmark parameters, about $2\%$ of the cloud is lost in the process. None of the floating resonances, with $n=1$ or $n=2$ states, becomes adiabatic within the interval of inclinations discussed above.
\vskip 2pt
Finally, in case the binary is formed at radii small enough to avoid constraints on the inclination coming from fine resonances, an interesting scenario opens up. The strongest floating Bohr resonance is $\ket{322}\overset{g=-1}\longrightarrow\ket{211}$, which becomes adiabatic, for benchmark parameters, for $\beta<\SI{155}{\degree}$.\footnote{Due to the weakness of the resonance compared to most the (hyper)fine ones, it is not possible to expand around $\beta=\pi$ and get a simple formula for the upper limit on the angle as function of the parameters. Nevertheless, a good approximation is given by the following cubic equation: $(\pi-\beta)^4+2.8(\pi-\beta)^6>0.056\times(10^5M\ped{c}q/M)^{1/2}$.} Among all possible scenarios we considered in Sections~\ref{sec:evolution-211} and~\ref{sec:evolution-322}, this is the only case where the binary's evolution in the Bohr region features a new phenomenon, beyond ionisation and non-adiabatic sinking resonances:~namely, an adiabatic floating resonance. The companion's motion continues to ionise the cloud while this resonance takes place, potentially changing $M\ped{c}$ significantly before its end. This is also the only floating resonance with the actual potential to partially move the cloud to a different state, rather than merely destroying it:~as can be seen in Figure~\ref{fig:tfloat_tdecay} (\emph{right panel}), the hierarchy $\Delta t\ped{float}\gg t\ped{decay}$ is not valid in the entire parameter space. Hence, depending on the parameters, when the resonance ends, the inspiral can either continue without the cloud, or with a cloud in a (decaying) $\ket{211}$ state and a reduced value of $M\ped{c}$. In the latter case, the discussion in Section~\ref{sec:evolution-211} applies from this point onwards.
\section{Observational Signatures}\label{sec_Legacy:observational-signatures}
The dynamics of the cloud-binary system are intricate and depend on the parameters. In Section~\ref{sec_Legacy:history}, we determined when the cloud is entirely destroyed in the early inspiral, when it loses some of its mass upon resonance breaking, and when it remains intact until the binary enters the Bohr region. There are thus two main ways the cloud can leave an imprint on the gravitational waveform:~(i)  modifications of the waveform due to interaction with the cloud, in case it is still present in the late stages of the inspiral (Section~\ref{sec:direct-signatures});~(ii) permanent consequences on the binary parameters left by a cloud destroyed early in the inspiral (Section~\ref{sec:indirect-evidence}). A partially destroyed cloud, left by a broken resonance, may be able to combine both kinds of signatures.
\subsection{Direct Signatures of the Cloud}\label{sec:direct-signatures}
As discussed extensively in Section~\ref{sec_Legacy:history}, the requirement that the cloud survives the hyperfine and fine resonances forces either the inclination angle to be within $\mathcal{O}(10^\circ)$ of a counter-rotating configuration or the binary to form at radii too small to ever excite those resonances. Then, most phenomena producing direct observational evidence of the cloud happen when the binary reaches the Bohr region. Here, ionisation takes over GW radiation as the primary mechanism of orbital energy loss. When $P\ped{ion}\gg P_\slab{gw}$, the evolution of the GW frequency $f_\slab{gw}$ approximately follows a universal shape~\cite{Baumann:2022pkl},
\begin{equation}
f_\slab{gw}(t)=\frac{\alpha^3}M\,f\left(\frac{M\ped{c}q\alpha^3}{M^2}t\right)\,,
\label{eqn:universal-f-ionisation}
\end{equation}
where the function $f$ can be explicitly determined from the shape of $P\ped{ion}$. This universal behaviour of $f_\slab{gw}(t)$ constitutes a direct evidence of the presence of the cloud.
\vskip 2pt
On top of this, sinking resonances can cause non-negligible upward ``jumps'' of $f_\slab{gw}$ due to their backreaction,\footnote{In this chapter, we only study the backreaction on the orbital parameters. When including the  backreaction on the geometry as well, the cloud's transitions could cause ``resonant'' features in the emitted GWs, see e.g.,~Figure~1 of~\cite{Duque:2023seg}.} even if they are strongly non-adiabatic. For a Bohr resonance $\ket{n_a\ell_am_a}\to\ket{n_b\ell_bm_b}$, they are located at
\begin{equation}
f_\slab{gw}\ap{res}=\frac{\SI{26}{mHz}}g\left(\frac{10^4M_\odot}{M}\right)\left(\frac\alpha{0.2}\right)^3\left(\frac1{n_a^2}-\frac1{n_b^2}\right)\,,
\label{eqn:position-resonances}
\end{equation}
where $g=m_b-m_a$, and thus fall inside the LISA band for benchmark parameters.\footnote{Formula~\eqref{eqn:position-resonances}, with $n_b\to\infty$, also describes the position of the $g$-th ``kink'' of the function $f$ appearing in~\eqref{eqn:universal-f-ionisation}, corresponding to the $g$-th discontinuity of $P\ped{ion}$ (see Figure~\ref{fig:ionisation-bohr-resonances}).}
\vskip 2pt
The amplitude of the jump can be computed explicitly from~\eqref{eqn:dimensionless-omega-evolution} (assuming quasi-circular orbits):
\begin{equation}
\Delta f_\slab{gw}=\frac{\SI{0.61}{mHz}}{\Delta m^{1/3}}\,\left(\frac{10^4M_\odot}{M}\right)\!\left(\frac{M\ped{c}/M}{0.01}\right)\!\left(\frac{10^{-3}}{q}\right)\!\left(\frac\alpha{0.2}\right)^3\!\left(\frac1{n_a^2}-\frac1{n_b^2}\right)^{4/3}\!\left(\frac{\abs{c_b}^2}{10^{-3}}\right)\,,
\label{eqn:jump-resonances}
\end{equation}
where the values of $\abs{c_b}^2$ and their dependence on the parameters are given in Figure~\ref{fig:sinking-from-211-and-322} and eq.~\eqref{eqn:scaling-cb2}. The increase in frequency comes with smaller, long-lived oscillations of the frequency, and with a slight increase of the eccentricity; both these effects have been shown in the \emph{right panels} of Figure~\ref{fig:eccentric-backreacted-resonance} for example parameters. The dephasing introduced by a single sinking resonance on top of the one coming from ionisation is $\Delta\Phi_\slab{gw}\approx\pi f_\slab{gw}\ap{res}\Delta f_\slab{gw}/\gamma$. This is of the order of thousands of radians, although the exact number can vary by a few orders of magnitude in different regions of the parameter space. Not only is this well above the expected LISA precision of $\Delta\Phi_\slab{gw}\sim2\pi$, but such a dephasing would happen in a very narrow frequency range, in contrast to most other environmental effects, including ionisation. This unique behaviour would aid parameter estimation by directly linking the cloud's parameters with $\Delta\Phi_\slab{gw}$ via~\eqref{eqn:position-resonances} and~\eqref{eqn:jump-resonances}, especially if multiple jumps are observed within one signal.
\vskip 2pt
As discussed in Sections~\ref{sec:evolution-211} and~\ref{sec:evolution-322}, the only cases where a floating resonance can be observed in the Bohr region require a binary formation at very small radii, so that all early resonances are skipped without a strict requirement on the inclination angle. Resonances of the type $\ket{n_a\ell_am_a}\to\ket{100}$ happen very late in the inspiral (see Figure~\ref{fig:ionisation-bohr-resonances}), where relativistic corrections are expected to be more important~\cite{Brito:2023pyl,Dyson:2025dlj} (as we will show in the next chapter). The only other floating Bohr resonance encountered in Section~\ref{sec_Legacy:history} is $\ket{322}\to\ket{211}$. This is an interesting case because it may not entirely destroy the cloud. The expected GW signal is a constant frequency $f_\slab{gw}$ given by eq.~\eqref{eqn:position-resonances}, for a total floating time of\footnote{This value assumes a quasi-circular co-rotating orbit. Moderate nonzero values of eccentricity or inclination introduce $\mathcal O(1)$ variations in $\Delta t\ped{float}$.}
\begin{equation}
\Delta t\ped{float}=\SI{5.8}{yrs}\,\left(\frac{M}{10^4M_\odot}\right)\left(\frac{10^{-3}}{q}\right)^{2}\left(\frac{0.2}{\alpha}\right)^{3}\,.
\label{eqn:deltatfloat-322-211}
\end{equation}
Although the cloud's mass is continuously reduced by ionisation while the resonance takes place, the value given in~\eqref{eqn:deltatfloat-322-211} remains independent of $M\ped{c}$ as long as it is large enough to guarantee $P\ped{ion}\gg P_\slab{gw}$. 
\subsection{Indirect Signatures:~Impact on Binary Parameters}\label{sec:indirect-evidence}
For sufficiently small orbital inclinations, as seen in Figures~\ref{fig:deltas211} and~\ref{fig:deltas322}, the cloud can be destroyed during one of the floating resonances in the early inspiral, to a level where it no longer affects the binary dynamics in an observable way. Then, by the time the system enters in band, its evolution is expected to follow the rules of vacuum General Relativity. Nevertheless, the binary still carries the marks of the previously existing boson cloud, and of the resonance that destroyed it. These are due to the backreaction on the orbit from that floating resonance, and come in the form of a change in the eccentricity and tilt of the inclination angle.
\begin{figure}[t!]
\centering
\includegraphics[width=\linewidth]{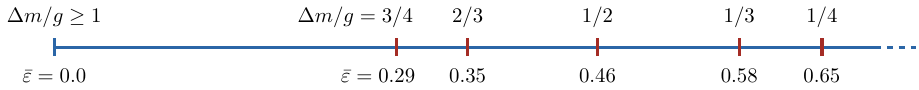}
\caption{Example values of the fixed point $\bar{\varepsilon}$ depending on $\Delta m/g$. Numbers can be found by solving eqs.~\eqref{eqn:dimensionless-eccentricity-evolution} and~\eqref{eqn:dimensionless-inclination-evolution} on a floating orbit.}
\label{fig:fixedpointsresonances}
\end{figure}
\vskip 2pt
While in Section~\ref{sec_Legacy:history} we could simplify the analysis by studying only the strongest resonance, the impact on the orbital parameters strongly depends on which overtone (i.e., which value of $g$) mediated the last adiabatic resonance encountered by the system.\footnote{If the system undergoes multiple floats, for example, because broken resonances leave a cloud massive enough to excite other adiabatic resonances, then the evolution of the eccentricity follows several nontrivial steps. Here, however, we focus on the last of those as it has the most direct observational consequences.} As shown in Figure~\ref{fig:streamplot_inclination-eccentricity}, the orbital parameters follow specific sets of trajectories on the $(\varepsilon,\beta)$-plane, until the resonance breaks or completes. While floating orbits \emph{always} tilt the inclination angle towards a co-rotating configuration, the eccentricity is forced towards a fixed point, whose value depends on $\Delta m/g$. Some examples of the value of this fixed point are shown in Figure~\ref{fig:fixedpointsresonances} for different values of $\Delta m / g$.
\vskip 2pt
Assuming that the resonance does not break prematurely, the distance travelled by the binary in the $(\varepsilon,\beta)$-plane depends on the parameter $D$ alone, introduced in~\eqref{eqn:D}, given by
\begin{equation}\label{eqn:D_full}
D=D_0\left(\frac{-g}2\right)^{\!2/3}\left(\frac{M\ped{c}/M}{10^{-2}}\right)\left(\frac{10^{-3}}{q}\right)\left(\frac\alpha{0.2}\right)\left(\frac{\tilde a}{0.5}\right)^{\!1/3}\,.
\end{equation}
For the two strongest hyperfine resonances from $\ket{211}$ ($\ket{322}$), the parameter $D_0$ assumes the values $3.30$ and $4.16$ ($1.28$ and $1.62$). Very roughly, the system gets $e^D$ times closer to the eccentricity fixed point than it was before the resonance started. Due to $D$ being inversely proportional to $q$, mainly intermediate or extreme mass ratio binaries change significantly their orbital parameters during a floating resonance. Examples of variations of the parameters during a floating orbit are reported in Figure~\ref{fig:examples-backreaction-orbit} for the resonances $\ket{211}\to\ket{21\,\minus1}$ and $\ket{322}\to\ket{320}$.
\begin{figure}[t!] 
\centering
\includegraphics[width=\linewidth]{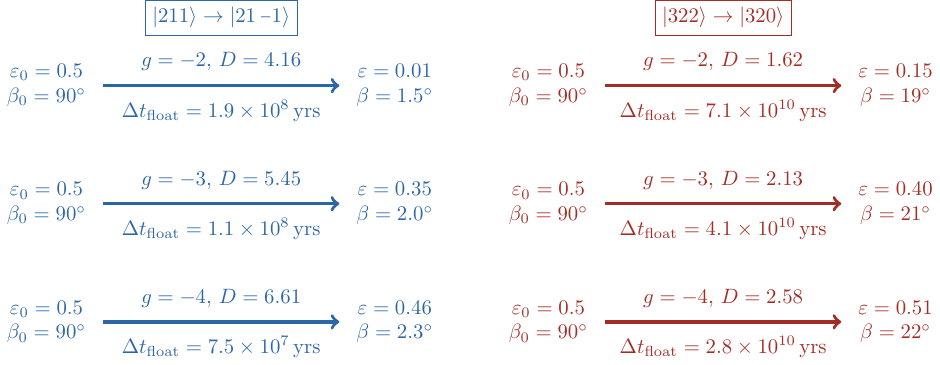}
\caption{Examples of backreaction on the eccentricity $\varepsilon$ and inclination $\beta$ during floating orbits that destroy the cloud entirely. We show the strongest resonance ($\Delta m/g=1$) and two overtones in each scenario, using the benchmark parameters. Each case is initialised with $\varepsilon_0=0.5$ and $\beta_0=\SI{90}{\degree}$ for illustrative purposes, but the final values of $\varepsilon$ and $\beta$ are very robust against the choice of different initial conditions. The final values of $\varepsilon$ and $\beta$, as well as $\Delta t\ped{float}$, are computed integrating numerically eqs.~\eqref{eqn:dimensionless-dressed-schrodinger},~\eqref{eqn:dimensionless-omega-evolution},~\eqref{eqn:dimensionless-eccentricity-evolution}, and~\eqref{eqn:dimensionless-inclination-evolution}. For benchmark parameters, the floating time of resonances from $\ket{322}$ exceeds the Hubble time; however, it strongly depends on the parameters, as shown in~\eqref{eqn:hyperfine-B-tfloat-exponent}.}
\label{fig:examples-backreaction-orbit}
\end{figure}
\vskip 2pt
\begin{figure}
\centering
\includegraphics[width=\linewidth]{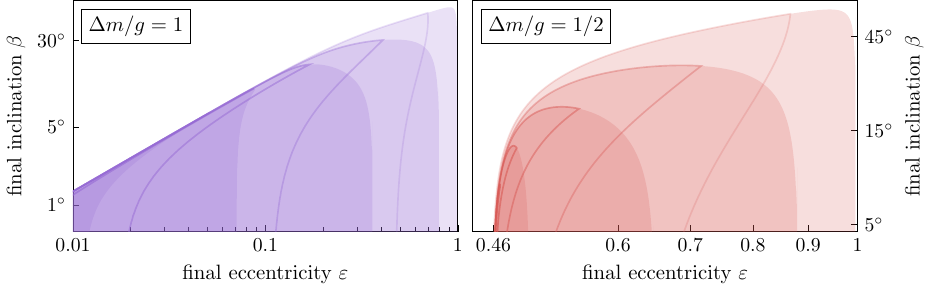}
\caption{The shaded regions show the possible values of eccentricity $\varepsilon$ and inclination $\beta$ at the completion of a floating resonance, starting from any initial values $\varepsilon_0$ and $\beta_0$, for different values of $D$. The values used, from the outermost to the innermost region, are $D=1,1.5,2,2.5,3$ (\emph{left panel}) and $D=1.8,2.6,3.4,4.2,5$ (\emph{right panel}); cf.~\eqref{eqn:D_full}. When the initial inclination is required to satisfy the conditions necessary to sustain the float, a smaller portion of each region is reachable. We enclosed in solid lines the reachable portions for $\beta_0\le\SI{128}{\degree}$ (\emph{left panel}) and $\beta_0\le\SI{142}{\degree}$ (\emph{right panel}), which are the thresholds for $\ket{211}\to\ket{21\,\minus1}$ and $\ket{211}\to\ket{210}$, for the reference parameters used in~\eqref{eqn:211_chi_1}.}
\label{fig:parameter-range} 
\end{figure}
Starting from generic initial conditions, we show in Figure~\ref{fig:parameter-range} the possible values of $\varepsilon$ and $\beta$ at the end of a floating resonance, as functions of $D$ and for two values of $\Delta m/g$. The float brings the orbit significantly close to the equatorial plane, even for large initial inclinations. An abundance of quasi-planar inspiral events can thus be indirect evidence for boson clouds. Whether the formation mechanisms of the binary, or other astrophysical processes, also lead to a natural preference for small inclinations is still subject to large uncertainties~\cite{Amaro-Seoane:2012lgq,Pan:2021ksp,Pan:2021oob}.
\vskip 2pt
Additionally, the eccentricity is suppressed by the main tones ($g=\Delta m$) and brought close to, or above, a nonzero fixed point by overtones ($g>\Delta m$). The latter scenario is especially interesting for binaries that are not dynamically captured, such as in the case of comparable mass ratios, because they are generally expected to be on quasi-circular orbits. The past interaction with a cloud can overturn this prediction. The float-induced high eccentricities are mitigated by the subsequent GW emission, but the binary will remain more eccentric than it would have been otherwise, even in late stages of the inspiral. The cloud-binary dynamics thus leave a unique mark on the binary parameters and opens up the possibility of performing a statistical test of the parameters of a large number of binary inspirals, and comparing them with the ones predicted from a suitable model of their formation channels.
\vskip 2pt
Lastly, we note that the extremely long floating time associated with some hyperfine or fine resonances can stop many binaries from getting in band at all, reducing the merger rate. For example, for our choice of benchmark parameters, the hyperfine resonances from the $\ket{322}$ state, shown in Figure~\ref{fig:examples-backreaction-orbit}, float for longer than the Hubble time.
\section{Summary and Outlook}\label{sec_Legacy:conclusions}
In the context of GW astronomy, binary BH environments have long been proposed as a laboratory for fundamental physics. One such example are gravitational atoms, or clouds of ultralight bosons produced by superradiance around spinning BHs. Compared to other kinds of environments, the phenomenology of gravitational atoms is extremely rich. The two most striking types of interaction between the binary and the cloud are resonant phenomena~\cite{Zhang:2018kib,Baumann:2018vus,Baumann:2019ztm} and friction effects~\cite{Zhang:2019eid,Baumann:2021fkf,Baumann:2022pkl,Tomaselli:2023ysb}, both of which leave very distinct signatures on the emitted gravitational waveform. This complexity is a blessing for the potential detection and identification~\cite{Cole:2022yzw} of such systems. However, it is a curse for the achievement of a complete characterisation of their evolution.
\vskip 2pt
Previous studies have described the effects on the waveform as function of the state of the cloud and of the binary configuration at the time of observation. These are, however, the final products of a complex series of cloud-binary interactions that characterise former phases of the inspiral. Despite a number of relevant studies~\cite{Takahashi:2021eso,Zhang:2018kib,Ding:2020bnl,Tong:2021whq,Du:2022trq,Tong:2022bbl,Fan:2023jjj,Takahashi:2023flk}, the combinations of cloud states and binary configurations compatible with this kind of evolution have not yet been determined.
\vskip 2pt
In this chapter, we finalise such a programme by systematically studying the chronological sequence of resonances encountered during the binary inspiral. We do so in the most general possible set of assumptions:~we allow for any value of the initial eccentricity, inclination and separation of the binary; at the same time, we keep the scaling with the cloud's parameters explicit, so to apply our results to the entire parameter space. Furthermore, we take into account the backreaction of the resonances on the orbit, and study how this impacts the behaviour of the resonances themselves. This aspect, as well as the impact of inclination and eccentricity on resonances (and vice versa), have never been studied before, and each of these novel results turns out to play a crucial role in our analysis. Finally, we perform explicitly the exercise of ``following'' the evolution of the system from the initial states most likely to be populated by superradiance, $\ket{211}$ and $\ket{322}$, until the merger, and then summarise the GW signatures compatible with the scenarios studied.
\vskip 2pt
In principle, one might have expected the evolution of the system to be extremely complicated. The S-matrix approach developed in~\cite{Baumann:2019ztm} suggests a tree of populated states branching more and more, every time a new resonance is encountered. In practice, however, we find that the hierarchy between the floating and decay timescales simplifies the picture dramatically.\footnote{As shown in Figure~\ref{fig:tfloat_tdecay} and eq.~\eqref{eqn:t-decay-t-float}, the separation of timescales is larger for smaller mass ratios. For equal-mass ratios $q\sim1$, the timescales could become comparable.} During every hyperfine or fine adiabatic resonance, the cloud loses mass until the resonance breaks, often to the point where it is no longer directly observable. The conclusion is then remarkably simple, and similar among the two cases studied explicitly. Only binaries close enough to a counter-rotating configuration, where these early resonances are very weak, are able to carry the cloud up to the point where it becomes observable through the effects of ionisation and a large number of weak resonances. Our detailed study of the nonlinear behaviour of resonances allows us to precisely quantify the angular interval of inclinations where this scenario is realised, see eqs.~\eqref{eqn:211_chi_1},~\eqref{eqn:322_chi_1}, and~\eqref{eqn:322_chi_2}, as well as Figures~\ref{fig:deltas211} and~\ref{fig:deltas322}.
\vskip 2pt
The early disappearance of the cloud for generic orbital configurations may seem to suggest that the chances of detecting ultralight bosons using binary systems is slim after all. It is certainly true that our work puts strict conditions for the direct observation of the cloud-binary interaction. To avoid all early resonances, either the initial separation must be small enough, or the orbital angular momentum must be approximately anti-aligned (within a tolerance interval whose size depends on the parameters) with the central BH's spin. The likelihood of either scenario depends on the astrophysical processes governing the formation of the binary system and is subject to large uncertainties (though see~\cite{Goodman:2002gv,Goodman:2003sf,Levin:2006uc,LISA:2022yao} and~\cite{Amaro-Seoane:2012jcd,Babak:2017tow} for the two cases, respectively). We note, however, that the event leading to the cloud destruction -- an adiabatic floating resonance -- necessarily exerts a strong backreaction on the binary's inclination and eccentricity, forcing them to evolve towards a fixed point. The possibility of inferring the past existence of a cloud from its \emph{legacy} left on the binary parameters is a new and exciting observational prospect for both ground-based GW detectors, such as LIGO-Virgo or the Einstein Telescope, and space-borne ones, such as LISA. Proper population studies will be needed to turn this prediction into a test of fundamental physics with the available and future data.
\vskip 2pt
The present chapter answers the main remaining open questions left on the phenomenology of gravitational atoms in binaries, within a certain set of assumptions. We neglected a number of subleading effects. Some of them, like the accretion onto the secondary~\cite{Baumann:2021fkf} and the cloud's self gravity~\cite{Ferreira:2017pth,Hannuksela:2018izj} can be straightforwardly included in the binary's evolution, and do not change qualitatively the picture drawn here. Others, like the backreaction of the cloud's decay in GWs~\cite{Cao:2023fyv}, the non-resonant overlap between growing and decaying states~\cite{Tong:2022bbl,Fan:2023jjj}, and the change in the BH's mass and spin due to the absorption of the cloud during a resonance (see footnote~\ref{fn:bh-parameters}) can potentially introduce new relevant features. But perhaps most importantly, we stuck to a non-relativistic analysis;~an assumption that will be relaxed in the next chapter.
\chapter{Extreme Mass Ratio Inspirals:~Perturbing the Environment in Kerr} \label{chap:inspirals_selfforce}
\begingroup
\renewcommand{\vec}[1]{\mathaccent"017E{#1}}
\vspace{-0.8cm}
\hfill \emph{Je n'aurai pas le temps, pas le temps}

\hfill \emph{M\^{e}me en courant}

\hfill \emph{Plus vite que le vent}

\hfill \emph{Plus vite que le temps}

\hfill \emph{M\^{e}me en volant}

\hfill \emph{Je n'aurai pas le temps, pas le temps}

\hfill \emph{De visiter toute l'immensit\'{e}}

\hfill \emph{D'un si grand univers}

\hfill \emph{Même en cent ans}

\hfill \emph{Je n'aurai pas le temps de tout faire}
\vskip 5pt

\hfill Michel Fugain, \emph{Je n'aurai pas le temps}
\vskip 35pt
\noindent The previous chapter explored how the cloud and the binary interact during the early phases of the inspiral -- well before the system enters the sensitivity band of GW detectors. The main conclusion is that resonances typically destroy the cloud, except when the binary is nearly counter-rotating relative to it [see Figure~\ref{fig:history_211} and eqs.~\eqref{eqn:211_chi_1},~\eqref{eqn:322_chi_1}, and~\eqref{eqn:322_chi_2}]. In such cases, the cloud can survive until the later stages of the inspiral, when the binary enters the strong gravity regime. At that point, \emph{relativistic effects} become important, and modelling the dynamics of both binary and its environment requires a different set of tools.
\vskip 2pt
In this chapter, I present the work of~\cite{Dyson:2025dlj}, which introduces the first self-consistent, fully relativistic calculation of a perturbation to a BH environment induced by an inspiralling secondary in the Kerr geometry. The framework is general and allows environmental effects to be incorporated in EMRI modelling in a fully relativistic way. As a representative and simple example, I consider again a superradiantly grown scalar cloud and analyse the response of the scalar field due the perturbation induced by the secondary. As will be shown, the field develops a distinctive spiralling wake trailing the secondary. Moreover, the fluxes emitted to infinity and through the horizon exhibit substantial deviations compared to the Newtonian and Schwarzschild case, with relative differences of tens of percent, even at relatively large binary separations. These results underscore the necessity of modelling environments relativistically, as well as relaxing the assumption of spherical symmetry and using the Kerr geometry as a background. Neglecting these aspects could lead to significant biases in data analysis with upcoming detectors.
\vskip 2pt
The structure of this chapter is as follows. In Section~\ref{sec_SF:Field_eq}, I establish the relativistic framework for generic environments, followed by Section~\ref{sec_SF:scalar_fields}, where I specialise to scalar fields. In Section~\ref{sec_SF:scalar_wake}, I solve for the field and examine the wake structure, while in Section~\ref{sec_SF:fluxes}, I analyse the fluxes to infinity and through the horizon, comparing them to the Schwarzschild and Newtonian cases. I conclude in Section~\ref{sec_SF:conclusions}, with additional technical details provided in Appendix~\ref{app:BHPT_Kerr}.
\section{Field Equations with Environments}\label{sec_SF:Field_eq}
We focus on astrophysical systems whose geometry is dominated by the Kerr spacetime, with perturbations arising from the (small) binary companion (or ``secondary'') and surrounding matter. We first outline our perturbation scheme for generic matter fields.
\vskip 2pt
The action for generic matter fields minimally coupled to gravity in the presence of a perturber with mass $m\ped{p}$ is given by
\begin{equation}\label{eq:action_generic}
\begin{aligned}
    S =\int\mathrm{d}^4 x \sqrt{-\mathbf{g}} \left(\frac{\bf R}{16 \pi} + \mathcal{L}^{\rm env}[ {\bf \Psi}]\!\right)- m\ped{p}\int \mathrm{d}\tau \sqrt{-\mathbf{g}_{\mu\nu}\mathbf{u}^\mu\mathbf{u}^\nu}\,,
\end{aligned}
\end{equation}
where $\mathcal{L}^{\rm env}$ is the Lagrangian of the matter field ${\bf \Psi}$ and the action of the point-particle encodes the curvature of the secondary BH using the \emph{skeletonised} source approach~\cite{Mathisson:1937zz,Dixon:2015vxa}. Here, $\mathbf{u^\mu}$ denotes the four-velocity of the secondary on some effective, regularised metric. Bold quantities denote fully nonlinear terms.
\vskip 2pt
By varying the action~\eqref{eq:action_generic} with respect to the matter field and metric, we obtain the field equations:
\begin{align}
    \mathcal{Q} [{\bf \Psi},{\bf g} ]&=0\label{eq:GenericEquationsOfMotionPsi_matter}\,, \\
   G_{\mu \nu}[ {\bf g}]&=  T^{\rm env}_{\mu\nu}[{\bf \Psi},{\bf g} ] +  T^{\rm p}_{\mu\nu}[\gamma] \label{eq:GenericEquationsOfMotionPsi_gravity}\,,
\end{align}
where $\gamma$ is the world-line of the secondary and $\mathcal{Q}$ is a generic nonlinear operator involving ${\bf \Psi}$ and ${\bf g}$.
\vskip 2pt
At leading order, the surrounding matter (henceforth, ``the environment'') is treated as a stationary solution on top of the Kerr geometry and is characterised by a total mass $M^{\rm env}$ and typical length $L^{\rm env}$. The stress-energy tensor of the matter field scales with the energy density as
\begin{equation}
\begin{aligned}
   T^{\rm env}_{\mu\nu} \sim \rho^{\rm env}_0 \sim \frac{M^{\rm env}}{(L^{\rm env})^3}\,.
\end{aligned}
\end{equation}
The natural expansion for the matter field then follows the characteristic density ratio, i.e.,
\begin{equation}\label{eq:EmviromentalExpansionParameter}
\begin{aligned}
   \epsilon^n = \frac{\rho_0^{\rm env}}{\rho_0^{\rm BH}} = \frac{M^{\rm env}} {(L^{\rm env})^3}\frac{L^3}{M} = \eta \left(\frac{L}{L^{\rm env}}\right)^3\,,
\end{aligned}
\end{equation}
where $M$ and $L$ denote the mass and length scale of the primary (Kerr) BH, respectively, and $\eta  \equiv M^{\rm env}/M$. In most astrophysical scenarios, this parameter is expected to be small. For example, taking $\rho_{\rm BH} \sim 10^{8}\left(10^6M_{\odot}/M\right)^2\,\mathrm{kg}/\mathrm{m}^3$ and referring to Figure~\ref{fig:taxonomy}, accretion disks around supermassive BHs typically have $\epsilon^n \sim 10^{-12}$.
\vskip 2pt
The exponent $n$ corresponds to the leading-order power of ${\bf \Psi}$ in the matter Lagrangian. For instance, $\mathcal{L}^{\rm env} = |\partial{\bf \Psi}|^2|{\bf \Psi}|^2+ |{\bf \Psi}|^6$ implies $n=4$. When $\epsilon\ll 1$, the matter field is then naturally expanded as ${\bf \Psi} = \epsilon \psi + \cdots$ where $\psi$ satisfies the matter field equations on Kerr~\eqref{eq:GenericEquationsOfMotionPsi_matter}. As usual, we define the mass ratio between the secondary and primary BH as $q \equiv m\ped{p}/M$. Finally, besides $q\ll1$ and  $\epsilon\ll1$, we do not require any scaling relation between $q$ and $\epsilon$:~they act as independent perturbative parameters. To track this dual expansion, we label quantities of some perturbative order $S^{(n,m)}$ as being associated with a perturbative coefficient $\sim \mathcal{O}(\epsilon^n q^m)$.
\vskip 2pt
In this framework, the gravitational and matter fields are expanded as follows:
\begin{align}
    \mathbf{g}_{\mu\nu}&= g_{\mu\nu}+\epsilon^n h^{(n,0)}_{\mu\nu}+q h^{(0,1)}_{\mu\nu}+ \epsilon^n  q  h^{(n,1)}_{\mu\nu}+ \epsilon^n q^2   h^{(n,2)}_{\mu\nu}+  q^2  h^{(0,2)}_{\mu\nu}+\cdots\,,\label{eq:expansion_metric}\\
    \mathbf{\Psi} &= \epsilon \psi^{(1,0)} +\epsilon  q  \psi^{(1,1)}+\epsilon  q^2 \psi^{(1,2)}+\cdots\label{eq:expansion_field}\,,
\end{align}
where $h^{(0,1)}_{\mu\nu}$ is the metric perturbation arising from a vacuum point-particle source. From the equations of motion, we can then generically expand the nonlinear operators as
\begin{equation}\label{eq:Gexpanded}
    G_{a b }[g_{\mu \nu } + h_{\mu \nu} ] = G_{ab}[ g_{\mu\nu}] + \delta G_{a b } [h_{\mu \nu }] + \delta^2 G_{a b } [h_{\mu \nu },h_{\mu \nu }] + \cdots\,,
\end{equation}
where we defined:
\begin{equation}
     \delta^m G_{ab} [h_{\mu \nu }]  = \frac{1}{m!} \frac{\mathrm{d}^m}{\mathrm{d}\lambda^m} G[g_{\mu\nu}+ \lambda h_{\mu\nu}]|_{\lambda=0}\,.
\end{equation}
Similarly, we expand $\mathcal{Q}$ as
\begin{equation}\label{eq:Qexpanded}
\begin{aligned}
    \mathcal{Q}[\psi+\bar{\psi},\ g_{\mu \nu }  +h_{\mu \nu} ] &= 
    \mathcal{Q}[\psi, g_{\mu\nu}]+\delta^{(1,0)}\mathcal{Q}[\bar{\psi},g_{\mu \nu }]  \\
    & + \delta^{(1,1)}\mathcal{Q}[\bar{\psi},h_{\mu \nu }]+ \delta^{(2,1)}\mathcal{Q}[\bar{\psi},\bar{\psi},h_{\mu \nu }] \\
    &+\delta^{(1,2)}\mathcal{Q}[\bar{\psi},h_{\mu \nu },h_{\mu \nu }] + \cdots\,,
\end{aligned}
\end{equation}
where
\begin{equation}
     \delta^{(n,m)} \mathcal{Q}[\bar{\psi}, h_{\mu \nu }] = \frac{1}{n! m!} \frac{\mathrm{d}^{n+m}}{\mathrm{d}\kappa^n \mathrm{d}\lambda^m} \left\{\mathcal{Q}[\psi+ \kappa \bar{\psi}, g_{\mu\nu}+ \lambda h_{\mu\nu}]\right\}|_{\lambda=0, \kappa=0}\,.
\end{equation}
Substituting the perturbed fields~\eqref{eq:expansion_metric}--\eqref{eq:expansion_field} into the expansions of eqs.~\eqref{eq:GenericEquationsOfMotionPsi_matter}--\eqref{eq:GenericEquationsOfMotionPsi_gravity}, the $\mathcal{O}(\epsilon q)$ perturbation to the environment satisfies:
\begin{equation}\label{eq:dynamicalperubation}
  \delta^{(1,0)}\mathcal{Q}[\psi^{(1,1)}, g_{\mu \nu }] = -  \delta^{(1,1)}\mathcal{Q}[\psi^{(1,0)},h^{(0,1)}_{\mu \nu }]\,.
\end{equation}
This expression captures the leading-order \emph{dynamical perturbation} from the secondary BH to the environment.  While the formulation above focuses on a single matter field, our framework naturally extends to multiple matter fields, enabling a direct application to more complex systems such as spin-1 fields or fluid environments. For example, consider a plasma described by an electromagnetic field $A_{\mu} = \epsilon A^{(1,0)}_{\mu} + \epsilon q A^{(1,1)}_{\mu}+\cdots$ and a fluid four-velocity $U_{\mu} = \epsilon U^{(1,0)}_{\mu} + \epsilon q U^{(1,1)}_{\mu}+\cdots$ In this case, a similar structure of equations arises (in a slight abuse of notation):
\begin{equation}
\begin{aligned}
  \delta^{(1,1,0)}\mathcal{W}_{\mu}[A_{\mu}^{(1,1)},U_{\mu}^{(1,0)}, g_{\mu \nu }] &= -  \delta^{(1,1,1)}\mathcal{W}_{\mu}[A_{\mu}^{(1,0)}, U_{\mu}^{(1,0)},h^{(0,1)}_{\mu \nu}]\,,\\
  \delta^{(1,1,0)}\mathcal{P}_{\mu}[U_{\mu}^{(1,1)},A_{\mu}^{(1,0)}, g_{\mu \nu }] &= -  \delta^{(1,1,1)}\mathcal{P}_{\mu}[U_{\mu}^{(1,0)}, A_{\mu}^{(1,0)},h^{(0,1)}_{\mu \nu}]\,.
\end{aligned}
\end{equation}
Applying our formalism to such scenarios is an exciting direction for future research.
\section{Scalar Fields}\label{sec_SF:scalar_fields}
The above framework is general and applies to any non-vacuum spacetime with a stress-energy tensor $T^{\rm env}_{\mu \nu}$ satisfying known equations of motion, provided that perturbations to the surrounding medium remain in the linear regime. We now specialise to the case of a massive scalar field around a Kerr BH, where superradiance has led to the formation of a bosonic cloud -- also known as a gravitational atom (see Section~\ref{BHenv_subsec:gatoms} for a discussion of its properties). The cloud is assumed to be in a quasi-stationary state with a spin given by~\eqref{eq:spinatsat} and residing in its dominant dipolar ground state, $\ket{n\ped{b}\ell\ped{b} m\ped{b}} = \ket{211}$.
\vskip 2pt
We consider the gravitational atom to be perturbed by a point-particle on an equatorial, circular orbit in the Kerr geometry. The matter Lagrangian for a massive scalar is given by
\begin{equation}\label{eq:action}
\begin{aligned}
\mathcal{L}^{\rm env}[{\bf \Phi}] =  \nabla_\nu {\bf \Phi} \nabla^\nu {\bf \Phi}^* - \mu^2 |{\bf \Phi}|^2\,,
\end{aligned}
\end{equation}
yielding $n=2$ and the equation of motion~\eqref{eq:GenericEquationsOfMotionPsi_matter}:
\begin{equation}
   \mathcal{Q}[{\bf \Phi}, {\bf g}_{\mu\nu}]=\frac{{1}}{\sqrt{- {\bf g}}} \partial_{\mu} (\sqrt{- {\bf g}}\, \partial^\mu {\bf \Phi}) - \mu^2 {\Phi}\,. 
\end{equation}
Following the previous section, we expand the scalar field as
\begin{equation}\label{eq:expansion_field2}
    \mathbf{\Phi} = \epsilon \phi^{(1,0)} + \epsilon q \phi^{(1,1)}+\epsilon  q^2 \phi^{(1,2)}+\cdots
\end{equation}
We take the background solution $\epsilon \phi^{(1,0)}$ to be that of a superradiant cloud, whose normalisation is set such that its total mass is given by $M\ped{c}$. The characteristic length scale of the cloud is given by its Bohr radius $L^{\rm env} = (\mu \alpha)^{-1} = \alpha^{-2}M$, yielding, $\epsilon = \alpha^3 \sqrt{M\ped{c}/M}$~\eqref{eq:EmviromentalExpansionParameter}. This expression defines our perturbative parameter and imposes the condition $\epsilon \ll 1$, provided that both $\alpha$ and $M_{\rm c}/M$ are sufficiently small. The expansions of the scalar equation of motion at leading dynamical order $\sim \mathcal{O}(\epsilon q) $, yields the following expressions:
\begin{align}
 \delta^{(1,0)}\mathcal{Q}[\phi^{(1,1)},g_{\mu \nu }] & =  \frac{ 1}{\sqrt{- g}} \partial_{\mu} (\sqrt{- g} \partial^\mu \phi^{(1,1)}) - \mu^2 \phi^{(1,1)} =\left( \Box- \mu^2\right) \phi^{(1,1)}\,, \label{eq:KG_pre}\\
 \delta^{(1,1)}\mathcal{Q}[\phi^{(1,0)},h^{(0,1)}_{\mu \nu }]  &= -
h^{\mu\nu}_{(0,1)}\nabla_\mu \nabla_\nu \phi^{(1,0)}-\frac{h^{(0,1)}}{2} \delta^{(1,0)}\mathcal{Q}[\phi^{(1,0)},g_{\mu \nu }] \nonumber \\&- \left( \nabla_\mu \bar{h}^{\mu\nu}_{(0,1)}\right) \left( \nabla_\nu \phi^{(1,0)} \right)\,.\label{eq:operators2}
\end{align}
Here, $\bar{h}_{\mu\nu}$ is the trace-reversed metric perturbation and from now on, all covariant derivatives and d'Alembertian operators are defined with respect to the background metric. As $\phi^{(1,0)}$ is a test-field solution on Kerr, the second term on the right-hand side of eq.~\eqref{eq:operators2} vanishes. Moreover, as we solve for the $\mathcal{O}(q)$ metric perturbation in Lorenz gauge, i.e., $\nabla_\mu \bar{h}^{\mu\nu}_{(0,1)} = 0$ (see~\cite{Dolan:2023enf,Dolan:2021ijg} and Appendix~\ref{app_BHPT_Kerr:LG}), the final term in eq.~\eqref{eq:operators2} also vanishes, resulting in the expression:
\begin{equation}\label{eq:rhs_lead}
\begin{aligned}
\delta^{(1,1)}\mathcal{Q}[\phi^{(1,0)},h^{(0,1)}_{\mu \nu }]  = -h_{(0,1)}^{\mu\nu}\nabla_\mu \nabla_\nu \phi^{(1,0)}\,. 
\end{aligned}
\end{equation}
Thus, from eqs.~\eqref{eq:dynamicalperubation},~\eqref{eq:KG_pre} and~\eqref{eq:rhs_lead}, the leading-order \emph{dynamical} perturbation from the secondary to the scalar field is governed by
\begin{equation}\label{eq:EOM_sourced}
\begin{aligned}
\left(\Box - \mu^2\right) \phi^{(1,1)} =  h_{(0,1)}^{\mu\nu}\nabla_\mu \nabla_\nu \phi^{(1,0)}\,.
\end{aligned}
\end{equation}
The specialisation to Lorenz gauge results in a source that diverges as $1/|\vec{r} - \vec{r} \ped{p}|$ at the position of the secondary. At the level of spheroidal harmonics, this leads to a source that is $C^0$ continuous, which makes it particularly well-suited for producing extended solutions across the entire domain.
\section{Scalar Field Wake}\label{sec_SF:scalar_wake}
Figure~\ref{fig:Wake_Profile} shows the wake profile of the matter field due to a secondary perturber on a prograde equatorial circular orbit.\footnote{An observant reader may note that this configuration is, strictly speaking, unrealistic. As discussed in Chapter~\ref{chap:legacy}, co-rotating inspirals disrupt the cloud early in the evolution through orbital resonances. Nevertheless, we consider co-rotating orbits here for simplicity and as an initial step, since the qualitative dynamical behaviour remains largely unchanged in the counter-rotating case, as established in a Newtonian analysis~\cite{Baumann:2021fkf}.} The system parameters are chosen deep in the relativistic regime, with orbital radius $r\ped{p}=3.5M$, $a=0.88M$ and $\alpha = 0.3$ (corresponding to a cloud whose density peaks at $r\sim 20M$). We solve eq.~\eqref{eq:EOM_sourced} by decomposing in a spheroidal harmonic basis and find a full solution that constitutes a rich wake structure in the azimuthal and equatorial planes. As the secondary scatters matter through the transfer of angular momentum, a \emph{low-density} trail is formed and a spiralling outwash causes matter flux to infinity. 
\vskip 2pt
We also analysed configurations at larger separations, where certain modes transition from radiative to bound configurations. At these radii, we observe complex changes in the cloud's morphology, including configurations where a low-density region forms in front of the secondary, with a \emph{high-density} region in its wake. As we always stay in a regime where $\Omega\ped{p}< \omega\ped{c}$, this contrasts the picture presented in studies on linear motion of BHs in homogeneous media~\cite{1999ApJ...513..252O,2007MNRAS.382..826B,Traykova:2021dua,Vicente:2022ivh,Traykova:2023qyv,Wang:2024cej,Dyson:2024qrq}, where a trail of over-density is found to form \emph{in front} of the secondary due to its relative velocity with respect to the background. Consequently, applying results from such studies
to the binary case would yield incorrect conclusions, highlighting the importance of addressing the full binary problem directly.
\begin{figure}[t!]
    \centering
    \includegraphics[scale = 0.7]{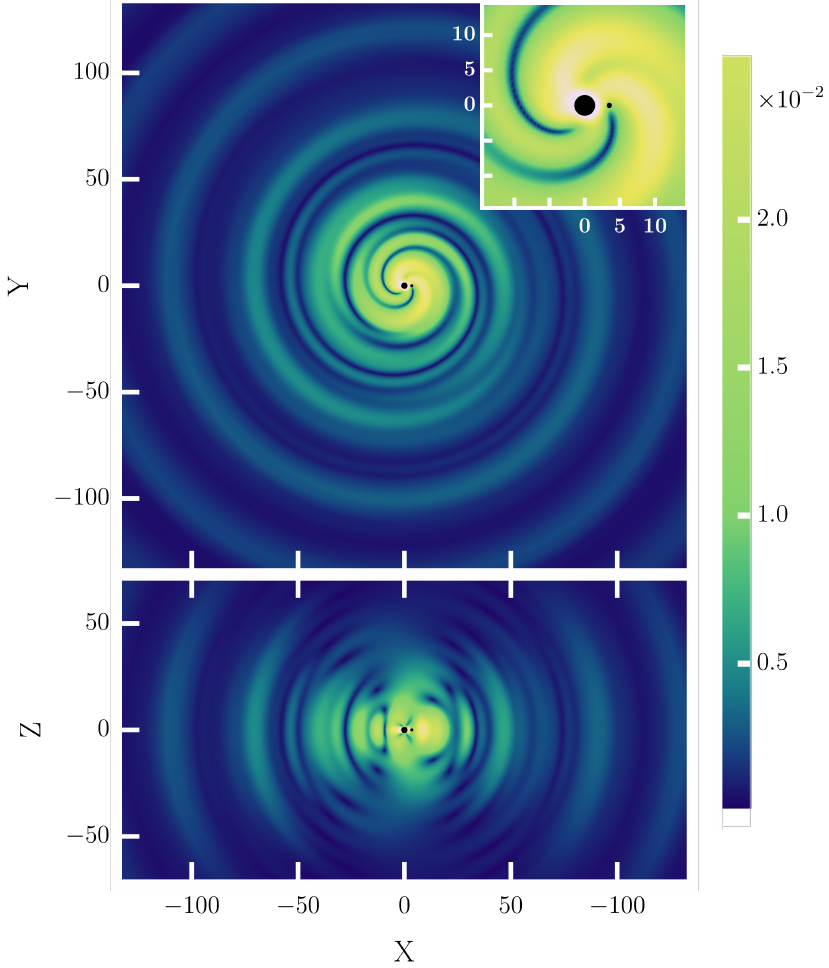}
    \caption{The absolute value of the perturbed scalar field $|\phi^{(1,1)}|$ for $\ell\geq 2$, taking $\alpha = 0.3$, $a = 0.88M$ and $r\ped{p} = 3.5M$. In the \emph{top panel}, we show an equatorial slice of the field solution, in which the $\hat{Z}$--axis is aligned with the BH spin. In the \emph{bottom panel}, we show an azimuthal slice of the field, where the secondary moves ``into the plane''.}
    \label{fig:Wake_Profile}
\end{figure}
\section{Radiative Energy Loss}\label{sec_SF:fluxes}
As the secondary orbits the central BH, its perturbation induces a transfer of energy and angular momentum to the scalar field and into GWs. These will in turn determine how the secondary evolves. It is clear from eqs.~\eqref{eq:expansion_metric}--\eqref{eq:expansion_field} that many dissipative and conservative effects will contribute to the change in orbital energy. We now derive the explicit form of the flux formulae in the case of boson clouds. For clarity, we suppress most of the perturbative indexing, though the order of each quantity should be clear from context.
\vskip 2pt
The (orbit-averaged) energy fluxes of the perturbed field to infinity and through the horizon can be calculated, respectively, as~\cite{Teukolsky:1973ha,Teukolsky:1974yv} 
\begin{equation}\label{eq:energy_fluxes}
\begin{aligned}
\dot E^{\Phi,\infty} &= -\lim_{r\to +\infty} r^2\int \mathrm{d}\Omega\, T^{\Phi}_{\mu r}K^{\mu}_{(t)}\,,\\ 
\dot E^{\Phi,\mathrm{H}} &= \lim_{r\to r_+} 2M r_+ \int \mathrm{d}\Omega\, T^{\Phi}_{\mu\nu}K^{\mu}_{(t)}l^{\nu}\,,
\end{aligned}
\end{equation}
where 
\begin{equation}\label{eq:stress}
T^{\Phi}_{\mu\nu}  = 2\nabla_{(\mu} \Phi \nabla_{\nu)} \Phi^* - g_{\mu\nu}  ( \nabla_{\sigma}\Phi\nabla^{\sigma}\Phi^* + \mu^2 |\Phi|^2 )\,.
\end{equation}
Here, $\mathrm{d}\Omega$ is the area element of the $2$--sphere, $l^{\mu}=\partial/\partial t+\Omega\ped{H}\partial/\partial \varphi$ is a null vector normal to the horizon, with $\Omega\ped{H}$ given in~\eqref{eq:OmegaH} and $K^{\mu}_{(t)}\equiv\partial/\partial t$, $K_{(\varphi)}^{\mu}\equiv\partial/\partial\varphi$ are the Killing vectors of the Kerr metric~\eqref{eq:metric_kerr}. Analogous expressions for angular momentum fluxes are obtained by swapping $K_{(t)}^{\mu} \rightarrow K_{(\varphi)}^{\mu}$ in eq.~\eqref{eq:energy_fluxes}.
We define the mass of the background boson cloud to be given by the volume integral on a spacelike slice of the time component of the stress-energy tensor. In particular, we define
\begin{equation}
    M\ped{c} = -\int_{r_+}^{\infty}  \int_{S^2} T_t^t[\phi^{(1,0)},g^{\text{Kerr}}]\,(r^2+a^2\cos^2\theta) \mathrm{d}r\,  \mathrm{d}\Omega\,.
\end{equation}
We decompose $\phi^{(1,1)}$ in a spheroidal harmonics basis, 
\begin{equation}
    \phi^{(1,1)} = \sum_{\ell m} \phi^{(1,1)}_{\ell m}(r)S_{\ell m}(\theta,\varphi,\gamma_{m\ped{g}})e^{- i (\Omega_{m\ped{g}} + \omega\ped{c}) t}\,,
\end{equation}
where $\gamma_{m\ped{g}} =  a \sqrt{(\Omega_{m\ped{g}}+\omega\ped{c})^2-\mu^2}$ and for circular orbits $\Omega_{m\ped{g}} = (m- m\ped{b})\Omega\ped{p} \equiv m\ped{g}\Omega\ped{p}$. The above equations yield expressions for the energy fluxes of individual modes:
\begin{equation}\label{eq:enery_flux_ind}
\begin{aligned}
    \dot E^{\Phi,\infty}_{\ell m} &= \lim_{r\to +\infty} r^2\Big\{2\,|\omega\ped{c} + \Omega_{m\ped{g}}|\,\mathrm{Re}\left[\sqrt{\left(\Omega_{m\ped{g}}+\omega\ped{c}\right)^{2}-\mu^2}\,\right]|\phi^{(1,1)}_{\ell m}|^{2} \Big\}\,,\\
    \dot E^{\Phi,\mathrm{H}}_{\ell m} &= \lim_{r\to r_+} 2M r\Big\{2\, \left(\omega\ped{c}+\Omega_{m\ped{g}}\right) \left(\omega\ped{c}+\Omega_{m\ped{g}} - m\Omega\ped{H}\right)|\phi^{(1,1)}_{\ell m}|^{2}\Big\}\,.
\end{aligned}
\end{equation}
Here, we have set the frequency at the superradiant threshold, i.e., $\omega = \omega\ped{c}$ and we remind the reader that  at leading-order $\phi^{(1,1)}_{\ell m} \propto 1/r$, when $r\to\infty$. 
\vskip 2pt
Similarly, the angular momentum fluxes are given by
\begin{equation}\label{eq:angmom_flux_ind}
\begin{aligned}
    \dot L^{\Phi,\infty}_{\ell m} &= \lim_{r\to +\infty} r^2\Big\{ 2m s_{m\ped{g}} \mathrm{Re}\left[\sqrt{\left(\Omega_{m\ped{g}}+\omega\ped{c}\right)^{2}-\mu^2}\right]|\phi^{(1,1)}_{\ell m}|^{2} \Big\}\,,\\
    \dot L^{\Phi,\mathrm{H}}_{\ell m} &= \lim_{r\to r_+} 2 M r\left\{ 2m\left(\omega\ped{c}+\Omega_{m\ped{g}} - m\Omega\ped{H}\right)|\phi^{(1,1)}_{\ell m}|^{2}\right\}\,,
\end{aligned}
\end{equation}
where $s_{m\ped{g}} \equiv \mathrm{sgn}\left(\omega\ped{c} + \Omega_{m\ped{g}}\right)$. While $\dot{E}^{\Phi,\,\infty/\mathrm{H}}$~\eqref{eq:enery_flux_ind} can be truly associated as a ``scalar flux'', the scalar perturbations also affect the cloud, which in turn impacts the evolution of the secondary. 
\vskip 2pt
For small mass ratios $q$, the secondary evolves adiabatically, meaning that the energy dissipated over one orbit is much smaller than the total orbital energy. Consequently, the evolution of the secondary can be constructed with a sequence of geodesics~\cite{Barack:2018yvs,Wardell:2021fyy,Chua:2020stf,Katz:2021yft,Hughes:2021exa}. As a first step towards understanding how the secondary moves from one geodesic to the next, we can consider the leading-order flux balance of the system:
\begin{equation}\label{eq:fluxbalance2}
\begin{aligned}
    \dot{E}\ped{orb} + \dot{M}\ped{c} &= -\dot{E}^{\scalebox{0.7}{$\mathrm{GW}$},\infty}-\dot{E}^{\scalebox{0.7}{$\mathrm{GW}$}, \mathrm{H}}-\dot{E}^{\Phi,\,\infty}-\dot{E}^{\Phi,\mathrm{H}}\,,\\
    \dot{L}\ped{orb} + \dot{S}\ped{c} &= -\dot{L}^{\scalebox{0.7}{$\mathrm{GW}$},\infty}-\dot{L}^{\scalebox{0.7}{$\mathrm{GW}$}, \mathrm{H}}-\dot{L}^{\Phi,\infty}-\dot{L}^{\Phi,\mathrm{H}}\,,
\end{aligned}
\end{equation}
where $S\ped{c}$ is the spin of the cloud. This balance equation allows one to evolve orbital parameters due to energy emission from the secondary and the environment. Importantly, this formula excludes the effects of conservative energy transfer between the orbit and bound states of the cloud. As such, it should be used with some caution until all contributions up to order $\mathcal{O}(\epsilon^2 q^2)$ (e.g., $h^{(2,1)}$) are fully understood. Even so, in the absence of a complete understanding, this equation provides a useful starting point for time-domain evolutions and generating relativistic waveforms for EMRIs with environments. Some insights into how such conservative transfer might occur in a relativistic setting have been studied in~\cite{Redondo-Yuste:2023snb}.
\vskip 2pt
To compute the rate at which the mass and spin of the cloud changes, we make use of the global $U(1)$-symmetry of the (complex) scalar field, whose conserved current implies the existence of a Noether charge $Q$:
\begin{equation}
    Q = \int_{\Sigma} \mathrm{d}^{3}x\sqrt{-g}\,j^0_{\Phi}\,,
\end{equation}
where $\Sigma$ is a space-like hypersurface and
\begin{equation}\label{KG_current}
    j^{\Phi}_{\mu} = -i\left(\Phi^{*}\partial_\mu\Phi-\Phi\partial_\mu\Phi^{*}\right)\,.
\end{equation}
The mass and spin of the cloud are then related to the cloud's Noether charge, i.e., $M\ped{c} = \omega\ped{c} Q$ and $S\ped{c} = m\ped{b}Q$, respectively. The rate of change of the scalar charge is
\begin{equation}\label{eq:charge_fluxes}
\begin{aligned}
\dot Q^{\Phi,\infty} &= -\lim_{r\to +\infty} r^2\int \mathrm{d}\Omega\, j^{\Phi}_{r}\,,\\ 
\dot Q^{\Phi,\mathrm{H}} &= \lim_{r\to r_+} 2M r \int \mathrm{d}\Omega\, j^{\Phi}_{\mu}l^{\mu}\,,
\end{aligned}
\end{equation}
leading to:
\begin{equation}\label{eq:Noether_ind}
\begin{aligned}
    \dot Q^{\Phi,\infty}_{\ell m} &= -\lim_{r\to +\infty} r^2\Big\{ 2\,s_{m\ped{g}}\, \mathrm{Re}\left[\sqrt{\left(\Omega_{m\ped{g}}+\omega\ped{c}\right)^{2}-\mu^2}\right]|\phi^{(1,1)}_{\ell m}|^{2} \Big\}\,,\\
    \dot Q^{\Phi,\mathrm{H}}_{\ell m} &= -\lim_{r\to r_+} 2 M r_+\left\{ 2\,\left(\omega\ped{c}+\Omega_{m\ped{g}} - m\Omega\ped{H}\right)|\phi^{(1,1)}_{\ell m}|^{2}\right\}\,.
\end{aligned}
\end{equation}
Through the Noether charge, we can then define the scalar energy and angular momentum \emph{power} as
\begin{equation}
\begin{aligned}
    \dot{E}^{\mathrm{s},\infty/\mathrm{H}} &= \dot{E}^{\Phi,\infty/\mathrm{H}} + \omega\ped{c} \dot{Q}^{\Phi,\infty/\mathrm{H}}\,,\\
    \dot{L}^{\mathrm{s},\infty/\mathrm{H}} &= \dot{L}^{\Phi,\infty/\mathrm{H}} + m\ped{b} \dot{Q}^{\Phi,\infty/\mathrm{H}}\,.
\end{aligned}
\end{equation}
Plugging in eqs.~\eqref{eq:enery_flux_ind}, \eqref{eq:angmom_flux_ind} and~\eqref{eq:Noether_ind} in the balance equation~\eqref{eq:fluxbalance2}, we find:
\begin{equation}\label{eq:scalar_flux}
\begin{aligned}
    \dot E^{\mathrm{s},\infty}_{\ell m} &= \lim_{r\to +\infty} r^2\Big\{ 2\,\Omega_{m\ped{g}}s_{m\ped{g}}\,\mathrm{Re}\left[\sqrt{\left(\Omega_{m\ped{g}}+\omega\ped{c}\right)^{2}-\mu^2}\right]|\phi^{(1,1)}_{\ell m}|^{2} \Big\}\,,\\
    \dot E^{\mathrm{s},\mathrm{H}}_{\ell m} &= \lim_{r\to r_+} 2 M r\left\{ 2\Omega_{m\ped{g}}\left(\omega\ped{c}+\Omega_{m\ped{g}} - m\Omega\ped{H}\right)|\phi^{(1,1)}_{\ell m}|^{2}\right\}\,.
\end{aligned}
\end{equation}
\begin{figure}[t!]
    \centering
    \includegraphics[width = \linewidth]{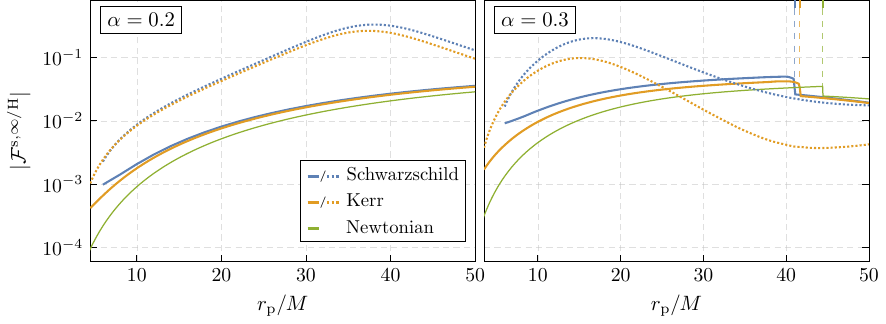}
    \caption{The total flux to infinity (solid lines) and through the horizon (dotted lines) considering a prograde orbit and $\alpha = 0.2$ (\emph{left panel}) or $\alpha = 0.3$ (\emph{right panel}). Note that the horizon fluxes are negative on the entire radial domain. The sharp features in the infinity flux in the \emph{right panel}, computed using eq.~\eqref{eq:discontinuities}, are marked by vertical dashed lines. Note the Schwarzschild results stop at the innermost stable circular orbit (ISCO) ($r\ped{p} = 6M$). We sum up to $\ell = 6$ (5) for the infinity (horizon) fluxes.}
    \label{fig:Fluxes_Inf_Hor}
\end{figure}
Finally, for the angular momentum, we have
\begin{equation}
\begin{aligned}
    \dot L^{\mathrm{s},\infty}_{\ell m} &= \lim_{r\to +\infty} r^2\Big\{ 2m\ped{g} s_{m\ped{g}}\,\mathrm{Re}\left[\sqrt{\left(\Omega_{m\ped{g}}+\omega\ped{c}\right)^{2}-\mu^2}\right]|\phi^{(1,1)}_{\ell m}|^{2} \Big\}\,,\\
    \dot L^{\mathrm{s},\mathrm{H}}_{\ell m} &= \lim_{r\to r_+} 2 M r\left\{ 2m\ped{g}\left(\omega\ped{c}+\Omega_{m\ped{g}} - m\Omega\ped{H}\right)|\phi^{(1,1)}_{\ell m}|^{2}\right\}\,,
\end{aligned}
\end{equation}
which satisfy
\begin{equation}\label{eq:energy_angmom}    
\dot{E}^{\mathrm{s},\infty/\mathrm{H}} = \Omega\ped{p} \dot{L}^{\mathrm{s},\infty/\mathrm{H}}\,.
\end{equation}
Consequently, the backreaction from the leading-order scalar fluxes onto the secondary evolves circular orbits into circular orbits, justifying the quasi-circular limit studied in this chapter. Furthermore, because of this relation~\eqref{eq:energy_angmom}, it is sufficient to look at the energy fluxes only.
\vskip 2pt
In a slight abuse of terminology, we define the \emph{scalar flux} of the environment as 
\begin{equation}\label{eq:scalar_fluxx}
\mathcal{F}^{\mathrm{s}, \infty/\mathrm{H}} \equiv \epsilon^{-2} q^{-2}\left( \dot{E}^{\Phi,\,\infty/\mathrm{H}}+ \dot{M}\ped{c}^{\infty/\mathrm{H}} \right)\,,
\end{equation}
and calculate its emission to infinity and through the horizon. In Figure~\ref{fig:Fluxes_Inf_Hor}, we show the scalar fluxes for $\alpha = 0.2$ (\emph{left panel}) and $\alpha = 0.3$ (\emph{right panel}), corresponding to highly spinning BHs with $a = 0.69M$ and $a = 0.88M$, respectively. For comparison, we also include the fluxes obtained in Schwarzschild\footnote{In Schwarzschild, massive scalar fields still settle on quasi-bound states, even though spin is essential for superradiance to occur. These states are decaying ($M\mathrm{Im}[\omega\ped{c}] < 0$) however, due to absorption at the horizon, preventing the cloud from achieving a stationary configuration. Following~\cite{Brito:2023pyl}, we explicitly ignore this, setting $M\mathrm{Im}[\omega\ped{c}]  = 0$.} and in the Newtonian regime~\cite{Baumann:2021fkf,Baumann:2022pkl,Tomaselli:2023ysb} (see Appendix~\ref{app_BHPT_Kerr:scalar_Newt}). 
\vskip 2pt
The main features in Figure~\ref{fig:Fluxes_Inf_Hor} are:
\begin{enumerate}
\item As the orbital separation between the binary components grows, effects related to spin of the primary become small for the flux to infinity, and the Schwarzschild and Kerr results become similar. At the same time, the Newtonian treatment converges towards the relativistic cases at large radii. It is worthwhile to compare two cases in more detail:~Kerr versus Schwarzschild, and ``our'' Schwarzschild results versus those from an earlier work~\cite{Brito:2023pyl}, which used a different method and gauge. We show the relative differences in the fluxes at infinity and at the horizon in Figure~\ref{fig:RelDiffFluxes_Inf_Hor}, using our Schwarzschild results as the reference.
\begin{figure}[t!]
    \centering
    \includegraphics[width = \linewidth]{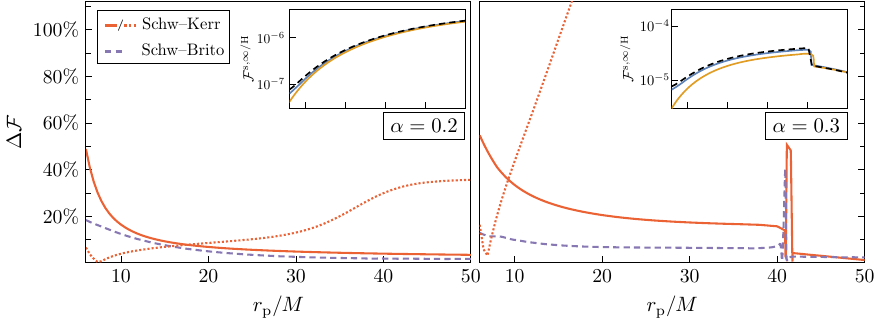}
    \caption{The relative error $\Delta \mathcal{F}$ in the total flux to infinity (solid lines) and through the horizon (dotted lines), taking our Schwarzschild data as a reference. We consider a prograde orbit and $\alpha = 0.2$ (\emph{left panel}) or $\alpha = 0.3$ (\emph{right panel}). The comparison is between our Kerr and Schwarzschild results [{\color{Mathematica4}{red}}], as well as our Schwarzschild results with those from an earlier work~\cite{Brito:2023pyl} [{\color{Mathematica5}{purple}}], which used a different gauge. In the inset, we show the actual flux $\mathcal{F}^{\mathrm{s}, \infty}$ in Schwarzschild/Kerr from our data [{\color{Mathematica1}{blue}}/{\color{Mathematica2}orange}] compared to the one from~\cite{Brito:2023pyl} [\textbf{black} dashed]. The horizontal axis in the inset is the same as in the main plot.}
    \label{fig:RelDiffFluxes_Inf_Hor}
\end{figure}
As expected, at large radii, the relative differences between the Schwarzschild and Kerr flux becomes small, dropping to a few percent at $r\ped{p} = 50M$. Closer to the ISCO, the differences increase, reaching approximately $50\%$. This emphasises the need to perform calculations in Kerr and not in Schwarzschild. When comparing our Schwarzschild results to those in~\cite{Brito:2023pyl}, we find relative differences up to $20\%$, due to gauge ambiguities, as we will discuss at the end of this section. In addition, mode-by-mode comparisons reveal notable differences between our results and~\cite{Brito:2023pyl}, particularly in the quadrupole.
\vskip 2pt
Importantly, the scalar flux we calculate~\eqref{eq:scalar_fluxx} does not describe the complete flux at $\mathcal{O}(\epsilon^2q^2)$. This is due to the fact that there are two additional gravitational contributions to the energy fluxes which we do not consider. One is due to the expansion of the Einstein operator, $\delta^2G[h^{(2,1)},h^{(0,1)} ]$, and arises as an additional term in the integrand of $\mathcal{F}^{\mathrm{s}, \infty/\mathrm{H}}_{(2,2)}$, which we do not calculate. The other terms comes from the $\mathcal{O}(\epsilon^2)$ correction to the orbital frequency, $\Omega\ped{p}^{(2,0)}$, from the background matter field. Thus, to calculate the full gauge invariant flux at $\mathcal{O}(\epsilon^2q^2) $ one requires the expression:
\begin{equation}\label{eq:gaugeinvariantflux}
\mathcal{F}^{ \infty/\mathrm{H}}_{(2,2)} \equiv \mathcal{F}^{\mathrm{s}, \infty/\mathrm{H}}_{(2,2)} + \frac{\mathrm{d}\mathcal{F}^{\scalebox{0.7}{$\mathrm{GW}$}, \infty/\mathrm{H}}_{(0,2)}}{\mathrm{d} \Omega\ped{p}^{\text{Kerr}}}\Omega\ped{p}^{(2,0)}\,,
\end{equation}
where $\mathcal{F}^{\mathrm{s}, \infty/\mathrm{H}}_{(2,2)}$ should be corrected to include the term arising from the expansion of the Einstein operator and $\mathcal{F}^{\scalebox{0.7}{$\mathrm{GW}$}, \infty/\mathrm{H}}_{(0,2)}$ is the first-order vacuum flux due to the secondary. Moreover, $\Omega\ped{p}^{\text{Kerr}}$ is the frequency in Kerr for a given orbital radius. The calculation of the $h^{(2,0)}$ and $h^{(2,1)}$ metric perturbations is necessary to calculate these additional terms and is still an open problem.
\item The horizon flux exceeds the infinity flux across most of the shown radial domain. It is dominated by the $(\ell,m) = (0,0)$ mode and always negative, indicating that the binary's orbit gains energy. This is due to a resonance between bound states of the cloud:~the initial $\ket{211}$ state resonates with $\ket{100}$, which has lower energy and angular momentum. This surplus is fed back into the orbit, potentially giving rise to a \emph{floating orbit}~\cite{Baumann:2019ztm,Tomaselli:2024dbw,Tomaselli:2024bdd}, where the binary's evolution is slowed down or even stalled for a period of time. For $\alpha = 0.3$, the $(\ell,m) = (2,2)$ mode becomes significant around $r\ped{p} \sim 50M$, nearly overtaking the $(0,0)$ contribution. Unlike the $(0,0)$ mode, it produces a positive horizon flux, inducing a \emph{sinking orbit}:~a period of accelerated inspiral. Such resonances are key observables for probing the cloud's properties, yet they are also efficient at depleting the cloud itself, as we saw in Chapter~\ref{chap:legacy}.
\item The \emph{right panel} reveals a sharp feature in the flux to infinity, consistent with earlier studies~\cite{Baumann:2021fkf,Baumann:2022pkl,Tomaselli:2023ysb,Brito:2023pyl}. These arise when a new mode starts contributing to the flux, specifically, occurring when
\begin{equation}\label{eq:discontinuities}
 r^{*,m}\ped{p} = \left(\frac{m-m\ped{b}}{(\mu-\text{Re}[\omega\ped{c}])M}-\tilde a\right)^{2/3}M\,.
\end{equation}
\vskip 2pt
Using eq.~\eqref{eq:discontinuities} and calculating $\omega\ped{c}$ with Leaver's method~\cite{Leaver:1985ax,Dolan:2007mj}, we find for $\alpha = 0.3$ that $r^{*,2}\ped{p}/M = 41.01, 41.66, 44.44$ in the Schwarzschild, Kerr and Newtonian case, respectively, in precise agreement with Figure~\ref{fig:Fluxes_Inf_Hor} (vertical dashed lines). Notably, sharp features are absent in the $\alpha = 0.2$ case as they occur at larger radii (e.g.~$r^{*,2}\ped{p} \sim 100M$). The physical origin of these features lies in the long-range nature of the gravitational potential, as detailed in Appendix D of~\cite{Baumann:2021fkf}. Close to $r^{*,m}\ped{p}$, the wavelength of the modes becomes extremely large, which requires the flux to be extracted far out. The small dip in the flux preceding the feature is thus merely a numerical artefact. In Figure~\ref{fig:Wake_Profile_SM}, we show an equatorial slice of the field solution just before (\emph{top panel)} and after (\emph{bottom panel)} $r^{*,2}\ped{p}$~\eqref{eq:discontinuities}. Indeed, the morphology of the cloud changes completely ``in a short time''. The reason being that $(\ell,m)=(2,2)$ mode of the field solution transitions from a radiative configuration -- with $\sim 1/r$ decay -- to a bound configuration that decays exponentially. This sharp transition in the field profile arises due to the single harmonic state configuration of the boson cloud background and will not be as prominent in environments with a more general harmonic dependence.
\begin{figure}[t!]
    \centering
    \includegraphics[scale = 0.7]{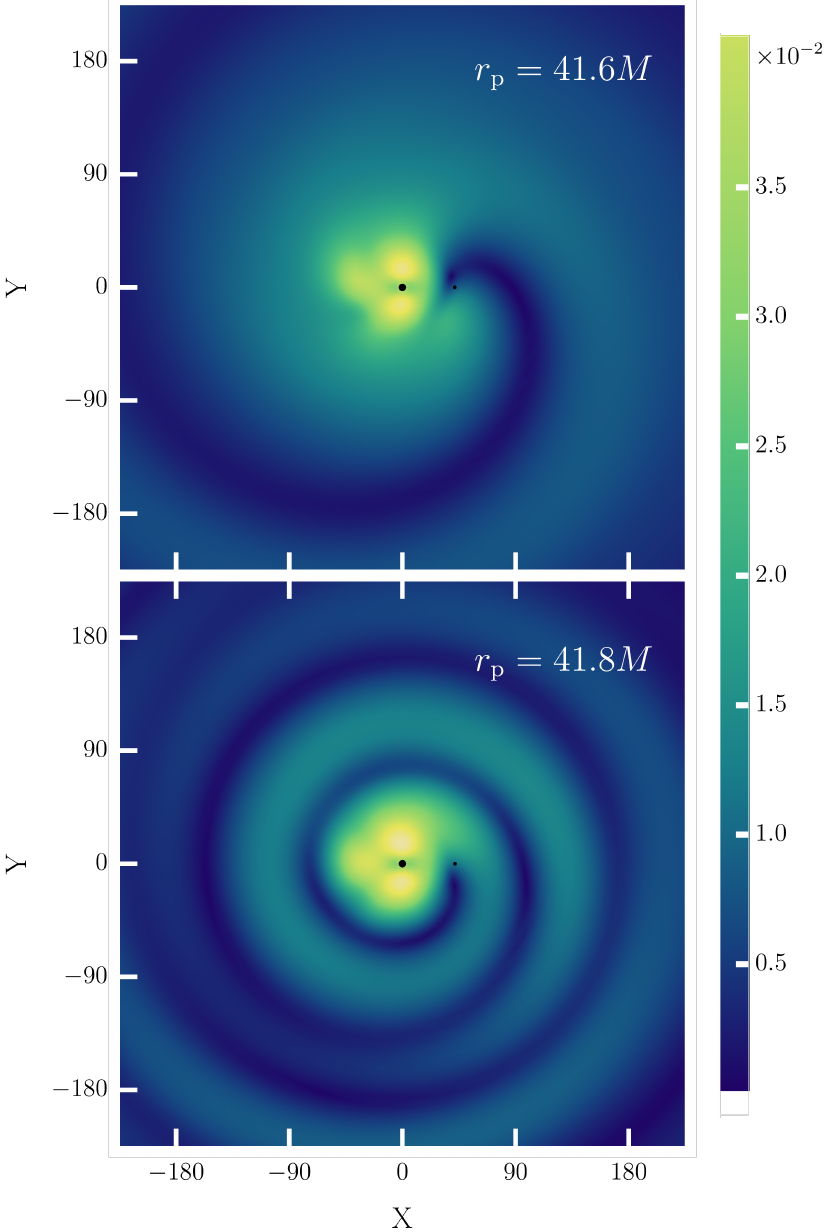}
    \caption{The absolute value of the perturbed scalar field $|\phi^{(1,1)}|$ for $\ell\geq 2$, taking $\alpha = 0.3$, $a = 0.88M$. In the \emph{top panel}, we show an equatorial slice of the field solution at $r\ped{p} = 41.6M$, in which the $\hat{Z}$--axis is aligned with the BH spin. In the \emph{bottom panel}, we show an equatorial slice at $r\ped{p} = 41.8M$.}
    \label{fig:Wake_Profile_SM}
\end{figure}
\item Consistent with previous studies~\cite{Baumann:2021fkf,Baumann:2022pkl,Brito:2023pyl,Duque:2023seg}, we observe that scalar fluxes tend to dominate over gravitational fluxes during the early inspiral stage. As the gravitational and scalar fluxes rely on independent perturbative parameters, i.e., $q$ and $\epsilon$, a general comparison with the results in Figure~\ref{fig:Fluxes_Inf_Hor} should not be made. However, an example case for a given value for $q$ and $\epsilon$ is shown in Figure~\ref{fig:RatioFluxes_Inf_Hor}. Since both the GW and scalar flux scale as $q^2$, this factor cancels out in their ratio. We show a case for which the cloud has obtained a maximum mass $\eta = M\ped{c}/M = 0.1$~\eqref{eq:massatsat}, with a typical EMRI mass ratio $q = 10^{-6}$ and $\alpha = 0.3$. The results show that scalar horizon fluxes dominate over the gravitational horizon fluxes in most of the inspiral, while the scalar fluxes to infinity will probably overtake the gravitational fluxes at larger radii. Finally, we note that the horizon flux is less relevant in spherically symmetric structures that were studied before~\cite{Brito:2023pyl,Duque:2023seg}. A possible reason is that for spherical structures the $(\ell,m)=(0,0)$ mode does not contribute to $\dot E^{\mathrm{s,H}}$, unlike for dipolar clouds. Since there is no angular barrier for $(\ell,m)=(0,0)$ modes, those are more ``easily'' absorbed at the horizon.
\begin{figure}[t!]
    \centering
    \includegraphics[scale = 1]{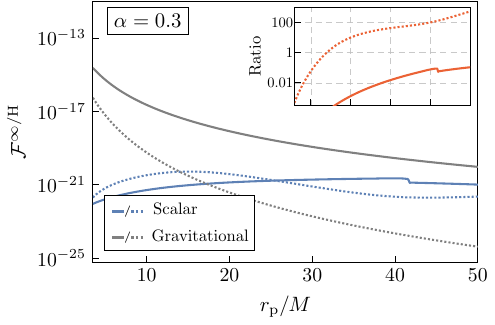}
    \caption{Fluxes to infinity and through the horizon for the scalar and gravitational case, including the perturbative prefactors and considering $q = 10^{-6}$, $\epsilon^2 = 0.1\alpha^3$, $\alpha = 0.3$. In the inset, we show the \emph{ratio} between the two, i.e., $\mathcal{F}\ped{s}/\mathcal{F}_{\scalebox{0.6}{$\mathrm{GW}$}}$. The horizontal axis of the inset is the same as in the main plot.}
    \label{fig:RatioFluxes_Inf_Hor}
\end{figure}
\end{enumerate}
\vskip 2pt
As can be seen in Figure~\ref{fig:RelDiffFluxes_Inf_Hor}, our results for the scalar flux in Schwarzschild are not in full agreement with previous work~\cite{Brito:2023pyl}, which used a different gauge. We find a discrepancy of up to $\sim 10\%$.\footnote{After completion of this thesis, a typo in~\cite{Brito:2023pyl} was identified. Correcting it reduces the discrepancy between our Schwarzschild results and those of~\cite{Brito:2023pyl} to below the sub-percent level across the domain in a mode-by-mode comparison.} A plausible source of this discrepancy lies in the nature of the background solution in Schwarzschild, which includes an exponentially decaying term $\sim e^{-\mathrm{Im}[\omega\ped{c}] t}$, where $\mathrm{Im}[\omega\ped{c} M]$ is $\sim 1-10\%$ the value of the unnormalised flux. This decaying behaviour makes the background ill-suited as a stationary state to perturb around in frequency domain. To address this, we follow~\cite{Brito:2023pyl} and set $\mathrm{Im}[\omega\ped{c} M] =0 $ by hand, which means the background solution is not an exact solution of the homogeneous Klein-Gordon equation. This approximation introduces gauge dependence in the asymptotic values of the perturbed scalar field and violates the conservation laws underpinning eq.~\eqref{eq:fluxbalance2}. This issue does not arise in Kerr where a stationary background solution can be found.
\section{Summary and Outlook}\label{sec_SF:conclusions}
In this chapter, we developed a framework to study BH environments in EMRIs. As an example case, we apply it to boson clouds and, for the first time, self-consistently compute the perturbation of an EMRI to the environment in the Kerr geometry. We have demonstrated the importance of performing these calculations in Kerr, by comparing them with fluxes in Schwarzschild. For less relativistic environments ($\alpha = 0.2$), where the density peaks at $\sim 50M$, we find relative differences of around 10\%, increasing to $50\%$ near the ISCO. In more relativistic environments ($\alpha = 0.3$), these differences are even more significant, attaining $30-100\%$ throughout the region where EMRIs are expected to enter the LISA band.
\vskip 2pt
Additionally, we solved for the field perturbations across the entire domain, revealing a rich wake structure induced by the secondary. Our results demonstrate how the morphology of the environment changes with the position of the secondary, highlighting the intricate and rich dynamics of these systems, which are linked to striking observational signatures with future GW detectors~\cite{Baumann:2018vus,Baumann:2019ztm,Baumann:2021fkf,Cole:2022yzw,Baumann:2022pkl,Tomaselli:2023ysb,Brito:2023pyl,Duque:2023seg,Tomaselli:2024bdd,Boskovic:2024fga,Tomaselli:2024dbw,Khalvati:2024tzz}. These results raise important questions about existing studies that use linear motion of BHs in a homogeneous medium as a proxy for dynamical friction in a binary inspiral (e.g., \cite{Kavanagh:2020cfn,Coogan:2021uqv,Dosopoulou:2023umg,Macedo:2013qea,Cardoso:2019rou}). Such approximations, or using Schwarzschild as a background are clearly inadequate and will lead to significant errors. Instead, the correct approach in perturbation theory is to use the framework developed in this chapter.
\vskip 2pt
We expect this work to serve as the starting point for self-consistent modelling of EMRIs and environments in Kerr. There are several directions that warrant further exploration in the future. For instance, applying our framework to the Navier-Stokes system would provide a crucial step towards understanding EMRI dynamics in accretion disks in the fully relativistic regime. Another key challenge still lies in calculating the conservative and dissipative effects of all field perturbations, and the slow-time contributions inherent to these systems have yet to be explored. A two-timescale analysis (see \cite{Hinderer:2008dm,Miller:2020bft}) will be necessary to understand how all of these contributions affect the binaries' orbital parameters.
\endgroup
\chapter{In the Grip of the Disk:~Dragging the Companion through an AGN}\label{chap:capture}
\vspace{-0.8cm}
\hfill \emph{Considerate la vostra semenza:}

\hfill \emph{fatti non foste a viver come bruti,}

\hfill \emph{ma per seguir virtute e canoscenza.}
\vskip 5pt

\hfill Dante Alighieri, \emph{Inferno}, Canto XXVI
\vskip 35pt
\noindent Much of this thesis has explored how BHs can act as powerful probes of new, ``exotic'' physics. But even if such physics does not exist, environments will still play a critical role in GW astrophysics. In this final chapter, I turn to one particularly compelling example -- an environment with direct observational support:~accretion disks in active galactic nuclei (AGN) (see Section~\ref{BHenv_sec:accretion_disks_AGN}). These disks may prove essential to GW observations, not only by imprinting themselves on the waveforms of compact-object binaries~\cite{Tanaka_2002,Yunes:2011ws,Kocsis:2011dr,Derdzinski:2018qzv,Duffell:2019uuk,Tagawa:2019osr,Derdzinski:2020wlw,Tagawa:2020qll,Pan:2021ksp,Pan:2021oob,Zwick:2021dlg,Speri:2022upm,Derdzinski:2022ltb,Garg:2022nko,Morton:2023wxg,Garg:2024zku,Copparoni:2025jhq}, but also by helping to form the binaries in the first place. In the standard ``loss-cone'' scenario, EMRIs form when stellar-mass objects are scattered onto tightly bound orbits via multi-body interactions with a surrounding stellar cluster~\cite{Amaro-Seoane:2012lgq,Babak:2017tow,Gair:2017ynp}. An alternative channel has emerged in gas-rich AGN environments, where the disk facilitates the \emph{capture} of a compact objects, potentially increasing the EMRI formation rate by orders of magnitude~\cite{Pan:2021oob,Pan:2021ksp,Wang:2022obu,Derdzinski:2022ltb}. The captured secondary typically follows a highly eccentric, generically inclined orbit, that intersects the disk twice per cycle. Over time, repeated interactions with the disk gradually align the orbit with the disk plane~\cite{Fabj:2020qqc, Nasim:2022rvl, 2023MNRAS.522.1763G, 2024MNRAS.528.4958W}, where subsequent gas-driven migration accelerates its inward drift~\cite{1979ApJ...233..857G, 1980ApJ...241..425G, Tanaka_2002, 2004ApJ...602..388T, Kocsis:2011dr,Speri:2022upm,Duque:2024mfw}. In some cases, compact objects may even form directly within the disk;~a process known as \emph{in-situ} formation~\cite{Derdzinski:2022ltb}.
\vskip 2pt
AGN disks have also been proposed as ``nurseries'' for BH binaries in the LIGO--Virgo--KAGRA band~\cite{Stone:2016wzz,Bartos:2016dgn,Leigh:2017wff,Mckernan:2017ssq,Secunda:2018kar,Tagawa:2019osr,Grobner:2020drr}. Realistic disk models suggest the existence of \emph{migration traps}:~regions where the torque exerted by the disk changes sign~\cite{Bellovary:2015ifg}. In such regions, objects at larger radii migrate inwards, while those at smaller radii migrate outwards, leading to a natural accumulation of compact objects. This process can facilitate hierarchical mergers within AGN disks~\cite{McKernan_2012,McKernan_2014,Mckernan:2017ssq,Yang:2019cbr,McKernan_2020,Gerosa:2021mno}. Notably, this mechanism has been invoked to explain events like GW190521, where one of the merging BHs lies in the pair instability gap~\cite{Toubiana:2020drf,Sberna:2022qbn}.
\vskip 2pt
The rich phenomenology of BHs in AGN disks highlights the need for a precise description of the drag forces acting on compact objects in generic orbits. While disk-satellite interactions in EMRIs have been studied extensively~\cite{1998MNRAS.293L...1V,1999A&A...352..452S,MacLeod:2019jxd,Fabj:2020qqc,Nasim:2022rvl, 2023MNRAS.522.1763G,2024MNRAS.528.4958W,Li:2025zgo}, several key aspects remain underexplored or require more accurate modelling. Many previous works approximate the orbital evolution by considering two scatterings at most and extrapolating those results over long timescales. Additionally, these studies often assume circular or highly symmetric orbits, which can obscure more complex behaviour and have led to conflicting conclusions in the literature.
\vskip 2pt
In this chapter, I present the results of~\cite{Spieksma:2025wex}, which introduces a novel framework for consistently tracking the evolution of a secondary through an arbitrary number of disk crossings. This approaches enables (i) the \emph{identification} of regions in parameter space where the disk drags the secondary to align with it within realistic timescales and (ii) the \emph{characterisation} of the orbit at the end of the drag process, as the secondary has aligned. This can serve as important input for source modelling with LISA. For example, in vacuum, EMRIs are expected to have moderate eccentricity when entering the LISA band~\cite{Hopman:2005vr,Amaro-Seoane:2012lgq,Babak:2017tow}. As it turns out, the presence of a disk changes this outcome. The framework developed here also reveals new dynamical behaviours for initially highly eccentric binaries and enables a systematic comparison with previous studies, clarifying existing discrepancies in the literature.
\vskip 2pt
This chapter is organised as follows. In Section~\ref{sec_Capture:setup}, I describe the setup, and introduce the orbital parameters necessary to describe the secondary. In Section~\ref{sec_Capture:scattering}, I describe the hydrodynamic drag and accretion in the disk. I then present the main results and comparisons with previous work in Section~\ref{sec_Capture:results}. Finally, I conclude in Section~\ref{sec_Capture:conclusions}.
\section{Binary System}\label{sec_Capture:setup}
\begin{figure}[t!]
    \centering
    \includegraphics[scale = 1.1]{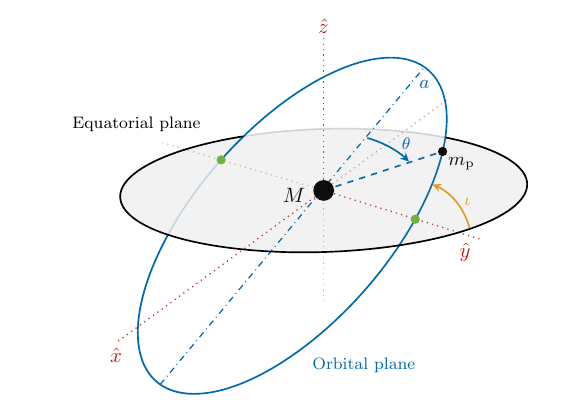}
    \caption{Schematic illustration of our setup. The system features an accretion disk in the equatorial plane, while the orbital plane [{\color{cornellBlue}{blue}}] is inclined by an angle $\iota$ and follows an eccentric trajectory with semi-major axis $a$. The angle between the secondary (with mass $m_{\rm p}$) and the semi-major axis is the true anomaly $\theta$. The primary (with mass $M$) resides at one of the focal points of the ellipse. The axes are oriented such that the secondary interacts with the disk -- thereby accreting matter or experiencing a drag -- at the designated \emph{scattering points}, located at $\vec{r} = (0, y, 0)$, and marked by green dots.}
    \label{fig:schematic_overview}
\end{figure}

We consider a binary system consisting of two BHs:~a non-spinning primary of mass $M$ and a secondary of mass $m_{\rm p}$, such that the mass ratio $q \equiv m_{\rm p}/M \ll 1$. The primary is surrounded by the AGN disk, which we define as the \emph{equatorial plane}. The secondary follows a generic orbit characterised by an inclination $\iota$\footnote{In this chapter, we denote the inclination by $\iota$, rather than the $\beta$ used in Chapter~\ref{chap:legacy} to avoid confusion with the $\beta$-disk model of the Shakura–Sunyaev disk.} relative to the disk, an eccentricity $\varepsilon$, and a semi-major axis $a$. A schematic illustration of this configuration is shown in Figure~\ref{fig:schematic_overview}.
Throughout this chapter, we use Cartesian coordinates, identifying the $\hat{z}$--axis as orthogonal to the equatorial plane.
\vskip 2pt
We focus on the binary just after capture, when the secondary is far away from the primary and follows a Keplerian orbit. The corresponding orbital energy and angular momentum are given by
\begin{equation}\label{eq:E&L}
E_{\rm orb} = -\frac{Mm_{\rm p}}{2a}\,,\quad
L_{\rm orb} = m_{\rm p}\sqrt{Ma(1-\varepsilon^2)}\,.
\end{equation}
The separation between the secondary and the primary, with the latter located at a focal point of the ellipse, reads
\begin{equation}\label{eq:r_nextscatter}
    R(t) = \frac{a(1-\varepsilon^2)}{1+\varepsilon\cos{\theta(t)}}\,.
\end{equation}
Here, $\theta$ denotes the \emph{true anomaly}, defined as the angle between the semi-major axis and the position of the secondary as it moves along the orbit (see Figure~\ref{fig:schematic_overview}). In what follows, we will focus exclusively on the secondary’s position at either the ascending or descending node. Consequently, we omit the explicit time dependence in~\eqref{eq:r_nextscatter} and refer to $R$ and $\theta$ simply as the radial distance and true anomaly evaluated at these nodes. Note that the true anomaly at the ascending node is commonly identified with the \emph{argument of periapsis} (modulo a minus sign). Inverting relation~\eqref{eq:r_nextscatter} and using eq.~\eqref{eq:E&L}, we find:
\begin{equation}\label{eq:theta}
    \cos{\theta} = \frac{1}{\varepsilon}\left(\frac{L_{\rm orb}^2}{m_{\rm p}^2M R}-1\right)\,.
\end{equation}
Conservation of mechanical energy determines the orbital velocity of the secondary:
\begin{equation}\label{eq:v_visvisa}
    v_{\rm orb}=\sqrt{M \left(\frac{2}{R}-\frac{1}{a}\right)}\,.
\end{equation}
Finally, the eccentricity can be expressed as
\begin{equation}\label{eq:eccentricity_prime}
    \varepsilon = \sqrt{1+\frac{2L^2_{\rm orb}E_{\rm orb}}{m_{\rm p}^3M^2}}\,.
\end{equation}
The above relations will be used later to update the orbital parameters after interactions between the disk and the secondary.
\section{Scattering Process}\label{sec_Capture:scattering}
Each time the secondary crosses the equatorial plane, its dynamics are influenced by two main effects. First, the secondary accretes matter from the disk, resulting in an exchange of energy and linear momentum that alters its orbit. Second, the disk exerts a gravitational drag on the secondary, commonly referred to as \emph{dynamical friction}~\cite{Chandrasekhar:1943ys}. We will account for both effects and develop a framework to determine (i) the orbital changes after each crossing and (ii) the location of the subsequent crossing, iterating this process self-consistently. 
\vskip 2pt
Without loss of generality, we align the $\hat{y}$--axis in the equatorial plane with the position of the first scattering point (marked by the green dot in Figure~\ref{fig:schematic_overview}). In Cartesian coordinates, this scattering occurs at $\vec{r}_{\rm d} = (0, y, 0)$. Since the particles in the disk follow circular Keplerian orbits in the equatorial plane,\footnote{In some accretion disk models, the motion may not be strictly Keplerian, e.g., in the Thompson et al.~model~\eqref{eq:vel_non-kep}. However, deviations from Keplerian motion do not affect our results appreciably.} their velocity at this point is $\vec{v}_{\rm d} =(v_{\rm d},0,0)$, where
\begin{equation}\label{eq:vel_disk}
    v_{\rm d} = \sqrt{\frac{M}{R}}\,.
\end{equation}
At the moment of scattering, the secondary's position coincides with that of the fluid, i.e., $\vec{r}_{\rm p} =\vec{r}_{\rm d}=(0,y,0)$. To fully specify the scattering conditions, we also need the secondary's velocity, which we denote as $\vec{v}_{\rm s} = (v_{\mathrm{s},x}, v_{\mathrm{s},y}, v_{\mathrm{s},z})$. The change in the orbit due to the scattering are then determined by imposing conservation of angular momentum and energy.
\subsection{Single Crossing}
We start by describing the effects of a single crossing of the secondary through the disk. First, we examine accretion, followed by dynamical friction, and finally the backreaction on the orbit. This provides the basis for understanding the cumulative effect of multiple crossings.
\subsubsection{Accretion}\label{sec:single_accretion}
As the secondary crosses the disk, it accretes mass and momentum. We model this interaction as a perfectly inelastic collision. Quantities associated with the secondary \emph{after} the disk passage are denoted with primes. Consider first the change in linear momentum. Since the disk lies in the $x-y$ plane and the fluid rotates in circular orbits, its velocity and position are orthogonal. Given that the position of the scattering point lies along the $\hat{y}$--axis, the fluid velocity aligns with the $\hat{x}$--axis. The momentum of a fluid element is then given by
\begin{equation}\label{eq:P_fluid}
\vec{P}_{\rm d} = m_{\rm d}\vec{v}_{\rm d} = m_{\rm d}(v_{\rm d},0,0)\,, 
\end{equation}
where $m_{\rm d}$ is the mass of the fluid element. For the secondary, the momenta before and after the interaction are expressed as 
\begin{equation}\label{eq:P_secondary}
\begin{aligned}
\vec{P}_{\rm p} &= m_{\rm p}(v_{\mathrm{s},x}, v_{\mathrm{s},y}, v_{\mathrm{s},z})\,,\\
\vec{P}'_{\rm p} &= m_{\rm p}(v_{\mathrm{s},x}, v_{\mathrm{s},y}, v_{\mathrm{s},z}) + \Delta m (v_{\rm d},0,0)\,,
\end{aligned}
\end{equation}
where $\Delta m$ represents the accreted mass. Consequently, the secondary's mass increases as
\begin{equation}\label{eq:m_prime}
m_{\rm p}' = m_{\rm p} + \Delta m\,.
\end{equation}
Before specifying $\Delta m$, we note that accretion effects can alternatively be described using an effective force. Rearranging eq.~\eqref{eq:P_secondary} yields
\begin{equation}
   \vec{P}'_{\rm p}- \vec{P}_{\rm p}= \Delta \vec{P}_{\rm p}=  \Delta m \vec{v}_{\rm d}= \dot{m}_{\rm p} \Delta t\,\vec{v}_{\rm d}\,,
\end{equation}
where the overhead dot denotes a time derivative and $\Delta t$ is the crossing time. Using the impulse-momentum theorem, which relates the change in momentum to the applied force over time, we can define an \emph{effective force}:
\begin{equation}
\vec{F}_{\rm acc}=\dot{m}_{\rm p}\vec{v}_{\rm d}\,.
\end{equation}
This expression coincides with the one used in~\cite{Barausse:2007dy}. 
\vskip 2pt
To calculate the accretion rate of the secondary as it passes through the disk, we consider \emph{Bondi-Hoyle accretion}, which describes the spherical accretion of material onto a BH that is moving through a medium. The effective accretion cross-section is given by the Bondi-Hoyle-Lyttleton formula~\cite{1939PCPS...35..405H,1944MNRAS.104..273B}:
\begin{equation}\label{eq:cross_section}
    \sigma_{\scalebox{0.65}{BHL}} = \frac{4\pi m_{\rm p}^{2}}{(c_{\rm s}^2 + v^2_{\rm rel})^{2}}\,.
\end{equation}
Here, $\vec{v}_{\rm rel} = \vec{v}_{\rm d} - \vec{v}_{\rm p}$ denotes the relative velocity between the disk and the secondary, and $c_{\rm s}$ is the sound speed in the disk. Using eq.~\eqref{eq:cross_section}, the mass accreted by the secondary during a single disk passage is given by
\begin{equation}\label{eq:Deltam}
    \Delta m = \int\dot{m}_{\rm p}\,\mathrm{d}t = \int\rho\,\sigma_{\scalebox{0.65}{BHL}} \sqrt{c_{\rm s}^2+v_{\rm rel}^2}\,\mathrm{d}t\,.
\end{equation}
For simplicity, we first consider a perpendicular crossing, such that $\iota = \pi/2$ and the radius $r$ remains constant. From eq.~\eqref{eq:Deltam}, the accreted mass is expressed as
\begin{equation}\label{eq:deltaM2}
    \Delta m = \int\!\rho(r,z)\,\sigma_{\scalebox{0.65}{BHL}} \sqrt{c_{\rm s}^2+v_{\rm rel}^2}\,\mathrm{d}z\frac{\mathrm{d}t}{\mathrm{d}z} =\int_{-H/2}^{H/2}\rho(r)\, \frac{4\pi m_{\rm p}^{2}}{(v^2_{\rm rel} + c_{\rm s}^2)^{3/2}v_z}\,\mathrm{d}z\,,
\end{equation}
where we use the piecewise profile from eq.~\eqref{eq:verticalpiecewise} and $\mathrm{d}z/\mathrm{d}t = v_{\mathrm{rel},z} = v_z$, since the disk velocity has no $\hat{z}$--component. The total accreted mass is then
\begin{equation}\label{eq:perpendicular_accreted_mass}
    \Delta m = H(r)\rho(r)\frac{4\pi m_{\rm p}^{2}}{(v^2_{\rm rel} + c_{\rm s}^2)^{3/2}}\frac{1}{v_z}\,.
\end{equation}
This result agrees with that of~\cite{2023MNRAS.522.1763G}.\footnote{If the disk's vertical profile is modelled using a Gaussian~\eqref{eq:verticalgaussian} instead of a piecewise function~\eqref{eq:verticalpiecewise}, the results differ by a factor of $\sqrt{2 \pi}$.}
\vskip 2pt
Consider then the secondary crossing the disk at a generic angle with a velocity $\mathcal{V}$ in the $y-z$ plane. The angle between the ``vertical'' $\hat{z}$--direction and the ``non-perpendicular'' $\hat{z}'$--direction of the secondary, is given by $\xi$ (i.e., $\xi = 0$ corresponds to the perpendicular case). It is important to note that $\xi$ is not the inclination of the orbit. The velocity component along the $\hat{z}$--axis is $v_z = \mathcal{V}\cos{\xi}$. The mass accreted during the passage is then given by
\begin{equation}\label{eq:Deltam_generic}
    \Delta m = \int\!\rho(r,z)\,\frac{4\pi m_{\rm p}^{2}}{(v^2_{\rm rel} + c_{\rm s}^2)^{3/2}}\, \mathrm{d}t = \int\!\rho(r,z)\,\frac{4\pi m_{\rm p}^{2}}{(v^2_{\rm rel} + c_{\rm s}^2)^{3/2}}\, \frac{1}{\mathcal{V}}\mathrm{d}z'\,,
\end{equation}
while the secondary travels a distance $H(r)/\cos{\xi}$ in the $\hat{z}$--direction. We can thus write:
\begin{equation}
    \Delta m = \int_0^{H/\cos{\xi}}\!\rho(r,z)\,\frac{4\pi m_{\rm p}^{2}}{(v^2_{\rm rel} + c_{\rm s}^2)^{3/2}}\, \frac{1}{\mathcal{V}}\mathrm{d}z'\,.
\end{equation}
Rotating back to the original plane, where $z' = z\cos{\xi}-y\sin{\xi}$, we obtain:
\begin{equation}\label{eq:nonperpendicular_accreted_mass}
\begin{aligned}
    \Delta m &= \int_0^{H}\!\rho(r,z)\,\frac{4\pi m_{\rm p}^{2}}{(v^2_{\rm rel} + c_{\rm s}^2)^{3/2}}\, \frac{1}{\mathcal{V}}\cos{\xi}\,\mathrm{d}z \\&+ \int_0^{H\sin{\xi}/\cos{\xi}}\!\rho(r,z)\,\frac{4\pi m_{\rm p}^{2}}{(v^2_{\rm rel} + c_{\rm s}^2)^{3/2}}\, \frac{1}{\mathcal{V}}\sin{\xi}\,\mathrm{d}y\,.
\end{aligned}
\end{equation}
In the special case of $\xi = 0$, the second integral vanishes, recovering the result from the perpendicular crossing case~\eqref{eq:perpendicular_accreted_mass}. On the other hand, when the orbit becomes equatorial, i.e., $\xi = \pi/2$, the path in the disk becomes ``infinite'', and the integral in eq.~\eqref{eq:nonperpendicular_accreted_mass} indeed diverges.
\vskip 2pt
Analytically solving eq.~\eqref{eq:nonperpendicular_accreted_mass} is generally challenging due to the spatial dependence of the fluid density. However, assuming the arc of the orbit inside the disk to be small, we can approximate the density to be constant. In this case, the integral above admits a simple analytical solution: 
\begin{equation}
    \Delta m = \frac{4\pi m_{\rm p}^{2} \rho H}{(v^2_{\rm rel} + c_{\rm s}^2)^{3/2}} \Big(\frac{\cos \xi}{\mathcal{V}}+\frac{\sin^2 \xi}{\mathcal{V} \cos \xi}\Big)=\frac{4\pi m_{\rm p}^{2} \rho H}{(v^2_{\rm rel} + c_{\rm s}^2)^{3/2}}\frac{1}{v_z}\,,
\end{equation}
where in the last equality, we used $\mathcal{V}=v_z/\cos \xi$. Remarkably, this expression matches the result for the perpendicular crossing~\eqref{eq:perpendicular_accreted_mass}. The actual accreted mass, however, will differ because the relative velocity between the secondary and the disk changes whenever the crossing is non-perpendicular. Finally, the result remains unchanged even if the secondary's velocity has a component along the $\hat{x}$--axis. In this case, the calculation only requires an additional rotation in the $x-y$ plane, which again leads to the same result.
\subsubsection{Dynamical friction}
Dynamical friction arises from the interaction between the secondary and the wake of particles affected by its motion, but not accreted by it. While one could model this using partially inelastic scatterings, we find it more convenient to adopt the force description introduced by Ostriker~\cite{1999ApJ...513..252O}, based on the impulse theorem. In this framework, the change in the secondary’s momentum is given by
\begin{equation}\label{eq:P_secondary_DF}
\begin{aligned}
\vec{P}'_{\rm p} &= \vec{P}_{\rm p} + \vec{F}_{\rm DF} \Delta t\,,\\
\vec{F}_{\rm DF} &= \frac{4 \pi m_{\rm p}^2\rho}{v_{\rm rel}^3} \vec{v}_{\rm rel}\,\mathcal{I}\left(v_{\mathrm{rel}} / c_{\rm s}\right)\,,
\end{aligned}
\end{equation}
where $\mathcal{I}\left(v_{\mathrm{rel}} / c_{\rm s}\right)$ is a dimensionless factor that depends on the \emph{Mach number}, $\mathcal{M} = v_{\rm rel}/c_{\rm s}$. This factor is defined as
\begin{equation}
    \mathcal{I}(\mathcal{M})= \begin{dcases}\frac{1}{2} \ln \left(1-\frac{1}{\mathcal{M}^2}\right)+\ln \Lambda \quad &\mathcal{M}>1\,, \\ \frac{1}{2} \ln \left(\frac{1+\mathcal{M}}{1-\mathcal{M}}\right)-\mathcal{M}\quad&\mathcal{M}<1\,.\end{dcases}
\end{equation}
When the secondary is far from the primary, the Mach number is large ($\mathcal{M} \gg 1$), and $\mathcal{I}$ becomes nearly independent of $\mathcal{M}$. To avoid the divergence at $\mathcal{M} = 1$, we adopt the following regularised form:
\begin{equation}
I(\mathcal{M})= \begin{dcases}\ln \Lambda \quad &\mathcal{M} \geq 1\,, \\ \min \left[\ln \Lambda\,, \frac{1}{2} \ln \left(\frac{1+\mathcal{M}}{1-\mathcal{M}}\right)-\mathcal{M}\right] \quad &\mathcal{M}<1\,.\end{dcases} 
\end{equation}
Here, $\ln \Lambda$ is the Coulomb logarithm~\cite{1987gady.book.....B}, which serves as a regulator that defines the effective size of the medium contributing to the gravitational drag on the secondary. Without this cutoff, the medium would generate an infinitely extended wake, leading to a divergent drag force. While the precise value of $\ln \Lambda$ depends on the properties of the medium and the secondary, numerical simulations of gas accretion onto BHs suggest that $\ln \Lambda \approx 3$ provides a good fit~\cite{2013MNRAS.429.3114C}.
\vskip 2pt
Two important considerations are worth highlighting. 
\begin{enumerate}
\item Both accretion and friction act as effective forces that modify the secondary's velocity after scattering. In principle, these forces would also cause a displacement of the secondary's position. However, while the change in velocity scales as $\propto \vec{F} \Delta t$, the change in position scales as $\propto \vec{F} \Delta t^2$. Since we always assume the arch of the orbit inside the disk to be small, the crossing time satisfies $\Delta{t}\ll 1$. Consequently, under the impulsive force approximation, the displacement is a next-to-leading order effect in $\Delta t$ and can thus be safely neglected. 
\item Both Bondi accretion and our model for dynamical friction implicitly assume an infinite medium.  If the Bondi radius becomes comparable to the medium size, these effects may be reduced~\cite{Vicente:2019ilr,2021ApJ...916...48D}. In our analysis, we rigorously verified that the disk's thickness remains significantly larger than the Bondi radius, thereby validating our assumptions.
\end{enumerate}
\subsubsection{Backreaction on the orbit}
Once the linear momentum of the secondary after scattering is determined, using either eq.~\eqref{eq:P_secondary} or~\eqref{eq:P_secondary_DF}, the corresponding angular momentum can be calculated as
\begin{equation}\label{eq:L_prime}
\vec{L}'_{\rm p} = \vec{r}_{\rm p} \times \vec{P}_{\rm p}' \,,
\end{equation}
where $\vec{r}_{\rm p} = (0,R,0) = (0,y,0)$. The updated orbital energy is then given by
\begin{equation}
    E'_{\rm orb} = \frac{|\vec{P}_{\rm p}'|^2}{2m'_{\rm p}} - \frac{M m'_{\rm p}}{|R|}\,.
\end{equation}
After each scattering event, we verify that the orbital energy is negative, ensuring the orbit remains bound.
\vskip 2pt
We then update the orbital parameters. The inclination is determined using
\begin{equation}\label{eq:i_prime}
    \iota' = \arccos{\left(\frac{-L_{\mathrm{s},z}'}{|\vec{L}'_{\rm p}|}\right)}\,,
\end{equation}
where $L_{\mathrm{s},z}'$ and $|\vec{L}'_{\rm p}|$ denote the $z$--component and the magnitude of the updated angular momentum vector~\eqref{eq:L_prime}, respectively. The orbital eccentricity is updated using the relation
\begin{equation}\label{eq:e_prime}
    \varepsilon' = \sqrt{\frac{-|\vec{L}'_{\rm p}|^2+a' m^{\prime\,2}_{\rm p} M}{a'M m_{\rm p}^{\prime\,2}}}\,,
\end{equation}
where the updated semi-major axis $a'$ is obtained by inverting the vis-viva equation~\eqref{eq:v_visvisa}:
\begin{equation}\label{eq:a_prime}
    a' = \frac{M R}{2M- R|\vec{v}'_{\rm p}|^2}\,,
\end{equation}
with the magnitude of the secondary's velocity given by $|\vec{v}_{\rm p}|=\sqrt{v^2_x+v^2_y+v^2_z}$. Finally, the true anomaly is updated using
\begin{equation}\label{eq:theta_prime}
    \theta' = \arccos{\left(\frac{a'(1-\varepsilon^{\prime\,2})-R}{\varepsilon'R}\right)}\,.
\end{equation}
%
\subsection{Multiple Crossings}
After the scattering, the secondary follows a new orbit in vacuum with the updated orbital parameters, starting from the first crossing point until it intersects the disk again. While one could numerically evolve the orbit to locate the next scattering point, we exploit the symmetry of the problem to determine it analytically. From Figure~\ref{fig:schematic_overview}, it follows that the next crossing point occurs at $\theta'' = \pi-\theta'$, where double primes denote the quantities at the subsequent scattering point. 
\vskip 2pt
By definition, the next crossing point must lie on both the disk and orbital planes. These planes intersect along the $\hat{y}$--axis, where both the primary, located at the origin, and the ``previous'' scattering point are.\footnote{Including the displacement induced by hydrodynamic drag would introduce a minor $\hat{x}$--axis component to the updated position. However, as discussed before, this effect is subleading and can be neglected.} Using eqs.~\eqref{eq:r_nextscatter} and~\eqref{eq:v_visvisa}, we compute the updated separation $|R''|$ and velocity magnitude $|\vec{v}''|$. Since the components of the angular momentum must be conserved individually, we use the conservation along the $\hat{x}$--axis ($\hat{z}$--axis) to find $v''_{z}$ ($v''_{x}$) as 
\begin{equation}\label{eq:vz_primeprime}
    v_z'' = \frac{R p'_z}{R''m_{\rm p}'}\,,\quad v_x'' = \frac{R p'_x}{R''m_{\rm p}'}\,.
\end{equation}
The final component, $v''_y$, is then found as 
\begin{equation}\label{eq:vy_primeprime}
    v_y'' = \sqrt{|v''|^2-v^{\prime\prime\,2}_x -v^{\prime\prime\,2}_z}\,.
\end{equation}
With all the updated quantities in hand, the new crossing position and velocity are given by $\vec{r}_{\rm new} = (0,-|R''|,0)$ and $\vec{v}_{\rm new} = (v_x'',v_y'',v_z'')$. This process is then repeated for subsequent crossings. 
\subsection{Initialisation}\label{subsec:ini}
At each step of our algorithm, the position $\vec{r}_{\rm p}$ and velocity $\vec{v}_{\rm p}$ of the secondary are updated, fully characterising the orbit. However, directly specifying the initial velocity vector $\vec{v}_{\mathrm{p},0}$ is not intuitive. To address this, we initialise the orbit using the standard orbital elements, from which we derive the initial velocity of the secondary.
\vskip 2pt
The position of the secondary is given by eq.~\eqref{eq:r_nextscatter}. From this, we compute the radial and transverse velocities as 
\begin{equation}\label{eq:vr_vrtheta}
    v_{r} = \sqrt{\frac{M}{a(1-\varepsilon^2)}}\,\varepsilon \sin{\theta}\,,\quad 
    v_{\theta} =\sqrt{\frac{M}{a(1-\varepsilon^2)}}\, (1+\varepsilon\cos{\theta})\,.
\end{equation}
The velocity of the secondary in the orbital plane is then:
\begin{equation}\label{eq:v_trans}
    \vec{v}_{\mathrm{p, orb}} = (v_r\cos{\theta} - v_{\theta}\sin{\theta}, v_r\sin{\theta} + v_{\theta}\cos{\theta}, 0)\,.
\end{equation}
To transform this velocity into the original coordinate system, we apply the rotation matrix:
\begin{equation}\label{eq:rotation_matrix}
    \vec{\mathcal{R}} = \vec{\mathcal{R}}_{\Omega}\vec{\mathcal{R}}_{\omega}\vec{\mathcal{R}}_{\iota}\,,
\end{equation}
where $\vec{\mathcal{R}}_{\Omega}$, $\vec{\mathcal{R}}_{\omega}$ and $\vec{\mathcal{R}}_{\iota}$ are the rotation matrices for the longitude of the ascending node ($\Omega$), the argument of the periapsis ($\omega$) and the inclination, respectively. Since we are only concerned with the position of the secondary as it crosses the equatorial plane, and we have aligned our axes accordingly, the longitude of the ascending node is given by $\Omega = \pi/2$. Thus, the first rotation matrix required for eq.~\eqref{eq:rotation_matrix} simplifies to:
\begin{equation}
    \vec{\mathcal{R}}_{\Omega} =    
    \begin{bmatrix}
    \cos{\Omega} & -\sin{\Omega} & 0\\
    \sin{\Omega} & \cos{\Omega} & 0\\
    0 & 0 & 1
    \end{bmatrix} = 
    \begin{bmatrix}
    0 & -1 & 0\\
    1 & 0 & 0\\
    0 & 0 & 1
    \end{bmatrix}\,.
\end{equation}
Furthermore, since the first scattering point occurs at the ascending node, the argument of the periapsis is given by $\omega = -\theta$. The corresponding rotation matrix is then:
\begin{equation}
    \vec{\mathcal{R}}_{\omega} =    
    \begin{bmatrix}
    \cos{\omega} & -\sin{\omega} & 0\\
    \sin{\omega} & \cos{\omega} & 0\\
    0 & 0 & 1
    \end{bmatrix} = 
    \begin{bmatrix}
    \cos{\theta} & \sin{\theta} & 0\\
    -\sin{\theta} & \cos{\theta} & 0\\
    0 & 0 & 1
    \end{bmatrix}\,.
\end{equation}
Finally, the rotation matrix for the inclination takes the form:
\begin{equation}
    \vec{\mathcal{R}}_{\iota} =    
    \begin{bmatrix}
    1 & 0 & 0\\
    0 & \cos{\iota} & -\sin{\iota}\\
    0 & \sin{\iota} & \cos{\iota}
    \end{bmatrix}\,.
\end{equation}
The initial velocity is then given by $\vec{v}_{\mathrm{p},0} = \vec{\mathcal{R}}\vec{v}_{\mathrm{p, orb}}$, making use of eqs.~\eqref{eq:vr_vrtheta}--\eqref{eq:rotation_matrix}. As a result, the initialisation of the algorithm requires the following parameters:~(i) the semi-major axis, $a_0$;~(ii) the eccentricity $\varepsilon_0$;~(iii) the true anomaly $\theta_0$;~(iv) the inclination $\iota_0$ and (v)~the initial position $\vec{r}_0$.
\section{Results}\label{sec_Capture:results}
We are now equipped to evolve the system over an arbitrary number of orbits. To understand the total time evolution of the orbit, we can estimate the effect of GW radiation reaction, which drives the system towards the plunge, ending the inspiral. The inspiral and orbital timescales can be approximated as
\begin{equation}
 t_{\rm{insp}}\approx \frac{r^4}{q M^3}\quad \text{and}\quad t_{\rm{orb}}\approx \sqrt{\frac{r^3}{M}}\,, 
\end{equation}
leading to an approximate number of orbits:
\begin{equation}\label{eq:Ninsp}
    N_{\rm orbits} \approx \frac{1}{q}\sqrt{\left(\frac{r}{M}\right)^5}\,.
\end{equation}
Thus, the number of orbits is inversely proportional to the mass ratio, i.e., $N_{\rm orbits} \sim q^{-1}$. The precise number of orbits over an EMRI lifetime is highly uncertain~\cite{Berry:2019wgg,Colpi:2024xhw}, especially in the presence of astrophysical environments. Therefore, we will consider a conservative scenario where $N_{\rm orbits} = q^{-1}$ and the secondary located at very large separations, such that we can adopt the adiabatic approximation and neglect the GW radiation reaction. 
\vskip 2pt
In the following, we primarily focus on dynamical friction and the Sirko-Goodman AGN model (see Section~\ref{BHenv_sec:accretion_disks_AGN}), using a set of benchmark parameters listed in Table~\ref{tab:benchmark}. A comparison of the Sirko-Goodman prescription with the Thompson et al.~model (see Figure~\ref{fig:densities_AGN}) is presented in Section~\ref{subsec_Capture:compareAGN}. We show that both models yield qualitatively similar results, suggesting that our conclusions are robust to the choice of disk model. Furthermore, in Section~\ref{subsec_Capture:compareeffect}, we examine the distinction between dynamical friction and accretion-driven drag. While both processes influence the orbital evolution in a similar manner, friction is typically slightly larger in magnitude. In fact, in the supersonic regime -- relevant for most of our parameter space -- the main difference arises from the Coulomb logarithm, which is an $\mathcal{O}(1)$ factor. Consequently, the influence of accretion is generally more adiabatic and less abrupt than that of friction. 
\begin{table}
    \centering
    \renewcommand{\arraystretch}{1.08}
    \begin{tabular}{|>{\centering\arraybackslash}p{1.2cm}|c|c|}
        \hline
        \multicolumn{3}{|c|}{Benchmark system} \\
        \hline
        \textbf{Symbol} & \textbf{Meaning} & \textbf{Value} \\
        \hline
        $M$ & BH mass & $10^7 M_{\odot}$ \\
        $q$ & Mass ratio & $10^{-4}$ \\
        $a$ & Semi-major axis & $10^{6}M$ \\
        $\iota$ & Inclination &  \\
        $\varepsilon$ & Eccentricity &  \\
        $\theta$ & True anomaly &  \\
        $\alpha\ped{visc} $ & Viscosity & $0.01$ \\
        $f_{\rm Edd}$ & Luminosity ratio & 0.5 \\
        $\eta$ & Radiative efficiency & 0.1 \\
        $X$ & Hydrogen abundance & 0.7 \\
        \hline
    \end{tabular}
    \caption{Benchmark system of parameters when using the Sirko-Goodman model (see Section~\ref{BHenv_sec:accretion_disks_AGN})~\cite{Sirko:2002ex,Gangardt:2024bic}. We will always model the inner disk according to the $\alpha$-prescription of the Shakura-Sunyaev disk~\eqref{eq:surface_density}.}
    \label{tab:benchmark}
\end{table}
\subsection{Evolution of the Inclination and Semi-Major Axis}
\subsubsection{Initial inclination}
\begin{figure}[t!]
    \centering
    \includegraphics[scale = 1.1]{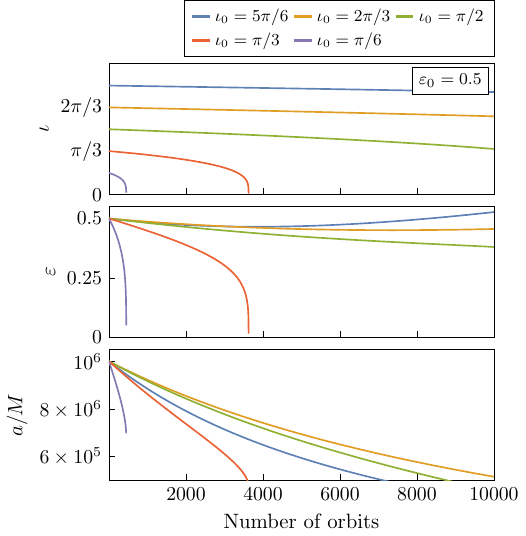}
    \caption{The impact of dynamical friction from repeated disk crossings on the evolution of the inclination, eccentricity and semi-major axis, for various initial inclinations. We use the benchmark parameters from Table~\ref{tab:benchmark}, setting $\varepsilon_0 = 0.5$, $a_0 = 10^6 M$ ($\sim 0.5\,\mathrm{pc}$) and $\theta_0 = \pi/3$. Increasing either the primary's mass $M$ or the semi-major axis enhances the magnitude of the effect.}
    \label{fig:General_SG_varyIncl_M1e7}
\end{figure}
First, we examine how the disk influences the inclination and semi-major axis as functions of the \emph{initial inclination} $\iota_0$. We consider both prograde ($\iota_0<\pi/2$) and retrograde ($\iota_0>\pi/2$) orbits. Figure~\ref{fig:General_SG_varyIncl_M1e7} shows the evolution of the orbital parameters. The setup uses the Sirko-Goodman model with $M = 10^7M_{\odot}$, $\varepsilon_0=0.5$, $a_0=10^6 M$ and $\theta_0=\pi/3$. As illustrated, interactions with the disk generally lead to a decrease in both the inclination and semi-major axis. The evolution of eccentricity is more complex and will be discussed in detail in the next section. The closer the initial inclination is to alignment with the disk ($\iota_0 \rightarrow 0$), the more pronounced the reduction in inclination is. For instance, a prograde orbit with $\iota_0 = \pi/6$ is fully dragged into the equatorial plane and circularised in fewer than 1000 orbits, while an orbit with $\iota_0=\pi/3$ achieves alignment in approximately $4000$ orbits. Using Kepler's third law and the updated semi-major axis after each orbit, we find that this corresponds to an alignment timescale of $T_{\rm align} \simeq 10^{5}$ and $10^{6}\,\mathrm{yrs}$, respectively. At higher inclinations, the effect diminishes:~while an orbit with $\iota_0 = \pi/2$ still experiences a noticeable change after $10^4$ orbits, the effect becomes negligible for $\iota_0 \gtrsim 2\pi/3$. Interestingly, the decrease in the semi-major axis does not follow the same trend and remains relevant for all different values of $\iota_0$, while the ordering between the curves changes. A possible explanation involves the two competing effects that influence the energy lost during each interaction with the disk:~(i) how much is the secondary immersed in the disk (which is maximised when the orbit is nearly co-planar, i.e., $\iota \approx 0$ or $\iota \approx \pi$) and~(ii) if the motion is aligned or anti-aligned with the disk’s rotation, i.e., whether the secondary moves \emph{with} or \emph{against} the flow of the gas in the disk. For example, consider the case of a nearly retrograde orbit with $\iota_0 = 5\pi/6$ [{\color{Mathematica1}{blue}}]. The secondary is almost fully embedded in the disk, maximising the interaction and thus the potential for energy loss. However, because it is moving against the direction of the disk, the relative velocity is large, which leads to a smaller drag force and thus less efficient energy extraction. Now compare this to a slightly less inclined orbit, e.g., $\iota_0 = 2\pi/3$ [{\color{Mathematica2}{orange}}]:~the secondary is less immersed in the disk, reducing the duration and intensity of each scattering event, but the alignment with the disk flow is more favourable for transferring energy. The resulting decrease in semi-major axis is thus determined by the interplay between these two factors.
\subsubsection{Initial eccentricity and true anomaly}
\begin{figure}[t!]
    \centering
    \includegraphics[scale = 1.1]{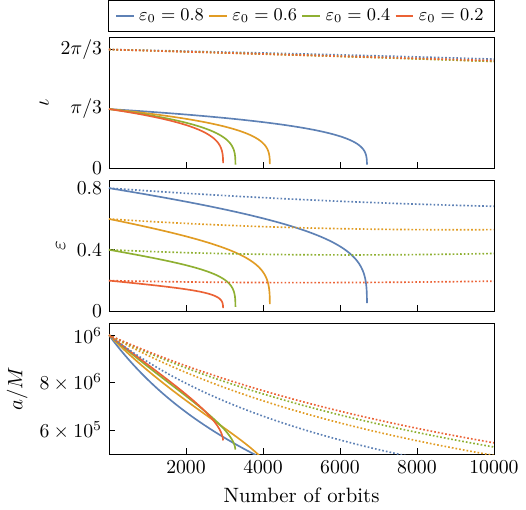}
    \caption{Similar configuration as in Figure~\ref{fig:General_SG_varyIncl_M1e7}, but now varying the initial eccentricity. We take $\iota_0 = \pi/3$ (solid) and $\iota_0 = 2\pi/3$ (dotted).}
    \label{fig:General_SG_varyEcc_M1e7}
\end{figure}

\begin{figure}[t!]
    \centering
    \includegraphics[scale = 1.1]{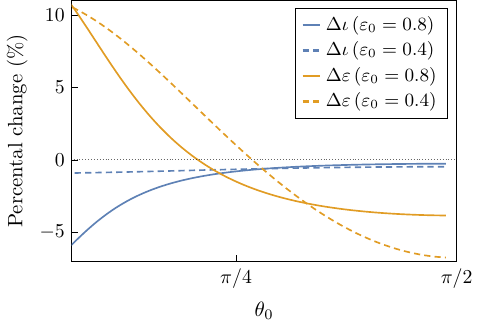}
    \caption{Fractional change in inclination and eccentricity after $1000$ orbits, for $\iota_0 = 5\pi/6$ and either $\varepsilon_0 = 0.8$ (solid) or $\varepsilon_0 = 0.4$ (dashed). The benchmark parameters used are listed in Table~\ref{tab:benchmark}.}
    \label{fig:theta_varying}
\end{figure}

We now investigate a similar setup, this time varying the \emph{initial eccentricity} $\varepsilon_0$ and the \emph{true anomaly} $\theta_0$. These two parameters are closely linked, and the system’s evolution depends strongly on their interplay. Since there is no physically motivated choice for the initial true anomaly, we will explore a range of values to characterise its influence. Together, $\varepsilon_0$ and $\theta_0$ determine the relative positioning of the ascending and descending nodes with respect to the BH, effectively controlling how \emph{symmetric} or \emph{asymmetric} the two disk crossings per orbit are, as illustrated in Figure~\ref{fig:schematic_overview}. When $\theta_0=\pi/2$, the nodes are equidistant from the BH, ensuring that each scattering happens in regions of the disk with the same density and local velocity. In contrast, for $\theta_0 = 0$ and $\varepsilon_0 \neq 0$, the scatterings take place in highly asymmetric regions of the disk [see eq.~\eqref{eq:r_nextscatter}]. 
\vskip 2pt
We begin by varying $\varepsilon_0$, fixing the inclination at $\iota_0 = \pi/3$ (solid lines) or $\iota_0 = 2\pi/3$ (dotted lines), and setting $\theta_0 = \pi/3$. The results are shown in Figure~\ref{fig:General_SG_varyEcc_M1e7}. As seen in the \emph{top panel}, the evolution of the inclination is strongly influenced by the initial eccentricity. Lower values of $\varepsilon_0$ lead to a more rapid decrease in inclination, while larger eccentricities (e.g., $\varepsilon_0 = 0.8$ [{\color{Mathematica1}blue}]) result in significantly longer alignment timescales. In contrast, the decay rates of the eccentricity and semi-major axis exhibit a weaker dependence on $\varepsilon_0$, especially for prograde orbits.
\vskip 2pt
Crucially, this behaviour can be completely reversed by changing the initial anomaly $\theta_0$. Figure~\ref{fig:theta_varying} shows the fractional change in inclination as a function of $\theta_0$ for two eccentricities, $\varepsilon_0 = 0.4$ and $\varepsilon_0 = 0.8$. When $\theta_0 \lesssim \pi/4$, higher eccentricity leads to \emph{larger decrease} in inclination -- the opposite trend compared to the previous setup. This highlights the intricate relationship between $\varepsilon_0$ and $\theta_0$, making it difficult to draw general conclusions about either one independently. For completeness, Figure~\ref{fig:theta_varying} also shows the evolution of the eccentricity. While the inclination consistently decreases, we observe regions where the eccentricity actually increases, as also evident in the \emph{centre panel} of Figure~\ref{fig:General_SG_varyIncl_M1e7}. A detailed explanation of this behaviour will be provided in the next section.
\vskip 2pt
So far, we have examined how orbital evolution depends on the initial conditions. It is crucial, however, to consider how this evolution behaves dynamically. The inclination and eccentricity do not decay linearly with the number of orbits. As shown in Figures~\ref{fig:General_SG_varyIncl_M1e7} and~\ref{fig:General_SG_varyEcc_M1e7}, their rates of change experience a sharp transition when the inclination falls below a critical threshold -- approximately $\iota \approx \pi/12$ for the parameters used here. This behaviour can be attributed to the increasing relevance of dynamical friction, $F_{\rm DF} \propto v_{\rm rel}^{-2}$, as the orbit aligns with the disk and becomes more circular. These results underscore the importance of a self-consistent framework for modelling the system’s evolution, rather than relying on timescale estimates based solely on the initial conditions.
%
\subsubsection{Masses}
The dependence of the system's evolution on the \emph{masses} of the primary and secondary is more straightforward. We find that the variation of the orbital parameters is inversely proportional to the mass of the secondary. However, since the number of orbits scales inversely with the mass ratio~\eqref{eq:Ninsp}, the overall change in the orbital parameters over the \emph{entire} inspiral remains independent on the secondary's mass. In contrast, increasing the mass of the primary increases the number of orbits, while simultaneously decreasing the disk's density. In this case, the former effect dominates, making the overall impact of the scatterings more significant for larger values of the primary's mass.
\subsection{Evolution of the Eccentricity}
\begin{figure}[t!]
    \centering
    \includegraphics[width=\linewidth]{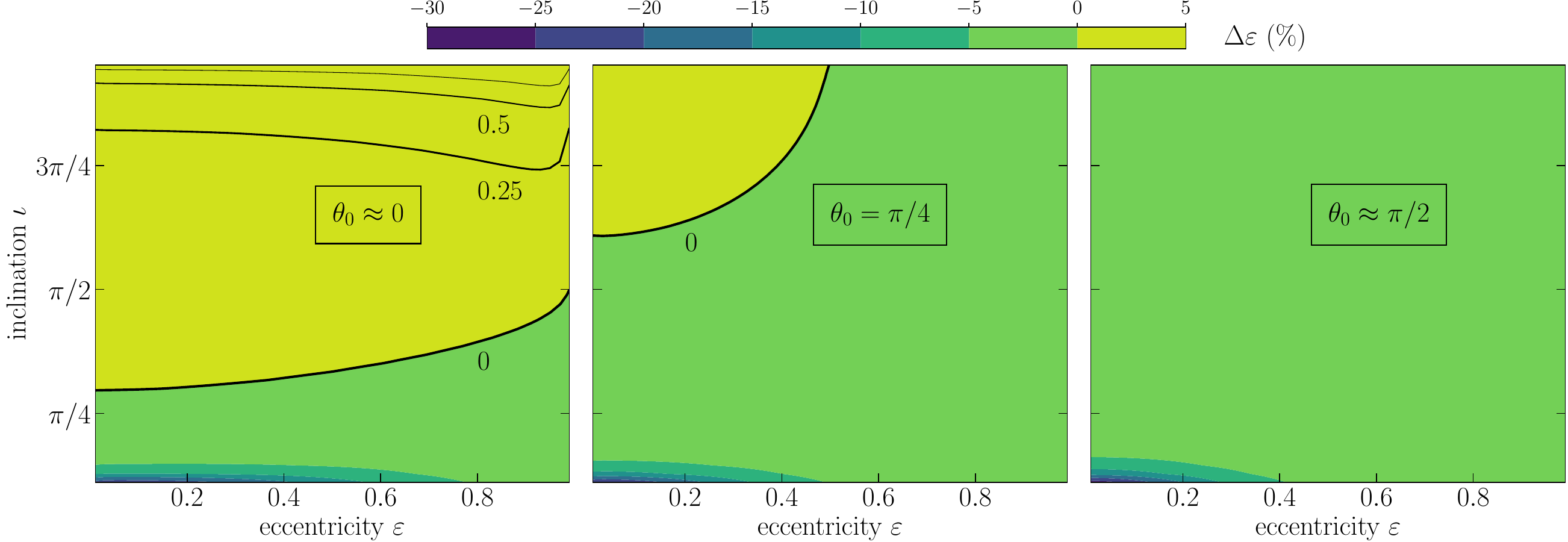}
    \caption{Contour plot of the fractional change in eccentricity, $\Delta \varepsilon$, after just 25 orbits in the Sirko-Goodman model with $M = 10^{7} M_{\odot}$. The \emph{left}, \emph{centre}, and \emph{right panels} correspond to initial values of the true anomaly of $\theta_0 \approx 0$, $\theta_0 = \pi/4$, and $\theta_0 \approx \pi/2$, respectively. Black contours mark regions of positive $\Delta \varepsilon = 0,\ 0.25,\ 0.5,\ 0.75$, with decreasing line thickness. In reality, the secondary continues to interact with the disk even after its orbit aligns with the disk plane. However, our model is no longer valid beyond this point. As a result, the total eccentricity reduction over the full inspiral due to the disk is much greater than what is shown in this figure.}
    \label{fig:contourplots}
\end{figure}

\begin{figure}[t!]
    \centering
    \includegraphics{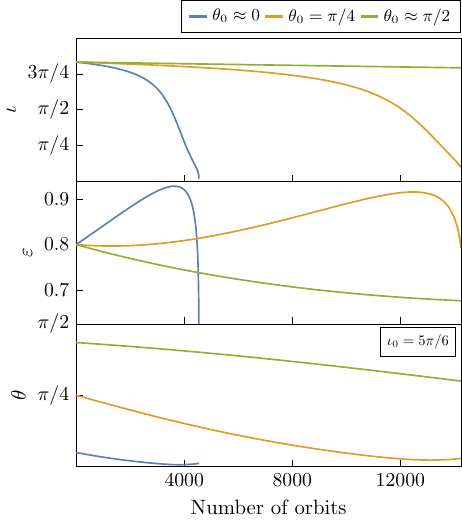}
    \caption{Using the same initial true anomaly values as in the contour plots, we show the evolution of a system with $\varepsilon_0 = 0.8$ and $\iota_0 = 5\pi/6$. The results reveal a clear and strong dependence of the eccentricity evolution on the initial choice of the true anomaly (\emph{centre panel}). In contrast, the inclination consistently decreases regardless of $\theta$ (\emph{top panel}). While the eccentricity behaviour can be complex and highly sensitive to initial conditions, the overall trend remains:~the system ultimately tends towards circularisation. Benchmark parameters are listed in Table~\ref{tab:benchmark}.}
    \label{fig:ecc_pumping}
\end{figure}

Thus far, our results have revealed a complex and often non-intuitive evolution of the eccentricity. In particular, Figures~\ref{fig:General_SG_varyIncl_M1e7} and~\ref{fig:theta_varying} highlight specific conditions under which the eccentricity can temporarily increase—a phenomenon we refer to as \emph{eccentricity pumping}. Similar behaviour has been observed in other astrophysical environments, such as circumbinary disks~\cite{Zrake:2020zkw,DOrazio:2021kob,Siwek:2023rlk,Tiede:2023dwq} and superradiant boson clouds~\cite{Tomaselli:2023ysb,Tomaselli:2024bdd,Tomaselli:2024dbw}.
\vskip 2pt
To better understand when eccentricity pumping occurs in our context, we study its dependence on the true anomaly. Figure~\ref{fig:contourplots} presents a contour plot of the fractional change in eccentricity, $\Delta \varepsilon$, after just 25 orbits. The three panels correspond to different initial values for the true anomaly:~$\theta_0 \approx 0$ (\emph{left}), $\theta_0 = \pi/4$ (\emph{centre}), and $\theta_0 \approx \pi/2$ (\emph{right}). These plots show that eccentricity pumping is most prominent when the nodes are maximally asymmetric relative to the disk, i.e., when $\theta \approx 0$. In this regime, large portions of parameter space experience a net increase in eccentricity. As $\theta$ increases, the pumping region shrinks (\emph{centre panel}), and it disappears entirely when the nodes are symmetric with respect to the BH, i.e., $\theta \approx \pi/2$ (\emph{right panel}). Notably, eccentricity pumping occurs predominantly at large inclinations.
\vskip 2pt
To explore the time evolution more directly, Figure~\ref{fig:ecc_pumping} shows the eccentricity evolution for three representative cases, corresponding to the same true anomaly values as the contour plots. For $\theta_0 \approx 0$ [{\color{Mathematica1}blue}], the eccentricity increases monotonically throughout the evolution up until the point where the secondary is nearly aligned and it starts decreasing rapidly. In contrast, for $\theta_0 \approx \pi/2$ [{\color{Mathematica3}{green}}], the eccentricity always decreases. The most intricate behaviour arises for $\theta_0 = \pi/4$ [{\color{Mathematica2}{orange}}]:~the system initially undergoes a slight decrease in eccentricity, but as it evolves, it enters a pumping region (as seen in the corresponding contour plot), causing $\varepsilon$ to increase before ultimately decaying again as the secondary comes close to alignment. This demonstrates that eccentricity can \emph{dynamically switch behaviour} over time, depending on how the system evolves through phase space. The underlying mechanism behind this evolution involves two competing effects:~while a decrease in inclination generally forces the system towards region of eccentricity decrease, a decrease in the true anomaly widens the available parameter space for eccentricity pumping (see Figure~\ref{fig:contourplots}). Importantly, we find that once the inclination drops below a critical threshold, the impact of the inclination becomes dominant and eccentricity always decreases (see Figure~\ref{fig:ecc_pumping}). Consequently, the system always evolves towards a circular, prograde orbit as its final state. Lastly, as we discussed in the previous section, alignment happens faster in regions of large eccentricity and small true anomaly. As shown in Figure~\ref{fig:ecc_pumping}, in many cases the system is pushed \emph{dynamically} towards such a configuration, with a rapid alignment as a result of it.
\vskip 2pt
At first glance, it may seem that interactions with the disk should always lead to \emph{circularisation} of the orbit. After all, drag dissipates orbital energy, which typically shrinks the orbit and reduces eccentricity. However, as we have seen, this intuition does not always hold:~under certain conditions, the orbit can instead become more eccentric. An intuitive explanation can be found by considering the geometry of the orbit and the direction and strength of the drag force at key points along it. 
\vskip 2pt
The change in eccentricity depends critically on where the drag is strongest, which in turn is determined by two factors:~(i) the local density of the disk at the scattering points, and~(ii) the relative velocity between the secondary and the gas in the disk. In our setup, the orbit intersects the disk at two points:~the ascending and descending nodes. The positions of these nodes relative to periapsis and apoapsis are set by the true anomaly $\theta$. When $\theta \approx 0$, the periapsis is close to one node and the apoapsis close to the other. If the density of the disk is asymmetric between these two points (i.e., when $\varepsilon \neq 0$), the resulting drag forces can differ significantly. In particular, when the drag is stronger near periapsis, the energy loss is concentrated there, leading to the apoapsis shrinking more rapidly than the periapsis, i.e., the system is \emph{circularising}. Conversely, if the drag is stronger near apoapsis, the opposite occurs and the periapsis shrinks faster, causing the orbit to undergo \emph{eccentricity pumping}. In addition, the orientation of the orbit with respect to the disk plays a role. For prograde orbits, the secondary moves with the disk. At periapsis, it is faster than the gas in the disk, so drag points backward and strongly damps the orbit, leading to circularisation. For retrograde orbits, the secondary moves opposite to the disk and drag acts in the prograde direction. But crucially, the magnitude of the drag is not symmetric:~it is weaker at periapsis (due to the high relative velocity) and stronger at apoapsis, leading to eccentricity pumping.
\vskip 2pt
In summary, the geometry of the system -- the inclination, eccentricity, and location of the periapsis (set by $\theta$) -- together determine whether eccentricity is damped or pumped. The results in Figure~\ref{fig:contourplots} confirm the intuitive picture outlined above:~eccentricity pumping is most efficient for orbits with high inclinations and small values of the true anomaly.
\subsection{Dynamics of Highly-Eccentric Orbits}
\begin{figure}[t!]
    \centering
    \includegraphics{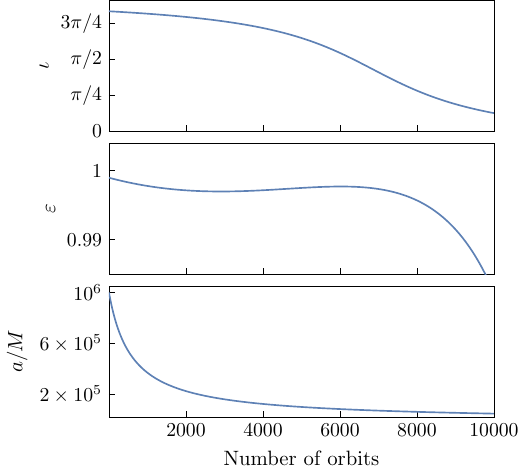}
    \caption{Evolution of the orbital parameters for an initially high eccentricity $\varepsilon_0 = 0.999$, and a nearly counter-rotating orbit $\iota_0 = 5\pi/6$, with $\theta_0 \approx 0$. While the eccentricity remains approximately constant, the orbit flips on relatively short timescales. Disk and binary parameters follow Table~\ref{tab:benchmark}.}
    \label{fig:high_ecc}
\end{figure}

EMRIs are expected to form with extremely high eccentricities, potentially reaching values as large as $\varepsilon \gtrsim 0.9999$~\cite{Amaro-Seoane:2012lgq}. As discussed in previous sections, orbits with high eccentricity and asymmetric nodes ($\theta \approx 0$) can experience significant changes in both inclination and eccentricity. In Figure~\ref{fig:high_ecc}, we explore such a scenario, revealing an intriguing evolution of the system. For a surprisingly large number of orbits ($\approx 8000$), the eccentricity remains nearly constant while the inclination undergoes a dramatic shift, transitioning from a nearly counter-rotating configuration to a co-rotating one. Meanwhile, the semi-major axis decreases significantly. This highlights the complex interplay between the eccentricity and inclination evolution in astrophysically realistic systems, and how their evolution can differ significantly.
\vskip 2pt
Note that in the highly-eccentric regime, GW emission can become significant. While our current setup neglects radiation reaction, this effect could, in principle, be included to extend the validity of the algorithm. That said, for the semi-major axis and eccentricity values considered here (specifically, $\varepsilon_0 < 0.999$ and $a_0 = 10^6 M$), the periapsis remains at radii larger than $1000M$, corresponding to inspiral timescales of millions of years for typical EMRI systems. As such, ignoring GW emission will not significantly impact the results in the parameter space we explore.
\subsection{Comparisons}
\subsubsection{AGN models}\label{subsec_Capture:compareAGN}
\begin{figure}[t!]
    \centering
    \includegraphics[scale = 1.1]{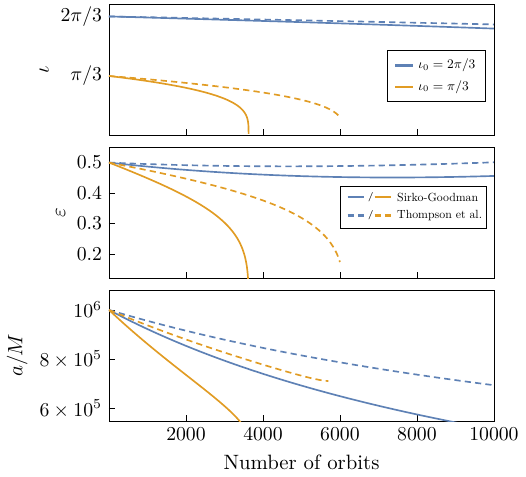}
    \caption{Comparison of the Sirko-Goodman (solid) and Thompson et al.~(dashed) models. We use $\varepsilon_0 = 0.5$, $\theta_0 = \pi/3$ and $\iota_0 = 2\pi/3$ [{\color{Mathematica1}{blue}}] or $\iota_0 = \pi/3$ [{\color{Mathematica2}{orange}}]. The benchmark parameters for the Sirko-Goodman model are taken from Table~\ref{tab:benchmark}, while the Thompson et al.~parameters are the same as in Figure~\ref{fig:densities_AGN}.}
    \label{fig:SG_Tho}
\end{figure}
So far, we have exclusively focused on the Sirko-Goodman model. This choice is motivated by the fact that, despite differences in disk structure and velocity profiles, both models yield comparable effects on the orbital parameters, with the magnitude of these effects being sensitive to specific parameter choices. To highlight their differences more clearly, Figure~\ref{fig:SG_Tho} illustrates the behaviour of the two models in the context of dynamical friction.
\vskip 2pt
As discussed in Section~\ref{BHenv_sec:accretion_disks_AGN}, the Thompson et al.~model incorporates non-Keplerian angular velocities~\eqref{eq:vel_non-kep}. However, we explicitly verified that this modification has only a marginal impact on the orbital evolution. In general, the Sirko-Goodman model tends to produce stronger effects, mainly due to its larger scale heights at the orbital separations we are interested in. Moreover, the gas density in the Thompson et al.~model is generally lower at smaller radii, as seen in Figure~\ref{fig:densities_AGN}.
\subsubsection{Accretion versus friction}\label{subsec_Capture:compareeffect}
\begin{figure}[t!]
    \centering
    \includegraphics[scale = 1.1]{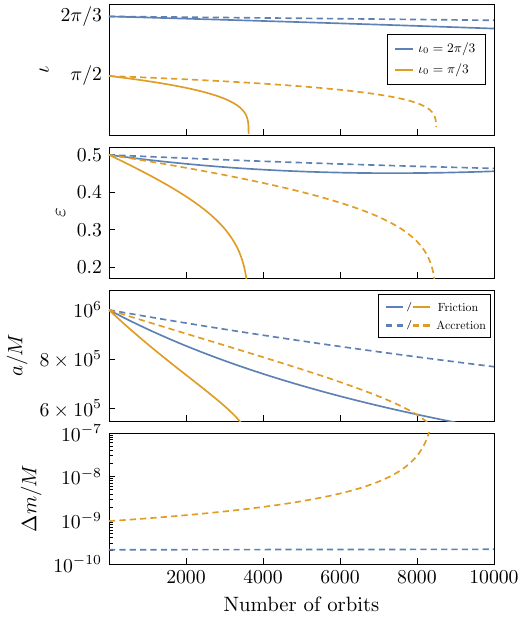}
    \caption{The impact of dynamical friction (solid) and accretion (dashed) on the evolution of inclination, eccentricity, and semi-major axis for $\iota_0 = 2\pi/3$ [{\color{Mathematica1}{blue}}] or $\iota_0 = \pi/3$ [{\color{Mathematica2}{orange}}]. All other parameters follow Figure~\ref{fig:SG_Tho}.}
    \label{fig:accretion_vs_friction}
\end{figure}

In the previous analysis, we neglected accretion in order to streamline the discussion. As illustrated in Figure~\ref{fig:accretion_vs_friction}, both accretion and dynamical friction act in qualitatively similar ways on the secondary’s orbit, exerting drag forces that alter its motion over time. While the overall impact of accretion is generally weaker compared to that of dynamical friction, it does introduce one qualitatively distinct effect:~the gradual increase in the mass of the secondary. Although this mass growth is typically subdominant in terms of its direct impact on orbital evolution, it can become significant near the point of alignment, as shown in Figure~\ref{fig:accretion_vs_friction} (\emph{bottom panel}). In such cases, the evolving mass of the secondary may feed back into the dynamical process, modifying the inspiral rate and potentially influencing the final configuration of the system.
\subsubsection{Previous work}
Our framework enables, for the first time, a self-consistent evolution of the orbital parameters during the early inspiral phase of an EMRI interacting with an AGN disk. Previous studies typically considered only a few scatterings, extrapolating timescales to infer long-term evolution, or applied restrictive assumptions (see e.g.,~\cite{1998MNRAS.293L...1V, 1999A&A...352..452S, MacLeod:2019jxd, Fabj:2020qqc, Nasim:2022rvl, 2023MNRAS.522.1763G, 2024MNRAS.528.4958W}). As a result, these studies often reported contradictory results.
\vskip 2pt
We find that the interactions between the secondary and the disk can significantly influence the evolution of the binary across a broad parameter space. This stands in contrast to earlier studies that argued dynamical friction and accretion would be negligible for compact objects~\cite{1998MNRAS.293L...1V, 1999A&A...352..452S, MacLeod:2019jxd}. The discrepancy is likely attributable to their assumption of extremely thin disks. Our findings align more closely with recent studies~\cite{Fabj:2020qqc,Nasim:2022rvl,2023MNRAS.522.1763G, 2024MNRAS.528.4958W}. 
\vskip 2pt
Nevertheless, key differences remain even among these more recent works. While in~\cite{Fabj:2020qqc, Nasim:2022rvl}, circular orbits were assumed, our results suggest that eccentricity plays a crucial role in the alignment process, significantly affecting timescales. We thus consider the assumption of circular orbits in these studies to be unrealistic. Furthermore, in~\cite{Nasim:2022rvl} a critical inclination angle was proposed beyond which the secondary would align into a fully retrograde orbit. Our simulations do not support this claim;~instead, we consistently find circular, prograde orbits as the final outcome. Eccentricity was included in~\cite{2023MNRAS.522.1763G}, which found that it can grow for retrograde orbits but is always damped in the prograde case. Our results partly challenge this picture:~eccentricity growth can occur even for prograde orbits across a significant portion of the parameter space (see Figure~\ref{fig:contourplots}). Similar to their study, we account for both accretion and dynamical friction, and we confirm their conclusion that accretion can be as important as friction, altering timescales by an $\mathcal{O}(1)$ factor. Our general conclusions share some qualitative similarities with the estimates in~\cite{2024MNRAS.528.4958W}, particularly regarding the evolution of the eccentricity with respect to the true anomaly. In line with their predictions, we observe that for highly asymmetric crossing points ($\theta \approx 0$), eccentricity increases across a large parameter space before quickly damping once the inclination reaches a critical value. Additionally, we independently confirm their prediction that for highly symmetric crossing points ($\theta \approx \pi/2$), eccentricity never grows, even for retrograde orbits.
\section{Summary and Outlook}\label{sec_Capture:conclusions}
In this chapter, we developed a new framework to study the orbital evolution of EMRIs embedded in AGN disks. Unlike previous approaches, our method enables the evolution of the system to be tracked over an arbitrary number of orbits, capturing both short-term interactions and long-term orbital changes in a computationally efficient way.
\vskip 2pt
Our simulations show that AGN disks strongly influence the evolution of EMRIs. As a general outcome, the secondary always aligns with the disk plane, usually in relatively short timescales. While the evolution of eccentricity is more intricate, we find that it, too, eventually damps as the system aligns. These results provide insight into the \emph{expected configuration} of EMRIs once they enter the sensitivity band of GW detectors and raise questions about the possibility of using residual eccentricity as a diagnostic for identifying EMRIs in accretion disks~\cite{Klein:2022rbf,Garg:2023lfg,Wang:2023tle,Romero-Shaw:2024klf,Samsing:2024syt,Duque:2024mfw}. We also uncover previously unexplored dynamics for highly eccentric binaries and perform a comparison with existing literature.
\vskip 2pt
This study opens new avenues for a comprehensive investigation of disk-satellite interactions in EMRIs. There are several directions that warrant further exploration in the future.
\begin{itemize}
\item Our framework can be readily adapted to track the evolution of stars embedded in AGN disks, where disk interactions are expected to be even more pronounced~\cite{Fabj:2020qqc,Nasim:2022rvl,2023MNRAS.522.1763G}.
\item By computing the semi-major axis at the end of the alignment process, our framework enables a systematic study of populations of massive stars and BHs in AGN disks.
\item Integrating our model with EMRI formation scenarios and gas-driven migration models would enable a coherent and chronological picture of EMRIs before they enter the LISA band -- from initial capture, through alignment to the onset of GW emission in the LISA band.
\end{itemize}
Ultimately, the strength of this framework lies in its versatility and computational efficiency, offering a powerful tool to explore the rich and complex dynamics of EMRIs in gas-rich environments. It can serve both as a cross-check for full numerical simulations and as a guide for identifying the most relevant regions of parameter space, aiding GW source modelling for future detectors.
\chapter{Conclusions and Outlook}\label{chap:Conclusions}
\vspace{-0.8cm}
\hfill \emph{The first principles of the Universe are atoms and empty space;}

\hfill \emph{everything else is merely thought to exist.}
\vskip 5pt

\hfill Diogenes La\"{e}rtius, \emph{Democritus} ({\selectlanguage{greek}Dhm\'okritos}), Book~IX
\vskip 35pt
\noindent The detection of gravitational waves has ushered in a new era of astrophysics. Gravity, once testable only in the weak-field regime, is now explored in its most extreme form. Black holes, once visible only through indirect electromagnetic observations, are now directly observed. And gravitational waves, once confined only to theory, are now tools for precise measurement.
\vskip 2pt
These developments are not just technological:~they are conceptual. Gravity is the one universal language of the Universe, and we finally have access to its dictionary. This has sparked excitement across gravitational, astro, and high-energy physics communities, in part driven by the question:~can gravitational waves reveal the fundamental constituents of the Universe? The environments surrounding black holes may hold part of the answer. With next-generation detectors pushing the boundaries of sensitivity and frequency coverage, there is an increasing interest in how these environments imprint themselves on gravitational-wave signals. This thesis contributes to that ongoing effort.
\vskip 2pt
One particularly striking scenario involves an environment black holes can generate through superradiance. This process enables ultralight bosonic fields to ``tap'' energy and angular momentum from rotating black holes, forming a dense cloud around them. In Chapters~\ref{chap:SR_Axionic} and~\ref{chap:in_medium_supp}, I explored the interactions between such clouds and the electromagnetic sector. While previous studies suggested the possibility of electromagnetic bursts of radiation, evolving the joint dynamics of the boson and electromagnetic field reveals a different outcome:~rather than bursts, a \emph{stationary} emission of light arises, where the black hole's rotational energy sustains continuous, monochromatic radiation. The detectability of this signal, however, depends on its ability to reach Earth. By studying astrophysical plasmas on curved spacetime, I concluded that even low-density plasmas can prevent most of the radiation from propagating -- a phenomenon known as \emph{in-medium suppression}. This alters the expected observational signatures from these systems and emphasises the importance of considering multiple, potentially coexisting matter configurations around black holes.
\vskip 2pt
In binary systems, environments can shape the dynamics and leave distinctive features in the emitted gravitational waves. \emph{In vacuum}, the final stage of a binary coalescence -- the ringdown -- provides a powerful probe of the properties of the remnant black hole. This naturally raises the question whether environments influence this outcome. One compelling scenario involves the interaction between charged black holes and astrophysical plasmas. In Chapter~\ref{chap:plasma_ringdown}, I showed that two key effects can arise:~(i) when plasma extends to the light ring, it alters the fundamental \emph{gravitational} quasi-normal mode of the remnant;~(ii) when plasma is localised further out -- as may occur if a binary decouples from its circumbinary disk late in the inspiral~\cite{Armitage:2002uu,Dittmann:2023dss,ONeill:2025bjd} -- electromagnetic ringdown modes can reflect off the plasma barrier, producing gravitational-wave echoes. While such scenarios yield distinct observational signatures, the question of their detectability still needs to be addressed. In Chapter~\ref{chap:BHspec}, I examined this issue in the context of dark matter halos. Focusing on the projected sensitivity of future detectors and adopting realistic astrophysical parameters, I found that the ringdown signal is effectively \emph{indistinguishable} from that in vacuum. As a result, analyses based solely on the ringdown phase remain reliable for inferring the remnant's mass and spin, which can, in turn, help identify any environmental effects that may have influenced the earlier stages of the coalescence.
\vskip 2pt
Although the ringdown can offer a clean probe of new physics, its duration is short-lived. In contrast, the inspiral phase lasts much longer -- particularly for extreme mass ratio binaries, which are key targets for next-generation detectors. These systems can remain in-band for years, allowing even subtle influences to accumulate to a detectable level. Consequently, much of the focus on black hole environments has centred on this phase. A remarkably rich phenomenology arises in the case of superradiant boson clouds. To fully understand the cloud-binary system in the late inspiral, when it enters the detector's sensitivity band, it is necessary to track its evolution from formation onwards and construct a coherent picture of the system’s history. A crucial mechanism shaping this evolution is the occurrence of orbital resonances between different bound states of the cloud. In Chapter~\ref{chap:legacy}, I generalised the framework of resonances to fully generic orbits, revealing a strong dependence on the binary's orbital parameters. In particular, if the binary is nearly counter-rotating, resonances are weak, and the cloud remains intact throughout the inspiral. This enables more accurate predictions of the system’s configuration when it enters the sensitivity band of detectors, aiding parameter estimation in the analysis of the gravitational-wave signal. Conversely, for binaries that are not close to counter-rotating, resonances lead to a complete absorption of the cloud by the central black hole. As this happens, the binary is driven towards a co-rotating configuration, while the eccentricity settles towards an attractor value that may be nonzero. Consequently, the cloud leaves a lasting \emph{imprint} on the binary and the resulting gravitational-wave signal.
\vskip 2pt
Should the cloud survive its early interaction with the binary, it eventually enters the relativistic regime. At this stage, new modelling techniques are required. In the vacuum case, the self-force method~\cite{Barack:2018yvs,Pound:2021qin} offers the most effective approach for extreme mass ratio systems. In Chapter~\ref{chap:inspirals_selfforce}, I extended this framework to non-vacuum spacetimes, making it possible to account for generic environments. Applying it to the cloud-binary system in the Kerr geometry, I explored its behaviour deep in the relativistic regime. As expected, Newtonian approximations and linear-motion estimates lead to large deviations. Perhaps surprisingly, even the differences with the Schwarzschild case prove significant, highlighting the critical role of black hole spin.
\vskip 2pt
Understanding the dynamics between binaries and their environment is crucial for characterising gravitational-wave signals, but it also raises questions about their formation. In particular, the origin of extreme mass ratio binaries remains largely uncertain~\cite{Amaro-Seoane:2012lgq}. One proposed mechanism involves the capture of compact objects in the accretion disks of active galactic nuclei. Once captured, the object typically follows a high-eccentricity orbit with a generic inclination, undergoing frequent scatterings with the disk well before reaching the sensitivity band of detectors. In Chapter~\ref{chap:capture}, I introduced a new framework to model these scatterings and explored the system across a wide variety of initial conditions. The results showed that the disk always drives the binary into the disk plane on relatively short timescales. The evolution of the eccentricity is more complex:~while it typically decreases, certain regions of parameter space allow it to grow. This intricate behaviour provides valuable insights into the binary's dynamics during the early inspiral, and could shed light on its formation.
\subsubsection{Outlook}
The study of black hole environments is broad. Addressing every open question regarding their modelling or impact on gravitational-wave astronomy is neither feasible nor particularly insightful. Instead, the following points represent, in my view, the key next steps for future research.
\vskip 2pt
\noindent \textbf{\emph{Modelling.}} 
\vspace{-0.15cm}
\begin{itemize}
\item Decades of research have produced a variety of models describing structures around black holes, particularly in the context of dark matter (see Section~\ref{BHenv_subsec:DMStruct}). With LISA on the horizon~\cite{Colpi:2024xhw}, a systematic assessment of these models is timely. This includes evaluating their stability, such as in the presence of hierarchical black hole mergers~\cite{Merritt:2002vj,Kavanagh:2018ggo}, and ensuring their consistency with astrophysical and cosmological constraints (see, e.g., Figure~\ref{fig:constraints}).
\item Different environments may interact not just with the black hole and the binary, but also with each other. One instance of this was discussed in Chapter~\ref{chap:SR_Axionic}, where boson clouds generate electromagnetic radiation, but surrounding plasma suppresses its propagation. Another example involves accretion disks, which can spin up black holes, potentially facilitating the formation of boson clouds~\cite{Sarmah:2024nst}. While modelling individual environments is already challenging, real astrophysical systems -- especially in galactic centres -- are likely to host multiple coexisting environments. Understanding how these influence each other requires a sufficiently flexible modelling framework.
\end{itemize}
\noindent \textbf{\emph{Formation.}} The process by which extreme mass ratio binaries form is still not well understood~\cite{Amaro-Seoane:2012lgq}. In vacuum, various mechanisms have been proposed, such as successive two-body encounters~\cite{Amaro-Seoane:2012lgq} or the Hill mechanism~\cite{1988Natur.331..687H}. Environments may influence the formation process by providing an additional channel for energy loss. This has been studied extensively for accretion disks and active galactic nuclei (see Chapter~\ref{chap:capture} and references therein), but other environments, like boson clouds, can play a similar role~\cite{Tomaselli:2023ysb}. A promising direction for future work is to better understand how environments shape binary formation and how they influence the expected EMRI rate for next-generation detectors.
\vskip 2pt 
\noindent \textbf{\emph{Dynamics.}} The evolution of binaries within astrophysical environments is a central focus of current research. Accurately modelling their dynamics requires a detailed understanding of effects such as accretion and dynamical friction in various matter configurations. Recently, increasing attention has been directed towards how environments influence the orbital parameters of binaries.
\vspace{-0.15cm}
\begin{itemize}
\item As discussed in Chapters~\ref{chap:legacy} and~\ref{chap:capture}, environments can dramatically impact the eccentricity and inclination of binaries. While those studies were conducted in the Newtonian regime, extending binary evolution to include generically inclined and eccentric orbits in the relativistic regime remains an open challenge. Even for vacuum Kerr, this problem has not been fully solved for extreme mass ratio binaries, making it a crucial avenue for future research.
\item Extending to generic orbits is not only necessary for a consistent evolution of the system, but it also opens the possibility of using orbital parameters as a probe of the environment itself;~for example, through a measurable residual eccentricity in the waveform~\cite{Klein:2022rbf,Garg:2023lfg,Wang:2023tle}. Various mechanisms, such as perturbations from third bodies~\cite{Gultekin:2004pm,Samsing:2013kua}, circumbinary disks~\cite{Zrake:2020zkw,DOrazio:2021kob,Siwek:2023rlk,Tiede:2023dwq}, and boson clouds~\cite{Tomaselli:2023ysb,Tomaselli:2024bdd,Tomaselli:2024dbw,Boskovic:2024fga} (see Figures~\ref{fig:streamplot_inclination-eccentricity}--\ref{fig:parameter-range}) all induce eccentricity, with the latter two even admitting nontrivial equilibrium points. While eccentricity alone may not suffice to distinguish between environments~\cite{Romero-Shaw:2024klf}, orbital inclination provides an additional diagnostic. Identifying a statistically significant preference for a particular inclination angle or eccentricity among detected sources could offer compelling evidence for the presence of an environment. Population studies aimed at uncovering such trends represent a promising area for further exploration.
\end{itemize}
\vskip 2pt
\noindent \textbf{\emph{Waveforms.}}
\vspace{-0.15cm}
\begin{itemize}
\item Numerical relativity can play an important role in the study of environments for two main reasons:~(i) it provides an essential cross-check for waveform computations, and (ii)~it enables the evolution of systems in a regime where the full nonlinear nature of Einstein’s equations becomes important, such as during merger.\footnote{For extreme mass ratio inspirals, the secondary simply plunges into the primary black hole, which can be modelled \emph{in vacuum} using semi-analytic techniques~\cite{Ori:2000zn,Apte:2019txp,Hughes:2019zmt,Becker:2024xdi,Kuchler:2024esj}. However, whether these can -- or need to -- be extended to include environments is unclear. Furthermore, numerical relativity could still be important for the (pre)-merger stage of intermediate mass ratio systems.} Due to prohibitively long timescales, numerical relativity is impractical for systems with small mass ratios. However, in vacuum, second-order self-force calculations have been successfully compared against numerical relativity for binaries with mass ratios in the range $\sim 0.01-0.1$~\cite{Wardell:2021fyy}. A similar comparison including environmental effects would be an important validation step. That said, numerical relativity in non-vacuum spacetimes still faces challenges related to the well-posedness of some systems and the ability to generate sufficiently accurate waveforms for detectors like LISA or Einstein Telescope~\cite{Purrer:2019jcp,Ferguson:2020xnm}.
\item For extreme mass ratio inspirals in vacuum, second-order self-force corrections are expected to be necessary to meet LISA's precision requirements~\cite{Hinderer:2008dm,Hughes:2016xwf}. When extending the self-force framework to include environmental effects, as outlined in Chapter~\ref{chap:inspirals_selfforce}, additional contributions arise. While we computed the leading-order perturbation to the scalar field, a fully self-consistent treatment requires incorporating all terms up to order $q^2$, such as the $\epsilon^2 q$--term~\eqref{eq:expansion_metric} of the metric perturbation. This would pave the way for constructing the first fully self-consistent waveforms for EMRIs in astrophysical environments -- an exciting prospect for upcoming work.
\end{itemize}
\vskip 2pt
\noindent \textbf{\emph{Data analysis.}} With accurate waveforms in hand, the signal must be properly interpreted. Three key aspects need to be considered:~(i) whether the signal is loud enough to be detected, that is, if its signal-to-noise ratio is high enough;~(ii) whether it is distinguishable from vacuum signals; and~(iii) whether the parameters of the system can be accurately extracted. In the context of environments, the following points deserve extra care.
\vspace{-0.15cm}
\begin{itemize}
\item Although a range of environments have been shown to influence gravitational-wave emission, comparisons are often made against vacuum only. To determine whether properties of the environment can be extracted from data, it is necessary to distinguish between different environments, not just between vacuum and non-vacuum scenarios. Preliminary Newtonian studies suggest this may be feasible~\cite{Cole:2022yzw}, but a more realistic (and relativistic) analysis is needed.
\item Additionally, various astrophysical phenomena can mimic environmental effects in gravitational-wave signals. This could introduce significant parameter biases and even lead to the misidentification of the compact object. A data-analysis-driven study is essential to understand to what degree all these effects are important. Some examples are:
\begin{itemize}
\item Possible modifications to General Relativity at high curvature scales, such as in dynamical Chern-Simons or Einstein-Gauss-Bonnet gravity.  These could induce ``beyond-GR'' effects, which alter binary mergers in the strong-field regime~\cite{Yunes:2009ke,Yunes:2013dva,Yunes:2016jcc,Okounkova:2017yby,Okounkova:2020rqw}, and can produce waveform deviations resembling those induced by matter configurations~\cite{Yuan:2024duo}. Distinguishing between these scenarios may require systematic tests, such as inspiral-merger-ringdown consistency checks, as beyond-GR deviations often become most pronounced during the merger and ringdown phases, where gravity is strongest. That said, small effects can also accumulate during the inspiral, particularly in long-lived systems such as EMRIs.
\item Third-body interactions, especially in galactic centres, where additional objects are likely present~\cite{Yunes:2010sm,Meiron:2016ipr,Zwick:2024yzh,Samsing:2024syt,Hendriks:2024zbu}.
\item Nonzero tidal Love numbers induced by environments, which impact waveforms at 5PN order~\cite{Baumann:2018vus,Cardoso:2019upw,DeLuca:2021ite,Cardoso:2021wlq,Brito:2023pyl,Cannizzaro:2024fpz}. Early studies have shown that neglecting these can indeed lead to significant biases~\cite{DeLuca:2025bph}.
\end{itemize}
\item Instrumental effects in detectors such as LISA further complicate parameter inference. Systematic uncertainties due to data gaps~\cite{Robson:2018jly,Baghi:2021tfd,Seoane:2021kkk,Burke:2025bun} and calibration inaccuracies may affect the precision with which astrophysical parameters can be extracted~\cite{Cornish:2005qw,Littenberg:2023xpl,Strub:2024kbe,Katz:2024oqg}.
\end{itemize}
\vskip 2pt
Beyond the specific points outlined above, realistic gravitational-wave source modelling requires going beyond toy models and crude approximations. Recent years have seen a promising shift from Newtonian models to fully relativistic treatments~\cite{Brito:2023pyl,Duque:2023seg,Khalvati:2024tzz,Dyson:2025dlj,Vicente:2025gsg}. Looking ahead, it is crucial to develop a systematic and flexible framework capable of capturing a variety of environments. Chapter~\ref{chap:inspirals_selfforce} marks an important step in this direction. The real challenge now lies in applying this framework to more complex scenarios, such as accretion disks, which will undoubtedly introduce various complications. Equally important is maintaining a strong focus on data-analysis considerations, not only to account for parameter degeneracies but also to avoid overly complex models that risk making the analysis impractical. 
\vskip 2pt
We are fortunate to be working in an era of gravitational physics where data is present, guiding us towards the next phase. The excitement surrounding this field continues to grow, with the study of black hole environments attracting attention not only within gravitational physics but also from other fields like high-energy physics. Realising its full potential, however, requires a careful approach:~astrophysics is inherently messy, while gravitational-wave physics demands precision. With proposals for new detectors moving forward~\cite{ET:2019dnz,Colpi:2024xhw}, now is the time to set up a robust pipeline for modelling EMRIs in astrophysical environments and integrating these efforts into waveform packages~\cite{Chua:2020stf,Katz:2021yft,Hughes:2021exa}. This will maximise the scientific return and ensure that the most astrophysically plausible scenarios are prioritised.
\vskip 2pt
Achieving this goal is undoubtedly a herculean task, but the rewards are undeniable:~gravitational waves have the potential to illuminate regions of the Universe that were once in the dark. The present thesis has sought to explore the wealth of observational signatures astrophysical environments may bring and to lay the groundwork for future exploration. Above all, I hope to have conveyed this message:~gravitational waves are cosmic messengers, carrying insights from some of the most enigmatic regions of our Universe. The challenge ahead is great -- but so too is the opportunity it presents.
\epilogue{Epilogue}{\input{Epilogue}}
\appendix
\addtocontents{toc}{\protect\setcounter{tocdepth}{1}}
\chapter{Numerical Relativity Simulations} \label{app:NR}
This appendix provides additional details on the numerical relativity simulations discussed in Chapter~\ref{chap:SR_Axionic}. Section~\ref{appNR_sec:freescalars} covers aspects specific to scalar fields, while Section~\ref{appNR_sec:waveextraction} explains the wave extraction process. Section~\ref{appNR_sec:Cauchy} describes the formulation of our problem as an initial-value problem. Section~\ref{appNR_sec:convergence} presents numerical convergence results, followed by Section~\ref{appNR_sec:higherorder}, which examines higher multipole contributions to the scalar and electromagnetic flux. Finally, Section~\ref{appNR_sec:selectionrules} outlines the selection rules that determine which multipoles are excited.
\section{Benchmarks for Evolution of Scalar Fields}\label{appNR_sec:freescalars}
The purpose of this appendix is to study in some detail the time evolution of free massive scalar fields in the vicinity of a Schwarzschild BH. Even though superradiance requires a spinning BH and thus the use of the Kerr metric, timescales are prohibitively large. Nevertheless, the main focus of Chapter~\ref{chap:SR_Axionic} is on physics related to the \emph{existence} of scalar clouds, more than to what caused them in the first place.
\vskip 2pt
As such, we mimic superradiant growth without the need of a spinning BH (see Section~\ref{appNR_subsec:artificial_super} below) and therefore we consider a Schwarzschild spacetime for simplicity. We still need to guarantee that, on the required timescales, a bound state exists, so that it can mimic well the true superradiant clouds. Fortunately, massive scalars around non-spinning BHs do settle on quasi-bound states which, while not unstable, have extremely large lifetimes. Thus, we want to show first of all that our numerical framework reproduces well such states.
\subsection{Bound States}\label{appNR_subsec:boundstates}
\begin{figure}[t!]
\centering
    \includegraphics[scale=0.45, trim = 0 0 0 0]{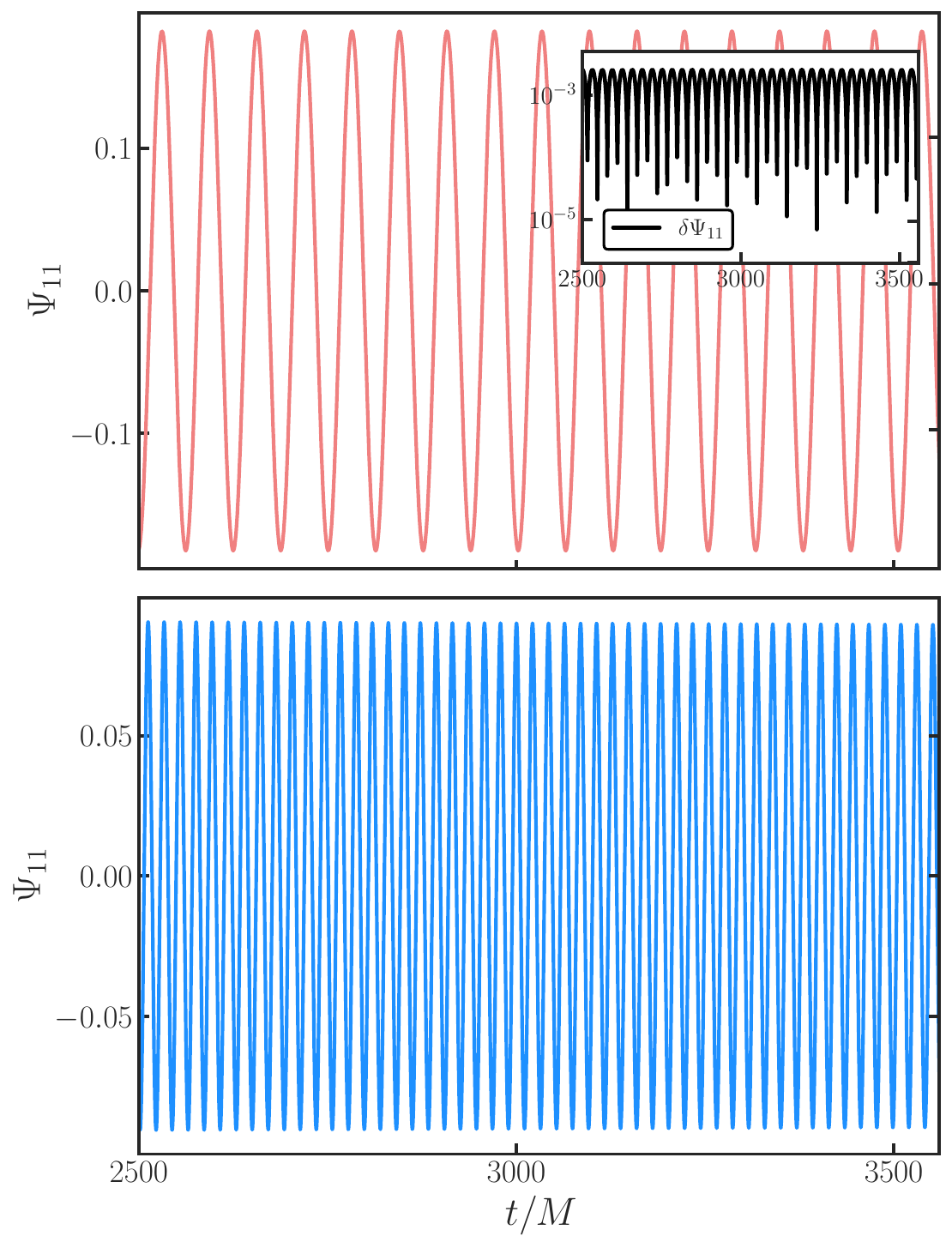}
    \caption{\emph{Top panel}:~The $\ell = m = 1$ component of a scalar field around a Schwarzschild BH. Figure shows fraction of the time evolution of the initial conditions from~\eqref{eq:scalarfieldKS}. The field is extracted at $r\ped{ex} = 100 M$ and $\mu M = 0.1$. Inset shows $\delta \Psi_{11}$, which is the difference between the numerical output and the theoretically predicted fundamental mode $\Psi^{\mathrm{Fund}}_{11} \sim \cos{(\omega_{\scalebox{0.55}{$\mathrm{R}$}} t)}e^{-\omega_{\scalebox{0.55}{$\mathrm{I}$}} t}$, where $\omega_{\scalebox{0.65}{$\mathrm{R}$}}$, $\omega_{\scalebox{0.55}{$\mathrm{I}$}} $ are the real and imaginary part of the eigenfrequency, respectively. These were independently computed using Leaver's method. \emph{Bottom panel}:~Same for $\mu M=0.3$ and extraction radius $r\ped{ex} = 40 M$. There is an apparent decay of the field on timescales shorter than those implied by the quasi-bound state decay. This effect is due to finite resolution, and its magnitude is small enough such that we can ignore it in our study.}
    \label{fig:scalarfieldmuM01}
\end{figure}

\begin{figure}[t!]
    \centering
    \includegraphics[scale=0.6, trim = 0 0 0 0]{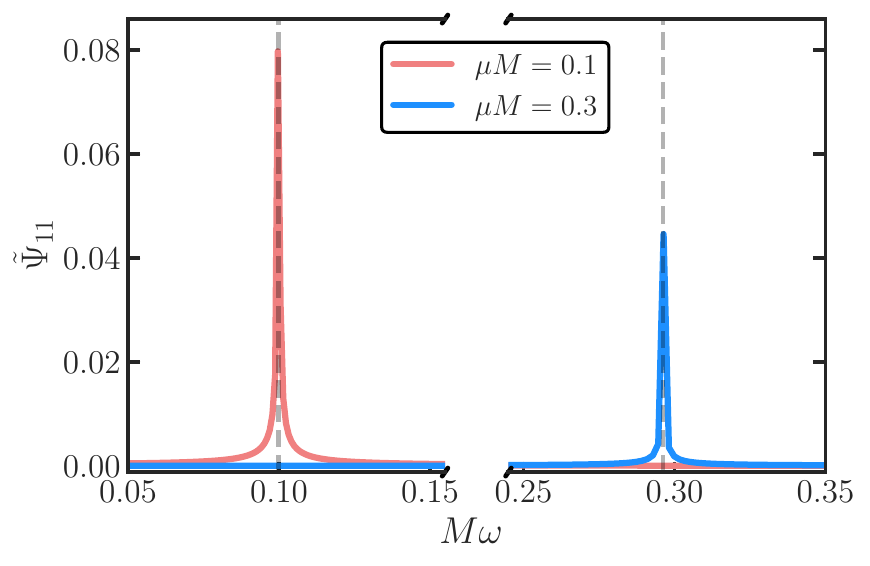}
    \caption{Fourier transform of the dipole component of the scalar field for $\mu M = 0.1$ and $\mu M = 0.3$ when the field is extracted at $r\ped{ex} = 100 M$ and $r\ped{ex} = 40 M$, respectively. Fourier transform is taken on the entire time evolution of Figure~\ref{fig:scalarfieldmuM01}. Dashed lines indicate the (real part of the) frequency of the fundamental mode for $\mu M = 0.1$ and $\mu M = 0.3$. Clearly, we are not triggering any overtones.}
    \label{fig:Fourierplots}
\end{figure}

The initial data whose time evolution we will study, are the quasi-bound states of a massive scalar field, which are solutions localised in the vicinity of the BH and prone to become unstable in the superradiance regime (if the BH is allowed to spin). There exist various methods to find such quasi-bound solutions, either by direct numerical integration or using continued fractions~\cite{Leaver:1985ax,Cardoso:2005vk,Dolan:2007mj,Berti:2009kk}. We will use Leaver's continued fraction approach~\cite{Leaver:1985ax}. It is crucial to have accurate solutions describing pure quasi-bound states, as deviations from such a pure state may trigger excitations of overtones, resulting in a beating pattern~\cite{Witek:2012tr}.
\vskip 2pt
In Boyer-Lindquist (BL) coordinates ($t_{\scalebox{0.55}{$\mathrm{BL}$}}$, $r_{\scalebox{0.55}{$\mathrm{BL}$}}$, $\theta_{\scalebox{0.55}{$\mathrm{BL}$}}$, $\varphi_{\scalebox{0.55}{$\mathrm{BL}$}}$), the scalar field bound state is given by [cf.~eq.~\eqref{eq:scala_ansatz}]\footnote{We include spin here for generality, although we evolve the scalar field in a Schwarzschild background.}
\begin{equation}\label{eq:scalarfieldBL}
\Psi_{\ell m}=e^{-i \omega t_{\scalebox{0.55}{$\mathrm{BL}$}}} e^{-i m \varphi_{\scalebox{0.55}{$\mathrm{BL}$}}} S_{\ell m}\left(\theta_{\scalebox{0.55}{$\mathrm{BL}$}}\right) R_{\ell m}\left(r_{\scalebox{0.55}{$\mathrm{BL}$}}\right)\,,
\end{equation}
where $S_{\ell m}(\theta_{\scalebox{0.55}{$\mathrm{BL}$}})$ are the spheroidal harmonics. In a Schwarzschild geometry, the angular dependence is fully captured by the familiar spherical harmonics 
$ Y_{\ell m}(\theta_{\scalebox{0.55}{$\mathrm{BL}$}}, \varphi_{\scalebox{0.55}{$\mathrm{BL}$}}) = S_{\ell m}(\theta_{\scalebox{0.55}{$\mathrm{BL}$}})e^{i m\varphi_{\scalebox{0.55}{$\mathrm{BL}$}}}$. The radial dependence is given by
\begin{equation}\label{eq:radialdep}
R_{\ell m}\left(r_{\scalebox{0.55}{$\mathrm{BL}$}}\right)=\left(r_{\scalebox{0.55}{$\mathrm{BL}$}}-r_{\scalebox{0.55}{$\mathrm{BL, +}$}}\right)^{-i \sigma}\left(r_{\scalebox{0.55}{$\mathrm{BL}$}}-r_{\scalebox{0.55}{$\mathrm{BL, -}$}}\right)^{i \sigma+\chi-1}e^{r_{\scalebox{0.55}{$\mathrm{BL}$}} q} \sum_{n=0}^{\infty} a_n\left(\frac{r_{\scalebox{0.55}{$\mathrm{BL}$}}-r_{\scalebox{0.55}{$\mathrm{BL, +}$}}}{r_{\scalebox{0.55}{$\mathrm{BL}$}}-r_{\scalebox{0.55}{$\mathrm{BL, -}$}}}\right)^n\,,
\end{equation}
where
\begin{equation}
\sigma=\frac{2 M r_{\scalebox{0.55}{$\mathrm{BL, +}$}}\left(\omega-\omega\ped{c}\right)}{r_{\scalebox{0.55}{$\mathrm{BL, +}$}}-r_{\scalebox{0.55}{$\mathrm{BL, -}$}}}\,, \quad q=\pm \sqrt{\mu^2-\omega^2}\,,\quad\chi=M\frac{\mu^2-2 \omega^2}{q}\,.
\end{equation}
Here, $r_{\scalebox{0.55}{$\mathrm{BL, \pm}$}} = M \pm \sqrt{M^{2}-a_{\scalebox{0.60}{$\mathrm{J}$}}^{2}}$ are the inner~$({\scalebox{0.85}{$\mathrm{-}$}})$ and outer~$({\scalebox{0.85}{$\mathrm{+}$}})$ horizon, $\omega\ped{c} = m \Omega\ped{H} = m a_{\scalebox{0.60}{$\mathrm{J}$}}/(2Mr_{\scalebox{0.55}{$\mathrm{BL, +}$}})$ is the critical superradiance frequency~\eqref{eq:env_SR_condition}, $a_{\scalebox{0.65}{$\mathrm{J}$}}$ is the spin of the BH and to obtain quasi-bound states, one should consider the minus sign in the expression for $q$. Since all the terms in these expressions are known in closed form, we only need to solve for the frequency of the mode of interest, $\omega$. This is found by solving the following condition for $\omega$:
\begin{equation}\label{eq:Leaver}
\beta_0 - \cfrac{\alpha_0 \gamma_{1}}{\beta_{1} - 
    \cfrac{\alpha_{1} \gamma_{2}}{\beta_{2} - 
    \cfrac{\alpha_{2} \gamma_{3}}{\beta_{3} - \dots}}} = 0\,,
\end{equation}
where all the coefficients can be found in e.g.,~\cite{Dolan:2007mj}. In~\eqref{eq:radialdep}, the amplitude of the scalar field is defined arbitrarily (as long as one neglects the backreaction of the field on the background geometry). Hence, we must choose a suitable normalisation. We will normalise the field by assigning a predetermined value to the maximum of the radial wave function. In previous works~\cite{Boskovic:2018lkj, Ikeda:2018nhb}, the hydrogenic approximation was used instead, where the wave function is defined as 
\begin{equation}
\Psi=\Psi_0 r_{\scalebox{0.55}{$\mathrm{BL}$}} M \mu^2 e^{-r_{\scalebox{0.55}{$\mathrm{BL}$}} M \mu^2 / 2} \cos \left(\varphi_{\scalebox{0.55}{$\mathrm{BL}$}}-\omega_{\scalebox{0.65}{$\mathrm{R}$}} t\right) \sin \theta_{\scalebox{0.55}{$\mathrm{BL}$}}\,,
\end{equation}
where $\omega_{\scalebox{0.65}{$\mathrm{R}$}}$ is the real part of the eigenfrequency. In order to allow for a direct comparison with those works, we relate our normalisation, the maximum value of the real part of the field, $\left(R_{\ell m}\right)\ped{max}$, to this parameter $\Psi_0$. They are related by
\begin{equation}\label{eq:normalisation}
    \left(R_{\ell m}\right)\ped{max} = \frac{4 \Psi_0 \sqrt{2\pi/3}}{e}\,,
\end{equation}
where the factor $\sqrt{2\pi/3}$ comes from the normalisation of the spherical harmonics $\ell = 1$ modes, and should be adapted accordingly for higher multipoles. We will introduce relevant quantities in terms of $\Psi_0$.
\vskip 2pt
For numerical purposes, BL coordinates are not ideal due to the coordinate singularity at the horizon. Therefore, we employ Kerr-Schild coordinates, which are horizon penetrating-coordinates~\cite{Witek:2012tr}. The coordinate transformation from BL to Kerr-Schild (KS) coordinates is given by
\begin{equation}\label{eq:coordBLKS}
\mathrm{d} t_{\scalebox{0.55}{$\mathrm{KS}$}}=\mathrm{d} t_{\scalebox{0.55}{$\mathrm{BL}$}}+\frac{2 M r_{\scalebox{0.55}{$\mathrm{BL}$}}}{\Delta} \mathrm{d} r_{\scalebox{0.55}{$\mathrm{BL}$}}\,, \ \mathrm{d} r_{\scalebox{0.55}{$\mathrm{KS}$}}=\mathrm{d} r_{\scalebox{0.55}{$\mathrm{BL}$}}\,,\ \mathrm{d} \theta_{\scalebox{0.55}{$\mathrm{KS}$}}=\mathrm{d} \theta_{\scalebox{0.55}{$\mathrm{BL}$}}\,, \ \mathrm{d} \varphi_{\scalebox{0.55}{$\mathrm{KS}$}}=\mathrm{d} \varphi_{\scalebox{0.55}{$\mathrm{BL}$}}+\frac{a_{\scalebox{0.60}{$\mathrm{J}$}}}{\Delta} \mathrm{d} r_{\scalebox{0.55}{$\mathrm{BL}$}}\,,
\end{equation}
where $\Delta \equiv r^{2} -2 M r + a^{2}_{\scalebox{0.60}{$\mathrm{J}$}}$. Using this coordinate transformation in~\eqref{eq:scalarfieldBL}, we can construct the bound state scalar field as
\begin{equation}\label{eq:scalarfieldKS}
\Psi_{\ell m}=e^{-i \omega t_{\scalebox{0.55}{$\mathrm{KS}$}}}\left(r_{\scalebox{0.55}{$\mathrm{KS}$}}-r_{\scalebox{0.55}{$\mathrm{KS, +}$}}\right)^P\left(r_{\scalebox{0.55}{$\mathrm{KS}$}}-r_{\scalebox{0.55}{$\mathrm{KS, -}$}}\right)^Q 
\left(\frac{r_{\scalebox{0.55}{$\mathrm{KS}$}}-r_{\scalebox{0.55}{$\mathrm{KS, +}$}}}{r_{{}\ped{KS}}-r_{\scalebox{0.55}{$\mathrm{KS, -}$}}}\right)^R Y_{\ell m}\left(\theta_{\scalebox{0.55}{$\mathrm{KS}$}}, \varphi_{\scalebox{0.55}{$\mathrm{KS}$}}\right) R_{\ell m}(r_{\scalebox{0.55}{$\mathrm{KS}$}})\,,
\end{equation}
where $P=2 i \omega M r_{\scalebox{0.45}{$\mathrm{KS, +}$}}/(r_{\scalebox{0.45}{$\mathrm{KS, +}$}}- r_{\scalebox{0.45}{$\mathrm{KS, -}$}})$, $Q=-2 i \omega M r_{\scalebox{0.45}{$\mathrm{KS, -}$}}/(r_{\scalebox{0.45}{$\mathrm{KS, +}$}}-r_{\scalebox{0.45}{$\mathrm{KS, -}$}})$, $R=i m a_{\scalebox{0.45}{$\mathrm{J}$}}/(r_{\scalebox{0.45}{$\mathrm{KS, -}$}}-r_{\scalebox{0.45}{$\mathrm{KS, +}$}})$. From now on, we use KS coordinates without the subscript unless otherwise stated. In our non-spinning BH case, $Q = R = 0$. The remaining extra term instead cancels exactly the divergence of the field at the BH horizon. We test our numerical setup by constructing the bound state initial configuration for scalar fields with mass couplings $\mu M = 0.1$ and $\mu M = 0.3$ and evolving them in a Schwarzschild background. 
\vskip 2pt
In Figure~\ref{fig:scalarfieldmuM01}, we show the non-vanishing multipolar component of the field for $\mu M = 0.1$ and $\mu M = 0.3$, where we only display a fraction of the time evolution such that individual oscillations are visible. For $\mu M = 0.1$, the scalar field is exceptionally stable on timescales longer than $5000M$. For $\mu M = 0.3$, there is a decrease in the amplitude of a few percent on those timescales, which does not have severe consequences. In fact, this problem can be resolved by increasing the spatial resolution. 
\vskip 2pt
As a last check, in Figure~\ref{fig:Fourierplots} the Fourier transform for both $\mu M = 0.1$ and $\mu M = 0.3$ is shown, and compared with the real part of the eigenfrequency of the fundamental mode. They are in excellent comparison, ensuring that we are not triggering any overtones.
\subsection{Artificial Superradiance}\label{appNR_subsec:artificial_super}
\begin{figure}[t!]
    \centering
    \includegraphics[scale=0.45, trim = 0 0 0 30]{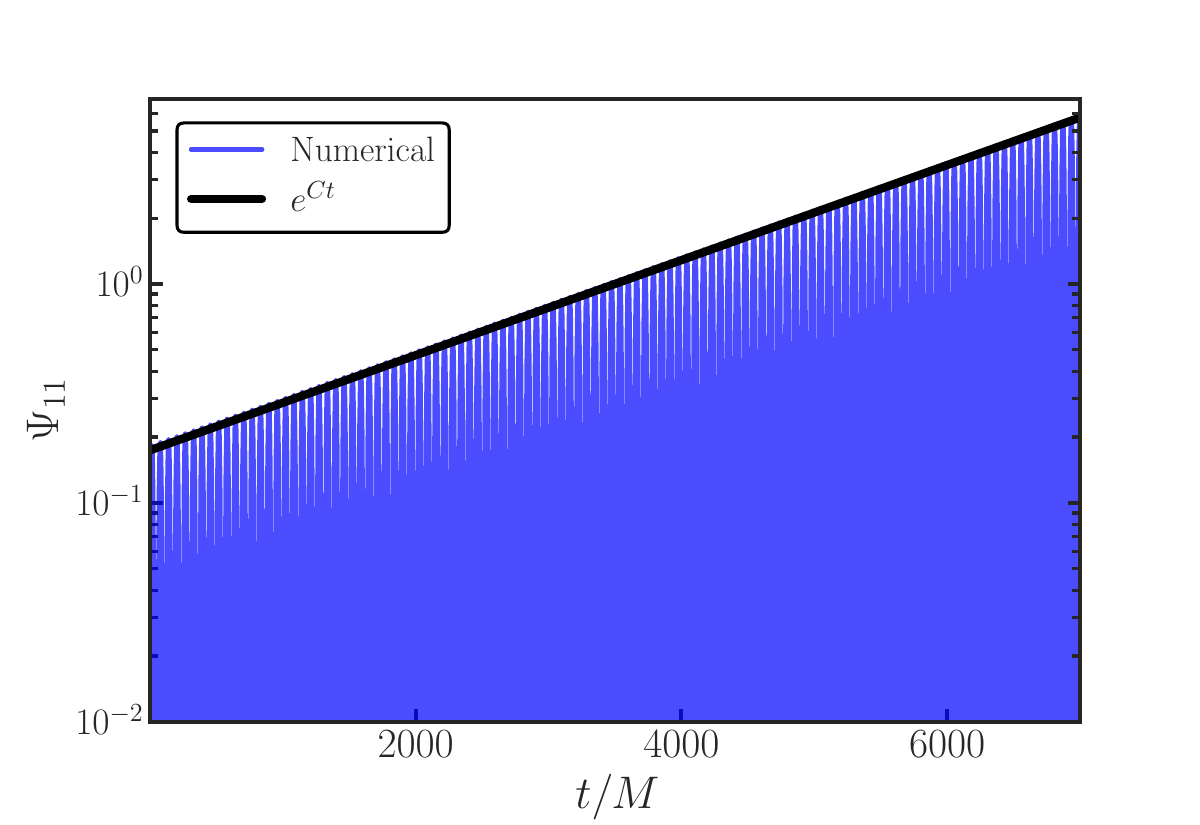}
    \caption{The time evolution of $\Psi_{11}$ extracted at $r\ped{ex} = 100 M$ with $C = 5 \times 10^{-4} M^{-1}$ and $\mu M = 0.1$. We show the growth of the scalar field from the initial conditions~\eqref{eq:scalarfieldKS}, using Zel'Dovich trick described in Chapter~\ref{chap:SR_Axionic}.}
    \label{fig:ASR}
\end{figure}

Studying superradiance for scalars is numerically challenging, since timescales for superradiant growth are very large. Fortunately, an effective superradiance-like instability can be introduced by adding a simple $C \partial \Psi/\partial t $ term to the Klein-Gordon equation as shown in~\eqref{eq:ASR}. This ``trick'' was first used by Zel'dovich~\cite{ZelDovich1971, ZelDovich1972, Cardoso:2015zqa} and it can mimic the correct description of many superradiance systems. The addition of this Lorentz-invariance-violating term causes an instability on a timescale of the order $1/C$, where we can tune $C$ to be within our numerical limits. For reference, we report the relevant timescales in our problem.
\vskip 2pt
\noindent \emph{Normal superradiance}:
\begin{equation}
    t_{\scalebox{0.65}{$\mathrm{SR}$}} \sim 48 \left(\frac{a_{\scalebox{0.65}{$\mathrm{J}$}}}{M}(\mu M)^{-9} \right) M\,, \quad \text{when} \quad \mu M \ll 1\,.
\end{equation}
\emph{Artificial superradiance}:
\begin{equation}
    t_{\scalebox{0.65}{$\mathrm{ASR}$}} \sim \frac{1}{C} M\,.
\end{equation}
\emph{Electromagnetic (EM) instability}:
\begin{equation}
    t_{\scalebox{0.65}{$\mathrm{EM}$}} \sim 10 k\ped{a}^{-1}\left(\frac{M}{M\ped{c}}\right)^{1/2}(\mu M)^{-3} = 5 (\mu M)^{-1} M\,,
\end{equation}
which is the EM instability timescale that was found in~\cite{Ikeda:2018nhb} and we used $k\ped{a} \geq 2 \sqrt{M/M\ped{c}}(\mu M)^{-2} M^{-1}$. 
\vskip 2pt
Accordingly, for a reasonable mass coupling of $\mu M = 0.1$, and while optimising superradiant growth with a maximally spinning BH, normal superradiance timescales are on 
the order of $t_{\scalebox{0.65}{$\mathrm{SR}$}} \sim 10^{10} M$. This should be compared to the EM instability, which is on $t_{\scalebox{0.65}{$\mathrm{EM}$}} \sim 50 M$. 
\vskip 2pt
To test whether we implemented the artificial superradiant growth in the correct way, we set $C = 5 \times 10^{-4} M^{-1}$ and evolve the scalar field. From Figure~\ref{fig:ASR}, we can see that artificial superradiance is correctly implemented in the code, as it leads to the desired exponential evolution of the field. 
\section{Wave Extraction}\label{appNR_sec:waveextraction}
From the simulations, we extract the radiated scalar and vector waves at some radius $r = r\ped{ex}$. For the scalar waves, the field $\Psi$ and its conjugated momentum $\Pi$ are projected onto spheres of constant coordinate radius using the spherical harmonics with spin weight $s\ped{w} = 0$:
\begin{equation}\label{eq:scalarextract}
\begin{aligned}
    \Psi_{\ell m}(t)&= \int \mathrm{d}\Omega\,\Psi(t, \theta, \varphi)\, {}_{0}\mkern-2mu Y_{\ell m}^{*}(\theta, \varphi), \\ \Pi_{\ell m}(t)&= \int \mathrm{d}\Omega\, \Pi(t, \theta, \varphi)\, {}_{0}\mkern-2mu Y_{\ell m}^{*}(\theta, \varphi)\,.
\end{aligned}
\end{equation}
To monitor the emitted EM (vector) waves, we use the Newman-Penrose formalism~\cite{Newman:1961qr}, in which the radiative degrees of freedom are given by complex scalars. For EM, these are defined as contractions between the Maxwell tensor and vectors of a null tetrad ($k^{\mu}, \ell^{\mu}, m^{\mu}, \Bar{m}^{\mu}$), where $k^\mu \ell_\mu=-m^\mu \bar{m}_\mu = -1$. The null tetrad itself is constructed from the orthonormal timelike vector $n^\mu$ and a Cartesian orthonormal basis $\left\{u^i, v^i, w^i\right\}$ on the spatial hypersurface. Asymptotically, the basis
vectors $\left\{u^i, v^i, w^i\right\}$ behave as the unit radial, polar and azimuthal vectors, respectively. For our purposes, the quantity of interest is the gauge-invariant Newman-Penrose scalar $\Phi_{2}$, which captures the outgoing EM radiation at infinity and is defined as
\begin{equation}\label{eq:NPscalar2}
    \Phi_{2} = F_{\mu \nu}\ell^{\mu}\Bar{m}^{\nu}\,,
\end{equation}
where $\ell^{\mu} = 1/\sqrt{2}\,(n^{\mu}-u^{\mu})$ and $\Bar{m}^{\mu} = 1/\sqrt{2}\,(v^{\mu}-iw^{\mu})$. Decomposing the Maxwell tensor gives
\begin{equation}
F_{\mu \nu}=n_\mu E_\nu-n_\nu E_\mu+D_\mu \mathcal{A}_\nu-D_\nu \mathcal{A}_\mu\,,
\end{equation}
where $E_{\mu} = F_{\mu\nu}n^{\nu}$ and $\mathcal{A}_\mu$ is the spatial part of the vector field $A_{\mu}$. The real and imaginary components of $\Phi_2$ are then given by
\begin{equation}
\begin{aligned}
\Phi_2^{\rm R} =&-\frac{1}{2}\left[E_i^{\rm R} v^i+u^i v^j\left(D_i \mathcal{A}_j^{\rm R}-D_j \mathcal{A}_i^{\rm R}\right) +E_i^{\rm I} w^i+u^i w^j\left(D_i \mathcal{A}_j^{\rm I}-D_j \mathcal{A}_i^{\rm I}\right)\right]\,, \\ 
\Phi_2^{\rm I} =&\,\frac{1}{2}\left[E_i^{\rm R} w^i+u^i w^j\left(D_i \mathcal{A}_j^{\rm R}-D_j \mathcal{A}_i^{\rm R}\right) -E_i^{\rm I} v^i-u^i v^j\left(D_i \mathcal{A}_j^{\rm I}-D_j \mathcal{A}_i^{\rm I}\right)\right]\,.
\end{aligned}
\end{equation}
Similar to the scalar case, we obtain the multipoles of $\Phi_{2}$ at a certain extraction radius $r\ped{ex}$, by projecting $\Phi_2$ onto the $s\ped{w}=-1$ spin-weighted spherical harmonics:
\begin{equation}\label{eq:NPextract}
\begin{aligned}
(\Phi^{\rm R}_{2})_{\ell m}(t) = \int 
\mathrm{d} \Omega&\left[\Phi_2^{\rm R}(t, \theta, \varphi)\> {}_{{\scalebox{0.65}{$-$}}1}\mkern-2mu Y_{\ell m}^{\rm R}(\theta, \varphi) +\Phi_2^{\rm I}(t, \theta, \varphi)\> {}_{{\scalebox{0.65}{$-$}}1}\mkern-2mu Y_{\ell m}^{\rm I}(\theta, \varphi)\right]\,, \\
(\Phi^{\rm I}_{2})_{\ell m}(t) = \int 
\mathrm{d} \Omega&\left[\Phi_2^{\rm I}(t, \theta, \varphi)\> {}_{{\scalebox{0.65}{$-$}}1}\mkern-2mu Y_{\ell m}^{\rm R}(\theta, \varphi) -\Phi_2^{\rm R}(t, \theta, \varphi)\> {}_{{\scalebox{0.65}{$-$}}1}\mkern-2mu Y_{\ell m}^{\rm I}(\theta, \varphi)\right]\,.
\end{aligned}
\end{equation}
In Chapter~\ref{chap:SR_Axionic}, we often show $|(\Phi_{2})_{\ell m}| = \sqrt{(\Phi_{2})^{*}_{\ell m}(\Phi_{2})_{\ell m}}$\,.
\section{Formulation as a Cauchy Problem}\label{appNR_sec:Cauchy}
We continue by formalising our equations of motion~\eqref{eq:SRAxion_evoleqns} as an (initial value) Cauchy problem and discussing the initial data.
\subsection{3+1 Decomposition}\label{appNR_subsec:DecompGeneral}
The equations of motion of our axion-photon-plasma system are given by~\eqref{eq:SRAxion_evoleqns}. We ignore the dynamics of gravity and solve the Klein-Gordon, Maxwell and plasma equations on a fixed spacetime background. In order to evolve the system in time, we use the standard 3+1 decomposition of the spacetime (see e.g.,~\cite{Alcubierre}). The metric then takes the following generic form:
\begin{equation}
    ds^2=-\alpha^2 \mathrm{d} t^2+\gamma_{i j}\left(\mathrm{d} x^i+\beta^i \mathrm{d} t\right)\left(\mathrm{d} x^j+\beta^j \mathrm{d} t\right)\,, 
\end{equation}
where $\alpha$ is the lapse function, $\beta^{i}$ is the shift vector and $\gamma_{ij}$ is the 3-metric on the spatial hypersurface. Furthermore, we introduce the scalar momentum as
\begin{equation}
    \Pi = -n^{\mu}\nabla_{\mu}\Psi\,, 
\end{equation}
where $n^{\mu}$ is the unit normal vector to the spatial hypersurface, which takes on the form $n^{\mu} = (1/\alpha, -\beta^{i}/\alpha)$. The vector field $A_{\mu}$ can be decomposed as
\begin{equation}
    A_{\mu} = \mathcal{A}_{\mu} + n_{\mu}A_{\varphi}\,,
\end{equation}
where
\begin{equation}  
\mathcal{A}_i=\gamma^j{ }_i A_j \quad \text{and} \quad A_\varphi=-n^\mu A_\mu\,.
\end{equation}
We also introduce the EM fields
\begin{equation} 
E^i=\gamma^i{ }_j F^{j \nu} n_\nu \quad \text{and} \quad B^i=\gamma^i{ }_j{ }^* \!F^{j \nu} n_\nu\,,
\end{equation}
which are defined with respect to an Eulerian observer.\footref{ft:Eulerian} As for the plasma quantities, the fluids' four velocities are decomposed as~\cite{Gourgoulhon:2007ue}
\begin{equation}\label{eq:fluidvel}
    u\ped{e}^\mu=\Gamma\ped{e}(n^\mu + \mathcal{U}^\mu)\,, \quad  u\ped{ion}^\mu=\Gamma\ped{ion}(n^\mu + \mathcal{V}^\mu)\,,
\end{equation}
where $\mathcal{U}^\mu$ and $\mathcal{V}^\mu$ are also defined with respect to an Eulerian observer. From the normalisation of the four velocities, the Lorentz factor is then:
\begin{equation}\label{eq:lorentzfactor}
    \Gamma\ped{e}=- n_\mu u\ped{e}^\mu=\frac{1}{\sqrt{1- \mathcal{U}_\mu \mathcal{U}^\mu}}\,, \quad \Gamma\ped{ion}=\frac{1}{\sqrt{1- \mathcal{V}_\mu \mathcal{V}^\mu}}\,.
\end{equation}
Note that even though in~\eqref{eq:fluidvel} and~\eqref{eq:lorentzfactor} the ion quantities are included for generality, we do not actually use them in Chapter~\ref{chap:SR_Axionic} as we ignore the oscillations of the ions [assumption (ii) in Section~\ref{sec_SR_Axionic:SRax_Plasma}]. Finally, we introduce the charge density as seen by an Eulerian observer as
\begin{equation}
\label{eq:rho}
    \rho=- n_\mu j^\mu\,.
\end{equation}
Since $j^\mu$ is the sum of the currents of the two fluids, we can express~\eqref{eq:rho} also as $\rho=\rho\ped{e}+\rho\ped{ion}$. 
\vskip 2pt
Using the above definitions, we obtain the following evolution equations for the full axion-photon-plasma system (for the decomposition of the momentum equation, we refer to Appendix~\ref{appNR_subsec:DecompMom}; for the EM part, see e.g.,~\cite{Alcubierre:2009ij}):
\begin{equation}
\begin{aligned}
\label{eq:3+1set}
\partial_t \Psi&= -\alpha \Pi+\mathcal{L}_\beta \Psi\,, \\
\partial_t \Pi&=\alpha\left(-D^2 \Psi+\mu^2 \Psi+K \Pi-2 k\ped{a} E^i B_i\right) -D^i \alpha D_i \Psi+\mathcal{L}_\beta \Pi\,, \\
\partial_t \mathcal{A}_i&=-\alpha\left(E_i+D_i A_\varphi\right)-A_\varphi D_i \alpha+\mathcal{L}_\beta \mathcal{A}_i\,, \\
\partial_t E^i&= \alpha K E^i-\alpha D_j\left(D^j \mathcal{A}^i-D^i \mathcal{A}^j\right)-\left(D^i \mathcal{A}^j-D^j \mathcal{A}^i\right) D_j \alpha + \alpha\left(D^{i}\mathcal{Z}-j^{i}\right) \\ &+2 k\ped{a}\alpha\left(B^i \Pi+\epsilon^{i j k} E_k D_j \Psi\right)+ \mathcal{L}_\beta E^i\,,\\
\partial_t A_\varphi&=-\mathcal{A}^i D_i \alpha+\alpha\left(K A_\varphi-D_i \mathcal{A}^i-\mathcal{Z}\right)+\mathcal{L}_\beta A_\varphi\,,
&\\
\partial_{t}(\Gamma\ped{e}\mathcal{U}_{i}) &= \alpha\left(\frac{q\ped{e}}{m\ped{e}}E_{i} + \epsilon_{ijk}\mathcal{U}^{j}B^{k} -\Gamma\ped{e} a_{i} -\mathcal{U}^{j} D_{j} \left(\Gamma\ped{e}\mathcal{U}_{i}\right)\right) +\mathcal{L}_\beta \Gamma\ped{e}\mathcal{U}_{i}\,,  \\
\partial_{t}\rho\ped{e} &= -D_{i}(\alpha j^{i}) +\alpha \rho\ped{e} K + \mathcal{L}_{\beta}\rho\ped{e}\,, \\
\partial_t \mathcal{Z}&= \alpha\left(D_iE^i+2 k\ped{a} B^i D_i \Psi -\rho \right)-\kappa \alpha \mathcal{Z}+\mathcal{L}_\beta \mathcal{Z}\,,
\end{aligned}
\end{equation}
where $\mathcal{L}_\beta$ is the Lie derivative and we have introduced a constraint damping variable $\mathcal{Z}$ to stabilise the numerical time evolution. Furthermore, we define $D_i$ as the covariant derivative with respect to $\gamma_{i j}$, the extrinsic curvature as $K_{i j}=\frac{1}{2 \alpha}\left[-\partial_t \gamma_{i j}+D_i \beta_j+D_j \beta_i\right]$ and $K$ as its trace. Note that due to assumption (ii), the evolution equations for the ions are absent.
\vskip 2pt
Finally, we get the following constraints:
\begin{equation}\label{eq:constrainteqn}
    \begin{aligned}
        D_{i}B^{i} &= 0\,, \\
        D_{i}E^{i} &= \rho -2k\ped{a}B_{i}D^{i}\Psi\,,\\
        (n^{\mu}+\mathcal{U}^{\mu})\nabla_{\mu}\Gamma\ped{e}&=\Gamma\ped{e}\, \mathcal{U}^{i}\mathcal{U}^{j}K_{ij}-\Gamma\ped{e}\, \mathcal{U}^{i}a_{i}-\frac{q\ped{e}}{m\ped{e}}E^{i}\mathcal{U}_{i}\,.
    \end{aligned}
\end{equation}
Upon ignoring the gravitational term in the momentum evolution equation [assumption (v) in Section~\ref{sec_SR_Axionic:SRax_Plasma}], this last constraint is trivially satisfied on the linear level.
\subsection{Background Metric}\label{appNR_subsec:backmetric}
As discussed in Appendix~\ref{appNR_subsec:boundstates}, we employ Kerr-Schild coordinates in our numerical setup to avoid the coordinate singularity at the horizon. These are related to Cartesian coordinates by
\begin{equation}
\begin{aligned}
&x=r \cos \varphi \sin \theta-a_{\scalebox{0.60}{$\mathrm{J}$}} \sin \varphi \sin \theta\,,\\
&y=r \sin \varphi \sin \theta+a_{\scalebox{0.60}{$\mathrm{J}$}} \cos \varphi \sin \theta\,,\\
&z=r \cos \theta\,.
\end{aligned}
\end{equation}
In these coordinates, the metric takes the following form:
\begin{equation}
ds^{2}=(\eta_{\mu\nu}+2Hl_{\mu}l_{\nu})\mathrm{d}x^{\mu}\mathrm{d}x^{\nu}\,,
\end{equation}
where
\begin{equation}
\begin{aligned}
H&=\frac{r^{3}M}{r^{4}+a_{\scalebox{0.60}{$\mathrm{J}$}}^{2}z^{2}}\,,\\
l_{\mu}&=\left(
1,\frac{rx+a_{\scalebox{0.60}{$\mathrm{J}$}}y}{r^{2}+a_{\scalebox{0.60}{$\mathrm{J}$}}^{2}},\frac{-a_{\scalebox{0.60}{$\mathrm{J}$}}x+ry}{r^{2}+a_{\scalebox{0.60}{$\mathrm{J}$}}^{2}},\frac{z}{r}
\right)\,,\\
r&=\left[\frac{1}{2}\left(x^{2}+y^{2}+z^{2}-a_{\scalebox{0.60}{$\mathrm{J}$}}^{2}+\sqrt{(x^{2}+y^{2}+z^{2})^{2}+4a_{\scalebox{0.60}{$\mathrm{J}$}}^{2}z^{2}}\,\right)\right]^{1/2}\,.
\end{aligned}
\end{equation}
Furthermore, we define
\begin{equation}
\begin{aligned}
\alpha&=\frac{1}{\sqrt{1+2 H}}\,, \\
\beta_i&=2 H l_i\,, \\
\gamma_{i j}&=\delta_{i j}+2 H l_i l_j\,, \\
K_{i j}&=\frac{\partial_i\left(H l_j\right)+\partial_j\left(H l_i\right)+2 H (l^{*})^k \partial_k\left(H l_i l_j\right)}{\sqrt{1+2 H}}\,,
\end{aligned}
\end{equation}
which are the lapse function, shift vector, spatial metric, and the extrinsic curvature, respectively.
\subsection{Evolution Without Plasma}\label{appNR_subsec:evolwithoutplasma}
Since the simulations \emph{with} and \emph{without} plasma have a slightly different structure, we separate these clearly in the following sections. First, we consider the full set of equations in the absence of plasma. These belong to the simulations $\mathcal{I}_{i}$ and $\mathcal{J}_{i}$ from Sections~\ref{sec_SR_Axionic:withoutSR} and~\ref{sec_SR_Axionic:withSR}. They are
\begin{equation}
\begin{aligned}
\partial_t \Psi&= -\alpha \Pi+\mathcal{L}_\beta \Psi\,, \\
\partial_t \Pi&=\alpha\left(-D^2 \Psi+\mu^2 \Psi+K \Pi-2 k\ped{a} E^i B_i\right) 
-D^i \alpha D_i \Psi+\mathcal{L}_\beta \Pi\,, \\
\partial_t \mathcal{A}_i&=-\alpha\left(E_i+D_i A_\varphi\right)-A_\varphi D_i \alpha+\mathcal{L}_\beta \mathcal{A}_i\,, \\
\partial_t E^i&= \alpha K E^i-\alpha D_j\left(D^j \mathcal{A}^i-D^i \mathcal{A}^j\right)-\left(D^i \mathcal{A}^j-D^j \mathcal{A}^i\right) D_j \alpha + \alpha D^i \mathcal{Z}\\ &+ 2 k\ped{a}\alpha\left(B^i \Pi+\epsilon^{i j k} E_k D_j \Psi\right) + \mathcal{L}_\beta E^i\,,\\
\partial_t A_\varphi&=-\mathcal{A}^i D_i \alpha+\alpha\left(K A_\varphi-D_i \mathcal{A}^i-\mathcal{Z}\right)+\mathcal{L}_\beta A_\varphi\,,
&\\
\partial_t \mathcal{Z}&= \alpha\left(D_iE^i+2 k\ped{a} B^i D_i \Psi \right)-\kappa \alpha \mathcal{Z}+\mathcal{L}_\beta \mathcal{Z}\,.
\end{aligned}
\end{equation}
These are the same as in~\cite{Boskovic:2018lkj, Ikeda:2018nhb}.
\subsubsection{Initial data}
To construct the initial data for our simulations, we must solve the constraint equations~\eqref{eq:constrainteqn}. By doing so on the initial time-slice, the Bianchi identity will ensure they are satisfied throughout the evolution. As explained in Appendix~\ref{appNR_subsec:boundstates}, we use Leaver's method to construct the scalar field bound state. For the electric field, we use initial data analogous to~\cite{Boskovic:2018lkj,Ikeda:2018nhb}. In particular, we choose the Gaussian profile defined in~\eqref{eq:InitialElectric}. 
\subsection{Evolution With Plasma}\label{appNR_subsec:evolPlasma}
In the simulations \emph{with} plasma, we linearise the axion-photon-plasma system due to the complexity of the problem, and we neglect ion perturbations. We express the perturbed quantities with an overhead bar, such that
\begin{equation}
\begin{aligned}
\mathcal{A}^{i}&=\mathcal{A}^{i}\ped{b}+\epsilon\bar{\mathcal{A}}_{i}\,, \quad E^{i}=E^{i}\ped{b}+\epsilon \bar{E}_{i}\,, \quad A_{\varphi}=A\ped{b,\varphi}+\epsilon \bar{A}_{\varphi}\,, \\ 
\mathcal{U}^{i}&=\mathcal{U}^{i}\ped{b}+\epsilon\, \bar{\mathcal{U}}_{i}\,, \quad \Gamma\ped{e}=1\,, \quad \rho=\rho_{\mathrm{b, e}}+\rho\ped{b, ion}+\epsilon\bar{\rho}\ped{e}\,, 
\end{aligned}
\end{equation}
where we denote background quantities with a subscript $\mathrm{b}$ and $\epsilon$ is the arbitrarily small parameter in the perturbation scheme. For simplicity, we consider a quasi-neutral, field-free background plasma, i.e.,~$E^{i}\ped{b}=A^{i}\ped{b}=A_{\mathrm{b},\varphi}=0$, and $\rho\ped{b, e}=-\rho\ped{b, ion}$. The problem at hand naturally introduces two distinct reference frames:~the Eulerian observer rest frame and the plasma rest frame. The relative velocity between the two is the background quantity $\mathcal{U}^i$. We consider a plasma co-moving with the Eulerian observer, such that the plasma is static in the spacetime foliation.
Since the background field of the electron charge density does not vanish, according to~\eqref{eq:3+1set} it should evolve as 
\begin{equation}\label{eq:bkgdensity} 
 \partial_{t}\rho\ped{b}=\alpha\rho\ped{b, e}K+ \mathcal{L}_\beta \rho\ped{b, e}\,.
\end{equation}
We are mainly interested in the evolution of this variable in a localised region of spacetime far away from the BH, i.e.,~the axion cloud, and thus the evolution of~\eqref{eq:bkgdensity} due to strong gravity terms is extremely slow compared to the linear system. Therefore, we neglect its evolution similarly to the gravitational influence on the evolution of the background velocity [assumption (v) in Section~\ref{sec_SR_Axionic:SRax_Plasma}].
\vskip 2pt
Before proceeding, there is one other subtlety. The plasma response to the perturbing EM field is proportional to the electron charge-to-mass ratio, which is extremely large $q\ped{e}/m\ped{e}\approx 10^{22}$.
Nevertheless, as we are linearising the system, and therefore neglecting the backreaction of the EM field onto the axion field, the amplitude of the former is arbitrary in our scheme. If we were to consider the full problem including backreaction instead, the amplitude of the axion field would clearly introduce a scale. Due to this freedom, we rescale the EM variables as
\begin{equation}
\hat{E}^{i}=\frac{q\ped{e}}{m\ped{e}}\bar{E}^{i}\,, \quad \hat{A}_{i}=\frac{q\ped{e}}{m\ped{e}}\bar{A}_{i}\,, \quad \hat{\mathcal{Z}}=\frac{q\ped{e}}{m\ped{e}}\mathcal{Z}\,.
\end{equation}
Then, we can write down the full set of equations \emph{including} the plasma as
\begin{equation}\label{eq:evoleqnplasma}
\begin{aligned}
\partial_t \Psi&= -\alpha \Pi+\mathcal{L}_\beta \Psi\,, \\
\partial_t \Pi&=\alpha\left(-D^2 \Psi+\mu^2 \Psi+K \Pi \right) -D^i \alpha D_i \Psi+\mathcal{L}_\beta \Pi\,, \\
\partial_t \hat{\mathcal{A}}_i&=-\alpha\left(\hat{E}_i+D_i \hat{A}_\varphi\right)-\hat{A}_\varphi D_i \alpha+\mathcal{L}_\beta \hat{\mathcal{A}}_i\,, \\
\partial_t \hat{E}^i&= \alpha K \hat{E}^i-\alpha D_j\left(D^j \hat{\mathcal{A}}^i-D^i \hat{\mathcal{A}}^j\right)-\left(D^i \hat{\mathcal{A}}^j-D^j \hat{\mathcal{A}}^i\right) D_j \alpha + \alpha\left(D^i \hat{\mathcal{Z}} - \omega\ped{p}^{2}\bar{\mathcal{U}}^{i}\right)\\ &+2 k\ped{a}\alpha\left(\hat{B}^i \Pi+\epsilon^{i j k} \hat{E}_k D_j \Psi\right) + \mathcal{L}_\beta \hat{E}^i\,,\\
\partial_t \hat{A}_{\varphi}&=-\hat{\mathcal{A}}^{i}D_{i}\alpha+\alpha(K\hat{\mathcal{A}}_{\varphi}-D_{i}\hat{\mathcal{A}}^i-\hat{\mathcal{Z}})+\mathcal{L}_{\beta}\hat{A}_{\varphi}\,,\\
\partial_t\bar{\mathcal{U}}_i&=\alpha\hat{E}_i+\mathcal{L}_{\beta}\bar{\mathcal{U}}_{i}\,,\\
\partial_t\bar{\omega}\ped{p}^{2}&=-D_{i}(\alpha\omega\ped{p}^{2}\bar{\mathcal{U}}^{i})+\alpha\bar{\omega}^{2}\ped{p}K+\mathcal{L}_{\beta}\bar{\omega}^{2}\ped{p}\,, \\
\partial_t \hat{\mathcal{Z}}&= \alpha\left(D_i \hat{E}^i+2 k\ped{a} \hat{B}^i D_i \Psi -\bar{\omega}^{2}\ped{p} \right)-\kappa \alpha \hat{\mathcal{Z}}+\mathcal{L}_\beta \hat{\mathcal{Z}}\,, 
\end{aligned}
\end{equation}
where $\omega\ped{p}^{2}$ is the plasma frequency, and $\bar{\omega}\ped{p}^{2}$ its perturbation:
\begin{equation}
\omega\ped{p}^{2}=\frac{q\ped{e}}{m\ped{e}}\rho\ped{b,e}\,,\quad
\bar{\omega}\ped{p}^{2}=\frac{q\ped{e}}{m\ped{e}}\bar{\rho}\ped{e}\,.
\end{equation}
Note that due to the rescaling, there is no charge-to-mass ratio of the electrons and the field equations are written only in terms of the plasma frequency, which is $\mathcal{O}(1/M)$. 
\vskip 2pt
As we detail in the following subsection, including a linearised fluid model in the equations of motion, causes the system~\eqref{eq:evoleqnplasma} to become ill-posed upon using a damping variable.\footnote{The ill-posedness originates from the linearisation of the fluid equation, whereas the fully nonlinear system of equations is strongly hyperbolic and thus well-posed~\cite{MUNZ2000484,Abgrall:2014rfb}} However, this damping variable is essential in constraining Gauss' law and without it, the simulations diverge for large EM values. As a resolution, we slightly adjust our equations by not including the perturbed plasma frequency, $\bar{\omega}\ped{p}$, in the evolution equation of the damping variable $\mathcal{Z}$, i.e.,
\begin{equation}
    \partial_t \hat{\mathcal{Z}}= \alpha\left(D_i \hat{E}^i+2 k\ped{a} \hat{B}^i D_i \Psi \right)-\kappa \alpha \hat{\mathcal{Z}}+\mathcal{L}_\beta \hat{\mathcal{Z}}\,.
\end{equation}
This is a minimal change as the perturbed plasma frequency does not enter any other evolution equation of the system in the linearised regime, yet it does restore the well-posedness of our setup. We justify this approach in two ways;~(i) we evolve the system \emph{with} and \emph{without} the damping variable for large plasma frequencies (where the EM values remain small) and we find excellent agreement between the two, and (ii) for small plasma frequencies, where the EM field is allowed to grow, the effects of the plasma are negligible, and therefore ignoring $\bar{\omega}\ped{p}$ leads to a subleading error compared to the EM values. 
\subsubsection{Initial data}
For the plasma part, we assume quasi-neutrality [assumption (iii) in Section~\ref{sec_SR_Axionic:SRax_Plasma}], i.e., $\rho = -n_{\mu}(e n\ped{e} u\ped{e}^{\mu} - Z e n\ped{ion} u\ped{ion}^{\mu}) = e n\ped{e} - Z e n\ped{ion} = 0$. As shown in~\eqref{eq:constrainteqn}, the constraint equation for the plasma is trivially satisfied on the linear level, and thus the initial data listed in the previous section solves all of our constraints. As for the electronic density, in principle, depending on the specific environment we are interested in, we can assume different spatial profiles~\cite{Dima:2020rzg, Wang:2022hra, Abramowicz2013}, which correspond to a space-dependent effective mass for the photon. However, the length scale of interest to us, i.e.,~the size of the axion cloud, is typically much shorter than length scale on which the effective mass varies. Hence, for simplicity, we assume a constant density plasma.
\subsection{Hyperbolicity of Fluid Model}\label{appNR_subsec:hyperbolicity}
The evolution equations~\eqref{eq:evoleqnplasma} are not strongly hyperbolic and therefore do not form a well-posed system. Consequently, the existence of a unique solution that depends continuously on the initial data is not guaranteed and any numerical approach is bound to fail. We now show explicitly that our system is not strongly hyperbolic.
\vskip 2pt
Based on~\cite{Hilditch:2013sba}, we introduce an arbitrary unit vector $s^{i}$ and consider the principal part of the system, i.e.,~we consider only the highest derivative terms from~\eqref{eq:evoleqnplasma}:
\begin{equation}
\begin{aligned}
    \partial_{t}\lbrack\partial_{s}^{2}\psi\rbrack&\sim -\alpha\partial_{s}\left(\partial_{s}\hat{A}_{\varphi}\right)+\beta^{s}\partial_{s}\lbrack\partial_{t}^{2}\psi\rbrack\,,\\
    \partial_{t}(\partial_{s}\hat{\mathcal{A}}_{A})&\sim -\alpha \partial_{s}\hat{E}_{A}+\beta^{s}\partial_{s}(\partial_{s}\hat{\mathcal{A}}_{A})\,,\\
    \partial_{t}(\partial_{s}\hat{A}_{\varphi})&\sim \beta^{s}(\partial_{s}\hat{A}_{\varphi})-\alpha\partial_{s}\lbrack\partial_{s}^{2}\psi\rbrack\,,\\
    \partial_{t}\hat{E}_{A}&\sim -\alpha\partial_{s}(\partial_{s}\hat{\mathcal{A}}_{A})+\beta^{s}\partial_{s}\hat{E}_{A}\,,\\
    \partial_{t}\hat{E}_{s}&\sim \alpha\partial_{s}\hat{\mathcal{Z}}+\beta^{s}\partial_{s}\hat{E}_{s}\,,\\
    \partial_{t}\hat{\mathcal{Z}}&\sim \alpha\partial_{s}\hat{E}^{s}+\beta^{s}\partial_{s}\hat{\mathcal{Z}}\,,\\
    \partial_{t}\bar{\mathcal{U}}_{s}&\sim\beta^{s}\partial_{s}\bar{\mathcal{U}}_{s}\,,\\
    \partial_{t}\bar{\mathcal{U}}_{A}&\sim\beta^{s}\partial_{s}\bar{\mathcal{U}}_{A}\,,\\
    \partial_{t}\bar{\omega}^{2}\ped{p}&\sim -\alpha\omega\ped{p}^{2}\partial_{s}\bar{\mathcal{U}}_{s}+\beta^{s}\partial_{s}\bar{\omega}\ped{p}^{2}\,,
\end{aligned}
\label{highest derivative terms}
\end{equation}
where the index $A$ denotes the component projected into the surface orthogonal to $s^{i}$, and $\lbrack\partial_{s}^{2}\psi\rbrack$ can be written as
\begin{equation}
\lbrack\partial_{s}^{2}\psi\rbrack=\partial_{s}\hat{\mathcal{A}}_{s}+\hat{\mathcal{Z}}\,.
\end{equation}
Defining the principal symbol of 
$(\lbrack\partial_{s}^{2}\psi\rbrack,\hat{A}_{\varphi})$, $(\hat{\mathcal{Z}},E_{s})$, $(\partial_{s}\bar{\mathcal{A}}_{A},\hat{E}_{A})$ and $(\bar{\mathcal{U}}_{s},\bar{\omega}\ped{p}^{2})$ as $P_{\mathcal{G}}$, $P_{\mathcal{C}}$, $P_{\mathcal{P}}$, and $P_{\mathcal{F}}$, respectively, we get:
\begin{equation}
\begin{aligned}
P_{\mathcal{G}}&=\begin{pmatrix}
\beta^{s}&-\alpha\\
-\alpha&\beta^{s}
\end{pmatrix}\,, \quad
&&P_{\mathcal{C}}=\begin{pmatrix}
\beta^{s}&\alpha\\
\alpha&\beta^{s}
\end{pmatrix}\,, \\
P_{\mathcal{P}}&=\begin{pmatrix}
\beta^{s}&-\alpha\\
-\alpha&\beta^{s}
\end{pmatrix}\,, \quad
&&P_{\mathcal{F}}=\begin{pmatrix}
\beta^{s}&0\\
-\alpha\omega\ped{p}^{2}&\beta^{s}
\end{pmatrix}\,.
\end{aligned}
\end{equation}
We see that the eigenvalue for $P_{\mathcal{F}}$ is degenerate with $\beta^{s}$ and the eigenvector is $(0,b)^{T}$, where $b$ is an arbitrary value. Therefore, the principal symbol does not have a complete set of eigenvectors, and the system is not strongly hyperbolic. If we ignore $\bar{\omega}^{2}\ped{p}$ in $\mathcal{Z}$, it becomes decoupled from Maxwell's equations, and the only relevant variable for the fluid part that remains, is the linearised four velocity $\bar{\mathcal{U}}_{i}$. As is shown in~\eqref{highest derivative terms}, the principal part for $\bar{\mathcal{U}}_{i}$ is just canonical the advection term.
\subsection{Momentum Equation}\label{appNR_subsec:DecompMom}
In order to include a plasma in our numerical setup, we need to apply the 3+1 decomposition to the momentum equation~\eqref{eq:SRAxion_evoleqns}. Even though this has been done before, e.g.,~in~\cite{Thorne1982}, it is not part of standard literature. Therefore, we do the decomposition explicitly here. Our starting point is the momentum equation for the electrons, given by (we drop the subscript ``$\mathrm{e}$'' here, since we only consider the momentum equation for the electrons)
\begin{equation}\label{eq:momeqn}
    u^{\nu}\nabla_{\nu}u^{\mu} = \frac{q}{m}F^{\mu\nu}u_{\nu}\,.
\end{equation}
The four-velocity can be written as $u^{\mu} = \Gamma(n^{\mu} + \mathcal{U}^{\mu})$, where $\mathcal{U}_{i}$ is tangent to $\Sigma_{t}$, the spatial hypersurface, such that $n^{\mu}\mathcal{U}_{\mu} = 0$ and $u^{\mu}u_{\mu} =-1$. Although we are interested in linear effects and therefore $\Gamma = 1$, we will derive the 3+1 equations in full generality and including this factor. 
\vskip 2pt
If we project~\eqref{eq:momeqn} using the projector operator $h^a_{\ b} = \delta^a_{\ b} + n^{a}n_{b}$ we obtain the evolution equation, if we project it onto $n^{a}$ we get the constraint equation. Let us start with the former.
\subsubsection{Evolution equation}
We start with the left-hand side (LHS) of~\eqref{eq:momeqn}:\footnote{We avoid writing the overall $\Gamma$ coming from $u^{\mu}$ on both sides of~\eqref{eq:momeqn}.}
\begin{equation}\label{eq:lhsmomeq}
\begin{aligned}             
    & h^\rho_{\  \mu}\left(n^{\nu}+\mathcal{U}^{\nu}\right)\nabla_{\nu}\left[\Gamma\left(n^{\mu}+\mathcal{U}^{\mu}\right)\right] =\\
    &\Gamma a^{\rho} +  \underbrace{n^{\nu}\nabla_{\nu}\left(\Gamma \mathcal{U}^{\rho}\right)  - \Gamma \mathcal{U}^{\mu}a_{\mu}n^{\rho}}_\text{I} + \underbrace{\Gamma\mathcal{U}^{\nu}\nabla_{\nu}n^{\rho}}_\text{II} +  \underbrace{h^{\rho}_{\mu}\mathcal{U}^{\nu}\nabla_{\nu}\left(\Gamma \mathcal{U}^{\mu}\right)}_\text{III}\,,
\end{aligned}
\end{equation}
where $a_{\mu} = n^{\nu}\nabla_{\nu}n_{\mu}$ is the acceleration of the Eulerian observer. Since this a projection onto the hypersurface and thus orthogonal to $n^{\mu}$, we focus on $\rho = i$, i.e.,~the spatial part. To write~\eqref{eq:lhsmomeq} in a more convenient form, we work out parts I, II and III separately in~\eqref{eq:partI},~\eqref{eq:partII},~\eqref{eq:partIII}. First, we define the useful relations~\cite{Thorne1982}:
\begin{equation}
\begin{aligned}
    D_{\tau}\mathcal{U}^{i} &= n^{\mu}\nabla_{\mu}\mathcal{U}^{i} - n^{i}a_{\mu}\mathcal{U}^{\mu}\,,\\
     \mathcal{L}_{t}\mathcal{U}^{i} &= \alpha (D_{\tau}\mathcal{U}^{i} + K^{i}_{j}\mathcal{U}^{j}) + \mathcal{L}_{\beta}\mathcal{U}^{i}\,,
\end{aligned}
\end{equation}
where $\mathcal{L}_t=\mathcal{L}_{\alpha n \scalebox{0.55}{$\mathrm{+}$} \beta}$. For part I, we then have
\begin{equation}\label{eq:partI}
n^{\mu} \nabla_{\mu}\left(\Gamma\mathcal{U}^{i}\right)- \Gamma n^{i}a_{\mu}\mathcal{U}^{\mu} = \frac{1}{\alpha}\Big[\mathcal{L}_t\left(\Gamma\mathcal{U}^{i}\right) -\mathcal{L}_{\beta}\left(\Gamma\mathcal{U}^{i}\right)\Big] - \Gamma K^{i}_{j}\mathcal{U}^{j}\,,
\end{equation}
for term II, we find
\begin{equation}\label{eq:partII}
\begin{aligned}
   \Gamma\mathcal{U}^{\nu}\nabla_{\nu}n^{i} = 
   \Gamma\mathcal{U}^{\nu}\left(-a^{i}n_{\nu} - K^{i}_{\nu}\right) = -\Gamma K^{i}_{j}\mathcal{U}^{j}\,,
\end{aligned}
\end{equation}
and finally, for term III, we have
\begin{equation}\label{eq:partIII}       h^{i}_{\mu}\mathcal{U}^{\nu}\nabla_{\nu}\left(\Gamma\mathcal{U}^{\mu}\right) = h^{i}_{\mu}h^{\nu}_{ \alpha}\mathcal{U}^{\alpha}\nabla_{\nu}\left(\Gamma\mathcal{U}^{\mu}\right) =\mathcal{U}^{\alpha}D_{\alpha}\left(\Gamma\mathcal{U}^{i}\right)\,.
\end{equation}
Combining these, we can write the LHS as
\begin{equation}\label{eq:evolLHS}
\Gamma a^{i}+\frac{1}{\alpha}\left[\mathcal{L}_t\left(\Gamma \mathcal{U}^{i}\right) -\mathcal{L}_{\beta}\left(\Gamma\mathcal{U}^{i}\right)\right]- \Gamma K^{i}_{j}\mathcal{U}^{j} - \Gamma K^{i}_{j}\mathcal{U}^{j}+ \mathcal{U}^{\alpha}D_{\alpha}\left(\Gamma\mathcal{U}^{i}\right)\,.
\end{equation}
For the right-hand side (RHS) of the momentum equation~\eqref{eq:momeqn}, we first write out the standard form of the decomposition of the Maxwell tensor (see, e.g.,~\cite{Alcubierre:2009ij}) and then project it onto the spatial hypersurface using the projector operator: 
\begin{equation}\label{eq:evolRHS}
\begin{aligned}
    &h^\rho_{\ \mu}\frac{q}{m}F^{\mu\nu}(n_{\nu}+\mathcal{U}_{\nu}) \\
    &=h^\rho_{\ \mu}\frac{q}{m}(\epsilon^{\alpha\mu\nu\sigma}n_{\alpha}B_{\sigma}\mathcal{U}_{\nu} + n^{\mu}E^{\nu}\mathcal{U}_{\nu} + E^{\mu}) \\ &=
    \frac{q}{m}\left( ^{(3)}\epsilon^{\rho\nu\sigma}\mathcal{U}_{\nu}B_{\sigma} + E^{\rho} \right)\,,
\end{aligned}
\end{equation}
where we used $E^{\nu}n_{\nu} = 0$, $\epsilon^{\alpha\mu\nu\sigma}n_{\alpha}n_{\nu} = 0$, and $\epsilon^{\alpha\mu\nu\sigma}n_{\alpha} =^{(3)}\epsilon^{\mu\nu\sigma}$.
\vskip 2pt
We can then piece together~\eqref{eq:evolLHS} and~\eqref{eq:evolRHS} to obtain
\begin{equation}
\mathcal{L}_t\left(\Gamma \mathcal{U}^{i}\right) -\mathcal{L}_\beta \left(\Gamma\mathcal{U}^{i}\right) + \alpha \mathcal{U}^{j} D_{j} \left(\Gamma\mathcal{U}^{i}\right)= \alpha\left(\frac{q}{m}\left(E^{i} + \epsilon^{ijk}\mathcal{U}_{j}B_{k}\right) - \Gamma a^{i}+2 \Gamma K^{ij}\mathcal{U}_j\right)\,.
\end{equation}
Finally, we apply the 3-metric tensor $\gamma_{ij}$ to the above equation to lower the index. To do so, we use the following identities~\cite{Thorne1982}:
\begin{equation}
\begin{aligned}
   \gamma_{ij} \mathcal{L}_t\left(\Gamma \mathcal{U}^{j}\right) &= \mathcal{L}_t \left(\Gamma\mathcal{U}_i \right)- \Gamma \mathcal{U}^j \mathcal{L}_t \gamma_{ij}\,, \\
    \Gamma \mathcal{U}^j \mathcal{L}_t \gamma_{ij} &= - 2 \alpha \Gamma \mathcal{U}^j  K_{ij}+ \Gamma \mathcal{U}^j \mathcal{L}_\beta \gamma_{ij}\,,   
\end{aligned}
\end{equation} 
from which we obtain:
\begin{equation}\label{eq:Uupieq}
\partial_t\left(\Gamma \mathcal{U}_{i}\right) -\mathcal{L}_\beta \left(\Gamma\mathcal{U}_{i}\right) + \alpha \mathcal{U}^{j} D_{j} \left(\Gamma\mathcal{U}_{i}\right)= \alpha\left(\frac{q}{m}\left(E_{i} + \epsilon_{ijk}\mathcal{U}^{j}B^{k}\right) - \Gamma a_{i}\right)\,.
\end{equation}
As explained in Appendix~\ref{sec_SR_Axionic:SRax_Plasma}, we simplify the plasma to make it more suitable to our numerical setup. By linearising this equation, which also implies $\Gamma\sim 1$, we are left with
\begin{equation}\label{eq:momeqnlinear}
    \partial_{t}\mathcal{U}_{i} = \alpha\left(\frac{q}{m}E_{i} - a_{i} \right) +\mathcal{L}_\beta \mathcal{U}_{i}\,.
\end{equation}
Finally, we also neglect gravity, which brings us to our final equation:
\begin{equation}
    \partial_{t}\mathcal{U}_{i} = \alpha\frac{q}{m}E_{i}+\mathcal{L}_\beta \mathcal{U}_{i}\,.
\end{equation}
\subsubsection{Constraint equation}
By projecting the momentum equation~\eqref{eq:momeqn} onto the timelike unit vector, $n^{\mu}$, we obtain the constraint equation. Again, we first show the LHS: 
\begin{equation}
n_\mu\left(n^{\nu}+\mathcal{U}^{\nu}\right)\nabla_{\nu}\left[\Gamma\left(n^{\mu}+\mathcal{U}^{\mu}\right)\right] = 
- \Gamma a_{\mu}\mathcal{U}^{\mu} - \Gamma \mathcal{U}^{\nu}\mathcal{U}^{\mu}\nabla_{\nu}n_{\mu}-n^\nu \partial_\nu \Gamma- \mathcal{U}^\nu \partial_\nu \Gamma\,,
\end{equation}
where we used that $n_{\mu}\nabla_{\nu}n^{\mu} = 0$. For the RHS, we have
\begin{equation}
\frac{q}{m}n_\mu(\epsilon^{\alpha\mu\nu\sigma}n_{\alpha}B_{\sigma}\mathcal{U}_{\nu} + n^{\mu}E^{\nu}\mathcal{U}_{\nu} + E^{\mu}) = -\frac{q}{m}E^{\nu}\mathcal{U}_{\nu}\,.
\end{equation}
Thus we end up with the following constraint equation:
\begin{equation}\label{eq:constrainMomeqn}
    \begin{aligned}
        n^\nu \partial_\nu \Gamma+ \mathcal{U}^\nu \partial_\nu \Gamma +\Gamma a_{\mu}\mathcal{U}^{\mu} + \Gamma \mathcal{U}^{\nu}\mathcal{U}^{\mu}K_{\mu \nu} = \frac{q}{m}E^{\nu}\mathcal{U}_{\nu}\,.
    \end{aligned}
\end{equation}
The fourth and fifth term are second-order and thus drop out in our linearised setup [assumption (i) in Section~\ref{sec_SR_Axionic:SRax_Plasma}]. Furthermore, we can neglect the third term since we ignore the gravity term [assumption~(v)]. As $\Gamma$ depends quadratically on the four-velocity, it must be $1$ in the linear theory, and therefore the constraint is trivially satisfied.
\section{Numerical Convergence}\label{appNR_sec:convergence}
\begin{figure}[t!]
    \centering
    \includegraphics[scale=0.45, trim = 0 0 0 20]{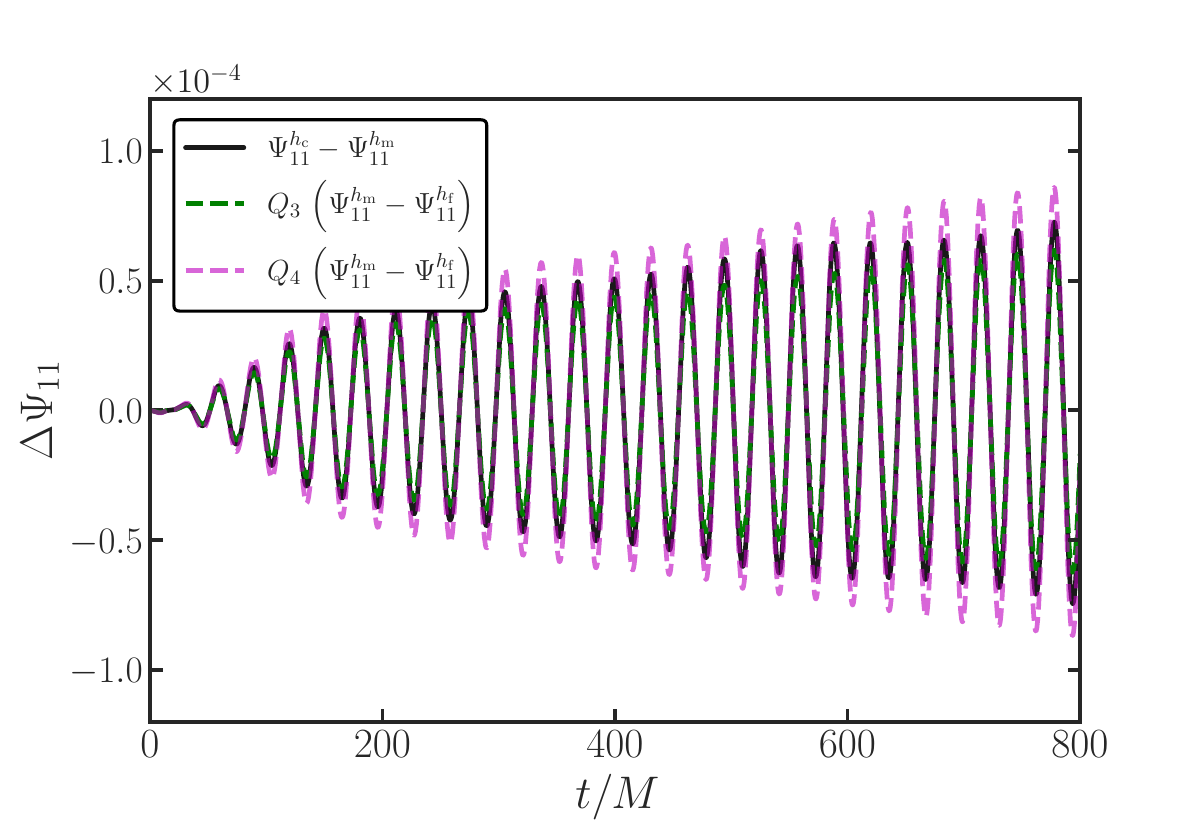}
    \caption{Convergence analysis of the $\ell = m = 1$ multipole of $\Psi$, extracted at $r\ped{ex} = 20M$ for $\mu M = 0.2$ and $C = 10^{-3}$. We show the expected result for third-order convergence [{\color{Mathematica3Dark}{green}}] ($Q_{3} = 5.64$), and fourth-order convergence [{\color{Mathematica5Light}{purple}}] ($Q_{4} = 7.85$).}
    \label{fig:convergencePhiC}
\end{figure}

In our numerical framework, we employ the method of lines, where spatial derivatives are approximated by a fourth-order accurate finite-difference scheme and we integrate using a fourth-order Runge-Kutta method. Furthermore, Kreiss-Oliger dissipation is applied to evolved quantities in order to suppress high-frequency modes that come from the boundaries between adjacent refinement regions. The numerical simulations are performed using the open source \textsc{Einstein Toolkit}~\cite{Loffler:2011ay, Zilhao:2013hia}. For the evolution of the scalar and vector field, we extent the \textsc{ScalarEvolve}~\cite{Bernard:2019nkv, Ikeda:2020xvt, Cunha:2017wao} and \textsc{ProcaEvolve} thorns~\cite{Zilhao:2015tya, Sanchis-Gual:2022zsr}, respectively. We use \textsc{Multipatch} to interpolate between different grids in our numerical domain~\cite{Pollney:2009yz, Reisswig:2012nc}. In particular, to connect the central Cartesian grid with the spherical wave zone. Additionally, \textsc{Carpet} communicates between refinement levels with second-order and fifth-order accuracy in time and space, respectively. The Courant number in all our simulations is $0.2$, such that the Courant–Friedrichs–Lewy condition is satisfied. To check whether our numerical results respect the required convergence, we evolve the same configuration with a coarse ($h\ped{c}$), medium ($h\ped{m}$) and fine ($h\ped{f}$) resolution. The convergence factor can then be calculated according to
\begin{equation}
Q_n=\frac{f_{h\ped{c}}-f_{h\ped{m}}}{f_{h\ped{m}}-f_{h\ped{f}}}=\frac{h\ped{c}^n-h\ped{m}^n}{h\ped{m}^n-h\ped{f}^n}\,,
\end{equation}
where $n$ is the expected convergence order. In our case, we take as the coarsest level $h\ped{c}=$ $1.8 M$, then $h\ped{m}=1.2 M$ and $h\ped{f}=1.0 M$. As can be seen in Figure~\ref{fig:convergencePhiC}, we obtain a convergence order between $3$ and $4$. We have performed similar tests for the other simulations (\emph{with} or \emph{without} $C$ and \emph{with} or \emph{without} the plasma) and we find similar conclusions.
\section{Higher Multipoles}\label{appNR_sec:higherorder}
\begin{figure}[t!]
\centering
    \includegraphics[scale=0.45, trim = 0 0 0 30]{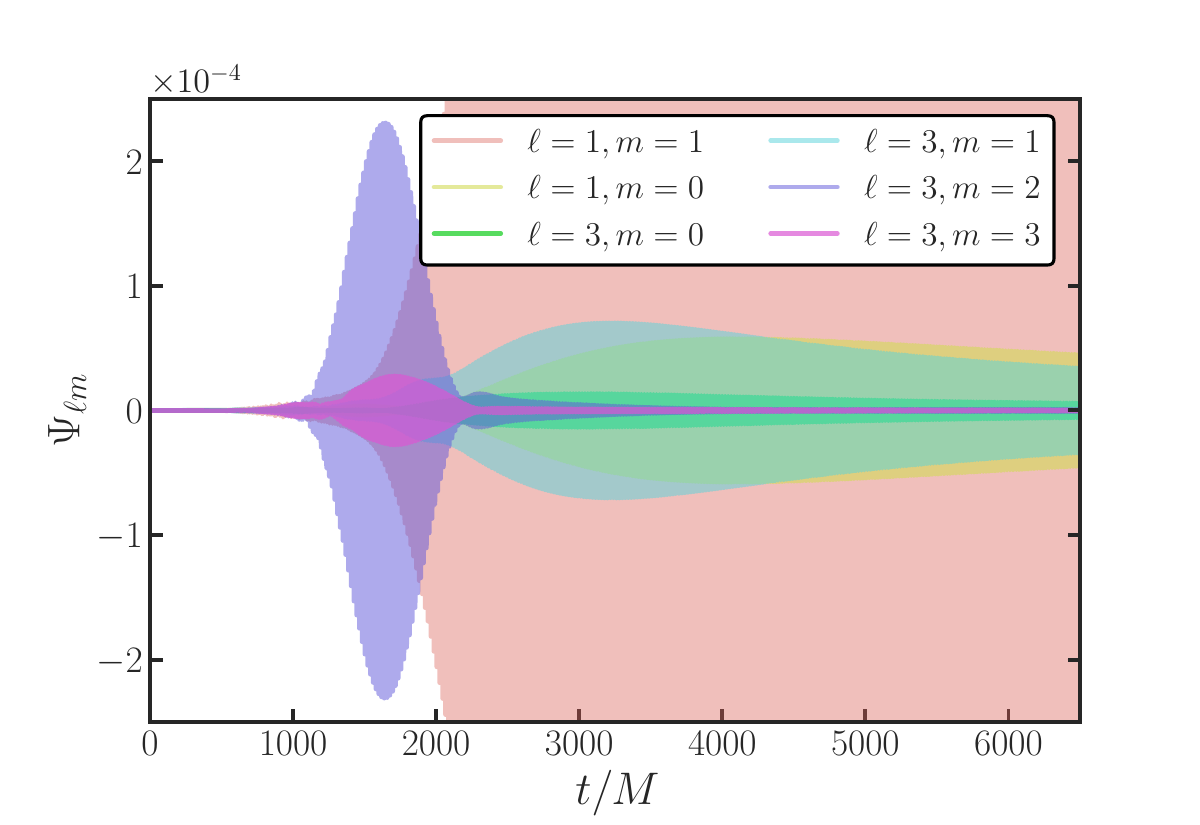}	
    	\includegraphics[scale=0.45, trim = 0 0 0 0]{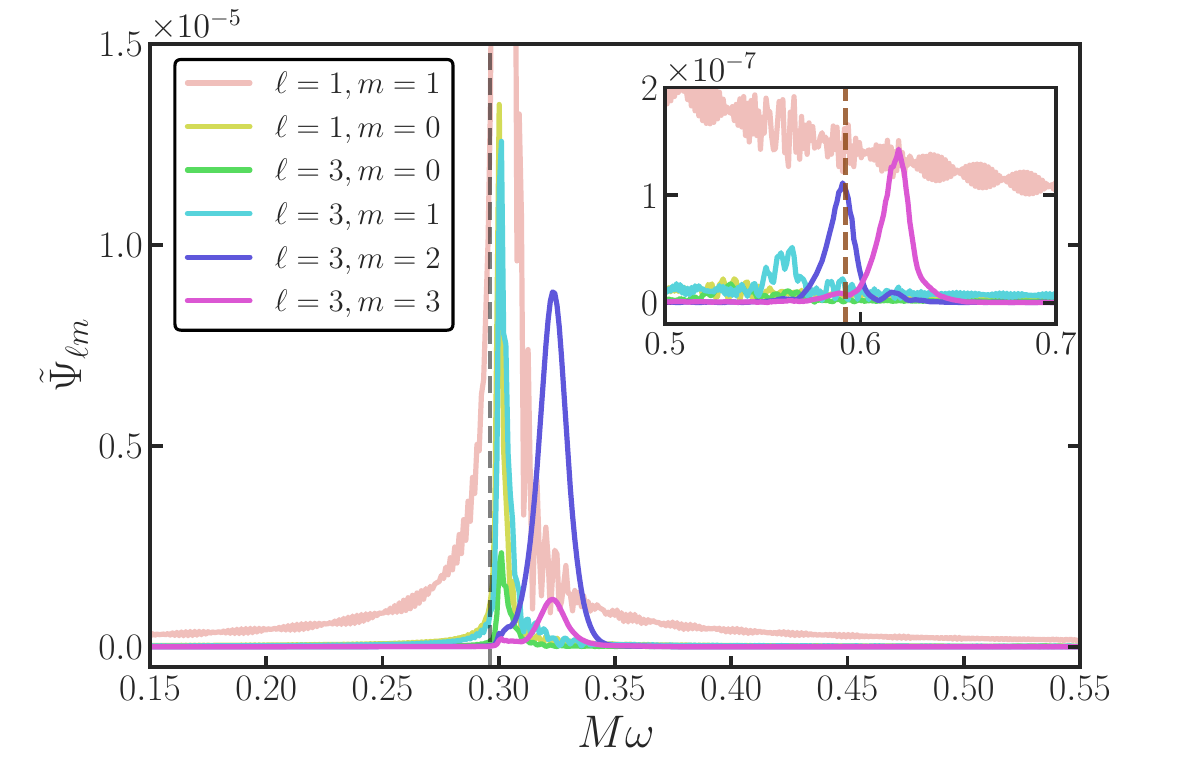}
    \caption{\emph{Top panel}:~Time evolution of various multipole modes of the scalar field in the supercritical case (simulation $\mathcal{I}_{3}$). The field is extracted at $r\ped{ex} = 400M$ and $\mu M = 0.3$. Interestingly, even $\ell$ modes are not excited, while odd $\ell$ modes are, which we explain in Appendix~\ref{appNR_sec:selectionrules}.
    \emph{Bottom panel}:~Fourier transform of the multipole modes shown in the \emph{top panel}. Dashed line denotes the frequency of the fundamental mode ($\omega_0$) [{\color{gray}{grey}}], while the peaks around $\omega = 2\omega_0$ [{\color{MathematicaBrown}{brown}}] (seen in the inset) originate from interactions between ``up-scattered'' photons ($\omega = 3\omega_0/2$) with ``normal'' photons ($\omega = \omega_0/2$).}   
    \label{fig:scalarfieldmodes}
\end{figure}
\begin{figure}[t!]
\centering
    \includegraphics[scale=0.45, trim = 0 0 0 0]{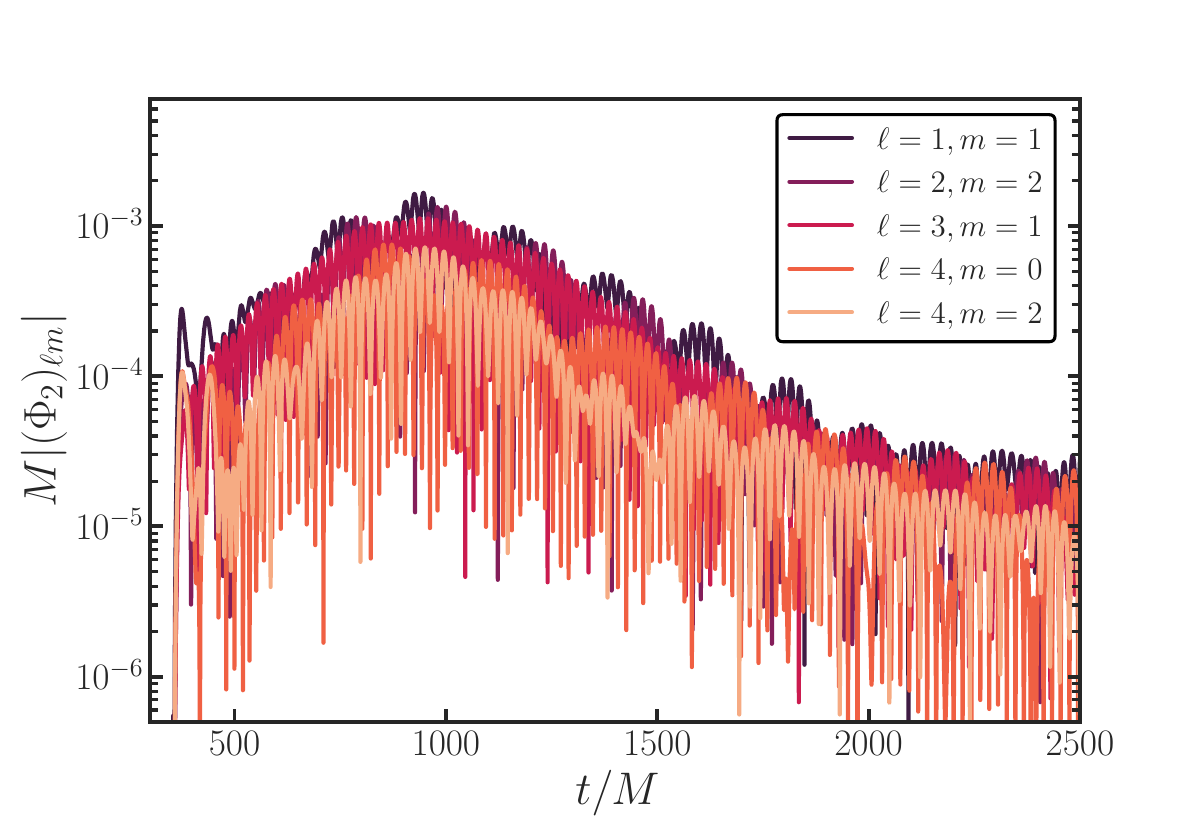}	
    	\includegraphics[scale=0.45, trim = -20 0 0 0]{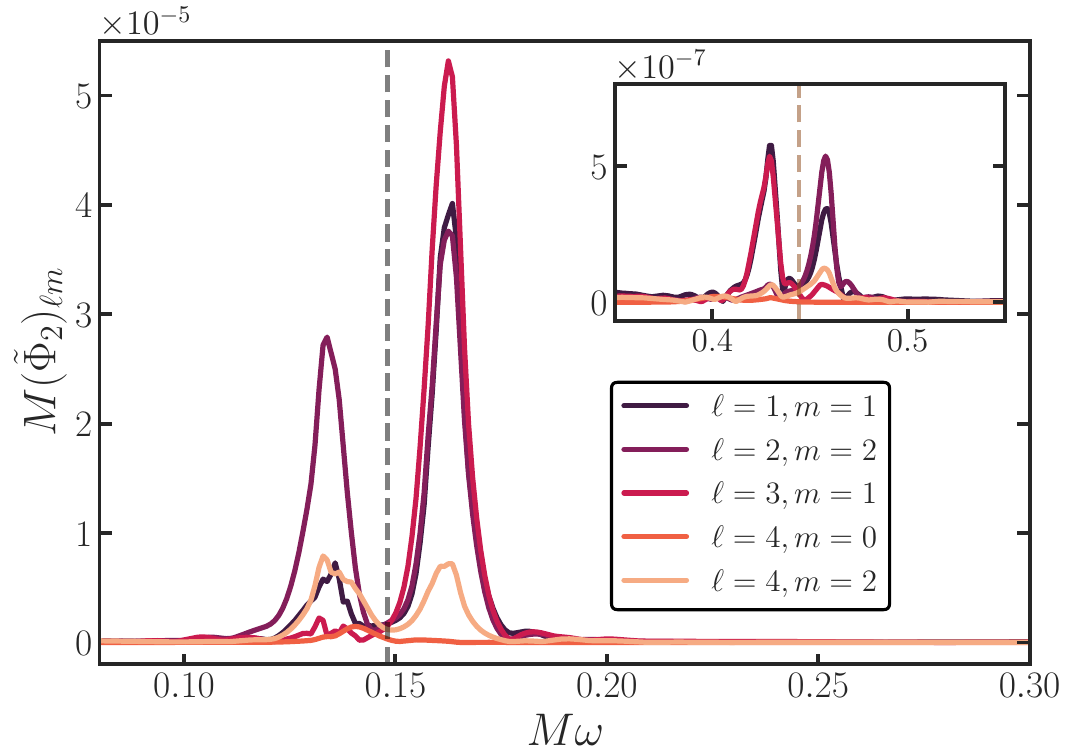}
    \caption{\emph{Top panel}:~Time evolution of various multipole
    modes of the Newman-Penrose scalar $|(\Phi_{2})_{\ell m}|$ in the supercritical regime. Considered simulation is $\mathcal{I}_{3}$. Field is extracted at $r\ped{ex} = 400M$ and $\mu M = 0.3$.
    \emph{Bottom panel}:~Fourier transform of the multipole modes shown in the \emph{top panel}. Dashed lines indicate the frequencies at $\omega_0/2$ [{\color{MathematicaBrown}{brown}}] and $3\omega_0/2$ [{\color{gray}{grey}}].}    \label{fig:EMfieldmodes}
\end{figure}

In Chapter~\ref{chap:SR_Axionic}, we have shown the dominant contribution coming from the dipole $\ell = m = 1$ mode (cf.~Figures~\ref{fig:BurstmuM03r20}--\ref{fig:BurstAxionmuM03r40100}). In general however, higher order multipoles are also produced. In Figures~\ref{fig:scalarfieldmodes} and~\ref{fig:EMfieldmodes}, we show a subset of those from the scalar and vector field, respectively. In both figures, we consider simulation $\mathcal{I}_{3}$ (see Table~\ref{tb:simulations}), where superradiance is turned off and we start in the supercritical regime. Three features are worth noting;~(i) only axion modes with odd $\ell$ can be produced from our initial data. An explanation for this selection rule is provided in Appendix~\ref{appNR_sec:selectionrules};~(ii) the Fourier transform of the vector field shows additional peaks with a frequency slightly lower than $\mu/2$ and two near $3\mu/2$. As discussed in Chapter~\ref{chap:SR_Axionic}, these should be interpreted as \emph{photon echoes} created by outwards travelling photons that interact with the axion cloud;~(iii) in Figure~\ref{fig:scalarfieldmodes}, we observe that some of these up-scattered photons can recombine with ``normal'' photons ($\omega \sim \mu/2$) to form axion waves with a frequency of twice the boson mass. 
\section{Selection Rules}\label{appNR_sec:selectionrules}
\begin{figure}[t!]
    \centering
    \includegraphics[scale=0.45, trim = 0 0 0 0]{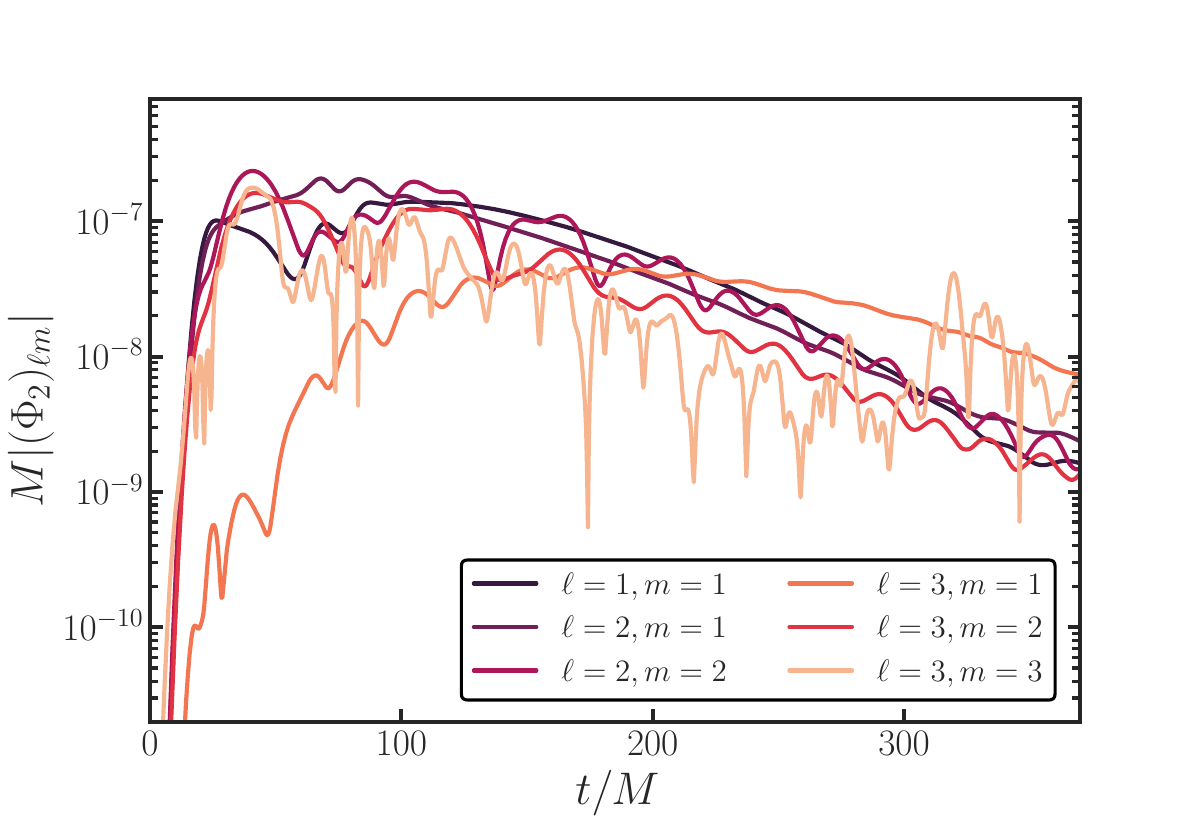}
    \caption{The time evolution of various multipole
    modes of the Newman-Penrose scalar $\Phi_{2}$ in the subcritical regime. The considered simulation is $\mathcal{I}_{2}$, where the field is extracted at $r\ped{ex} = 20M$ and $\mu M = 0.3$. Each of the curves has been rescaled according to the order found in~\eqref{eq:selectionrules}.}
    \label{fig:EMfieldSelectionrules}
\end{figure}

Using spherical harmonics, the equations of motion can be decomposed, and, as we will show, allow us to predict which modes are excited from the axionic coupling. This approach yields a consistency check of our simulations in the case where superradiant growth is absent. Using the electric field $E_{i}$ and the magnetic field $B_{i}$, the Maxwell equations can be written as 
\begin{equation}
\begin{aligned}
    \partial_{t}E^{i}&
    =\alpha
    KE^{i}+\beta^{j}\partial_{j}E^{i}-E^{j}\partial_{j}\beta^{i}-\epsilon^{ijk}D_{j}(\alpha B_{k})
    +2k\ped{a}\alpha\left(
    \epsilon^{ijk} E_{k}D_{j}\Psi +B^{i}n^{\alpha}\partial_{\alpha}\Psi
    \right)\,,\\
    \partial_{t}B^{i}&=\beta^{j}\partial_{j}B^{i}-B^{j}\partial_{j}\beta^{i}+\alpha KB^{i}+\epsilon^{ijk}D_{j}(\alpha E_{k})\,,
\end{aligned}
\end{equation}
where $\epsilon^{ijk}=-\frac{1}{\sqrt{\gamma}}E^{ijk}$ and $E^{ijk}$ is the totally anti-symmetric tensor with $E^{123}=1$. We focus on a Schwarzschild BH which has the following metric:
\begin{equation}
ds^{2}=-f(r)\mathrm{d}t^{2}+\frac{1}{f(r)}(r)\mathrm{d}r^{2}+r^{2}\hat{\gamma}_{AB}\mathrm{d}x^{A}\mathrm{d}x^{B}\,,
\end{equation}
where $f(r)=1-2M/r$, and $\hat{\gamma}_{AB}\mathrm{d}x^{A}\mathrm{d}x^{B}=\mathrm{d}\theta^{2}+\sin^{2}\theta \mathrm{d}\varphi^{2}$. From the spacetime symmetry, the electric and magnetic field can be decomposed using the scalar spherical harmonics $Y_{\ell m}(\theta,\varphi)$ and the vector spherical harmonics $\hat{\nabla}_{A}Y_{\ell m}$ and $V_{\ell m,A}=\hat{\epsilon}^{B}_{A}\hat{\nabla}_{B}Y_{\ell m}$. Here, $\hat{\epsilon}_{AB}$ is the anti-symmetric tensor with $\hat{\epsilon}_{\theta\varphi}=\sin\theta$. Using these harmonics functions, we expand the electric, magnetic, and scalar field as follows:
\begin{equation}
\begin{aligned}
E_{r}&=\sum_{\ell m}\mathcal{E}_{\ell m,r}Y_{\ell m}+{\rm c.c.}\,,\\
E_{A}&=\sum_{\ell m}\left\{
\mathcal{E}_{\ell m,S}\frac{\hat{\nabla}_{A}Y_{\ell m}}{\sqrt{\ell(\ell+1)}}+\mathcal{E}_{\ell m,V}V_{\ell m,A}+{\rm c.c.}
\right\}\,,\\
B_{r}&=\sum_{\ell m}\mathcal{B}_{\ell m,r}Y_{\ell m}+{\rm c.c.}\,,\\
B_{A}&=\sum_{\ell m}\left\{
\mathcal{B}_{\ell m,S}\frac{\hat{\nabla}_{A}Y_{\ell m}}{\sqrt{\ell (\ell +1)}}+\mathcal{B}_{\ell m,V}V_{\ell m,A}+{\rm c.c.}
\right\}\,,\\
\Psi&=\sum_{\ell m}\Psi_{\ell m}Y_{\ell m}+{\rm c.c.}\,,
\end{aligned}
\end{equation}
where $A = \{\theta, \varphi\}$ and $\mathcal{E}_{\ell m,r}$, $\mathcal{E}_{\ell m,S}$, $\mathcal{E}_{\ell m,V}$, $\mathcal{B}_{\ell m,r}$, $\mathcal{B}_{\ell m,S}$, $\mathcal{B}_{\ell m,V}$, and $\Psi_{\ell m}$ are all coefficients that depend on $t$ and $r$ only. Time or space derivatives are denoted with a dot or prime, respectively. Since both the scalar and vector spherical harmonics are orthogonal functions, the Maxwell equations with axionic coupling can be decomposed. To simplify the notation, we define $\boldsymbol{\Lambda} = (\ell m, \ell' m', \ell'' m'')$, so that the coefficients for the electric field can be written as
\begin{equation}
    \begin{aligned}
     \dot{\mathcal{E}}_{\ell m,r}&=-\frac{\ell (\ell +1)}{r^{2}}\mathcal{B}_{\ell m,V}-\frac{2k\ped{a}}{\sqrt{f}r^{2}}\sum_{\substack{\ell'm'\\ \ell''m''}}\sum_{I=1}^{4}C^{(r,I)}_{\boldsymbol{\Lambda} }X_{(I),\boldsymbol{\Lambda} }\,,\\
    \dot{\mathcal{E}}_{\ell m,S}&=-\frac{\sqrt{\ell(\ell+1)}}{2}(f'\mathcal{B}_{\ell m,V}+2f\mathcal{B}'_{\ell m,V})-2k\ped{a}\sqrt{\ell(\ell+1)} \sum_{\substack{\ell'm'\\ \ell''m''}}\sum_{I=1}^{4}C^{(S,I)}_{\boldsymbol{\Lambda}}X_{(I),\boldsymbol{\Lambda} }\,,\\
    \dot{\mathcal{E}}_{\ell m,V}&=-f\mathcal{B}_{\ell m,r}+\frac{f'\mathcal{B}_{\ell m,S}+2f\mathcal{B}'_{\ell m,S}}{2\sqrt{\ell(\ell+1)}}-2k\ped{a} \sum_{\substack{\ell'm'\\ \ell''m''}}\sum_{I=1}^{4}C^{(V,I)}_{\boldsymbol{\Lambda} }X_{(I),\boldsymbol{\Lambda} }\,.
    \label{Maxeqspherical1}
    \end{aligned}
\end{equation}
Then, we proceed with the coefficients for the magnetic field:
\begin{equation}
    \begin{aligned}
    \dot{\mathcal{B}}_{\ell m,r}&=\frac{\ell(\ell+1)}{r^{2}}\mathcal{E}_{\ell m,V}\,,\\
    \dot{\mathcal{B}}_{\ell m,S}&=\frac{\sqrt{\ell(\ell+1)}}{2}\left(f'\mathcal{E}_{\ell m,V}+2f\mathcal{E}'_{\ell m,V}\right)\,,\\
    \dot{\mathcal{B}}_{\ell m,V}&=f\mathcal{E}_{\ell m,r}-\frac{f'\mathcal{E}_{\ell m,S}+2f\mathcal{E}'_{\ell m,S}}{2\sqrt{\ell(\ell+1)}}\,,
    \label{Maxeqspherical2}
    \end{aligned}
\end{equation}
and the coefficients for the scalar field:
\begin{equation}
    \begin{aligned}
    \ddot{\Psi}_{\ell m}&=-f\left(\frac{2}{r}+f'\right)\Psi_{\ell m}'-f^{2}\Psi''_{\ell m}+f\left(
    \frac{\ell(\ell+1)}{r^{2}}+\mu^{2}
    \right)\Psi_{\ell m}-2k\ped{a}f\\
    &\times \sum_{\substack{\ell'm'\\ \ell''m''}}\sum_{I=1}^{4}C^{(\Psi,I)}_{\boldsymbol{\Lambda}}X_{(I),\boldsymbol{\Lambda} }\,,
    \label{Maxeqspherical3}
    \end{aligned}
\end{equation}
where $X_{(I),\boldsymbol{\Lambda}}$ is found by
\begin{equation}
\begin{aligned}
    X_{(1),\boldsymbol{\Lambda}}&=\int \mathrm{d}^{2}\Omega\, Y^{\ast}_{\ell m}Y_{\ell''m''}Y_{\ell'm'}\,,\\
    X_{(2),\boldsymbol{\Lambda}}&=\int \mathrm{d}^{2}\Omega\,  Y^{\ast}_{\ell m}Y^{\ast}_{\ell''m''}Y_{\ell'm'}\,,\\
    X_{(3),\boldsymbol{\Lambda}}&=\int \mathrm{d}^{2}\Omega\,  Y^{\ast}_{\ell m}\hat{\nabla}^{A}Y_{\ell''m''}V_{\ell'm',A}\,,\\
    X_{(4),\boldsymbol{\Lambda}}&=\int \mathrm{d}^{2}\Omega\,  Y^{\ast}_{\ell m}\hat{\nabla}^{A}Y^{\ast}_{\ell''m''}V_{\ell'm',A}\,.
\end{aligned}
\end{equation}
Upon defining the prefactor
\begin{equation}    
Q_{\ell\ell'\ell''}=\frac{\ell(\ell+1)+\ell'(\ell'+1)-\ell''(\ell''+1)}{2}\,,
\end{equation}
we find for $C^{(r,I)}$:
\begin{equation}
    \begin{aligned}
     C^{(r,1)}_{\boldsymbol{\Lambda}}&=Q_{\ell''\ell'\ell}
    \mathcal{E}_{\ell''m'',V}\Psi_{\ell'm'} -r^{2}\mathcal{B}_{\ell''m'',r}\dot{\Psi}_{\ell'm'}\,,\\
    C^{(r,2)}_{\boldsymbol{\Lambda}}&=Q_{\ell''\ell'\ell}
    \mathcal{E}_{\ell''m'',V}^{\ast}\Psi_{\ell'm'} -r^{2}\mathcal{B}_{\ell''m'',r}^{\ast}\dot{\Psi}_{\ell'm'}\,,\\
    C^{(r,3)}_{\boldsymbol{\Lambda}}&=\frac{\mathcal{E}_{\ell ''m'',S}\Psi_{\ell 'm'}}{\sqrt{\ell''(\ell''+1)}}\,,\\
    C^{(r,4)}_{\boldsymbol{\Lambda}}&=\frac{\mathcal{E}^{\ast}_{\ell''m'',S}\Psi_{\ell 'm'}}{\sqrt{\ell''(\ell''+1)}}\,,
    \end{aligned}
    \label{coeff Cr}
\end{equation}
for $C^{(S,I)}$:
\begin{equation}
    \begin{aligned}
    C^{(S,1)}_{\boldsymbol{\Lambda}}&=-Q_{\ell''\ell\ell'}\left(f\mathcal{E}_{\ell''m'',V}\Psi_{\ell'm'}'+\frac{\mathcal{B}_{\ell''m'',S}\dot{\Psi}_{\ell'm'}}{\sqrt{\ell''(\ell''+1)}}\right)\,,\\
    C^{(S,2)}_{\boldsymbol{\Lambda}}&=-Q_{\ell''\ell\ell'}\left(f\mathcal{E}^{\ast}_{\ell''m'',V}\Psi_{\ell'm'}'+\frac{\mathcal{B}^{\ast}_{\ell''m'',S}\dot{\Psi}_{\ell'm'}}{\sqrt{\ell''(\ell''+1)}}\right)\,,\\
    C^{(S,3)}_{\boldsymbol{\Lambda}}&=f\mathcal{E}_{\ell''m'',r}\Psi_{\ell'm'}
    -\frac{f}{\sqrt{\ell''(\ell''+1)}}\mathcal{E}_{\ell''m'',S}\Psi_{\ell'm'}'+\mathcal{B}_{\ell''m'',V}\dot{\Psi}_{\ell'm'}\,,\\
    C^{(S,4)}_{\boldsymbol{\Lambda}}&=
    f\mathcal{E}^{\ast}_{\ell''m''m,r}\Psi_{\ell'm'}
    -\frac{f}{\sqrt{\ell''(\ell''+1)}}\mathcal{E}^{\ast}_{\ell''m'',S}\Psi_{\ell'm'}'+\mathcal{B}^{\ast}_{\ell''m'',V}\dot{\Psi}_{\ell'm'}\,,
    \end{aligned}
     \label{coeff CS}
\end{equation}
for $C^{(V,I)}$:
\begin{equation}
    \begin{aligned}
    C^{(V,1)}_{\boldsymbol{\Lambda}}&=-
    Q_{\ell'\ell\ell''}f\mathcal{E}_{\ell''m'',r}\Psi_{\ell'm'} +
    Q_{\ell''\ell\ell'}\!\left(
    \frac{f}{\sqrt{\ell''(\ell''+1)}}\mathcal{E}_{\ell''m'',S}\Psi'_{\ell'm'}-\mathcal{B}_{\ell''m'',V}\dot{\Psi}_{\ell'm'}
    \right)\,,\\
    C^{(V,2)}_{\boldsymbol{\Lambda}}&=
    -    Q_{\ell'\ell\ell''}f\mathcal{E}^{\ast}_{\ell''m'',r}\Psi_{\ell'm'}+
    Q_{\ell''\ell\ell'}\!\left(
    \frac{f}{\sqrt{\ell''(\ell''+1)}}\mathcal{E}_{\ell''m'',S}\Psi_{\ell'm'}'-\mathcal{B}_{\ell''m''}\dot{\Psi}_{\ell'm'}
    \right)\,,\\
    C^{(V,3)}_{\boldsymbol{\Lambda}}&=-\left(f\mathcal{E}_{\ell''m'',V}\Psi_{\ell'm'}'+\frac{\mathcal{B}_{\ell''m'',S}\dot{\Psi}_{\ell'm'}}{\sqrt{\ell''(\ell''+1)}}\right)\,,\\
    C^{(V,4)}_{\boldsymbol{\Lambda}}&=-\left(f\mathcal{E}^{\ast}_{\ell''m'',V}\Psi_{\ell'm'}'+\frac{\mathcal{B}^{\ast}_{\ell''m'',S}\dot{\Psi}_{\ell'm'}}{\sqrt{\ell''(\ell''+1)}}\right)\,,
    \end{aligned}
     \label{coeff CV}
\end{equation}
and finally for $C^{(\Psi,I)}$:
\begin{equation}
\begin{aligned}
&C^{(\Psi,1)}_{\boldsymbol{\Lambda}}=f\mathcal{E}_{\ell'm',r}\mathcal{B}_{\ell''m'',r}+\frac{Q_{\ell''\ell'\ell}}{r^{2}}  
\left(
\frac{\mathcal{E}_{\ell'm',S}\mathcal{B}_{\ell''m''S,}}{\sqrt{\ell'\ell''(\ell'+1)(\ell''+1)}}+\mathcal{E}_{\ell'm',V}\mathcal{B}_{\ell''m'',V}
\right)\,,\\
&C^{(\Psi,2)}_{\boldsymbol{\Lambda}}=f\mathcal{E}_{\ell'm',r}\mathcal{B}^{\ast}_{\ell''m'',r}
\frac{Q_{\ell''\ell'\ell}}{r^{2}} \left(
\frac{\mathcal{E}_{\ell'm'S}\mathcal{B}^{\ast}_{\ell''m'',S}}{\sqrt{\ell'\ell''(\ell'+1)(\ell''+1)}}+\mathcal{E}_{\ell'm',V}\mathcal{B}^{\ast}_{\ell''m'',V}
\right)\,,\\
&C^{(\Psi,3)}_{\boldsymbol{\Lambda}}=\frac{1}{r^{2}}\left(\mathcal{E}_{\ell'm',V}\frac{\mathcal{B}_{\ell''m'',S}}{\sqrt{\ell''(\ell''+1)}}
\frac{\mathcal{E}_{\ell'm'S}}{\sqrt{\ell'(\ell'+1)}}\mathcal{B}_{\ell''m''V}
\right)\,,\\
&C^{(\Psi,4)}_{\boldsymbol{\Lambda}}=\frac{1}{r^{2}}\left(
\mathcal{E}_{\ell'm',V}\frac{\mathcal{B}^{\ast}_{\ell''m'',S}}{r^{2}\sqrt{\ell''(\ell''+1)}}
\frac{\mathcal{E}_{\ell'm'S}}{\sqrt{\ell'(\ell'+1)}}\mathcal{B}_{\ell'm'V}^{\ast}
\right)
\,.
\end{aligned}
\label{coeff CPsi}
\end{equation}
In our simulations, we monitor the Newman-Penrose variable $\Phi_{2}$, which has decomposed coefficients defined by
\begin{equation}
\begin{aligned}
(\Phi_{2})_{\ell m}=\frac{\sqrt{\ell(\ell+1)}}{2r}\left\{
-\left(\mathcal{B}_{\ell m,V}+\frac{\mathcal{E}_{\ell m,S}}{\sqrt{\ell(\ell+1)}}
\right)
+i\left(
-\mathcal{E}_{\ell m,V}+\frac{\mathcal{B}_{\ell m,S}}{\sqrt{\ell(\ell+1)}}
\right)\right\}\,.
\end{aligned}
\end{equation}
The non-vanishing components of our initial data (see Appendix~\ref{appNR_subsec:evolwithoutplasma}) are 
\begin{equation}
\begin{aligned}
    \Psi_{1,\pm 1}&\sim \Psi_{0}\,,\\
    \mathcal{E}_{10,V}(t=0,r)&=\frac{1}{2}\sqrt{\frac{\pi}{3}}E^{\varphi}(r)\,,
\end{aligned}
\label{eq:initalSelection}
\end{equation}
where $E^{\varphi}(r)$ is defined in~\eqref{eq:InitialElectric}. Since this is a perturbative approach, we focus on the subcritical regime and assume $E^{\varphi}(r)$ is order $\mathcal{O}(\epsilon)$. Using the above equations, we can obtain the order of each mode of $(\Phi_{2})_{\ell m}$ as
\begin{equation}
    \begin{aligned}
    (\Phi_{2})_{1,0}&\sim \mathcal{O}(\epsilon)\,,\\
    (\Phi_{2})_{1,\pm 1}&\sim \mathcal{O}(k\ped{a}\Psi_{0}\epsilon)\,,\\
    (\Phi_{2})_{2,\pm 1}&\sim \mathcal{O}(k\ped{a}\Psi_{0}\epsilon)\,,\\
    (\Phi_{2})_{2,\pm 2}&\sim \mathcal{O}((k\ped{a}\Psi_{0})^{2}\epsilon)\,,\\
    (\Phi_{2})_{3,\pm 3}&\sim \mathcal{O}((k\ped{a}\Psi_{0})^{3}\epsilon)\,.
    \label{eq:selectionrules}  
    \end{aligned}
\end{equation}
In Figure~\ref{fig:EMfieldSelectionrules}, we show $|(\Phi_{2})_{\ell m}|$ in the subcritical regime $(\mathcal{I}_{2})$, and rescale all multipoles according to~\eqref{eq:selectionrules}. As can be seen, using the rescaling, all curves are on the same order, demonstrating that our simulations show consistent behaviour.
\vskip 2pt
As alluded to in Section~\ref{sec_SR_Axionic:withoutSR}, from our initial data only odd $\ell$ scalar multipoles can be produced. We can now proof this. From $Y_{\ell m}(\pi-\theta,\varphi+\pi)=(-1)^{\ell}Y_{\ell m}(\theta,\varphi)$, we find that $X_{(1),\boldsymbol{\Lambda}}$ and $X_{(2),\boldsymbol{\Lambda}}$ are nonzero when $\ell+\ell'+\ell''$ is an \emph{even} number, and $X_{(3),\boldsymbol{\Lambda}}$ and $X_{(4),\boldsymbol{\Lambda}}$ are nonzero when $\ell+\ell'+\ell''$ is an \emph{odd} number.
\vskip 2pt
Then, eqs.~\eqref{coeff Cr}--\eqref{coeff CV} show that the non-vanishing modes of our initial data, $\Psi_{\ell m}$ and $\mathcal{E}_{\ell m,V}$~\eqref{eq:initalSelection} with odd $\ell$ can \emph{only} excite $\mathcal{E}_{\ell m,r}$ and $\mathcal{E}_{\ell m,S}$ with even $\ell$, while it excites $\mathcal{E}_{\ell m,V}$ with odd $\ell$. Next,~\eqref{Maxeqspherical2} implies that the non-vanishing component of $\mathcal{E}_{\ell m,r}$ and $\mathcal{E}_{\ell m,S}$ with even $\ell$ excites $\mathcal{B}_{\ell m,V}$ with even $\ell$, while the non-vanishing component of $\mathcal{E}_{\ell m,V}$ with odd $\ell$ excites $\mathcal{B}_{\ell m,r}$ and $\mathcal{B}_{\ell m,S}$ with odd $\ell$. 
\vskip 2pt
Finally,~\eqref{coeff CPsi} shows that the non-vanishing component of $\mathcal{E}_{\ell m,r}$, $\mathcal{E}_{\ell m,S}$, and $\mathcal{B}_{\ell m,V}$ with even $\ell$, and $\mathcal{E}_{\ell m,V}$, $\mathcal{B}_{\ell m,r}$, and $\mathcal{B}_{\ell m,S}$ with odd $\ell$ \emph{only} excites $\Psi_{\ell m}$ with odd $\ell$. Therefore, the non-vanishing components of the simulation starting from initial data~\eqref{eq:initalSelection}, only excite $\Psi_{\ell m}$ with odd $\ell$. These results are consistent with our simulations, see Figure~\ref{fig:scalarfieldmodes}. 
\chapter{Mathieu Equation}\label{app:Mathieu}
Most of the results in Chapter~\ref{chap:SR_Axionic} can be interpreted through the Mathieu equation. In its simplest form, the \emph{Mathieu functions} are solutions to
\begin{equation}
\frac{\dd^2 y}{\dd x^2} + (a + b\cos{2x})y = 0\,,\nonumber
\end{equation}
where $a$ and $b$ are real-valued constants. A striking feature of these solutions is the presence of instabilities whenever $a = n^2/4$ for $n \in \mathbb{N}$ and $b \neq 0$~\cite{bender78:AMM}. Notably, in Lorenz gauge, the Maxwell equations for the coupled axion-photon system~\eqref{eq:SRAxion_evoleqns} in flat spacetime reduce to a Mathieu-like form. In this appendix, I show that this structure persists even when accounting for a superradiantly growing cloud (Section~\ref{app_Mathieu_sec:SR}) or a background plasma (Section~\ref{app_Mathieu_sec:plasmaMathieu}). Consequently, the well-understood properties of Mathieu functions can provide valuable insights into the dynamics of the full system in curved spacetime.
\section{Superradiant Mathieu Equation}\label{app_Mathieu_sec:SR}
By solving the superradiant Mathieu equation~\eqref{eq:modifiedMathieu} numerically for different values of $C$, we were able to find a growth rate~\eqref{eq:modifiedgrowthrate} for the electromagnetic field in flat spacetime, while assuming a homogeneous axion condensate. Remarkably, this estimate is accurate in describing the super-exponential growth of the electromagnetic field in presence of superradiance, even when considering the full setup on a Schwarzschild background (including the finite-size effects of the cloud with $\lambda\ped{esc}$). In this appendix, we show a few examples of the numerical solutions to~\eqref{eq:modifiedMathieu} and we use a \emph{multiple-scale method} to derive the growth rate analytically.
\begin{figure}[t!]
    \centering
    \includegraphics[scale=0.45, trim = 0 0 0 0]{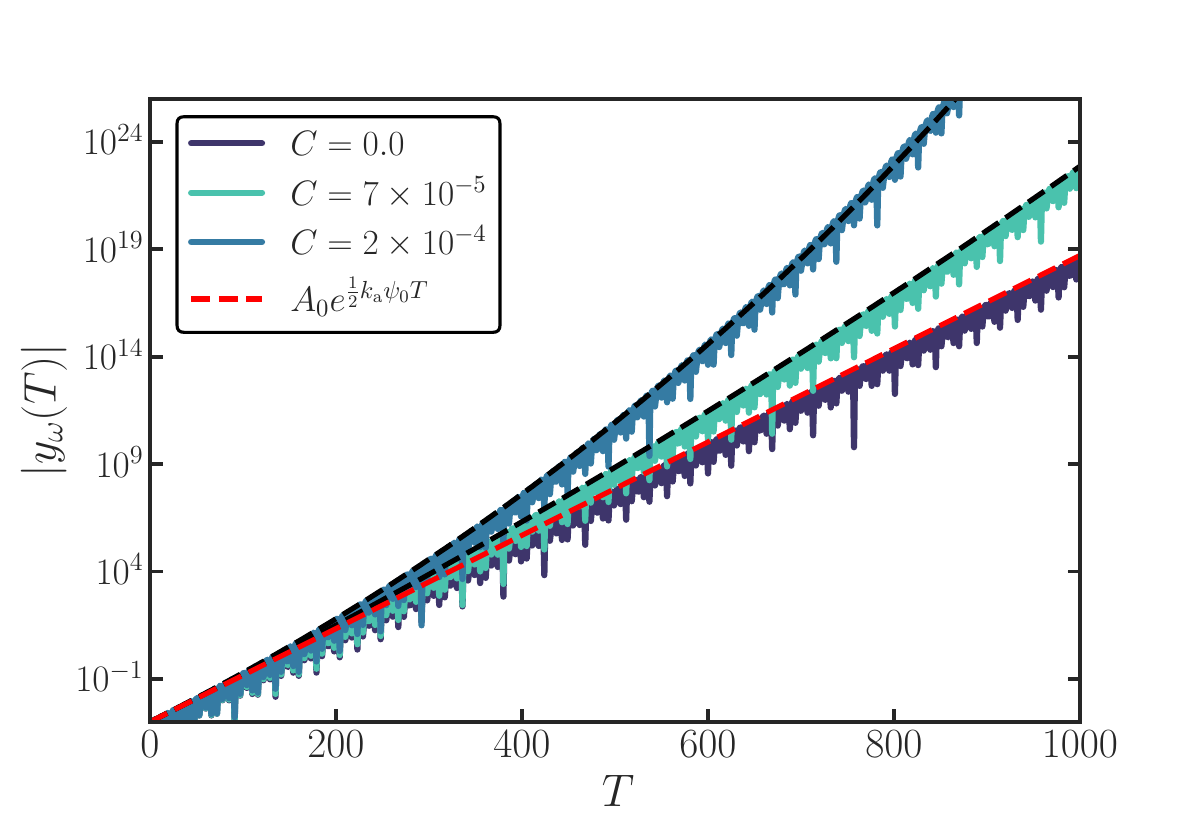}
    \caption{Numerical solutions to the superradiant Mathieu equation~\eqref{eq:modifiedMathieu} for different values of $C$. Horizontal axis shows the rescaled time $T = \mu t$. Dashed lines show the ``standard'' Mathieu growth rate [{\color{Mathematica4}{red}}], and the analytic growth rate from~\eqref{eq:modifiedgrowthrate} [\textbf{black}]. Chosen parameters are $\mu =0.2, p_z=0.1, k\ped{a}\psi_0 =0.1$.}
    \label{fig:mathieuwithC}
\end{figure}
\vskip 2pt
Figure~\ref{fig:mathieuwithC} shows various numerical solutions to the superradiant Mathieu equation~\eqref{eq:modifiedMathieu}. For $C=0$, the solution is well-described by the standard Mathieu growth rate (red dashed line). For nonzero values of $C$ instead, the standard Mathieu prediction becomes inaccurate and the numerical solutions are well fitted by the super-exponential growth rate~\eqref{eq:modifiedgrowthrate} (black dashed lines).
\subsubsection{Multiple-scale analysis}
Regular perturbation theory fails to describe certain problems at late times due to the appearance of \emph{secular terms}, which introduce non-uniformities between consecutive orders of the perturbation series. An example is the Mathieu equation, where a \emph{multiple-scale analysis} provides a more suitable framework (see, e.g.,~\cite{bender78:AMM}). In the following, we demonstrate its effectiveness in solving the superradiant Mathieu equation and thereby provide an analytical explanation for the numerically fitted growth rate~\eqref{eq:modifiedgrowthrate}. 
\vskip 2pt
Consider the superradiant Mathieu equation~\eqref{eq:modifiedMathieu}, where we expand the exponential as
\begin{equation}\label{eq:multiscalemathieu}
\frac{\mathrm{d}^{2}y}{\mathrm{d}T^{2}} + (b + 2 \delta (1+C T)\cos T)\,y=0\,,
\end{equation}
assuming $\delta$ is small. We introduce two timescales, a \emph{fast} timescale $T$, and a \emph{slow} timescale $\mathcal{T}=\delta T$, treated as independent variables. Promoting the solution to depend on both, i.e.,~$y(T)\rightarrow y(T,\mathcal{T})$ and expanding as $y=y_0(T,\mathcal{T})+ \delta y_1(T,\mathcal{T})$, with $b=b_0+\delta b_1$, the additional freedom from $\mathcal{T}$ allows us to eliminate secular terms, extending the solution's validity to longer times compared to a standard perturbative approach. Assuming the superradiant term to be small as well, i.e.,~$C=\delta \mathfrak{C}$, we introduce yet another timescale, the \emph{very slow} one, $\mathfrak{T}=\delta^2 T$. Expanding further, $b=b_0+\delta b_1 + \delta^2 b_2$, we obtain:
\begin{equation}\label{eq:multiscalefull}
     \frac{\mathrm{d}^{2}y}{\mathrm{d}T^{2}} +\left(b + \delta (b_1 + 2 \cos T)+\delta^2 (b_2 + 2 \mathfrak{C} T \cos T)\right)\,y=0\,.
\end{equation}
In the spirit of the multiple-scale method, we promote $y$ to depend on \emph{all the timescales} as independent variables and then expand, i.e.,~$y(T,\mathcal{T}, \mathfrak{T})=y_0 (T,\mathcal{T}, \mathfrak{T})+\delta y_1 (T,\mathcal{T}, \mathfrak{T})+\delta^2 y_2(T,\mathcal{T}, \mathfrak{T})$. 
\vskip 2pt
We can now consider~\eqref{eq:multiscalefull} order by order. At zeroth order, we have:
\begin{equation}
    \partial^2_T y_0 + b_0 y_0=0\,,
\end{equation}
where $b_0 = 1/4$ at the inset of the first unstable Mathieu band. At this order, we obtain the solution: $y_0=A(\mathcal{T}, \mathfrak{T}) e^{i T/2}+ \mathrm{c.c.}$ Hence, the solution at the fast timescale $T$ just describes the harmonic behaviour. At first-order, we have
\begin{equation}
    \partial^2_T y_1+\frac{1}{4}y_1=-2\partial_T \partial_\mathcal{T}y_0 -(b_1+2 \cos T)\,y_0\,.
\end{equation}
The right-hand side contains secular terms. However, we can use the extra dependence of $y_0$ on the slow timescale to remove it, namely by requiring $i \partial_\mathcal{T}A(\mathcal{T}, \mathfrak{T})=-b_1 A(\mathcal{T}, \mathfrak{T})+A^*(\mathcal{T}, \mathfrak{T})$. Solving this leads to the dependence of the zeroth-order solution on the slow timescale, i.e.,~$y_0=A(\mathfrak{T})e^{\sqrt{1-b_1^2}\mathcal{T}} e^{i T/2}+\mathrm{c.c.}$ Similarly, the dependence on the very slow timescale can be used to eliminate secular terms in the second-order equation.\footnote{As these computations become quite cumbersome, we do not report them here.} Following this procedure, we obtain for the zeroth-order solution: 
\begin{equation}
y_0\approx e^{\sqrt{1-b_1^2}\mathcal{T}} e^{i T/2}e^{ \mathfrak{C} T\mathfrak{T}}+\mathrm{c.c.}    
\end{equation} 
At sufficiently large times ($T \gg 1/\mathfrak{C}$), the growth rate is thus dominated by $e^{\mathfrak{C}T \mathfrak{T}}=e^{\delta C T^2}$. 
\vskip 2pt
Finally, by comparing~\eqref{eq:multiscalemathieu} with~\eqref{eq:modifiedMathieu}, we identify $\delta=p_z\psi_0 k\ped{a}/\mu^2$ and $p_z=\mu/2$, such that, after rescaling the physical time $t=T/\mu$, the growth rate reads
\begin{equation}
    e^{\lambda t}\,,\!\!\quad \!\text{with}\quad\! \lambda t = \frac{\mu}{2} k\ped{a} \psi_0 C t^2\,,
\end{equation}
which is the dominant growth rate at late times we found in~\eqref{eq:modifiedgrowthrateFirst} (up to a factor of 2).
\vskip 2pt
The multiple-scale method thus produces a solution with three timescales;~(i) the \emph{fast} timescale that corresponds to the harmonic oscillations with a frequency at half the boson mass, (ii) the \emph{slow} timescale belonging to the standard Mathieu growth rate, and (iii) the \emph{very slow} timescale which originates from the super-exponential growth induced by superradiance and becomes dominant at late times. In conclusion, the Mathieu equation provides us, once again, with a simple analytical explanation to the behaviour of the full system.
\section{Plasma Mathieu Equation}\label{app_Mathieu_sec:plasmaMathieu}
We continue by studying the axion-photon-plasma system in flat spacetime. This analysis closely follows~\cite{SenPlasma}, yet now in the context of a Mathieu-like equation. Furthermore, we generalise their work by including a momentum equation rather than assuming Ohm's law. 
\vskip 2pt
The starting point is the equations of motion~\eqref{eq:SRAxion_evoleqns} in Minkowski. As in Section~\ref{sec_SR_Axionic:analyticalgrow}, we take the wave vector to point along the $\hat{z}$--direction, i.e.,~$\vec{p}=(0,0,p_z)$, adopt the electromagnetic \emph{ansatz}~\eqref{eq:MathieuEM}, and consider a homogeneous axion condensate defined as
\begin{equation}
    \Psi=\frac{1}{2}(\psi_0 e^{-i \mu t}+ \psi_0^* e^{i \mu t})\,.
\end{equation}
By linearising~\eqref{eq:SRAxion_evoleqns}, one can straightforwardly solve the momentum equation and find the velocity of the electrons with respect to the electromagnetic field. Once again, due to their large inertia, we neglect the perturbations of the ions and treat them as a neutralising background. Since the longitudinal and transverse modes are decoupled in linear photon-plasma theory in flat spacetime, we focus on the transverse sector, and obtain
\begin{equation}
    u^k=-\frac{q\ped{e}}{m\ped{e}}\alpha^k e^{i (\vec{p} \cdot \vec{x} - \omega t)}\,,
\end{equation}
where $k=\hat{x},\hat{y}$ denote the transverse directions. This expression can now be inserted into the current on the right-hand side of Maxwell's equations, namely $j^k=q\ped{e} n\ped{e} u\ped{e}^k$, leading to two decoupled equations for the transverse polarisations $\alpha^k$. To proceed, we redefine the fields as $y_k =e^{i \omega t}\alpha_k$, rescale time via $T=\mu t$, and project onto a circular polarisation basis $e_\pm$ so that $y=y_\omega e_\pm$. In this form, we recover the Mathieu equation in the presence of plasma as~\cite{hertzberg:2018zte}
\begin{equation}\label{eq:PlasmaMathieu}
    \partial_T^2 y_\omega +\frac{1}{\mu^{2}}\Big(p_z^2+\omega\ped{p}^2 -2 \mu  p_z \psi_0 k\ped{a}\text{sin}T\Big)y_\omega=0\,.
\end{equation}
One can readily see that in the absence of plasma, i.e.,~$\omega\ped{p}=0$, the vacuum Mathieu equation is recovered~\cite{Boskovic:2018lkj}.
\begin{figure}[t!]
\centering
    \includegraphics[scale=0.45, trim = 0 0 0 0]{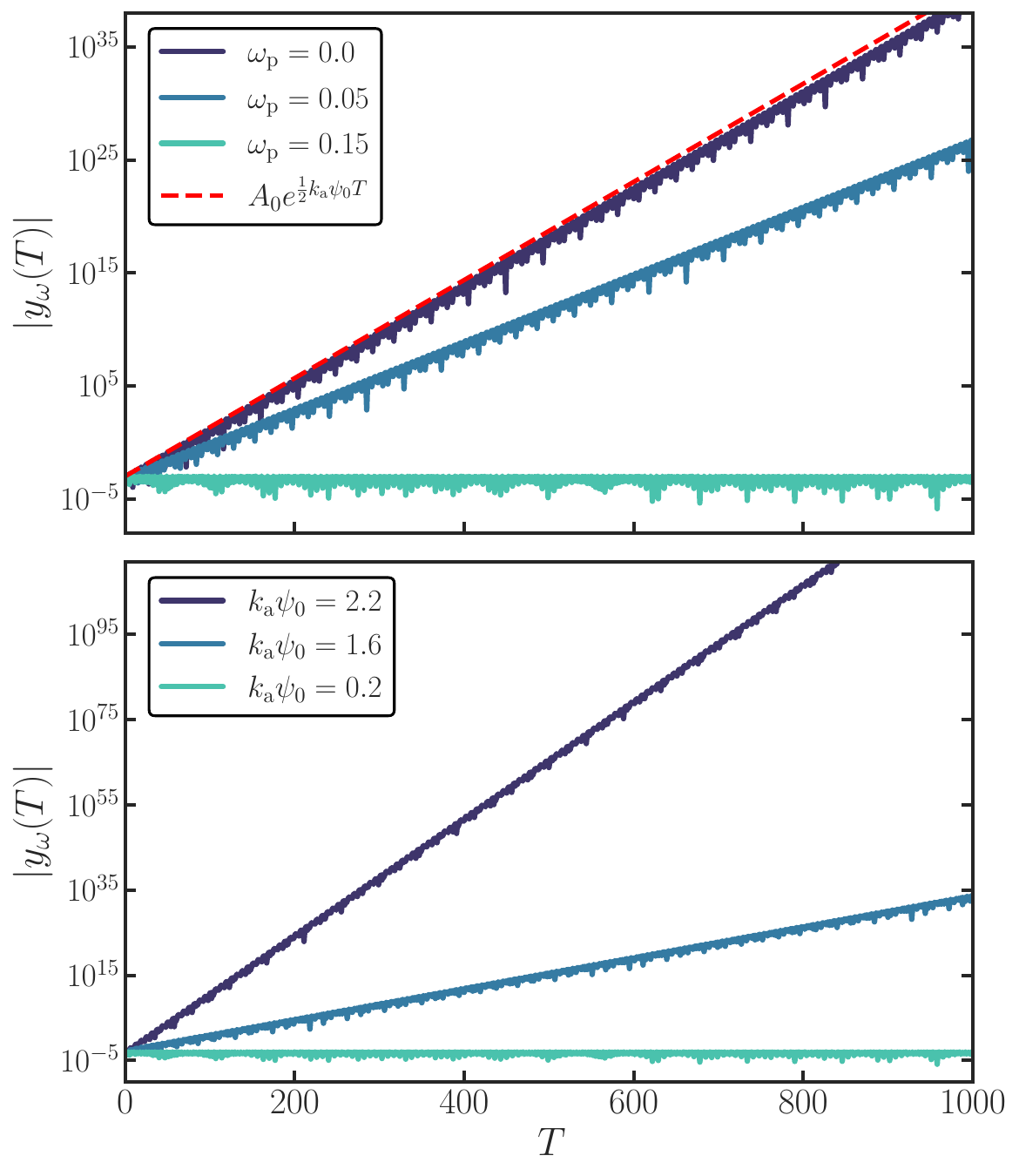}
    \caption{\emph{Top panel}:~Solutions of the plasma Mathieu equation~\eqref{eq:PlasmaMathieu} for different values of the plasma frequency. We take $\mu=0.2$, $p_z=0.1$ and $k\ped{a}\psi_0 =0.2$.
    \emph{Bottom panel}:~Similar setup as above, yet now the axionic coupling is varied and $\omega\ped{p} = 0.15 > \mu /2$. The other parameters are the same as above. As can be seen, for large values of $k\ped{a}\psi_0$ the instability is restored.}
    \label{fig:mathieuplasma}
\end{figure}
\vskip 2pt
The \emph{top panel} of Figure~\ref{fig:mathieuplasma} shows numerical solutions to the plasma Mathieu equation~\eqref{eq:PlasmaMathieu} for different values of the plasma frequency. For $\omega\ped{p} = 0$, the solution develops an instability that is well-described by the analytic solution of the vacuum Mathieu equation [{\color{Mathematica4}{red}} dashed]. By increasing $\omega\ped{p}$, the growth rate of the instability becomes smaller as the interval of the momentum corresponding to the instability shrinks until the solution becomes stable when $\omega\ped{p}\geq\mu/2$. We find good agreement with the instability interval predicted in eq.~(19) from~\cite{SenPlasma}, by exploring a wide region of the parameter space. Interestingly, even when $\omega\ped{p}\geq\mu/2$, the instability band can be widened by increasing the value of $k\ped{a} \psi_0$, making it possible to restore the instability even for dense plasmas. This effect can be seen in the \emph{bottom panel} of Figure~\ref{fig:mathieuplasma}:~when fixing $\omega\ped{p}=0.15$ (with $\mu = 0.2$), the instability is restored for large enough values of $k\ped{a} \psi_0$. 
\subsubsection{Band analysis}
\begin{figure}[t!]
    \centering
    \includegraphics[scale=0.45, trim = 0 0 0 0]{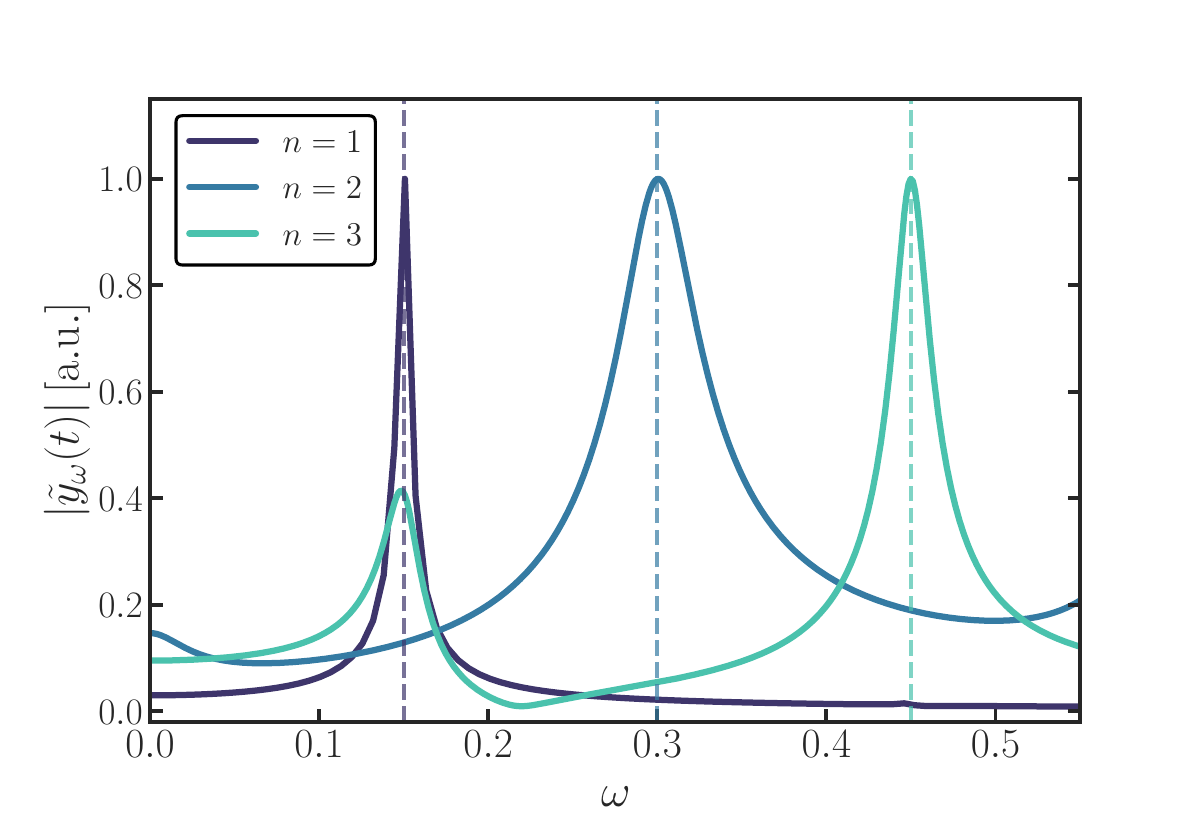}
    \caption{The Fourier transforms of three numerical solutions to the plasma Mathieu equation~\eqref{eq:PlasmaMathieu} with $\mu=0.3$ and $p_z=0.15$. The peaks have been arbitrarily normalised and we revert $t = \mu^{-1}T$ for the Fourier transform. The considered parameters for the plasma frequency and the axionic coupling for $n = 1, 2, 3$ are $\omega\ped{p}=0, 0.2, 0.43$ and $k\ped{a} \psi_0 = 0.02, 2, 2.8$, respectively. The dashed lines indicate $n \mu/2$ and show that each of the solutions lies in a different instability band.}
    \label{fig:bandsmathieuplasma}
\end{figure}
In~\cite{SenPlasma}, the maximum growth rate in the low-coupling regime is found when the condition $p_z^2 + \omega\ped{p}^2 = \mu^2/4$ holds. Although this result lacks an immediate interpretation in~\cite{SenPlasma}, our reformulation of the system in terms of the Mathieu equation offers a straightforward explanation. From~\eqref{eq:PlasmaMathieu}, it follows directly that the instability bands are located where
\begin{equation}\label{eq:plasmabands}
 p_z^2 + \omega\ped{p}^2= n^2 \, \frac{\mu^2}{4}\,,\: \text{with} \quad n \in \mathbb{N}\,.   
\end{equation}
Thus, the maximum growth rate found in~\cite{SenPlasma} can be interpreted as the first, dominant instability band, corresponding to $n = 1$. As long as the couplings are sufficiently low, the photon dispersion relation remains unaffected by the condensate, so that $\omega^2 \approx p_z^2 + \omega\ped{p}^2$, with $\omega$ the photon frequency. Therefore, similar to the vacuum case, the instability bands correspond to frequencies that are multiples of $\mu/2$. 
\vskip 2pt
For sufficiently large plasma frequencies, specifically when $\omega\ped{p} \geq \mu/2$, the condition~\eqref{eq:plasmabands} can no longer be satisfied for $n = 1$. However, crucially, it remains possible to satisfy this condition for $n > 1$, corresponding to the excitation of higher instability bands and thereby restoring the instability.\footnote{Higher-order bands of the Mathieu equation are narrower than the first, reducing the available parameter space for an instability. Nonetheless, these bands broaden at large couplings $k\ped{a}\psi_0$, making it possible to trigger an efficient instability even for $n>1$.} Figure~\ref{fig:bandsmathieuplasma} illustrates this behaviour. We numerically solve~\eqref{eq:PlasmaMathieu} and take its Fourier transform. The dark-blue curve, peaking at $\mu/2$, corresponds to $\omega\ped{p} = 0$ and moderate couplings $k\ped{a} \psi_0 = 0.02$, and exhibits the typical parametric resonance of the first band. When the plasma frequency is increased to $\omega\ped{p} = 0.2 > \mu/2$, the first instability band can no longer be excited. However, by increasing the axionic coupling to $k\ped{a} \psi_0 = 2$, the instability reappears in the second band at frequency $\mu$ [{\color{mediumblue1}{blue}}]. This pattern continues for even larger plasma frequencies:~for example, with $\omega\ped{p} = 0.43$ and $k\ped{a} \psi_0 = 2.8$, the third instability band is excited at $\omega = 3\mu/2$ [{\color{tealgreen1}{turquoise}}]. In all three cases, the chosen values of $\omega\ped{p}$ approximately satisfy condition~\eqref{eq:plasmabands} for $n = 1$, $2$, and $3$, respectively.
\chapter{Black Hole Perturbation Theory}\label{app:BHPT}
Understanding BHs and their dynamics requires a variety of theoretical tools, each suited to different aspects of the problem. A particularly successful approach involves describing the spacetime as a small deviation from a known exact solution, an approach referred to as \emph{black hole perturbation theory}. This framework has been widely applied in modelling systems such as EMRIs~\cite{Barack:2018yvs,Pound:2021qin} and in studying the QNMs of BHs~\cite{Kokkotas:1999bd,Nollert:1999ji,Berti:2009kk,Konoplya:2011qq}.
\vskip 2pt
This appendix establishes the theoretical foundation for BH perturbation theory in stationary, spherically symmetric spacetimes at the linear level. This approach has been central to the analyses in Chapters~\ref{chap:in_medium_supp},~\ref{chap:plasma_ringdown}, and~\ref{chap:BHspec}. Section~\ref{appBHPT_sec:perturbations} introduces the perturbative setup. Section~\ref{appBHPT_sec:evoleq} presents the evolution equations governing BH perturbations in vacuum, outlines how QNMs are extracted from these equations, and highlights specific considerations in different environments. Section~\ref{appBHPT_sec:ID} then addresses various choices of initial data for time-domain evolutions, and Section~\ref{appBHPT_sec:num_frame} describes the numerical framework, and includes convergence tests for the code.
\section{Perturbations}\label{appBHPT_sec:perturbations}
Black hole perturbation theory offers a framework for studying small deviations from exact solutions of Einstein’s field equations~\eqref{eq:EFE}. The goal is to understand how small disturbances -- such as a passing wave, a falling particle, or a companion star -- evolve in a given background metric. 
\vskip 2pt
The main idea is to expand relevant quantities as 
\begin{equation}
X = X^{(0)} + \epsilon\,\delta\!X + \mathcal{O}(\epsilon^{2})\,,
\end{equation}
where $X$ can represent tensorial quantities such as the metric $g_{\alpha\beta}$, as well as vector or scalar quantities. Here, the superscript $(0)$ denotes background quantities and $\epsilon$ is a bookkeeping parameter controlling the perturbation order. Given the spherical symmetry of the background, all perturbations can be decomposed into irreducible representations of the rotation group $\mathrm{SO}(3)$. This allows us to express the angular dependence of the perturbations through a multipolar expansion.
\vskip 2pt
The study of gravitational perturbations begins by expanding the metric:
\begin{equation}
g_{\mu \nu}=g_{\alpha \beta}^{(0)}+\epsilon\, \delta g_{\alpha \beta}\,,
\end{equation}
where $\delta g_{\alpha \beta} \ll 1$ is the \emph{metric perturbation}. Since $\delta g_{\alpha \beta}$ is a rank--$2$ tensor, it has ten independent components. The key to simplifying the analysis lies in separating the angular dependence of $\delta g_{\alpha \beta}$, which was first done by Regge and Wheeler in 1957~\cite{ReggeWheeler}. For a Schwarzschild background, this can be done using tensor spherical harmonics~\cite{NIST:DLMF}, reducing the problem to a set of perturbation functions that depend only on the radial coordinate $r$ and time $t$. 
\vskip 2pt
These perturbations fall into two distinct classes based on their transformation under parity ($\theta \rightarrow \pi - \theta$, $\varphi \rightarrow \varphi+\pi$):
\begin{itemize}
\item \emph{Axial} (or \emph{odd-parity)} perturbations, which acquire a factor of $(-1)^{\ell +1}$,
\item \emph{Polar} (or \emph{even-parity)} perturbations, which acquire a factor of $(-1)^{\ell}$.
\end{itemize}
This classification simplifies the equations significantly, as perturbations of different parity decouple. The ten metric components then fall into seven (three) independent functions for polar (axial) perturbations. Furthermore, perturbations with different harmonic indices $\ell$ are independent, meaning that for each $\ell$, we are left with two separate sets of equations -- one for each parity sector -- which completely characterise the linear response of the system. The gravitational perturbations can thus be decomposed as
\begin{equation}\label{eq:BHPT_metricPert}
\delta g_{\alpha \beta}(t,r,\theta,\varphi)= \delta g_{\alpha \beta}^{\rm axial}(t,r,\theta,\varphi)+\delta g_{\alpha \beta}^{\rm polar}(t,r,\theta,\varphi)\,.
\end{equation}
The explicit expressions for these components in terms of tensor spherical harmonics are lengthy and can be found in e.g.,~\cite{Sago:2002fe}.
\vskip 2pt
The diffeomorphism invariance of General Relativity allows for infinitesimal coordinate transformations,
\begin{equation}
\bar{x}^{\alpha} = x^{\alpha} + \xi^{\alpha}(x)\,, \quad \delta g_{\alpha \beta} \rightarrow \delta g'_{\alpha \beta} = \delta g_{\alpha \beta} -2 \nabla_{(\alpha} \xi_{\beta)}\,,
\end{equation}
where the four functions composing $ \xi^{\alpha}(x)$ can be used to eliminate certain components of the metric perturbations. A common and convenient choice is the \emph{Regge--Wheeler gauge}, which removes specific components involving higher-order angular derivatives, reducing the number of independent functions to two for the axial sector and four for the polar sector. It is then possible to derive a decoupled wave equation for a ``master variable'', which encapsulates all the dynamics. In the Regge--Wheeler gauge, the metric perturbation~\eqref{eq:BHPT_metricPert} takes the form
\begin{align}
\label{eq:Regge-Wheeler-gauge}
\renewcommand{\arraystretch}{1.25}
\setlength\arraycolsep{2pt}
\delta g_{\alpha \beta}\!=\!\sum_{\ell m}\begin{pNiceArray}{cc|cc}[margin = 0.cm]
\Block[fill=blue!10,rounded-corners]{2-2}{}
H^{\ell m}_0Y^{\ell m} & H^{\ell m}_1Y^{\ell m} & \Block[fill=red!10,rounded-corners]{2-2}{}
        h^{\ell m}_0 S^{\ell m}_{\theta} & h^{\ell m}_0 S^{\ell m}_{\varphi} \\
H^{\ell m}_1Y^{\ell m} & H_2^{\ell m}Y^{\ell m} & h^{\ell m}_1S^{\ell m}_{\theta} & h^{\ell m}_1 S^{\ell m}_{\varphi} \\
\hline
\Block[fill=red!10,rounded-corners]{2-2}{}
* & * & \Block[fill=blue!10,rounded-corners]{2-2}{}
        r^2 K^{\ell m}Y^{\ell m} & 0 \\
* & * & 0 & r^{2}\sin^{2}{\theta} K^{\ell m}Y^{\ell m} \\
\end{pNiceArray}e^{-i \omega t}\,.
\end{align}
Here, the red and blue blocks correspond to the axial $(\delta g_{\alpha \beta}^{\rm axial})$ and polar part $(\delta g_{\alpha \beta}^{\rm polar})$, respectively. To keep the notation streamlined, we omit explicit dependence on $(t,r,\theta,\varphi)$ and define the axial vector harmonics as $(S^{\ell m}_{\theta}, S^{\ell m}_{\varphi})=(-\partial_{\varphi}Y^{\ell m}/\sin{\theta},\sin{\theta}\partial_{\theta}Y^{\ell m})$. 
\vskip 2pt
Throughout this thesis, we frequently consider the presence of additional fundamental fields, which can be decomposed into spherical harmonics of the appropriate type. For example, a generic vector field $v_{\alpha}$ can be expanded in terms of vector spherical harmonics as
\begin{equation}\label{eq:vector_expansions}
    \delta v_\alpha =\frac{1}{r}\sum_{i=1}^4\sum_{\ell m}c_iv_{i}^{\ell m}  Z_\alpha^{(i) \ell m} e^{-i \omega t}\,,
\end{equation}
where the coefficients $c_1=c_2=1$ and $c_3=c_4=1/\sqrt{\ell(\ell+1)}$. The vector spherical harmonics $Z_\alpha^{\ell m}$ are defined in e.g.,~\cite{Rosa:2011my}.  Similarly, scalar quantities such as $P$ are expanded as 
\begin{equation}\label{eq:scalar_decomposition}
\delta P = \sum_{\ell m} P^{\ell m} Y^{\ell m} e^{-i \omega t}\,, 
\end{equation}
where $Y_{\ell m}$ denote the standard scalar spherical harmonics. 
\section{Evolution Equations}\label{appBHPT_sec:evoleq}
Inserting the decomposed perturbations [eqs.~\eqref{eq:Regge-Wheeler-gauge}--\eqref{eq:scalar_decomposition}] into the linearised field equations leads to a system of coupled second-order differential equations. Through an appropriate field redefinition, one can obtain a single master equation governing the evolution of perturbations. For the axial gravitational sector, that is is the Regge--Wheeler equation with the vector-type gravitational variable $\Psi\ped{RW}$~\cite{ReggeWheeler} that consists of the axial perturbation functions $h_0$ and $h_1$. For the polar sector, it is the Zerilli equation~\cite{Zerilli:1970se,Zerilli:1970wzz} with the scalar-type gravitational variable $\Psi\ped{Z}$ that consists of the polar perturbation functions $H_0$, $H_1$, $H_2$ and $K$. The evolution equations take the form of a wave equation with an effective potential that depends on the background metric, i.e., 
\begin{equation}\label{eq:evol_perturb_eq_time}
\left(-\frac{\partial^2}{\partial t^2}+\frac{\partial^2}{\partial r_*^2} - V_{\mathrm{RW}/\mathrm{Z}}\right)\Psi_{\mathrm{RW}/\mathrm{Z}} = \mathcal{S}_{\mathrm{RW}/\mathrm{Z}}\,,
\end{equation}
where $r_*$ is the tortoise coordinate and $\mathcal{S}_{\mathrm{RW}/\mathrm{Z}}$ are source terms that depend on the energy-momentum tensor. The potentials in eq.~\eqref{eq:evol_perturb_eq_time} are given by
\begin{equation}\label{eq:potentials_RW_Z}
\begin{aligned}
V\ped{RW} &= \left(1-\frac{2M}{r}\right)\left[\frac{\ell(\ell+1)}{r^2}-\frac{6M}{r^3}\right]\,,\\
V\ped{Z} &= \frac{2}{r^3}\left(1-\frac{2M}{r}\right)\frac{9M^3+3\Lambda^2Mr^2+\Lambda^2(1+\Lambda)r^3+9M^2\Lambda r}{(3M + \Lambda r)^2}\,,
\end{aligned}
\end{equation}
where $\Lambda = (\ell-1)(\ell+2)/2$. Despite their different forms, the Regge--Wheeler and Zerilli equations are \emph{isospectral}, meaning they share the same QNM frequencies~\cite{Chandrasekhar:1975zza,MTB}.
\vskip 2pt
The source term $\mathcal{S}_{\mathrm{RW}/\mathrm{Z}}$ describes the object that excites the spacetime perturbations. One of the widely-used approaches is that of a ``point-particle'' that represents e.g.,~a much smaller, secondary BH. This secondary can be assumed to move along geodesics with its gravitational backreaction accounted for by the energy and angular momentum loss due to GWs~\cite{PhysRevD.50.3816,Hughes:1999bq,Mino:2003yg}. This is the central objective of the self-force programme (see Section~\ref{GravAstro_subsec:binary_insp}).
\subsubsection{Quasi-normal modes}
Another key application of BH perturbation theory is modelling the ringdown phase of a BH coalescence (see Section~\ref{GravAstro_subsec:ringdown}). Here, we briefly outline how QNM frequencies are computed in the frequency domain.
\vskip 2pt
As shown in~\eqref{eq:evol_perturb_eq_time}, perturbations of the background spacetime satisfy a wave equation for a master variable $\Psi$, which encodes the radiative degrees of freedom of the gravitational field~\cite{ReggeWheeler,Zerilli:1970se,Teukolsky:1974yv,MTB}. Equation~\eqref{eq:evol_perturb_eq_time} can be solved either in the time domain, by evolving wavepackets scattered off the BH potential~\cite{Vishveshwara:1970zz}, or in the frequency domain, where one extracts the characteristic modes of the system~\cite{Chandrasekhar:1975zza,Leaver:1986gd}. Written in terms of the ``standard'' radial coordinate $r$ and using again $\Psi$ in a slight abuse of notation, the master equation for the Regge--Wheeler potential~\eqref{eq:potentials_RW_Z} in the frequency domain reads:
\begin{equation}\label{eq:evol_perturb_eq_freq}
r(r-r_+)\frac{\dd^2\Psi}{\dd r^2} + \frac{\dd \Psi}{\dd r} - \left[\ell(\ell+1) - \frac{3}{r} - \frac{\omega^2 r^3}{r-r_+}\right]\Psi = 0\,,
\end{equation}
where in Schwarzschild $r_+ = 2M$. QNMs correspond to solutions of this equation satisfying boundary conditions of purely ingoing waves at the BH horizon and purely outgoing waves at infinity:
\begin{equation}\label{eq:QNM_boundary}
\Psi\sim\begin{cases}
e^{-i\omega r_*} \sim e^{-i\omega\ln{(r-r_+)}} \sim (r-r_+)^{-i\omega}  \:& r\to r_+\,,\\
e^{i\omega r_*}\sim e^{i\omega(r+\ln{r})} \sim r^{i\omega}e^{i\omega r} \:& r\to\infty\,.
\end{cases}
\end{equation}
Similar to constructing the bound states of scalar fields around BHs (see Appendix~\ref{appNR_subsec:boundstates}), one of the most effective methods for solving this problem and computing the QNM frequencies, is Leaver’s continued fraction method~\cite{Leaver:1985ax,Leaver:1986,Leaver:1986gd}. It begins by expanding the solution as
\begin{equation}
\Psi = (r-r_+)^{-i\omega} r^{2i\omega}e^{i\omega(r-r_+)}\sum_{n=0}^{\infty}b_n\left(\frac{r-r_+}{r}\right)^n\,,
\end{equation}
where the prefactor is chosen to incorporate the QNM boundary conditions~\eqref{eq:QNM_boundary}. Substituting this \emph{ansatz} into eq.~\eqref{eq:evol_perturb_eq_freq} leads to a three-term recurrence relation for the expansion coefficients: 
\begin{equation}\label{eq:recurrence}
\begin{aligned}
\alpha_0b_1+\beta_0 b_0 &= 0\,,\\
\alpha_n b_{n+1}+\beta_n b_n+\gamma_n b_{n-1} &= 0\,, \quad n = 1,2,\ldots\,,
\end{aligned}
\end{equation}
where $\alpha_n$, $\beta_n$, $\gamma_n$ depend on $\ell$ and $\omega$. Given any initial values $b_0$ and $b_1$, one can iteratively solve~\eqref{eq:recurrence} for $b_n$ and thereby find a solution for $\Psi$. In general, for arbitrary $\omega$, the series divergences as $n\to \infty$. However, QNM frequencies correspond to special values of $\omega$ for which the recurrence relation~\eqref{eq:recurrence} admits a \emph{minimal solution}. Pincherle's theorem~\cite{Pincherele,Gautschi:1967cat} ensures that such minimal solutions exist (i.e., that the series is convergent and the QNM boundary conditions are satisfied) if $\omega$ satisfies a continued fraction equation~\eqref{eq:Leaver}. This method is particularly powerful because it converges rapidly, even for high overtones, and has been successfully applied to Schwarzschild and Kerr spacetimes~\cite{Nollert:1993zz,Onozawa:1996ux,Berti:2003jh,Berti:2004um}.
\vskip 2pt
Having outlined the theoretical framework of BH perturbation theory, we now turn to the specific cases considered in this thesis.
\subsection{Specifics for Plasma}\label{appBHPT_subsec:plasma}
In Chapters~\ref{chap:in_medium_supp} and~\ref{chap:plasma_ringdown}, we study plasmas surrounding BHs using perturbation theory.  Given their shared features, we outline the core features of the framework here. For clarity, the discussion focuses on plasmas around charged BHs. Extensions to cases involving couplings to a scalar field (Section~\ref{sec_IMS:chargedBHs}) or uncharged BHs (Section~\ref{sec_IMS:darkphotons}) follow straightforwardly and are omitted here to avoid unnecessary distractions.
\vskip 2pt
Starting from the background Reissner-Nordstr\"{o}m metric~\eqref{eq:RN_Sol}, we linearise the field equations [eqs.~\eqref{eq:IMS_evoleqns} or~\eqref{eq:einstein_maxwell}] to first-order perturbations. Unlike vacuum Reissner-Nordstr\"{o}m~\cite{Zerilli:1970se,Zerilli:1970wzz}, where one only the gravitational and electromagnetic field are perturbed, the presence of plasma requires us to account for perturbations in the fluid quantities. Given the large hierarchy between the mass of the electrons and ions, we ignore perturbations on the latter and treat them as a stationary, neutralising background.
\subsubsection{Axial sector}
As we will see, the evolution of the perturbations in the axial sector can be written as a set of two coupled wave equations. This procedure is aided by the existence of a simple relation between the axial components of the plasma velocity and the electromagnetic field, found by perturbing the angular components of the momentum equation~\eqref{eq:momentum_continuity}:
\begin{equation}\label{eq:momentumcons}
v_{4}=-\frac{e}{m\ped{e}}u_{4}\,,
\end{equation}
where $v_{i}$ are the coefficients in the expansion of the four-velocity of the electrons while $u_{i}$ are the coefficients from the electromagnetic field~\eqref{eq:vector_expansions}. The above equation simply corresponds to the conservation of the generalised transverse momentum of a charged particle with impulse $\vec{P}$ in an electromagnetic field $\vec{A}$, i.e., $\partial_t (\vec{P}+ e \vec{A})=0$. Given that the axial sector is composed solely of transverse modes~\cite{Baryakhtar:2017ngi}, this very simple relation holds for our purposes. Moreover, by considering only axial quantities, the continuity equation~\eqref{eq:momentum_continuity} and the radial component of the momentum equation show that $n^{\ell m}\ped{e}=P^{\ell m}\ped{t}=0$, which is a consequence of variations in the fluid density and pressure being longitudinal -- and thus strictly polar -- degrees of freedom. Fluctuations of the fluid's stress-energy tensor affect the axial Einstein equations via terms proportional to $\omega\ped{p} m\ped{e}/e$ and $\omega\ped{p} ^2 m\ped{e}^2/e^2$. Given that in astrophysical systems of interest $\omega\ped{p}=\mathcal{O}(1/M)$, the large charge-to-mass ratio of the electron ($e/m\ped{e} \approx 10^{22}$) drastically suppresses the impact of the fluid on the gravitational sector, which we thus neglect. 
\vskip 2pt
This leaves us with three degrees of freedom, $h_0$, $h_1$ and $u_4$, for which there are three coupled equations (two of which are gravitational and one that is electromagnetic). The first gravitational equation comes from the $t\!-\!\theta$ and $t\!-\!\varphi$ components of Einstein equation, while the second one can be derived from the $r\!-\!\theta$ and $r\!-\!\varphi$ components. The Einstein equations are not modified by plasma and one can solve the latter to obtain:\footnote{In some of the computations below, it is convenient to perform computations in the frequency domain by assuming a sinusoidal time dependence, and switch back to the time domain once the master equations are obtained. Note that, as the system is stationary, it is possible to go back and forth between them simply through $\partial/\partial t \leftrightarrow -i \omega$. }
\begin{equation}
    \frac{\partial h_0}{\partial r}= 2 \frac{h_0}{r}-\frac{4 Q u_{4}}{\lambda r^2}+\frac{i\Big(-2+\lambda-r^2\omega^2/f\Big)}{r\omega}\Psi\ped{RW}\,,
\end{equation}
where $f$ is the $g_{tt}$ component of the metric~\eqref{eq:RN_Sol}, $\lambda=\ell(\ell+1)$, and we defined the Regge--Wheeler ``master variable'' $\Psi\ped{RW}= f h_1 /r$. We can then use this expression in the first gravitational equation, which reduces to a simple relation between $h_0$ and $\Psi\ped{RW}$:
\begin{equation}
    h_0=\frac{f}{-i \omega}\frac{\partial (r\Psi\ped{RW})}{\partial r}\,.
\end{equation}
As $h_0$ is now decoupled from the system, we are able to obtain two coupled equations for the ``master variables'' $\Psi\ped{RW}$ and $u_4$ as
\begin{equation}
\label{eq:wavelikeRW}
\begin{aligned}
&\frac{\partial^{2}\Psi\ped{RW}}{\partial r^{2}}-\frac{2(Q^2- M r)}{r^{3}f}\frac{\partial\Psi\ped{RW}}{\partial r} - \frac{4iQ\omega}{r^{3}\lambda f}u_{4} + \left(\frac{\omega^{2}}{f^{2}} - \frac{r^2 \lambda -6 M r + 4 Q^2}{r^{4}f}  \right)\Psi\ped{RW}  = 0\,,\\[6pt]
&\frac{\partial^{2}u_{4}}{\partial r^{2}} - \frac{2(Q^{2}-Mr)}{r^{3}f}\frac{\partial u_4}{\partial r} +\frac{iQ\lambda(\lambda-2)}{r^{3}\omega f}\Psi\ped{RW} + \frac{r^{4}\omega^{2} - (4 Q^2+r^2\lambda+\omega\ped{p}^2 r^4)f}{r^{4}f^{2}}u_4 = 0\,.
\end{aligned}
\end{equation}
As advertised, the full system has thus reduced to a set of coupled ordinary differential equations for $\Psi\ped{RW}$ and $u_{4}$. To study the dynamics emerging from these equations, we transform them back to the time domain. Furthermore, to acquire a wave-like form of the equations, we consider the Moncrief master variable $\Psi\ped{M}$ instead of the Regge--Wheeler one. They are related through~\cite{Moncrief:1974ng,PhysRevD.9.2707,Moncrief:1975sb}:
\begin{equation}
\label{eq:Moncrief}
\Psi\ped{M}= \frac{2 i \Psi\ped{RW}}{\omega}\,.
\end{equation}
Using this substitution, we can rewrite~\eqref{eq:wavelikeRW} as wave equations:
\begin{equation}
\label{eq:wavelike-eqn}
\begin{aligned}
\hat{\mathcal{L}}\Psi\ped{M} &=\Bigg(\frac{4 Q^4}{r^6}+ \frac{Q^2(-14 M + r (4+\lambda))}{r^5} + \left(1-\frac{2M}{r}\right)\left[\frac{\lambda}{r^{2}}-\frac{6M}{r^{3}}
\right]\Bigg)\Psi\ped{M}-\frac{8 Q f}{r^3 \lambda} u_{4}\,, \\
\hat{\mathcal{L}}u_{4}&=f\left(\omega\ped{p}^2 + \frac{\lambda}{r^{2}} + \frac{4 Q^2}{r^4}\right) u_{4}-\frac{(\ell-1)\lambda(\ell+2) Q f}{2 r^3}\Psi\ped{M}\,,
\end{aligned}
\end{equation}
where the differential operator is defined as $\hat{\mathcal{L}} = \partial^{2}/\partial r_*^{2} - \partial^{2}/\partial t^{2}$, with the tortoise coordinate satisfying $\mathrm{d}r_*/\mathrm{d}r = f^{-1}$. In the limit where the charge $Q\rightarrow 0$, the system decouples:~the equation for $\Psi\ped{M}$ simplifies to the Regge--Wheeler equation~\eqref{eq:potentials_RW_Z}, while the equation for $u_4$ matches the axial electromagnetic mode in a Schwarzschild background with plasma~\cite{Cannizzaro:2020uap}.
\subsubsection{Polar sector}
In contrast to the axial sector, fluid and electromagnetic perturbations are coupled in the polar sector. This is due the presence of longitudinal, electrostatic modes as well as gravitational fluctuations in the momentum equation~\eqref{eq:momentum_continuity}. Additionally, the polar density and pressure fluctuations $\delta n\ped{e}^{\ell m}$, $\delta P\ped{t}^{\ell m}$ do not vanish, while the perturbed velocity does not decouple trivially. There is thus no equivalent to eq.~\eqref{eq:momentumcons}, which complicates the procedure.
\vskip 2pt
In order to study the polar sector consistently, one should evolve the fluid variables along with the gravitational and electromagnetic fields and close the system with a suitable equation of state, as done in e.g., Chapter~\ref{chap:BHspec}. We do not undertake that exercise here yet rather show that, similar to the axial sector, electromagnetic modes in the polar sector are dressed with an effective mass given by $\omega\ped{p}$. Since the results of Chapters~\ref{chap:in_medium_supp} and~\ref{chap:plasma_ringdown} are centred around the presence of such a mass, there is no need to repeat the same detailed study in the polar sector as we did in the axial sector.
\vskip 2pt
In the following, we briefly sketch the procedure to show the presence of an effective mass $\omega\ped{p}$, in the polar sector. To obtain the electromagnetic master equation, one should consider the perturbation of the normalisation of the four velocity, $\delta(v_\mu v^\mu)=0$ and solve for $v_1$ in terms of metric perturbations. This relation can then be substituted in the $t-r$ component of the Einstein equations, which gives $v_2$ in terms of metric perturbations. Finally, from the angular components of the momentum equation, one can obtain $v_3$ in terms of metric, electromagnetic and density perturbations. Next, consider the polar Maxwell's equations, and in particular the radial one -- which we shall denote $\mathcal{M}_r$ -- and a combination of the angular ones, given by $\mathcal{M}_\theta+\mathcal{M}_\varphi/\text{sin}^2\theta$. In analogy to previous work~\cite{Zerilli:1970se, Pani:2018flj}, the electromagnetic master relation can be derived from these equations. Following Zerilli, we find it convenient to use perturbations of the field strength $\delta F_{\mu\nu}=\partial_\mu\,\delta A_\nu-\partial_\nu\,\delta A_\mu$, rather than the electromagnetic potential $\delta A^\mu$, as a dynamical variable. We choose a gauge such that $u_3=0$, while the other components of the potential are expressed as
\begin{equation}
\begin{aligned}
    u_1&= r\,\delta \bar{F}_{02}\,, \quad    u_2= r f\, \delta\bar{F}_{12}\,,\\
    \frac{\partial u_1}{\partial r}&= r\,\delta \bar{F}_{01}+\delta \bar{F}_{02}-i r \omega\,\delta \bar{F}_{02}\,,
\end{aligned}
\end{equation}
where $\delta \bar{F}_{\mu\nu}$ is the angle-independent part of $\delta F_{\mu\nu}$. Using this relation, one can solve $\mathcal{M}_r$ and $\mathcal{M}_\theta+\mathcal{M}_\varphi/\text{sin}^2\theta$ to obtain $\delta \bar{F}_{01}$ and $\delta \bar{F}_{02}$, respectively. These solutions can then be used in the homogeneous Maxwell equation,
\begin{equation}
    \delta\bar{F}_{01}=\frac{ \partial\,\delta \bar{F}_{02}}{\partial r}+i \omega\,\delta \bar{F}_{12}\,,
\end{equation}
to obtain a second-order differential equation for $\delta \bar{F}_{12}$. At large radii and neglecting the coupling with metric and density perturbations induced by the fluid, one finds the standard dispersion relation of a transverse electromagnetic mode in a plasma dressed by an effective mass given by the plasma frequency $\omega^2= k^2+\omega\ped{p}^2$.
\subsection{Specifics for Galactic Environments}\label{appBHPT_subsec:galactic}
In Chapter~\ref{chap:BHspec}, we investigate the ringdown of BHs within a galactic environment using perturbation theory for non-vacuum spacetimes~\cite{Barack:2018yvs,Pound:2021qin,Cardoso:2021wlq,Cardoso:2022whc,Duque:2023seg}. This involves perturbing the metric and the anisotropic fluid characterising the halo at linear order in the small mass ratio $q = m\ped{p}/M$. As the background is again spherically symmetric, the perturbations for the axial and polar sector decouple. Our focus here lies on the polar sector, where the matter and gravitational perturbations are coupled, potentially giving rise to a richer dynamics. 
\vskip 2pt
Details on the derivations of the equations of motion governing the evolution of the perturbations can be found in~\cite{Cardoso:2022whc}. Schematically, they are given by a set of 3 wave-like equations
\begin{equation}\label{eqn:wavelikeeqn}
\hat{\mathcal{L}}\vec{\phi} = \hat{A}\vec{\phi}_{,r_{*}} + \hat{A}\vec{\phi} + \vec{S}^{\rm p}\,,
\end{equation}
where $\vec{\phi} = (S, K, \delta \rho)$, $S$ and $K$ are functions representing perturbations of the metric and $\delta \rho$ is the perturbation of the matter density profile. Moreover, $\mathcal{L}_{v} = v^{2}\partial^{2}/\partial r_*^{2} - \partial^{2}/\partial t^{2}$ is a wave operator and $\hat{\mathcal{L}}\vec{\phi} = \left( \mathcal{L}_1 \phi_1,  \mathcal{L}_1 \phi_2,  \mathcal{L}_{c\ped{s_{r}}} \phi_3 \right)$. The coefficient matrices of the homogeneous part $\hat{A}$ and $\hat{B}$ are the same as in~\cite{Cardoso:2022whc}, with the difference in the setup being the source term $\vec{S}^{\rm p}$, representing the radial plunge of the small body.
\vskip 2pt
From the evolved variables $\vec{\phi}$~\eqref{eqn:wavelikeeqn}, we can construct the Zerilli-Moncrief~\cite{Zerilli:1970wzz} function in the near-vacuum region as
\begin{equation}
\mathcal{Z}_{\ell m}=\frac{r}{n+1}\left[K(r)+\frac{A(r)}{n+3(M+M\ped{H})/r} \left(H_2(r)-r \frac{\partial K}{\partial r}\right)\right]\,,
\end{equation}
where $n = \ell(\ell+1)/2 - 1$ and $H_2(r)$ can be found in eq.~(47) in~\cite{Cardoso:2022whc}. This function controls the radiative degrees of freedom of the gravitational field and is related to the GWs polarisations by
\begin{equation}
\left(h_{+}-i h_{\times}\right)_{\ell m}=\frac{1}{r} \sqrt{\frac{(\ell+2) !}{(\ell-2) !}} \mathcal{Z}_{\ell m} \> {}_{{\scalebox{0.65}{$-$}}2}\mkern-2mu Y_{\ell m}(\theta, \varphi)+\mathcal{O}\left(\frac{1}{r^2}\right)\,,
\end{equation}
where ${}_{{\scalebox{0.65}{$-$}}2}\mkern-2mu Y_{\ell m}(\theta, \varphi)$ are the spin-weighted spherical harmonics with $s = 2$. For the radial plunges we consider in Chapter~\ref{chap:BHspec}, the cross-polarisation $h_{\times}$ is zero and the GW radiation is fully determined by $h_+$.
\section{Initial Data}\label{appBHPT_sec:ID}
Historically, the first study of BH perturbations in the time domain was carried out by Vishveshwara, who examined the scattering of Gaussian wavepackets at BHs~\cite{Vishveshwara:1970zz}. This approach remains relevant in cases where the specific nature of the perturbation is not of interest, and one simply seeks to understand how the system evolves once initialised. The clear advantage of this method is the absence of a source term on the right-hand side of the evolution equation~\eqref{eq:evol_perturb_eq_time}. We have used this approach in Chapters~\ref{chap:in_medium_supp} and~\ref{chap:plasma_ringdown}. However, a more physically realistic setup involves perturbations induced by a plunging particle, which we describe below.
\subsubsection{A plunging particle}
In Chapter~\ref{chap:BHspec}, we studied the response of the BH immersed in a halo by having a secondary BH plunging into the primary one. The secondary is represented by a point-particle with world-line $x\ped{p}^\mu(\tau)$, four-velocity $u\ped{p}^\mu = \mathrm{d}x\ped{p}^\mu/\mathrm{d}\tau$ and stress-energy tensor:
\begin{equation}
T\ped{p}^{\mu \nu}=m\ped{p} \int u\ped{p}^\mu u\ped{p}^\nu \frac{\delta^{(4)}\left(x^\mu-x\ped{p}^\mu(\tau)\right)}{\sqrt{-g}} \mathrm{d} \tau\,,
\end{equation}
where $\tau$ is the proper time related to coordinate time via
\begin{equation}
\frac{\mathrm{d}t}{\mathrm{d}\tau} = \frac{E\ped{p}}{\sqrt{A(r\ped{p})}}\,.
\end{equation}
Here, $E\ped{p}$ is the energy of the point-particle~\cite{Cardoso:2022whc} and $A(r)$ the $g_{tt}$ component of the metric, see eq.~\eqref{eqn:line_element}. We do not consider backreaction on the orbit due to GW emission, i.e., we assume the particle follows geodesic radial motion, determined by 
\begin{equation}\label{eqn:plunge}
\frac{\mathrm{d}r\ped{p}}{\mathrm{d}t} = -\sqrt{A(r\ped{p})B(r\ped{p})}\sqrt{1 - \frac{A(r\ped{p})}{E\ped{p}^2}}\,,
\end{equation}
where $B(r) = 1 - 2m(r)/r$. Since we take the plunge to be along the radial $\hat{z}$--direction, such that $\theta\ped{p} = \varphi\ped{p} = 0$, only $m = 0$ modes are excited.
\vskip 2pt
Following the same procedure as~\cite{Cardoso:2022whc}, we find $\vec{S}^{\rm p} = q \left(S_1^{\rm p}, S_2^{\rm p}, S_3^{\rm p}  \right)$ with coefficients:
\begin{equation}
\begin{aligned}
S_1^{\rm p} &=-\frac{8\sqrt{\pi}}{E\ped{p}} \sqrt{2\ell+1}\frac{A}{r^3}\left(E\ped{p}^2 - A\right) \sqrt{A B}\,\delta(r-r\ped{p}(t))\,,\\
S_2^{\rm p} &= \, -\frac{4\sqrt{\pi}}{E\ped{p}} \sqrt{2\ell+1}\frac{A}{r^2}\sqrt{AB}\, \delta(r-r\ped{p}(t))\,,\\
S_3^{\rm p} &= -\frac{2\sqrt{\pi}}{E\ped{p}} \sqrt{2\ell+1}\frac{\rho + 2 P\ped{t}}{r^2}\left(2E\ped{p}^2-A\right) \sqrt{AB}\, \delta(r-r\ped{p}(t))\,,     
\end{aligned}
\end{equation}
where we suppressed dependencies on $r$. Numerically, the Dirac delta representing the point-particle is approximated by a smoothed Gaussian-like distribution.
\begin{equation}\label{eq:pp_num}
\delta(r-r\ped{p}(t)) =\frac {\exp\left[-(r-r\ped{p}(t))^2/2\lambda\ped{p}^2 \right]}{\sqrt{2\pi}\lambda\ped{p}}\,,
\end{equation}
where $\lambda\ped{p}$ needs to be sufficiently small to ensure numerical convergence as $\lambda\ped{p} \rightarrow 0$. Finally, when initialising with a point-particle, we typically apply a window function to $\vec{S}^{\rm p}$ to start the particle smoothly, and reduce the initial junk radiation.  
\section{Numerical Framework}\label{appBHPT_sec:num_frame}
The evolution equations of the master variables described above are too complex to admit analytic solutions. Consequently, we rely on numerical methods. In Chapters~\ref{chap:in_medium_supp},~\ref{chap:plasma_ringdown} and~\ref{chap:BHspec}, we employ the \emph{two-step Lax-Wendroff algorithm} with second-order finite differences, a well-suited approach for solving ``wave-like'' partial differential equations in the time domain~\cite{Krivan:1997hc,Pazos_valos_2005,Sundararajan:2007jg,Zenginoglu:2011zz,Cardoso:2021vjq,Zenginoglu:2012us,Cardoso:2022whc}. We will first provide an overview of the algorithm and the implementation in Section~\ref{appBHPT_subsec:impl} and then show a convergence test of our code in Section~\ref{appBHPT_subsec:convergence}. 
\subsection{Implementation}\label{appBHPT_subsec:impl}
The first step is to discretise the continuous evolution equations. Assuming spherical symmetry, we expand in spherical harmonics, which removes the angular dependence and reduces the problem to a single spatial dimension. Let $t$ denote the time coordinate and $x$ the radial coordinate. Under these assumptions, the homogeneous component of the equation takes the general form of a second-order partial differential equation:
\begin{equation}\label{eq:Lax_evol}
\partial^2_{t}\Psi = \left[A^{tx}\partial_t\partial_x + A^{xx}\partial_x^2+B^t\partial_t+B^{x}\partial_x+C\right]\Psi\,,
\end{equation}
where the coefficients $A^{ij}$, $B^{k}$ and $C$ have been normalised by dividing through $-A^{tt}$. These coefficients depend solely on the radial coordinate $x$, and their values can be read off from the equations of motion. This formulation can be straightforwardly generalised to systems involving two or more coupled wave equations by promoting the coefficients to matrices.
\vskip 2pt
To favour a stable numerical evolution, it proves useful to rewrite the second-order equation as a system of two coupled first-order equations~\cite{Krivan:1997hc}. This is done by introducing an auxiliary variable, often referred to as the \emph{conjugate momentum}. It is defined as
\begin{equation}
\Pi \equiv (\partial_t+b\,\partial_x)\Psi\,,\quad \text{where}\quad b \equiv -\frac{A^{tx}+\sqrt{(A^{tx})^2+4A^{xx}}}{2}\,.
\end{equation}
Including a source term $\boldsymbol{T}$, the evolution equation~\eqref{eq:Lax_evol} can then be written in a first-order matrix form as
\begin{equation}\label{eq:Lax_evol2}
\partial_{t}\boldsymbol{u} + \boldsymbol{M}\cdot\partial_x\boldsymbol{u}+\boldsymbol{A}\cdot\boldsymbol{u}= \boldsymbol{T}\,,
\end{equation}
where the solution vector is:
\begin{equation}
\boldsymbol{u} \equiv \{\Psi\ped{R}, \Psi\ped{I}, \Pi\ped{R}, \Pi\ped{I}\}\,.
\end{equation}
Here, $\mathrm{R}$ and $\mathrm{I}$ represent the real and imaginary parts of the field, respectively. The structure of the matrices $\boldsymbol{M}$ and $\boldsymbol{A}$ is:
\begin{equation}
\boldsymbol{M} = \begin{pmatrix}
b & 0 & 0 & 0\\
0 & b & 0 &0\\
m_{31} & m_{32} & -b & 0\\
-m_{32} & m_{31} & 0 & -b
\end{pmatrix}\,,\quad  
\boldsymbol{A} = \begin{pmatrix}
0 & 0 & -1 & 0\\
0 & 0 & 0 & -1\\
a_{31} & a_{32} & a_{33} & a_{34}\\
-a_{32} & a_{31} & -a_{34} & a_{33}
\end{pmatrix}\,.
\end{equation}
The coefficients entering these matrices depend on the specific evolution equations (for e.g., the Teukolsky equation, see~\cite{Krivan:1997hc,Sundararajan:2007jg}). In our case, these coefficients are generally quite simple, or even zero, as they have already been extensively manipulated [see, e.g.,~\eqref{eq:wavelike-eqn}].
\vskip 2pt
We employ a time-explicit numerical scheme based on the Lax–Wendroff finite-difference method. To apply this scheme, we first recast eq.~\eqref{eq:Lax_evol2} in the form of an advection equation:
\begin{equation}
\left(\partial_{t} + \boldsymbol{D}\cdot\partial_{x}\right)\boldsymbol{u} = \boldsymbol{S}\,, 
\end{equation}
where
\begin{equation}
\boldsymbol{D} = \text{diag}(b,b,-b,-b)\quad \text{and} \quad
\boldsymbol{S} = -(\boldsymbol{M}-\boldsymbol{D})\cdot \partial_{x}\boldsymbol{u}-\boldsymbol{A}\cdot \boldsymbol{u}+\boldsymbol{T}\,.
\end{equation}
Here, the matrices $\boldsymbol{M}$, $\boldsymbol{A}$, and $\boldsymbol{D}$ do not depend on time. This equation is then discretised on a uniform one-dimensional grid with spatial resolution $\delta x$. Each time step proceeds in two stages:~first, an intermediate solution is computed at the midpoints between grid cells
\begin{equation}
\boldsymbol{u}^{n+1/2}_{i+1/2} = \frac{1}{2}\left(\boldsymbol{u}^{n}_{i+1}+ \boldsymbol{u}^{n}_{i}\right) -\frac{\delta t}{2} \left[\frac{1}{\delta x} \boldsymbol{D}^{n}_{i+1/2} \left(\boldsymbol{u}^{n}_{i+1}-\mathbf{u}^{n}_{i}\right)-\boldsymbol{S}^{n}_{i+1/2}\right]\,,
\end{equation}
where $\delta t$ denotes the time step. Radial derivatives are approximated using second-order centred differences evaluated at grid points $i$ and $i+1$. The algebraic terms in $\boldsymbol{D}^{n}_{i+1/2}$ and $\boldsymbol{S}^{n}_{i+1/2}$ are computed by averaging the corresponding values at $i$ and $i+1$. In the second stage, the intermediate solution is used to advance the solution to the next time step:
\begin{equation}
\boldsymbol{u}^{n+1}_{i} =  \boldsymbol{u}^{n}_{i} -\delta t \left[\frac{1}{\delta x} \boldsymbol{D}^{n+1/2}_{i} \left(\boldsymbol{u}^{n+1/2}_{i+1/2}-\boldsymbol{u}^{n+1/2}_{i-1/2}\right)-\boldsymbol{S}^{n+1/2}_{i}\right]\,.
\end{equation}
Here, the centred radial differences and averages are taken on the values $\boldsymbol{u}^{n+1/2}_{i+1/2}$ and $\boldsymbol{u}^{n+1/2}_{i-1/2}$.
\vskip 2pt
Simulations in Schwarzschild coordinates $(t,r,\theta,\varphi)$ present numerically difficulties because of the coordinate singularity at the BH horizon. To get around this, we use the tortoise coordinate $r_*$ (defining $x = r_*$), which effectively sends the horizon to $r_* \rightarrow -\infty$. The computational domain is then extended far enough outwards, such that any artificial effect from the boundary cannot influence the dynamics at the radius of extraction. This allows us to make an arbitrary choice for the boundary conditions on $\Psi$ and $\Pi$, which we take to be zero. 
\vskip 2pt
The numerical grid is uniformly spaced in the tortoise coordinate, and we typically choose $\lambda\ped{p} \approx 4 \delta x$~\eqref{eq:pp_num}. To avoid divergences near the horizon, we excise any matter distribution in that region and verify that the qualitative behaviour of the solution remains unaffected.
\subsection{Convergence Test}\label{appBHPT_subsec:convergence}
\begin{figure}[t!]
    \centering
    \includegraphics[scale=1]{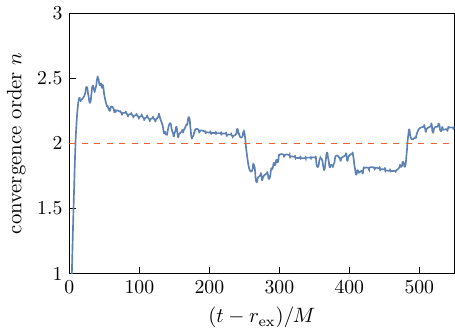}
    \caption{Convergence order as measured with $\mathrm{d}x/M =  \{0.1, 0.05, 0.025\}$. The result is consistent with the expected second-order convergence rate.}
    \label{fig:convergence}
\end{figure}

To validate the implementation of our numerical scheme, we perform a convergence test. In Figure~\ref{fig:convergence}, we show the Cauchy convergence order, defined as 
\begin{equation}
n = \log_{2}\left(\frac{||\Psi_{2h} - \Psi_{h}||_{2}}{||\Psi_{h}-\Psi_{h/2}||_{2}}\right)\,,
\end{equation}
where a moving average is applied to smooth the resulting curve. The observed convergence rate is consistent with the second-order accuracy of the finite-difference scheme employed. Most simulations are performed using the resolution $\Psi_{h}$, corresponding to a spatial grid spacing of $\mathrm{d}x = 0.05M$. The time step is set to $\mathrm{d}t = 0.5 \mathrm{d}x$, ensuring the Courant–Friedrichs–Lewy condition is satisfied throughout.
\chapter{Axionic Instabilities in Flat Spacetime}\label{app:flatinstab}
\renewcommand{\theequation}{\thechapter.\arabic{equation}}
\numberwithin{equation}{chapter} 
Background electromagnetic fields in electro-vacuum charged BHs can give rise to axionic instabilities. However, the presence of plasma may significantly modify this behaviour, as discussed in Chapter~\ref{chap:in_medium_supp}. To support that discussion, this appendix presents a simplified analysis in flat spacetime.
\vskip 2pt
Consider eqs.~\eqref{eq:IMS_evoleqns} with $\sin\chi_0=0$ and a background ($\mathrm{b}$) system with a constant electric field along the $\hat{z}$--direction:
\begin{equation}\label{eq:constant_electric}
    A^{\rm b}_\mu=\left(z E_z,0,0,0\right)\,.
\end{equation}
Moreover, we assume the presence of ions, such that the plasma is globally neutral. Under the influence of the electric field~\eqref{eq:constant_electric}, electrons are subject to a constant acceleration. From the momentum equation~\eqref{eq:momentum_continuity}, one can easily infer the four velocity, which reads
\begin{equation}\label{eq:four_vel_fl}
    v^{\rm b}_\mu=\left(\frac{\sqrt{m\ped{e}^2+e^2 E_z^2 t^2}}{m\ped{e}},0,0,\frac{e E_z t}{m\ped{e}}\right) \,,
\end{equation}
such that $v^{\rm b}_\mu v^{\mathrm{b},\mu}=-1$. Likewise, ions feel a force in the opposite direction along the $\hat{z}$--axis. From the continuity equation~\eqref{eq:momentum_continuity}, one can then infer the density profile of the fluid:
\begin{equation}\label{eq:density_profile}
    n^{\rm b}=\frac{n^{(0)}}{\sqrt{m\ped{e}^2+ e^2 E_z^2 t^2}}\,,
\end{equation}
where $n^{(0)}$ is an integration constant. Given the time dependence of the four-velocity~\eqref{eq:four_vel_fl}, the system is not stationary. To obtain a full solution, one should thus perform a consistent evolution in the time domain. However, by assuming that the electric field is weak with respect to the particle's inertia, i.e., $e E_z\ll m\ped{e}$, we can approximate the plasma as static on timescales $t\approx \mathcal{O}(m\ped{e}/e E_z)$ and use standard frequency-domain methods. Moreover, in this approximation, the density is stationary and constant in space~\eqref{eq:density_profile}. Finally, as the ion's inertia is much larger than the electron's one, the same considerations hold.
\vskip 2pt
We now perturb the fields in the frequency domain as
\begin{equation}
\begin{aligned}
     \Psi & \sim \epsilon \bar{\psi} e^{-i (\omega t - k_i x^i)}\,, \\
    A_\mu & \sim A^{\rm b}_\mu+ \epsilon \bar{A}_\mu  e^{-i (\omega t - k_i x^i)}\,, \\ 
    v_\mu &\sim v^{\rm b}_\mu+ \epsilon \bar{v}_\mu  e^{-i (\omega t - k_i x^i)}\,, \\
    n\ped{e} &\sim  n^{\rm b} + \epsilon \bar{n} e^{-i (\omega t - k_i x^i)} \,,   
\end{aligned}
\end{equation}
where $k_i$ is the wave vector in Fourier space and perturbed variables are marked with an overhead bar. Jointly solving the Maxwell and momentum equations in terms of $\bar{\psi}$ and $\bar{A}_z$ then yields
\begin{equation}\label{eq:flateqs}
\begin{aligned}
    \bar{A}_0 &= -\frac{\bar{A}_z \omega}{k_z}\,,\\
    \bar{A}_x &= \frac{k_x \bar{A}_z}{k_z}+\frac{2 i E_z k_y k\ped{a} \bar{\psi}}{\omega\ped{p}^2+ k^2-\omega^2}\,, \\ 
    \bar{A}_y &= \frac{k_y \bar{A}_z}{k_z}-\frac{2 i E_z k_x k\ped{a} \bar{\psi}}{\omega\ped{p}^2+ k^2-\omega^2}\,, \\
    \bar{v}_0 &= 0\,,\\
    \bar{v}_x &= -\frac{2 i E_z k_y e k\ped{a} \bar{\psi}}{m\ped{e}(\omega\ped{p}^2+ k^2-\omega^2)}\,, \\
    \bar{v}_y &= \frac{2 i E_z k_x e k\ped{a} \bar{\psi}}{m\ped{e}(\omega\ped{p}^2+ k^2-\omega^2)}\,, \\
    \bar{v}_z &= 0\,,    
\end{aligned}
\end{equation}
\begin{figure}[t!]
    \centering
    \includegraphics[scale=1]{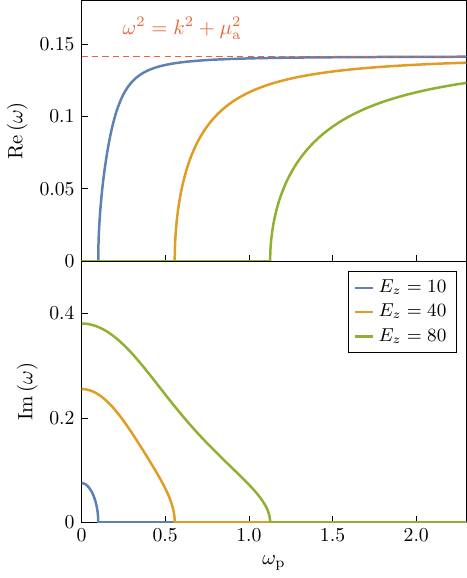}
    \caption{Spectrum of the axionic modes as a function of the plasma frequency for three different choices of the electric field strength. Increasing the plasma frequency stabilises the system, as modes turn from imaginary to real, and the frequency approaches the axion vacuum dispersion relation. The parameters in this plot are $k\ped{a}=0.01$, $k^2=0.1$ and $\mu\ped{a}=0.1$ in arbitrary units.}
    \label{fig:flatmodes}
\end{figure}
where $k^2=k_i k^i$ and we defined the plasma frequency of the background plasma $\omega\ped{p}^2=e^2 n^{\rm b}/m\ped{e}$. Since $k_\mu \bar{v}^\mu=0$, the electromagnetic perturbation is purely transverse. As a result, the continuity equation implies that density perturbations, which are longitudinal quantities, vanish identically, i.e., $\bar{n}=0$. 
\vskip 2pt
Finally, we plug these relations back in the Klein-Gordon equation. As the latter is sourced by the combination $k_y \bar{A}_x-k_x \bar{A}_y$, it follows from eq.~\eqref{eq:flateqs} that $\bar{A}_z$ vanishes, and we are left with a decoupled expression for $\bar{\psi}$, which reads
\begin{equation}
 \left(-\omega^2+k^2+\mu\ped{a}^2-\frac{4 E_z^2 (k_x^2+k_y^2)k\ped{a}^2}{-\omega^2+k^2+\omega\ped{p}^2} \right)\bar{\psi}=0\,.
\end{equation}
The dispersion relation is then found as
\begin{equation}
    \omega^2=\frac{1}{2}\left(2k^2+\mu\ped{a}^2+\omega\ped{p}^2 \pm \sqrt{(\mu\ped{a}^2-\omega\ped{p}^2)^2+16 E_z^2 k^2 k\ped{a}^2}\right)\,.
\end{equation}
Clearly, the electric field has a critical threshold above which an instability arises. This threshold is given by
\begin{equation}
    E_z^{\rm crit}=\frac{\sqrt{k^2+\mu\ped{a}^2}\sqrt{k^2+\omega\ped{p}^2}}{2 k\ped{a} k}\,.
\end{equation}
When $E_z>E_z^{\rm crit}$, the system admits purely imaginary, unstable modes. For $\omega\ped{p} \rightarrow 0$, one correctly obtains the vacuum threshold as found in~\cite{Boskovic:2018lkj}. Importantly, in the limit $\omega\ped{p} \gg \mu\ped{a}$, the threshold increases, as the axion-photon conversion angle drops~\eqref{eq:angle-axion-photon}. 
\vskip 2pt
Figure~\ref{fig:flatmodes} shows the real and imaginary part of the axionic frequency as a function of the plasma frequency for different values of the electric field. In all three cases, the electric field is above the threshold value for low plasma frequencies, and the modes are purely imaginary. By increasing the plasma frequency, the system stabilises as the modes become purely real while they approach the vacuum dispersion relation of the axion. This is exactly the impact of \emph{in-medium suppression}, as in the limit $\omega\ped{p} \gg \mu\ped{a}$ the axion effectively decouples from the photon. For small electric fields [{\color{Mathematica1}{blue}}], this happens precisely at the threshold $\omega\ped{p}=\mu\ped{a}$.
\renewcommand{\theequation}{\thechapter.\thesection.\arabic{equation}}
\numberwithin{equation}{section}
\chapter{Dark Photon Basis}\label{app:DPbasis}
In Chapter~\ref{chap:in_medium_supp}, I studied the dark photon in the \emph{interaction basis}. Depending on the context, however, other choices -- such as the \emph{mass basis} -- can be more convenient. This appendix is devoted to that purpose. In Section~\ref{appDPbasis:basis}, I derive the relevant equations of motion in the mass basis, while in Section~\ref{appDPbasis:BHPT}, I discuss how to apply black hole perturbation theory in that basis.
\section{Basis Choices}\label{appDPbasis:basis}
Kinetic mixing between the Standard Model and dark photon is described by the Lagrangian (similar to Chapter~\ref{chap:in_medium_supp}, dark photon quantities are denoted with a prime): 
\begin{equation}
    \mathcal{L}=\mathcal{L}\ped{EM}+\mathcal{L}\ped{{Proca}}+\frac{1}{2}\sin \chi_0 F'_{\mu\nu}F^{\mu\nu}\,.
\end{equation}
To make the equations more tractable, the mixing term can be removed through a field redefinition. Two different choices are possible. Redefining $A'_\mu \rightarrow A'_\mu+\sin\chi_0 A_\mu$ leads to the \emph{interaction basis}, in which the Lagrangian takes the form:
\begin{equation}\label{eq:lagrangian_interaction}
\mathcal{L}\ped{inter}=-\frac{1}{4}\left(F_{\mu\nu}F^{\mu\nu}+F'_{\mu\nu}F'^{ \mu\nu}\right)-\frac{\mu_{\gamma '}^2}{2} A'^\mu A'_\mu - \mu_{\gamma'}^2\sin\chi_0 A'_\mu A^\mu+ j^\mu A_\mu \,.
\end{equation}
In this basis, the fields are directly coupled, as clearly shown in~\eqref{eq:IMS_evoleqns}, while only the visible field $A_\mu$ couples with the Standard Model current $j_\mu$. This is the basis we use in Chapter~\ref{chap:in_medium_supp}.
\vskip 2pt
An alternative choice is $A_\mu \rightarrow A_\mu+\sin\chi_0 A'_\mu$, which removes the kinetic mixing is and leads to the \emph{mass basis} with a Lagrangian:
\begin{equation}\label{eq:lagrangian_mass}
\mathcal{L}\ped{mass}=-\frac{1}{4}\left(F_{\mu\nu}F^{\mu\nu}+F'_{\mu\nu}F'^{ \mu\nu}\right)-\frac{\mu_{\gamma '}^2}{2} A'^\mu A'_\mu + j^\mu (A_\mu+ \sin \chi_0 A'_\mu)\,.
\end{equation}
The corresponding field equations are given by
\begin{equation}\label{eq:masseqs}
\begin{aligned}
    \nabla_\nu F^{\mu \nu} &=j^\mu\,, \\
    \nabla_\nu F'^{\mu \nu} &=-\mu_{\gamma'}^2 A'^\mu+ \sin \chi_0 j^\mu\,,
\end{aligned}
\end{equation}
while the momentum equation in the Einstein cluster setup reads:
\begin{equation}
    \nabla^\nu T^{\rm p}_{\mu\nu}=e n\ped{e} (F_{\mu\nu}+ \sin \chi_0 F'_{\mu\nu})\,.
\end{equation}
In this case, the two fields are not directly coupled, yet both of them are coupled to the electrons. Note that, in this basis $A^\mu$ is not the visible photon, i.e., the one accelerating charged particles. Instead, that role is played by the combination $A\ped{{obs}}=A_\mu+\sin\chi_0 A'_\mu$. 
\vskip 2pt
As it should, the physics in the two bases is equivalent, and one can simply choose a preferred basis depending on the problem at hand. 
\section{Black Hole Perturbation Theory in the Mass Basis}\label{appDPbasis:BHPT}
For completeness, we briefly outline the computations presented in Chapter~\ref{chap:in_medium_supp}, now performed in the mass basis. This also facilitates the comparison with previous work~\cite{Caputo:2021efm}. By performing the multipolar decomposition of the momentum equation, we obtain the following relation between the axial four-velocity and the electromagnetic and dark photon axial fields:
\begin{equation}\label{eq:DP_massbasis}
v_{4}=-\frac{e}{m\ped{e}}(u^{\rm mb}_{4}+ \sin \chi_0 u_4^{\rm mb} {}')\,,
\end{equation}
where the superscript ``$\mathrm{mb}$'' refers to mass basis quantities. From eq.~\eqref{eq:DP_massbasis}, it is immediately clear that, in this basis, the dark photon field affects the motion of charged particles. Expanding Eq~\eqref{eq:masseqs} yields a set of coupled partial differential equations:
\begin{equation}
\label{eq:wavelike-eqn_DP_mass}
\begin{aligned}
\hat{\mathcal{L}}u^{\rm mb}_{4}&=f\left(\omega\ped{p}^2 + \frac{\lambda}{r^{2}}\right) u^{\rm mb}_{4}+f \omega\ped{p}^2 \sin{\chi_0}\,u^{\rm mb}_{4} {}'\,,
 \\
\hat{\mathcal{L}}u^{\rm mb}_{4} {}'&=f\left(\mu_{\gamma'}^2 + \frac{\lambda}{r^{2}}\right) u^{\rm mb}_{4} {}'+f \omega\ped{p}^2 \sin{\chi_0}\,u^{\rm mb}_4\,.
\end{aligned}
\end{equation}
The visible photon is then given by $u_4=u^{\rm mb}_4+ \sin \chi_0 u^{\rm mb}_4 {}'$. Thus, one can perform the same computations as in Chapter~\ref{chap:in_medium_supp} in the mass basis, comparing the results through this linear combination.
\chapter{Resonances in Gravitational Atoms}\label{app:GA}
This appendix presents additional results related to the study of resonances in boson clouds from Chapter~\ref{chap:legacy}. Section~\ref{appGA_sec:hyperfine-angular-momentum} discusses the assumption of total angular momentum conservation, while Section~\ref{appGA_sec:breaKING} provides a more general treatment of resonance breaking. In Section~\ref{appGA_sec:ion-at-resonance}, I show that the contribution from ionisation at resonance frequencies can be safely neglected. Section~\ref{appGA_sec:211_200} examines a special resonance mediated solely by the dipole of the gravitational perturbation. Finally, Section~\ref{appGA_sec:variables} summarises the variables used throughout Chapter~\ref{chap:legacy}.
\section{Hyperfine Resonances and Angular Momentum}\label{appGA_sec:hyperfine-angular-momentum}
A nonzero BH spin is responsible for the existence of the hyperfine energy splitting~\eqref{eq:BHenv_eigenenergy}, as it breaks the spherical symmetry of the background spacetime. At the same time, we study the backreaction of resonances (hyperfine or not) on the orbit in the Newtonian approximation, assuming the conservation of the total angular momentum, which leads to eqs.~\eqref{eqn:dimensionless-omega-evolution},~\eqref{eqn:dimensionless-eccentricity-evolution}, and~\eqref{eqn:dimensionless-inclination-evolution}. This methodology might appear as fundamentally inconsistent, so let us inspect it more closely.
\vskip 2pt
The weak-field approximation of the Kerr metric, which is valid at large distances, reads
\begin{equation}
ds^2=-\left(1-\frac{2M}r\right)\dd t^2+\left(1+\frac{2M}r\right)\dd r^2+r^2(\dd\theta^2+\sin^2\theta\dd\phi^2)-\tilde aM\frac{4M}r\sin^2\theta\dd t\dd\phi\,.
\end{equation}
The last term is known to give rise to the \emph{Lense-Thirring precession}, as the equation of motion of a scalar particle can be put in the form
\begin{equation}
\frac{\dd^2\vec r}{\dd t^2}=-\frac{M}{r^3}\vec r+4\frac{\dd\vec r}{\dd t}\times\vec B\,,
\end{equation}
where the \emph{gravito-magnetic field} $\vec B$ is related to the BH spin as
\begin{equation}
\vec B=\vec\nabla\times\vec A\,,\qquad\vec A=-\frac{\vec J\times\vec r}{2r^3}\,,\qquad\vec J=\tilde aM^2\hat z\,.
\end{equation}
The corresponding Hamiltonian is
\begin{equation}
H=\frac{(\vec p-4\mu\vec A)^2}{2\mu}-\frac{\mu M}r\approx\frac{\vec p^2}{2\mu}-\frac\alpha{r}+\frac{2\tilde aM^2}{r^3}L_z\,,
\label{eqn:hamiltonian-gravitomagnetism}
\end{equation}
where $\mu=\alpha/M$ is the mass of the particle. We can immediately check that the last term in~\eqref{eqn:hamiltonian-gravitomagnetism} gives rise to the expected hyperfine splitting,
\begin{equation}
\braket{n\ell m|H|n\ell m}=2\tilde aM^2m\Braket{n\ell m|\frac1{r^3}|n\ell m}=2\tilde aM^2m\,\frac{(\mu \alpha)^3}{n^3\ell(\ell+1/2)(\ell+1)}\,,
\end{equation}
which perfectly matches the last term in~\eqref{eq:BHenv_eigenenergy}.
\vskip 2pt
The orbital angular momentum $\vec L=\vec r\times\vec p$ evolves as
\begin{equation}
\frac{\dd \vec L}{\dd t}=i[H,\vec L]=\frac{2}{r^3}\vec J\times\vec L\,,
\end{equation}
which is the expected Lense-Thirring precession. Applying this equation to the cloud-binary system gives rise to two additional terms on the right-hand sides of~\eqref{eqn:Lz-balance} and~\eqref{eqn:Lx-balance}, corresponding to the Lense-Thirring precession of the cloud (which vanishes in most cases, as $\vec S\ped{c}\parallel\vec J$ even during a transition, as we will see below) and of the binary. This precession is, however, \emph{parametrically small}. None of the other terms in~\eqref{eqn:Lz-balance} and~\eqref{eqn:Lx-balance} depend on the BH spin $\tilde a$, even in the case of hyperfine resonances, where the energy splitting is proportional to $\tilde a$. Not only for realistic parameters is this precession extremely slow, but it also does not disrupt the approach in Chapter~\ref{chap:legacy}, as~\eqref{eqn:Lx-balance} can be simply replaced by the analogous equation for the (precessing) equatorial projection of the angular momentum.
\vskip 2pt
Having justified the use of the conservation of total angular momentum, there is another potentially worrying aspect of the breaking of spherical symmetry, that has to do with the spin of the cloud when it is in a mixed state, for example during a transition. As long as the Hamiltonian is spherically symmetric, $\ket{n\ell m}$ are guaranteed to be eigenstates of the scalar field's orbital angular momentum $\vec L$. Its matrix elements are given by $L_z\ket{n\ell m}=m\ket{n\ell m}$ and, in the Condon-Shortley convention, $L_\pm\ket{n\ell m}=\sqrt{\ell(\ell+1)-m(m\pm1)}\ket{n\ell,m\pm1}$, where $L_\pm=L_x\pm iL_y$. If the cloud is in a mixed state of the form $\ket{\psi}=c_a\ket{n_a\ell_am_a}+c_b\ket{n_b\ell_bm_b}$, then the $z$--component of its angular momentum is $m_a\abs{c_a}^2+m_b\abs{c_b}^2$, while the equatorial components vanish unless $\ell_a=\ell_b$ and $\abs{m_a-m_b}=1$.
\vskip 2pt
Remarkably, all the previous results still hold for the Hamiltonian~\eqref{eqn:hamiltonian-gravitomagnetism}. That is because the perturbation $\sim L_z/r^3$ is diagonal on the basis $\ket{\ell m}$, only mixing states with different $n$. Even though the spacetime is not spherically symmetric, the angular structure of the eigenstates is unchanged. The equations in Chapter~\ref{chap:legacy} then do not need any modification, except for the case of hyperfine transitions with $\abs{\Delta m}=1$. For a hyperfine transition with $m_b=m_a-1$, careful computation (in the Schr\"{o}dinger, not dressed, frame) of the equatorial components of $\vec S\ped{c}$ shows that eq.~\eqref{eqn:Lx-balance} would need to be corrected with a term
\begin{equation}
\frac{\dd S_{\text{c},x}}{\dd\tau}\sim\frac{B}{g}\sqrt{\ell(\ell+1)-m_a(m_a-1)}\sqrt{Z}(\abs{c_b}^2-\abs{c_a}^2)\sin(C\tau/3)\,.
\end{equation}
This is a fast oscillating term that averages to zero on timescales much shorter than the evolution of the orbital parameters and the duration of the resonance. We can thus safely ignore it in Chapter~\ref{chap:legacy}.
\section{General Resonance Breaking}\label{appGA_sec:breaKING}
The phenomenon of resonance breaking was discussed in Section~\ref{sec:resonance-breaking} in the simplified scenario where only one of the following quantities is allowed to vary at a time:~the eccentricity $\varepsilon$, the Landau-Zener parameter $Z$, or the cloud's mass $M\ped{c}$. We derive here the result in the general case. Taking the time derivative of~\eqref{eqn:dimensionless-omega-evolution-Gamma}, we find
\begin{equation}
\begin{split}
\frac{\dd^2\omega}{\dd\tau^2}&=\frac{\dd f(\varepsilon)}{\dd\tau}+B\frac{\dd^2\abs{c_a}^2}{\dd\tau^2}=\frac{\dd f(\varepsilon)}{\dd\tau}+B\biggl(\frac{\dd^2c_a^*}{\dd\tau^2}c_a+2\frac{\dd c_a^*}{\dd\tau}\frac{\dd c_a}{\dd\tau}+c_a^*\frac{\dd^2c_a}{\dd\tau^2}\biggr)\\
&=\frac{\dd f(\varepsilon)}{\dd\tau}-2ZB(\abs{c_a}^2-\abs{c_b}^2)+\biggl(\frac{1}{2Z}\frac{\dd Z}{\dd\tau}-\Gamma\biggr)\biggl(\frac{\dd\omega}{\dd\tau}-f(\varepsilon)\biggr)\\&+\omega\sqrt ZB(c_a^*c_b+c_ac_b^*)\,,
\end{split}
\label{eqn:breaKING-harmonic-oscillator}
\end{equation}
where the second and third line are obtained by repeated use of the Schr\"{o}dinger equation~\eqref{eqn:schrodinger-Gamma} together with~\eqref{eqn:dimensionless-omega-evolution-Gamma}. Under the assumption that all coefficients appearing above evolve slowly during a floating orbit, eq.~\eqref{eqn:breaKING-harmonic-oscillator} has the structure of a damped harmonic oscillator, with solution
\begin{equation}
\omega=\frac{\frac{\dd f(\varepsilon)}{\dd\tau}-\frac{f(\varepsilon)}{2Z}\frac{\dd Z}{\dd\tau}+\Gamma-2ZB(\abs{c_a}^2-\abs{c_b}^2)}{-\sqrt{Z}B(c_a^*c_b+c_ac_b^*)}+\text{damped oscillatory terms}\,.
\label{eqn:breaKING-harmonic-oscillator-solution}
\end{equation}
The resonance breaks whenever $c_b^*c_a+c_a^*c_b=0$. By direct application of the Schr\"{o}dinger equation, we find
\begin{equation}
\sqrt Z\frac\dd{\dd\tau}(c_a^*c_b+c_ac_b^*)=-\omega\frac{\dd\abs{c_a}^2}{\dd\tau}-\Gamma\sqrt Z(c_a^*c_b+c_b^*c_a)\,.
\label{eqn:breaKING-dcacbcbcadt}
\end{equation}
Substituting the non-oscillatory term of~\eqref{eqn:breaKING-harmonic-oscillator-solution} in eq.~\eqref{eqn:breaKING-dcacbcbcadt}, we arrive at an equation for the sole unknown $c_a^*c_b+c_b^*c_a$:
\begin{equation}
\begin{split}
&\frac{ZB}2\left(\!\frac\dd{\dd\tau}+2\Gamma\!\right)\!(c_a^*c_b+c_b^*c_a)^2\\&=\left(\frac{\dd f(\varepsilon)}{\dd\tau}-\frac{f(\varepsilon)}{2Z}\frac{\dd Z}{\dd\tau}+f(\varepsilon)\Gamma-2ZB(\abs{c_a}^2-\abs{c_b}^2)\!\right)\frac{\dd\abs{c_a}^2}{\dd\tau}\,.
\end{split}
\label{eqn:breaKING-master}
\end{equation}
Remarkably, the evolution of the eccentricity, the variation of the Landau-Zener parameter and the decay of the cloud contribute additively to~\eqref{eqn:breaKING-master}, each with its own term. In realistic cases, $\Gamma$ is large enough to force the population of state $\ket{b}$ to reach a saturation value $\abs{c_b}^2=f(\varepsilon)/(2\Gamma B)$, which is usually small enough to be neglected in~\eqref{eqn:breaKING-master}. The point of resonance breaking, then, only involves the population left in the initial state, $\abs{c_a}^2$.
\section{Ionisation at Resonance}\label{appGA_sec:ion-at-resonance}
The expressions for the ionisation rate and power derived in~\cite{Baumann:2021fkf} are valid under the assumption that the frequency $\Omega$ of the perturbation is away from any bound-to-bound state resonance. Here, we relax this assumption by computing the new term contributing at resonance and showing that its effect is ultimately negligible. To allow for an easy match with the notations of~\cite{Baumann:2021fkf}, we denote the state initially populated by $\ket{b}$ and any other bound state by $\ket{a}$.
\vskip 2pt
Ignoring couplings between different continuum states (as justified in Appendix~A2 of~\cite{Baumann:2021fkf}), the Hamiltonian of the gravitational atom reads,
\begin{equation}
\mathcal{H}=\sum_{b} \epsilon_{b}\ket{b}\!\bra{b} + \sum_{a \neq b} \eta_{ab}(t) \ket{a}\!\bra{b} + \sum_{K} \epsilon_{K}\ket{K}\!\bra{K}+\sum_{K, b}\left[\eta_{Kb}(t)\ket{K}\!\bra{b}+\text{h.c.}\right]\,,
\end{equation}
where $\ket{K}\equiv\ket{k\ell m}$ is a continuum state multi-index, $\epsilon_K\approx k^2/(2\mu)$, and the couplings $\eta_{ab}(t)$ and $\eta_{Kb}(t)$ are the matrix elements of the perturbation~\eqref{eqn:V_star}. As in~\cite{Baumann:2021fkf}, by integrating out the continuum the Schr\"{o}dinger equation can be recast in the following form:\footnote{All quantities here depend on time, either through fast oscillatory terms, due to the evolving phase $\varphi_*$, or through the slow frequency chirp. For ease of notation, we will only explicitly write the time dependence of terms falling in the first class.}
\begin{equation}
\label{eq:IonResoccupation}
i \frac{\dd c_{b}}{\dd t}=\mathcal{E}_{b}c_{b}(t)+\sum_{a \neq b}\left[\eta_{ba}(t) e^{i(\epsilon_b-\epsilon_a) t}+\mathcal{E}_{ba}(t)\right] c_{a}(t)\,,
\end{equation}
where we define the \emph{induced couplings},
\begin{equation}
\label{eq:inducedcouplingdef}
\mathcal{E}_{ba}(t)\equiv-i \int_{-\infty}^{t} \dd t^{\prime} \sum_{K} \eta^{*}_{K b}(t) \eta_{K a}(t^{\prime}) e^{-i(\epsilon_K-\epsilon_b)t+i(\epsilon_K-\epsilon_a)t'}\,,
\end{equation}
and the \emph{induced energies} $\mathcal E_b\equiv\mathcal E_{bb}$. The first term in~\eqref{eq:IonResoccupation} controls the ionisation of state $\ket{b}$, while the first term in the parenthesis is responsible for the $\ket{b}\to\ket{a}$ resonance. The last term, which is the focus here, is a coupling between $\ket{b}$ and $\ket{a}$ induced via the interaction with the continuum. Because $\mathcal{E}_{ba}(t)$ oscillates very rapidly unless $(m_b-m_a)\dot\varphi_*=\epsilon_b-\epsilon_a$, the parenthesis in~\eqref{eq:IonResoccupation} can be neglected altogether whenever the system is not actively on resonance.
\vskip 2pt
Let us study what happens when this is the case instead. The same saddle-point approximation done in~\cite{Baumann:2021fkf} can be applied to the case $a\ne b$, arriving to
\begin{equation}
\mathcal E_{ba}(t)=e^{i(\epsilon_b-\epsilon_a)t-i(m_b-m_a)\varphi_*(t)}\sum_{\ell m}\left[-\frac{i\mu\,\eta_{Kb}^{*\floq{g_b}}\eta_{Ka}^\floq{g_a}}{2k_*^\floq{g_a}}\,\Theta\left((k_*^\floq{g_a})^2\right)\right]\,.
\end{equation}
Here, we defined $g_a=m-m_a$, evaluated $\ket{K}$ at $k_*^\floq{g_a}=\sqrt{2\mu((m-m_a)\dot\varphi_*+\varepsilon_a)}$, and expanded the bound-continuum coupling in its Floquet components, $\eta_{Ka}=\eta_{Ka}^\floq{g_a}e^{i(m-m_a)t}$ (and similarly for $a\leftrightarrow b$). To understand the effect of the induced coupling $\mathcal E_{ba}$, we can temporarily set $\eta_{ba}=0$ and write~\eqref{eq:IonResoccupation} as
\begin{equation}
\frac{\dd\abs{c_b}^2}{\dd t}=\sum_{a}\sum_{\ell m}\frac{\mu}{k_*^\floq{g_a}}\Theta\left((k_*^\floq{g_a})^2\right)\Re\left[e^{i(\epsilon_b-\epsilon_a)t-i(m_b-m_a)\varphi_*(t)}\eta_{Kb}^{*\floq{g_b}}\eta_{Ka}^\floq{g_a}c_b^*(t)c_a(t)\right]\,.
\label{eqn:ionisation+mixedcoupling}
\end{equation}
Here, the term with $a=b$ reproduces the ionisation term $\mathcal{E}_{b}c_{b}(t)$ in~\eqref{eq:IonResoccupation}. Moreover, the evolution of state $\ket{a}$ is determined by the same formula, swapping $b\leftrightarrow a$. For $a\ne b$, however, this operation transforms the term in brackets into its complex conjugate, so its real part stays unchanged. We thus see that the induced coupling $\mathcal E_{ba}$ does \emph{not} contribute to a $\ket{b}\to\ket{a}$ transition alongside $\eta_{ba}$, as one might have expected from~\eqref{eq:IonResoccupation} and as was speculated in~\cite{Baumann:2021fkf}. Instead, both $\abs{c_b}^2$ and $\abs{c_a}^2$ experience an \emph{identical} depletion (in addition to ionisation) or recombination, depending on the sign of the real part appearing in~\eqref{eqn:ionisation+mixedcoupling};~both cases are possible.
\vskip 2pt
We have validated the previous results by comparing them to an explicit numerical integration of the Schr\"{o}dinger equation, with the continuum states modelled as a large set of discrete states, quadratically spaced in energy. By tuning the parameters to make the impact of the induced coupling clearly visible, we found that~\eqref{eqn:ionisation+mixedcoupling} gives, indeed, a very accurate description of the evolution of the populations around the resonance. In Chapter~\ref{chap:legacy}, in particular for Bohr resonances, we are mainly concerned with the correction from the induced coupling to a naive approach where the contributions of ionisation and the resonance are simply summed up. To determine its importance, we assume for simplicity that $\eta_{ab}=0$, $\abs{c_b}^2=1$ and $\abs{c_a}^2=0$ at $t=-\infty$ and employ a (further) saddle-point approximation in~\eqref{eq:IonResoccupation} around the time $t_0$ such that $\dot\varphi_*=\Omega_0=(\epsilon_b-\epsilon_a)/(m_b-m_a)$. The population at $t=+\infty$ is then
\begin{equation}
\abs{c_a}^2=\frac{2\pi}{\abs{m_b-m_a}\gamma}\,\abs*{\sum_{\ell m}\frac{\mu\,\eta_{Kb}^{*\floq{g_b}}\eta_{Ka}^\floq{g_a}}{2k_*^\floq{g_b}}\Theta\left((k_*^\floq{g_b})^2\right)}^2\,,
\label{eqn:final-saddle-point}
\end{equation}
where the couplings and $k_*^\floq{g_b}$ have to be evaluated at $\Omega=\Omega_0$. Similar to an argument already developed in~\cite{Baumann:2021fkf}, this quantity $\abs{c_a}^2$ is $\mathcal O(q^3\alpha^4)$, and it has to compete with the $\eta^2/\gamma\sim\mathcal O(q\alpha^2)$ contributions due to the direct coupling $\abs{\eta_{ba}}^2/\gamma$. Once again, we have validated~\eqref{eqn:final-saddle-point} by comparing it to a direct numerical integration of the Schr\"{o}dinger equation and evaluated it for a typical Bohr resonance, finding a final population of $\mathcal O(10^{-11})$. We conclude that simply adding the steady deoccupation introduced by ionisation on top of the resonant transition studied in Chapter~\ref{chap:legacy} is a good approximation for our purposes.
\section{$\ket{211}\to\ket{200}$ Resonance}\label{appGA_sec:211_200}
The strength of the fine resonance $\ket{211}\to\ket{200}$ has anomalous scaling with parameters, due to the dipole $\ell_*=1$ being entirely responsible for the coupling and the binary separation falling partially inside the region where ionisation dominates over GW emission. We therefore determine the angle $\delta_2$, such that for $\pi-\delta_2<\beta\le\pi$ the resonance is non-adiabatic, as
\begin{equation}
\delta_2=\SI{6.7}{\degree}\,\left(\frac{10^{-2}}{M\ped{c}/M}\right)^{1/4}\left(\frac{q}{10^{-3}}\right)^{1/8} \left(\frac{\alpha}{0.2}\right)^{7/8}f(\varepsilon_0)^{3/8}F(\alpha,M\ped{c})\,,
\label{eqn:211-delta2}
\end{equation}
where the formula holds for small $\delta_2$, and the function $F(\alpha,M\ped{c})$ is calculated numerically and shown in Figure~\ref{fig:211_200_F(alpha,Mc)}.
\begin{figure} 
\centering
\includegraphics[scale = 0.9]{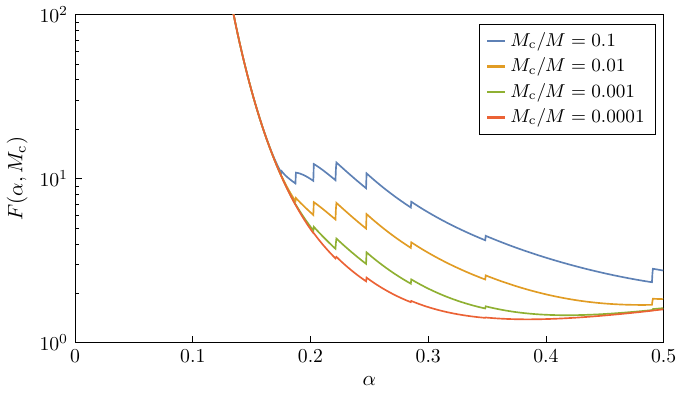}
\caption{Function $F(\alpha,M\ped{c})$ appearing in eq.~\eqref{eqn:211-delta2}, which defines the angular interval $\delta_2$ around a counter-rotating orbit where the resonance $\ket{211}\to\ket{200}$ is not adiabatic.}
\label{fig:211_200_F(alpha,Mc)}
\end{figure}
\section{Summary of Resonance Variables}\label{appGA_sec:variables}
\begin{center}
\begin{tabular}{clc}
\toprule
\textbf{Symbol} & \textbf{Meaning} & \textbf{Reference} \\
\midrule
$\varepsilon$ & Binary eccentricity & \\
$\beta$ & Binary inclination & \\
$g$ & Overtone number &~\eqref{eqn:eta-circular} \\
$\gamma$ & Frequency chirp rate induced by GWs/ionisation &~\eqref{eqn:gamma_gws} \\
$\tau$ & Dimensionless time &~\eqref{eqn:coefficients} \\
$\omega$ & Dimensionless frequency &~\eqref{eqn:coefficients} \\
$Z$ & Landau-Zener parameter &~\eqref{eqn:coefficients} \\
$B$ & Backreaction of a resonance &~\eqref{eqn:BC} \\
$C$ & Inertia of $\varepsilon$ and $\beta$ w.r.t.~resonance backreaction &~\eqref{eqn:BC} \\
$D$ & Distance parameter, $D=B/C$ &~\eqref{eqn:D} \\
$\Gamma$ & Dimensionless decay width of the final state &~\eqref{eqn:schrodinger-Gamma}\\
\bottomrule
\end{tabular}
\end{center}
\chapter{Perturbation Theory with Environments}\label{app:BHPT_Kerr}
In Chapter~\ref{chap:inspirals_selfforce}, the perturbation of a secondary BH on an environment in the Kerr geometry was studied. In this appendix, I report on some of the technical details related to the gauge choice (Section~\ref{app_BHPT_Kerr:LG}), the scalar fluxes in the Newtonian regime (Section~\ref{app_BHPT_Kerr:scalar_Newt}) and the numerical procedure (Section~\ref{app_BHPT_Kerr:num_proc}).
\section{Lorenz Gauge Metric Perturbations}\label{app_BHPT_Kerr:LG}
Diffeomorphism invariance of General Relativity becomes \emph{gauge freedom} in perturbation theory. Under a change of coordinates $x^\mu \rightarrow x^\mu + \epsilon X^\mu(x)$, the linear perturbation of a tensor $\mathbf{T}$ transforms as $\delta T \rightarrow \delta T - \pounds_X T$, where $\pounds_X T$ is the Lie derivative of $T$ on the background;~consequently, the linear metric perturbation~\eqref{eq:expansion_metric} transforms as $h_{\mu \nu} \rightarrow h_{\mu \nu} - 2 \nabla_{(\mu} X_{\nu)}$.
\vskip 2pt
For many calculations, it is convenient to exploit this gauge freedom by working in Lorenz gauge, defined by
\begin{equation}
\nabla^\mu \bar{h}_{\mu \nu}^{(n,m)} = 0\,, 
\end{equation}
where $\bar{h}_{\mu \nu} = h_{\mu \nu} - \frac{1}{2} g_{\mu \nu} h$ is the trace-reversed metric perturbation, and $h\equiv h^\alpha{}_\alpha$ is the trace (indices $(n,m)$ omitted for clarity). The covariant derivative $\nabla_\mu$ is defined on the background spacetime $g_{\mu \nu}$. Then the perturbed Einstein equations become a system of hyperbolic equations, 
\begin{equation}\label{eq:LEE} 
\Box \bar{h}^{(n,m)}_{\mu \nu} + 2 \tensor{R}{^\alpha _\mu ^\beta _\nu} \bar{h}^{(n,m)}_{\alpha \beta}  
= S^{(n,m)}_{\mu\nu}\,,    
\end{equation}
where the source term $S^{(n,m)}_{\mu\nu}$ depends on metric and field perturbations of lower orders.
\vskip 2pt
Recent work~\cite{Dolan:2021ijg, Dolan:2023enf, Wardell:2024yoi} has developed a prescription for constructing the Lorenz-gauge metric perturbation on the Kerr background~\eqref{eq:metric_kerr} from scalar variables that satisfy decoupled, separable equations;~more specifically, sourced Teukolsky equations of spin-2, spin-1 and spin-0 types \cite{Teukolsky:1973ha}. In Chapter~\ref{chap:inspirals_selfforce}, we make use the implementation from~\cite{Dolan:2023enf} to compute the metric perturbation of a point-like body on a circular equatorial orbit around a rotating BH in Lorenz gauge.
\section{Scalar Fluxes in the Newtonian Regime}\label{app_BHPT_Kerr:scalar_Newt}
As a consistency check of our results in Section~\ref{sec_SF:fluxes}, we compared the relativistic scalar fluxes with those in the Newtonian regime. In the latter case, the study of dynamical friction in gravitational atoms in the Newtonian regime has been referred to as \emph{ionisation}~\cite{Baumann:2021fkf}, due to analogy with atomic physics (see~\cite{Tomaselli:2023ysb} for a thorough comparison between ``classic'' dynamical friction and ionisation). Below, we briefly summarise the ionisation process and refer to~\cite{Baumann:2021fkf} for details.
\vskip 2pt
In the language of quantum mechanics, ionisation describes the transfer of the cloud from its bound state $\ket{n\ped{b}\ell\ped{b} m\ped{b}}$~\eqref{eqn:BHenv_eigenstates}, to any unbound state $\ket{k \ell m}$, where $k$ represents the wavenumber. This process is governed by the coupling strength between these states, defined by the matrix element: 
\begin{equation}
    \eta = \bra{k\ell m}V\ket{n\ped{b}\ell\ped{b} m\ped{b}}\,,
\end{equation}
where $V$ is the gravitational perturbation of the secondary, which is expressed as a multipole expansion of the Newtonian potential~\eqref{eqn:V_star}. 
\vskip 2pt
The \emph{ionisation power} is then found by summing over all the unbound states. On circular, equatorial orbits, it is given by 
\begin{equation}
    P\ped{ion} = \frac{M\ped{c}}{\mu}\sum_{\ell m} \Omega_{m\ped{g}} |\eta(k_*)|^{2}\Theta(k_*^2)\,,
\end{equation}
where $k_* = -\mu \alpha^2/(2n\ped{b}^2)\pm \Omega_{m\ped{g}}$ and $\Theta$ is the Heaviside step function. This quantity is equivalent to the energy flux to infinity, $P\ped{ion} \equiv \dot{E}^{\mathrm{s}, \infty}$, and the quantity shown in Figure~\ref{fig:Fluxes_Inf_Hor}.
\clearpage
\section{Numerical Procedure and Validation}\label{app_BHPT_Kerr:num_proc}
Solving eq.~\eqref{eq:EOM_sourced} in Chapter~\ref{chap:inspirals_selfforce} is a nontrivial task. Here, we outline the key steps involved, along with the consistency checks we performed.
\subsubsection{Metric data} 
To construct the metric perturbation $h^{(0,1)}_{\mu\nu}$ in Lorenz gauge, we adapt the \textsc{Mathematica} notebook developed in~\cite{Dolan:2021ijg, Dolan:2023enf, Wardell:2024yoi} to a modular \textsc{Mathematica} package suitable for exploring large parameter spaces. The package outputs spin-weighted spherical harmonic data on a tortoise coordinate grid. The data is then summed over $\ell$--modes up to $\ell\ped{max} = 18$, constructing $m$--mode data on a two-dimensional $(r_*,\theta)$ grid. This step is essential for generating the source term in eq.~\eqref{eq:dynamicalperubation}, as it circumvents the problem of infinite mode couplings, which would otherwise require a solution such as outlined in~\cite{Spiers:2024src}.
\vskip 2pt
To check our results, the two-dimensional $m$--mode components of $h^{(0,1)}_{\mu\nu}$ are numerically reprojected onto spin-weighted spherical harmonics in the Schwarzschild limit. The resulting data is then compared against the first-order Schwarzschild Lorenz-gauge data used by the \textsc{Multiscale Self-Force collaboration}~\cite{Warburton:2021kwk,Wardell:2021fyy}. We find agreement at the level of machine precision across all modes and radii, with one exception:~the $(\ell,m)=(1,0)$ mode. This discrepancy is well-understood and originates from different completion choices between the two approaches. The correction term, given by eqs.~(D3a)--(D3b) in~\cite{Miller:2020bft}, resolves this discrepancy, achieving machine precision for all $m$--modes. Importantly, the difference in the $(\ell, m) = (1,0)$ mode only affects the $m = 1$ modes of the perturbed scalar field. As the background field is in a single-state configuration with $(\ell\ped{b}, m\ped{b}) = (1,1)$, this mode cannot contribute to the fluxes at infinity. 
\subsubsection{Background field} 
The background scalar field $\phi^{(1,0)}$ is constructed using Leaver's method~\cite{Leaver:1985ax,Dolan:2007mj}. For a given value of the boson mass $\mu$, we construct the radial profile at the threshold frequency $\omega\ped{c}$ of a pure $(\ell\ped{b},m\ped{b}) = (1,1)$ harmonic state. We then build the field profile on the same $(r_*,\theta)$ grid as the $m$--mode $h^{(0,1)}$ data. 
\subsubsection{Derivatives} 
To calculate the derivatives of the scalar field, i.e., $\nabla_\mu\nabla_\nu \phi^{(1,0)}$, we build a two-dimensional covariant derivative operator using the method of splines. As a consistency check, we contract this quantity with the background metric and confirm that $g_{\text{Kerr}}^{\mu\nu}\nabla_\mu\nabla_\nu \phi^{(1,0)} = \mu^2 \phi^{(1,0)}$ within machine precision. 
\vskip 2pt
In the Schwarzschild case, the absence of a stationary threshold configuration of the cloud leads to oscillatory behaviour at the horizon, where the field oscillates as $e^{- i\omega r_*}$. This contrasts the Kerr case, where the field goes to a constant at the horizon (as $\omega \rightarrow \Omega\ped{H}$). To ensure the robustness of our Schwarzschild results, we recompute all quantities using ingoing Eddington-Finkelstein coordinates, which do give regular derivatives at the horizon. We find that our results remain unchanged.
\subsubsection{Source construction} 
With the $m$--mode tensors $h_{(0,1)}^{\mu\nu}$ and $\nabla_\mu\nabla_\nu \phi^{(1,0)}$ validated, we contract them to form the
$m$--mode source term for eq.~\eqref{eq:dynamicalperubation}. This source term is projected onto spheroidal harmonics (or spherical harmonics in the Schwarzschild case) using the same projection routines applied earlier.
\subsubsection{Finding solutions} 
\begin{figure}[t!]
    \centering
    \includegraphics[scale = 1]{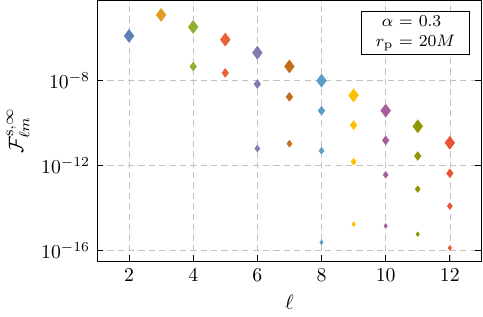}
    \caption{Contributions from different $\ell$--modes of the flux to infinity in Kerr for $\alpha = 0.3$, with the secondary on a prograde orbit at $r\ped{p} = 20 M$. Due to selection rules, modes with opposite parity are zero, i.e., when $\ell$ is odd and $m$ is even or vice versa. Additionally, all $m = 0$ and $m = 1$ modes do not contribute to the flux to infinity. The different sized diamonds show the contribution from the modes that are not zero, where the higher the $m$, the larger contribution. For example, for $\ell = 8$, we show $m = 8, 6, 4, 2$. Fluxes through the horizon follow a similar trend.}
    \label{fig:flux_l_mode_convergence}
\end{figure}

\begin{figure}[t!]
    \centering
    \includegraphics[scale = 1]{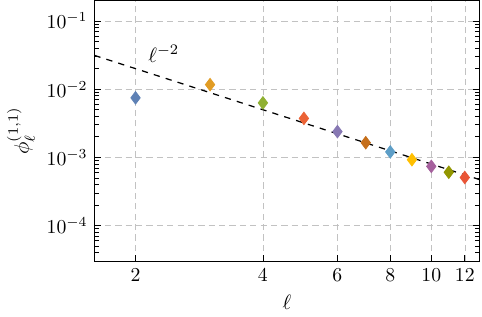}
   \caption{We show the contribution from different $\ell$--modes of the scalar field perturbation evaluated on the orbital radius of the secondary ($r\ped{p} = 20 M$). They fall off with the expected rate, $\ell^{-2}$, indicated by the black dashed line.}
    \label{fig:flux_l_mode_convergence_particle}
\end{figure}

Using the radial source functions, we solve the radial Klein-Gordon equation using a variation of parameters approach. We first build the ``$\mathrm{In}$'' and ``$\mathrm{Up}$'' solution numerically as those which solve the homogeneous Klein-Gordon equation with boundary conditions given by
\begin{equation}
\begin{aligned}
\lim_{r\rightarrow r_+}R_{\mathrm{In}} &= e^{- i  (\Omega_{m\ped{g}}+\omega\ped{c} - m \Omega\ped{H})r^*}\,,\\
\lim_{r\rightarrow \infty}R_{\mathrm{Up}}  &= \frac{e^{ i  k_{m\ped{g}}  r^*} r^{\frac{i \mu^2}{ k_{m\ped{g}}}}}{\sqrt{a^2+r^2}}\,,
\end{aligned}
\end{equation}
where $k_{m\ped{g}} = \sqrt{(\Omega_{m\ped{g}}+\omega\ped{c} )^2-\mu^2}$. Taking the spheroidal projection of the source, $S^{(1,1)}_{\ell m}$,  we directly solve for the scalar field perturbations as,
\begin{equation}
    \phi^{(1,1)}_{\ell m}(r) = C^{\mathrm{In}}_{\ell m}(r) R^{\mathrm{In}}_{\ell m}(r) + C^{\mathrm{Up}}_{\ell m}(r) R^{\mathrm{Up}}_{\ell m}(r)\,,
\end{equation}
where 
\begin{equation}
\begin{aligned}
    C^{\mathrm{In}}_{\ell m}(r) &= \int_{r}^\infty \frac{ R^{\mathrm{Up}}_{\ell m} S^{(1,1)}_{\ell m}}{\mathcal{W}_0}\mathrm{d}r'\,,\\
    C^{\text{Up}}_{\ell m}(r) &= \int_{r_+}^r \frac{ R^{\mathrm{In}}_{\ell m} S^{(1,1)}_{\ell m}}{\mathcal{W}_0}\mathrm{d}r'\,, 
\end{aligned}
\end{equation}
and $\mathcal{W}_0$ is the constant Wronskian coefficient given by
\begin{equation}
   \frac{ {\mathcal{W}_0}}{\Delta}  = R^{\mathrm{In}}_{\ell m} \frac{ \mathrm{d} R^{\mathrm{Up}}_{\ell m} }{\mathrm{d}r}- R^{\mathrm{Up}}_{\ell m} \frac{\mathrm{d} R^{\mathrm{In}}_{\ell m} }{\mathrm{d}r}\,.
\end{equation}
Here, $\Delta= r^2 -2 M r + a^2$ is the standard Kerr quantity. We explicitly verify that the behaviour of the solution at both boundaries aligns with the boundary conditions of the homogeneous $\textit{In}$ and $\textit{Up}$ solutions. To cross-check our code, we insert our source data in a solver from an independent implementation~\cite{Brito:2023pyl}, finding consistent results.
\vskip 2pt
The solution of the Klein-Gordon equation gives $\phi^{(1,1)}_{\ell m}$, which we sum over $\ell$ and $m$ to reconstruct the field profile, which is shown in Figures~\ref{fig:Wake_Profile} and~\ref{fig:Wake_Profile_SM}. The asymptotes of these extended solutions are then extracted to obtain the input of the flux formulae~\eqref{eq:scalar_flux}, generating the results shown in Figure~\ref{fig:Fluxes_Inf_Hor}. In that figure, we sum up to $\ell = 6$, ensuring that the flux increment remains below $1\%$ across the considered radial domain. In the Newtonian case instead, the computational cost is much lower, allowing us to easily sum up to $\ell = 10$, which pushes the flux increment below $0.01\%$. At larger radii than we are showing in this Chapter~\ref{chap:inspirals_selfforce}, more modes might be required to obtain accurate results, which poses a computational challenge in the relativistic regime. Therefore, to compute the total flux across a large radial domain, e.g., for waveform modelling in packages like \textsc{FEW}~\cite{Chua:2020stf,Katz:2021yft,Hughes:2021exa}, a smooth interpolation between both approaches might be required.
\vskip 2pt
In Figure~\ref{fig:flux_l_mode_convergence}, we show the amplitude of the fluxes mode-by-mode. They follow the expected trend:~the \emph{main} $\ell = m$ contribution decays exponentially with increasing $\ell$. Interestingly, Figure~\ref{fig:flux_l_mode_convergence} also shows that \emph{subleading} modes with $\ell \neq m$ can have a non-negligible contributions. For example, the $(\ell,m) = (4,2)$ mode is larger than the $(\ell,m) = (8,8)$ mode.
\vskip 2pt
We show a similar plot in Figure~\ref{fig:flux_l_mode_convergence_particle}, where we define $\phi^{(1,1)}_{\ell} = \sum_{m=-\ell}^{\ell}\phi^{(1,1)}_{\ell m}$ and extract the field at the position of the secondary (the point at which the field is most irregular). We find that the field perturbation is finite and continuous at the particle with the $\ell$--modes of the perturbation constituting a convergent sequence decaying as $\ell^{-2}$.
\vskip 2pt
\backmatter
\selectlanguage{english}
\newpage
\phantomsection
\bibliographystyle{utphys}
\bibliography{thesis}
\end{document}

%% file: Colofon.tex
\noindent This thesis has been accomplished at the Niels Bohr Institute (NBI) of the University of Copenhagen. It was supported by the VILLUM Foundation (grant no.\ VIL37766), the DNRF Chair program (grant no.\ DNRF162) by the Danish National Research Foundation and the European Union’s H2020 ERC Advanced Grant “Black holes: gravitational engines of discovery” grant agreement no.\ Gravitas–101052587.

\vfill
\par\vskip 12cm

\noindent\copyright{} \textbf{Thomas Frederik Martinus Spieksma} 2025\\[0.5em]
\noindent \emph{Exploring Black Hole Environments}\\[1em]
   \begin{minipage}[b]{\textwidth}
All rights reserved. Without limiting the rights under copyright reserved above, no part of this book may be reproduced, stored in or introduced into a retrieval system, or transmitted, in any form or by any means (electronic, mechanical, photocopying, recording or otherwise) without the written permission of both the copyright owner and the author of the book.
      \end{minipage}    
\vfill
      

%% file: Poem.tex
{\it \hspace{-0.65cm} What joy it is, when out at sea the stormwinds are lashing the waters, to gaze from the shore at the heavy stress some other man is enduring! Not that anyone's afflictions are in themselves a source of delight;~but to realise from what troubles you yourself are free is joy indeed. What joy, again, to watch opposing hosts marshalled on the field of battle when you have yourself no part in their peril! But this is the greatest joy of all:~to stand aloof in a quiet citadel, stoutly fortified by the teaching of the wise, and to gaze down from that elevation on others wandering aimlessly in a vain search for the way of life, pitting their wits one against another, disputing for precedence, struggling night and day with unstinted effort to scale the pinnacles of wealth and power. O joyless hearts of men! O minds without vision! How dark and dangerous the life in which this tiny span is lived away! Do you not see that nature is clamouring for two things only, a body free from pain, a mind released from worry and fear for the enjoyment of pleasurable sensations?
\vskip 2pt
So we find that the requirements of our bodily nature are few indeed, no more than is necessary to banish pain. To heap pleasure upon pleasure may height men's enjoyment at times. But what matter if there are no golden images of youths about the house, holding flaming torches in their right hands to illumine banquets prolonged into the night? What matter if the hall does not sparkle with silver and gleam with gold, and no carved and gilded rafters ring to the music of the lute? Nature does not miss these luxuries when men recline in company on the soft grass by a running stream under the branches of a tall tree and refresh their bodies pleasurably at small expense. Better still if the weather smiles upon them and the season of the year stipples the green herbage with flowers. Burning fevers flee no swifter from your body if you toss under figured counterpanes and coverlets of crimson than if you must lie in rude home-spun.\vskip 20pt}
\vskip 2pt
\hfill Lucretius, \emph{De Rerum Natura}, Book II\hspace{-0.34cm}

%% file: Sammenfatning2.tex
The past decade has transformed our ability to observe the Universe. Via gravitational waves, merging black holes and neutron stars can now be directly detected, offering unprecedented opportunities to test General Relativity and explore astrophysics in a fundamentally new way. Driven by this breakthrough, the next generation of detectors is being developed to observe a wider range of sources with greater precision. This marks the beginning of a new era in gravitational-wave astronomy:~leveraging black holes as probes of new physics.
\vskip 2pt
Despite their reputation, black holes are remarkably simple macroscopic objects, described by only a few parameters. This simplicity enables highly accurate predictions of black hole mergers, with any deviation potentially signalling new physics -- from modifications to General Relativity to the presence of surrounding matter. Such prospects motivate the study of \emph{black hole environments}. As two black holes in a binary spiral towards each other, they interact dynamically with their environment, accreting material or experiencing drag forces that alter their motion and, in turn, the gravitational waves they emit. Detecting these effects requires careful modelling of both the binary and the environment. This thesis addresses various aspects of this challenge, with implications for future detectors such as LISA or Einstein Telescope.
\vskip 2pt
When one compact object is much lighter than the other, so-called \emph{extreme mass ratio binaries}, its motion becomes especially sensitive to the surrounding environment, influencing the binary's evolution from formation all the way to merger. While various types of environments exist, in this thesis, I focus on the main ones:~gaseous media such as plasma and accretion disks, as well as dark matter structures. A particularly striking scenario involves black hole superradiance, a process in which a black hole transfers some of its energy and angular momentum to a bosonic field, forming a dense ``boson cloud''. These clouds occupy quantised bound states similar to those in the hydrogen atom. The superradiant process, relevant for bosons with masses in the range $10^{-20}$ to $10^{-10}\,\mathrm{eV}$, relies solely on gravitational interactions, offering a unique opportunity to explore the weak-coupling and ultralight regime of particle physics.
\vskip 2pt
In the first part of this thesis, I investigate these ultralight bosons and their interactions with the electromagnetic sector. I show that, as a boson cloud grows via superradiance, its interaction with photons drives the system towards a stationary state, where energy extracted from the black hole is steadily converted into monochromatic electromagnetic radiation -- a compelling observational channel. I also assess whether this radiation can reach us, given the presence of matter throughout the Universe. I find that astrophysical plasmas can significantly suppress the conversion from bosons to photons, allowing light to propagate only for strong couplings.
\vskip 2pt
I then turn to black hole binaries, focusing first on the ringdown phase. Plasma may also play a role here if black holes carry charge. I demonstrate how plasma can modify the fundamental quasi-normal mode, or even induce echoes in the gravitational-wave signal. Similarly distinct signatures may occur in galactic dark matter halos. To evaluate their relevance, I investigate whether such halos affect the ringdown in a realistic data-analysis context. My findings suggest that, for both present and upcoming detectors, the ringdown signal remains indistinguishable from that in vacuum.
\vskip 2pt
In contrast to the ringdown phase, the inspiral lasts much longer, allowing effects from the environment to accumulate in the waveform. Returning to the study of boson clouds, I examine the early inspiral, where resonances between bound states can occur. A systematic exploration of the system’s ``history'' reveals that when the binary and cloud are nearly counter-rotating, resonances are ineffective, and the cloud persists into the late inspiral, where it may directly influence the waveform. For most orbital configurations, however, the cloud is absorbed by the black hole, leaving imprints on the binary’s eccentricity and inclination, thereby opening an indirect observational window. I also explore the early inspiral phase in the context of accretion disks. Here too, the environment shapes the orbital evolution by aligning the binary's orbit with the disk plane, while eccentricity may either increase or decrease. These effects are crucial for accurately modelling gravitational-wave sources.
\vskip 2pt
As the binary approaches merger, its motion becomes highly relativistic. In the case of extreme mass ratio inspirals, the self-force approach offers the most suitable framework for modelling this regime. I extend this method to account for environments and apply it to boson clouds, highlighting the limitations of Newtonian and Schwarzschild approximations. This framework paves the way for studying generic black hole environments in Kerr spacetime.
\vskip 2pt
This thesis explores the rich observational signatures produced by black hole environments, with implications for both gravitational physics and (astro)particle physics. A key challenge lies in determining whether these signatures can be identified in realistic data and how precisely the system parameters can be inferred. With this in mind, gravitational waves may yet unlock some of the deepest mysteries of our Universe.
\newpage
\noindent \textbf{Dansk overs\ae ttelse}
\vskip 2pt
\noindent Det seneste \aa rti har transformeret vores evne til at observere Universet. Kolliderende sorte huller og neutronstjerner kan nu observeres direkte via gravitationsb\o lger, hvilket giver hidtil usete muligheder for at teste den generelle relativitetsteori og udforske astrofysik p\aa{} en fundamentalt ny m\aa de. Drevet af dette gennembrud udvikles den n\ae ste generation af detektorer til at observere en bredere vifte af kilder med st\o rre pr\ae cision. Dette markerer begyndelsen p\aa{} en ny \ae ra inden for gravitationsb\o lgeastronomi:~udnyttelse af sorte huller som detektorer for ny fysik.
\vskip 2pt
Trods deres ry er sorte huller bem\ae rkelsesv\ae rdigt simple makroskopiske objekter, beskrevet af kun f\aa{} parametre. Deres simple beskrivelse muligg\o r meget n\o jagtige forudsigelser af kollisioner mellem af sorte huller, hvor enhver afvigelse potentielt signalerer ny fysik, s\aa som modificeringer af den generelle relativitetsteori eller tilstedev\ae relsen af omgivende stof.  S\aa danne perspektiver motiverer studiet af sorte hullers omgivelser. N\aa r et bin\ae rt system af to sorte huller er i bev\ae gelse mod hinanden i en spiral-bane, interagerer de dynamisk med deres milj\o{}, tilegner sig materiale eller oplever modstandskr\ae fter fra omgivelserne, der \ae ndrer deres bev\ae gelse og dermed de gravitationsb\o lger de udsender. M\aa ling af disse effekter kræver præcis modellering af b\aa de det binære system og milj\o et. Denne afhandling unders\o ger forskellige aspekter af denne udfordring, med implikationer for fremtidige detektorer s\aa som LISA eller Einstein Telescope.
\vskip 2pt
N\aa r et kompakt objekt i et binært system er meget lettere end det andet, et s\aa kaldt \emph{extreme-mass-ratio} bin\ae rt system, bliver dets bevægelse s\ae rligt f\o lsom over for omgivelserne, hvilket p\aa virker det bin\ae re systems udvikling;~fra dannelse og hele vejen til kollision. I denne afhandling fokuserer jeg p\aa{} de vigtigste typer af astrofysiske milj\o er:~gasformige medier s\aa som plasma og akkretionsskiver, samt m\o rke stofstrukturer. Et s\ae rligt sl\aa ende scenario involverer sorte hullers superradians, en proces, hvor et sort hul overf\o rer noget af sin energi og angul\ae rt moment til et bosonisk felt og danner en t\ae t ``bosonsky''. Disse skyer indtager kvantiserede bundne tilstande svarende til dem i hydrogenatomet. Superradiansprocessen, som er relevant for bosoner med masser i intervallet $10^{-20}$ til $10^{-10}\,\mathrm{eV}$, er udelukkende afh\ae ngig af gravitationelle interaktioner, hvilket giver en unik mulighed for at udforske partikelfysikkens svage koblings- og ultralette regime. 
\vskip 2pt
I den f\o rste del af denne afhandling unders\o ger jeg disse ultralette bosoner og deres interaktioner med den elektromagnetiske sektor. Jeg viser, at i processen hvor en bosonsky vokser via superradians-f\ae nomenet, driver dens interaktion med fotoner systemet mod en station\ae r tilstand, hvor energi udvundet fra det sorte hul kontinuerligt omdannes til monokromatisk elektromagnetisk str\aa ling -- en brugbar observationskilde for f\ae nomenet. Jeg unders\o ger ogs\aa{}, om denne str\aa ling kan n\aa{} os, givet tilstedeværelsen af stof i hele Universet. Jeg konkluderer, at astrofysiske plasmaer kan undertrykke omdannelsen fra bosoner til fotoner betydeligt, hvilket g\o r, at lyset kun kan udbredes ved st\ae rke koblinger. 
\vskip 2pt
Derefter unders\o ger jeg sorte huller i bin\ae re systemer og fokuserer f\o rst p\aa{} fasen kaldet \emph{ringdown}. Plasma kan ogs\aa{} spille en rolle her, hvis sorte huller b\ae rer ladning. Jeg demonstrerer, hvordan plasma kan \ae ndre den grundl\ae ggende s\aa kaldte kvasinormale tilstand eller endda inducere ekkoer i gravitationsb\o lgesignalet. Tilsvarende kan forskellige signaturer forekomme i galaktiske m\o rkstofhaloer. For at evaluere deres relevans unders\o ger jeg, om s\aa danne haloer p\aa virker ringdown fasen i en realistisk dataanalysekontekst. Mine resultater tyder p\aa{}, at ringdown-signalet, for b\aa de nuv\ae rende og kommende detektorer, forbliver umuligt at skelne fra signalet i vakuum.
\vskip 2pt
I mods\ae tning til ringdown fasen varer inspiralfasen, hvor de sorte huller bev\ae ger sig mod hinanden, meget l\ae ngere, hvilket tillader effekter fra omgivelserne at akkumulere i b\o lgeformen. Jeg vender her tilbage til at studere boson skyer, hvor jeg unders\o ger den tidlige inspiralfase, hvor resonanser mellem bundne tilstande kan forekomme. En systematisk udforskning af systemets ``historie'' viser, at n\aa r det bin\ae re system og boson-skyen er n\ae sten modsat rotererende, er resonanserne ineffektive, og skyen overlever ind i den sene inspiralfase, hvor den direkte kan p\aa virke b\o lgeformen. For de fleste orbitale konfigurationer absorberes skyen dog af det sorte hul, hvilket efterlader aftryk p\aa{} det bin\ae re systems excentricitet og h\ae ldning, hvilket \aa bner et indirekte observationsvindue. Jeg udforsker ogs\aa{} den tidlige inspiralfase i kontekst af akkretionsskiver. Ogs\aa{} her p\aa virker milj\o et orbitaludviklingen af det bin\ae re system, s\aa ledes at banen p\aa virkes i retning af at v\ae re sammenfaldende skiveplanet, mens excentriciteten af systemet enten kan \o ges eller mindskes. Det er afg\o rende at forst\aa{} disse effekter for at opn\aa{} en pr\ae cis modellering af gravitationsb\o lgekilder. 
\vskip 2pt
Efterh\aa nden som det bin\ae re system n\ae rmer sig kollision, bliver dens bev\ae gelse meget relativistisk. I tilf\ae lde af extreme-mass-ratio systemer er den s\aa kaldte selvkrafttilgang den mest passende metode til modellering af dette regime. Jeg udvider denne metode til at tage h\o jde for milj\o er og anvender den p\aa{} bosonskyer og p\aa peger begr\ae nsningerne ved Newtonske og Schwarzschild-approksimationer af systemet. Denne ramme baner vejen for at studere generiske sort-hul milj\o er i Kerr-rumtiden.
\vskip 2pt
Denne afhandling udforsker de rige observationssignaturer, der produceres af sorte hullers milj\o er, med implikationer for b\aa de gravitationsfysik og (astro)partikelfysik. En central udfordring ligger i at bestemme om disse signaturer kan identificeres i realistiske data, og hvor pr\ae cist systemparametrene kan udledes.  Ved at l\o se disse udfordringer har gravitationsb\o lger potentiale til at bidrage til opklaringen af nogle af de dybeste mysterier i vores Univers.
\vskip 2pt
\hfill \textit{Oversat af Asta Heinesen}

%% file: Basedon.tex
{\scshape This thesis is based on, and partially consists of reprints of, the following publications:} \vskip 20pt
\renewcommand{\ttdefault}{cmtt} 

\begin{itemize}[]
 
{\fontfamily{cmr}\selectfont

\item [\cite{Spieksma:2023vwl}]

T.~F.~M.~Spieksma, E.~Cannizzaro, T.~Ikeda, V.~Cardoso, and Y.~Chen, ``Superradiance:~Axionic couplings and plasma effects,'' \href{https://journals.aps.org/prd/abstract/10.1103/PhysRevD.108.063013}{\textit{Physical Review D} \textbf{108} no.~6, (2023) 063013}, \href{https://arxiv.org/abs/2306.16447}{\ttfamily arXiv:2306.16447 [gr-qc]}. \vskip 5pt

Presented in Chapter~{\bf \begingroup\hypersetup{linkcolor=black}\ref{chap:SR_Axionic}\endgroup}. \vskip 10pt

\item [\cite{Cannizzaro:2024hdg}]

E.~Cannizzaro and T.~F.~M.~Spieksma, ``Phenomenology of ultralight bosons around compact objects:~In-medium suppression,'' \href{https://journals.aps.org/prd/abstract/10.1103/PhysRevD.110.084021}{\textit{Physical Review D} \textbf{110} no.~8, (2024) 084021}, \href{https://arxiv.org/abs/2406.17016}{\ttfamily arXiv:2406.17016 [hep-ph]}. \vskip 5pt

Presented in Chapter~{\bf \begingroup\hypersetup{linkcolor=black}\ref{chap:in_medium_supp}\endgroup}. \vskip 10pt

\item [\cite{Cannizzaro:2024yee}]

E.~Cannizzaro, T.~F.~M.~Spieksma, V.~Cardoso, and T.~Ikeda, ``Impact of a plasma on the relaxation of black holes,'' \href{https://journals.aps.org/prd/abstract/10.1103/PhysRevD.110.L021302}{\textit{Physical Review D (Letters)} \textbf{110} no.~2, (2024) L021302}, \href{https://arxiv.org/abs/2405.05315}{\ttfamily arXiv:2405.05315 [gr-qc]}. \vskip 5pt

Presented in Chapter~{\bf \begingroup\hypersetup{linkcolor=black}\ref{chap:plasma_ringdown}\endgroup}. \vskip 10pt

\item [\cite{Spieksma:2024voy}] 

T.~F.~M.~Spieksma, V.~Cardoso, G.~Carullo, M.~D.~Rocca, and F.~Duque, ``Black Hole Spectroscopy in Environments:~Detectability Prospects,'' \href{https://doi.org/10.1103/PhysRevLett.134.081402}{\textit{Physical Review Letters} \textbf{134} no.~8, (2025) 081402}, \href{https://arxiv.org/abs/2409.05950}{\ttfamily arXiv:2409.05950 [gr-qc]}.  \vskip 5pt

Presented in Chapter~{\bf \begingroup\hypersetup{linkcolor=black}\ref{chap:BHspec}\endgroup}. \vskip 10pt

\newpage

\item [\cite{Tomaselli:2024bdd}]

G.~M.~Tomaselli, T.~F.~M.~Spieksma, and G.~Bertone, ``Resonant history of gravitational atoms in black hole binaries,'' \href{https://doi.org/10.1103/PhysRevD.110.064048}{\textit{Physical Review D} \textbf{110} no.~6, (2024) 064048}, \href{https://arxiv.org/abs/2403.03147}{\ttfamily arXiv:2403.03147 [gr-qc]}. \vskip 5pt

\textit{and} \vskip 5pt

\item [\cite{Tomaselli:2024dbw}]

G.~M.~Tomaselli, T.~F.~M.~Spieksma, and G.~Bertone, ``Legacy of Boson Clouds on Black Hole Binaries,'' \href{https://doi.org/10.1103/PhysRevLett.133.121402}{\textit{Physical Review Letters} \textbf{133} no.~12, (2024) 121402}, \href{https://arxiv.org/abs/2407.12908}{\ttfamily arXiv:2407.12908 [gr-qc]}. \vskip 5pt

Presented in Chapter~{\bf \begingroup\hypersetup{linkcolor=black}\ref{chap:legacy}\endgroup}. \vskip 10pt

\item [\cite{Dyson:2025dlj}] 

C.~Dyson, T.~F.~M.~Spieksma, R.~Brito, M.~van de Meent, and S.~Dolan, ``Environmental Effects in Extreme-Mass-Ratio Inspirals:~Perturbations to the Environment in Kerr Spacetimes,'' \href{https://doi.org/10.1103/PhysRevLett.134.211403}{\textit{Physical Review Letters} \textbf{134} no.~21, (2025) 211403}, \href{https://arxiv.org/abs/2501.09806}{\ttfamily arXiv:2501.09806 [gr-qc]}.  \vskip 5pt

Presented in Chapter~{\bf \begingroup\hypersetup{linkcolor=black}\ref{chap:inspirals_selfforce}\endgroup}. \vskip 10pt

\item [\cite{Spieksma:2025wex}] 

T.~F.~M.~Spieksma and E.~Cannizzaro, ``In the grip of the disk:~Dragging the companion through an AGN,'' \href{https://arxiv.org/abs/2504.08033}{\ttfamily arXiv:2504.08033 [astro-ph.GA]}.  \vskip 5pt

Presented in Chapter~{\bf \begingroup\hypersetup{linkcolor=black}\ref{chap:capture}\endgroup}.

}

\end{itemize}

\newpage

{\scshape During the completion of his PhD, T.~F.~M.~Spieksma was also an author of the following publications:} \vskip 20pt

\begin{itemize}[]

{\fontfamily{cmr}\selectfont

\item [\cite{Cole:2022yzw}]

P.~S.~Cole, G.~Bertone, A.~Coogan, D.~Gaggero, T.~Karydas, B.~J.~Kavanagh, T.~F.~M.~Spieksma, and G.~M.~Tomaselli, ``Distinguishing environmental effects on binary black hole gravitational waveforms,'' \href{http://dx.doi.org/10.1038/s41550-023-01990-2}{\textit{Nature Astronomy} \textbf{7} no.~8, (2023) 943--950}, \href{https://arxiv.org/abs/2211.01362}{\ttfamily arXiv:2211.01362 [gr-qc]}. \vskip 5pt

\item [\cite{Tomaselli:2023ysb}]

G.~M.~Tomaselli, T.~F.~M.~Spieksma, and G.~Bertone, ``Dynamical friction in gravitational atoms,'' \href{http://dx.doi.org/10.1088/1475-7516/2023/07/070}{\textit{Journal of Cosmology and Astroparticle Physics} \textbf{07} (2023) 070}, \href{https://arxiv.org/abs/2305.15460}{\ttfamily arXiv:2305.15460 [gr-qc]}. \vskip 5pt

\item [\cite{Cardoso:2024jme}]

V.~Cardoso, G. Carullo, M.~De Amicis, F.~Duque, T.~Katagiri, D.~Pere\~niguez, J.~Redondo-Yuste, T.~F.~M.~Spieksma, and Z.~Zhong, ``Hushing black holes:~Tails in dynamical spacetimes,'' \href{https://doi.org/10.1103/PhysRevD.109.L121502}{\textit{Physical Review D (Letters)} \textbf{109} no.~12, (2024) L121502}, \href{https://arxiv.org/abs/2405.12290}{\ttfamily arXiv:2405.12290 [gr-qc]}. \vskip 5pt

\item [\cite{Spieksma:2025sda}] 

T.~F.~M.~Spieksma and E.~Cannizzaro, ``Axion dissipation in conductive media and neutron star superradiance,''\href{http://dx.doi.org/10.1088/1475-7516/2025/06/028}{\textit{Journal of Cosmology and Astroparticle Physics} \textbf{06} (2025) 028}, \href{https://arxiv.org/abs/2503.19978}{\ttfamily arXiv:2503.19978 [hep-ph]}.  \vskip 5pt

}

\end{itemize}

\newpage

%% file: Acknowledgements.tex
First and foremost, I want to thank my supervisor, Vitor Cardoso. It’s been an absolute pleasure working with you over the past three years. I’ve learned so much from your approach to science, been inspired by your many ideas, and often was amazed by your incredible ``intuition for physics''. There were countless times I showed you my work and you could spot something was off, just by feel. It has pushed me to always triple-check the physics behind my results. But beyond the science, what truly made these years special is how much fun we had together. Social boundaries pretty much vanished after the first few months, and most of our ``meetings'' resulted in us hanging out, over coffee, beers or lying on the couch and chatting. Sometimes about physics, sometimes about life. You gave me the chance to travel the world for science and I’m grateful we got to share meals in Italy, Portugal, Japan, and many other places. Vitor, thank you, it has been an incredible time.
\vskip 4pt
Vitor wasn’t just a great supervisor, he’s also built the most fun and dynamic research group I’ve seen anywhere during my PhD (though I might be a little biased). At the heart of that success is Julie, and I’m sure anyone in the Strong group, or anyone who’s visited, would agree. Julie, on a personal note, you’ve also become a friend. Being the first two people to arrive on the upper floor, our morning coffee chats became a cherished tradition and it set me up for the day ahead. Organising social events for the group with you has been very fun and rewarding. I’ll truly miss our morning moments of peace.
\vskip 4pt
To the rest of the Strong group:~it has been one hell of a ride building this group together from the ground up. We all arrived at the same time, none of us spoke Danish, and none of us really knew how anything worked around here. Conor, David, Gregorio, Jaime and Jose, especially, thank you for the amazing times we’ve shared. For the countless beers on the N{\o}rrebro bridge, for the Hangaren evenings, for not taking life too seriously, for the science we did talk about, for the incredible food we shared, for the time spent abroad, and above all, for being absolutely brilliant scientists and wonderful friends. I’ll admit, it’s a little sad that we’re all heading off to different corners of the world, but that’s just how life goes. I hope we’ll keep seeing each other, you're the ones who made this Copenhagen story special.
\vskip 4pt
Writing this thesis would not have been possible without the input and guidance of my collaborators. Rightfully, the first person to mention is Enrico. In fact, calling you a ``collaborator'' doesn’t quite capture it. You’ve become a close friend over the years. We started working together on my very first day in Copenhagen and basically never stopped. At first, it was the three of us including Vitor, but by now, we've become a well-oiled tandem. I probably talk to you more than anyone else on this planet, about science but also about football, tennis and life. It is rare to have most of your meetings with a friend, and something I truly appreciate. You’ve made me feel at home in Rome and in Lisbon, and I know we’ll keep working together, keep bullshitting about life, and now and then, even have a serious conversation. Thank you, Enrico, for the scientific collaboration, for your friendship, and for teaching me an important life lesson:~``\emph{una carbonara senza guanciale è come la Juventus senza~Del~Piero}.'' 
\vskip 4pt
One of the important lessons I've learned during my PhD is that, when it comes to physics, you’ll always find yourself surrounded by Mediterranean people. Francisco, you gave me the honorary title of ``most Mediterranean Northern-European person'', of which I'm still incredibly proud. I'm happy to realise you’ve become a good friend, from sharing music to simply having a great time whenever we see each other. I hope we may have many more of those moments. A special thank you also goes out to the people in Amsterdam, with whom I’ve continued to collaborate throughout my PhD. In particular, Gimmy, you’ve taught me so much, not only about physics, but about doing science. Your deep understanding of physics, your rigour, your eye for detail and your way of communicating science are all things I truly admire. It’s not hard to find traces of your influence throughout my work. I’m grateful we had the chance to collaborate so closely over the years and I hope we’ll continue. I also want to thank Gianfranco, for setting me on the path towards a PhD and always bringing a bird’s-eye view to the projects we worked on together. And finally, a big thank you to Katy, for making my stay in London possible and giving me the warmest of welcomes. I genuinely enjoyed my time there (especially the arXiv bingo), and I appreciate the space you gave me to focus on this thesis while still collaborating. I didn’t just learn about \textsc{GRChombo} -- I also took away a lot from your relaxed and balanced approach to science. A heartfelt thanks as well to all the QMUL folks;~I felt at home from the very start.
\vskip 4pt
Living in Copenhagen has been nothing short of a hipster's dream. I will not fully proclaim myself as one, but I’ve truly enjoyed what this city has to offer. It’s a serene place in a chaotic world. To everyone who made this place even more memorable, to my roommates, to Jakob and Lisa, thank you. Copenhagen is deeply ingrained in me. 
\vskip 4pt
Leaving Amsterdam while all my friends were still there wasn't easy. I've been lucky enough to return semi-regularly, and it has always felt special. To the beloved \textsc{AMICI OPTIMI}, who by now feel more like extended family. For always welcoming me with open arms when I visit, for always making time, for giving me a home away from home, and for visiting so often in Copenhagen. We all know what we mean to each other;~no more words are needed. To Floris and Quint, for inspiring me time and again. Whenever I'm with you, I take away something new:~a fresh perspective, a deeper appreciation for life. Our yearly summer trips in nature are a much-needed pause from the crazy busy world, and they recharge me for the year ahead. May we keep that tradition going for many years to come. To my other friends in the Netherlands, Hidde, Joost, Joris, Laurens, and Oscar. We try to see each other as much as life allows us to, both in the Netherlands and Copenhagen. You're incredible friends, and I'm amazed by how well we stay in touch. That we may continue to do so. 
\vskip 4pt
Of course, always save the best for last. No note of thanks would be complete without including the most important people:~my parents and my sister. Thank you for giving me the foundation to stand where I stand, for providing support during difficult times, for checking in, for offering advice when I need it, for giving me the biggest and most unwavering safety net in the world, and above all, for giving me unconditional love. Thank you for it all, I could not ask for more.

%% file: Prologue.tex
\vspace{-0.207cm}
It has always seemed arbitrary to me, the way the brain stores memories. Some make sense -- celebrating your birthday or going on vacation -- while others feel inexplicably preserved:~flashes of a morning walk to school or home-made sandwiches from my mom on a summer day. Why those memories? Were they more important than others? Do I even control this process?
\vskip 2pt
One of my fondest childhood memories is sitting on the couch with my dad, watching the Czech stop-motion animation \emph{Buurman \& Buurman}. I believe its popularity never spread much beyond Czechia and the Netherlands. The show follows two men, Pat and Mat, as they handle everyday problems in their own -- let's call it ``practical'' -- way. Everything always goes wrong, yet their spirit, optimism, and endurance carries them through. It is something I hope I appreciated, even unconsciously, as a child. I certainly do now.
\vskip 2pt
In many ways, memories of Pat and Mat kept rushing back to me during my PhD. Tackling problems I had no idea how to solve demanded all of the qualities above:~spirit, optimism and endurance. Talking to my collaborators, who often were as confused as I was, became a constant. There were moments of relief when I thought I'd finally cracked the problem, only to step back, apply a bit of common sense, and realise it still doesn't add up. Over time, I've come to understand that the dam always bursts. It may resist for a long time, but enough energy, time and determination will break it down. 
\vskip 2pt
Doing a PhD in gravitational physics, especially at this moment in time, has been special. The pace at which this field is advancing is nothing short of extraordinary. Since that landmark detection nearly a decade ago, gravitational-wave astronomy has evolved from a hopeful concept into a weekly reality. These waves arrive to us as cosmic messengers, carrying information from distant events we could never witness before. Already, we test the known and challenge the expected, with our excitement limited only by imagination. Gravity’s universal nature ensures that whatever is out there will find its way to us through these waves. Our (not-so) simple task is to decipher their message.
\vskip 2pt
It is a great joy to be part of that whirlwind. It is a great pleasure to dedicate time and energy to studying the Universe and becoming familiar with objects so humanly unfamiliar. It is a great privilege to study objects with no opinions, no political stances;~things that simply \emph{are}. 
\vskip 2pt
And yet, this experience is not just personal. Many others are deeply drawn to this subject. Maybe it's no surprise. While reading this, your body is vibrating through the gravitational waves that pass by. You're being bombarded by millions of neutrinos from explosions of massive stars elsewhere in the Universe. There is, and always has been, a unique connection between humans and the Cosmos. Our curiosity towards it is only natural. It reflects a universal feature of human nature, regardless of race, religion or belief:~our desire to understand the outside world.
\vskip 2pt
This drive has ultimately shaped my journey for the last three years, which, I've learned, both gives and takes. It's now, when I've reached my destination, that I can make up the balance. In less than a decade, gravitational-wave astronomy has grown from nonexistence into a thriving field. As with all rapid progress, it brings with it a host of unsolved problems and challenges. Still, in the face of these obstacles -- both within science and beyond, in an increasingly uncertain world -- I remain an advocate for positivity. The future is bright. With this thesis, I hope to pass on some of that excitement and positivity to you. 
\vskip 2pt
A Je To.
\vskip 10pt
\hfill \emph{From the attic of the Niels Bohr Institute,}

\hfill \emph{Copenhagen, Summer of 2025}

\hfill \emph{Thomas}

\newpage

%% file: Epilogue.tex
\vspace{-0.207cm}
{\it  Thousands of years ago, the first man\blackasteriskfootnote{It could just as well have been a woman -- the sentiment applies equally.} \!discovered how to make fire. He was probably burned at the stake he had taught his brothers to light. He was considered an evildoer who had dealt with a demon mankind dreaded. But thereafter men had fire to keep them warm, to cook their food, to light their caves. He had left them a gift they had not conceived and he had lifted darkness off the earth. Centuries later, the first man invented the wheel. He was probably torn on the rack he had taught his brothers to build. He was considered a transgressor who ventured into forbidden territory. But thereafter, men could travel past any horizon. He had left them a gift they had not conceived and he had opened the roads of the world. 
\vskip 2pt
That man, the unsubmissive and first, stands in the opening chapter of every legend mankind has recorded about is beginning. Prometheus was chained to a rock and torn by vultures---because he had stolen the fire of the gods. Adam was condemned to suffer---because he had eaten the fruit of the tree of knowledge. Whatever the legend, somewhere in the shadows of its memory mankind knew that its glory began with one and that that one paid for his courage.
\vskip 2pt
Throughout the centuries there were men who took first steps down new roads armed with nothing but their own vision. Their goals differed, but they all had this in common:~that the step was first, the road new, the vision unburrowed, and the response they received---hatred. The great creators---the thinkers, the artists, the scientists, the inventors---stood alone against the men of their time. Every great new thought was opposed. Every great new invention was denounced. The first motor was considered foolish. The airplane was considered impossible. The power loom was considered vicious. Anesthesia was considered sinful. But the men of unborrowed vision went ahead. They fought, they suffered and they paid. But they won.\vskip 10pt}
\vskip 4pt
\hfill Howard Roark
\vskip 2pt
\hfill by Ayn Rand, \emph{The Fountainhead} \hspace{-0.335cm}

%% file: PhD_Thesis_TS.bbl
\providecommand{\href}[2]{#2}\begingroup\raggedright\begin{thebibliography}{100}

\bibitem{Spieksma:2023vwl}
T.~F.~M. Spieksma, E.~Cannizzaro, T.~Ikeda, V.~Cardoso, and Y.~Chen,
  ``{Superradiance: Axionic couplings and plasma effects},''
  \href{http://dx.doi.org/10.1103/PhysRevD.108.063013}{{\em Phys. Rev. D}
  {\bfseries 108} no.~6, (2023) 063013},
  \href{http://arxiv.org/abs/2306.16447}{{\ttfamily arXiv:2306.16447 [gr-qc]}}.

\bibitem{Cannizzaro:2024hdg}
E.~Cannizzaro and T.~F.~M. Spieksma, ``{Phenomenology of ultralight bosons
  around compact objects: In-medium suppression},''
  \href{http://dx.doi.org/10.1103/PhysRevD.110.084021}{{\em Phys. Rev. D}
  {\bfseries 110} no.~8, (2024) 084021},
  \href{http://arxiv.org/abs/2406.17016}{{\ttfamily arXiv:2406.17016
  [hep-ph]}}.

\bibitem{Cannizzaro:2024yee}
E.~Cannizzaro, T.~F.~M. Spieksma, V.~Cardoso, and T.~Ikeda, ``{Impact of a
  plasma on the relaxation of black holes},''
  \href{http://dx.doi.org/10.1103/PhysRevD.110.L021302}{{\em Phys. Rev. D}
  {\bfseries 110} no.~2, (2024) L021302},
  \href{http://arxiv.org/abs/2405.05315}{{\ttfamily arXiv:2405.05315 [gr-qc]}}.

\bibitem{Spieksma:2024voy}
T.~F.~M. Spieksma, V.~Cardoso, G.~Carullo, M.~Della~Rocca, and F.~Duque,
  ``{Black Hole Spectroscopy in Environments: Detectability Prospects},''
  \href{http://dx.doi.org/10.1103/PhysRevLett.134.081402}{{\em Phys. Rev.
  Lett.} {\bfseries 134} no.~8, (2025) 081402},
  \href{http://arxiv.org/abs/2409.05950}{{\ttfamily arXiv:2409.05950 [gr-qc]}}.

\bibitem{Tomaselli:2024bdd}
G.~M. Tomaselli, T.~F.~M. Spieksma, and G.~Bertone, ``{Resonant history of
  gravitational atoms in black hole binaries},''
  \href{http://dx.doi.org/10.1103/PhysRevD.110.064048}{{\em Phys. Rev. D}
  {\bfseries 110} no.~6, (2024) 064048},
  \href{http://arxiv.org/abs/2403.03147}{{\ttfamily arXiv:2403.03147 [gr-qc]}}.

\bibitem{Tomaselli:2024dbw}
G.~M. Tomaselli, T.~F.~M. Spieksma, and G.~Bertone, ``{Legacy of Boson Clouds
  on Black Hole Binaries},''
  \href{http://dx.doi.org/10.1103/PhysRevLett.133.121402}{{\em Phys. Rev.
  Lett.} {\bfseries 133} no.~12, (2024) 121402},
  \href{http://arxiv.org/abs/2407.12908}{{\ttfamily arXiv:2407.12908 [gr-qc]}}.

\bibitem{Dyson:2025dlj}
C.~Dyson, T.~F.~M. Spieksma, R.~Brito, M.~van~de Meent, and S.~Dolan,
  ``{Environmental effects in extreme mass ratio inspirals: perturbations to
  the environment in Kerr},'' \href{http://arxiv.org/abs/2501.09806}{{\ttfamily
  arXiv:2501.09806 [gr-qc]}}.

\bibitem{Spieksma:2025wex}
T.~F.~M. Spieksma and E.~Cannizzaro, ``{In the grip of the disk: dragging the
  companion through an AGN},''
  \href{http://arxiv.org/abs/2504.08033}{{\ttfamily arXiv:2504.08033
  [astro-ph.GA]}}.

\bibitem{Cole:2022yzw}
P.~S. Cole, G.~Bertone, A.~Coogan, D.~Gaggero, T.~Karydas, B.~J. Kavanagh,
  T.~F.~M. Spieksma, and G.~M. Tomaselli, ``{Distinguishing environmental
  effects on binary black hole gravitational waveforms},''
  \href{http://dx.doi.org/10.1038/s41550-023-01990-2}{{\em Nature Astron.}
  {\bfseries 7} no.~8, (2023) 943--950},
  \href{http://arxiv.org/abs/2211.01362}{{\ttfamily arXiv:2211.01362 [gr-qc]}}.

\bibitem{Tomaselli:2023ysb}
G.~M. Tomaselli, T.~F.~M. Spieksma, and G.~Bertone, ``{Dynamical friction in
  gravitational atoms},''
  \href{http://dx.doi.org/10.1088/1475-7516/2023/07/070}{{\em JCAP} {\bfseries
  07} (2023) 070}, \href{http://arxiv.org/abs/2305.15460}{{\ttfamily
  arXiv:2305.15460 [gr-qc]}}.

\bibitem{Cardoso:2024jme}
V.~Cardoso, G.~Carullo, M.~De~Amicis, F.~Duque, T.~Katagiri, D.~Pere\~niguez,
  J.~Redondo-Yuste, T.~F.~M. Spieksma, and Z.~Zhong, ``{Hushing black holes:
  Tails in dynamical spacetimes},''
  \href{http://dx.doi.org/10.1103/PhysRevD.109.L121502}{{\em Phys. Rev. D}
  {\bfseries 109} no.~12, (2024) L121502},
  \href{http://arxiv.org/abs/2405.12290}{{\ttfamily arXiv:2405.12290 [gr-qc]}}.

\bibitem{Spieksma:2025sda}
T.~F.~M. Spieksma and E.~Cannizzaro, ``{Axion dissipation in conductive media
  and neutron star superradiance},''
  \href{http://arxiv.org/abs/2503.19978}{{\ttfamily arXiv:2503.19978
  [hep-ph]}}.

\bibitem{Remillard:2006fc}
R.~A. Remillard and J.~E. McClintock, ``{X-ray Properties of Black-Hole
  Binaries},''
  \href{http://dx.doi.org/10.1146/annurev.astro.44.051905.092532}{{\em Ann.
  Rev. Astron. Astrophys.} {\bfseries 44} (2006) 49--92},
  \href{http://arxiv.org/abs/astro-ph/0606352}{{\ttfamily
  arXiv:astro-ph/0606352}}.

\bibitem{Ghez:2008ms}
A.~M. Ghez {\em et~al.}, ``{Measuring Distance and Properties of the Milky
  Way's Central Supermassive Black Hole with Stellar Orbits},''
  \href{http://dx.doi.org/10.1086/592738}{{\em Astrophys. J.} {\bfseries 689}
  (2008) 1044--1062}, \href{http://arxiv.org/abs/0808.2870}{{\ttfamily
  arXiv:0808.2870 [astro-ph]}}.

\bibitem{2009ApJ...692.1075G}
S.~{Gillessen}, F.~{Eisenhauer}, S.~{Trippe}, T.~{Alexander}, R.~{Genzel},
  F.~{Martins}, and T.~{Ott}, ``{Monitoring Stellar Orbits Around the Massive
  Black Hole in the Galactic Center},''
  \href{http://dx.doi.org/10.1088/0004-637X/692/2/1075}{{\em Astroph. J.}
  {\bfseries 692} no.~2, (Feb., 2009) 1075--1109},
  \href{http://arxiv.org/abs/0810.4674}{{\ttfamily arXiv:0810.4674
  [astro-ph]}}.

\bibitem{LIGOScientific:2016aoc}
{\bfseries LIGO Scientific, Virgo} Collaboration, B.~P. Abbott {\em et~al.},
  ``{Observation of Gravitational Waves from a Binary Black Hole Merger},''
  \href{http://dx.doi.org/10.1103/PhysRevLett.116.061102}{{\em Phys. Rev.
  Lett.} {\bfseries 116} no.~6, (2016) 061102},
  \href{http://arxiv.org/abs/1602.03837}{{\ttfamily arXiv:1602.03837 [gr-qc]}}.

\bibitem{EventHorizonTelescope:2019dse}
{\bfseries Event Horizon Telescope} Collaboration, K.~Akiyama {\em et~al.},
  ``{First M87 Event Horizon Telescope Results. I. The Shadow of the
  Supermassive Black Hole},''
  \href{http://dx.doi.org/10.3847/2041-8213/ab0ec7}{{\em Astrophys. J. Lett.}
  {\bfseries 875} (2019) L1}, \href{http://arxiv.org/abs/1906.11238}{{\ttfamily
  arXiv:1906.11238 [astro-ph.GA]}}.

\bibitem{EventHorizonTelescope:2022wkp}
{\bfseries Event Horizon Telescope} Collaboration, K.~Akiyama {\em et~al.},
  ``{First Sagittarius A* Event Horizon Telescope Results. I. The Shadow of the
  Supermassive Black Hole in the Center of the Milky Way},''
  \href{http://dx.doi.org/10.3847/2041-8213/ac6674}{{\em Astrophys. J. Lett.}
  {\bfseries 930} no.~2, (2022) L12},
  \href{http://arxiv.org/abs/2311.08680}{{\ttfamily arXiv:2311.08680
  [astro-ph.HE]}}.

\bibitem{Blandford:1977ds}
R.~D. Blandford and R.~L. Znajek, ``{Electromagnetic extractions of energy from
  Kerr black holes},'' \href{http://dx.doi.org/10.1093/mnras/179.3.433}{{\em
  Mon. Not. Roy. Astron. Soc.} {\bfseries 179} (1977) 433--456}.

\bibitem{King:2003ix}
A.~King, ``{Black holes, galaxy formation, and the M\_BH - sigma relation},''
  \href{http://dx.doi.org/10.1086/379143}{{\em Astrophys. J. Lett.} {\bfseries
  596} (2003) L27--L30},
  \href{http://arxiv.org/abs/astro-ph/0308342}{{\ttfamily
  arXiv:astro-ph/0308342}}.

\bibitem{Volonteri:2010wz}
M.~Volonteri, ``{Formation of Supermassive Black Holes},''
  \href{http://dx.doi.org/10.1007/s00159-010-0029-x}{{\em Astron. Astrophys.
  Rev.} {\bfseries 18} (2010) 279--315},
  \href{http://arxiv.org/abs/1003.4404}{{\ttfamily arXiv:1003.4404
  [astro-ph.CO]}}.

\bibitem{Fabian:2012xr}
A.~C. Fabian, ``{Observational Evidence of AGN Feedback},''
  \href{http://dx.doi.org/10.1146/annurev-astro-081811-125521}{{\em Ann. Rev.
  Astron. Astrophys.} {\bfseries 50} (2012) 455--489},
  \href{http://arxiv.org/abs/1204.4114}{{\ttfamily arXiv:1204.4114
  [astro-ph.CO]}}.

\bibitem{Kormendy:2013dxa}
J.~Kormendy and L.~C. Ho, ``{Coevolution (Or Not) of Supermassive Black Holes
  and Host Galaxies},''
  \href{http://dx.doi.org/10.1146/annurev-astro-082708-101811}{{\em Ann. Rev.
  Astron. Astrophys.} {\bfseries 51} (2013) 511--653},
  \href{http://arxiv.org/abs/1304.7762}{{\ttfamily arXiv:1304.7762
  [astro-ph.CO]}}.

\bibitem{LIGOScientific:2018mvr}
{B.~Abbott {\it et al.}~(LIGO/Virgo Collaboration)}, ``{GWTC-1: A
  Gravitational-Wave Transient Catalog of Compact Binary Mergers Observed by
  LIGO and Virgo during the First and Second Observing Runs},''
  \href{http://dx.doi.org/10.1103/PhysRevX.9.031040}{{\em Phys. Rev.}
  {\bfseries X9} (2019) 031040},
\href{http://arxiv.org/abs/1811.12907}{{\ttfamily arXiv:1811.12907
  [astro-ph.HE]}}.

\bibitem{LIGOScientific:2020ibl}
{\bfseries LIGO Scientific, Virgo} Collaboration, R.~Abbott {\em et~al.},
  ``{GWTC-2: Compact Binary Coalescences Observed by LIGO and Virgo During the
  First Half of the Third Observing Run},''
  \href{http://dx.doi.org/10.1103/PhysRevX.11.021053}{{\em Phys. Rev. X}
  {\bfseries 11} (2021) 021053},
  \href{http://arxiv.org/abs/2010.14527}{{\ttfamily arXiv:2010.14527 [gr-qc]}}.

\bibitem{KAGRA:2021vkt}
{\bfseries KAGRA, VIRGO, LIGO Scientific} Collaboration, R.~Abbott {\em
  et~al.}, ``{GWTC-3: Compact Binary Coalescences Observed by LIGO and Virgo
  during the Second Part of the Third Observing Run},''
  \href{http://dx.doi.org/10.1103/PhysRevX.13.041039}{{\em Phys. Rev. X}
  {\bfseries 13} no.~4, (2023) 041039},
  \href{http://arxiv.org/abs/2111.03606}{{\ttfamily arXiv:2111.03606 [gr-qc]}}.

\bibitem{LIGOScientific:2019fpa}
{\bfseries LIGO Scientific, Virgo} Collaboration, B.~P. Abbott {\em et~al.},
  ``{Tests of General Relativity with the Binary Black Hole Signals from the
  LIGO-Virgo Catalog GWTC-1},''
  \href{http://dx.doi.org/10.1103/PhysRevD.100.104036}{{\em Phys. Rev. D}
  {\bfseries 100} no.~10, (2019) 104036},
  \href{http://arxiv.org/abs/1903.04467}{{\ttfamily arXiv:1903.04467 [gr-qc]}}.

\bibitem{LIGOScientific:2020tif}
{\bfseries LIGO Scientific, Virgo} Collaboration, R.~Abbott {\em et~al.},
  ``{Tests of general relativity with binary black holes from the second
  LIGO-Virgo gravitational-wave transient catalog},''
  \href{http://dx.doi.org/10.1103/PhysRevD.103.122002}{{\em Phys. Rev. D}
  {\bfseries 103} no.~12, (2021) 122002},
  \href{http://arxiv.org/abs/2010.14529}{{\ttfamily arXiv:2010.14529 [gr-qc]}}.

\bibitem{LIGOScientific:2021sio}
{\bfseries LIGO Scientific, VIRGO, KAGRA} Collaboration, R.~Abbott {\em
  et~al.}, ``{Tests of General Relativity with GWTC-3},''
  \href{http://arxiv.org/abs/2112.06861}{{\ttfamily arXiv:2112.06861 [gr-qc]}}.

\bibitem{LIGOScientific:2017vwq}
{\bfseries LIGO Scientific, Virgo} Collaboration, B.~P. Abbott {\em et~al.},
  ``{GW170817: Observation of Gravitational Waves from a Binary Neutron Star
  Inspiral},'' \href{http://dx.doi.org/10.1103/PhysRevLett.119.161101}{{\em
  Phys. Rev. Lett.} {\bfseries 119} no.~16, (2017) 161101},
  \href{http://arxiv.org/abs/1710.05832}{{\ttfamily arXiv:1710.05832 [gr-qc]}}.

\bibitem{LIGOScientific:2017ync}
{\bfseries LIGO Scientific, Virgo, Fermi GBM, INTEGRAL, IceCube, AstroSat
  Cadmium Zinc Telluride Imager Team, IPN, Insight-Hxmt, ANTARES, Swift, AGILE
  Team, 1M2H Team, Dark Energy Camera GW-EM, DES, DLT40, GRAWITA, Fermi-LAT,
  ATCA, ASKAP, Las Cumbres Observatory Group, OzGrav, DWF (Deeper Wider Faster
  Program), AST3, CAASTRO, VINROUGE, MASTER, J-GEM, GROWTH, JAGWAR,
  CaltechNRAO, TTU-NRAO, NuSTAR, Pan-STARRS, MAXI Team, TZAC Consortium, KU,
  Nordic Optical Telescope, ePESSTO, GROND, Texas Tech University, SALT Group,
  TOROS, BOOTES, MWA, CALET, IKI-GW Follow-up, H.E.S.S., LOFAR, LWA, HAWC,
  Pierre Auger, ALMA, Euro VLBI Team, Pi of Sky, Chandra Team at McGill
  University, DFN, ATLAS Telescopes, High Time Resolution Universe Survey,
  RIMAS, RATIR, SKA South Africa/MeerKAT} Collaboration, B.~P. Abbott {\em
  et~al.}, ``{Multi-messenger Observations of a Binary Neutron Star Merger},''
  \href{http://dx.doi.org/10.3847/2041-8213/aa91c9}{{\em Astrophys. J. Lett.}
  {\bfseries 848} no.~2, (2017) L12},
  \href{http://arxiv.org/abs/1710.05833}{{\ttfamily arXiv:1710.05833
  [astro-ph.HE]}}.

\bibitem{Cowperthwaite:2017dyu}
P.~Cowperthwaite {\em et~al.}, ``{The Electromagnetic Counterpart of the Binary
  Neutron Star Merger LIGO/Virgo GW170817. II. UV, Optical, and Near-infrared
  Light Curves and Comparison to Kilonova Models},''
  \href{http://dx.doi.org/10.3847/2041-8213/aa8fc7}{{\em Astrophys. J. Lett.}
  {\bfseries 848} (2017) L17},
  \href{http://arxiv.org/abs/1710.05840}{{\ttfamily arXiv:1710.05840
  [astro-ph.HE]}}.

\bibitem{Troja:2017nqp}
E.~Troja {\em et~al.}, ``{The X-ray Counterpart to the Gravitational Wave Event
  GW 170817},'' \href{http://dx.doi.org/10.1038/nature24290}{{\em Nature}
  {\bfseries 551} (2017) 71--74},
  \href{http://arxiv.org/abs/1710.05433}{{\ttfamily arXiv:1710.05433
  [astro-ph.HE]}}.

\bibitem{Kasen:2017sxr}
D.~Kasen, B.~Metzger, J.~Barnes, E.~Quataert, and E.~Ramirez-Ruiz, ``{Origin of
  the heavy elements in binary neutron-star mergers from a gravitational wave
  event},'' \href{http://dx.doi.org/10.1038/nature24453}{{\em Nature}
  {\bfseries 551} (2017) 80}, \href{http://arxiv.org/abs/1710.05463}{{\ttfamily
  arXiv:1710.05463 [astro-ph.HE]}}.

\bibitem{Pian:2017gtc}
E.~Pian {\em et~al.}, ``{Spectroscopic identification of r-process
  nucleosynthesis in a double neutron star merger},''
  \href{http://dx.doi.org/10.1038/nature24298}{{\em Nature} {\bfseries 551}
  (2017) 67--70}, \href{http://arxiv.org/abs/1710.05858}{{\ttfamily
  arXiv:1710.05858 [astro-ph.HE]}}.

\bibitem{LIGOScientific:2017zic}
{\bfseries LIGO Scientific, Virgo, Fermi-GBM, INTEGRAL} Collaboration, B.~P.
  Abbott {\em et~al.}, ``{Gravitational Waves and Gamma-rays from a Binary
  Neutron Star Merger: GW170817 and GRB 170817A},''
  \href{http://dx.doi.org/10.3847/2041-8213/aa920c}{{\em Astrophys. J. Lett.}
  {\bfseries 848} no.~2, (2017) L13},
  \href{http://arxiv.org/abs/1710.05834}{{\ttfamily arXiv:1710.05834
  [astro-ph.HE]}}.

\bibitem{LIGOScientific:2017adf}
{\bfseries LIGO Scientific, Virgo, 1M2H, Dark Energy Camera GW-E, DES, DLT40,
  Las Cumbres Observatory, VINROUGE, MASTER} Collaboration, B.~P. Abbott {\em
  et~al.}, ``{A gravitational-wave standard siren measurement of the Hubble
  constant},'' \href{http://dx.doi.org/10.1038/nature24471}{{\em Nature}
  {\bfseries 551} no.~7678, (2017) 85--88},
  \href{http://arxiv.org/abs/1710.05835}{{\ttfamily arXiv:1710.05835
  [astro-ph.CO]}}.

\bibitem{Sadeghian:2013laa}
L.~Sadeghian, F.~Ferrer, and C.~M. Will, ``{Dark Matter Distributions around
  Massive Black Holes: A General Relativistic Analysis},''
  \href{http://dx.doi.org/10.1103/PhysRevD.88.063522}{{\em Phys. Rev. D}
  {\bfseries 88} (2013) 063522},
  \href{http://arxiv.org/abs/1305.2619}{{\ttfamily arXiv:1305.2619
  [astro-ph.GA]}}.

\bibitem{Colpi:2024xhw}
M.~Colpi {\em et~al.}, ``{LISA Definition Study Report},''
  \href{http://arxiv.org/abs/2402.07571}{{\ttfamily arXiv:2402.07571
  [astro-ph.CO]}}.

\bibitem{Cornish:2005qw}
N.~J. Cornish and J.~Crowder, ``{LISA data analysis using MCMC methods},''
  \href{http://dx.doi.org/10.1103/PhysRevD.72.043005}{{\em Phys. Rev. D}
  {\bfseries 72} (2005) 043005},
  \href{http://arxiv.org/abs/gr-qc/0506059}{{\ttfamily arXiv:gr-qc/0506059}}.

\bibitem{Littenberg:2023xpl}
T.~B. Littenberg and N.~J. Cornish, ``{Prototype global analysis of LISA data
  with multiple source types},''
  \href{http://dx.doi.org/10.1103/PhysRevD.107.063004}{{\em Phys. Rev. D}
  {\bfseries 107} no.~6, (2023) 063004},
  \href{http://arxiv.org/abs/2301.03673}{{\ttfamily arXiv:2301.03673 [gr-qc]}}.

\bibitem{Barausse:2014tra}
E.~Barausse, V.~Cardoso, and P.~Pani, ``{Can environmental effects spoil
  precision gravitational-wave astrophysics?},''
  \href{http://dx.doi.org/10.1103/PhysRevD.89.104059}{{\em Phys. Rev. D}
  {\bfseries 89} no.~10, (2014) 104059},
  \href{http://arxiv.org/abs/1404.7149}{{\ttfamily arXiv:1404.7149 [gr-qc]}}.

\bibitem{Bartos:2016dgn}
I.~Bartos, B.~Kocsis, Z.~Haiman, and S.~M\'arka, ``{Rapid and Bright
  Stellar-mass Binary Black Hole Mergers in Active Galactic Nuclei},''
  \href{http://dx.doi.org/10.3847/1538-4357/835/2/165}{{\em Astrophys. J.}
  {\bfseries 835} no.~2, (2017) 165},
  \href{http://arxiv.org/abs/1602.03831}{{\ttfamily arXiv:1602.03831
  [astro-ph.HE]}}.

\bibitem{Stone:2016wzz}
N.~C. Stone, B.~D. Metzger, and Z.~Haiman, ``{Assisted inspirals of stellar
  mass black holes embedded in AGN discs: solving the \textquoteleft{}final au
  problem\textquoteright{}},''
  \href{http://dx.doi.org/10.1093/mnras/stw2260}{{\em Mon. Not. Roy. Astron.
  Soc.} {\bfseries 464} no.~1, (2017) 946--954},
  \href{http://arxiv.org/abs/1602.04226}{{\ttfamily arXiv:1602.04226
  [astro-ph.GA]}}.

\bibitem{Tagawa:2019osr}
H.~Tagawa, Z.~Haiman, and B.~Kocsis, ``{Formation and Evolution of Compact
  Object Binaries in AGN Disks},''
  \href{http://dx.doi.org/10.3847/1538-4357/ab9b8c}{{\em Astrophys. J.}
  {\bfseries 898} no.~1, (2020) 25},
  \href{http://arxiv.org/abs/1912.08218}{{\ttfamily arXiv:1912.08218
  [astro-ph.GA]}}.

\bibitem{Pan:2021ksp}
Z.~Pan and H.~Yang, ``{Formation Rate of Extreme Mass Ratio Inspirals in Active
  Galactic Nuclei},'' \href{http://dx.doi.org/10.1103/PhysRevD.103.103018}{{\em
  Phys. Rev. D} {\bfseries 103} no.~10, (2021) 103018},
  \href{http://arxiv.org/abs/2101.09146}{{\ttfamily arXiv:2101.09146
  [astro-ph.HE]}}.

\bibitem{Derdzinski:2022ltb}
A.~Derdzinski and L.~Mayer, ``{In situ extreme mass ratio inspirals via
  subparsec formation and migration of stars in thin, gravitationally unstable
  AGN discs},'' \href{http://dx.doi.org/10.1093/mnras/stad749}{{\em Mon. Not.
  Roy. Astron. Soc.} {\bfseries 521} no.~3, (2023) 4522--4543},
  \href{http://arxiv.org/abs/2205.10382}{{\ttfamily arXiv:2205.10382
  [astro-ph.GA]}}.

\bibitem{Bertone:2004pz}
G.~Bertone, D.~Hooper, and J.~Silk, ``{Particle Dark Matter: Evidence,
  Candidates and Constraints},''
  \href{http://dx.doi.org/10.1016/j.physrep.2004.08.031}{{\em Phys. Rept.}
  {\bfseries 405} (2005) 279--390},
  \href{http://arxiv.org/abs/hep-ph/0404175}{{\ttfamily arXiv:hep-ph/0404175}}.

\bibitem{Cirelli:2024ssz}
M.~Cirelli, A.~Strumia, and J.~Zupan, ``{Dark Matter},''
  \href{http://arxiv.org/abs/2406.01705}{{\ttfamily arXiv:2406.01705
  [hep-ph]}}.

\bibitem{Navarro:1995iw}
J.~F. Navarro, C.~S. Frenk, and S.~D.~M. White, ``{The Structure of cold dark
  matter halos},'' \href{http://dx.doi.org/10.1086/177173}{{\em Astrophys. J.}
  {\bfseries 462} (1996) 563--575},
  \href{http://arxiv.org/abs/astro-ph/9508025}{{\ttfamily
  arXiv:astro-ph/9508025}}.

\bibitem{Gondolo:1999ef}
P.~Gondolo and J.~Silk, ``{Dark Matter Annihilation at the Galactic Center},''
  \href{http://dx.doi.org/10.1103/PhysRevLett.83.1719}{{\em Phys. Rev. Lett.}
  {\bfseries 83} (1999) 1719--1722},
  \href{http://arxiv.org/abs/astro-ph/9906391}{{\ttfamily
  arXiv:astro-ph/9906391}}.

\bibitem{Ferrer:2017xwm}
F.~Ferrer, A.~M. da~Rosa, and C.~M. Will, ``{Dark Matter Spikes in the Vicinity
  of Kerr Black Holes},''
  \href{http://dx.doi.org/10.1103/PhysRevD.96.083014}{{\em Phys. Rev. D}
  {\bfseries 96} (2017) 083014},
  \href{http://arxiv.org/abs/1707.06302}{{\ttfamily arXiv:1707.06302
  [astro-ph.CO]}}.

\bibitem{Hu:2000ke}
W.~Hu, R.~Barkana, and A.~Gruzinov, ``{Cold and Fuzzy Dark Matter},''
  \href{http://dx.doi.org/10.1103/PhysRevLett.85.1158}{{\em Phys. Rev. Lett.}
  {\bfseries 85} (2000) 1158--1161},
  \href{http://arxiv.org/abs/astro-ph/0003365}{{\ttfamily
  arXiv:astro-ph/0003365}}.

\bibitem{Marsh:2015xka}
D.~J.~E. Marsh, ``{Axion Cosmology},''
  \href{http://dx.doi.org/10.1016/j.physrep.2016.06.005}{{\em Phys. Rept.}
  {\bfseries 643} (2016) 1--79},
  \href{http://arxiv.org/abs/1510.07633}{{\ttfamily arXiv:1510.07633
  [astro-ph.CO]}}.

\bibitem{Hui:2016ltb}
L.~Hui, J.~Ostriker, S.~Tremaine, and E.~Witten, ``{Ultralight Scalars as
  Cosmological Dark Matter},''
  \href{http://dx.doi.org/10.1103/PhysRevD.95.043541}{{\em Phys. Rev. D}
  {\bfseries 95} (2017) 043541},
\href{http://arxiv.org/abs/1610.08297}{{\ttfamily arXiv:1610.08297
  [astro-ph.CO]}}.

\bibitem{Peccei:1977hh}
R.~D. Peccei and H.~R. Quinn, ``{CP Conservation in the Presence of
  Instantons},''
\href{http://dx.doi.org/10.1103/PhysRevLett.38.1440}{{\em Phys. Rev. Lett.}
  {\bfseries 38} (1977) 1440--1443}.

\bibitem{Weinberg:1977ma}
S.~Weinberg, ``{A New Light Boson?},''
\href{http://dx.doi.org/10.1103/PhysRevLett.40.223}{{\em Phys. Rev. Lett.}
  {\bfseries 40} (1978) 223--226}.

\bibitem{Wilczek:1977pj}
F.~Wilczek, ``{Problem of Strong P and T Invariance in the Presence of
  Instantons},''
\href{http://dx.doi.org/10.1103/PhysRevLett.40.279}{{\em Phys. Rev. Lett.}
  {\bfseries 40} (1978) 279--282}.

\bibitem{ZelDovich1971}
Y.~B. {Zel'Dovich}, ``{Generation of Waves by a Rotating Body},'' {\em Soviet
  Journal of Experimental and Theoretical Physics Letters} {\bfseries 14}
  (Aug., 1971) 180.

\bibitem{ZelDovich1972}
Y.~B. {Zel'Dovich}, ``{Amplification of Cylindrical Electromagnetic Waves
  Reflected from a Rotating Body},'' {\em Soviet Journal of Experimental and
  Theoretical Physics} {\bfseries 35} (Jan., 1972) 1085.

\bibitem{Starobinsky:1973aij}
A.~A. Starobinsky, ``{Amplification of waves reflected from a rotating ``black
  hole''\,},'' {\em Sov. Phys. JETP} {\bfseries 37} no.~1, (1973) 28--32.

\bibitem{Arvanitaki:2009fg}
A.~Arvanitaki, S.~Dimopoulos, S.~Dubovsky, N.~Kaloper, and J.~March-Russell,
  ``{String Axiverse},''
  \href{http://dx.doi.org/10.1103/PhysRevD.81.123530}{{\em Phys. Rev. D}
  {\bfseries 81} (2010) 123530},
  \href{http://arxiv.org/abs/0905.4720}{{\ttfamily arXiv:0905.4720 [hep-th]}}.

\bibitem{Brito:2015oca}
R.~Brito, V.~Cardoso, and P.~Pani, ``{Superradiance},''
  \href{http://dx.doi.org/10.1007/978-3-319-19000-6}{{\em Lect. Notes Phys.}
  {\bfseries 906} (2015) 1},
\href{http://arxiv.org/abs/1501.06570}{{\ttfamily arXiv:1501.06570 [gr-qc]}}.

\bibitem{Baumann:2018vus}
D.~Baumann, H.~S. Chia, and R.~A. Porto, ``{Probing Ultralight Bosons with
  Binary Black Holes},''
  \href{http://dx.doi.org/10.1103/PhysRevD.99.044001}{{\em Phys. Rev. D}
  {\bfseries 99} no.~4, (2019) 044001},
  \href{http://arxiv.org/abs/1804.03208}{{\ttfamily arXiv:1804.03208 [gr-qc]}}.

\bibitem{Baumann:2019ztm}
D.~Baumann, H.~S. Chia, R.~A. Porto, and J.~Stout, ``{Gravitational Collider
  Physics},'' \href{http://dx.doi.org/10.1103/PhysRevD.101.083019}{{\em Phys.
  Rev. D} {\bfseries 101} (2020) 083019},
  \href{http://arxiv.org/abs/1912.04932}{{\ttfamily arXiv:1912.04932 [gr-qc]}}.

\bibitem{Baumann:2021fkf}
D.~Baumann, G.~Bertone, J.~Stout, and G.~M. Tomaselli, ``{Ionization of
  gravitational atoms},''
  \href{http://dx.doi.org/10.1103/PhysRevD.105.115036}{{\em Phys. Rev. D}
  {\bfseries 105} no.~11, (2022) 115036},
  \href{http://arxiv.org/abs/2112.14777}{{\ttfamily arXiv:2112.14777 [gr-qc]}}.

\bibitem{ET:2019dnz}
{\bfseries ET} Collaboration, M.~Maggiore {\em et~al.}, ``{Science Case for the
  Einstein Telescope},''
  \href{http://dx.doi.org/10.1088/1475-7516/2020/03/050}{{\em JCAP} {\bfseries
  03} (2020) 050}, \href{http://arxiv.org/abs/1912.02622}{{\ttfamily
  arXiv:1912.02622 [astro-ph.CO]}}.

\bibitem{Khalvati:2024tzz}
H.~Khalvati, A.~Santini, F.~Duque, L.~Speri, J.~Gair, H.~Yang, and R.~Brito,
  ``{Impact of relativistic waveforms in LISA\textquoteright{}s science
  objectives with extreme-mass-ratio inspirals},''
  \href{http://dx.doi.org/10.1103/PhysRevD.111.082010}{{\em Phys. Rev. D}
  {\bfseries 111} no.~8, (2025) 082010},
  \href{http://arxiv.org/abs/2410.17310}{{\ttfamily arXiv:2410.17310 [gr-qc]}}.

\bibitem{Einstein:1915ca}
A.~Einstein, ``{The Field Equations of Gravitation},'' {\em Sitzungsber.
  Preuss. Akad. Wiss. Berlin (Math. Phys. )} {\bfseries 1915} (1915) 844--847.

\bibitem{Einstein:1916vd}
A.~Einstein, ``{The foundation of the general theory of relativity.},''
  \href{http://dx.doi.org/10.1002/andp.19163540702}{{\em Annalen Phys.}
  {\bfseries 49} no.~7, (1916) 769--822}.

\bibitem{Misner:1973prb}
C.~W. Misner, K.~S. Thorne, and J.~A. Wheeler, {\em {Gravitation}}.
\newblock W. H. Freeman, San Francisco, 1973.

\bibitem{Wald:1984rg}
R.~M. Wald,
  \href{http://dx.doi.org/10.7208/chicago/9780226870373.001.0001}{{\em {General
  Relativity}}}.
\newblock Chicago Univ. Pr., Chicago, USA, 1984.

\bibitem{Schutz:1985jx}
B.~F. Schutz, \href{http://dx.doi.org/10.1017/CBO9780511984181}{{\em {A first
  course in general relativity}}}.
\newblock Cambridge Univ. Pr., Cambridge, UK, 1985.

\bibitem{Schwarzschild1916}
K.~{Schwarzschild}, ``{{\"U}ber das Gravitationsfeld eines Massenpunktes nach
  der Einsteinschen Theorie},'' {\em Sitzungsberichte der K{\"o}niglich
  Preu{\ss}ischen Akademie der Wissenschaften (Berlin} (Jan., 1916) 189--196.

\bibitem{Droste1917}
J.~{Droste}, ``{The Field of a Single Centre in Einstein's Theory of
  Gravitation, and the Motion of a Particle in That Field},'' {\em Koninklijke
  Nederlandse Akademie van Wetenschappen Proceedings Series B Physical
  Sciences} {\bfseries 19} (Jan, 1917) 197--215.

\bibitem{Finkelstein:1958zz}
D.~Finkelstein, ``{Past-Future Asymmetry of the Gravitational Field of a Point
  Particle},'' \href{http://dx.doi.org/10.1103/PhysRev.110.965}{{\em Phys.
  Rev.} {\bfseries 110} (1958) 965--967}.

\bibitem{Birkhoff}
G.~D. {Birkhoff} and R.~E. {Langer}, {\em {Relativity and Modern Physics}}.
\newblock Harvard University Press, 1923.

\bibitem{Jebsen}
J.~T. {Jebsen}, ``{On the General Spherically Symmetric Solutions of Einstein's
  Gravitational Equations in Vacuo.},'' {\em Arkiv for Matematik, Astronomi och
  Fysik} {\bfseries 15} (Jan., 1921) 18.

\bibitem{Einstein1916mercury}
A.~Einstein, ``{Die Grundlage der allgemeinen Relativit{\"a}tstheorie},''
  \href{http://dx.doi.org/10.1002/andp.19163540702}{{\em Annalen der Physik}
  {\bfseries 354} (1916) 769--822}.

\bibitem{Dyson:1920cwa}
F.~W. Dyson, A.~S. Eddington, and C.~Davidson, ``{A Determination of the
  Deflection of Light by the Sun's Gravitational Field, from Observations Made
  at the Total Eclipse of May 29, 1919},''
  \href{http://dx.doi.org/10.1098/rsta.1920.0009}{{\em Phil. Trans. Roy. Soc.
  Lond. A} {\bfseries 220} (1920) 291--333}.

\bibitem{Kerr1963}
R.~P. Kerr, ``{Gravitational Field of a Spinning Mass as an Example of
  Algebraically Special Metrics},''
  \href{http://dx.doi.org/10.1103/PhysRevLett.11.237}{{\em Phys. Rev. Lett.}
  {\bfseries 11} (Sep, 1963) 237--238}.

\bibitem{Newman:1961qr}
E.~Newman and R.~Penrose, ``{An Approach to gravitational radiation by a method
  of spin coefficients},'' \href{http://dx.doi.org/10.1063/1.1724257}{{\em J.
  Math. Phys.} {\bfseries 3} (1962) 566--578}.

\bibitem{BL1967}
R.~H. Boyer and R.~W. Lindquist, ``{Maximal Analytic Extension of the Kerr
  Metric},'' \href{http://dx.doi.org/10.1063/1.1705193}{{\em Journal of
  Mathematical Physics} {\bfseries 8} (1967) 265--281}.

\bibitem{Stephani_Kramer_MacCallum_Hoenselaers_Herlt_2003}
H.~Stephani, D.~Kramer, M.~MacCallum, C.~Hoenselaers, and E.~Herlt, {\em Exact
  Solutions of Einstein’s Field Equations}.
\newblock Cambridge Monographs on Mathematical Physics. Cambridge University
  Press, 2~ed., 2003.

\bibitem{Israel:1967wq}
W.~Israel, ``{Event horizons in static vacuum space-times},''
  \href{http://dx.doi.org/10.1103/PhysRev.164.1776}{{\em Phys. Rev.} {\bfseries
  164} (1967) 1776--1779}.

\bibitem{Carter:1971zc}
B.~Carter, ``{Axisymmetric Black Hole Has Only Two Degrees of Freedom},''
\href{http://dx.doi.org/10.1103/PhysRevLett.26.331}{{\em Phys. Rev. Lett.}
  {\bfseries 26} (1971) 331}.

\bibitem{Robinson:1975bv}
D.~Robinson, ``{Uniqueness of the Kerr black hole},''
\href{http://dx.doi.org/10.1103/PhysRevLett.34.905}{{\em Phys. Rev. Lett.}
  {\bfseries 34} (1975) 905--906}.

\bibitem{Preskill:1984gd}
J.~Preskill, ``{Magnetic monopoles},''
  \href{http://dx.doi.org/10.1146/annurev.ns.34.120184.002333}{{\em Ann. Rev.
  Nucl. Part. Sci.} {\bfseries 34} (1984) 461--530}.

\bibitem{Gibbons:1975kk}
G.~W. Gibbons, ``{Vacuum Polarization and the Spontaneous Loss of Charge by
  Black Holes},'' \href{http://dx.doi.org/10.1007/BF01609829}{{\em Commun.
  Math. Phys.} {\bfseries 44} (1975) 245--264}.

\bibitem{1969ApJ...157..869G}
P.~{Goldreich} and W.~H. {Julian}, ``{Pulsar Electrodynamics},''
  \href{http://dx.doi.org/10.1086/150119}{{\em Astrophys. J.} {\bfseries 157}
  (Aug., 1969) 869}.

\bibitem{Eardley:1975kp}
D.~M. Eardley and W.~H. Press, ``{Astrophysical processes near black holes},''
  \href{http://dx.doi.org/10.1146/annurev.aa.13.090175.002121}{{\em Ann. Rev.
  Astron. Astrophys.} {\bfseries 13} (1975) 381--422}.

\bibitem{Reissner:1916cle}
H.~Reissner, ``{\"Uber die Eigengravitation des elektrischen Feldes nach der
  Einsteinschen Theorie},''
  \href{http://dx.doi.org/10.1002/andp.19163550905}{{\em Annalen Phys.}
  {\bfseries 355} no.~9, (1916) 106--120}.

\bibitem{1918KNAB...20.1238N}
G.~{Nordstr{\"o}m}, ``{On the Energy of the Gravitation field in Einstein's
  Theory},'' {\em Koninklijke Nederlandse Akademie van Wetenschappen
  Proceedings Series B Physical Sciences} {\bfseries 20} (Jan., 1918)
  1238--1245.

\bibitem{Newman:1965my}
E.~T. Newman, R.~Couch, K.~Chinnapared, A.~Exton, A.~Prakash, and R.~Torrence,
  ``{Metric of a Rotating, Charged Mass},''
  \href{http://dx.doi.org/10.1063/1.1704351}{{\em J. Math. Phys.} {\bfseries 6}
  (1965) 918--919}.

\bibitem{MTB}
S.~Chandrasekhar, {\em The Mathematical Theory of Black Holes}.
\newblock Oxford University Press, New York, 1983.

\bibitem{Lens-Thirring1918}
J.~{Lense} and H.~{Thirring}, ``{{\"U}ber den Einflu{\ss} der Eigenrotation der
  Zentralk{\"o}rper auf die Bewegung der Planeten und Monde nach der
  Einsteinschen Gravitationstheorie},'' {\em Physikalische Zeitschrift}
  {\bfseries 19} (Jan., 1918) 156.

\bibitem{Penrose:1971uk}
R.~Penrose and R.~M. Floyd, ``{Extraction of rotational energy from a black
  hole},'' \href{http://dx.doi.org/10.1038/physci229177a0}{{\em Nature}
  {\bfseries 229} (1971) 177--179}.

\bibitem{Bardeen:1973gs}
J.~M. Bardeen, B.~Carter, and S.~W. Hawking, ``{The Four laws of black hole
  mechanics},'' \href{http://dx.doi.org/10.1007/BF01645742}{{\em Commun. Math.
  Phys.} {\bfseries 31} (1973) 161--170}.

\bibitem{Christodoulou:1970}
D.~Christodoulou, ``{Reversible and Irreversible Transformations in Black-Hole
  Physics},''
\href{http://dx.doi.org/10.1103/PhysRevLett.25.1596}{{\em Phys. Rev. Lett.}
  {\bfseries 25} (1970) 1596}.

\bibitem{Giacconi:1962zz}
R.~Giacconi, H.~Gursky, F.~R. Paolini, and B.~B. Rossi, ``{Evidence for x Rays
  From Sources Outside the Solar System},''
  \href{http://dx.doi.org/10.1103/PhysRevLett.9.439}{{\em Phys. Rev. Lett.}
  {\bfseries 9} (1962) 439--443}.

\bibitem{1967ApJ...148L.119G}
R.~{Giacconi}, P.~{Gorenstein}, H.~{Gursky}, and J.~R. {Waters}, ``{An X-Ray
  Survey of the Cygnus Region},'' \href{http://dx.doi.org/10.1086/180028}{{\em
  Astrophys. J. Lett.} {\bfseries 148} (June, 1967) L119}.

\bibitem{1972Natur.235...37W}
B.~L. {Webster} and P.~{Murdin}, ``{Cygnus X-1-a Spectroscopic Binary with a
  Heavy Companion ?},'' \href{http://dx.doi.org/10.1038/235037a0}{{\em Nature}
  {\bfseries 235} no.~5332, (Jan., 1972) 37--38}.

\bibitem{2016ApJ...830...17B}
A.~{Boehle}, A.~M. {Ghez}, R.~{Sch{\"o}del}, L.~{Meyer}, S.~{Yelda},
  S.~{Albers}, G.~D. {Martinez}, E.~E. {Becklin}, T.~{Do}, J.~R. {Lu},
  K.~{Matthews}, M.~R. {Morris}, B.~{Sitarski}, and G.~{Witzel}, ``{An Improved
  Distance and Mass Estimate for Sgr A* from a Multistar Orbit Analysis},''
  \href{http://dx.doi.org/10.3847/0004-637X/830/1/17}{{\em Astroph. J.}
  {\bfseries 830} no.~1, (Oct., 2016) 17},
  \href{http://arxiv.org/abs/1607.05726}{{\ttfamily arXiv:1607.05726
  [astro-ph.GA]}}.

\bibitem{2021NatRP...3..732V}
M.~{Volonteri}, M.~{Habouzit}, and M.~{Colpi}, ``{The origins of massive black
  holes},'' \href{http://dx.doi.org/10.1038/s42254-021-00364-9}{{\em Nature
  Reviews Physics} {\bfseries 3} no.~11, (Sept., 2021) 732--743},
  \href{http://arxiv.org/abs/2110.10175}{{\ttfamily arXiv:2110.10175
  [astro-ph.GA]}}.

\bibitem{Maggiore:2007ulw}
M.~Maggiore,
  \href{http://dx.doi.org/10.1093/acprof:oso/9780198570745.001.0001}{{\em
  {Gravitational Waves. Vol. 1: Theory and Experiments}}}.
\newblock Oxford University Press, 2007.

\bibitem{Maggiore:2018sht}
M.~Maggiore, \href{http://dx.doi.org/10.1093/oso/9780198570899.001.0001}{{\em
  {Gravitational Waves. Vol. 2: Astrophysics and Cosmology}}}.
\newblock Oxford University Press, 2018.

\bibitem{Einstein1918}
A.~{Einstein}, ``{{\"U}ber Gravitationswellen},'' {\em Sitzungsberichte der
  K{\"o}niglich Preu{\ss}ischen Akademie der Wissenschaften (Berlin} (Jan.,
  1918) 154--167.

\bibitem{EinsteinRosen1937}
A.~Einstein and N.~Rosen, ``{On Gravitational Waves},''
  \href{http://dx.doi.org/10.1016/S0016-0032(37)90583-0}{{\em J. Franklin
  Inst.} {\bfseries 223} (1937) 43--54}.

\bibitem{Hulse:1974eb}
R.~Hulse and J.~Taylor, ``{Discovery of a Pulsar in a Binary System},''
\href{http://dx.doi.org/10.1086/181708}{{\em Astrophys. J.} {\bfseries 195}
  (1975) L51}.

\bibitem{Taylor1982}
J.~H. {Taylor} and J.~M. {Weisberg}, ``{A New Test of General Relativity -
  Gravitational Radiation and the Binary Pulsar PSR 1913+16},''
  \href{http://dx.doi.org/10.1086/159690}{{\em Astrophys. J.} {\bfseries 253}
  (Feb., 1982) 908--920}.

\bibitem{Purrer:2019jcp}
M.~P\"urrer and C.-J. Haster, ``{Gravitational waveform accuracy requirements
  for future ground-based detectors},''
  \href{http://dx.doi.org/10.1103/PhysRevResearch.2.023151}{{\em Phys. Rev.
  Res.} {\bfseries 2} no.~2, (2020) 023151},
  \href{http://arxiv.org/abs/1912.10055}{{\ttfamily arXiv:1912.10055 [gr-qc]}}.

\bibitem{LISAConsortiumWaveformWorkingGroup:2023arg}
{\bfseries LISA Consortium Waveform Working Group} Collaboration, N.~Afshordi
  {\em et~al.}, ``{Waveform Modelling for the Laser Interferometer Space
  Antenna},'' \href{http://arxiv.org/abs/2311.01300}{{\ttfamily
  arXiv:2311.01300 [gr-qc]}}.

\bibitem{Peters:1963ux}
P.~C. Peters and J.~Mathews, ``{Gravitational radiation from point masses in a
  Keplerian orbit},'' \href{http://dx.doi.org/10.1103/PhysRev.131.435}{{\em
  Phys. Rev.} {\bfseries 131} (1963) 435--439}.

\bibitem{Peters:1964zz}
P.~Peters, ``{Gravitational Radiation and the Motion of Two Point Masses},''
  \href{http://dx.doi.org/10.1103/PhysRev.136.B1224}{{\em Phys. Rev.}
  {\bfseries 136} (1964) B1224--B1232}.

\bibitem{Blanchet:2023bwj}
L.~Blanchet, G.~Faye, Q.~Henry, F.~Larrouturou, and D.~Trestini,
  ``{Gravitational-Wave Phasing of Quasicircular Compact Binary Systems to the
  Fourth-and-a-Half Post-Newtonian Order},''
  \href{http://dx.doi.org/10.1103/PhysRevLett.131.121402}{{\em Phys. Rev.
  Lett.} {\bfseries 131} no.~12, (2023) 121402},
  \href{http://arxiv.org/abs/2304.11185}{{\ttfamily arXiv:2304.11185 [gr-qc]}}.

\bibitem{Wardell:2021fyy}
B.~Wardell, A.~Pound, N.~Warburton, J.~Miller, L.~Durkan, and A.~Le~Tiec,
  ``{Gravitational Waveforms for Compact Binaries from Second-Order Self-Force
  Theory},'' \href{http://dx.doi.org/10.1103/PhysRevLett.130.241402}{{\em Phys.
  Rev. Lett.} {\bfseries 130} no.~24, (2023) 241402},
  \href{http://arxiv.org/abs/2112.12265}{{\ttfamily arXiv:2112.12265 [gr-qc]}}.

\bibitem{Bardeen1972}
J.~M. {Bardeen}, W.~H. {Press}, and S.~A. {Teukolsky}, ``{Rotating Black Holes:
  Locally Nonrotating Frames, Energy Extraction, and Scalar Synchrotron
  Radiation},'' \href{http://dx.doi.org/10.1086/151796}{{\em Astrophys. J.}
  {\bfseries 178} (Dec., 1972) 347--370}.

\bibitem{PhysRevD.50.3816}
C.~Cutler, D.~Kennefick, and E.~Poisson, ``Gravitational radiation reaction for
  bound motion around a schwarzschild black hole,''
  \href{http://dx.doi.org/10.1103/PhysRevD.50.3816}{{\em Phys. Rev. D}
  {\bfseries 50} (Sep, 1994) 3816--3835}.

\bibitem{Hughes:1999bq}
S.~A. Hughes, ``{The Evolution of circular, nonequatorial orbits of Kerr black
  holes due to gravitational wave emission},''
  \href{http://dx.doi.org/10.1103/PhysRevD.65.069902}{{\em Phys. Rev. D}
  {\bfseries 61} no.~8, (2000) 084004},
  \href{http://arxiv.org/abs/gr-qc/9910091}{{\ttfamily arXiv:gr-qc/9910091}}.
  [Erratum: Phys.Rev.D 63, 049902 (2001), Erratum: Phys.Rev.D 65, 069902
  (2002), Erratum: Phys.Rev.D 67, 089901 (2003), Erratum: Phys.Rev.D 78, 109902
  (2008), Erratum: Phys.Rev.D 90, 109904 (2014)].

\bibitem{Mino:2003yg}
Y.~Mino, ``{Perturbative approach to an orbital evolution around a supermassive
  black hole},'' \href{http://dx.doi.org/10.1103/PhysRevD.67.084027}{{\em Phys.
  Rev. D} {\bfseries 67} (2003) 084027},
  \href{http://arxiv.org/abs/gr-qc/0302075}{{\ttfamily arXiv:gr-qc/0302075}}.

\bibitem{Barack:2018yvs}
L.~Barack and A.~Pound, ``{Self-force and Radiation Reaction in General
  Relativity},'' \href{http://dx.doi.org/10.1088/1361-6633/aae552}{{\em Rept.
  Prog. Phys.} {\bfseries 82} (2019) 016904},
  \href{http://arxiv.org/abs/1805.10385}{{\ttfamily arXiv:1805.10385 [gr-qc]}}.

\bibitem{Pound:2021qin}
A.~Pound and B.~Wardell, ``{Black hole perturbation theory and gravitational
  self-force},'' \href{http://arxiv.org/abs/2101.04592}{{\ttfamily
  arXiv:2101.04592 [gr-qc]}}.

\bibitem{Hinderer:2008dm}
T.~Hinderer and E.~E. Flanagan, ``{Two Timescale Analysis of Extreme Mass Ratio
  Inspirals in Kerr. I. Orbital Motion},''
  \href{http://dx.doi.org/10.1103/PhysRevD.78.064028}{{\em Phys. Rev. D}
  {\bfseries 78} (2008) 064028},
  \href{http://arxiv.org/abs/0805.3337}{{\ttfamily arXiv:0805.3337 [gr-qc]}}.

\bibitem{Hughes:2016xwf}
S.~A. Hughes, \href{http://dx.doi.org/10.1142/9789813226609_0208}{``{Adiabatic
  and post-adiabatic approaches to extreme mass ratio inspiral},''} in {\em
  {14th Marcel Grossmann Meeting on Recent Developments in Theoretical and
  Experimental General Relativity, Astrophysics, and Relativistic Field
  Theories}}, vol.~2, pp.~1953--1959.
\newblock 2017.
\newblock \href{http://arxiv.org/abs/1601.02042}{{\ttfamily arXiv:1601.02042
  [gr-qc]}}.

\bibitem{Buonanno:1998gg}
A.~Buonanno and T.~Damour, ``{Effective One-body Approach to General
  Relativistic Two-body Dynamics},''
  \href{http://dx.doi.org/10.1103/PhysRevD.59.084006}{{\em Phys. Rev. D}
  {\bfseries 59} (1999) 084006},
  \href{http://arxiv.org/abs/gr-qc/9811091}{{\ttfamily arXiv:gr-qc/9811091}}.

\bibitem{Buonanno:2000ef}
A.~Buonanno and T.~Damour, ``{Transition from Inspiral to Plunge in Binary
  Black Hole Coalescences},''
  \href{http://dx.doi.org/10.1103/PhysRevD.62.064015}{{\em Phys. Rev. D}
  {\bfseries 62} (2000) 064015},
  \href{http://arxiv.org/abs/gr-qc/0001013}{{\ttfamily arXiv:gr-qc/0001013}}.

\bibitem{Taracchini:2013rva}
A.~Taracchini {\em et~al.}, ``{Effective-one-body Model for Black-hole Binaries
  with Generic Mass Ratios and Spins},''
  \href{http://dx.doi.org/10.1103/PhysRevD.89.061502}{{\em Phys. Rev. D}
  {\bfseries 89} (2014) 061502},
  \href{http://arxiv.org/abs/1311.2544}{{\ttfamily arXiv:1311.2544 [gr-qc]}}.

\bibitem{Gourgoulhon:2007ue}
E.~Gourgoulhon, ``{3+1 formalism and bases of numerical relativity},''
  \href{http://arxiv.org/abs/gr-qc/0703035}{{\ttfamily arXiv:gr-qc/0703035}}.

\bibitem{Alcubierre}
M.~Alcubierre,
  \href{http://dx.doi.org/10.1093/acprof:oso/9780199205677.001.0001}{{\em
  {Introduction to 3+1 Numerical Relativity}}}.
\newblock Oxford University Press, 2008.

\bibitem{Baumgarte:2010ndz}
T.~W. Baumgarte and S.~L. Shapiro,
  \href{http://dx.doi.org/10.1017/CBO9781139193344}{{\em {Numerical Relativity:
  Solving Einstein's Equations on the Computer}}}.
\newblock Cambridge University Press, 2010.

\bibitem{Arnowitt:1959ah}
R.~L. Arnowitt, S.~Deser, and C.~W. Misner, ``{Dynamical Structure and
  Definition of Energy in General Relativity},''
  \href{http://dx.doi.org/10.1103/PhysRev.116.1322}{{\em Phys. Rev.} {\bfseries
  116} (1959) 1322--1330}.

\bibitem{Pretorius:2005gq}
F.~Pretorius, ``{Evolution of Binary Black Hole Spacetimes},''
  \href{http://dx.doi.org/10.1103/PhysRevLett.95.121101}{{\em Phys. Rev. Lett.}
  {\bfseries 95} (2005) 121101},
  \href{http://arxiv.org/abs/gr-qc/0507014}{{\ttfamily arXiv:gr-qc/0507014}}.

\bibitem{Baker:2005vv}
J.~G. Baker, J.~Centrella, D.-I. Choi, M.~Koppitz, and J.~van Meter,
  ``{G}ravitational wave extraction from an inspiraling configuration of
  merging black holes,''
  \href{http://dx.doi.org/10.1103/PhysRevLett.96.111102}{{\em Phys. Rev. Lett.}
  {\bfseries 96} (2006) 111102},
\href{http://arxiv.org/abs/gr-qc/0511103}{{\ttfamily arXiv:gr-qc/0511103}}.

\bibitem{Campanelli:2005dd}
M.~Campanelli, C.~Lousto, P.~Marronetti, and Y.~Zlochower, ``{Accurate
  Evolutions of Orbiting Black-hole Binaries Without Excision},''
  \href{http://dx.doi.org/10.1103/PhysRevLett.96.111101}{{\em Phys. Rev. Lett.}
  {\bfseries 96} (2006) 111101},
  \href{http://arxiv.org/abs/gr-qc/0511048}{{\ttfamily arXiv:gr-qc/0511048}}.

\bibitem{Kidder:2000yq}
L.~E. Kidder, M.~A. Scheel, S.~A. Teukolsky, E.~D. Carlson, and G.~B. Cook,
  ``{Black Hole Evolution by Spectral Methods},''
  \href{http://dx.doi.org/10.1103/PhysRevD.62.084032}{{\em Phys. Rev. D}
  {\bfseries 62} (2000) 084032},
  \href{http://arxiv.org/abs/gr-qc/0005056}{{\ttfamily arXiv:gr-qc/0005056}}.

\bibitem{Loffler:2011ay}
F.~Loffler {\em et~al.}, ``{The Einstein Toolkit: A Community Computational
  Infrastructure for Relativistic Astrophysics},''
  \href{http://dx.doi.org/10.1088/0264-9381/29/11/115001}{{\em Class. Quant.
  Grav.} {\bfseries 29} (2012) 115001},
  \href{http://arxiv.org/abs/1111.3344}{{\ttfamily arXiv:1111.3344 [gr-qc]}}.

\bibitem{Moesta:2013dna}
P.~M{\"o}sta, B.~C. Mundim, J.~A. Faber, R.~Haas, S.~C. Noble, T.~Bode,
  F.~L{\"o}ffler, C.~D. Ott, C.~Reisswig, and E.~Schnetter, ``{GRHydro: A New
  Open Source General-relativistic Magnetohydrodynamics Code for the Einstein
  Toolkit},'' \href{http://dx.doi.org/10.1088/0264-9381/31/1/015005}{{\em
  Class. Quant. Grav.} {\bfseries 31} (2014) 015005},
  \href{http://arxiv.org/abs/1304.5544}{{\ttfamily arXiv:1304.5544 [gr-qc]}}.

\bibitem{Brugmann:2008zz}
B.~Bruegmann, J.~A. Gonzalez, M.~Hannam, S.~Husa, U.~Sperhake, and W.~Tichy,
  ``{Calibration of Moving Puncture Simulations},''
  \href{http://dx.doi.org/10.1103/PhysRevD.77.024027}{{\em Phys. Rev. D}
  {\bfseries 77} (2008) 024027},
  \href{http://arxiv.org/abs/gr-qc/0610128}{{\ttfamily arXiv:gr-qc/0610128}}.

\bibitem{Clough:2015sqa}
K.~Clough, P.~Figueras, H.~Finkel, M.~Kunesch, E.~A. Lim, and
  S.~Tunyasuvunakool, ``{GRChombo : Numerical Relativity with Adaptive Mesh
  Refinement},'' \href{http://dx.doi.org/10.1088/0264-9381/32/24/245011}{{\em
  Class. Quant. Grav.} {\bfseries 32} (2015) 24},
  \href{http://arxiv.org/abs/1503.03436}{{\ttfamily arXiv:1503.03436 [gr-qc]}}.

\bibitem{Jani:2016wkt}
K.~Jani, J.~Healy, J.~A. Clark, L.~London, P.~Laguna, and D.~Shoemaker,
  ``{Georgia Tech Catalog of Gravitational Waveforms},''
  \href{http://dx.doi.org/10.1088/0264-9381/33/20/204001}{{\em Class. Quant.
  Grav.} {\bfseries 33} (2016) 204001},
  \href{http://arxiv.org/abs/1605.03204}{{\ttfamily arXiv:1605.03204 [gr-qc]}}.

\bibitem{Healy:2019jyf}
J.~Healy, C.~O. Lousto, J.~Lange, R.~O'Shaughnessy, Y.~Zlochower, and
  M.~Campanelli, ``{Second RIT Binary Black Hole Simulations Catalog and Its
  Application to Gravitational Waves Parameter Estimation},''
  \href{http://dx.doi.org/10.1103/PhysRevD.100.024021}{{\em Phys. Rev. D}
  {\bfseries 100} (2019) 024021},
  \href{http://arxiv.org/abs/1901.02553}{{\ttfamily arXiv:1901.02553 [gr-qc]}}.

\bibitem{Boyle:2019kee}
M.~Boyle {\em et~al.}, ``{The SXS Collaboration Catalog of Binary Black Hole
  Simulations},'' \href{http://dx.doi.org/10.1088/1361-6382/ab34e2}{{\em Class.
  Quant. Grav.} {\bfseries 36} (2019) 195006},
  \href{http://arxiv.org/abs/1904.04831}{{\ttfamily arXiv:1904.04831 [gr-qc]}}.

\bibitem{Vishveshwara:1970zz}
C.~V. Vishveshwara, ``{Scattering of Gravitational Radiation by a Schwarzschild
  Black-hole},'' \href{http://dx.doi.org/10.1038/227936a0}{{\em Nature}
  {\bfseries 227} (1970) 936--938}.

\bibitem{Press:1971wr}
W.~H. Press, ``{Long Wave Trains of Gravitational Waves from a Vibrating Black
  Hole},'' \href{http://dx.doi.org/10.1086/180849}{{\em Astrophys. J. Lett.}
  {\bfseries 170} (1971) L105--L108}.

\bibitem{Chandrasekhar:1975zza}
S.~Chandrasekhar and S.~L. Detweiler, ``{The Quasi-normal Modes of the
  Schwarzschild Black Hole},''
  \href{http://dx.doi.org/10.1098/rspa.1975.0112}{{\em Proc. Roy. Soc. Lond. A}
  {\bfseries A344} (1975) 441--452}.

\bibitem{ReggeWheeler}
T.~Regge and J.~A. Wheeler, ``{Stability of a Schwarzschild Singularity},''
  \href{http://dx.doi.org/10.1103/PhysRev.108.1063}{{\em Phys. Rev.} {\bfseries
  108} (Nov, 1957) 1063--1069}.

\bibitem{Zerilli:1970se}
F.~J. Zerilli, ``{Effective potential for even parity Regge-Wheeler
  gravitational perturbation equations},''
  \href{http://dx.doi.org/10.1103/PhysRevLett.24.737}{{\em Phys. Rev. Lett.}
  {\bfseries 24} (1970) 737--738}.

\bibitem{Zerilli:1970wzz}
F.~J. Zerilli, ``{Gravitational field of a particle falling in a schwarzschild
  geometry analyzed in tensor harmonics},''
  \href{http://dx.doi.org/10.1103/PhysRevD.2.2141}{{\em Phys. Rev. D}
  {\bfseries 2} (1970) 2141--2160}.

\bibitem{Teukolsky:1972my}
S.~A. Teukolsky, ``{Rotating black holes - separable wave equations for
  gravitational and electromagnetic perturbations},''
  \href{http://dx.doi.org/10.1103/PhysRevLett.29.1114}{{\em Phys. Rev. Lett.}
  {\bfseries 29} (1972) 1114--1118}.

\bibitem{Teukolsky:1973ha}
S.~A. Teukolsky, ``{Perturbations of a rotating black hole. I - Fundamental
  equations for gravitational electromagnetic and neutrino field
  perturbations},'' \href{http://dx.doi.org/10.1086/152444}{{\em Astrophys. J.}
  {\bfseries 185} (1973) 635--647}.

\bibitem{Press:1973zz}
W.~H. Press and S.~A. Teukolsky, ``{Perturbations of a Rotating Black Hole. II.
  Dynamical Stability of the Kerr Metric},''
  \href{http://dx.doi.org/10.1086/152445}{{\em Astrophys. J.} {\bfseries 185}
  (1973) 649--674}.

\bibitem{Kokkotas:1999bd}
K.~D. Kokkotas and B.~G. Schmidt, ``{Quasinormal modes of stars and black
  holes},'' \href{http://dx.doi.org/10.12942/lrr-1999-2}{{\em Living Rev. Rel.}
  {\bfseries 2} (1999) 2},
\href{http://arxiv.org/abs/gr-qc/9909058}{{\ttfamily arXiv:gr-qc/9909058
  [gr-qc]}}.

\bibitem{Nollert:1999ji}
H.-P. Nollert, ``{Quasinormal modes: the characteristic `sound' of black holes
  and neutron stars},''
  \href{http://dx.doi.org/10.1088/0264-9381/16/12/201}{{\em Class. Quant.
  Grav.} {\bfseries 16} (1999) R159--R216}.

\bibitem{Berti:2009kk}
E.~Berti, V.~Cardoso, and A.~O. Starinets, ``{Quasinormal modes of black holes
  and black branes},''
  \href{http://dx.doi.org/10.1088/0264-9381/26/16/163001}{{\em Class. Quant.
  Grav.} {\bfseries 26} (2009) 163001},
\href{http://arxiv.org/abs/0905.2975}{{\ttfamily arXiv:0905.2975 [gr-qc]}}.

\bibitem{Konoplya:2011qq}
R.~A. Konoplya and A.~Zhidenko, ``{Quasinormal modes of black holes: From
  astrophysics to string theory},''
  \href{http://dx.doi.org/10.1103/RevModPhys.83.793}{{\em Rev. Mod. Phys.}
  {\bfseries 83} (2011) 793--836},
  \href{http://arxiv.org/abs/1102.4014}{{\ttfamily arXiv:1102.4014 [gr-qc]}}.

\bibitem{Leaver:1986gd}
E.~W. Leaver, ``{Spectral decomposition of the perturbation response of the
  Schwarzschild geometry},''
  \href{http://dx.doi.org/10.1103/PhysRevD.34.384}{{\em Phys. Rev. D}
  {\bfseries 34} (1986) 384--408}.

\bibitem{Price:1971fb}
R.~H. Price, ``{Nonspherical perturbations of relativistic gravitational
  collapse. 1. Scalar and gravitational perturbations},''
  \href{http://dx.doi.org/10.1103/PhysRevD.5.2419}{{\em Phys. Rev. D}
  {\bfseries 5} (1972) 2419--2438}.

\bibitem{Gundlach:1993tp}
C.~Gundlach, R.~H. Price, and J.~Pullin, ``{Late time behavior of stellar
  collapse and explosions: 1. Linearized perturbations},''
  \href{http://dx.doi.org/10.1103/PhysRevD.49.883}{{\em Phys. Rev. D}
  {\bfseries 49} (1994) 883--889},
  \href{http://arxiv.org/abs/gr-qc/9307009}{{\ttfamily arXiv:gr-qc/9307009}}.

\bibitem{Hintz:2020roc}
P.~Hintz, ``{A Sharp Version of Price\textquoteright{}s Law for Wave Decay on
  Asymptotically Flat Spacetimes},''
  \href{http://dx.doi.org/10.1007/s00220-021-04276-8}{{\em Commun. Math. Phys.}
  {\bfseries 389} no.~1, (2022) 491--542},
  \href{http://arxiv.org/abs/2004.01664}{{\ttfamily arXiv:2004.01664
  [math.AP]}}.

\bibitem{GRITJHU}
E.~Berti, {\em {\it \href{https://pages.jh.edu/~eberti2/ringdown/}{Ringdown}}}.

\bibitem{Lim:2019xrb}
H.~Lim, G.~Khanna, A.~Apte, and S.~A. Hughes, ``{Exciting Black Hole Modes via
  Misaligned Coalescences: II. The Mode Content of Late-time Coalescence
  Waveforms},'' \href{http://dx.doi.org/10.1103/PhysRevD.100.084032}{{\em Phys.
  Rev. D} {\bfseries 100} (2019) 084032},
  \href{http://arxiv.org/abs/1901.05902}{{\ttfamily arXiv:1901.05902 [gr-qc]}}.

\bibitem{Dreyer:2003bv}
O.~Dreyer, B.~J. Kelly, B.~Krishnan, L.~S. Finn, D.~Garrison, and
  R.~Lopez-Aleman, ``{Black Hole Spectroscopy: Testing General Relativity
  through Gravitational Wave Observations},''
  \href{http://dx.doi.org/10.1088/0264-9381/21/4/003}{{\em Class. Quant. Grav.}
  {\bfseries 21} (2004) 787--804},
  \href{http://arxiv.org/abs/gr-qc/0309007}{{\ttfamily arXiv:gr-qc/0309007}}.

\bibitem{Berti:2005ys}
E.~Berti, V.~Cardoso, and C.~M. Will, ``{On gravitational-wave spectroscopy of
  massive black holes with the space interferometer LISA},''
  \href{http://dx.doi.org/10.1103/PhysRevD.73.064030}{{\em Phys. Rev.}
  {\bfseries D73} (2006) 064030},
\href{http://arxiv.org/abs/gr-qc/0512160}{{\ttfamily arXiv:gr-qc/0512160
  [gr-qc]}}.

\bibitem{Baibhav:2023clw}
V.~Baibhav, M.~H.-Y. Cheung, E.~Berti, V.~Cardoso, G.~Carullo, R.~Cotesta,
  W.~Del~Pozzo, and F.~Duque, ``{Agnostic black hole spectroscopy: Quasinormal
  mode content of numerical relativity waveforms and limits of validity of
  linear perturbation theory},''
  \href{http://dx.doi.org/10.1103/PhysRevD.108.104020}{{\em Phys. Rev. D}
  {\bfseries 108} no.~10, (2023) 104020},
  \href{http://arxiv.org/abs/2302.03050}{{\ttfamily arXiv:2302.03050 [gr-qc]}}.

\bibitem{Detweiler:1977gy}
S.~L. Detweiler, ``{Resonant oscillations of a rapidly rotating black hole},''
  \href{http://dx.doi.org/10.1098/rspa.1977.0005}{{\em Proc. Roy. Soc. Lond. A}
  {\bfseries 352} (1977) 381--395}.

\bibitem{LIGOScientific:2016lio}
{\bfseries LIGO Scientific, Virgo} Collaboration, B.~P. Abbott {\em et~al.},
  ``{Tests of general relativity with GW150914},''
  \href{http://dx.doi.org/10.1103/PhysRevLett.116.221101}{{\em Phys. Rev.
  Lett.} {\bfseries 116} no.~22, (2016) 221101},
  \href{http://arxiv.org/abs/1602.03841}{{\ttfamily arXiv:1602.03841 [gr-qc]}}.
  [Erratum: Phys.Rev.Lett. 121, 129902 (2018)].

\bibitem{Berti:2015itd}
E.~Berti {\em et~al.}, ``{Testing General Relativity with Present and Future
  Astrophysical Observations},''
  \href{http://dx.doi.org/10.1088/0264-9381/32/24/243001}{{\em Class. Quant.
  Grav.} {\bfseries 32} (2015) 243001},
\href{http://arxiv.org/abs/1501.07274}{{\ttfamily arXiv:1501.07274 [gr-qc]}}.

\bibitem{Yunes:2016jcc}
N.~Yunes, K.~Yagi, and F.~Pretorius, ``{Theoretical Physics Implications of the
  Binary Black-Hole Mergers GW150914 and GW151226},''
  \href{http://dx.doi.org/10.1103/PhysRevD.94.084002}{{\em Phys. Rev.}
  {\bfseries D94} no.~8, (2016) 084002},
\href{http://arxiv.org/abs/1603.08955}{{\ttfamily arXiv:1603.08955 [gr-qc]}}.

\bibitem{Will:2014kxa}
C.~M. Will, ``{The Confrontation between General Relativity and Experiment},''
  \href{http://dx.doi.org/10.12942/lrr-2014-4}{{\em Living Rev. Rel.}
  {\bfseries 17} (2014) 4},
\href{http://arxiv.org/abs/1403.7377}{{\ttfamily arXiv:1403.7377 [gr-qc]}}.

\bibitem{Cardoso:2019rvt}
V.~Cardoso and P.~Pani, ``{Testing the nature of dark compact objects: a status
  report},'' \href{http://dx.doi.org/10.1007/s41114-019-0020-4}{{\em Living
  Rev. Rel.} {\bfseries 22} no.~1, (2019) 4},
  \href{http://arxiv.org/abs/1904.05363}{{\ttfamily arXiv:1904.05363 [gr-qc]}}.

\bibitem{GWPlotter}
R.~Cole, C.~Moore, and C.~Berry, {\em {\it
  \href{https://github.com/robsci/GWplotter/tree/master}{GWplotter}}}.

\bibitem{TheLIGOScientific:2014jea}
{J.~Aasi~{\it et al.} (LIGO Scientific Collaboration)}, ``{Advanced LIGO},''
  \href{http://dx.doi.org/10.1088/0264-9381/32/7/074001}{{\em Class. Quant.
  Grav.} {\bfseries 32} (2015) 074001},
\href{http://arxiv.org/abs/1411.4547}{{\ttfamily arXiv:1411.4547 [gr-qc]}}.

\bibitem{TheVirgo:2014hva}
{F.~Acernese {\it et al.}~(VIRGO Collaboration)}, ``{Advanced Virgo: a
  Second-Generation Interferometric Gravitational Wave Detector},''
  \href{http://dx.doi.org/10.1088/0264-9381/32/2/024001}{{\em Class. Quant.
  Grav.} {\bfseries 32} (2015) 024001},
\href{http://arxiv.org/abs/1408.3978}{{\ttfamily arXiv:1408.3978 [gr-qc]}}.

\bibitem{KAGRA:2020tym}
{\bfseries KAGRA} Collaboration, T.~Akutsu {\em et~al.}, ``{Overview of KAGRA:
  Detector design and construction history},''
  \href{http://dx.doi.org/10.1093/ptep/ptaa125}{{\em PTEP} {\bfseries 2021}
  no.~5, (2021) 05A101}, \href{http://arxiv.org/abs/2005.05574}{{\ttfamily
  arXiv:2005.05574 [physics.ins-det]}}.

\bibitem{LIGOScientific:2020kqk}
{\bfseries LIGO Scientific, Virgo} Collaboration, R.~Abbott {\em et~al.},
  ``{Population Properties of Compact Objects from the Second LIGO-Virgo
  Gravitational-Wave Transient Catalog},''
  \href{http://dx.doi.org/10.3847/2041-8213/abe949}{{\em Astrophys. J. Lett.}
  {\bfseries 913} no.~1, (2021) L7},
  \href{http://arxiv.org/abs/2010.14533}{{\ttfamily arXiv:2010.14533
  [astro-ph.HE]}}.

\bibitem{LIGOScientific:2018cki}
{\bfseries LIGO Scientific, Virgo} Collaboration, B.~P. Abbott {\em et~al.},
  ``{GW170817: Measurements of neutron star radii and equation of state},''
  \href{http://dx.doi.org/10.1103/PhysRevLett.121.161101}{{\em Phys. Rev.
  Lett.} {\bfseries 121} no.~16, (2018) 161101},
  \href{http://arxiv.org/abs/1805.11581}{{\ttfamily arXiv:1805.11581 [gr-qc]}}.

\bibitem{Unnikrishnan:2013qwa}
C.~S. Unnikrishnan, ``{IndIGO and LIGO-India: Scope and plans for gravitational
  wave research and precision metrology in India},''
  \href{http://dx.doi.org/10.1142/S0218271813410101}{{\em Int. J. Mod. Phys. D}
  {\bfseries 22} (2013) 1341010},
  \href{http://arxiv.org/abs/1510.06059}{{\ttfamily arXiv:1510.06059
  [physics.ins-det]}}.

\bibitem{Saleem:2021iwi}
M.~Saleem {\em et~al.}, ``{The science case for LIGO-India},''
  \href{http://dx.doi.org/10.1088/1361-6382/ac3b99}{{\em Class. Quant. Grav.}
  {\bfseries 39} no.~2, (2022) 025004},
  \href{http://arxiv.org/abs/2105.01716}{{\ttfamily arXiv:2105.01716 [gr-qc]}}.

\bibitem{Unnikrishnan:2023uou}
C.~S. Unnikrishnan, ``{LIGO-India: A decadal assessment on its scope,
  relevance, progress and future},''
  \href{http://dx.doi.org/10.1142/S0218271824500251}{{\em Int. J. Mod. Phys. D}
  {\bfseries 33} no.~05n06, (2024) 2450025},
  \href{http://arxiv.org/abs/2301.07522}{{\ttfamily arXiv:2301.07522
  [astro-ph.IM]}}.

\bibitem{Ackley:2020atn}
K.~Ackley {\em et~al.}, ``{Neutron Star Extreme Matter Observatory: A
  kilohertz-band gravitational-wave detector in the global network},''
  \href{http://dx.doi.org/10.1017/pasa.2020.39}{{\em Publ. Astron. Soc.
  Austral.} {\bfseries 37} (2020) e047},
  \href{http://arxiv.org/abs/2007.03128}{{\ttfamily arXiv:2007.03128
  [astro-ph.HE]}}.

\bibitem{Punturo:2010zz}
M.~Punturo {\em et~al.}, ``{The Einstein Telescope: A third-generation
  gravitational wave observatory},''
  \href{http://dx.doi.org/10.1088/0264-9381/27/19/194002}{{\em Class. Quant.
  Grav.} {\bfseries 27} (2010) 194002}.

\bibitem{Sathyaprakash:2012jk}
B.~Sathyaprakash {\em et~al.}, ``{Scientific Objectives of Einstein
  Telescope},'' \href{http://dx.doi.org/10.1088/0264-9381/29/12/124013,
  10.1088/0264-9381/30/7/079501}{{\em Class. Quant. Grav.} {\bfseries 29}
  (2012) 124013}, \href{http://arxiv.org/abs/1206.0331}{{\ttfamily
  arXiv:1206.0331 [gr-qc]}}.
[Erratum: Class. Quant. Grav. {\bf 30} (2013) 079501].

\bibitem{Reitze:2019iox}
D.~Reitze {\em et~al.}, ``{Cosmic Explorer: The U.S. Contribution to
  Gravitational-Wave Astronomy Beyond LIGO},'' {\em Bull. Am. Astron. Soc.}
  {\bfseries 51} (7, 2019) 035,
  \href{http://arxiv.org/abs/1907.04833}{{\ttfamily arXiv:1907.04833
  [astro-ph.IM]}}.

\bibitem{Evans:2021gyd}
M.~Evans {\em et~al.}, ``{A Horizon Study for Cosmic Explorer: Science,
  Observatories, and Community},''
  \href{http://arxiv.org/abs/2109.09882}{{\ttfamily arXiv:2109.09882
  [astro-ph.IM]}}.

\bibitem{LISA:2017pwj}
{\bfseries LISA} Collaboration, P.~Amaro-Seoane {\em et~al.}, ``{Laser
  Interferometer Space Antenna},''
  \href{http://arxiv.org/abs/1702.00786}{{\ttfamily arXiv:1702.00786
  [astro-ph.IM]}}.

\bibitem{TianQin:2015yph}
{\bfseries TianQin} Collaboration, J.~Luo {\em et~al.}, ``{TianQin: a
  space-borne gravitational wave detector},''
  \href{http://dx.doi.org/10.1088/0264-9381/33/3/035010}{{\em Class. Quant.
  Grav.} {\bfseries 33} no.~3, (2016) 035010},
  \href{http://arxiv.org/abs/1512.02076}{{\ttfamily arXiv:1512.02076
  [astro-ph.IM]}}.

\bibitem{TianQin:2020hid}
{\bfseries TianQin} Collaboration, J.~Mei {\em et~al.}, ``{The TianQin project:
  current progress on science and technology},''
  \href{http://dx.doi.org/10.1093/ptep/ptaa114}{{\em PTEP} {\bfseries 2021}
  no.~5, (2021) 05A107}, \href{http://arxiv.org/abs/2008.10332}{{\ttfamily
  arXiv:2008.10332 [gr-qc]}}.

\bibitem{Gong:2021any}
X.~{Gong}, S.~{Xu}, S.~{Gui}, S.~{Huang}, and Y.-K. {Lau},
  \href{http://dx.doi.org/10.1007/978-981-15-4702-7_24-1}{``{Mission Design for
  the TAIJI Mission and Structure Formation in Early Universe},''} in {\em
  Handbook of Gravitational Wave Astronomy}, C.~{Bambi}, S.~{Katsanevas}, and
  K.~D. {Kokkotas}, eds., p.~24.
\newblock Springer Nature, 2021.

\bibitem{Crowder:2005nr}
J.~Crowder and N.~J. Cornish, ``{Beyond LISA: Exploring Future Gravitational
  Wave Missions},'' \href{http://dx.doi.org/10.1103/PhysRevD.72.083005}{{\em
  Phys. Rev. D} {\bfseries 72} (2005) 083005},
  \href{http://arxiv.org/abs/gr-qc/0506015}{{\ttfamily arXiv:gr-qc/0506015}}.

\bibitem{Kawamura:2011zz}
S.~Kawamura {\em et~al.}, ``{The Japanese Space Gravitational Wave Antenna:
  DECIGO},''
\href{http://dx.doi.org/10.1088/0264-9381/28/9/094011}{{\em Class. Quant.
  Grav.} {\bfseries 28} (2011) 094011}.

\bibitem{Kuns:2019upi}
K.~A. Kuns, H.~Yu, Y.~Chen, and R.~X. Adhikari, ``{Astrophysics and Cosmology
  with a Deci-hertz Gravitational-wave Detector: TianGO},''
  \href{http://arxiv.org/abs/1908.06004}{{\ttfamily arXiv:1908.06004 [gr-qc]}}.

\bibitem{NANOGrav:2023gor}
{\bfseries NANOGrav} Collaboration, G.~Agazie {\em et~al.}, ``{The NANOGrav 15
  yr Data Set: Evidence for a Gravitational-wave Background},''
  \href{http://dx.doi.org/10.3847/2041-8213/acdac6}{{\em Astrophys. J. Lett.}
  {\bfseries 951} no.~1, (2023) L8},
  \href{http://arxiv.org/abs/2306.16213}{{\ttfamily arXiv:2306.16213
  [astro-ph.HE]}}.

\bibitem{Sirko:2002ex}
E.~Sirko and J.~Goodman, ``{Spectral energy distributions of selfgravitating
  QSO discs},'' \href{http://dx.doi.org/10.1046/j.1365-8711.2003.06431.x}{{\em
  Mon. Not. Roy. Astron. Soc.} {\bfseries 341} (2003) 501},
  \href{http://arxiv.org/abs/astro-ph/0209469}{{\ttfamily
  arXiv:astro-ph/0209469}}.

\bibitem{1973A&A....24..337S}
N.~I. {Shakura} and R.~A. {Sunyaev}, ``{Black holes in binary systems.
  Observational appearance.},''
  \href{http://dx.doi.org/10.1007/978-94-010-2585-0_13}{{\em Astron.
  Astrophys.} {\bfseries 24} (Jan., 1973) 337--355}.

\bibitem{Bonetti:2018tpf}
M.~Bonetti, A.~Sesana, F.~Haardt, E.~Barausse, and M.~Colpi, ``{Post-Newtonian
  evolution of massive black hole triplets in galactic nuclei \textendash{} IV.
  Implications for LISA},'' \href{http://dx.doi.org/10.1093/mnras/stz903}{{\em
  Mon. Not. Roy. Astron. Soc.} {\bfseries 486} no.~3, (2019) 4044--4060},
  \href{http://arxiv.org/abs/1812.01011}{{\ttfamily arXiv:1812.01011
  [astro-ph.GA]}}.

\bibitem{Yang:2017aht}
H.~Yang and M.~Casals, ``{General Relativistic Dynamics of an Extreme
  Mass-Ratio Binary interacting with an External Body},''
  \href{http://dx.doi.org/10.1103/PhysRevD.96.083015}{{\em Phys. Rev. D}
  {\bfseries 96} no.~8, (2017) 083015},
  \href{http://arxiv.org/abs/1704.02022}{{\ttfamily arXiv:1704.02022 [gr-qc]}}.

\bibitem{Wen:2002km}
L.~Wen, ``{On the eccentricity distribution of coalescing black hole binaries
  driven by the Kozai mechanism in globular clusters},''
  \href{http://dx.doi.org/10.1086/378794}{{\em Astrophys. J.} {\bfseries 598}
  (2003) 419--430}, \href{http://arxiv.org/abs/astro-ph/0211492}{{\ttfamily
  arXiv:astro-ph/0211492}}.

\bibitem{Blaes:2002cs}
O.~Blaes, M.~H. Lee, and A.~Socrates, ``{The kozai mechanism and the evolution
  of binary supermassive black holes},''
  \href{http://dx.doi.org/10.1086/342655}{{\em Astrophys. J.} {\bfseries 578}
  (2002) 775--786}, \href{http://arxiv.org/abs/astro-ph/0203370}{{\ttfamily
  arXiv:astro-ph/0203370}}.

\bibitem{Miller:2002pg}
M.~C. Miller and D.~P. Hamilton, ``{Four-body effects in globular cluster black
  hole coalescence},'' \href{http://dx.doi.org/10.1086/341788}{{\em Astrophys.
  J.} {\bfseries 576} (2002) 894},
  \href{http://arxiv.org/abs/astro-ph/0202298}{{\ttfamily
  arXiv:astro-ph/0202298}}.

\bibitem{Amaro-Seoane:2011rdr}
P.~Amaro-Seoane, P.~Brem, J.~Cuadra, and P.~J. Armitage, ``{The butterfly
  effect in the extreme-mass ratio inspiral problem},''
  \href{http://dx.doi.org/10.1088/2041-8205/744/2/L20}{{\em Astrophys. J.
  Lett.} {\bfseries 744} (2012) L20},
  \href{http://arxiv.org/abs/1108.5174}{{\ttfamily arXiv:1108.5174
  [astro-ph.CO]}}.

\bibitem{2013MNRAS.431.2155N}
S.~{Naoz}, W.~M. {Farr}, Y.~{Lithwick}, F.~A. {Rasio}, and J.~{Teyssandier},
  ``{Secular dynamics in hierarchical three-body systems},''
  \href{http://dx.doi.org/10.1093/mnras/stt302}{{\em Mon. Not. Roy. Astron.
  Soc.} {\bfseries 431} no.~3, (May, 2013) 2155--2171},
  \href{http://arxiv.org/abs/1107.2414}{{\ttfamily arXiv:1107.2414
  [astro-ph.EP]}}.

\bibitem{Kozai}
Y.~{Kozai}, ``{Secular Perturbations of Asteroids with High Inclination and
  Eccentricity},'' \href{http://dx.doi.org/10.1086/108790}{{\em Astron. J.}
  {\bfseries 67} (Nov., 1962) 591--598}.

\bibitem{LIDOV1962719}
M.~Lidov, ``{The Evolution of Orbits of Artificial Satellites of Planets under
  the Action of Gravitational Perturbations of External Bodies},''
  \href{http://dx.doi.org/https://doi.org/10.1016/0032-0633(62)90129-0}{{\em
  Planetary and Space Science} {\bfseries 9} (1962) 719 -- 759}.

\bibitem{Ezquiaga:2020gdt}
J.~M. Ezquiaga, D.~E. Holz, W.~Hu, M.~Lagos, and R.~M. Wald, ``{Phase effects
  from strong gravitational lensing of gravitational waves},''
  \href{http://dx.doi.org/10.1103/PhysRevD.103.064047}{{\em Phys. Rev. D}
  {\bfseries 103} no.~6, (2021) 064047},
  \href{http://arxiv.org/abs/2008.12814}{{\ttfamily arXiv:2008.12814 [gr-qc]}}.

\bibitem{Fang:2019mui}
Y.~{Fang}, X.~{Chen}, and Q.-G. {Huang}, ``{Impact of a Spinning Supermassive
  Black Hole on the Orbit and Gravitational Waves of a Nearby Compact
  Binary},'' \href{http://dx.doi.org/10.3847/1538-4357/ab510e}{{\em Astrophys.
  J.} {\bfseries 887} no.~2, (Dec., 2019) 210},
  \href{http://arxiv.org/abs/1908.01443}{{\ttfamily arXiv:1908.01443
  [astro-ph.HE]}}.

\bibitem{Roy:2024rhe}
S.~Roy and R.~Vicente, ``{Compact binary coalescences in dense gaseous
  environments can pose as ones in vacuum},''
  \href{http://dx.doi.org/10.1103/PhysRevD.111.084037}{{\em Phys. Rev. D}
  {\bfseries 111} no.~8, (2025) 084037},
  \href{http://arxiv.org/abs/2410.16388}{{\ttfamily arXiv:2410.16388 [gr-qc]}}.

\bibitem{Brito:2014wla}
R.~Brito, V.~Cardoso, and P.~Pani, ``{Black Holes as Particle Detectors:
  Evolution of Superradiant Instabilities},''
  \href{http://dx.doi.org/10.1088/0264-9381/32/13/134001}{{\em Class. Quant.
  Grav.} {\bfseries 32} (2015) 134001},
\href{http://arxiv.org/abs/1411.0686}{{\ttfamily arXiv:1411.0686 [gr-qc]}}.

\bibitem{East:2017ovw}
W.~East and F.~Pretorius, ``{Superradiant Instability and Backreaction of
  Massive Vector Fields around Kerr Black Holes},''
  \href{http://dx.doi.org/10.1103/PhysRevLett.119.041101}{{\em Phys. Rev.
  Lett.} {\bfseries 119} (2017) 041101},
\href{http://arxiv.org/abs/1704.04791}{{\ttfamily arXiv:1704.04791 [gr-qc]}}.

\bibitem{Herdeiro:2021znw}
C.~A.~R. Herdeiro, E.~Radu, and N.~M. Santos, ``{A bound on energy extraction
  (and hairiness) from superradiance},''
  \href{http://dx.doi.org/10.1016/j.physletb.2021.136835}{{\em Phys. Lett. B}
  {\bfseries 824} (2022) 136835},
  \href{http://arxiv.org/abs/2111.03667}{{\ttfamily arXiv:2111.03667 [gr-qc]}}.

\bibitem{Zwicky:1933gu}
F.~Zwicky, ``{Die Rotverschiebung von extragalaktischen Nebeln},''
  \href{http://dx.doi.org/10.1007/s10714-008-0707-4}{{\em Helv. Phys. Acta}
  {\bfseries 6} (1933) 110--127}.

\bibitem{Smith:1936mlg}
S.~Smith, ``{The Mass of the Virgo Cluster},''
  \href{http://dx.doi.org/10.1086/143697}{{\em Astrophys. J.} {\bfseries 83}
  (1936) 23--30}.

\bibitem{1937AnLun...6....1H}
E.~{Holmberg}, ``{A Study of Double and Multiple Galaxies Together with
  Inquiries into some General Metagalactic Problems},'' {\em Annals of the
  Observatory of Lund} {\bfseries 6} (Jan., 1937) 1--173.

\bibitem{Rubin:1970zza}
V.~C. Rubin and J.~Ford, W.Kent, ``{Rotation of the Andromeda Nebula from a
  Spectroscopic Survey of Emission Regions},''
  \href{http://dx.doi.org/10.1086/150317}{{\em Astrophys. J.} {\bfseries 159}
  (1970) 379--403}.

\bibitem{Freeman:1970mx}
K.~C. Freeman, ``{On the disks of spiral and SO Galaxies},''
  \href{http://dx.doi.org/10.1086/150474}{{\em Astrophys. J.} {\bfseries 160}
  (1970) 811}.

\bibitem{Ostriker:1973uit}
J.~P. Ostriker and P.~J.~E. Peebles, ``{A Numerical Study of the Stability of
  Flattened Galaxies: or, can Cold Galaxies Survive?},''
  \href{http://dx.doi.org/10.1086/152513}{{\em Astrophys. J.} {\bfseries 186}
  (1973) 467--480}.

\bibitem{Roberts1975}
M.~S. {Roberts} and R.~N. {Whitehurst}, ``{The Rotation Curve and Geometry of
  M31 at Large Galactocentric Distances.},''
  \href{http://dx.doi.org/10.1086/153889}{{\em Astrophys. J.} {\bfseries 201}
  (Oct., 1975) 327--346}.

\bibitem{Rubin:1980zd}
V.~Rubin, N.~Thonnard, and J.~Ford, W.K., ``{Rotational Properties of 21 SC
  Galaxies with a Large Range of Luminosities and Radii, from NGC 4605 /R =
  4kpc/ to UGC 2885 /R = 122 kpc/},''
  \href{http://dx.doi.org/10.1086/158003}{{\em Astrophys. J.} {\bfseries 238}
  (1980) 471}.

\bibitem{Persic:1995ru}
M.~Persic, P.~Salucci, and F.~Stel, ``{The Universal Rotation Curve of Spiral
  Galaxies: 1. The Dark Matter Connection},''
  \href{http://dx.doi.org/10.1093/mnras/278.1.27}{{\em Mon. Not. Roy. Astron.
  Soc.} {\bfseries 281} (1996) 27},
  \href{http://arxiv.org/abs/astro-ph/9506004}{{\ttfamily
  arXiv:astro-ph/9506004}}.

\bibitem{Iocco:2015xga}
F.~Iocco, M.~Pato, and G.~Bertone, ``{Evidence for Dark Matter in the Inner
  Milky Way},'' \href{http://dx.doi.org/10.1038/nphys3237}{{\em Nature Phys.}
  {\bfseries 11} (2015) 245--248},
  \href{http://arxiv.org/abs/1502.03821}{{\ttfamily arXiv:1502.03821
  [astro-ph.GA]}}.

\bibitem{2013ApJS..208...19H}
G.~{Hinshaw}, D.~{Larson}, E.~{Komatsu}, D.~N. {Spergel}, C.~L. {Bennett},
  J.~{Dunkley}, M.~R. {Nolta}, M.~{Halpern}, R.~S. {Hill}, N.~{Odegard},
  L.~{Page}, K.~M. {Smith}, J.~L. {Weiland}, B.~{Gold}, N.~{Jarosik},
  A.~{Kogut}, M.~{Limon}, S.~S. {Meyer}, G.~S. {Tucker}, E.~{Wollack}, and
  E.~L. {Wright}, ``{Nine-year Wilkinson Microwave Anisotropy Probe (WMAP)
  Observations: Cosmological Parameter Results},''
  \href{http://dx.doi.org/10.1088/0067-0049/208/2/19}{{\em Astrophys. J. Supp.}
  {\bfseries 208} no.~2, (Oct., 2013) 19},
  \href{http://arxiv.org/abs/1212.5226}{{\ttfamily arXiv:1212.5226
  [astro-ph.CO]}}.

\bibitem{Planck:2018vyg}
{\bfseries Planck} Collaboration, N.~Aghanim {\em et~al.}, ``{Planck 2018
  results. VI. Cosmological parameters},''
  \href{http://dx.doi.org/10.1051/0004-6361/201833910}{{\em Astron. Astrophys.}
  {\bfseries 641} (2020) A6}, \href{http://arxiv.org/abs/1807.06209}{{\ttfamily
  arXiv:1807.06209 [astro-ph.CO]}}. [Erratum: Astron.Astrophys. 652, C4
  (2021)].

\bibitem{2004IAUS..220..439H}
H.~{Hoekstra}, H.~K.~C. {Yee}, and M.~D. {Gladders},
  \href{http://dx.doi.org/10.48550/arXiv.astro-ph/0310756}{``{Properties of
  galaxy dark matter halos from weak lensing},''} in {\em Dark Matter in
  Galaxies}, S.~{Ryder}, D.~{Pisano}, M.~{Walker}, and K.~{Freeman}, eds.,
  vol.~220 of {\em IAU Symposium}, p.~439.
\newblock July, 2004.
\newblock \href{http://arxiv.org/abs/astro-ph/0310756}{{\ttfamily
  arXiv:astro-ph/0310756 [astro-ph]}}.

\bibitem{2006ApJ...648L.109C}
D.~{Clowe}, M.~{Brada{\v{c}}}, A.~H. {Gonzalez}, M.~{Markevitch}, S.~W.
  {Randall}, C.~{Jones}, and D.~{Zaritsky}, ``{A Direct Empirical Proof of the
  Existence of Dark Matter},'' \href{http://dx.doi.org/10.1086/508162}{{\em
  Astrophys. J. Lett.} {\bfseries 648} no.~2, (Sept., 2006) L109--L113},
  \href{http://arxiv.org/abs/astro-ph/0608407}{{\ttfamily
  arXiv:astro-ph/0608407 [astro-ph]}}.

\bibitem{2010MNRAS.404...60R}
B.~A. {Reid}, W.~J. {Percival}, D.~J. {Eisenstein}, L.~{Verde}, D.~N.
  {Spergel}, R.~A. {Skibba}, N.~A. {Bahcall}, T.~{Budavari}, J.~A. {Frieman},
  M.~{Fukugita}, J.~R. {Gott}, J.~E. {Gunn}, {\v{Z}}.~{Ivezi{\'c}}, G.~R.
  {Knapp}, R.~G. {Kron}, R.~H. {Lupton}, T.~A. {McKay}, A.~{Meiksin}, R.~C.
  {Nichol}, A.~C. {Pope}, D.~J. {Schlegel}, D.~P. {Schneider}, C.~{Stoughton},
  M.~A. {Strauss}, A.~S. {Szalay}, M.~{Tegmark}, M.~S. {Vogeley}, D.~H.
  {Weinberg}, D.~G. {York}, and I.~{Zehavi}, ``{Cosmological constraints from
  the clustering of the Sloan Digital Sky Survey DR7 luminous red galaxies},''
  \href{http://dx.doi.org/10.1111/j.1365-2966.2010.16276.x}{{\em Mon. Not. Roy.
  Astron. Soc.} {\bfseries 404} no.~1, (May, 2010) 60--85},
  \href{http://arxiv.org/abs/0907.1659}{{\ttfamily arXiv:0907.1659
  [astro-ph.CO]}}.

\bibitem{2011ApJS..192...18K}
E.~{Komatsu}, K.~M. {Smith}, J.~{Dunkley}, C.~L. {Bennett}, B.~{Gold},
  G.~{Hinshaw}, N.~{Jarosik}, D.~{Larson}, M.~R. {Nolta}, L.~{Page}, D.~N.
  {Spergel}, M.~{Halpern}, R.~S. {Hill}, A.~{Kogut}, M.~{Limon}, S.~S. {Meyer},
  N.~{Odegard}, G.~S. {Tucker}, J.~L. {Weiland}, E.~{Wollack}, and E.~L.
  {Wright}, ``{Seven-year Wilkinson Microwave Anisotropy Probe (WMAP)
  Observations: Cosmological Interpretation},''
  \href{http://dx.doi.org/10.1088/0067-0049/192/2/18}{{\em Astrophys. J. Supp.}
  {\bfseries 192} no.~2, (Feb., 2011) 18},
  \href{http://arxiv.org/abs/1001.4538}{{\ttfamily arXiv:1001.4538
  [astro-ph.CO]}}.

\bibitem{Cyburt:2004cq}
R.~H. Cyburt, ``{Primordial nucleosynthesis for the new cosmology: Determining
  uncertainties and examining concordance},''
  \href{http://dx.doi.org/10.1103/PhysRevD.70.023505}{{\em Phys. Rev. D}
  {\bfseries 70} (2004) 023505},
  \href{http://arxiv.org/abs/astro-ph/0401091}{{\ttfamily
  arXiv:astro-ph/0401091}}.

\bibitem{Steigman:2007xt}
G.~Steigman, ``{Primordial Nucleosynthesis in the Precision Cosmology Era},''
  \href{http://dx.doi.org/10.1146/annurev.nucl.56.080805.140437}{{\em Ann. Rev.
  Nucl. Part. Sci.} {\bfseries 57} (2007) 463--491},
  \href{http://arxiv.org/abs/0712.1100}{{\ttfamily arXiv:0712.1100
  [astro-ph]}}.

\bibitem{Iocco:2008va}
F.~Iocco, G.~Mangano, G.~Miele, O.~Pisanti, and P.~D. Serpico, ``{Primordial
  Nucleosynthesis: from precision cosmology to fundamental physics},''
  \href{http://dx.doi.org/10.1016/j.physrep.2009.02.002}{{\em Phys. Rept.}
  {\bfseries 472} (2009) 1--76},
  \href{http://arxiv.org/abs/0809.0631}{{\ttfamily arXiv:0809.0631
  [astro-ph]}}.

\bibitem{Fields:2011zzb}
B.~D. Fields, ``{The primordial lithium problem},''
  \href{http://dx.doi.org/10.1146/annurev-nucl-102010-130445}{{\em Ann. Rev.
  Nucl. Part. Sci.} {\bfseries 61} (2011) 47--68},
  \href{http://arxiv.org/abs/1203.3551}{{\ttfamily arXiv:1203.3551
  [astro-ph.CO]}}.

\bibitem{Cyburt:2015mya}
R.~H. Cyburt, B.~D. Fields, K.~A. Olive, and T.-H. Yeh, ``{Big Bang
  Nucleosynthesis: 2015},''
  \href{http://dx.doi.org/10.1103/RevModPhys.88.015004}{{\em Rev. Mod. Phys.}
  {\bfseries 88} (2016) 015004},
  \href{http://arxiv.org/abs/1505.01076}{{\ttfamily arXiv:1505.01076
  [astro-ph.CO]}}.

\bibitem{Peebles:1970ag}
P.~J.~E. Peebles and J.~T. Yu, ``{Primeval adiabatic perturbation in an
  expanding universe},'' \href{http://dx.doi.org/10.1086/150713}{{\em
  Astrophys. J.} {\bfseries 162} (1970) 815--836}.

\bibitem{Bond:1980ha}
J.~R. Bond, G.~Efstathiou, and J.~Silk, ``{Massive Neutrinos and the Large
  Scale Structure of the Universe},''
  \href{http://dx.doi.org/10.1103/PhysRevLett.45.1980}{{\em Phys. Rev. Lett.}
  {\bfseries 45} (1980) 1980--1984}.

\bibitem{Hu:2001bc}
W.~Hu and S.~Dodelson, ``{Cosmic Microwave Background Anisotropies},''
  \href{http://dx.doi.org/10.1146/annurev.astro.40.060401.093926}{{\em Ann.
  Rev. Astron. Astrophys.} {\bfseries 40} (2002) 171--216},
  \href{http://arxiv.org/abs/astro-ph/0110414}{{\ttfamily
  arXiv:astro-ph/0110414}}.

\bibitem{Carroll:2000fy}
S.~M. Carroll, ``{The Cosmological constant},''
  \href{http://dx.doi.org/10.12942/lrr-2001-1}{{\em Living Rev. Rel.}
  {\bfseries 4} (2001) 1},
  \href{http://arxiv.org/abs/astro-ph/0004075}{{\ttfamily
  arXiv:astro-ph/0004075}}.

\bibitem{Copeland:2006wr}
E.~J. Copeland, M.~Sami, and S.~Tsujikawa, ``{Dynamics of dark energy},''
  \href{http://dx.doi.org/10.1142/S021827180600942X}{{\em Int. J. Mod. Phys. D}
  {\bfseries 15} (2006) 1753--1936},
  \href{http://arxiv.org/abs/hep-th/0603057}{{\ttfamily arXiv:hep-th/0603057}}.

\bibitem{1990ApJ...356..359H}
L.~{Hernquist}, ``{An Analytical Model for Spherical Galaxies and Bulges},''
  \href{http://dx.doi.org/10.1086/168845}{{\em Astrophys. J.} {\bfseries 356}
  (June, 1990) 359}.

\bibitem{King:1962wi}
I.~King, ``{The structure of star clusters. I. An Empirical density law},''
  \href{http://dx.doi.org/10.1086/108756}{{\em Astron. J.} {\bfseries 67}
  (1962) 471}.

\bibitem{1965TrAlm...5...87E}
J.~{Einasto}, ``{On the Construction of a Composite Model for the Galaxy and on
  the Determination of the System of Galactic Parameters},'' {\em Trudy
  Astrofizicheskogo Instituta Alma-Ata} {\bfseries 5} (Jan., 1965) 87--100.

\bibitem{Jaffe:1983iv}
W.~Jaffe, ``{A Simple model for the distribution of light in spherical
  galaxies},'' {\em Mon. Not. Roy. Astron. Soc.} {\bfseries 202} (1983)
  995--999.

\bibitem{Kravtsov:1997dp}
A.~V. Kravtsov, A.~A. Klypin, J.~S. Bullock, and J.~R. Primack, ``{The Cores of
  dark matter dominated galaxies: Theory versus observations},''
  \href{http://dx.doi.org/10.1086/305884}{{\em Astrophys. J.} {\bfseries 502}
  (1998) 48}, \href{http://arxiv.org/abs/astro-ph/9708176}{{\ttfamily
  arXiv:astro-ph/9708176}}.

\bibitem{Moore:1999gc}
B.~Moore, T.~R. Quinn, F.~Governato, J.~Stadel, and G.~Lake, ``{Cold collapse
  and the core catastrophe},''
  \href{http://dx.doi.org/10.1046/j.1365-8711.1999.03039.x}{{\em Mon. Not. Roy.
  Astron. Soc.} {\bfseries 310} (1999) 1147--1152},
  \href{http://arxiv.org/abs/astro-ph/9903164}{{\ttfamily
  arXiv:astro-ph/9903164}}.

\bibitem{Cardoso:2021wlq}
V.~Cardoso, K.~Destounis, F.~Duque, R.~P. Macedo, and A.~Maselli, ``{Black
  holes in galaxies: Environmental impact on gravitational-wave generation and
  propagation},'' \href{http://dx.doi.org/10.1103/PhysRevD.105.L061501}{{\em
  Phys. Rev. D} {\bfseries 105} no.~6, (2022) L061501},
  \href{http://arxiv.org/abs/2109.00005}{{\ttfamily arXiv:2109.00005 [gr-qc]}}.

\bibitem{Cardoso:2022whc}
V.~Cardoso, K.~Destounis, F.~Duque, R.~Panosso~Macedo, and A.~Maselli,
  ``{Gravitational Waves from Extreme-Mass-Ratio Systems in Astrophysical
  Environments},'' \href{http://dx.doi.org/10.1103/PhysRevLett.129.241103}{{\em
  Phys. Rev. Lett.} {\bfseries 129} no.~24, (2022) 241103},
  \href{http://arxiv.org/abs/2210.01133}{{\ttfamily arXiv:2210.01133 [gr-qc]}}.

\bibitem{Einstein_cluster}
A.~Einstein, ``{On a stationary system with spherical symmetry consisting of
  many gravitating masses},'' \href{http://dx.doi.org/10.2307/1968902}{{\em
  Annals Math.} {\bfseries 40} (1939) 922--936}.

\bibitem{Einstein_cluster_2}
A.~{Geralico}, F.~{Pompi}, and R.~{Ruffini},
  \href{http://dx.doi.org/10.1142/S2010194512006356}{``{On Einstein
  Clusters},''} in {\em International Journal of Modern Physics Conference
  Series}, vol.~12 of {\em International Journal of Modern Physics Conference
  Series}, pp.~146--173.
\newblock Jan., 2012.

\bibitem{Konoplya:2022hbl}
R.~A. Konoplya and A.~Zhidenko, ``{Solutions of the Einstein Equations for a
  Black Hole Surrounded by a Galactic Halo},''
  \href{http://dx.doi.org/10.3847/1538-4357/ac76bc}{{\em Astrophys. J.}
  {\bfseries 933} no.~2, (2022) 166},
  \href{http://arxiv.org/abs/2202.02205}{{\ttfamily arXiv:2202.02205 [gr-qc]}}.

\bibitem{Figueiredo:2023gas}
E.~Figueiredo, A.~Maselli, and V.~Cardoso, ``{Black holes surrounded by generic
  dark matter profiles: Appearance and gravitational-wave emission},''
  \href{http://dx.doi.org/10.1103/PhysRevD.107.104033}{{\em Phys. Rev. D}
  {\bfseries 107} no.~10, (2023) 104033},
  \href{http://arxiv.org/abs/2303.08183}{{\ttfamily arXiv:2303.08183 [gr-qc]}}.

\bibitem{Speeney:2024mas}
N.~Speeney, E.~Berti, V.~Cardoso, and A.~Maselli, ``{Black holes surrounded by
  generic matter distributions: Polar perturbations and energy flux},''
  \href{http://dx.doi.org/10.1103/PhysRevD.109.084068}{{\em Phys. Rev. D}
  {\bfseries 109} no.~8, (2024) 084068},
  \href{http://arxiv.org/abs/2401.00932}{{\ttfamily arXiv:2401.00932 [gr-qc]}}.

\bibitem{Kim:2004tc}
Y.-R. Kim and R.~A.~C. Croft, ``{Gravitational redshifts in simulated galaxy
  clusters},'' \href{http://dx.doi.org/10.1086/383218}{{\em Astrophys. J.}
  {\bfseries 607} (2004) 164--174},
  \href{http://arxiv.org/abs/astro-ph/0402047}{{\ttfamily
  arXiv:astro-ph/0402047}}.

\bibitem{Gondolo:2000pn}
P.~Gondolo, ``{Either neutralino dark matter or cuspy dark halos},''
  \href{http://dx.doi.org/10.1016/S0370-2693(00)01218-1}{{\em Phys. Lett. B}
  {\bfseries 494} (2000) 181--186},
  \href{http://arxiv.org/abs/hep-ph/0002226}{{\ttfamily arXiv:hep-ph/0002226}}.

\bibitem{Ullio:2001fb}
P.~Ullio, H.~Zhao, and M.~Kamionkowski, ``{A Dark matter spike at the galactic
  center?},'' \href{http://dx.doi.org/10.1103/PhysRevD.64.043504}{{\em Phys.
  Rev. D} {\bfseries 64} (2001) 043504},
  \href{http://arxiv.org/abs/astro-ph/0101481}{{\ttfamily
  arXiv:astro-ph/0101481}}.

\bibitem{Bertone:2001jv}
G.~Bertone, G.~Sigl, and J.~Silk, ``{Astrophysical limits on massive dark
  matter},'' \href{http://dx.doi.org/10.1046/j.1365-8711.2001.04650.x}{{\em
  Mon. Not. Roy. Astron. Soc.} {\bfseries 326} (2001) 799--804},
  \href{http://arxiv.org/abs/astro-ph/0101134}{{\ttfamily
  arXiv:astro-ph/0101134}}.

\bibitem{Merritt:2002vj}
D.~Merritt, M.~Milosavljevic, L.~Verde, and R.~Jimenez, ``{Dark Matter Spikes
  and Annihilation Radiation from the Galactic Center},''
  \href{http://dx.doi.org/10.1103/PhysRevLett.88.191301}{{\em Phys. Rev. Lett.}
  {\bfseries 88} (2002) 191301},
  \href{http://arxiv.org/abs/astro-ph/0201376}{{\ttfamily
  arXiv:astro-ph/0201376}}.

\bibitem{Bertone:2005hw}
G.~Bertone and D.~Merritt, ``{Time-dependent models for dark matter at the
  Galactic Center},'' \href{http://dx.doi.org/10.1103/PhysRevD.72.103502}{{\em
  Phys. Rev. D} {\bfseries 72} (2005) 103502},
  \href{http://arxiv.org/abs/astro-ph/0501555}{{\ttfamily
  arXiv:astro-ph/0501555}}.

\bibitem{Merritt:2006mt}
D.~Merritt, S.~Harfst, and G.~Bertone, ``{Collisionally Regenerated Dark Matter
  Structures in Galactic Nuclei},''
  \href{http://dx.doi.org/10.1103/PhysRevD.75.043517}{{\em Phys. Rev. D}
  {\bfseries 75} (2007) 043517},
  \href{http://arxiv.org/abs/astro-ph/0610425}{{\ttfamily
  arXiv:astro-ph/0610425}}.

\bibitem{Bertone:2005xz}
G.~Bertone, A.~R. Zentner, and J.~Silk, ``{A new signature of dark matter
  annihilations: gamma-rays from intermediate-mass black holes},''
  \href{http://dx.doi.org/10.1103/PhysRevD.72.103517}{{\em Phys. Rev. D}
  {\bfseries 72} (2005) 103517},
  \href{http://arxiv.org/abs/astro-ph/0509565}{{\ttfamily
  arXiv:astro-ph/0509565}}.

\bibitem{Zhao:2005zr}
H.-S. Zhao and J.~Silk, ``{Mini-dark halos with intermediate mass black
  holes},'' \href{http://dx.doi.org/10.1103/PhysRevLett.95.011301}{{\em Phys.
  Rev. Lett.} {\bfseries 95} (2005) 011301},
  \href{http://arxiv.org/abs/astro-ph/0501625}{{\ttfamily
  arXiv:astro-ph/0501625}}.

\bibitem{2009PhRvL.103p1301B}
T.~{Bringmann}, J.~{Lavalle}, and P.~{Salati}, ``{Intermediate Mass Black Holes
  and Nearby Dark Matter Point Sources: A Critical Reassessment},''
  \href{http://dx.doi.org/10.1103/PhysRevLett.103.161301}{{\em Phys. Rev.
  Lett.} {\bfseries 103} no.~16, (Oct., 2009) 161301},
  \href{http://arxiv.org/abs/0902.3665}{{\ttfamily arXiv:0902.3665
  [astro-ph.CO]}}.

\bibitem{Kavanagh:2018ggo}
B.~J. Kavanagh, D.~Gaggero, and G.~Bertone, ``{Merger rate of a subdominant
  population of primordial black holes},''
  \href{http://dx.doi.org/10.1103/PhysRevD.98.023536}{{\em Phys. Rev. D}
  {\bfseries 98} no.~2, (2018) 023536},
  \href{http://arxiv.org/abs/1805.09034}{{\ttfamily arXiv:1805.09034
  [astro-ph.CO]}}.

\bibitem{Bertone:2024wbn}
G.~Bertone, A.~R. A.~C. Wierda, D.~Gaggero, B.~J. Kavanagh, M.~Volonteri, and
  N.~Yoshida, ``{Dark Matter Mounds: towards a realistic description of dark
  matter overdensities around black holes},''
  \href{http://arxiv.org/abs/2404.08731}{{\ttfamily arXiv:2404.08731
  [astro-ph.CO]}}.

\bibitem{Kaup1968}
D.~J. Kaup, ``{Klein-Gordon Geon},''
  \href{http://dx.doi.org/10.1103/PhysRev.172.1331}{{\em Phys. Rev.} {\bfseries
  172} (Aug, 1968) 1331--1342}.

\bibitem{Ruffini1969}
R.~Ruffini and S.~Bonazzola, ``{Systems of Self-Gravitating Particles in
  General Relativity and the Concept of an Equation of State},''
  \href{http://dx.doi.org/10.1103/PhysRev.187.1767}{{\em Phys. Rev.} {\bfseries
  187} (Nov, 1969) 1767--1783}.

\bibitem{Breit:1983nr}
J.~D. Breit, S.~Gupta, and A.~Zaks, ``{Cold Bose Stars},''
\href{http://dx.doi.org/10.1016/0370-2693(84)90764-0}{{\em Phys. Lett.}
  {\bfseries 140B} (1984) 329--332}.

\bibitem{Colpi1986}
M.~Colpi, S.~L. Shapiro, and I.~Wasserman, ``{Boson Stars: Gravitational
  Equilibria of Self-Interacting Scalar Fields},''
  \href{http://dx.doi.org/10.1103/PhysRevLett.57.2485}{{\em Phys. Rev. Lett.}
  {\bfseries 57} (Nov, 1986) 2485--2488}.

\bibitem{Schunck:2003kk}
F.~E. Schunck and E.~W. Mielke, ``{General Relativistic Boson Stars},''
  \href{http://dx.doi.org/10.1088/0264-9381/20/20/201}{{\em Class. Quant.
  Grav.} {\bfseries 20} (2003) R301--R356},
  \href{http://arxiv.org/abs/0801.0307}{{\ttfamily arXiv:0801.0307
  [astro-ph]}}.

\bibitem{Liebling:2012fv}
S.~L. Liebling and C.~Palenzuela, ``{Dynamical Boson Stars},''
  \href{http://dx.doi.org/10.12942/lrr-2012-6, 10.1007/s41114-017-0007-y}{{\em
  Living Rev. Rel.} {\bfseries 15} (2012) 6},
\href{http://arxiv.org/abs/1202.5809}{{\ttfamily arXiv:1202.5809 [gr-qc]}}.

\bibitem{Seidel1991}
E.~Seidel and W.-M. Suen, ``{Oscillating Soliton Stars},''
  \href{http://dx.doi.org/10.1103/PhysRevLett.66.1659}{{\em Phys. Rev. Lett.}
  {\bfseries 66} (Apr, 1991) 1659--1662}.

\bibitem{Copeland:1995fq}
E.~J. Copeland, M.~Gleiser, and H.~R. Muller, ``{Oscillons: Resonant
  Configurations during Bubble Collapse},''
  \href{http://dx.doi.org/10.1103/PhysRevD.52.1920}{{\em Phys. Rev.} {\bfseries
  D52} (1995) 1920--1933},
\href{http://arxiv.org/abs/hep-ph/9503217}{{\ttfamily arXiv:hep-ph/9503217
  [hep-ph]}}.

\bibitem{Braaten:2015eeu}
E.~Braaten, A.~Mohapatra, and H.~Zhang, ``{Dense Axion Stars},''
  \href{http://dx.doi.org/10.1103/PhysRevLett.117.121801}{{\em Phys. Rev.
  Lett.} {\bfseries 117} (2016) 121801},
  \href{http://arxiv.org/abs/1512.00108}{{\ttfamily arXiv:1512.00108
  [hep-ph]}}.

\bibitem{Lee1987-2}
R.~Friedberg, T.~D. Lee, and Y.~Pang, ``{Mini-soliton Stars},''
  \href{http://dx.doi.org/10.1103/PhysRevD.35.3640}{{\em Phys. Rev. D}
  {\bfseries 35} (Jun, 1987) 3640--3657}.

\bibitem{Lee1987-3}
R.~Friedberg, T.~D. Lee, and Y.~Pang, ``{Scalar Soliton Stars and Black
  Holes},'' \href{http://dx.doi.org/10.1103/PhysRevD.35.3658}{{\em Phys. Rev.
  D} {\bfseries 35} (Jun, 1987) 3658--3677}.

\bibitem{Lee1987-4}
T.~D. Lee and Y.~Pang, ``{Fermion Soliton Stars and Black Holes},''
  \href{http://dx.doi.org/10.1103/PhysRevD.35.3678}{{\em Phys. Rev. D}
  {\bfseries 35} (Jun, 1987) 3678--3694}.

\bibitem{Lynn:1988rb}
B.~W. Lynn, ``{Q-Stars},''
\href{http://dx.doi.org/10.1016/0550-3213(89)90352-0}{{\em Nucl. Phys.}
  {\bfseries B321} (1989) 465}.

\bibitem{Ozel:2016oaf}
F.~{\"O}zel and P.~Freire, ``{Masses, Radii, and the Equation of State of
  Neutron Stars},''
  \href{http://dx.doi.org/10.1146/annurev-astro-081915-023322}{{\em Ann. Rev.
  Astron. Astrophys.} {\bfseries 54} (2016) 401--440},
\href{http://arxiv.org/abs/1603.02698}{{\ttfamily arXiv:1603.02698
  [astro-ph.HE]}}.

\bibitem{Bianchi:1990mha}
M.~Bianchi, D.~Grasso, and R.~Ruffini, ``{Jeans Mass of a Cosmological Coherent
  Scalar Field},'' {\em Astron. Astrophys.} {\bfseries 231} (1990) 301--308.

\bibitem{Seidel:1993zk}
E.~Seidel and W.-M. Suen, ``{Formation of Solitonic Stars through Gravitational
  Cooling},'' \href{http://dx.doi.org/10.1103/PhysRevLett.72.2516}{{\em Phys.
  Rev. Lett.} {\bfseries 72} (1994) 2516--2519},
  \href{http://arxiv.org/abs/gr-qc/9309015}{{\ttfamily arXiv:gr-qc/9309015}}.

\bibitem{Brito:2015yga}
R.~Brito, V.~Cardoso, and H.~Okawa, ``{Accretion of dark matter by stars},''
  \href{http://dx.doi.org/10.1103/PhysRevLett.115.111301}{{\em Phys. Rev.
  Lett.} {\bfseries 115} no.~11, (2015) 111301},
  \href{http://arxiv.org/abs/1508.04773}{{\ttfamily arXiv:1508.04773 [gr-qc]}}.

\bibitem{DiGiovanni:2018bvo}
F.~Di~Giovanni, N.~Sanchis-Gual, C.~A.~R. Herdeiro, and J.~A. Font,
  ``{Dynamical formation of Proca stars and quasistationary solitonic
  objects},'' \href{http://dx.doi.org/10.1103/PhysRevD.98.064044}{{\em Phys.
  Rev. D} {\bfseries 98} no.~6, (2018) 064044},
  \href{http://arxiv.org/abs/1803.04802}{{\ttfamily arXiv:1803.04802 [gr-qc]}}.

\bibitem{Narain:2006kx}
G.~Narain, J.~Schaffner-Bielich, and I.~N. Mishustin, ``{Compact stars made of
  fermionic dark matter},''
  \href{http://dx.doi.org/10.1103/PhysRevD.74.063003}{{\em Phys. Rev. D}
  {\bfseries 74} (2006) 063003},
  \href{http://arxiv.org/abs/astro-ph/0605724}{{\ttfamily
  arXiv:astro-ph/0605724}}.

\bibitem{Kouvaris:2015rea}
C.~Kouvaris and N.~G. Nielsen, ``{Asymmetric Dark Matter Stars},''
  \href{http://dx.doi.org/10.1103/PhysRevD.92.063526}{{\em Phys. Rev. D}
  {\bfseries 92} no.~6, (2015) 063526},
  \href{http://arxiv.org/abs/1507.00959}{{\ttfamily arXiv:1507.00959
  [hep-ph]}}.

\bibitem{Gresham:2018rqo}
M.~I. Gresham and K.~M. Zurek, ``{Asymmetric Dark Stars and Neutron Star
  Stability},'' \href{http://dx.doi.org/10.1103/PhysRevD.99.083008}{{\em Phys.
  Rev. D} {\bfseries 99} no.~8, (2019) 083008},
  \href{http://arxiv.org/abs/1809.08254}{{\ttfamily arXiv:1809.08254
  [astro-ph.CO]}}.

\bibitem{Chang:2018bgx}
J.~H. Chang, D.~Egana-Ugrinovic, R.~Essig, and C.~Kouvaris, ``{Structure
  Formation and Exotic Compact Objects in a Dissipative Dark Sector},''
  \href{http://dx.doi.org/10.1088/1475-7516/2019/03/036}{{\em JCAP} {\bfseries
  03} (2019) 036}, \href{http://arxiv.org/abs/1812.07000}{{\ttfamily
  arXiv:1812.07000 [hep-ph]}}.

\bibitem{HawkingPBH}
S.~Hawking, ``{Gravitationally Collapsed Objects of Very Low Mass},''
  \href{http://dx.doi.org/10.1093/mnras/152.1.75}{{\em Mon. Not. Roy. Astron.
  Soc.} {\bfseries 152} (04, 1971) 75--78}.

\bibitem{Chapline:1975ojl}
G.~F. Chapline, ``{Cosmological effects of primordial black holes},''
  \href{http://dx.doi.org/10.1038/253251a0}{{\em Nature} {\bfseries 253}
  no.~5489, (1975) 251--252}.

\bibitem{Carr:2016drx}
B.~Carr, F.~Kuhnel, and M.~Sandstad, ``{Primordial Black Holes as Dark
  Matter},'' \href{http://dx.doi.org/10.1103/PhysRevD.94.083504}{{\em Phys.
  Rev. D} {\bfseries 94} no.~8, (2016) 083504},
  \href{http://arxiv.org/abs/1607.06077}{{\ttfamily arXiv:1607.06077
  [astro-ph.CO]}}.

\bibitem{Zeldovich:1967lct}
Y.~B. Zel'dovich and I.~D. Novikov, ``{The Hypothesis of Cores Retarded during
  Expansion and the Hot Cosmological Model},'' {\em Sov. Astron.} {\bfseries
  10} (1967) 602.

\bibitem{Carr:1974nx}
B.~J. Carr and S.~W. Hawking, ``{Black holes in the early Universe},''
  \href{http://dx.doi.org/10.1093/mnras/168.2.399}{{\em Mon. Not. Roy. Astron.
  Soc.} {\bfseries 168} (1974) 399--415}.

\bibitem{1975ApJ...201....1C}
B.~J. {Carr}, ``{The primordial black hole mass spectrum.},''
  \href{http://dx.doi.org/10.1086/153853}{{\em Astrophys. J.} {\bfseries 201}
  (Oct., 1975) 1--19}.

\bibitem{Tolman1939}
R.~C. Tolman, ``{Static Solutions of Einstein's Field Equations for Spheres of
  Fluid},'' \href{http://dx.doi.org/10.1103/PhysRev.55.364}{{\em Phys. Rev.}
  {\bfseries 55} (Feb, 1939) 364--373}.

\bibitem{Carr:2009jm}
B.~J. Carr, K.~Kohri, Y.~Sendouda, and J.~Yokoyama, ``{New cosmological
  constraints on primordial black holes},''
  \href{http://dx.doi.org/10.1103/PhysRevD.81.104019}{{\em Phys. Rev. D}
  {\bfseries 81} (2010) 104019},
  \href{http://arxiv.org/abs/0912.5297}{{\ttfamily arXiv:0912.5297
  [astro-ph.CO]}}.

\bibitem{EROS-2:2006ryy}
{\bfseries EROS-2} Collaboration, P.~Tisserand {\em et~al.}, ``{Limits on the
  Macho Content of the Galactic Halo from the EROS-2 Survey of the Magellanic
  Clouds},'' \href{http://dx.doi.org/10.1051/0004-6361:20066017}{{\em Astron.
  Astrophys.} {\bfseries 469} (2007) 387--404},
  \href{http://arxiv.org/abs/astro-ph/0607207}{{\ttfamily
  arXiv:astro-ph/0607207}}.

\bibitem{2013MNRAS.435.1582C}
S.~{Calchi Novati}, S.~{Mirzoyan}, P.~{Jetzer}, and G.~{Scarpetta},
  ``{Microlensing towards the SMC: a new analysis of OGLE and EROS results},''
  \href{http://dx.doi.org/10.1093/mnras/stt1402}{{\em {Mon. Not. Roy. Astron.
  Soc.}} {\bfseries 435} no.~2, (Oct., 2013) 1582--1597},
  \href{http://arxiv.org/abs/1308.4281}{{\ttfamily arXiv:1308.4281
  [astro-ph.GA]}}.

\bibitem{Niikura:2017zjd}
H.~Niikura {\em et~al.}, ``{Microlensing constraints on primordial black holes
  with Subaru/HSC Andromeda observations},''
  \href{http://dx.doi.org/10.1038/s41550-019-0723-1}{{\em Nature Astron.}
  {\bfseries 3} no.~6, (2019) 524--534},
  \href{http://arxiv.org/abs/1701.02151}{{\ttfamily arXiv:1701.02151
  [astro-ph.CO]}}.

\bibitem{Zumalacarregui:2017qqd}
M.~Zumalacarregui and U.~Seljak, ``{Limits on stellar-mass compact objects as
  dark matter from gravitational lensing of type Ia supernovae},''
  \href{http://dx.doi.org/10.1103/PhysRevLett.121.141101}{{\em Phys. Rev.
  Lett.} {\bfseries 121} no.~14, (2018) 141101},
  \href{http://arxiv.org/abs/1712.02240}{{\ttfamily arXiv:1712.02240
  [astro-ph.CO]}}.

\bibitem{Ali-Haimoud:2016mbv}
Y.~Ali-Ha\"\i{}moud and M.~Kamionkowski, ``{Cosmic microwave background limits
  on accreting primordial black holes},''
  \href{http://dx.doi.org/10.1103/PhysRevD.95.043534}{{\em Phys. Rev. D}
  {\bfseries 95} no.~4, (2017) 043534},
  \href{http://arxiv.org/abs/1612.05644}{{\ttfamily arXiv:1612.05644
  [astro-ph.CO]}}.

\bibitem{Poulin:2017bwe}
V.~Poulin, P.~D. Serpico, F.~Calore, S.~Clesse, and K.~Kohri, ``{CMB bounds on
  disk-accreting massive primordial black holes},''
  \href{http://dx.doi.org/10.1103/PhysRevD.96.083524}{{\em Phys. Rev. D}
  {\bfseries 96} no.~8, (2017) 083524},
  \href{http://arxiv.org/abs/1707.04206}{{\ttfamily arXiv:1707.04206
  [astro-ph.CO]}}.

\bibitem{Giddings:1992hh}
S.~B. Giddings, ``{Black holes and massive remnants},''
  \href{http://dx.doi.org/10.1103/PhysRevD.46.1347}{{\em Phys. Rev. D}
  {\bfseries 46} (1992) 1347--1352},
  \href{http://arxiv.org/abs/hep-th/9203059}{{\ttfamily arXiv:hep-th/9203059}}.

\bibitem{Posada:2018agb}
C.~Posada and C.~Chirenti, ``{On the radial stability of ultra compact
  Schwarzschild stars beyond the Buchdahl limit},''
  \href{http://dx.doi.org/10.1088/1361-6382/ab0526}{{\em Class. Quant. Grav.}
  {\bfseries 36} (2019) 065004},
  \href{http://arxiv.org/abs/1811.09589}{{\ttfamily arXiv:1811.09589 [gr-qc]}}.

\bibitem{Einstein:1935tc}
A.~Einstein and N.~Rosen, ``{The Particle Problem in the General Theory of
  Relativity},'' \href{http://dx.doi.org/10.1103/PhysRev.48.73}{{\em Phys.
  Rev.} {\bfseries 48} (1935) 73--77}.

\bibitem{Raidal:2018eoo}
M.~Raidal, S.~Solodukhin, V.~Vaskonen, and H.~Veerm\"ae, ``{Light Primordial
  Exotic Compact Objects as All Dark Matter},''
  \href{http://dx.doi.org/10.1103/PhysRevD.97.123520}{{\em Phys. Rev. D}
  {\bfseries 97} no.~12, (2018) 123520},
  \href{http://arxiv.org/abs/1802.07728}{{\ttfamily arXiv:1802.07728
  [astro-ph.CO]}}.

\bibitem{Deliyergiyev:2019vti}
M.~Deliyergiyev, A.~Del~Popolo, L.~Tolos, M.~Le~Delliou, X.~Lee, and F.~Burgio,
  ``{Dark compact objects: an extensive overview},''
  \href{http://dx.doi.org/10.1103/PhysRevD.99.063015}{{\em Phys. Rev. D}
  {\bfseries 99} no.~6, (2019) 063015},
  \href{http://arxiv.org/abs/1903.01183}{{\ttfamily arXiv:1903.01183 [gr-qc]}}.

\bibitem{Arcadi:2017kky}
G.~Arcadi, M.~Dutra, P.~Ghosh, M.~Lindner, Y.~Mambrini, M.~Pierre, S.~Profumo,
  and F.~S. Queiroz, ``{The Waning of the WIMP? A Review of Models, Searches,
  and Constraints},''
  \href{http://dx.doi.org/10.1140/epjc/s10052-018-5662-y}{{\em Eur. Phys. J. C}
  {\bfseries 78} (2018) 203}, \href{http://arxiv.org/abs/1703.07364}{{\ttfamily
  arXiv:1703.07364 [hep-ph]}}.

\bibitem{Srednicki:1988ce}
M.~Srednicki, R.~Watkins, and K.~A. Olive, ``{Calculations of Relic Densities
  in the Early Universe},''
  \href{http://dx.doi.org/10.1016/0550-3213(88)90099-5}{{\em Nucl. Phys. B}
  {\bfseries 310} (1988) 693}.

\bibitem{Gondolo:1990dk}
P.~Gondolo and G.~Gelmini, ``{Cosmic Abundances of Stable Particles: Improved
  Analysis},'' \href{http://dx.doi.org/10.1016/0550-3213(91)90438-4}{{\em Nucl.
  Phys. B} {\bfseries 360} (1991) 145--179}.

\bibitem{Akerib:2016vxi}
{D.S.~ Akerib {\it et al.}~(LUX Collaboration)}, ``{Results from a Search for
  Dark Matter in the Complete LUX Exposure},''
  \href{http://dx.doi.org/10.1103/PhysRevLett.118.021303}{{\em Phys. Rev.
  Lett.} {\bfseries 118} (2017) 021303},
  \href{http://arxiv.org/abs/1608.07648}{{\ttfamily arXiv:1608.07648
  [astro-ph.CO]}}.

\bibitem{Cui:2017nnn}
{X.~Cui {\it et al.}~(PandaX-II Collaboration)}, ``{Dark Matter Results From
  54-Ton-Day Exposure of PandaX-II Experiment},''
  \href{http://dx.doi.org/10.1103/PhysRevLett.119.181302}{{\em Phys. Rev.
  Lett.} {\bfseries 119} (2017) 181302},
  \href{http://arxiv.org/abs/1708.06917}{{\ttfamily arXiv:1708.06917
  [astro-ph.CO]}}.

\bibitem{Aprile:2018dbl}
{E.~Aprile {\it et al.}~(XENON Collaboration)}, ``{Dark Matter Search Results
  from a One Ton-Year Exposure of XENON1T},''
  \href{http://dx.doi.org/10.1103/PhysRevLett.121.111302}{{\em Phys. Rev.
  Lett.} {\bfseries 121} (2018) 111302},
  \href{http://arxiv.org/abs/1805.12562}{{\ttfamily arXiv:1805.12562
  [astro-ph.CO]}}.

\bibitem{XENON:2024wpa}
{\bfseries XENON} Collaboration, E.~Aprile {\em et~al.}, ``{The XENONnT dark
  matter experiment},''
  \href{http://dx.doi.org/10.1140/epjc/s10052-024-12982-5}{{\em Eur. Phys. J.
  C} {\bfseries 84} no.~8, (2024) 784},
  \href{http://arxiv.org/abs/2402.10446}{{\ttfamily arXiv:2402.10446
  [physics.ins-det]}}.

\bibitem{Schumann:2019eaa}
M.~Schumann, ``{Direct Detection of WIMP Dark Matter: Concepts and Status},''
  \href{http://dx.doi.org/10.1088/1361-6471/ab2ea5}{{\em J. Phys. G} {\bfseries
  46} (2019) 103003}, \href{http://arxiv.org/abs/1903.03026}{{\ttfamily
  arXiv:1903.03026 [astro-ph.CO]}}.

\bibitem{FermiLAT2009}
W.~B. e.~a. {Atwood}, ``{The Large Area Telescope on the Fermi Gamma-Ray Space
  Telescope Mission},''
  \href{http://dx.doi.org/10.1088/0004-637X/697/2/1071}{{\em Astrophys. J.}
  {\bfseries 697} (June, 2009) 1071--1102},
  \href{http://arxiv.org/abs/0902.1089}{{\ttfamily arXiv:0902.1089
  [astro-ph.IM]}}.

\bibitem{Adriani:2008zr}
{O.~ Adriani {\it et al.}~(PAMELA Collaboration)}, ``{An Anomalous Positron
  Abundance in Cosmic Rays with Energies 1.5-100 GeV},''
  \href{http://dx.doi.org/10.1038/nature07942}{{\em Nature} {\bfseries 458}
  (2009) 607--609}, \href{http://arxiv.org/abs/0810.4995}{{\ttfamily
  arXiv:0810.4995 [astro-ph]}}.

\bibitem{Gaskins:2016cha}
J.~M. Gaskins, ``{A Review of Indirect Searches for Particle Dark Matter},''
  \href{http://dx.doi.org/10.1080/00107514.2016.1175160}{{\em Contemp. Phys.}
  {\bfseries 57} (2016) 496--525},
  \href{http://arxiv.org/abs/1604.00014}{{\ttfamily arXiv:1604.00014
  [astro-ph.HE]}}.

\bibitem{Shi:1998km}
X.-D. Shi and G.~M. Fuller, ``{A New dark matter candidate: Nonthermal sterile
  neutrinos},'' \href{http://dx.doi.org/10.1103/PhysRevLett.82.2832}{{\em Phys.
  Rev. Lett.} {\bfseries 82} (1999) 2832--2835},
  \href{http://arxiv.org/abs/astro-ph/9810076}{{\ttfamily
  arXiv:astro-ph/9810076}}.

\bibitem{Laine:2008pg}
M.~Laine and M.~Shaposhnikov, ``{Sterile neutrino dark matter as a consequence
  of nuMSM-induced lepton asymmetry},''
  \href{http://dx.doi.org/10.1088/1475-7516/2008/06/031}{{\em JCAP} {\bfseries
  06} (2008) 031}, \href{http://arxiv.org/abs/0804.4543}{{\ttfamily
  arXiv:0804.4543 [hep-ph]}}.

\bibitem{Boyarsky:2009ix}
A.~Boyarsky, O.~Ruchayskiy, and M.~Shaposhnikov, ``{The Role of sterile
  neutrinos in cosmology and astrophysics},''
  \href{http://dx.doi.org/10.1146/annurev.nucl.010909.083654}{{\em Ann. Rev.
  Nucl. Part. Sci.} {\bfseries 59} (2009) 191--214},
  \href{http://arxiv.org/abs/0901.0011}{{\ttfamily arXiv:0901.0011 [hep-ph]}}.

\bibitem{Drewes:2016upu}
M.~Drewes {\em et~al.}, ``{A White Paper on keV Sterile Neutrino Dark
  Matter},'' \href{http://dx.doi.org/10.1088/1475-7516/2017/01/025}{{\em JCAP}
  {\bfseries 01} (2017) 025}, \href{http://arxiv.org/abs/1602.04816}{{\ttfamily
  arXiv:1602.04816 [hep-ph]}}.

\bibitem{Bulbul:2014sua}
E.~Bulbul, M.~Markevitch, A.~Foster, R.~K. Smith, M.~Loewenstein, and S.~W.
  Randall, ``{Detection of An Unidentified Emission Line in the Stacked X-ray
  spectrum of Galaxy Clusters},''
  \href{http://dx.doi.org/10.1088/0004-637X/789/1/13}{{\em Astrophys. J.}
  {\bfseries 789} (2014) 13}, \href{http://arxiv.org/abs/1402.2301}{{\ttfamily
  arXiv:1402.2301 [astro-ph.CO]}}.

\bibitem{Boyarsky:2014jta}
A.~Boyarsky, O.~Ruchayskiy, D.~Iakubovskyi, and J.~Franse, ``{Unidentified Line
  in X-Ray Spectra of the Andromeda Galaxy and Perseus Galaxy Cluster},''
  \href{http://dx.doi.org/10.1103/PhysRevLett.113.251301}{{\em Phys. Rev.
  Lett.} {\bfseries 113} (2014) 251301},
  \href{http://arxiv.org/abs/1402.4119}{{\ttfamily arXiv:1402.4119
  [astro-ph.CO]}}.

\bibitem{2015MNRAS.450.2143J}
T.~{Jeltema} and S.~{Profumo}, ``{Discovery of a 3.5 keV line in the Galactic
  Centre and a critical look at the origin of the line across astronomical
  targets},'' \href{http://dx.doi.org/10.1093/mnras/stv768}{{\em Mon. Not. Roy.
  Astron. Soc.} {\bfseries 450} no.~2, (June, 2015) 2143--2152},
  \href{http://arxiv.org/abs/1408.1699}{{\ttfamily arXiv:1408.1699
  [astro-ph.HE]}}.

\bibitem{Dessert:2018qih}
C.~Dessert, N.~L. Rodd, and B.~R. Safdi, ``{The dark matter interpretation of
  the 3.5-keV line is inconsistent with blank-sky observations},''
  \href{http://dx.doi.org/10.1126/science.aaw3772}{{\em Science} {\bfseries
  367} no.~6485, (2020) 1465--1467},
  \href{http://arxiv.org/abs/1812.06976}{{\ttfamily arXiv:1812.06976
  [astro-ph.CO]}}.

\bibitem{Silich:2021sra}
E.~M. Silich, K.~Jahoda, L.~Angelini, P.~Kaaret, A.~Zajczyk, D.~M. LaRocca,
  R.~Ringuette, and J.~Richardson, ``{A Search for the 3.5 keV Line from the
  Milky Way\textquoteright{}s Dark Matter Halo with HaloSat},''
  \href{http://dx.doi.org/10.3847/1538-4357/ac043b}{{\em Astrophys. J.}
  {\bfseries 916} no.~1, (2021) 2},
  \href{http://arxiv.org/abs/2105.12252}{{\ttfamily arXiv:2105.12252
  [astro-ph.HE]}}.

\bibitem{Goldberg:1986nk}
H.~Goldberg and L.~J. Hall, ``{A New Candidate for Dark Matter},''
  \href{http://dx.doi.org/10.1016/0370-2693(86)90731-8}{{\em Phys. Lett. B}
  {\bfseries 174} (1986) 151}.

\bibitem{DeRujula:1989fe}
A.~De~Rujula, S.~L. Glashow, and U.~Sarid, ``{Charged Dark Matter},''
  \href{http://dx.doi.org/10.1016/0550-3213(90)90227-5}{{\em Nucl. Phys. B}
  {\bfseries 333} (1990) 173--194}.

\bibitem{Cheung:2007ut}
K.~Cheung and T.-C. Yuan, ``{Hidden Fermion as Milli-charged Dark Matter in
  Stueckelberg Z-prime Model},''
  \href{http://dx.doi.org/10.1088/1126-6708/2007/03/120}{{\em JHEP} {\bfseries
  03} (2007) 120}, \href{http://arxiv.org/abs/hep-ph/0701107}{{\ttfamily
  arXiv:hep-ph/0701107}}.

\bibitem{Feldman:2007wj}
D.~Feldman, Z.~Liu, and P.~Nath, ``{The Stueckelberg Z-prime Extension with
  Kinetic Mixing and Milli-Charged Dark Matter From the Hidden Sector},''
  \href{http://dx.doi.org/10.1103/PhysRevD.75.115001}{{\em Phys. Rev. D}
  {\bfseries 75} (2007) 115001},
  \href{http://arxiv.org/abs/hep-ph/0702123}{{\ttfamily arXiv:hep-ph/0702123}}.

\bibitem{Dimopoulos:1989hk}
S.~Dimopoulos, D.~Eichler, R.~Esmailzadeh, and G.~D. Starkman, ``{Getting a
  Charge Out of Dark Matter},''
  \href{http://dx.doi.org/10.1103/PhysRevD.41.2388}{{\em Phys. Rev. D}
  {\bfseries 41} (1990) 2388}.

\bibitem{Davidson:2000hf}
S.~Davidson, S.~Hannestad, and G.~Raffelt, ``{Updated Bounds on Millicharged
  Particles},'' \href{http://dx.doi.org/10.1088/1126-6708/2000/05/003}{{\em
  JHEP} {\bfseries 05} (2000) 003},
  \href{http://arxiv.org/abs/hep-ph/0001179}{{\ttfamily arXiv:hep-ph/0001179}}.

\bibitem{Essig:2013lka}
R.~Essig {\em et~al.}, ``{Working Group Report: New Light Weakly Coupled
  Particles},'' in {\em {Community Summer Study 2013: Snowmass on the
  Mississippi (CSS2013) Minneapolis, MN, USA, July 29-August 6, 2013}}.
\newblock 2013.
\newblock
\href{http://arxiv.org/abs/1311.0029}{{\ttfamily arXiv:1311.0029 [hep-ph]}}.
\newblock

\bibitem{Berlin:2018bsc}
A.~Berlin, N.~Blinov, G.~Krnjaic, P.~Schuster, and N.~Toro, ``{Dark Matter,
  Millicharges, Axion and Scalar Particles, Gauge Bosons, and Other New Physics
  with LDMX},'' \href{http://dx.doi.org/10.1103/PhysRevD.99.075001}{{\em Phys.
  Rev. D} {\bfseries 99} (2019) 075001},
  \href{http://arxiv.org/abs/1807.01730}{{\ttfamily arXiv:1807.01730
  [hep-ph]}}.

\bibitem{Holdom:1986eq}
B.~Holdom, ``{Searching for $\epsilon$ Charges and a New U(1)},''
  \href{http://dx.doi.org/10.1016/0370-2693(86)90470-3}{{\em Phys. Lett. B}
  {\bfseries 178} (1986) 65--70}.

\bibitem{Okun:1982xi}
L.~Okun, ``{Limits of Electrodynamics: Paraphotons?},'' {\em Sov. Phys. JETP}
  {\bfseries 56} (1982) 502.

\bibitem{Galison:1983pa}
P.~Galison and A.~Manohar, ``{Two Z's or Not Two Z's?},''
  \href{http://dx.doi.org/10.1016/0370-2693(84)91161-4}{{\em Phys. Lett. B}
  {\bfseries 136} (1984) 279--283}.

\bibitem{Caputo:2021eaa}
A.~Caputo, A.~J. Millar, C.~A.~J. O'Hare, and E.~Vitagliano, ``{Dark photon
  limits: A handbook},''
  \href{http://dx.doi.org/10.1103/PhysRevD.104.095029}{{\em Phys. Rev. D}
  {\bfseries 104} no.~9, (2021) 095029},
  \href{http://arxiv.org/abs/2105.04565}{{\ttfamily arXiv:2105.04565
  [hep-ph]}}.

\bibitem{Aad:2012tfa}
{G.~Aad {\it et al.}~(ATLAS Collaboration)}, ``{Observation of a New Particle
  in the Search for the Standard Model Higgs Boson with the ATLAS Detector at
  the LHC},'' \href{http://dx.doi.org/10.1016/j.physletb.2012.08.020}{{\em
  Phys. Lett.} {\bfseries B716} (2012) 1--29},
\href{http://arxiv.org/abs/1207.7214}{{\ttfamily arXiv:1207.7214 [hep-ex]}}.

\bibitem{Chatrchyan:2012xdj}
{S.~Chatrchyan {\it et al.}~(CMS Collaboration)}, ``{Observation of a New Boson
  at a Mass of 125 GeV with the CMS Experiment at the LHC},''
  \href{http://dx.doi.org/10.1016/j.physletb.2012.08.021}{{\em Phys. Lett.}
  {\bfseries B716} (2012) 30--61},
\href{http://arxiv.org/abs/1207.7235}{{\ttfamily arXiv:1207.7235 [hep-ex]}}.

\bibitem{Tanabashi:2018oca}
{M.~ Tanabashi {\it et al.}~(Particle Data Group)}, ``{Review of Particle
  Physics},'' \href{http://dx.doi.org/10.1103/PhysRevD.98.030001}{{\em Phys.
  Rev. D} {\bfseries 98} (2018) 030001}.

\bibitem{Riess:1998cb}
{A.~Reiss {\it et al.}~(Supernova Search Team Collaboration)}, ``{Observational
  Evidence from Supernovae for an Accelerating Universe and a Cosmological
  Constant},'' \href{http://dx.doi.org/10.1086/300499}{{\em Astron. J.}
  {\bfseries 116} (1998) 1009--1038},
  \href{http://arxiv.org/abs/astro-ph/9805201}{{\ttfamily
  arXiv:astro-ph/9805201}}.

\bibitem{Perlmutter:1998np}
{S.~Perlmutter {\it et al.}~(Supernova Cosmology Project Collaboration)},
  ``{Measurements of $\Omega$ and $\Lambda$ from 42 High Redshift
  Supernovae},'' \href{http://dx.doi.org/10.1086/307221}{{\em Astrophys. J.}
  {\bfseries 517} (1999) 565--586},
  \href{http://arxiv.org/abs/astro-ph/9812133}{{\ttfamily
  arXiv:astro-ph/9812133}}.

\bibitem{Weinberg:1988cp}
S.~Weinberg, ``{The Cosmological Constant Problem},''
\href{http://dx.doi.org/10.1103/RevModPhys.61.1}{{\em Rev. Mod. Phys.}
  {\bfseries 61} (1989) 1--23}.

\bibitem{Padmanabhan:2002ji}
T.~Padmanabhan, ``{Cosmological constant: The Weight of the Vacuum},''
  \href{http://dx.doi.org/10.1016/S0370-1573(03)00120-0}{{\em Phys. Rept.}
  {\bfseries 380} (2003) 235--320},
  \href{http://arxiv.org/abs/hep-th/0212290}{{\ttfamily arXiv:hep-th/0212290}}.

\bibitem{Baker:2006ts}
C.~Baker {\em et~al.}, ``{An Improved Experimental Limit on the Electric Dipole
  Moment of the Neutron},''
  \href{http://dx.doi.org/10.1103/PhysRevLett.97.131801}{{\em Phys. Rev. Lett.}
  {\bfseries 97} (2006) 131801},
  \href{http://arxiv.org/abs/hep-ex/0602020}{{\ttfamily arXiv:hep-ex/0602020}}.

\bibitem{Afach:2015sja}
J.~M. Pendlebury {\em et~al.}, ``{Revised Experimental Upper Limit on the
  Electric Dipole Moment of the Neutron},''
  \href{http://dx.doi.org/10.1103/PhysRevD.92.092003}{{\em Phys. Rev. D}
  {\bfseries 92} (2015) 092003},
  \href{http://arxiv.org/abs/1509.04411}{{\ttfamily arXiv:1509.04411
  [hep-ex]}}.

\bibitem{Srednicki:1985xd}
M.~Srednicki, ``{Axion Couplings to Matter. 1. CP Conserving Parts},''
  \href{http://dx.doi.org/10.1016/0550-3213(85)90054-9}{{\em Nucl. Phys. B}
  {\bfseries 260} (1985) 689--700}.

\bibitem{diCortona:2015ldu}
G.~Grilli~di Cortona, E.~Hardy, J.~Pardo~Vega, and G.~Villadoro, ``{The QCD
  Axion, Precisely},'' \href{http://dx.doi.org/10.1007/JHEP01(2016)034}{{\em
  JHEP} {\bfseries 01} (2016) 034},
  \href{http://arxiv.org/abs/1511.02867}{{\ttfamily arXiv:1511.02867
  [hep-ph]}}.

\bibitem{Kim:1979if}
J.~E. Kim, ``{Weak Interaction Singlet and Strong CP Invariance},''
  \href{http://dx.doi.org/10.1103/PhysRevLett.43.103}{{\em Phys. Rev. Lett.}
  {\bfseries 43} (1979) 103}.

\bibitem{Shifman:1979if}
M.~A. Shifman, A.~Vainshtein, and V.~I. Zakharov, ``{Can Confinement Ensure
  Natural CP Invariance of Strong Interactions?},''
  \href{http://dx.doi.org/10.1016/0550-3213(80)90209-6}{{\em Nucl. Phys. B}
  {\bfseries 166} (1980) 493--506}.

\bibitem{Dine:1981rt}
M.~Dine, W.~Fischler, and M.~Srednicki, ``{A Simple Solution to the Strong CP
  Problem with a Harmless Axion},''
  \href{http://dx.doi.org/10.1016/0370-2693(81)90590-6}{{\em Phys. Lett. B}
  {\bfseries 104} (1981) 199--202}.

\bibitem{Zhitnitsky:1980tq}
A.~Zhitnitsky, ``{On Possible Suppression of the Axion Hadron Interactions.},''
  {\em Sov. J. Nucl. Phys.} {\bfseries 31} (1980) 260.

\bibitem{Acharya:2010zx}
B.~S. Acharya, K.~Bobkov, and P.~Kumar, ``{An M Theory Solution to the Strong
  CP Problem and Constraints on the Axiverse},''
  \href{http://dx.doi.org/10.1007/JHEP11(2010)105}{{\em JHEP} {\bfseries 11}
  (2010) 105}, \href{http://arxiv.org/abs/1004.5138}{{\ttfamily arXiv:1004.5138
  [hep-th]}}.

\bibitem{Cicoli:2012sz}
M.~Cicoli, M.~Goodsell, and A.~Ringwald, ``{The Type IIB String Axiverse and
  its Low-energy Phenomenology},''
  \href{http://dx.doi.org/10.1007/JHEP10(2012)146}{{\em JHEP} {\bfseries 10}
  (2012) 146}, \href{http://arxiv.org/abs/1206.0819}{{\ttfamily arXiv:1206.0819
  [hep-th]}}.

\bibitem{Green:1987mn}
M.~B. Green, J.~H. Schwarz, and E.~Witten, {\em {Superstring theory. Vol. 2:
  Loop amplitudes, anomalies and phenomenology}}.
\newblock Cambridge Monographs on Mathematical Physics. Cambridge University
  Press, 7, 1988.

\bibitem{Svrcek:2006yi}
P.~Svrcek and E.~Witten, ``{Axions In String Theory},''
  \href{http://dx.doi.org/10.1088/1126-6708/2006/06/051}{{\em JHEP} {\bfseries
  06} (2006) 051},
\href{http://arxiv.org/abs/hep-th/0605206}{{\ttfamily arXiv:hep-th/0605206
  [hep-th]}}.

\bibitem{Dine:2007zp}
M.~Dine, \href{http://dx.doi.org/10.1017/9781009290883}{{\em {Supersymmetry and
  String Theory : Beyond the Standard Model}}}.
\newblock Oxford University Press, 2016.

\bibitem{AxionLimits}
C.~O'Hare, {\em {\it
  \href{https://cajohare.github.io/AxionLimits/}{AxionLimits}}}.

\bibitem{Hui:2021tkt}
L.~Hui, ``{Wave Dark Matter},''
  \href{http://dx.doi.org/10.1146/annurev-astro-120920-010024}{{\em Ann. Rev.
  Astron. Astrophys.} {\bfseries 59} (2021) 247--289},
  \href{http://arxiv.org/abs/2101.11735}{{\ttfamily arXiv:2101.11735
  [astro-ph.CO]}}.

\bibitem{Weinberg:2013aya}
D.~H. Weinberg, J.~S. Bullock, F.~Governato, R.~Kuzio~de Naray, and A.~H.~G.
  Peter, ``{Cold dark matter: controversies on small scales},''
  \href{http://dx.doi.org/10.1073/pnas.1308716112}{{\em Proc. Nat. Acad. Sci.}
  {\bfseries 112} (2015) 12249--12255},
  \href{http://arxiv.org/abs/1306.0913}{{\ttfamily arXiv:1306.0913
  [astro-ph.CO]}}.

\bibitem{Klypin:1999uc}
A.~A. Klypin, A.~V. Kravtsov, O.~Valenzuela, and F.~Prada, ``{Where are the
  missing Galactic satellites?},'' \href{http://dx.doi.org/10.1086/307643}{{\em
  Astrophys. J.} {\bfseries 522} (1999) 82--92},
  \href{http://arxiv.org/abs/astro-ph/9901240}{{\ttfamily
  arXiv:astro-ph/9901240}}.

\bibitem{Boylan-Kolchin:2011qkt}
M.~Boylan-Kolchin, J.~S. Bullock, and M.~Kaplinghat, ``{Too big to fail? The
  puzzling darkness of massive Milky Way subhaloes},''
  \href{http://dx.doi.org/10.1111/j.1745-3933.2011.01074.x}{{\em Mon. Not. Roy.
  Astron. Soc.} {\bfseries 415} (2011) L40},
  \href{http://arxiv.org/abs/1103.0007}{{\ttfamily arXiv:1103.0007
  [astro-ph.CO]}}.

\bibitem{Flores:1994gz}
R.~A. Flores and J.~R. Primack, ``{Observational and theoretical constraints on
  singular dark matter halos},'' \href{http://dx.doi.org/10.1086/187350}{{\em
  Astrophys. J. Lett.} {\bfseries 427} (1994) L1--4},
  \href{http://arxiv.org/abs/astro-ph/9402004}{{\ttfamily
  arXiv:astro-ph/9402004}}.

\bibitem{Irsic:2017yje}
V.~Ir\v{s}i\v{c}, M.~Viel, M.~G. Haehnelt, J.~S. Bolton, and G.~D. Becker,
  ``{First constraints on fuzzy dark matter from Lyman-$\alpha$ forest data and
  hydrodynamical simulations},''
  \href{http://dx.doi.org/10.1103/PhysRevLett.119.031302}{{\em Phys. Rev.
  Lett.} {\bfseries 119} no.~3, (2017) 031302},
  \href{http://arxiv.org/abs/1703.04683}{{\ttfamily arXiv:1703.04683
  [astro-ph.CO]}}.

\bibitem{Rogers:2020ltq}
K.~K. Rogers and H.~V. Peiris, ``{Strong Bound on Canonical Ultralight Axion
  Dark Matter from the Lyman-Alpha Forest},''
  \href{http://dx.doi.org/10.1103/PhysRevLett.126.071302}{{\em Phys. Rev.
  Lett.} {\bfseries 126} no.~7, (2021) 071302},
  \href{http://arxiv.org/abs/2007.12705}{{\ttfamily arXiv:2007.12705
  [astro-ph.CO]}}.

\bibitem{PhysRevD.58.064014}
J.~D. Bekenstein and M.~Schiffer, ``The many faces of superradiance,''
  \href{http://dx.doi.org/10.1103/PhysRevD.58.064014}{{\em Phys. Rev. D}
  {\bfseries 58} (Aug, 1998) 064014}.

\bibitem{PhysRev.93.99}
R.~H. Dicke, ``Coherence in spontaneous radiation processes,''
  \href{http://dx.doi.org/10.1103/PhysRev.93.99}{{\em Phys. Rev.} {\bfseries
  93} (Jan, 1954) 99--110}.

\bibitem{2017NatPh..13..833T}
T.~{Torres}, S.~{Patrick}, A.~{Coutant}, M.~{Richartz}, E.~W. {Tedford}, and
  S.~{Weinfurtner}, ``{Rotational superradiant scattering in a vortex flow},''
  \href{http://dx.doi.org/10.1038/nphys4151}{{\em Nature Physics} {\bfseries
  13} no.~9, (Sept., 2017) 833--836}.

\bibitem{Bekenstein:1973mi}
J.~Bekenstein, ``{Extraction of Energy and Charge from a Black Hole},''
\href{http://dx.doi.org/10.1103/PhysRevD.7.949}{{\em Phys. Rev.} {\bfseries D7}
  (1973) 949--953}.

\bibitem{Starobinskil:1974nkd}
A.~A. Starobinsky and S.~M. Churilov, ``{Amplification of electromagnetic and
  gravitational waves scattered by a rotating "black hole"},'' {\em Sov. Phys.
  JETP} {\bfseries 65} no.~1, (1974) 1--5.

\bibitem{Unruh:1974bw}
W.~Unruh, ``{Second Quantization in the Kerr Metric},''
  \href{http://dx.doi.org/10.1103/PhysRevD.10.3194}{{\em Phys. Rev. D}
  {\bfseries 10} (1974) 3194--3205}.

\bibitem{1992ApJ...386..455H}
K.~{Hirotani}, M.~{Takahashi}, S.-Y. {Nitta}, and A.~{Tomimatsu}, ``{Accretion
  in a Kerr Black Hole Magnetosphere: Energy and Angular Momentum Transport
  between the Magnetic Field and the Matter},''
  \href{http://dx.doi.org/10.1086/171031}{{\em Astrophys. J.} {\bfseries 386}
  (Feb., 1992) 455}.

\bibitem{1999Sci...284..115V}
M.~H.~P.~M. {van Putten}, ``{Superradiance in a Torus Magnetosphere Around a
  Black Hole},'' \href{http://dx.doi.org/10.1126/science.284.5411.115}{{\em
  Science} {\bfseries 284} (Apr., 1999) 115}.

\bibitem{Carter:1968}
B.~Carter, ``{Hamilton-Jacobi and Schr\"odinger Separable Solutions of
  Einstein's Equations},''
{\em Commun. Math. Phys.} {\bfseries 10} (1968) 280.

\bibitem{Brill:1972xj}
D.~Brill, P.~Chrzanowski, M.~Pereira, E.~Fackerell, and J.~Ipser, ``{Solution
  of the Scalar Wave Equation in a Kerr Background by Separation of
  Variables},''
\href{http://dx.doi.org/10.1103/PhysRevD.5.1913}{{\em Phys. Rev.} {\bfseries
  D5} (1972) 1913--1915}.

\bibitem{Teukolsky:1974yv}
S.~A. Teukolsky and W.~H. Press, ``{Perturbations of a rotating black hole. III
  - Interaction of the hole with gravitational and electromagnet ic
  radiation},'' \href{http://dx.doi.org/10.1086/153180}{{\em Astrophys. J.}
  {\bfseries 193} (1974) 443--461}.

\bibitem{Press:1972zz}
W.~H. Press and S.~A. Teukolsky, ``{Floating Orbits, Superradiant Scattering
  and the Black-hole Bomb},''
\href{http://dx.doi.org/10.1038/238211a0}{{\em Nature} {\bfseries 238} (1972)
  211--212}.

\bibitem{Arvanitaki:2010sy}
A.~Arvanitaki and S.~Dubovsky, ``{Exploring the String Axiverse with Precision
  Black Hole Physics},''
  \href{http://dx.doi.org/10.1103/PhysRevD.83.044026}{{\em Phys. Rev. D}
  {\bfseries 83} (2011) 044026},
\href{http://arxiv.org/abs/1004.3558}{{\ttfamily arXiv:1004.3558 [hep-th]}}.

\bibitem{Baumann:2019eav}
D.~Baumann, H.~S. Chia, J.~Stout, and L.~ter Haar, ``{The Spectra of
  Gravitational Atoms},''
  \href{http://dx.doi.org/10.1088/1475-7516/2019/12/006}{{\em JCAP} {\bfseries
  12} (2019) 006}, \href{http://arxiv.org/abs/1908.10370}{{\ttfamily
  arXiv:1908.10370 [gr-qc]}}.

\bibitem{Detweiler:1980uk}
S.~Detweiler, ``{Klein-Gordon Equation and Rotating Black Holes},''
\href{http://dx.doi.org/10.1103/PhysRevD.22.2323}{{\em Phys. Rev.} {\bfseries
  D22} (1980) 2323--2326}.

\bibitem{Hui:2022sri}
L.~Hui, Y.~T.~A. Law, L.~Santoni, G.~Sun, G.~M. Tomaselli, and E.~Trincherini,
  ``{Black hole superradiance with dark matter accretion},''
  \href{http://dx.doi.org/10.1103/PhysRevD.107.104018}{{\em Phys. Rev. D}
  {\bfseries 107} no.~10, (2023) 104018},
  \href{http://arxiv.org/abs/2208.06408}{{\ttfamily arXiv:2208.06408 [gr-qc]}}.

\bibitem{Yoshino:2013ofa}
H.~Yoshino and H.~Kodama, ``{Gravitational radiation from an axion cloud around
  a black hole: Superradiant phase},''
  \href{http://dx.doi.org/10.1093/ptep/ptu029}{{\em PTEP} {\bfseries 2014}
  (2014) 043E02}, \href{http://arxiv.org/abs/1312.2326}{{\ttfamily
  arXiv:1312.2326 [gr-qc]}}.

\bibitem{Brito:2017zvb}
R.~Brito {\em et~al.}, ``{Gravitational Wave Searches for Ultralight Bosons
  with LIGO and LISA},''
  \href{http://dx.doi.org/10.1103/PhysRevD.96.064050}{{\em Phys. Rev. D}
  {\bfseries 96} (2017) 064050},
\href{http://arxiv.org/abs/1706.06311}{{\ttfamily arXiv:1706.06311 [gr-qc]}}.

\bibitem{Baryakhtar:2020gao}
M.~Baryakhtar, M.~Galanis, R.~Lasenby, and O.~Simon, ``{Black hole
  superradiance of self-interacting scalar fields},''
  \href{http://dx.doi.org/10.1103/PhysRevD.103.095019}{{\em Phys. Rev. D}
  {\bfseries 103} no.~9, (2021) 095019},
  \href{http://arxiv.org/abs/2011.11646}{{\ttfamily arXiv:2011.11646
  [hep-ph]}}.

\bibitem{Chia:2020dye}
H.~S. Chia, {\em {Probing Particle Physics with Gravitational Waves}}.
\newblock PhD thesis, Amsterdam U., 2020.
\newblock \href{http://arxiv.org/abs/2012.09167}{{\ttfamily arXiv:2012.09167
  [hep-ph]}}.

\bibitem{Tomaselli:2024faa}
G.~M. Tomaselli, {\em {Gravitational atoms and black hole binaries}}.
\newblock PhD thesis, Amsterdam U., 2024.
\newblock \href{http://arxiv.org/abs/2412.12526}{{\ttfamily arXiv:2412.12526
  [gr-qc]}}.

\bibitem{Chen_plasma}
F.~F. Chen, \href{http://dx.doi.org/10.1007/978-3-319-22309-4}{{\em
  {Introduction to Plasma Physics and Controlled Fusion}}}.
\newblock Springer Cham, 12, 2015.

\bibitem{Fitz_plasma}
R.~Fitzpatrick, \href{http://dx.doi.org/10.1201/9781003268253}{{\em {Plasma
  Physics: An Introduction}}}.
\newblock CRC Press, 08, 2014.

\bibitem{Saintonge:2022tfq}
A.~Saintonge and B.~Catinella, ``{The cold interstellar medium of galaxies in
  the Local Universe},''
  \href{http://dx.doi.org/10.1146/annurev-astro-021022-043545}{{\em Ann. Rev.
  Astron. Astrophys.} {\bfseries 60} (2022) 319--361},
  \href{http://arxiv.org/abs/2202.00690}{{\ttfamily arXiv:2202.00690
  [astro-ph.GA]}}.

\bibitem{BirdsallLangdon}
C.~K. Birdsall and A.~B. Langdon, {\em Plasma physics via computer simulation}.
\newblock Taylor and Francis, New York, 2005.

\bibitem{Arber_2015}
T.~D. Arber, K.~Bennett, C.~S. Brady, A.~Lawrence-Douglas, M.~G. Ramsay, N.~J.
  Sircombe, P.~Gillies, R.~G. Evans, H.~Schmitz, A.~R. Bell, and C.~P. Ridgers,
  ``Contemporary particle-in-cell approach to laser-plasma modelling,''
  \href{http://dx.doi.org/10.1088/0741-3335/57/11/113001}{{\em Plasma Physics
  and Controlled Fusion} {\bfseries 57} no.~11, (Sep, 2015) 113001}.

\bibitem{2012ASSL..391.....S}
B.~V. {Somov}, \href{http://dx.doi.org/10.1007/978-1-4614-4283-7}{{\em {Plasma
  Astrophysics, Part I: Fundamentals and Practice}}}, vol.~391.
\newblock Springer Nature, 2012.

\bibitem{osti_5612158}
J.~P. Freidberg, \href{http://dx.doi.org/10.1017/CBO9780511795046}{{\em {Ideal
  magnetohydrodynamics}}}.
\newblock Cambridge University Press, 07, 2014.

\bibitem{1991JPlPh..45..135K}
W.~L. {Kruer}, ``{The Physics of Laser Plasma Interactions},''
  \href{http://dx.doi.org/10.1017/S0022377800015555}{{\em Journal of Plasma
  Physics} {\bfseries 45} no.~1, (Jan., 1991) 135}.

\bibitem{1995ppai.book.....D}
R.~O. {Dendy}, {\em {Plasma Physics: An Introductory Course}}.
\newblock Cambridge University Press, 1995.

\bibitem{Cannizzaro:2020uap}
E.~Cannizzaro, A.~Caputo, L.~Sberna, and P.~Pani, ``{Plasma-photon interaction
  in curved spacetime I: formalism and quasibound states around nonspinning
  black holes},'' \href{http://dx.doi.org/10.1103/PhysRevD.103.124018}{{\em
  Phys. Rev. D} {\bfseries 103} (2021) 124018},
  \href{http://arxiv.org/abs/2012.05114}{{\ttfamily arXiv:2012.05114 [gr-qc]}}.

\bibitem{Cannizzaro:2021zbp}
E.~Cannizzaro, A.~Caputo, L.~Sberna, and P.~Pani, ``{Plasma-photon interaction
  in curved spacetime. II. Collisions, thermal corrections, and superradiant
  instabilities},'' \href{http://dx.doi.org/10.1103/PhysRevD.104.104048}{{\em
  Phys. Rev. D} {\bfseries 104} no.~10, (2021) 104048},
  \href{http://arxiv.org/abs/2107.01174}{{\ttfamily arXiv:2107.01174 [gr-qc]}}.

\bibitem{1981AA....96..293B}
R.~A. {Breuer} and J.~{Ehlers}, ``{Propagation of electromagnetic waves through
  magnetized plasmas in arbitrary gravitational fields},'' {\em Astron.
  Astrophys.} {\bfseries 96} no.~1-2, (Mar., 1981) 293--295.

\bibitem{Novikov:1973kta}
I.~D. Novikov and K.~S. Thorne, ``{Astrophysics and black holes},'' in {\em
  {Les Houches Summer School of Theoretical Physics}: {Black Holes}},
  pp.~343--550.
\newblock 1973.

\bibitem{Kocsis:2011dr}
B.~Kocsis, N.~Yunes, and A.~Loeb, ``{Observable Signatures of EMRI Black Hole
  Binaries Embedded in Thin Accretion Disks},''
  \href{http://dx.doi.org/10.1103/PhysRevD.86.049907}{{\em Phys. Rev. D}
  {\bfseries 84} (2011) 024032},
  \href{http://arxiv.org/abs/1104.2322}{{\ttfamily arXiv:1104.2322
  [astro-ph.GA]}}.

\bibitem{Speri:2022upm}
L.~Speri, A.~Antonelli, L.~Sberna, S.~Babak, E.~Barausse, J.~R. Gair, and M.~L.
  Katz, ``{Probing Accretion Physics with Gravitational Waves},''
  \href{http://dx.doi.org/10.1103/PhysRevX.13.021035}{{\em Phys. Rev. X}
  {\bfseries 13} no.~2, (2023) 021035},
  \href{http://arxiv.org/abs/2207.10086}{{\ttfamily arXiv:2207.10086 [gr-qc]}}.

\bibitem{1990agn..conf.....B}
R.~D. {Blandford}, H.~{Netzer}, L.~{Woltjer}, T.~J.~L. {Courvoisier}, and
  M.~{Mayor}, eds., {\em {Active Galactic Nuclei}}.
\newblock Jan., 1990.

\bibitem{2013peag.book.....N}
H.~{Netzer}, {\em {The Physics and Evolution of Active Galactic Nuclei}}.
\newblock Cambridge University Press, 2013.

\bibitem{1991ApJ...376..214B}
S.~A. {Balbus} and J.~F. {Hawley}, ``{A Powerful Local Shear Instability in
  Weakly Magnetized Disks. I. Linear Analysis},''
  \href{http://dx.doi.org/10.1086/170270}{{\em Astrophys. J.} {\bfseries 376}
  (July, 1991) 214}.

\bibitem{1964SPhD....9..195Z}
Y.~B. {Zel'dovich}, ``{The Fate of a Star and the Evolution of Gravitational
  Energy Upon Accretion},'' {\em Soviet Physics Doklady} {\bfseries 9} (Sept.,
  1964) 195.

\bibitem{1964ApJ...140..796S}
E.~E. {Salpeter}, ``{Accretion of Interstellar Matter by Massive Objects.},''
  \href{http://dx.doi.org/10.1086/147973}{{\em Astrophys. J.} {\bfseries 140}
  (Aug., 1964) 796--800}.

\bibitem{2015ARAA..53..365N}
H.~{Netzer}, ``{Revisiting the Unified Model of Active Galactic Nuclei},''
  \href{http://dx.doi.org/10.1146/annurev-astro-082214-122302}{{\em Annual
  Review of Astronomy and Astrophysics} {\bfseries 53} (Aug., 2015) 365--408},
  \href{http://arxiv.org/abs/1505.00811}{{\ttfamily arXiv:1505.00811
  [astro-ph.GA]}}.

\bibitem{2012MNRAS.425..460M}
B.~{McKernan}, K.~E.~S. {Ford}, W.~{Lyra}, and H.~B. {Perets}, ``{Intermediate
  mass black holes in AGN discs - I. Production and growth},''
  \href{http://dx.doi.org/10.1111/j.1365-2966.2012.21486.x}{{\em Mon. Not. Roy.
  Astron. Soc.} {\bfseries 425} no.~1, (Sept., 2012) 460--469},
  \href{http://arxiv.org/abs/1206.2309}{{\ttfamily arXiv:1206.2309
  [astro-ph.GA]}}.

\bibitem{2011MNRAS.417L.103M}
B.~{McKernan}, K.~E.~S. {Ford}, W.~{Lyra}, H.~B. {Perets}, L.~M. {Winter}, and
  T.~{Yaqoob}, ``{On rapid migration and accretion within discs around
  supermassive black holes},''
  \href{http://dx.doi.org/10.1111/j.1745-3933.2011.01132.x}{{\em Mon. Not. Roy.
  Astron. Soc.} {\bfseries 417} no.~1, (Oct., 2011) L103--L107},
  \href{http://arxiv.org/abs/1108.1787}{{\ttfamily arXiv:1108.1787
  [astro-ph.CO]}}.

\bibitem{Yang:2019cbr}
Y.~Yang {\em et~al.}, ``{Hierarchical Black Hole Mergers in Active Galactic
  Nuclei},'' \href{http://dx.doi.org/10.1103/PhysRevLett.123.181101}{{\em Phys.
  Rev. Lett.} {\bfseries 123} no.~18, (2019) 181101},
  \href{http://arxiv.org/abs/1906.09281}{{\ttfamily arXiv:1906.09281
  [astro-ph.HE]}}.

\bibitem{Secunda:2018kar}
A.~Secunda, J.~Bellovary, M.-M. Mac~Low, K.~E. Saavik~Ford, B.~McKernan,
  N.~Leigh, W.~Lyra, and Z.~S\'andor, ``{Orbital Migration of Interacting
  Stellar Mass Black Holes in Disks around Supermassive Black Holes},''
  \href{http://dx.doi.org/10.3847/1538-4357/ab20ca}{{\em Astrophys. J.}
  {\bfseries 878} no.~2, (2019) 85},
  \href{http://arxiv.org/abs/1807.02859}{{\ttfamily arXiv:1807.02859
  [astro-ph.HE]}}.

\bibitem{Fabj:2020qqc}
G.~Fabj, S.~S. Nasim, F.~Caban, K.~E.~S. Ford, B.~McKernan, and J.~M.
  Bellovary, ``{Aligning nuclear cluster orbits with an active galactic nucleus
  accretion disc},'' \href{http://dx.doi.org/10.1093/mnras/staa3004}{{\em Mon.
  Not. Roy. Astron. Soc.} {\bfseries 499} no.~2, (2020) 2608--2616},
  \href{http://arxiv.org/abs/2006.11229}{{\ttfamily arXiv:2006.11229
  [astro-ph.GA]}}.

\bibitem{Gerosa:2021mno}
D.~Gerosa and M.~Fishbach, ``{Hierarchical mergers of stellar-mass black holes
  and their gravitational-wave signatures},''
  \href{http://dx.doi.org/10.1038/s41550-021-01398-w}{{\em Nature Astron.}
  {\bfseries 5} no.~8, (2021) 749--760},
  \href{http://arxiv.org/abs/2105.03439}{{\ttfamily arXiv:2105.03439
  [astro-ph.HE]}}.

\bibitem{Santini:2023ukl}
A.~Santini, D.~Gerosa, R.~Cotesta, and E.~Berti, ``{Black-hole mergers in
  disklike environments could explain the observed q-$\chi$eff correlation},''
  \href{http://dx.doi.org/10.1103/PhysRevD.108.083033}{{\em Phys. Rev. D}
  {\bfseries 108} no.~8, (2023) 083033},
  \href{http://arxiv.org/abs/2308.12998}{{\ttfamily arXiv:2308.12998
  [astro-ph.HE]}}.

\bibitem{Whitehead:2023hmh}
H.~Whitehead, C.~Rowan, T.~Boekholt, and B.~Kocsis, ``{Gas assisted binary
  black hole formation in AGN discs},''
  \href{http://dx.doi.org/10.1093/mnras/stae1430}{{\em Mon. Not. Roy. Astron.
  Soc.} {\bfseries 531} no.~4, (2024) 4656--4680},
  \href{http://arxiv.org/abs/2309.11561}{{\ttfamily arXiv:2309.11561
  [astro-ph.GA]}}.

\bibitem{Vaccaro:2023cwr}
M.~P. Vaccaro, M.~Mapelli, C.~P\'erigois, D.~Barone, M.~C. Artale,
  M.~Dall'Amico, G.~Iorio, and S.~Torniamenti, ``{Impact of gas hardening on
  the population properties of hierarchical black hole mergers in active
  galactic nucleus disks},''
  \href{http://dx.doi.org/10.1051/0004-6361/202348509}{{\em Astron. Astrophys.}
  {\bfseries 685} (2024) A51},
  \href{http://arxiv.org/abs/2311.18548}{{\ttfamily arXiv:2311.18548
  [astro-ph.HE]}}.

\bibitem{1972ApJ...173..431W}
J.~R. {Wilson}, ``{Numerical Study of Fluid Flow in a Kerr Space},''
  \href{http://dx.doi.org/10.1086/151434}{{\em Astrophys. J.} {\bfseries 173}
  (Apr., 1972) 431}.

\bibitem{1991MNRAS.250..581F}
S.~A.~E.~G. {Falle}, ``{Self-similar jets.},''
  \href{http://dx.doi.org/10.1093/mnras/250.3.581}{{\em Mon. Not. Roy. Astron.
  Soc.} {\bfseries 250} (June, 1991) 581--596}.

\bibitem{Schartmann:2005pe}
M.~Schartmann, K.~Meisenheimer, M.~Camenzind, S.~Wolf, and T.~Henning,
  ``{Towards a physical model of dust tori in active galactic nuclei -
  Radiative transfer calculations for a hydrostatic torus model},''
  \href{http://dx.doi.org/10.1051/0004-6361:20042363}{{\em Astron. Astrophys.}
  {\bfseries 437} (2005) 861},
  \href{http://arxiv.org/abs/astro-ph/0504105}{{\ttfamily
  arXiv:astro-ph/0504105}}.

\bibitem{Schartmann:2008qb}
M.~Schartmann, K.~Meisenheimer, M.~Camenzind, S.~Wolf, K.~R.~W. Tristram, and
  T.~Henning, ``{Three-dimensional radiative transfer models of clumpy tori in
  Seyfert galaxies},'' \href{http://dx.doi.org/10.1051/0004-6361:20078907}{{\em
  Astron. Astrophys.} {\bfseries 482} (2008) 67},
  \href{http://arxiv.org/abs/0802.2604}{{\ttfamily arXiv:0802.2604
  [astro-ph]}}.

\bibitem{2009ApJ...702...63W}
K.~{Wada}, P.~P. {Papadopoulos}, and M.~{Spaans}, ``{Molecular Gas Disk
  Structures Around Active Galactic Nuclei},''
  \href{http://dx.doi.org/10.1088/0004-637X/702/1/63}{{\em Astrophys. J.}
  {\bfseries 702} no.~1, (Sept., 2009) 63--74},
  \href{http://arxiv.org/abs/0906.5444}{{\ttfamily arXiv:0906.5444
  [astro-ph.GA]}}.

\bibitem{Husko:2022uwx}
F.~Hu\v{s}ko and C.~G. Lacey, ``{Active galactic nuclei jets simulated with
  smoothed particle hydrodynamics},''
  \href{http://dx.doi.org/10.1093/mnras/stad450}{{\em Mon. Not. Roy. Astron.
  Soc.} {\bfseries 520} no.~4, (2023) 5090--5109},
  \href{http://arxiv.org/abs/2205.08884}{{\ttfamily arXiv:2205.08884
  [astro-ph.GA]}}.

\bibitem{Jiang:2019bxn}
Y.-F. {Jiang}, O.~{Blaes}, J.~M. {Stone}, and S.~W. {Davis}, ``{Global
  Radiation Magnetohydrodynamic Simulations of sub-Eddington Accretion Disks
  around Supermassive Black Holes},''
  \href{http://dx.doi.org/10.3847/1538-4357/ab4a00}{{\em Astrophys. J.}
  {\bfseries 885} no.~2, (Nov., 2019) 144},
  \href{http://arxiv.org/abs/1904.01674}{{\ttfamily arXiv:1904.01674
  [astro-ph.HE]}}.

\bibitem{Page:1974he}
D.~N. Page and K.~S. Thorne, ``{Disk-Accretion onto a Black Hole. Time-Averaged
  Structure of Accretion Disk},'' \href{http://dx.doi.org/10.1086/152990}{{\em
  Astrophys. J.} {\bfseries 191} (1974) 499--506}.

\bibitem{1995ApJ...450..508R}
H.~{Riffert} and H.~{Herold}, ``{Relativistic Accretion Disk Structure
  Revisited},'' \href{http://dx.doi.org/10.1086/176161}{{\em Astrophys. J.}
  {\bfseries 450} (Sept., 1995) 508}.

\bibitem{2012MNRAS.420..684P}
R.~F. {Penna}, A.~{S{\k{a}}dowski}, and J.~C. {McKinney}, ``{Thin-disc theory
  with a non-zero-torque boundary condition and comparisons with
  simulations},''
  \href{http://dx.doi.org/10.1111/j.1365-2966.2011.20084.x}{{\em Mon. Not. Roy.
  Astron. Soc.} {\bfseries 420} no.~1, (Feb., 2012) 684--698},
  \href{http://arxiv.org/abs/1110.6556}{{\ttfamily arXiv:1110.6556
  [astro-ph.HE]}}.

\bibitem{Abramowicz2013}
M.~A. Abramowicz and P.~C. Fragile, ``{Foundations of Black Hole Accretion Disk
  Theory},'' \href{http://dx.doi.org/10.12942/lrr-2013-1}{{\em Living Rev.
  Rel.} {\bfseries 16} (2013) 1},
  \href{http://arxiv.org/abs/1104.5499}{{\ttfamily arXiv:1104.5499
  [astro-ph.HE]}}.

\bibitem{1978ApJ...221..652P}
T.~{Piran}, ``{The role of viscosity and cooling mechanisms in the stability of
  accretion disks.},'' \href{http://dx.doi.org/10.1086/156069}{{\em Astrophys.
  J.} {\bfseries 221} (Apr., 1978) 652--660}.

\bibitem{1974ApJ...187L...1L}
A.~P. {Lightman} and D.~M. {Eardley}, ``{Black Holes in Binary Systems:
  Instability of Disk Accretion},''
  \href{http://dx.doi.org/10.1086/181377}{{\em Astrophys. J.} {\bfseries 187}
  (Jan., 1974) L1}.

\bibitem{1976MNRAS.175..613S}
N.~I. {Shakura} and R.~A. {Sunyaev}, ``{A theory of the instability of disk
  accretion on to black holes and the variability of binary X-ray sources,
  galactic nuclei and quasars.},''
  \href{http://dx.doi.org/10.1093/mnras/175.3.613}{{\em Mon. Not. Roy. Astron.
  Soc.} {\bfseries 175} (June, 1976) 613--632}.

\bibitem{1977A&A....59..111B}
G.~S. {Bisnovatyi-Kogan} and S.~I. {Blinnikov}, ``{Disk accretion onto a black
  hole at subcritical luminosity},'' {\em Astron. Astrophys.} {\bfseries 59}
  (July, 1977) 111--125.

\bibitem{1981ApJ...247...19S}
P.~J. {Sakimoto} and F.~V. {Coroniti}, ``{Accretion disk models for QSOs and
  active galactic nuclei - The role of magnetic viscosity},''
  \href{http://dx.doi.org/10.1086/159005}{{\em Astrophys. J.} {\bfseries 247}
  (July, 1981) 19--31}.

\bibitem{2010ApJ...713...52D}
S.~W. {Davis}, J.~M. {Stone}, and M.~E. {Pessah}, ``{Sustained
  Magnetorotational Turbulence in Local Simulations of Stratified Disks with
  Zero Net Magnetic Flux},''
  \href{http://dx.doi.org/10.1088/0004-637X/713/1/52}{{\em Astrophys. J.}
  {\bfseries 713} no.~1, (Apr., 2010) 52--65},
  \href{http://arxiv.org/abs/0909.1570}{{\ttfamily arXiv:0909.1570
  [astro-ph.HE]}}.

\bibitem{2012A&A...545A.115K}
I.~{Kotko} and J.~P. {Lasota}, ``{The viscosity parameter {\ensuremath{\alpha}}
  and the properties of accretion disc outbursts in close binaries},''
  \href{http://dx.doi.org/10.1051/0004-6361/201219618}{{\em Astron. Astrophys.}
  {\bfseries 545} (Sept., 2012) A115},
  \href{http://arxiv.org/abs/1209.0017}{{\ttfamily arXiv:1209.0017
  [astro-ph.SR]}}.

\bibitem{Nouri:2023nss}
F.~H. Nouri and A.~Janiuk, ``{Viscous torque in turbulent magnetized active
  galactic nucleus accretion disks and its effects on the gravitational waves
  of extreme mass ratio inspirals},''
  \href{http://dx.doi.org/10.1051/0004-6361/202348796}{{\em Astron. Astrophys.}
  {\bfseries 687} (2024) A184},
  \href{http://arxiv.org/abs/2309.06028}{{\ttfamily arXiv:2309.06028
  [astro-ph.HE]}}.

\bibitem{1981ARA&A..19..137P}
J.~E. {Pringle}, ``{Accretion discs in astrophysics},''
  \href{http://dx.doi.org/10.1146/annurev.aa.19.090181.001033}{{\em Annual
  review of astronomy and astrophysics} {\bfseries 19} (Jan., 1981) 137--162}.

\bibitem{Thompson:2005mf}
T.~A. Thompson, E.~Quataert, and N.~Murray, ``{Radiation pressure supported
  starburst disks and AGN fueling},''
  \href{http://dx.doi.org/10.1086/431923}{{\em Astrophys. J.} {\bfseries 630}
  (2005) 167--185}, \href{http://arxiv.org/abs/astro-ph/0503027}{{\ttfamily
  arXiv:astro-ph/0503027}}.

\bibitem{Gangardt:2024bic}
D.~Gangardt, A.~A. Trani, C.~Bonnerot, and D.~Gerosa, ``{pAGN: the one-stop
  solution for AGN disc modelling},''
  \href{http://dx.doi.org/10.1093/mnras/stae1117}{{\em Mon. Not. Roy. Astron.
  Soc.} {\bfseries 530} no.~4, (2024) 3689--3705},
  \href{http://arxiv.org/abs/2403.00060}{{\ttfamily arXiv:2403.00060
  [astro-ph.HE]}}.

\bibitem{1964ApJ...139.1217T}
A.~{Toomre}, ``{On the gravitational stability of a disk of stars.},''
  \href{http://dx.doi.org/10.1086/147861}{{\em Astrophys. J.} {\bfseries 139}
  (May, 1964) 1217--1238}.

\bibitem{Ayzenberg:2023hfw}
D.~Ayzenberg {\em et~al.}, ``{Fundamental Physics Opportunities with the
  Next-Generation Event Horizon Telescope},''
  \href{http://arxiv.org/abs/2312.02130}{{\ttfamily arXiv:2312.02130
  [astro-ph.HE]}}.

\bibitem{EventHorizonTelescope:2021bee}
{\bfseries Event Horizon Telescope} Collaboration, K.~Akiyama {\em et~al.},
  ``{First M87 Event Horizon Telescope Results. VII. Polarization of the
  Ring},'' \href{http://dx.doi.org/10.3847/2041-8213/abe71d}{{\em Astrophys. J.
  Lett.} {\bfseries 910} no.~1, (2021) L12},
  \href{http://arxiv.org/abs/2105.01169}{{\ttfamily arXiv:2105.01169
  [astro-ph.HE]}}.

\bibitem{EventHorizonTelescope:2022xqj}
{\bfseries Event Horizon Telescope} Collaboration, K.~Akiyama {\em et~al.},
  ``{First Sagittarius A* Event Horizon Telescope Results. VI. Testing the
  Black Hole Metric},'' \href{http://dx.doi.org/10.3847/2041-8213/ac6756}{{\em
  Astrophys. J. Lett.} {\bfseries 930} no.~2, (2022) L17},
  \href{http://arxiv.org/abs/2311.09484}{{\ttfamily arXiv:2311.09484
  [astro-ph.HE]}}.

\bibitem{Nampalliwar:2021tyz}
S.~Nampalliwar, S.~Kumar, K.~Jusufi, Q.~Wu, M.~Jamil, and P.~Salucci,
  ``{Modeling the Sgr A* Black Hole Immersed in a Dark Matter Spike},''
  \href{http://dx.doi.org/10.3847/1538-4357/ac05cc}{{\em Astrophys. J.}
  {\bfseries 916} no.~2, (2021) 116},
  \href{http://arxiv.org/abs/2103.12439}{{\ttfamily arXiv:2103.12439
  [astro-ph.HE]}}.

\bibitem{Ezquiaga:2020dao}
J.~M. Ezquiaga and M.~Zumalac\'arregui, ``{Gravitational wave lensing beyond
  general relativity: birefringence, echoes and shadows},''
  \href{http://dx.doi.org/10.1103/PhysRevD.102.124048}{{\em Phys. Rev. D}
  {\bfseries 102} no.~12, (2020) 124048},
  \href{http://arxiv.org/abs/2009.12187}{{\ttfamily arXiv:2009.12187 [gr-qc]}}.

\bibitem{2013A&A...554A..36L}
T.~{Lacroix} and J.~{Silk}, ``{Constraining the distribution of dark matter at
  the Galactic centre using the high-resolution Event Horizon Telescope},''
  \href{http://dx.doi.org/10.1051/0004-6361/201220753}{{\em Astron. Astrophys.}
  {\bfseries 554} (June, 2013) A36},
  \href{http://arxiv.org/abs/1211.4861}{{\ttfamily arXiv:1211.4861
  [astro-ph.GA]}}.

\bibitem{Profumo:2006im}
S.~Profumo and K.~Sigurdson, ``{The Shadow of Dark Matter},''
  \href{http://dx.doi.org/10.1103/PhysRevD.75.023521}{{\em Phys. Rev. D}
  {\bfseries 75} (2007) 023521},
  \href{http://arxiv.org/abs/astro-ph/0611129}{{\ttfamily
  arXiv:astro-ph/0611129}}.

\bibitem{Yang:2023tip}
Y.~Yang, D.~Liu, A.~\"Ovg\"un, G.~Lambiase, and Z.-W. Long, ``{Black hole
  surrounded by the pseudo-isothermal dark matter halo},''
  \href{http://dx.doi.org/10.1140/epjc/s10052-024-12412-6}{{\em Eur. Phys. J.
  C} {\bfseries 84} no.~1, (2024) 63},
  \href{http://arxiv.org/abs/2308.05544}{{\ttfamily arXiv:2308.05544 [gr-qc]}}.

\bibitem{Liu:2023oab}
D.~Liu, Y.~Yang, Z.~Xu, and Z.-W. Long, ``{Modeling the black holes surrounded
  by a dark matter halo in the galactic center of M87},''
  \href{http://dx.doi.org/10.1140/epjc/s10052-024-12492-4}{{\em Eur. Phys. J.
  C} {\bfseries 84} no.~2, (2024) 136},
  \href{http://arxiv.org/abs/2307.13553}{{\ttfamily arXiv:2307.13553 [gr-qc]}}.

\bibitem{Davoudiasl:2019nlo}
H.~Davoudiasl and P.~B. Denton, ``{Ultralight Boson Dark Matter and Event
  Horizon Telescope Observations of M87*},''
  \href{http://dx.doi.org/10.1103/PhysRevLett.123.021102}{{\em Phys. Rev.
  Lett.} {\bfseries 123} no.~2, (2019) 021102},
  \href{http://arxiv.org/abs/1904.09242}{{\ttfamily arXiv:1904.09242
  [astro-ph.CO]}}.

\bibitem{Saha:2022hcd}
A.~K. Saha, P.~Parashari, T.~N. Maity, A.~Dubey, S.~Bouri, and R.~Laha,
  ``{Bounds on ultralight bosons from the Event Horizon Telescope observation
  of Sgr A$^*$},''
  \href{http://dx.doi.org/10.1140/epjc/s10052-024-13239-x}{{\em Eur. Phys. J.
  C} {\bfseries 84} no.~9, (2024) 901},
  \href{http://arxiv.org/abs/2208.03530}{{\ttfamily arXiv:2208.03530
  [astro-ph.HE]}}.

\bibitem{GRAVITY:2020gka}
{\bfseries GRAVITY} Collaboration, R.~Abuter {\em et~al.}, ``{Detection of the
  Schwarzschild precession in the orbit of the star S2 near the Galactic centre
  massive black hole},''
  \href{http://dx.doi.org/10.1051/0004-6361/202037813}{{\em Astron. Astrophys.}
  {\bfseries 636} (2020) L5}, \href{http://arxiv.org/abs/2004.07187}{{\ttfamily
  arXiv:2004.07187 [astro-ph.GA]}}.

\bibitem{Zakharov:2007fj}
A.~F. Zakharov, A.~A. Nucita, F.~De~Paolis, and G.~Ingrosso, ``{Apoastron Shift
  Constraints on Dark Matter Distribution at the Galactic Center},''
  \href{http://dx.doi.org/10.1103/PhysRevD.76.062001}{{\em Phys. Rev. D}
  {\bfseries 76} (2007) 062001},
  \href{http://arxiv.org/abs/0707.4423}{{\ttfamily arXiv:0707.4423
  [astro-ph]}}.

\bibitem{Lacroix:2018zmg}
T.~Lacroix, ``{Dynamical constraints on a dark matter spike at the Galactic
  Centre from stellar orbits},''
  \href{http://dx.doi.org/10.1051/0004-6361/201832652}{{\em Astron. Astrophys.}
  {\bfseries 619} (2018) A46},
  \href{http://arxiv.org/abs/1801.01308}{{\ttfamily arXiv:1801.01308
  [astro-ph.GA]}}.

\bibitem{Shen:2023kkm}
Z.-Q. Shen, G.-W. Yuan, C.-Z. Jiang, Y.-L.~S. Tsai, Q.~Yuan, and Y.-Z. Fan,
  ``{Exploring dark matter spike distribution around the Galactic centre with
  stellar orbits},'' \href{http://dx.doi.org/10.1093/mnras/stad3282}{{\em Mon.
  Not. Roy. Astron. Soc.} {\bfseries 527} no.~2, (2023) 3196--3207},
  \href{http://arxiv.org/abs/2303.09284}{{\ttfamily arXiv:2303.09284
  [astro-ph.GA]}}.

\bibitem{2022NatSR..1215258C}
M.~H. {Chan}, C.~M. {Lee}, and C.~W. {Yu}, ``{Investigating the nature of mass
  distribution surrounding the Galactic supermassive black hole},''
  \href{http://dx.doi.org/10.1038/s41598-022-18946-7}{{\em Scientific Reports}
  {\bfseries 12} (Sept., 2022) 15258},
  \href{http://arxiv.org/abs/2208.12471}{{\ttfamily arXiv:2208.12471
  [astro-ph.GA]}}.

\bibitem{GRAVITY:2023cjt}
{\bfseries GRAVITY} Collaboration, A.~Foschi {\em et~al.}, ``{Using the motion
  of S2 to constrain scalar clouds around Sgr A*},''
  \href{http://dx.doi.org/10.1093/mnras/stad1939}{{\em Mon. Not. Roy. Astron.
  Soc.} {\bfseries 524} no.~1, (2023) 1075--1086},
  \href{http://arxiv.org/abs/2306.17215}{{\ttfamily arXiv:2306.17215
  [astro-ph.GA]}}.

\bibitem{Zuriaga-Puig:2023imf}
J.~Zuriaga-Puig, V.~Gammaldi, D.~Gaggero, T.~Lacroix, and M.~A.
  S\'anchez-Conde, ``{Multi-TeV dark matter density in the inner Milky Way
  halo: spectral and dynamical constraints},''
  \href{http://dx.doi.org/10.1088/1475-7516/2023/11/063}{{\em JCAP} {\bfseries
  11} (2023) 063}, \href{http://arxiv.org/abs/2307.06823}{{\ttfamily
  arXiv:2307.06823 [astro-ph.HE]}}.

\bibitem{John:2023knt}
I.~John, R.~K. Leane, and T.~Linden, ``{Dark matter scattering constraints from
  observations of stars surrounding Sgr A*},''
  \href{http://dx.doi.org/10.1103/PhysRevD.109.123041}{{\em Phys. Rev. D}
  {\bfseries 109} no.~12, (2024) 123041},
  \href{http://arxiv.org/abs/2311.16228}{{\ttfamily arXiv:2311.16228
  [astro-ph.HE]}}.

\bibitem{Lechien:2023psa}
T.~Lechien, G.~Hei\ss{}el, J.~Grover, and D.~Izzo, ``{Dark matter
  reconstruction from stellar orbits in the Galactic centre},''
  \href{http://dx.doi.org/10.1051/0004-6361/202347738}{{\em Astron. Astrophys.}
  {\bfseries 686} (2024) A179},
  \href{http://arxiv.org/abs/2308.09170}{{\ttfamily arXiv:2308.09170
  [astro-ph.GA]}}.

\bibitem{Arvanitaki:2014wva}
A.~Arvanitaki, M.~Baryakhtar, and X.~Huang, ``{Discovering the QCD Axion with
  Black Holes and Gravitational Waves},''
  \href{http://dx.doi.org/10.1103/PhysRevD.91.084011}{{\em Phys. Rev.}
  {\bfseries D91} no.~8, (2015) 084011},
\href{http://arxiv.org/abs/1411.2263}{{\ttfamily arXiv:1411.2263 [hep-ph]}}.

\bibitem{Yoshino:2014wwa}
H.~Yoshino and H.~Kodama, ``{Probing the string axiverse by gravitational waves
  from Cygnus X-1},'' \href{http://dx.doi.org/10.1093/ptep/ptv067}{{\em PTEP}
  {\bfseries 2015} no.~6, (2015) 061E01},
  \href{http://arxiv.org/abs/1407.2030}{{\ttfamily arXiv:1407.2030 [gr-qc]}}.

\bibitem{Arvanitaki:2016qwi}
A.~Arvanitaki, M.~Baryakhtar, S.~Dimopoulos, S.~Dubovsky, and R.~Lasenby,
  ``{Black Hole Mergers and the QCD Axion at Advanced LIGO},''
  \href{http://dx.doi.org/10.1103/PhysRevD.95.043001}{{\em Phys. Rev. D}
  {\bfseries 95} (2017) 043001},
\href{http://arxiv.org/abs/1604.03958}{{\ttfamily arXiv:1604.03958 [hep-ph]}}.

\bibitem{Baryakhtar:2017ngi}
M.~Baryakhtar, R.~Lasenby, and M.~Teo, ``{Black Hole Superradiance Signatures
  of Ultralight Vectors},''
  \href{http://dx.doi.org/10.1103/PhysRevD.96.035019}{{\em Phys. Rev.}
  {\bfseries D96} (2017) 035019},
\href{http://arxiv.org/abs/1704.05081}{{\ttfamily arXiv:1704.05081 [hep-ph]}}.

\bibitem{Brito:2017wnc}
R.~Brito, S.~Ghosh, E.~Barausse, E.~Berti, V.~Cardoso, I.~Dvorkin, A.~Klein,
  and P.~Pani, ``{Stochastic and resolvable gravitational waves from ultralight
  bosons},'' \href{http://dx.doi.org/10.1103/PhysRevLett.119.131101}{{\em Phys.
  Rev. Lett.} {\bfseries 119} no.~13, (2017) 131101},
  \href{http://arxiv.org/abs/1706.05097}{{\ttfamily arXiv:1706.05097 [gr-qc]}}.

\bibitem{Ng:2020jqd}
K.~K.~Y. Ng, M.~Isi, C.-J. Haster, and S.~Vitale, ``{Multiband
  gravitational-wave searches for ultralight bosons},''
  \href{http://dx.doi.org/10.1103/PhysRevD.102.083020}{{\em Phys. Rev. D}
  {\bfseries 102} no.~8, (2020) 083020},
  \href{http://arxiv.org/abs/2007.12793}{{\ttfamily arXiv:2007.12793 [gr-qc]}}.

\bibitem{Tsukada:2018mbp}
L.~Tsukada, T.~Callister, A.~Matas, and P.~Meyers, ``{First search for a
  stochastic gravitational-wave background from ultralight bosons},''
  \href{http://dx.doi.org/10.1103/PhysRevD.99.103015}{{\em Phys. Rev. D}
  {\bfseries 99} no.~10, (2019) 103015},
  \href{http://arxiv.org/abs/1812.09622}{{\ttfamily arXiv:1812.09622
  [astro-ph.HE]}}.

\bibitem{Palomba:2019vxe}
C.~Palomba {\em et~al.}, ``{Direct constraints on ultra-light boson mass from
  searches for continuous gravitational waves},''
  \href{http://dx.doi.org/10.1103/PhysRevLett.123.171101}{{\em Phys. Rev.
  Lett.} {\bfseries 123} (2019) 171101},
  \href{http://arxiv.org/abs/1909.08854}{{\ttfamily arXiv:1909.08854
  [astro-ph.HE]}}.

\bibitem{LIGOScientific:2021rnv}
{\bfseries LIGO Scientific, Virgo, KAGRA} Collaboration, R.~Abbott {\em
  et~al.}, ``{All-sky search for gravitational wave emission from scalar boson
  clouds around spinning black holes in LIGO O3 data},''
  \href{http://dx.doi.org/10.1103/PhysRevD.105.102001}{{\em Phys. Rev. D}
  {\bfseries 105} no.~10, (2022) 102001},
  \href{http://arxiv.org/abs/2111.15507}{{\ttfamily arXiv:2111.15507
  [astro-ph.HE]}}.

\bibitem{Yuan:2022bem}
C.~Yuan, Y.~Jiang, and Q.-G. Huang, ``{Constraints on an ultralight scalar
  boson from Advanced LIGO and Advanced Virgo\textquoteright{}s first three
  observing runs using the stochastic gravitational-wave background},''
  \href{http://dx.doi.org/10.1103/PhysRevD.106.023020}{{\em Phys. Rev. D}
  {\bfseries 106} no.~2, (2022) 023020},
  \href{http://arxiv.org/abs/2204.03482}{{\ttfamily arXiv:2204.03482
  [astro-ph.CO]}}.

\bibitem{Stott:2018opm}
M.~J. Stott and D.~J.~E. Marsh, ``{Black hole spin constraints on the mass
  spectrum and number of axionlike fields},''
  \href{http://dx.doi.org/10.1103/PhysRevD.98.083006}{{\em Phys. Rev. D}
  {\bfseries 98} no.~8, (2018) 083006},
  \href{http://arxiv.org/abs/1805.02016}{{\ttfamily arXiv:1805.02016
  [hep-ph]}}.

\bibitem{Fernandez:2019qbj}
N.~Fernandez, A.~Ghalsasi, and S.~Profumo, ``{Superradiance and the Spins of
  Black Holes from LIGO and X-ray binaries},''
  \href{http://arxiv.org/abs/1911.07862}{{\ttfamily arXiv:1911.07862
  [hep-ph]}}.

\bibitem{Ng:2019jsx}
K.~K.~Y. Ng, O.~A. Hannuksela, S.~Vitale, and T.~G.~F. Li, ``{Searching for
  ultralight bosons within spin measurements of a population of binary black
  hole mergers},'' \href{http://dx.doi.org/10.1103/PhysRevD.103.063010}{{\em
  Phys. Rev. D} {\bfseries 103} no.~6, (2021) 063010},
  \href{http://arxiv.org/abs/1908.02312}{{\ttfamily arXiv:1908.02312 [gr-qc]}}.

\bibitem{Stott:2020gjj}
M.~J. Stott, ``{Ultralight Bosonic Field Mass Bounds from Astrophysical Black
  Hole Spin},'' \href{http://arxiv.org/abs/2009.07206}{{\ttfamily
  arXiv:2009.07206 [hep-ph]}}.

\bibitem{Ng:2020ruv}
K.~K.~Y. Ng, S.~Vitale, O.~A. Hannuksela, and T.~G.~F. Li, ``{Constraints on
  Ultralight Scalar Bosons within Black Hole Spin Measurements from the
  LIGO-Virgo GWTC-2},''
  \href{http://dx.doi.org/10.1103/PhysRevLett.126.151102}{{\em Phys. Rev.
  Lett.} {\bfseries 126} no.~15, (2021) 151102},
  \href{http://arxiv.org/abs/2011.06010}{{\ttfamily arXiv:2011.06010 [gr-qc]}}.

\bibitem{Mehta:2021pwf}
V.~M. Mehta, M.~Demirtas, C.~Long, D.~J.~E. Marsh, L.~McAllister, and M.~J.
  Stott, ``{Superradiance in string theory},''
  \href{http://dx.doi.org/10.1088/1475-7516/2021/07/033}{{\em JCAP} {\bfseries
  07} (2021) 033}, \href{http://arxiv.org/abs/2103.06812}{{\ttfamily
  arXiv:2103.06812 [hep-th]}}.

\bibitem{Wen:2021yhz}
S.~Wen, P.~G. Jonker, N.~C. Stone, and A.~I. Zabludoff, ``{Mass, Spin, and
  Ultralight Boson Constraints from the Intermediate-mass Black Hole in the
  Tidal Disruption Event 3XMM J215022.4\textendash{}055108},''
  \href{http://dx.doi.org/10.3847/1538-4357/ac00b5}{{\em Astrophys. J.}
  {\bfseries 918} no.~2, (2021) 46},
  \href{http://arxiv.org/abs/2104.01498}{{\ttfamily arXiv:2104.01498
  [astro-ph.HE]}}.

\bibitem{Hoof:2024quk}
S.~Hoof, D.~J.~E. Marsh, J.~Sisk-Reyn\'es, J.~H. Matthews, and C.~Reynolds,
  ``{Getting More Out of Black Hole Superradiance: a Statistically Rigorous
  Approach to Ultralight Boson Constraints},''
  \href{http://arxiv.org/abs/2406.10337}{{\ttfamily arXiv:2406.10337
  [hep-ph]}}.

\bibitem{Pani:2012bp}
P.~Pani, V.~Cardoso, L.~Gualtieri, E.~Berti, and A.~Ishibashi, ``{Perturbations
  of Slowly Rotating Black Holes: Massive Vector Fields in the Kerr Metric},''
  \href{http://dx.doi.org/10.1103/PhysRevD.86.104017}{{\em Phys. Rev.}
  {\bfseries D86} (2012) 104017},
\href{http://arxiv.org/abs/1209.0773}{{\ttfamily arXiv:1209.0773 [gr-qc]}}.

\bibitem{Pani:2012vp}
P.~Pani, V.~Cardoso, L.~Gualtieri, E.~Berti, and A.~Ishibashi, ``{Black hole
  bombs and photon mass bounds},''
  \href{http://dx.doi.org/10.1103/PhysRevLett.109.131102}{{\em Phys. Rev.
  Lett.} {\bfseries 109} (2012) 131102},
  \href{http://arxiv.org/abs/1209.0465}{{\ttfamily arXiv:1209.0465 [gr-qc]}}.

\bibitem{Cardoso:2018tly}
V.~Cardoso, {\'O}.~Dias, G.~Hartnett, M.~Middleton, P.~Pani, and J.~Santos,
  ``{Constraining the Mass of Dark Photons and Axion-Like Particles Through
  Black Hole Superradiance},''
  \href{http://dx.doi.org/10.1088/1475-7516/2018/03/043}{{\em JCAP} {\bfseries
  1803} (2018) 043},
\href{http://arxiv.org/abs/1801.01420}{{\ttfamily arXiv:1801.01420 [gr-qc]}}.

\bibitem{Brito:2013wya}
R.~Brito, V.~Cardoso, and P.~Pani, ``{Massive spin-2 fields on black hole
  spacetimes: Instability of the Schwarzschild and Kerr solutions and bounds on
  the graviton mass},''
  \href{http://dx.doi.org/10.1103/PhysRevD.88.023514}{{\em Phys. Rev. D}
  {\bfseries 88} no.~2, (2013) 023514},
  \href{http://arxiv.org/abs/1304.6725}{{\ttfamily arXiv:1304.6725 [gr-qc]}}.

\bibitem{Rosa:2017ury}
J.~a.~G. Rosa and T.~W. Kephart, ``{Stimulated Axion Decay in Superradiant
  Clouds around Primordial Black Holes},''
  \href{http://dx.doi.org/10.1103/PhysRevLett.120.231102}{{\em Phys. Rev.
  Lett.} {\bfseries 120} no.~23, (2018) 231102},
  \href{http://arxiv.org/abs/1709.06581}{{\ttfamily arXiv:1709.06581 [gr-qc]}}.

\bibitem{Boskovic:2018lkj}
M.~Boskovic, R.~Brito, V.~Cardoso, T.~Ikeda, and H.~Witek, ``{Axionic
  instabilities and new black hole solutions},''
  \href{http://dx.doi.org/10.1103/PhysRevD.99.035006}{{\em Phys. Rev.}
  {\bfseries D99} no.~3, (2019) 035006},
\href{http://arxiv.org/abs/1811.04945}{{\ttfamily arXiv:1811.04945 [gr-qc]}}.

\bibitem{Ikeda:2018nhb}
T.~Ikeda, R.~Brito, and V.~Cardoso, ``{Blasts of Light from Axions},''
  \href{http://dx.doi.org/10.1103/PhysRevLett.122.081101}{{\em Phys. Rev.
  Lett.} {\bfseries 122} no.~8, (2019) 081101},
  \href{http://arxiv.org/abs/1811.04950}{{\ttfamily arXiv:1811.04950 [gr-qc]}}.

\bibitem{Miller:2025yyx}
A.~L. Miller, ``{Gravitational wave probes of particle dark matter: a
  review},'' \href{http://arxiv.org/abs/2503.02607}{{\ttfamily arXiv:2503.02607
  [astro-ph.HE]}}.

\bibitem{Olejak:2025hcx}
A.~Olejak, J.~Stegmann, S.~E. de~Mink, R.~Valli, R.~Sari, and S.~Justham,
  ``{Supermassive black holes stripping a subgiant star down to its helium
  core: a new type of multi-messenger source for LISA},''
  \href{http://arxiv.org/abs/2503.21995}{{\ttfamily arXiv:2503.21995
  [astro-ph.HE]}}.

\bibitem{1980ApJ...241..425G}
P.~{Goldreich} and S.~{Tremaine}, ``{Disk-satellite interactions.},''
  \href{http://dx.doi.org/10.1086/158356}{{\em Astrophys. J.} {\bfseries 241}
  (Oct., 1980) 425--441}.

\bibitem{Yunes:2011ws}
N.~Yunes, B.~Kocsis, A.~Loeb, and Z.~Haiman, ``{Imprint of Accretion
  Disk-Induced Migration on Gravitational Waves from Extreme Mass Ratio
  Inspirals},'' \href{http://dx.doi.org/10.1103/PhysRevLett.107.171103}{{\em
  Phys. Rev. Lett.} {\bfseries 107} (2011) 171103},
  \href{http://arxiv.org/abs/1103.4609}{{\ttfamily arXiv:1103.4609
  [astro-ph.CO]}}.

\bibitem{Derdzinski:2020wlw}
A.~Derdzinski, D.~D'Orazio, P.~Duffell, Z.~Haiman, and A.~MacFadyen,
  ``{Evolution of gas disc\textendash{}embedded intermediate mass ratio
  inspirals in the $LISA$ band},''
  \href{http://dx.doi.org/10.1093/mnras/staa3976}{{\em Mon. Not. Roy. Astron.
  Soc.} {\bfseries 501} no.~3, (2021) 3540--3557},
  \href{http://arxiv.org/abs/2005.11333}{{\ttfamily arXiv:2005.11333
  [astro-ph.HE]}}.

\bibitem{Chandrasekhar:1943ys}
S.~Chandrasekhar, ``{Dynamical Friction. I. General Considerations: the
  Coefficient of Dynamical Friction},''
  \href{http://dx.doi.org/10.1086/144517}{{\em Astrophys. J.} {\bfseries 97}
  (1943) 255}.

\bibitem{1943ApJ....97..263C}
S.~{Chandrasekhar}, ``{Dynamical Friction. II. The Rate of Escape of Stars from
  Clusters and the Evidence for the Operation of Dynamical Friction.},''
  \href{http://dx.doi.org/10.1086/144518}{{\em Astroph. J.} {\bfseries 97}
  (Mar., 1943) 263}.

\bibitem{1943ApJ....98...54C}
S.~{Chandrasekhar}, ``{Dynamical Friction. III. a More Exact Theory of the Rate
  of Escape of Stars from Clusters.},''
  \href{http://dx.doi.org/10.1086/144544}{{\em Astroph. J.} {\bfseries 98}
  (July, 1943) 54}.

\bibitem{Macedo:2013jja}
C.~F.~B. Macedo, P.~Pani, V.~Cardoso, and L.~C.~B. Crispino, ``{Astrophysical
  signatures of boson stars: quasinormal modes and inspiral resonances},''
  \href{http://dx.doi.org/10.1103/PhysRevD.88.064046}{{\em Phys. Rev. D}
  {\bfseries 88} no.~6, (2013) 064046},
  \href{http://arxiv.org/abs/1307.4812}{{\ttfamily arXiv:1307.4812 [gr-qc]}}.

\bibitem{Macedo:2013qea}
C.~F.~B. Macedo, P.~Pani, V.~Cardoso, and L.~C.~B. Crispino, ``{Into the lair:
  gravitational-wave signatures of dark matter},''
  \href{http://dx.doi.org/10.1088/0004-637X/774/1/48}{{\em Astrophys. J.}
  {\bfseries 774} (2013) 48}, \href{http://arxiv.org/abs/1302.2646}{{\ttfamily
  arXiv:1302.2646 [gr-qc]}}.

\bibitem{Traykova:2021dua}
D.~Traykova, K.~Clough, T.~Helfer, E.~Berti, P.~G. Ferreira, and L.~Hui,
  ``{Dynamical friction from scalar dark matter in the relativistic regime},''
  \href{http://dx.doi.org/10.1103/PhysRevD.104.103014}{{\em Phys. Rev. D}
  {\bfseries 104} no.~10, (2021) 103014},
  \href{http://arxiv.org/abs/2106.08280}{{\ttfamily arXiv:2106.08280 [gr-qc]}}.

\bibitem{Vicente:2022ivh}
R.~Vicente and V.~Cardoso, ``{Dynamical friction of black holes in ultralight
  dark matter},'' \href{http://dx.doi.org/10.1103/PhysRevD.105.083008}{{\em
  Phys. Rev. D} {\bfseries 105} no.~8, (2022) 083008},
  \href{http://arxiv.org/abs/2201.08854}{{\ttfamily arXiv:2201.08854 [gr-qc]}}.

\bibitem{Geroch:1970cd}
R.~Geroch, ``{Multipole Moments. II. Curved Space},''
\href{http://dx.doi.org/10.1063/1.1665427}{{\em J. Math. Phys.} {\bfseries 11}
  (1970) 2580}.

\bibitem{Hansen:1974zz}
R.~Hansen, ``{Multipole Moments of Stationary Spacetimes},''
\href{http://dx.doi.org/10.1063/1.1666501}{{\em J. Math. Phys.} {\bfseries 15}
  (1974) 46}.

\bibitem{Thorne:1980ru}
K.~Thorne, ``{Multipole Expansions of Gravitational Radiation},''
\href{http://dx.doi.org/10.1103/RevModPhys.52.299}{{\em Rev. Mod. Phys.}
  {\bfseries 52} (1980) 299--339}.

\bibitem{DeLuca:2021ite}
V.~De~Luca and P.~Pani, ``{Tidal deformability of dressed black holes and tests
  of ultralight bosons in extended mass ranges},''
  \href{http://dx.doi.org/10.1088/1475-7516/2021/08/032}{{\em JCAP} {\bfseries
  08} (2021) 032}, \href{http://arxiv.org/abs/2106.14428}{{\ttfamily
  arXiv:2106.14428 [gr-qc]}}.

\bibitem{Binnington:2009bb}
T.~Binnington and E.~Poisson, ``{Relativistic Theory of Tidal Love Numbers},''
  \href{http://dx.doi.org/10.1103/PhysRevD.80.084018}{{\em Phys. Rev. D}
  {\bfseries 80} (2009) 084018},
\href{http://arxiv.org/abs/0906.1366}{{\ttfamily arXiv:0906.1366 [gr-qc]}}.

\bibitem{Damour:2009vw}
T.~Damour and A.~Nagar, ``{Relativistic Tidal Properties of Neutron Stars},''
  \href{http://dx.doi.org/10.1103/PhysRevD.80.084035}{{\em Phys. Rev. D}
  {\bfseries 80} (2009) 084035},
\href{http://arxiv.org/abs/0906.0096}{{\ttfamily arXiv:0906.0096 [gr-qc]}}.

\bibitem{Gurlebeck:2015xpa}
N.~G{\"u}rlebeck, ``{No-hair Theorem for Black Holes in Astrophysical
  Environments},'' \href{http://dx.doi.org/10.1103/PhysRevLett.114.151102}{{\em
  Phys. Rev. Lett.} {\bfseries 114} (2015) 151102},
\href{http://arxiv.org/abs/1503.03240}{{\ttfamily arXiv:1503.03240 [gr-qc]}}.

\bibitem{Landry:2015zfa}
P.~Landry and E.~Poisson, ``{Tidal Deformation of a Slowly Rotating Material
  Body. External Metric},''
  \href{http://dx.doi.org/10.1103/PhysRevD.91.104018}{{\em Phys. Rev. D}
  {\bfseries 91} (2015) 104018},
\href{http://arxiv.org/abs/1503.07366}{{\ttfamily arXiv:1503.07366 [gr-qc]}}.

\bibitem{Pani:2015hfa}
P.~Pani, L.~Gualtieri, A.~Maselli, and V.~Ferrari, ``{Tidal Deformations of a
  Spinning Compact Object},''
  \href{http://dx.doi.org/10.1103/PhysRevD.92.024010}{{\em Phys. Rev.}
  {\bfseries D92} (2015) 024010},
\href{http://arxiv.org/abs/1503.07365}{{\ttfamily arXiv:1503.07365 [gr-qc]}}.

\bibitem{Chia:2020yla}
H.~S. Chia, ``{Tidal deformation and dissipation of rotating black holes},''
  \href{http://dx.doi.org/10.1103/PhysRevD.104.024013}{{\em Phys. Rev. D}
  {\bfseries 104} no.~2, (2021) 024013},
  \href{http://arxiv.org/abs/2010.07300}{{\ttfamily arXiv:2010.07300 [gr-qc]}}.

\bibitem{Cardoso:2019upw}
V.~Cardoso and F.~Duque, ``{Environmental effects in gravitational-wave
  physics: Tidal deformability of black holes immersed in matter},''
  \href{http://dx.doi.org/10.1103/PhysRevD.101.064028}{{\em Phys. Rev. D}
  {\bfseries 101} no.~6, (2020) 064028},
  \href{http://arxiv.org/abs/1912.07616}{{\ttfamily arXiv:1912.07616 [gr-qc]}}.

\bibitem{Brito:2023pyl}
R.~Brito and S.~Shah, ``{Extreme mass-ratio inspirals into black holes
  surrounded by scalar clouds},''
  \href{http://dx.doi.org/10.1103/PhysRevD.108.084019}{{\em Phys. Rev. D}
  {\bfseries 108} no.~8, (2023) 084019},
  \href{http://arxiv.org/abs/2307.16093}{{\ttfamily arXiv:2307.16093 [gr-qc]}}.

\bibitem{Cannizzaro:2024fpz}
E.~Cannizzaro, V.~De~Luca, and P.~Pani, ``{Tidal deformability of black holes
  surrounded by thin accretion disks},''
  \href{http://dx.doi.org/10.1103/PhysRevD.110.123004}{{\em Phys. Rev. D}
  {\bfseries 110} no.~12, (2024) 123004},
  \href{http://arxiv.org/abs/2408.14208}{{\ttfamily arXiv:2408.14208
  [astro-ph.HE]}}.

\bibitem{CanevaSantoro:2023aol}
G.~Caneva~Santoro, S.~Roy, R.~Vicente, M.~Haney, O.~J. Piccinni, W.~Del~Pozzo,
  and M.~Martinez, ``{First Constraints on Compact Binary Environments from
  LIGO-Virgo Data},''
  \href{http://dx.doi.org/10.1103/PhysRevLett.132.251401}{{\em Phys. Rev.
  Lett.} {\bfseries 132} no.~25, (2024) 251401},
  \href{http://arxiv.org/abs/2309.05061}{{\ttfamily arXiv:2309.05061 [gr-qc]}}.

\bibitem{Ikeda:2020xvt}
T.~Ikeda, L.~Bernard, V.~Cardoso, and M.~Zilh\~ao, ``{Black hole binaries and
  light fields: Gravitational molecules},''
  \href{http://dx.doi.org/10.1103/PhysRevD.103.024020}{{\em Phys. Rev. D}
  {\bfseries 103} no.~2, (2021) 024020},
  \href{http://arxiv.org/abs/2010.00008}{{\ttfamily arXiv:2010.00008 [gr-qc]}}.

\bibitem{Bamber:2022pbs}
J.~Bamber, J.~C. Aurrekoetxea, K.~Clough, and P.~G. Ferreira, ``{Black hole
  merger simulations in wave dark matter environments},''
  \href{http://dx.doi.org/10.1103/PhysRevD.107.024035}{{\em Phys. Rev. D}
  {\bfseries 107} no.~2, (2023) 024035},
  \href{http://arxiv.org/abs/2210.09254}{{\ttfamily arXiv:2210.09254 [gr-qc]}}.

\bibitem{Aurrekoetxea:2023jwk}
J.~C. Aurrekoetxea, K.~Clough, J.~Bamber, and P.~G. Ferreira, ``{Effect of Wave
  Dark Matter on Equal Mass Black Hole Mergers},''
  \href{http://dx.doi.org/10.1103/PhysRevLett.132.211401}{{\em Phys. Rev.
  Lett.} {\bfseries 132} no.~21, (2024) 211401},
  \href{http://arxiv.org/abs/2311.18156}{{\ttfamily arXiv:2311.18156 [gr-qc]}}.

\bibitem{Aurrekoetxea:2024cqd}
J.~C. Aurrekoetxea, J.~Marsden, K.~Clough, and P.~G. Ferreira,
  ``{Self-interacting scalar dark matter around binary black holes},''
  \href{http://dx.doi.org/10.1103/PhysRevD.110.083011}{{\em Phys. Rev. D}
  {\bfseries 110} no.~8, (2024) 083011},
  \href{http://arxiv.org/abs/2409.01937}{{\ttfamily arXiv:2409.01937 [gr-qc]}}.

\bibitem{Tomaselli:2024ojz}
G.~M. Tomaselli, ``{Scattering of wave dark matter by supermassive black
  holes},'' \href{http://dx.doi.org/10.1103/PhysRevD.111.063075}{{\em Phys.
  Rev. D} {\bfseries 111} no.~6, (2025) 063075},
  \href{http://arxiv.org/abs/2501.00090}{{\ttfamily arXiv:2501.00090 [gr-qc]}}.

\bibitem{Tanaka_2002}
H.~Tanaka, T.~Takeuchi, and W.~R. Ward, ``Three-dimensional interaction between
  a planet and an isothermal gaseous disk. i. corotation and lindblad torques
  and planet migration,'' \href{http://dx.doi.org/10.1086/324713}{{\em
  Astrophys. J.} {\bfseries 565} no.~2, (Feb, 2002) 1257}.

\bibitem{Barausse:2007dy}
E.~Barausse and L.~Rezzolla, ``{The Influence of the hydrodynamic drag from an
  accretion torus on extreme mass-ratio inspirals},''
  \href{http://dx.doi.org/10.1103/PhysRevD.77.104027}{{\em Phys. Rev. D}
  {\bfseries 77} (2008) 104027},
  \href{http://arxiv.org/abs/0711.4558}{{\ttfamily arXiv:0711.4558 [gr-qc]}}.

\bibitem{Derdzinski:2018qzv}
A.~M. Derdzinski, D.~D'Orazio, P.~Duffell, Z.~Haiman, and A.~MacFadyen,
  ``{Probing gas disc physics with LISA: simulations of an intermediate mass
  ratio inspiral in an accretion disc},''
  \href{http://dx.doi.org/10.1093/mnras/stz1026}{{\em Mon. Not. Roy. Astron.
  Soc.} {\bfseries 486} no.~2, (2019) 2754--2765},
  \href{http://arxiv.org/abs/1810.03623}{{\ttfamily arXiv:1810.03623
  [astro-ph.HE]}}. [Erratum: Mon.Not.Roy.Astron.Soc. 489, 4860--4861 (2019)].

\bibitem{Duffell:2019uuk}
P.~C. Duffell, D.~D'Orazio, A.~Derdzinski, Z.~Haiman, A.~MacFadyen, A.~L.
  Rosen, and J.~Zrake, ``{Circumbinary Disks: Accretion and Torque as a
  Function of Mass Ratio and Disk Viscosity},''
  \href{http://dx.doi.org/10.3847/1538-4357/abab95}{{\em Astrophys. J.}
  {\bfseries 901} no.~1, (2020) 25},
  \href{http://arxiv.org/abs/1911.05506}{{\ttfamily arXiv:1911.05506
  [astro-ph.SR]}}.

\bibitem{Pan:2021oob}
Z.~Pan, Z.~Lyu, and H.~Yang, ``{Wet extreme mass ratio inspirals may be more
  common for spaceborne gravitational wave detection},''
  \href{http://dx.doi.org/10.1103/PhysRevD.104.063007}{{\em Phys. Rev. D}
  {\bfseries 104} no.~6, (2021) 063007},
  \href{http://arxiv.org/abs/2104.01208}{{\ttfamily arXiv:2104.01208
  [astro-ph.HE]}}.

\bibitem{Zwick:2021dlg}
L.~Zwick, A.~Derdzinski, M.~Garg, P.~R. Capelo, and L.~Mayer, ``{Dirty
  waveforms: multiband harmonic content of gas-embedded gravitational wave
  sources},'' \href{http://dx.doi.org/10.1093/mnras/stac299}{{\em Mon. Not.
  Roy. Astron. Soc.} {\bfseries 511} no.~4, (2022) 6143--6159},
  \href{http://arxiv.org/abs/2110.09097}{{\ttfamily arXiv:2110.09097
  [astro-ph.HE]}}.

\bibitem{Garg:2022nko}
M.~Garg, A.~Derdzinski, L.~Zwick, P.~R. Capelo, and L.~Mayer, ``{The imprint of
  gas on gravitational waves from LISA intermediate-mass black hole
  binaries},'' \href{http://dx.doi.org/10.1093/mnras/stac2711}{{\em Mon. Not.
  Roy. Astron. Soc.} {\bfseries 517} (2022) 1339--1354},
  \href{http://arxiv.org/abs/2206.05292}{{\ttfamily arXiv:2206.05292
  [astro-ph.GA]}}.

\bibitem{Tiede:2023cje}
C.~Tiede, D.~J. D'Orazio, L.~Zwick, and P.~C. Duffell, ``{Disk-induced Binary
  Precession: Implications for Dynamics and Multimessenger Observations of
  Black Hole Binaries},''
  \href{http://dx.doi.org/10.3847/1538-4357/ad2613}{{\em Astrophys. J.}
  {\bfseries 964} no.~1, (2024) 46},
  \href{http://arxiv.org/abs/2312.01805}{{\ttfamily arXiv:2312.01805
  [astro-ph.HE]}}.

\bibitem{Zwick:2024yzh}
L.~Zwick, C.~Tiede, A.~A. Trani, A.~Derdzinski, Z.~Haiman, D.~J. D'Orazio, and
  J.~Samsing, ``{Novel category of environmental effects on gravitational waves
  from binaries perturbed by periodic forces},''
  \href{http://dx.doi.org/10.1103/PhysRevD.110.103005}{{\em Phys. Rev. D}
  {\bfseries 110} no.~10, (2024) 103005},
  \href{http://arxiv.org/abs/2405.05698}{{\ttfamily arXiv:2405.05698 [gr-qc]}}.

\bibitem{Garg:2024oeu}
M.~Garg, A.~Derdzinski, S.~Tiwari, J.~Gair, and L.~Mayer, ``{Measuring
  eccentricity and gas-induced perturbation from gravitational waves of LISA
  massive black hole binaries},''
  \href{http://dx.doi.org/10.1093/mnras/stae1764}{{\em Mon. Not. Roy. Astron.
  Soc.} {\bfseries 532} no.~4, (2024) 4060--4074},
  \href{http://arxiv.org/abs/2402.14058}{{\ttfamily arXiv:2402.14058
  [astro-ph.GA]}}.

\bibitem{Garg:2024yrs}
M.~Garg, C.~Tiede, and D.~J. D'Orazio, ``{Accretion-mediated
  spin\textendash{}eccentricity correlations in LISA massive black hole
  binaries},'' \href{http://dx.doi.org/10.1093/mnras/stae2357}{{\em Mon. Not.
  Roy. Astron. Soc.} {\bfseries 534} no.~4, (2024) 3705--3712},
  \href{http://arxiv.org/abs/2405.04411}{{\ttfamily arXiv:2405.04411
  [astro-ph.HE]}}.

\bibitem{Garg:2024zku}
M.~Garg, A.~Franchini, A.~Lupi, M.~Bonetti, and L.~Mayer, ``{Gas-induced
  perturbations on the gravitational wave in-spiral of live post-Newtonian LISA
  massive black hole binaries},''
  \href{http://arxiv.org/abs/2410.17305}{{\ttfamily arXiv:2410.17305
  [astro-ph.HE]}}.

\bibitem{Duque:2024mfw}
F.~Duque, S.~Kejriwal, L.~Sberna, L.~Speri, and J.~Gair, ``{Constraining
  accretion physics with gravitational waves from eccentric extreme-mass-ratio
  inspirals},'' \href{http://dx.doi.org/10.1103/PhysRevD.111.084006}{{\em Phys.
  Rev. D} {\bfseries 111} no.~8, (2025) 084006},
  \href{http://arxiv.org/abs/2411.03436}{{\ttfamily arXiv:2411.03436 [gr-qc]}}.

\bibitem{Copparoni:2025jhq}
L.~Copparoni, L.~Speri, L.~Sberna, A.~Derdzinski, and E.~Barausse, ``{The
  implications of stochastic gas torques for asymmetric binaries in the LISA
  band},'' \href{http://arxiv.org/abs/2502.10087}{{\ttfamily arXiv:2502.10087
  [gr-qc]}}.

\bibitem{Tagawa:2020qll}
H.~Tagawa, B.~Kocsis, Z.~Haiman, I.~Bartos, K.~Omukai, and J.~Samsing,
  ``{Mass-gap Mergers in Active Galactic Nuclei},''
  \href{http://dx.doi.org/10.3847/1538-4357/abd555}{{\em Astrophys. J.}
  {\bfseries 908} no.~2, (2021) 194},
  \href{http://arxiv.org/abs/2012.00011}{{\ttfamily arXiv:2012.00011
  [astro-ph.HE]}}.

\bibitem{Morton:2023wxg}
S.~L. Morton, S.~Rinaldi, A.~Torres-Orjuela, A.~Derdzinski, M.~P. Vaccaro, and
  W.~Del~Pozzo, ``{GW190521: A binary black hole merger inside an active
  galactic nucleus?},''
  \href{http://dx.doi.org/10.1103/PhysRevD.108.123039}{{\em Phys. Rev. D}
  {\bfseries 108} no.~12, (2023) 123039},
  \href{http://arxiv.org/abs/2310.16025}{{\ttfamily arXiv:2310.16025 [gr-qc]}}.

\bibitem{Eda:2013gg}
K.~Eda, Y.~Itoh, S.~Kuroyanagi, and J.~Silk, ``{New Probe of Dark-Matter
  Properties: Gravitational Waves from an Intermediate-Mass Black Hole Embedded
  in a Dark-Matter Minispike},''
  \href{http://dx.doi.org/10.1103/PhysRevLett.110.221101}{{\em Phys. Rev.
  Lett.} {\bfseries 110} no.~22, (2013) 221101},
  \href{http://arxiv.org/abs/1301.5971}{{\ttfamily arXiv:1301.5971 [gr-qc]}}.

\bibitem{Eda:2014kra}
K.~Eda, Y.~Itoh, S.~Kuroyanagi, and J.~Silk, ``{Gravitational waves as a probe
  of dark matter minispikes},''
  \href{http://dx.doi.org/10.1103/PhysRevD.91.044045}{{\em Phys. Rev. D}
  {\bfseries 91} no.~4, (2015) 044045},
  \href{http://arxiv.org/abs/1408.3534}{{\ttfamily arXiv:1408.3534 [gr-qc]}}.

\bibitem{Yue:2018vtk}
X.-J. Yue, W.-B. Han, and X.~Chen, ``{Dark matter: an efficient catalyst for
  intermediate-mass-ratio-inspiral events},''
  \href{http://dx.doi.org/10.3847/1538-4357/ab06f6}{{\em Astrophys. J.}
  {\bfseries 874} no.~1, (2019) 34},
  \href{http://arxiv.org/abs/1802.03739}{{\ttfamily arXiv:1802.03739 [gr-qc]}}.

\bibitem{Hannuksela:2019vip}
O.~A. Hannuksela, K.~C.~Y. Ng, and T.~G.~F. Li, ``{Extreme dark matter tests
  with extreme mass ratio inspirals},''
  \href{http://dx.doi.org/10.1103/PhysRevD.102.103022}{{\em Phys. Rev. D}
  {\bfseries 102} no.~10, (2020) 103022},
  \href{http://arxiv.org/abs/1906.11845}{{\ttfamily arXiv:1906.11845
  [astro-ph.CO]}}.

\bibitem{Kavanagh:2020cfn}
B.~J. Kavanagh, D.~A. Nichols, G.~Bertone, and D.~Gaggero, ``{Detecting dark
  matter around black holes with gravitational waves: Effects of dark-matter
  dynamics on the gravitational waveform},''
  \href{http://dx.doi.org/10.1103/PhysRevD.102.083006}{{\em Phys. Rev. D}
  {\bfseries 102} no.~8, (2020) 083006},
  \href{http://arxiv.org/abs/2002.12811}{{\ttfamily arXiv:2002.12811 [gr-qc]}}.

\bibitem{Coogan:2021uqv}
A.~Coogan, G.~Bertone, D.~Gaggero, B.~J. Kavanagh, and D.~A. Nichols,
  ``{Measuring the dark matter environments of black hole binaries with
  gravitational waves},''
  \href{http://dx.doi.org/10.1103/PhysRevD.105.043009}{{\em Phys. Rev. D}
  {\bfseries 105} no.~4, (2022) 043009},
  \href{http://arxiv.org/abs/2108.04154}{{\ttfamily arXiv:2108.04154 [gr-qc]}}.

\bibitem{Dai:2021olt}
N.~Dai, Y.~Gong, T.~Jiang, and D.~Liang, ``{Intermediate mass-ratio inspirals
  with dark matter minispikes},''
  \href{http://dx.doi.org/10.1103/PhysRevD.106.064003}{{\em Phys. Rev. D}
  {\bfseries 106} no.~6, (2022) 064003},
  \href{http://arxiv.org/abs/2111.13514}{{\ttfamily arXiv:2111.13514 [gr-qc]}}.

\bibitem{Speeney:2022ryg}
N.~Speeney, A.~Antonelli, V.~Baibhav, and E.~Berti, ``{Impact of relativistic
  corrections on the detectability of dark-matter spikes with gravitational
  waves},'' \href{http://dx.doi.org/10.1103/PhysRevD.106.044027}{{\em Phys.
  Rev. D} {\bfseries 106} no.~4, (2022) 044027},
  \href{http://arxiv.org/abs/2204.12508}{{\ttfamily arXiv:2204.12508 [gr-qc]}}.

\bibitem{Becker:2022wlo}
N.~Becker and L.~Sagunski, ``{Comparing accretion disks and dark matter spikes
  in intermediate mass ratio inspirals},''
  \href{http://dx.doi.org/10.1103/PhysRevD.107.083003}{{\em Phys. Rev. D}
  {\bfseries 107} no.~8, (2023) 083003},
  \href{http://arxiv.org/abs/2211.05145}{{\ttfamily arXiv:2211.05145 [gr-qc]}}.

\bibitem{Berezhiani:2023vlo}
L.~Berezhiani, G.~Cintia, V.~De~Luca, and J.~Khoury, ``{Dynamical friction in
  dark matter superfluids: The evolution of black hole binaries},''
  \href{http://dx.doi.org/10.1088/1475-7516/2024/06/024}{{\em JCAP} {\bfseries
  06} (2024) 024}, \href{http://arxiv.org/abs/2311.07672}{{\ttfamily
  arXiv:2311.07672 [astro-ph.CO]}}.

\bibitem{Karydas:2024fcn}
T.~K. Karydas, B.~J. Kavanagh, and G.~Bertone, ``{Sharpening the dark matter
  signature in gravitational waveforms.~I.~Accretion and eccentricity
  evolution},'' \href{http://dx.doi.org/10.1103/PhysRevD.111.063070}{{\em Phys.
  Rev. D} {\bfseries 111} no.~6, (2025) 063070},
  \href{http://arxiv.org/abs/2402.13053}{{\ttfamily arXiv:2402.13053 [gr-qc]}}.

\bibitem{Kavanagh:2024lgq}
B.~J. Kavanagh, T.~K. Karydas, G.~Bertone, P.~Di~Cintio, and M.~Pasquato,
  ``{Sharpening the dark matter signature in gravitational waveforms. II.
  Numerical simulations},''
  \href{http://dx.doi.org/10.1103/PhysRevD.111.063071}{{\em Phys. Rev. D}
  {\bfseries 111} no.~6, (2025) 063071},
  \href{http://arxiv.org/abs/2402.13762}{{\ttfamily arXiv:2402.13762 [gr-qc]}}.

\bibitem{Gliorio:2025cbh}
S.~Gliorio, E.~Berti, A.~Maselli, and N.~Speeney, ``{Extreme mass ratio
  inspirals in dark matter halos: dynamics and distinguishability of halo
  models},'' \href{http://arxiv.org/abs/2503.16649}{{\ttfamily arXiv:2503.16649
  [gr-qc]}}.

\bibitem{Ferreira:2017pth}
M.~C. Ferreira, C.~F.~B. Macedo, and V.~Cardoso, ``{Orbital fingerprints of
  ultralight scalar fields around black holes},''
  \href{http://dx.doi.org/10.1103/PhysRevD.96.083017}{{\em Phys. Rev. D}
  {\bfseries 96} no.~8, (2017) 083017},
  \href{http://arxiv.org/abs/1710.00830}{{\ttfamily arXiv:1710.00830 [gr-qc]}}.

\bibitem{Kim:2022mdj}
H.~Kim, A.~Lenoci, I.~Stomberg, and X.~Xue, ``{Adiabatically compressed wave
  dark matter halo and intermediate-mass-ratio inspirals},''
  \href{http://dx.doi.org/10.1103/PhysRevD.107.083005}{{\em Phys. Rev. D}
  {\bfseries 107} no.~8, (2023) 083005},
  \href{http://arxiv.org/abs/2212.07528}{{\ttfamily arXiv:2212.07528
  [astro-ph.GA]}}.

\bibitem{Buehler:2022tmr}
R.~Buehler and V.~Desjacques, ``{Dynamical friction in fuzzy dark matter:
  Circular orbits},'' \href{http://dx.doi.org/10.1103/PhysRevD.107.023516}{{\em
  Phys. Rev. D} {\bfseries 107} no.~2, (2023) 023516},
  \href{http://arxiv.org/abs/2207.13740}{{\ttfamily arXiv:2207.13740
  [astro-ph.CO]}}.

\bibitem{Traykova:2023qyv}
D.~Traykova, R.~Vicente, K.~Clough, T.~Helfer, E.~Berti, P.~G. Ferreira, and
  L.~Hui, ``{Relativistic drag forces on black holes from scalar dark matter
  clouds of all sizes},''
  \href{http://dx.doi.org/10.1103/PhysRevD.108.L121502}{{\em Phys. Rev. D}
  {\bfseries 108} no.~12, (2023) L121502},
  \href{http://arxiv.org/abs/2305.10492}{{\ttfamily arXiv:2305.10492 [gr-qc]}}.

\bibitem{Wang:2024cej}
Z.~Wang, T.~Helfer, D.~Traykova, K.~Clough, and E.~Berti, ``{Gravitational
  Magnus effect from scalar dark matter},''
  \href{http://dx.doi.org/10.1103/PhysRevD.110.024009}{{\em Phys. Rev. D}
  {\bfseries 110} no.~2, (2024) 024009},
  \href{http://arxiv.org/abs/2402.07977}{{\ttfamily arXiv:2402.07977 [gr-qc]}}.

\bibitem{Dyson:2024qrq}
C.~Dyson, J.~Redondo-Yuste, M.~van~de Meent, and V.~Cardoso, ``{Relativistic
  aerodynamics of spinning black holes},''
  \href{http://dx.doi.org/10.1103/PhysRevD.109.104038}{{\em Phys. Rev. D}
  {\bfseries 109} no.~10, (2024) 104038},
  \href{http://arxiv.org/abs/2402.07981}{{\ttfamily arXiv:2402.07981 [gr-qc]}}.

\bibitem{Zhang:2018kib}
J.~Zhang and H.~Yang, ``{Gravitational floating orbits around hairy black
  holes},'' \href{http://dx.doi.org/10.1103/PhysRevD.99.064018}{{\em Phys. Rev.
  D} {\bfseries 99} no.~6, (2019) 064018},
  \href{http://arxiv.org/abs/1808.02905}{{\ttfamily arXiv:1808.02905 [gr-qc]}}.

\bibitem{Zhang:2019eid}
J.~Zhang and H.~Yang, ``{Dynamic Signatures of Black Hole Binaries with
  Superradiant Clouds},''
  \href{http://dx.doi.org/10.1103/PhysRevD.101.043020}{{\em Phys. Rev. D}
  {\bfseries 101} no.~4, (2020) 043020},
  \href{http://arxiv.org/abs/1907.13582}{{\ttfamily arXiv:1907.13582 [gr-qc]}}.

\bibitem{Takahashi:2021eso}
T.~Takahashi and T.~Tanaka, ``{Axion clouds may survive the perturbative tidal
  interaction over the early inspiral phase of black hole binaries},''
  \href{http://dx.doi.org/10.1088/1475-7516/2021/10/031}{{\em JCAP} {\bfseries
  10} (2021) 031}, \href{http://arxiv.org/abs/2106.08836}{{\ttfamily
  arXiv:2106.08836 [gr-qc]}}.

\bibitem{Baumann:2022pkl}
D.~Baumann, G.~Bertone, J.~Stout, and G.~M. Tomaselli, ``{Sharp Signals of
  Boson Clouds in Black Hole Binary Inspirals},''
  \href{http://dx.doi.org/10.1103/PhysRevLett.128.221102}{{\em Phys. Rev.
  Lett.} {\bfseries 128} no.~22, (2022) 221102},
  \href{http://arxiv.org/abs/2206.01212}{{\ttfamily arXiv:2206.01212 [gr-qc]}}.

\bibitem{Takahashi:2023flk}
T.~Takahashi, H.~Omiya, and T.~Tanaka, ``{Evolution of binary systems
  accompanying axion clouds in extreme mass ratio inspirals},''
  \href{http://dx.doi.org/10.1103/PhysRevD.107.103020}{{\em Phys. Rev. D}
  {\bfseries 107} no.~10, (2023) 103020},
  \href{http://arxiv.org/abs/2301.13213}{{\ttfamily arXiv:2301.13213 [gr-qc]}}.

\bibitem{Duque:2023seg}
F.~Duque, C.~F.~B. Macedo, R.~Vicente, and V.~Cardoso, ``{Extreme-Mass-Ratio
  Inspirals in Ultralight Dark Matter},''
  \href{http://dx.doi.org/10.1103/PhysRevLett.133.121404}{{\em Phys. Rev.
  Lett.} {\bfseries 133} no.~12, (2024) 121404},
  \href{http://arxiv.org/abs/2312.06767}{{\ttfamily arXiv:2312.06767 [gr-qc]}}.

\bibitem{Boskovic:2024fga}
M.~Bo\v{s}kovi\'c, M.~Koschnitzke, and R.~A. Porto, ``{Signatures of Ultralight
  Bosons in the Orbital Eccentricity of Binary Black Holes},''
  \href{http://dx.doi.org/10.1103/PhysRevLett.133.121401}{{\em Phys. Rev.
  Lett.} {\bfseries 133} no.~12, (2024) 121401},
  \href{http://arxiv.org/abs/2403.02415}{{\ttfamily arXiv:2403.02415 [gr-qc]}}.

\bibitem{Palenzuela:2006wp}
C.~Palenzuela, I.~Olabarrieta, L.~Lehner, and S.~L. Liebling, ``{Head-on
  collisions of boson stars},''
  \href{http://dx.doi.org/10.1103/PhysRevD.75.064005}{{\em Phys. Rev. D}
  {\bfseries 75} (2007) 064005},
  \href{http://arxiv.org/abs/gr-qc/0612067}{{\ttfamily arXiv:gr-qc/0612067}}.

\bibitem{Palenzuela:2017kcg}
C.~Palenzuela, P.~Pani, M.~Bezares, V.~Cardoso, L.~Lehner, and S.~Liebling,
  ``{Gravitational Wave Signatures of Highly Compact Boson Star Binaries},''
  \href{http://dx.doi.org/10.1103/PhysRevD.96.104058}{{\em Phys. Rev. D}
  {\bfseries 96} no.~10, (2017) 104058},
  \href{http://arxiv.org/abs/1710.09432}{{\ttfamily arXiv:1710.09432 [gr-qc]}}.

\bibitem{Helfer:2018vtq}
T.~Helfer, E.~A. Lim, M.~A.~G. Garcia, and M.~A. Amin, ``{Gravitational Wave
  Emission from Collisions of Compact Scalar Solitons},''
  \href{http://dx.doi.org/10.1103/PhysRevD.99.044046}{{\em Phys. Rev. D}
  {\bfseries 99} no.~4, (2019) 044046},
  \href{http://arxiv.org/abs/1802.06733}{{\ttfamily arXiv:1802.06733 [gr-qc]}}.

\bibitem{Bezares:2018qwa}
M.~Bezares and C.~Palenzuela, ``{Gravitational Waves from Dark Boson Star
  Binary Mergers},'' \href{http://dx.doi.org/10.1088/1361-6382/aae87c}{{\em
  Class. Quant. Grav.} {\bfseries 35} (2018) 234002},
\href{http://arxiv.org/abs/1808.10732}{{\ttfamily arXiv:1808.10732 [gr-qc]}}.

\bibitem{Clough:2018exo}
K.~Clough, T.~Dietrich, and J.~C. Niemeyer, ``{Axion star collisions with black
  holes and neutron stars in full 3D numerical relativity},''
  \href{http://dx.doi.org/10.1103/PhysRevD.98.083020}{{\em Phys. Rev. D}
  {\bfseries 98} no.~8, (2018) 083020},
  \href{http://arxiv.org/abs/1808.04668}{{\ttfamily arXiv:1808.04668 [gr-qc]}}.

\bibitem{Zwick:2022dih}
L.~Zwick, P.~R. Capelo, and L.~Mayer, ``{Priorities in gravitational waveforms
  for future space-borne detectors: vacuum accuracy or environment?},''
  \href{http://dx.doi.org/10.1093/mnras/stad707}{{\em Mon. Not. Roy. Astron.
  Soc.} {\bfseries 521} no.~3, (2023) 4645--4651},
  \href{http://arxiv.org/abs/2209.04060}{{\ttfamily arXiv:2209.04060 [gr-qc]}}.

\bibitem{Garg:2024qxq}
M.~Garg, L.~Sberna, L.~Speri, F.~Duque, and J.~Gair, ``{Systematics in tests of
  general relativity using LISA massive black hole binaries},''
  \href{http://dx.doi.org/10.1093/mnras/stae2605}{{\em Mon. Not. Roy. Astron.
  Soc.} {\bfseries 535} no.~4, (2024) 3283--3292},
  \href{http://arxiv.org/abs/2410.02910}{{\ttfamily arXiv:2410.02910
  [astro-ph.GA]}}.

\bibitem{SenPlasma}
S.~Sen, ``{Plasma effects on lasing of a uniform ultralight axion
  condensate},'' \href{http://dx.doi.org/10.1103/PhysRevD.98.103012}{{\em Phys.
  Rev. D} {\bfseries 98} no.~10, (2018) 103012},
  \href{http://arxiv.org/abs/1805.06471}{{\ttfamily arXiv:1805.06471
  [hep-ph]}}.

\bibitem{Cardoso:2020nst}
V.~Cardoso, W.-D. Guo, C.~F.~B. Macedo, and P.~Pani, ``{The tune of the
  Universe: the role of plasma in tests of strong-field gravity},''
  \href{http://dx.doi.org/10.1093/mnras/stab404}{{\em Mon. Not. Roy. Astron.
  Soc.} {\bfseries 503} no.~1, (2021) 563--573},
  \href{http://arxiv.org/abs/2009.07287}{{\ttfamily arXiv:2009.07287 [gr-qc]}}.

\bibitem{Blas:2020kaa}
D.~Blas and S.~J. Witte, ``{Quenching Mechanisms of Photon Superradiance},''
  \href{http://dx.doi.org/10.1103/PhysRevD.102.123018}{{\em Phys. Rev. D}
  {\bfseries 102} no.~12, (2020) 123018},
  \href{http://arxiv.org/abs/2009.10075}{{\ttfamily arXiv:2009.10075
  [hep-ph]}}.

\bibitem{Caputo:2021efm}
A.~Caputo, S.~J. Witte, D.~Blas, and P.~Pani, ``{Electromagnetic signatures of
  dark photon superradiance},''
  \href{http://dx.doi.org/10.1103/PhysRevD.104.043006}{{\em Phys. Rev. D}
  {\bfseries 104} no.~4, (2021) 043006},
  \href{http://arxiv.org/abs/2102.11280}{{\ttfamily arXiv:2102.11280
  [hep-ph]}}.

\bibitem{Siemonsen:2022ivj}
N.~Siemonsen, C.~Mondino, D.~Egana-Ugrinovic, J.~Huang, M.~Baryakhtar, and
  W.~E. East, ``{Dark photon superradiance: Electrodynamics and multimessenger
  signals},'' \href{http://dx.doi.org/10.1103/PhysRevD.107.075025}{{\em Phys.
  Rev. D} {\bfseries 107} no.~7, (2023) 075025},
  \href{http://arxiv.org/abs/2212.09772}{{\ttfamily arXiv:2212.09772
  [astro-ph.HE]}}.

\bibitem{Kaplan:2015fuy}
D.~E. Kaplan and R.~Rattazzi, ``{Large field excursions and approximate
  discrete symmetries from a clockwork axion},''
  \href{http://dx.doi.org/10.1103/PhysRevD.93.085007}{{\em Phys. Rev. D}
  {\bfseries 93} no.~8, (2016) 085007},
  \href{http://arxiv.org/abs/1511.01827}{{\ttfamily arXiv:1511.01827
  [hep-ph]}}.

\bibitem{Farina:2016tgd}
M.~Farina, D.~Pappadopulo, F.~Rompineve, and A.~Tesi, ``{The photo-philic QCD
  axion},'' \href{http://dx.doi.org/10.1007/JHEP01(2017)095}{{\em JHEP}
  {\bfseries 01} (2017) 095}, \href{http://arxiv.org/abs/1611.09855}{{\ttfamily
  arXiv:1611.09855 [hep-ph]}}.

\bibitem{Sokolov:2021ydn}
A.~V. Sokolov and A.~Ringwald, ``{Photophilic hadronic axion from heavy
  magnetic monopoles},'' \href{http://dx.doi.org/10.1007/JHEP06(2021)123}{{\em
  JHEP} {\bfseries 06} (2021) 123},
  \href{http://arxiv.org/abs/2104.02574}{{\ttfamily arXiv:2104.02574
  [hep-ph]}}.

\bibitem{Sokolov:2022fvs}
A.~V. Sokolov and A.~Ringwald, ``{Electromagnetic Couplings of Axions},''
  \href{http://arxiv.org/abs/2205.02605}{{\ttfamily arXiv:2205.02605
  [hep-ph]}}.

\bibitem{Yoshino:2012kn}
H.~Yoshino and H.~Kodama, ``{Bosenova collapse of axion cloud around a rotating
  black hole},'' \href{http://dx.doi.org/10.1143/PTP.128.153}{{\em Prog. Theor.
  Phys.} {\bfseries 128} (2012) 153--190},
  \href{http://arxiv.org/abs/1203.5070}{{\ttfamily arXiv:1203.5070 [gr-qc]}}.

\bibitem{Omiya:2022gwu}
H.~Omiya, T.~Takahashi, T.~Tanaka, and H.~Yoshino, ``{Impact of multiple modes
  on the evolution of self-interacting axion condensate around rotating black
  holes},'' \href{http://dx.doi.org/10.1088/1475-7516/2023/06/016}{{\em JCAP}
  {\bfseries 06} (2023) 016}, \href{http://arxiv.org/abs/2211.01949}{{\ttfamily
  arXiv:2211.01949 [gr-qc]}}.

\bibitem{Chia:2022udn}
H.~S. Chia, C.~Doorman, A.~Wernersson, T.~Hinderer, and S.~Nissanke,
  ``{Self-interacting gravitational atoms in the strong-gravity regime},''
  \href{http://dx.doi.org/10.1088/1475-7516/2023/04/018}{{\em JCAP} {\bfseries
  04} (2023) 018}, \href{http://arxiv.org/abs/2212.11948}{{\ttfamily
  arXiv:2212.11948 [gr-qc]}}.

\bibitem{Zouros:1979iw}
T.~J.~M. Zouros and D.~M. Eardley, ``{Instabilities of massive scalar
  perturbations of a rotating black hole},''
  \href{http://dx.doi.org/10.1016/0003-4916(79)90237-9}{{\em Annals Phys.}
  {\bfseries 118} (1979) 139--155}.

\bibitem{Cardoso:2005vk}
V.~Cardoso and S.~Yoshida, ``{Superradiant Instabilities of Rotating Black
  Branes and Strings},''
  \href{http://dx.doi.org/10.1088/1126-6708/2005/07/009}{{\em JHEP} {\bfseries
  07} (2005) 009},
\href{http://arxiv.org/abs/hep-th/0502206}{{\ttfamily arXiv:hep-th/0502206
  [hep-th]}}.

\bibitem{Dolan:2007mj}
S.~Dolan, ``{Instability of the Massive Klein-Gordon Field on the Kerr
  Spacetime},'' \href{http://dx.doi.org/10.1103/PhysRevD.76.084001}{{\em Phys.
  Rev.} {\bfseries D76} (2007) 084001},
\href{http://arxiv.org/abs/0705.2880}{{\ttfamily arXiv:0705.2880 [gr-qc]}}.

\bibitem{Cardoso:2015zqa}
V.~Cardoso, R.~Brito, and J.~L. Rosa, ``{Superradiance in stars},''
  \href{http://dx.doi.org/10.1103/PhysRevD.91.124026}{{\em Phys. Rev. D}
  {\bfseries 91} no.~12, (2015) 124026},
  \href{http://arxiv.org/abs/1505.05509}{{\ttfamily arXiv:1505.05509 [gr-qc]}}.

\bibitem{KrallTrivelpiece1973}
N.~Krall and A.~Trivelpiece, {\em Principles of Plasma Physics}.
\newblock International series in pure and applied physics. McGraw-Hill, 1973.

\bibitem{Kephart1995}
T.~W. Kephart and T.~J. Weiler, ``Stimulated radiation from axion cluster
  evolution,'' \href{http://dx.doi.org/10.1103/PhysRevD.52.3226}{{\em Phys.
  Rev. D} {\bfseries 52} (Sep, 1995) 3226--3238}.

\bibitem{hertzberg:2018zte}
M.~P. Hertzberg and E.~D. Schiappacasse, ``{Dark Matter Axion Clump Resonance
  of Photons},'' \href{http://dx.doi.org/10.1088/1475-7516/2018/11/004}{{\em
  JCAP} {\bfseries 11} (2018) 004},
  \href{http://arxiv.org/abs/1805.00430}{{\ttfamily arXiv:1805.00430
  [hep-ph]}}.

\bibitem{Carenza:2019vzg}
P.~Carenza, A.~Mirizzi, and G.~Sigl, ``{Dynamical evolution of axion
  condensates under stimulated decays into photons},''
  \href{http://dx.doi.org/10.1103/PhysRevD.101.103016}{{\em Phys. Rev. D}
  {\bfseries 101} no.~10, (2020) 103016},
  \href{http://arxiv.org/abs/1911.07838}{{\ttfamily arXiv:1911.07838
  [hep-ph]}}.

\bibitem{Dima:2020rzg}
A.~Dima and E.~Barausse, ``{Numerical investigation of plasma-driven
  superradiant instabilities},''
  \href{http://dx.doi.org/10.1088/1361-6382/ab9ce0}{{\em Class. Quant. Grav.}
  {\bfseries 37} no.~17, (2020) 175006},
  \href{http://arxiv.org/abs/2001.11484}{{\ttfamily arXiv:2001.11484 [gr-qc]}}.

\bibitem{Ferriere:2001rg}
K.~M. Ferriere, ``{The interstellar environment of our galaxy},''
  \href{http://dx.doi.org/10.1103/RevModPhys.73.1031}{{\em Rev. Mod. Phys.}
  {\bfseries 73} (2001) 1031--1066},
  \href{http://arxiv.org/abs/astro-ph/0106359}{{\ttfamily
  arXiv:astro-ph/0106359}}.

\bibitem{burns2019farside}
J.~Burns, G.~Hallinan, J.~Lux, A.~Romero-Wolf, T.-C. Chang, J.~Kocz, J.~Bowman,
  R.~MacDowall, J.~Kasper, R.~Bradley, M.~Anderson, and D.~Rapetti, ``Farside:
  A low radio frequency interferometric array on the lunar farside,''
  \href{http://arxiv.org/abs/1907.05407}{{\ttfamily arXiv:1907.05407
  [astro-ph.IM]}}.

\bibitem{Carroll:1989vb}
S.~M. Carroll, G.~B. Field, and R.~Jackiw, ``{Limits on a Lorentz and Parity
  Violating Modification of Electrodynamics},''
  \href{http://dx.doi.org/10.1103/PhysRevD.41.1231}{{\em Phys. Rev. D}
  {\bfseries 41} (1990) 1231}.

\bibitem{Harari:1992ea}
D.~Harari and P.~Sikivie, ``{Effects of a Nambu-Goldstone boson on the
  polarization of radio galaxies and the cosmic microwave background},''
  \href{http://dx.doi.org/10.1016/0370-2693(92)91363-E}{{\em Phys. Lett. B}
  {\bfseries 289} (1992) 67--72}.

\bibitem{Chen:2019fsq}
Y.~Chen, J.~Shu, X.~Xue, Q.~Yuan, and Y.~Zhao, ``{Probing Axions with Event
  Horizon Telescope Polarimetric Measurements},''
  \href{http://dx.doi.org/10.1103/PhysRevLett.124.061102}{{\em Phys. Rev.
  Lett.} {\bfseries 124} no.~6, (2020) 061102},
  \href{http://arxiv.org/abs/1905.02213}{{\ttfamily arXiv:1905.02213
  [hep-ph]}}.

\bibitem{Chen:2021lvo}
Y.~Chen, Y.~Liu, R.-S. Lu, Y.~Mizuno, J.~Shu, X.~Xue, Q.~Yuan, and Y.~Zhao,
  ``{Stringent axion constraints with Event Horizon Telescope polarimetric
  measurements of M87$^{*}$},''
  \href{http://dx.doi.org/10.1038/s41550-022-01620-3}{{\em Nature Astron.}
  {\bfseries 6} no.~5, (2022) 592--598},
  \href{http://arxiv.org/abs/2105.04572}{{\ttfamily arXiv:2105.04572
  [hep-ph]}}.

\bibitem{Chen:2022oad}
Y.~Chen, C.~Li, Y.~Mizuno, J.~Shu, X.~Xue, Q.~Yuan, Y.~Zhao, and Z.~Zhou,
  ``{Birefringence tomography for axion cloud},''
  \href{http://dx.doi.org/10.1088/1475-7516/2022/09/073}{{\em JCAP} {\bfseries
  09} (2022) 073}, \href{http://arxiv.org/abs/2208.05724}{{\ttfamily
  arXiv:2208.05724 [hep-ph]}}.

\bibitem{CAST:2017uph}
{\bfseries CAST} Collaboration, V.~Anastassopoulos {\em et~al.}, ``{New CAST
  Limit on the Axion-Photon Interaction},''
  \href{http://dx.doi.org/10.1038/nphys4109}{{\em Nature Phys.} {\bfseries 13}
  (2017) 584--590}, \href{http://arxiv.org/abs/1705.02290}{{\ttfamily
  arXiv:1705.02290 [hep-ex]}}.

\bibitem{Hoof:2022xbe}
S.~Hoof and L.~Schulz, ``{Updated constraints on axion-like particles from
  temporal information in supernova SN1987A gamma-ray data},''
  \href{http://dx.doi.org/10.1088/1475-7516/2023/03/054}{{\em JCAP} {\bfseries
  03} (2023) 054}, \href{http://arxiv.org/abs/2212.09764}{{\ttfamily
  arXiv:2212.09764 [hep-ph]}}.

\bibitem{KawDawson1970}
P.~{Kaw} and J.~{Dawson}, ``{Relativistic Nonlinear Propagation of Laser Beams
  in Cold Overdense Plasmas},'' \href{http://dx.doi.org/10.1063/1.1692942}{{\em
  Physics of Fluids} {\bfseries 13} no.~2, (Feb., 1970) 472--481}.

\bibitem{1971PhRvL..27.1342M}
C.~{Max} and F.~{Perkins}, ``{Strong Electromagnetic Waves in Overdense
  Plasmas},'' \href{http://dx.doi.org/10.1103/PhysRevLett.27.1342}{{\em Phys.
  Rev. Lett.} {\bfseries 27} no.~20, (Nov., 1971) 1342--1345}.

\bibitem{Cannizzaro:2023ltu}
E.~Cannizzaro, F.~Corelli, and P.~Pani, ``{Nonlinear photon-plasma interaction
  and the black hole superradiant instability},''
  \href{http://dx.doi.org/10.1103/PhysRevD.109.023007}{{\em Phys. Rev. D}
  {\bfseries 109} no.~2, (2024) 023007},
  \href{http://arxiv.org/abs/2306.12490}{{\ttfamily arXiv:2306.12490 [gr-qc]}}.

\bibitem{Battye:2023oac}
R.~A. Battye, M.~J. Keith, J.~I. McDonald, S.~Srinivasan, B.~W. Stappers, and
  P.~Weltevrede, ``{Searching for time-dependent axion dark matter signals in
  pulsars},'' \href{http://dx.doi.org/10.1103/PhysRevD.108.063001}{{\em Phys.
  Rev. D} {\bfseries 108} no.~6, (2023) 063001},
  \href{http://arxiv.org/abs/2303.11792}{{\ttfamily arXiv:2303.11792
  [astro-ph.CO]}}.

\bibitem{Huang:2018lxq}
F.~P. Huang, K.~Kadota, T.~Sekiguchi, and H.~Tashiro, ``{Radio telescope search
  for the resonant conversion of cold dark matter axions from the magnetized
  astrophysical sources},''
  \href{http://dx.doi.org/10.1103/PhysRevD.97.123001}{{\em Phys. Rev. D}
  {\bfseries 97} no.~12, (2018) 123001},
  \href{http://arxiv.org/abs/1803.08230}{{\ttfamily arXiv:1803.08230
  [hep-ph]}}.

\bibitem{Hook:2018iia}
A.~Hook, Y.~Kahn, B.~R. Safdi, and Z.~Sun, ``{Radio Signals from Axion Dark
  Matter Conversion in Neutron Star Magnetospheres},''
  \href{http://dx.doi.org/10.1103/PhysRevLett.121.241102}{{\em Phys. Rev.
  Lett.} {\bfseries 121} no.~24, (2018) 241102},
  \href{http://arxiv.org/abs/1804.03145}{{\ttfamily arXiv:1804.03145
  [hep-ph]}}.

\bibitem{Leroy:2019ghm}
M.~Leroy, M.~Chianese, T.~D.~P. Edwards, and C.~Weniger, ``{Radio Signal of
  Axion-Photon Conversion in Neutron Stars: A Ray Tracing Analysis},''
  \href{http://dx.doi.org/10.1103/PhysRevD.101.123003}{{\em Phys. Rev. D}
  {\bfseries 101} no.~12, (2020) 123003},
  \href{http://arxiv.org/abs/1912.08815}{{\ttfamily arXiv:1912.08815
  [hep-ph]}}.

\bibitem{Witte:2020rvb}
S.~J. Witte, S.~Rosauro-Alcaraz, S.~D. McDermott, and V.~Poulin, ``{Dark photon
  dark matter in the presence of inhomogeneous structure},''
  \href{http://dx.doi.org/10.1007/JHEP06(2020)132}{{\em JHEP} {\bfseries 06}
  (2020) 132}, \href{http://arxiv.org/abs/2003.13698}{{\ttfamily
  arXiv:2003.13698 [astro-ph.CO]}}.

\bibitem{Witte:2021arp}
S.~J. Witte, D.~Noordhuis, T.~D.~P. Edwards, and C.~Weniger, ``{Axion-photon
  conversion in neutron star magnetospheres: The role of the plasma in the
  Goldreich-Julian model},''
  \href{http://dx.doi.org/10.1103/PhysRevD.104.103030}{{\em Phys. Rev. D}
  {\bfseries 104} no.~10, (2021) 103030},
  \href{http://arxiv.org/abs/2104.07670}{{\ttfamily arXiv:2104.07670
  [hep-ph]}}.

\bibitem{Burrage:2023zvk}
C.~Burrage, P.~G.~S. Fernandes, R.~Brito, and V.~Cardoso, ``{Spinning black
  holes with axion hair},''
  \href{http://dx.doi.org/10.1088/1361-6382/acf9d6}{{\em Class. Quant. Grav.}
  {\bfseries 40} no.~20, (2023) 205021},
  \href{http://arxiv.org/abs/2306.03662}{{\ttfamily arXiv:2306.03662 [gr-qc]}}.

\bibitem{PhysRevD.37.1237}
G.~Raffelt and L.~Stodolsky, ``Mixing of the photon with low-mass particles,''
  \href{http://dx.doi.org/10.1103/PhysRevD.37.1237}{{\em Phys. Rev. D}
  {\bfseries 37} (Mar, 1988) 1237--1249}.

\bibitem{Raffelt:1996wa}
G.~Raffelt, {\em {Stars as Laboratories for Fundamental Physics}}.
\newblock University of Chicago Press, Chicago, IL, 1996.

\bibitem{Mirizzi:2006zy}
A.~Mirizzi, G.~G. Raffelt, and P.~D. Serpico, ``{Photon-axion conversion in
  intergalactic magnetic fields and cosmological consequences},''
  \href{http://dx.doi.org/10.1007/978-3-540-73518-2_7}{{\em Lect. Notes Phys.}
  {\bfseries 741} (2008) 115--134},
  \href{http://arxiv.org/abs/astro-ph/0607415}{{\ttfamily
  arXiv:astro-ph/0607415}}.

\bibitem{Redondo:2008aa}
J.~Redondo, ``{Helioscope Bounds on Hidden Sector Photons},''
  \href{http://dx.doi.org/10.1088/1475-7516/2008/07/008}{{\em JCAP} {\bfseries
  07} (2008) 008}, \href{http://arxiv.org/abs/0801.1527}{{\ttfamily
  arXiv:0801.1527 [hep-ph]}}.

\bibitem{An:2013yfc}
H.~An, M.~Pospelov, and J.~Pradler, ``{New stellar constraints on dark
  photons},'' \href{http://dx.doi.org/10.1016/j.physletb.2013.07.008}{{\em
  Phys. Lett. B} {\bfseries 725} (2013) 190--195},
  \href{http://arxiv.org/abs/1302.3884}{{\ttfamily arXiv:1302.3884 [hep-ph]}}.

\bibitem{Wald:1974np}
R.~M. Wald, ``{Black hole in a uniform magnetic field},''
  \href{http://dx.doi.org/10.1103/PhysRevD.10.1680}{{\em Phys. Rev. D}
  {\bfseries 10} (1974) 1680--1685}.

\bibitem{Komissarov:2021vks}
S.~S. Komissarov, ``{Electrically charged black holes and the
  Blandford\textendash{}Znajek mechanism},''
  \href{http://dx.doi.org/10.1093/mnras/stab2686}{{\em Mon. Not. Roy. Astron.
  Soc.} {\bfseries 512} no.~2, (2022) 2798--2805},
  \href{http://arxiv.org/abs/2108.08161}{{\ttfamily arXiv:2108.08161
  [astro-ph.HE]}}.

\bibitem{Cardoso:2016olt}
V.~Cardoso, C.~F.~B. Macedo, P.~Pani, and V.~Ferrari, ``{Black holes and
  gravitational waves in models of minicharged dark matter},''
  \href{http://dx.doi.org/10.1088/1475-7516/2016/05/054}{{\em JCAP} {\bfseries
  05} (2016) 054}, \href{http://arxiv.org/abs/1604.07845}{{\ttfamily
  arXiv:1604.07845 [hep-ph]}}. [Erratum: JCAP 04, E01 (2020)].

\bibitem{Jaeckel:2012mjv}
J.~Jaeckel, ``{A force beyond the Standard Model - Status of the quest for
  hidden photons},'' {\em Frascati Phys. Ser.} {\bfseries 56} (2012) 172--192,
  \href{http://arxiv.org/abs/1303.1821}{{\ttfamily arXiv:1303.1821 [hep-ph]}}.

\bibitem{Siemonsen:2022yyf}
N.~Siemonsen, T.~May, and W.~E. East, ``{Modeling the black hole superradiance
  gravitational waveform},''
  \href{http://dx.doi.org/10.1103/PhysRevD.107.104003}{{\em Phys. Rev. D}
  {\bfseries 107} no.~10, (2023) 104003},
  \href{http://arxiv.org/abs/2211.03845}{{\ttfamily arXiv:2211.03845 [gr-qc]}}.

\bibitem{Feng:2022evy}
J.~C. Feng, S.~Chakraborty, and V.~Cardoso, ``{Shielding a charged black
  hole},'' \href{http://dx.doi.org/10.1103/PhysRevD.107.044050}{{\em Phys. Rev.
  D} {\bfseries 107} no.~4, (2023) 044050},
  \href{http://arxiv.org/abs/2211.05261}{{\ttfamily arXiv:2211.05261 [gr-qc]}}.

\bibitem{PhysRevLett.51.1415}
P.~Sikivie, ``Experimental tests of the "invisible" axion,''
  \href{http://dx.doi.org/10.1103/PhysRevLett.51.1415}{{\em Phys. Rev. Lett.}
  {\bfseries 51} (Oct, 1983) 1415--1417}.

\bibitem{Pani:2013wsa}
P.~Pani, E.~Berti, and L.~Gualtieri, ``{Scalar, Electromagnetic and
  Gravitational Perturbations of Kerr-Newman Black Holes in the Slow-Rotation
  Limit},'' \href{http://dx.doi.org/10.1103/PhysRevD.88.064048}{{\em Phys. Rev.
  D} {\bfseries 88} (2013) 064048},
  \href{http://arxiv.org/abs/1307.7315}{{\ttfamily arXiv:1307.7315 [gr-qc]}}.

\bibitem{Berti:2005eb}
E.~Berti and K.~D. Kokkotas, ``{Quasinormal modes of Kerr-Newman black holes:
  Coupling of electromagnetic and gravitational perturbations},''
  \href{http://dx.doi.org/10.1103/PhysRevD.71.124008}{{\em Phys. Rev. D}
  {\bfseries 71} (2005) 124008},
  \href{http://arxiv.org/abs/gr-qc/0502065}{{\ttfamily arXiv:gr-qc/0502065}}.

\bibitem{An:2023mvf}
H.~An, S.~Ge, and J.~Liu, ``{Solar Radio Emissions and Ultralight Dark
  Matter},'' \href{http://dx.doi.org/10.3390/universe9030142}{{\em Universe}
  {\bfseries 9} no.~3, (2023) 142},
  \href{http://arxiv.org/abs/2304.01056}{{\ttfamily arXiv:2304.01056
  [hep-ph]}}.

\bibitem{McDonald:2023shx}
J.~I. McDonald and S.~J. Witte, ``{Generalized ray tracing for axions in
  astrophysical plasmas},''
  \href{http://dx.doi.org/10.1103/PhysRevD.108.103021}{{\em Phys. Rev. D}
  {\bfseries 108} no.~10, (2023) 103021},
  \href{http://arxiv.org/abs/2309.08655}{{\ttfamily arXiv:2309.08655
  [hep-ph]}}.

\bibitem{Zerilli:1974ai}
F.~J. Zerilli, ``{Perturbation analysis for gravitational and electromagnetic
  radiation in a reissner-nordstroem geometry},''
  \href{http://dx.doi.org/10.1103/PhysRevD.9.860}{{\em Phys. Rev. D} {\bfseries
  9} (1974) 860--868}.

\bibitem{Zenginoglu:2011zz}
A.~Zenginoglu and G.~Khanna, ``{Null infinity waveforms from extreme-mass-ratio
  inspirals in Kerr spacetime},''
  \href{http://dx.doi.org/10.1103/PhysRevX.1.021017}{{\em Phys. Rev. X}
  {\bfseries 1} (2011) 021017},
  \href{http://arxiv.org/abs/1108.1816}{{\ttfamily arXiv:1108.1816 [gr-qc]}}.

\bibitem{Krivan:1997hc}
W.~Krivan, P.~Laguna, P.~Papadopoulos, and N.~Andersson, ``{Dynamics of
  perturbations of rotating black holes},''
  \href{http://dx.doi.org/10.1103/PhysRevD.56.3395}{{\em Phys. Rev. D}
  {\bfseries 56} (1997) 3395--3404},
  \href{http://arxiv.org/abs/gr-qc/9702048}{{\ttfamily arXiv:gr-qc/9702048}}.

\bibitem{Pazos_valos_2005}
E.~Pazos-Avalos and C.~O. Lousto, ``{Numerical integration of the Teukolsky
  equation in the time domain},''
  \href{http://dx.doi.org/10.1103/PhysRevD.72.084022}{{\em Phys. Rev. D}
  {\bfseries 72} (2005) 084022},
  \href{http://arxiv.org/abs/gr-qc/0409065}{{\ttfamily arXiv:gr-qc/0409065}}.

\bibitem{Zenginoglu:2012us}
A.~Zenginoglu, G.~Khanna, and L.~M. Burko, ``{Intermediate behavior of Kerr
  tails},'' \href{http://dx.doi.org/10.1007/s10714-014-1672-8}{{\em Gen. Rel.
  Grav.} {\bfseries 46} (2014) 1672},
\href{http://arxiv.org/abs/1208.5839}{{\ttfamily arXiv:1208.5839 [gr-qc]}}.

\bibitem{Cardoso:2021vjq}
V.~Cardoso, F.~Duque, and G.~Khanna, ``{Gravitational tuning forks and
  hierarchical triple systems},''
  \href{http://dx.doi.org/10.1103/PhysRevD.103.L081501}{{\em Phys. Rev. D}
  {\bfseries 103} no.~8, (2021) L081501},
  \href{http://arxiv.org/abs/2101.01186}{{\ttfamily arXiv:2101.01186 [gr-qc]}}.

\bibitem{Dubovsky:2015cca}
S.~Dubovsky and G.~Hern\'andez-Chifflet, ``{Heating up the Galaxy with Hidden
  Photons},'' \href{http://dx.doi.org/10.1088/1475-7516/2015/12/054}{{\em JCAP}
  {\bfseries 1512} no.~12, (2015) 054},
\href{http://arxiv.org/abs/1509.00039}{{\ttfamily arXiv:1509.00039 [hep-ph]}}.

\bibitem{East:2017mrj}
W.~East, ``{Superradiant Instability of Massive Vector Fields around Spinning
  Black Holes in the Relativistic Regime},''
  \href{http://dx.doi.org/10.1103/PhysRevD.96.024004}{{\em Phys. Rev.}
  {\bfseries D96} (2017) 024004},
\href{http://arxiv.org/abs/1705.01544}{{\ttfamily arXiv:1705.01544 [gr-qc]}}.

\bibitem{Prabhu:2021zve}
A.~Prabhu, ``{Axion production in pulsar magnetosphere gaps},''
  \href{http://dx.doi.org/10.1103/PhysRevD.104.055038}{{\em Phys. Rev. D}
  {\bfseries 104} no.~5, (2021) 055038},
  \href{http://arxiv.org/abs/2104.14569}{{\ttfamily arXiv:2104.14569
  [hep-ph]}}.

\bibitem{Noordhuis:2022ljw}
D.~Noordhuis, A.~Prabhu, S.~J. Witte, A.~Y. Chen, F.~Cruz, and C.~Weniger,
  ``{Novel Constraints on Axions Produced in Pulsar Polar-Cap Cascades},''
  \href{http://dx.doi.org/10.1103/PhysRevLett.131.111004}{{\em Phys. Rev.
  Lett.} {\bfseries 131} no.~11, (2023) 111004},
  \href{http://arxiv.org/abs/2209.09917}{{\ttfamily arXiv:2209.09917
  [hep-ph]}}.

\bibitem{Noordhuis:2023wid}
D.~Noordhuis, A.~Prabhu, C.~Weniger, and S.~J. Witte, ``{Axion Clouds around
  Neutron Stars},'' \href{http://dx.doi.org/10.1103/PhysRevX.14.041015}{{\em
  Phys. Rev. X} {\bfseries 14} no.~4, (2024) 041015},
  \href{http://arxiv.org/abs/2307.11811}{{\ttfamily arXiv:2307.11811
  [hep-ph]}}.

\bibitem{Caputo:2023cpv}
A.~Caputo, S.~J. Witte, A.~A. Philippov, and T.~Jacobson, ``{Pulsar Nulling and
  Vacuum Radio Emission from Axion Clouds},''
  \href{http://dx.doi.org/10.1103/PhysRevLett.133.161001}{{\em Phys. Rev.
  Lett.} {\bfseries 133} no.~16, (2024) 161001},
  \href{http://arxiv.org/abs/2311.14795}{{\ttfamily arXiv:2311.14795
  [hep-ph]}}.

\bibitem{Brito:2014nja}
R.~Brito, V.~Cardoso, and P.~Pani, ``{Superradiant instability of black holes
  immersed in a magnetic field},''
  \href{http://dx.doi.org/10.1103/PhysRevD.89.104045}{{\em Phys. Rev. D}
  {\bfseries 89} no.~10, (2014) 104045},
  \href{http://arxiv.org/abs/1405.2098}{{\ttfamily arXiv:1405.2098 [gr-qc]}}.

\bibitem{Day:2019bbh}
F.~V. Day and J.~I. McDonald, ``{Axion superradiance in rotating neutron
  stars},'' \href{http://dx.doi.org/10.1088/1475-7516/2019/10/051}{{\em JCAP}
  {\bfseries 10} (2019) 051}, \href{http://arxiv.org/abs/1904.08341}{{\ttfamily
  arXiv:1904.08341 [hep-ph]}}.

\bibitem{deFreitasPacheco:2023hpb}
J.~A. de~Freitas~Pacheco, E.~Kiritsis, M.~Lucca, and J.~Silk, ``{Quasiextremal
  primordial black holes are a viable dark matter candidate},''
  \href{http://dx.doi.org/10.1103/PhysRevD.107.123525}{{\em Phys. Rev. D}
  {\bfseries 107} no.~12, (2023) 123525},
  \href{http://arxiv.org/abs/2301.13215}{{\ttfamily arXiv:2301.13215
  [astro-ph.CO]}}.

\bibitem{Alonso-Monsalve:2023brx}
E.~Alonso-Monsalve and D.~I. Kaiser, ``{Primordial Black Holes with QCD Color
  Charge},'' \href{http://dx.doi.org/10.1103/PhysRevLett.132.231402}{{\em Phys.
  Rev. Lett.} {\bfseries 132} no.~23, (2024) 231402},
  \href{http://arxiv.org/abs/2310.16877}{{\ttfamily arXiv:2310.16877
  [hep-ph]}}.

\bibitem{Bai:2019zcd}
Y.~Bai and N.~Orlofsky, ``{Primordial Extremal Black Holes as Dark Matter},''
  \href{http://dx.doi.org/10.1103/PhysRevD.101.055006}{{\em Phys. Rev. D}
  {\bfseries 101} no.~5, (2020) 055006},
  \href{http://arxiv.org/abs/1906.04858}{{\ttfamily arXiv:1906.04858
  [hep-ph]}}.

\bibitem{Kritos:2021nsf}
K.~Kritos and J.~Silk, ``{Mergers of maximally charged primordial black
  holes},'' \href{http://dx.doi.org/10.1103/PhysRevD.105.063011}{{\em Phys.
  Rev. D} {\bfseries 105} no.~6, (2022) 063011},
  \href{http://arxiv.org/abs/2109.09769}{{\ttfamily arXiv:2109.09769 [gr-qc]}}.

\bibitem{Holdom:1985ag}
B.~Holdom, ``{Two U(1)'s and Epsilon Charge Shifts},''
  \href{http://dx.doi.org/10.1016/0370-2693(86)91377-8}{{\em Phys. Lett. B}
  {\bfseries 166} (1986) 196--198}.

\bibitem{Dubovsky:2003yn}
S.~L. Dubovsky, D.~S. Gorbunov, and G.~I. Rubtsov, ``{Narrowing the window for
  millicharged particles by CMB anisotropy},''
  \href{http://dx.doi.org/10.1134/1.1675909}{{\em JETP Lett.} {\bfseries 79}
  (2004) 1--5}, \href{http://arxiv.org/abs/hep-ph/0311189}{{\ttfamily
  arXiv:hep-ph/0311189}}.

\bibitem{Sigurdson:2004zp}
K.~Sigurdson, M.~Doran, A.~Kurylov, R.~R. Caldwell, and M.~Kamionkowski,
  ``{Dark-matter electric and magnetic dipole moments},''
  \href{http://dx.doi.org/10.1103/PhysRevD.70.083501}{{\em Phys. Rev. D}
  {\bfseries 70} (2004) 083501},
  \href{http://arxiv.org/abs/astro-ph/0406355}{{\ttfamily
  arXiv:astro-ph/0406355}}. [Erratum: Phys.Rev.D 73, 089903 (2006)].

\bibitem{Gies:2006ca}
H.~Gies, J.~Jaeckel, and A.~Ringwald, ``{Polarized Light Propagating in a
  Magnetic Field as a Probe of Millicharged Fermions},''
  \href{http://dx.doi.org/10.1103/PhysRevLett.97.140402}{{\em Phys. Rev. Lett.}
  {\bfseries 97} (2006) 140402},
  \href{http://arxiv.org/abs/hep-ph/0607118}{{\ttfamily arXiv:hep-ph/0607118}}.

\bibitem{Gies:2006hv}
H.~Gies, J.~Jaeckel, and A.~Ringwald, ``{Accelerator Cavities as a Probe of
  Millicharged Particles},''
  \href{http://dx.doi.org/10.1209/epl/i2006-10356-5}{{\em EPL} {\bfseries 76}
  (2006) 794--800}, \href{http://arxiv.org/abs/hep-ph/0608238}{{\ttfamily
  arXiv:hep-ph/0608238}}.

\bibitem{Burrage:2009yz}
C.~Burrage, J.~Jaeckel, J.~Redondo, and A.~Ringwald, ``{Late time CMB
  anisotropies constrain mini-charged particles},''
  \href{http://dx.doi.org/10.1088/1475-7516/2009/11/002}{{\em JCAP} {\bfseries
  11} (2009) 002}, \href{http://arxiv.org/abs/0909.0649}{{\ttfamily
  arXiv:0909.0649 [astro-ph.CO]}}.

\bibitem{Ahlers:2009kh}
M.~Ahlers, ``{The Hubble diagram as a probe of mini-charged particles},''
  \href{http://dx.doi.org/10.1103/PhysRevD.80.023513}{{\em Phys. Rev. D}
  {\bfseries 80} (2009) 023513},
  \href{http://arxiv.org/abs/0904.0998}{{\ttfamily arXiv:0904.0998 [hep-ph]}}.

\bibitem{McDermott:2010pa}
S.~D. McDermott, H.-B. Yu, and K.~M. Zurek, ``{Turning off the Lights: How Dark
  is Dark Matter?},'' \href{http://dx.doi.org/10.1103/PhysRevD.83.063509}{{\em
  Phys. Rev. D} {\bfseries 83} (2011) 063509},
  \href{http://arxiv.org/abs/1011.2907}{{\ttfamily arXiv:1011.2907 [hep-ph]}}.

\bibitem{Dolgov:2013una}
A.~D. Dolgov, S.~L. Dubovsky, G.~I. Rubtsov, and I.~I. Tkachev, ``{Constraints
  on millicharged particles from Planck data},''
  \href{http://dx.doi.org/10.1103/PhysRevD.88.117701}{{\em Phys. Rev. D}
  {\bfseries 88} no.~11, (2013) 117701},
  \href{http://arxiv.org/abs/1310.2376}{{\ttfamily arXiv:1310.2376 [hep-ph]}}.

\bibitem{Haas:2014dda}
A.~Haas, C.~S. Hill, E.~Izaguirre, and I.~Yavin, ``{Looking for milli-charged
  particles with a new experiment at the LHC},''
  \href{http://dx.doi.org/10.1016/j.physletb.2015.04.062}{{\em Phys. Lett. B}
  {\bfseries 746} (2015) 117--120},
  \href{http://arxiv.org/abs/1410.6816}{{\ttfamily arXiv:1410.6816 [hep-ph]}}.

\bibitem{Khalil:2018aaj}
M.~Khalil, N.~Sennett, J.~Steinhoff, J.~Vines, and A.~Buonanno, ``{Hairy binary
  black holes in Einstein-Maxwell-dilaton theory and their effective-one-body
  description},'' \href{http://dx.doi.org/10.1103/PhysRevD.98.104010}{{\em
  Phys. Rev. D} {\bfseries 98} no.~10, (2018) 104010},
  \href{http://arxiv.org/abs/1809.03109}{{\ttfamily arXiv:1809.03109 [gr-qc]}}.

\bibitem{Caputo:2019tms}
A.~Caputo, L.~Sberna, M.~Frias, D.~Blas, P.~Pani, L.~Shao, and W.~Yan,
  ``{Constraints on millicharged dark matter and axionlike particles from
  timing of radio waves},''
  \href{http://dx.doi.org/10.1103/PhysRevD.100.063515}{{\em Phys. Rev. D}
  {\bfseries 100} no.~6, (2019) 063515},
  \href{http://arxiv.org/abs/1902.02695}{{\ttfamily arXiv:1902.02695
  [astro-ph.CO]}}.

\bibitem{Gupta:2021rod}
P.~K. Gupta, T.~F.~M. Spieksma, P.~T.~H. Pang, G.~Koekoek, and C.~V.~D. Broeck,
  ``{Bounding dark charges on binary black holes using gravitational waves},''
  \href{http://dx.doi.org/10.1103/PhysRevD.104.063041}{{\em Phys. Rev. D}
  {\bfseries 104} no.~6, (2021) 063041},
  \href{http://arxiv.org/abs/2107.12111}{{\ttfamily arXiv:2107.12111 [gr-qc]}}.

\bibitem{Carullo:2021oxn}
G.~Carullo, D.~Laghi, N.~K. Johnson-McDaniel, W.~Del~Pozzo, O.~J.~C. Dias,
  M.~Godazgar, and J.~E. Santos, ``{Constraints on Kerr-Newman black holes from
  merger-ringdown gravitational-wave observations},''
  \href{http://dx.doi.org/10.1103/PhysRevD.105.062009}{{\em Phys. Rev. D}
  {\bfseries 105} no.~6, (2022) 062009},
  \href{http://arxiv.org/abs/2109.13961}{{\ttfamily arXiv:2109.13961 [gr-qc]}}.

\bibitem{Fiorillo:2024upk}
D.~F.~G. Fiorillo and E.~Vitagliano, ``{Self-Interacting Dark Sectors in
  Supernovae Can Behave as a Relativistic Fluid},''
  \href{http://dx.doi.org/10.1103/PhysRevLett.133.251004}{{\em Phys. Rev.
  Lett.} {\bfseries 133} no.~25, (2024) 251004},
  \href{http://arxiv.org/abs/2404.07714}{{\ttfamily arXiv:2404.07714
  [hep-ph]}}.

\bibitem{Bah:2020ogh}
I.~Bah and P.~Heidmann, ``{Topological Stars and Black Holes},''
  \href{http://dx.doi.org/10.1103/PhysRevLett.126.151101}{{\em Phys. Rev.
  Lett.} {\bfseries 126} no.~15, (2021) 151101},
  \href{http://arxiv.org/abs/2011.08851}{{\ttfamily arXiv:2011.08851
  [hep-th]}}.

\bibitem{Bah:2021owp}
I.~Bah and P.~Heidmann, ``{Smooth bubbling geometries without supersymmetry},''
  \href{http://dx.doi.org/10.1007/JHEP09(2021)128}{{\em JHEP} {\bfseries 09}
  (2021) 128}, \href{http://arxiv.org/abs/2106.05118}{{\ttfamily
  arXiv:2106.05118 [hep-th]}}.

\bibitem{Leaver:1990zz}
E.~W. Leaver, ``{Quasinormal modes of Reissner-Nordstrom black holes},''
  \href{http://dx.doi.org/10.1103/PhysRevD.41.2986}{{\em Phys. Rev. D}
  {\bfseries 41} (1990) 2986--2997}.

\bibitem{Armitage:2005xq}
P.~J. Armitage and P.~Natarajan, ``{Eccentricity of supermassive black hole
  binaries coalescing from gas rich mergers},''
  \href{http://dx.doi.org/10.1086/497108}{{\em Astrophys. J.} {\bfseries 634}
  (2005) 921--928}, \href{http://arxiv.org/abs/astro-ph/0508493}{{\ttfamily
  arXiv:astro-ph/0508493}}.

\bibitem{Artymowicz:1994bw}
P.~Artymowicz and S.~H. Lubow, ``{Dynamics of binary-disk interaction. 1:
  Resonances and disk gap sizes},''
  \href{http://dx.doi.org/10.1086/173679}{{\em Astrophys. J.} {\bfseries 421}
  (1994) 651--667}.

\bibitem{Grobner:2020drr}
M.~Gr\"obner, W.~Ishibashi, S.~Tiwari, M.~Haney, and P.~Jetzer, ``{Binary black
  hole mergers in AGN accretion discs: gravitational wave rate density
  estimates},'' \href{http://dx.doi.org/10.1051/0004-6361/202037681}{{\em
  Astron. Astrophys.} {\bfseries 638} (2020) A119},
  \href{http://arxiv.org/abs/2005.03571}{{\ttfamily arXiv:2005.03571
  [astro-ph.GA]}}.

\bibitem{Farris:2014zjo}
B.~D. Farris, P.~Duffell, A.~I. MacFadyen, and Z.~Haiman, ``{Binary Black Hole
  Accretion During Inspiral and Merger},''
  \href{http://dx.doi.org/10.1093/mnrasl/slu184}{{\em Mon. Not. Roy. Astron.
  Soc.} {\bfseries 447} no.~1, (2015) L80--L84},
  \href{http://arxiv.org/abs/1409.5124}{{\ttfamily arXiv:1409.5124
  [astro-ph.HE]}}.

\bibitem{Ishibashi:2020zzy}
W.~Ishibashi and M.~Gr\"obner, ``{Evolution of binary black holes in AGN
  accretion discs: Disc-binary interaction and gravitational wave emission},''
  \href{http://dx.doi.org/10.1051/0004-6361/202037799}{{\em Astron. Astrophys.}
  {\bfseries 639} (2020) A108},
  \href{http://arxiv.org/abs/2006.07407}{{\ttfamily arXiv:2006.07407
  [astro-ph.GA]}}.

\bibitem{2013MNRAS.436.2997D}
D.~J. {D'Orazio}, Z.~{Haiman}, and A.~{MacFadyen}, ``{Accretion into the
  central cavity of a circumbinary disc},''
  \href{http://dx.doi.org/10.1093/mnras/stt1787}{{\em Mon. Not. Roy. Astron.
  Soc.} {\bfseries 436} no.~4, (Dec., 2013) 2997--3020},
  \href{http://arxiv.org/abs/1210.0536}{{\ttfamily arXiv:1210.0536
  [astro-ph.GA]}}.

\bibitem{2017AA...598A..43C}
H.~{Canovas}, A.~{Hardy}, A.~{Zurlo}, Z.~{Wahhaj}, M.~R. {Schreiber},
  A.~{Vigan}, E.~{Villaver}, J.~{Olofsson}, G.~{Meeus}, F.~{M\'{e}nard},
  C.~{Caceres}, L.~A. {Cieza}, and A.~{Garufi}, ``{Constraining the mass of the
  planet(s) sculpting a disk cavity. The intriguing case of 2MASS
  J16042165-2130284},''
  \href{http://dx.doi.org/10.1051/0004-6361/201629145}{{\em Astron. Astrophys.}
  {\bfseries 598} (Feb., 2017) A43},
  \href{http://arxiv.org/abs/1606.07087}{{\ttfamily arXiv:1606.07087
  [astro-ph.SR]}}.

\bibitem{Alonso-Monsalve:2023jfq}
E.~Alonso-Monsalve and D.~I. Kaiser, ``{Debye screening of non-Abelian plasmas
  in curved spacetimes},''
  \href{http://dx.doi.org/10.1103/PhysRevD.108.125010}{{\em Phys. Rev. D}
  {\bfseries 108} no.~12, (2023) 125010},
  \href{http://arxiv.org/abs/2309.15385}{{\ttfamily arXiv:2309.15385
  [hep-ph]}}.

\bibitem{Moncrief:1974ng}
V.~Moncrief, ``{Stability of Reissner-Nordstrom black holes},''
  \href{http://dx.doi.org/10.1103/PhysRevD.10.1057}{{\em Phys. Rev. D}
  {\bfseries 10} (1974) 1057--1059}.

\bibitem{PhysRevD.9.2707}
V.~Moncrief, ``Odd-parity stability of a reissner-nordstr\"om black hole,''
  \href{http://dx.doi.org/10.1103/PhysRevD.9.2707}{{\em Phys. Rev. D}
  {\bfseries 9} (May, 1974) 2707--2709}.

\bibitem{Moncrief:1975sb}
V.~Moncrief, ``{Gauge-invariant perturbations of Reissner-Nordstrom black
  holes},'' \href{http://dx.doi.org/10.1103/PhysRevD.12.1526}{{\em Phys. Rev.
  D} {\bfseries 12} (1975) 1526--1537}.

\bibitem{Lingetti:2022psy}
G.~Lingetti, E.~Cannizzaro, and P.~Pani, ``{Superradiant instabilities by
  accretion disks in scalar-tensor theories},''
  \href{http://dx.doi.org/10.1103/PhysRevD.106.024007}{{\em Phys. Rev. D}
  {\bfseries 106} no.~2, (2022) 024007},
  \href{http://arxiv.org/abs/2204.09335}{{\ttfamily arXiv:2204.09335 [gr-qc]}}.

\bibitem{Cardoso:2016rao}
V.~Cardoso, E.~Franzin, and P.~Pani, ``{Is the gravitational-wave ringdown a
  probe of the event horizon?},''
  \href{http://dx.doi.org/10.1103/PhysRevLett.117.089902,
  10.1103/PhysRevLett.116.171101}{{\em Phys. Rev. Lett.} {\bfseries 116}
  no.~17, (2016) 171101}, \href{http://arxiv.org/abs/1602.07309}{{\ttfamily
  arXiv:1602.07309 [gr-qc]}}.
[Erratum: Phys. Rev. Lett.117,no.8,089902(2016)].

\bibitem{Cardoso:2016oxy}
V.~Cardoso, S.~Hopper, C.~F.~B. Macedo, C.~Palenzuela, and P.~Pani,
  ``{Gravitational-wave signatures of exotic compact objects and of quantum
  corrections at the horizon scale},''
  \href{http://dx.doi.org/10.1103/PhysRevD.94.084031}{{\em Phys. Rev. D}
  {\bfseries 94} no.~8, (2016) 084031},
  \href{http://arxiv.org/abs/1608.08637}{{\ttfamily arXiv:1608.08637 [gr-qc]}}.

\bibitem{Oshita:2018fqu}
N.~Oshita and N.~Afshordi, ``{Probing microstructure of black hole spacetimes
  with gravitational wave echoes},''
  \href{http://dx.doi.org/10.1103/PhysRevD.99.044002}{{\em Phys. Rev. D}
  {\bfseries 99} no.~4, (2019) 044002},
  \href{http://arxiv.org/abs/1807.10287}{{\ttfamily arXiv:1807.10287 [gr-qc]}}.

\bibitem{Wang:2019rcf}
Q.~Wang, N.~Oshita, and N.~Afshordi, ``{Echoes from Quantum Black Holes},''
  \href{http://dx.doi.org/10.1103/PhysRevD.101.024031}{{\em Phys. Rev. D}
  {\bfseries 101} no.~2, (2020) 024031},
  \href{http://arxiv.org/abs/1905.00446}{{\ttfamily arXiv:1905.00446 [gr-qc]}}.

\bibitem{Ferrari:2000sr}
V.~Ferrari and K.~D. Kokkotas, ``{Scattering of particles by neutron stars:
  Time evolutions for axial perturbations},''
  \href{http://dx.doi.org/10.1103/PhysRevD.62.107504}{{\em Phys. Rev. D}
  {\bfseries 62} (2000) 107504},
  \href{http://arxiv.org/abs/gr-qc/0008057}{{\ttfamily arXiv:gr-qc/0008057}}.

\bibitem{Pani:2018flj}
P.~Pani and V.~Ferrari, ``{On gravitational-wave echoes from neutron-star
  binary coalescences},''
  \href{http://dx.doi.org/10.1088/1361-6382/aacb8f}{{\em Class. Quant. Grav.}
  {\bfseries 35} no.~15, (2018) 15LT01},
  \href{http://arxiv.org/abs/1804.01444}{{\ttfamily arXiv:1804.01444 [gr-qc]}}.

\bibitem{Buoninfante:2019swn}
L.~Buoninfante and A.~Mazumdar, ``{Nonlocal star as a blackhole mimicker},''
  \href{http://dx.doi.org/10.1103/PhysRevD.100.024031}{{\em Phys. Rev. D}
  {\bfseries 100} no.~2, (2019) 024031},
  \href{http://arxiv.org/abs/1903.01542}{{\ttfamily arXiv:1903.01542 [gr-qc]}}.

\bibitem{Buoninfante:2019teo}
L.~Buoninfante, A.~Mazumdar, and J.~Peng, ``{Nonlocality amplifies echoes},''
  \href{http://dx.doi.org/10.1103/PhysRevD.100.104059}{{\em Phys. Rev. D}
  {\bfseries 100} no.~10, (2019) 104059},
  \href{http://arxiv.org/abs/1906.03624}{{\ttfamily arXiv:1906.03624 [gr-qc]}}.

\bibitem{Delhom:2019btt}
A.~Delhom, C.~F.~B. Macedo, G.~J. Olmo, and L.~C.~B. Crispino, ``{Absorption by
  black hole remnants in metric-affine gravity},''
  \href{http://dx.doi.org/10.1103/PhysRevD.100.024016}{{\em Phys. Rev. D}
  {\bfseries 100} no.~2, (2019) 024016},
  \href{http://arxiv.org/abs/1906.06411}{{\ttfamily arXiv:1906.06411 [gr-qc]}}.

\bibitem{Zhang:2017jze}
J.~Zhang and S.-Y. Zhou, ``{Can the graviton have a large mass near black
  holes?},'' \href{http://dx.doi.org/10.1103/PhysRevD.97.081501}{{\em Phys.
  Rev. D} {\bfseries 97} no.~8, (2018) 081501},
  \href{http://arxiv.org/abs/1709.07503}{{\ttfamily arXiv:1709.07503 [gr-qc]}}.

\bibitem{Cardoso:2017cqb}
V.~Cardoso and P.~Pani, ``{Tests for the existence of black holes through
  gravitational wave echoes},''
  \href{http://dx.doi.org/10.1038/s41550-017-0225-y}{{\em Nat. Astron.}
  {\bfseries 1} no.~9, (2017) 586--591},
\href{http://arxiv.org/abs/1709.01525}{{\ttfamily arXiv:1709.01525 [gr-qc]}}.

\bibitem{Pereniguez:2023wxf}
D.~Pere\~niguez, ``{Black hole perturbations and electric-magnetic duality},''
  \href{http://dx.doi.org/10.1103/PhysRevD.108.084046}{{\em Phys. Rev. D}
  {\bfseries 108} no.~8, (2023) 084046},
  \href{http://arxiv.org/abs/2302.10942}{{\ttfamily arXiv:2302.10942 [gr-qc]}}.

\bibitem{Dyson:2023ujk}
C.~Dyson and D.~Pere\~niguez, ``{Magnetic black holes: From Thomson dipoles to
  the Penrose process and cosmic censorship},''
  \href{http://dx.doi.org/10.1103/PhysRevD.108.084064}{{\em Phys. Rev. D}
  {\bfseries 108} no.~8, (2023) 084064},
  \href{http://arxiv.org/abs/2306.15751}{{\ttfamily arXiv:2306.15751 [gr-qc]}}.

\bibitem{Nollert:1996rf}
H.-P. Nollert, ``{About the significance of quasinormal modes of black
  holes},'' \href{http://dx.doi.org/10.1103/PhysRevD.53.4397}{{\em Phys. Rev.
  D} {\bfseries 53} (1996) 4397--4402},
  \href{http://arxiv.org/abs/gr-qc/9602032}{{\ttfamily arXiv:gr-qc/9602032}}.

\bibitem{Nollert:1998ys}
H.-P. Nollert and R.~H. Price, ``{Quantifying excitations of quasinormal mode
  systems},'' \href{http://dx.doi.org/10.1063/1.532698}{{\em J. Math. Phys.}
  {\bfseries 40} (1999) 980--1010},
  \href{http://arxiv.org/abs/gr-qc/9810074}{{\ttfamily arXiv:gr-qc/9810074}}.

\bibitem{Leung:1997was}
P.~T. Leung, Y.~T. Liu, W.~M. Suen, C.~Y. Tam, and K.~Young, ``{Quasinormal
  modes of dirty black holes},''
  \href{http://dx.doi.org/10.1103/PhysRevLett.78.2894}{{\em Phys. Rev. Lett.}
  {\bfseries 78} (1997) 2894--2897},
  \href{http://arxiv.org/abs/gr-qc/9903031}{{\ttfamily arXiv:gr-qc/9903031}}.

\bibitem{Jaramillo:2021tmt}
J.~L. Jaramillo, R.~Panosso~Macedo, and L.~A. Sheikh, ``{Gravitational Wave
  Signatures of Black Hole Quasinormal Mode Instability},''
  \href{http://dx.doi.org/10.1103/PhysRevLett.128.211102}{{\em Phys. Rev.
  Lett.} {\bfseries 128} no.~21, (2022) 211102},
  \href{http://arxiv.org/abs/2105.03451}{{\ttfamily arXiv:2105.03451 [gr-qc]}}.

\bibitem{Jaramillo:2020tuu}
J.~L. Jaramillo, R.~Panosso~Macedo, and L.~Al~Sheikh, ``{Pseudospectrum and
  Black Hole Quasinormal Mode Instability},''
  \href{http://dx.doi.org/10.1103/PhysRevX.11.031003}{{\em Phys. Rev. X}
  {\bfseries 11} no.~3, (2021) 031003},
  \href{http://arxiv.org/abs/2004.06434}{{\ttfamily arXiv:2004.06434 [gr-qc]}}.

\bibitem{Cheung:2022rbm}
M.~H.-Y. Cheung {\em et~al.}, ``{Nonlinear Effects in Black Hole Ringdown},''
  \href{http://dx.doi.org/10.1103/PhysRevLett.130.081401}{{\em Phys. Rev.
  Lett.} {\bfseries 130} no.~8, (2023) 081401},
  \href{http://arxiv.org/abs/2208.07374}{{\ttfamily arXiv:2208.07374 [gr-qc]}}.

\bibitem{Konoplya:2022pbc}
R.~A. Konoplya and A.~Zhidenko, ``{First few overtones probe the event horizon
  geometry},'' \href{http://dx.doi.org/10.1016/j.jheap.2024.10.015}{{\em JHEAp}
  {\bfseries 44} (2024) 419--426},
  \href{http://arxiv.org/abs/2209.00679}{{\ttfamily arXiv:2209.00679 [gr-qc]}}.

\bibitem{Konoplya:2022hll}
R.~A. Konoplya, A.~F. Zinhailo, J.~Kunz, Z.~Stuchlik, and A.~Zhidenko,
  ``{Quasinormal ringing of regular black holes in asymptotically safe gravity:
  the importance of overtones},''
  \href{http://dx.doi.org/10.1088/1475-7516/2022/10/091}{{\em JCAP} {\bfseries
  10} (2022) 091}, \href{http://arxiv.org/abs/2206.14714}{{\ttfamily
  arXiv:2206.14714 [gr-qc]}}.

\bibitem{Cardoso:2024mrw}
V.~Cardoso, S.~Kastha, and R.~Panosso~Macedo, ``{Physical significance of the
  black hole quasinormal mode spectra instability},''
  \href{http://dx.doi.org/10.1103/PhysRevD.110.024016}{{\em Phys. Rev. D}
  {\bfseries 110} no.~2, (2024) 024016},
  \href{http://arxiv.org/abs/2404.01374}{{\ttfamily arXiv:2404.01374 [gr-qc]}}.

\bibitem{Yang:2024vor}
Y.~Yang, Z.-F. Mai, R.-Q. Yang, L.~Shao, and E.~Berti, ``{Spectral instability
  of black holes: Relating the frequency domain to the time domain},''
  \href{http://dx.doi.org/10.1103/PhysRevD.110.084018}{{\em Phys. Rev. D}
  {\bfseries 110} no.~8, (2024) 084018},
  \href{http://arxiv.org/abs/2407.20131}{{\ttfamily arXiv:2407.20131 [gr-qc]}}.

\bibitem{Ianniccari:2024ysv}
A.~Ianniccari, A.~J. Iovino, A.~Kehagias, P.~Pani, G.~Perna, D.~Perrone, and
  A.~Riotto, ``{Deciphering the Instability of the Black Hole Ringdown
  Quasinormal Spectrum},''
  \href{http://dx.doi.org/10.1103/PhysRevLett.133.211401}{{\em Phys. Rev.
  Lett.} {\bfseries 133} no.~21, (2024) 211401},
  \href{http://arxiv.org/abs/2407.20144}{{\ttfamily arXiv:2407.20144 [gr-qc]}}.

\bibitem{Berti:2022xfj}
E.~Berti, V.~Cardoso, M.~H.-Y. Cheung, F.~Di~Filippo, F.~Duque, P.~Martens, and
  S.~Mukohyama, ``{Stability of the fundamental quasinormal mode in time-domain
  observations against small perturbations},''
  \href{http://dx.doi.org/10.1103/PhysRevD.106.084011}{{\em Phys. Rev. D}
  {\bfseries 106} no.~8, (2022) 084011},
  \href{http://arxiv.org/abs/2205.08547}{{\ttfamily arXiv:2205.08547 [gr-qc]}}.

\bibitem{Leong:2023nuk}
S.~H.~W. Leong, J.~Calder\'on~Bustillo, M.~Gracia-Linares, and P.~Laguna,
  ``{Detectability of dense-environment effects on black-hole mergers: The
  scalar field case, higher-order ringdown modes, and parameter biases},''
  \href{http://dx.doi.org/10.1103/PhysRevD.108.124079}{{\em Phys. Rev. D}
  {\bfseries 108} no.~12, (2023) 124079},
  \href{http://arxiv.org/abs/2308.03250}{{\ttfamily arXiv:2308.03250 [gr-qc]}}.

\bibitem{Redondo-Yuste:2023ipg}
J.~Redondo-Yuste, D.~Pere\~niguez, and V.~Cardoso, ``{Ringdown of a dynamical
  spacetime},'' \href{http://dx.doi.org/10.1103/PhysRevD.109.044048}{{\em Phys.
  Rev. D} {\bfseries 109} no.~4, (2024) 044048},
  \href{http://arxiv.org/abs/2312.04633}{{\ttfamily arXiv:2312.04633 [gr-qc]}}.

\bibitem{DeLuca:2024uju}
V.~De~Luca, G.~Franciolini, and A.~Riotto, ``{Flea on the elephant: Tidal Love
  numbers in subsolar primordial black hole searches},''
  \href{http://dx.doi.org/10.1103/PhysRevD.110.104041}{{\em Phys. Rev. D}
  {\bfseries 110} no.~10, (2024) 104041},
  \href{http://arxiv.org/abs/2408.14207}{{\ttfamily arXiv:2408.14207 [gr-qc]}}.

\bibitem{Quinlan:1994ed}
G.~D. Quinlan, L.~Hernquist, and S.~Sigurdsson, ``{Models of Galaxies with
  Central Black Holes: Adiabatic Growth in Spherical Galaxies},''
  \href{http://dx.doi.org/10.1086/175295}{{\em Astrophys. J.} {\bfseries 440}
  (1995) 554--564}, \href{http://arxiv.org/abs/astro-ph/9407005}{{\ttfamily
  arXiv:astro-ph/9407005}}.

\bibitem{Schoepe:2017cvt}
A.~Schoepe, D.~Hilditch, and M.~Bugner, ``{Revisiting Hyperbolicity of
  Relativistic Fluids},''
  \href{http://dx.doi.org/10.1103/PhysRevD.97.123009}{{\em Phys. Rev. D}
  {\bfseries 97} no.~12, (2018) 123009},
  \href{http://arxiv.org/abs/1712.09837}{{\ttfamily arXiv:1712.09837 [gr-qc]}}.

\bibitem{Datta:2023zmd}
S.~Datta, ``{Black holes immersed in dark matter: Energy condition and sound
  speed},'' \href{http://dx.doi.org/10.1103/PhysRevD.109.104042}{{\em Phys.
  Rev. D} {\bfseries 109} no.~10, (2024) 104042},
  \href{http://arxiv.org/abs/2312.01277}{{\ttfamily arXiv:2312.01277 [gr-qc]}}.

\bibitem{Albanesi:2023bgi}
S.~Albanesi, S.~Bernuzzi, T.~Damour, A.~Nagar, and A.~Placidi, ``{Faithful
  effective-one-body waveform of small-mass-ratio coalescing black hole
  binaries: The eccentric, nonspinning case},''
  \href{http://dx.doi.org/10.1103/PhysRevD.108.084037}{{\em Phys. Rev. D}
  {\bfseries 108} no.~8, (2023) 084037},
  \href{http://arxiv.org/abs/2305.19336}{{\ttfamily arXiv:2305.19336 [gr-qc]}}.

\bibitem{DeAmicis:2024not}
M.~De~Amicis, S.~Albanesi, and G.~Carullo, ``{Inspiral-inherited ringdown
  tails},'' \href{http://dx.doi.org/10.1103/PhysRevD.110.104005}{{\em Phys.
  Rev. D} {\bfseries 110} no.~10, (2024) 104005},
  \href{http://arxiv.org/abs/2406.17018}{{\ttfamily arXiv:2406.17018 [gr-qc]}}.

\bibitem{Islam:2024vro}
T.~Islam, G.~Faggioli, G.~Khanna, S.~E. Field, M.~van~de Meent, and
  A.~Buonanno, ``{Phenomenology and origin of late-time tails in eccentric
  binary black hole mergers},''
  \href{http://arxiv.org/abs/2407.04682}{{\ttfamily arXiv:2407.04682 [gr-qc]}}.

\bibitem{Robson:2018ifk}
T.~Robson, N.~J. Cornish, and C.~Liu, ``{The construction and use of LISA
  sensitivity curves},'' \href{http://dx.doi.org/10.1088/1361-6382/ab1101}{{\em
  Class. Quant. Grav.} {\bfseries 36} no.~10, (2019) 105011},
  \href{http://arxiv.org/abs/1803.01944}{{\ttfamily arXiv:1803.01944
  [astro-ph.HE]}}.

\bibitem{Owen:1995tm}
B.~J. Owen, ``{Search templates for gravitational waves from inspiraling
  binaries: Choice of template spacing},''
  \href{http://dx.doi.org/10.1103/PhysRevD.53.6749}{{\em Phys. Rev. D}
  {\bfseries 53} (1996) 6749--6761},
  \href{http://arxiv.org/abs/gr-qc/9511032}{{\ttfamily arXiv:gr-qc/9511032}}.

\bibitem{Flanagan:1997kp}
E.~E. Flanagan and S.~A. Hughes, ``{Measuring gravitational waves from binary
  black hole coalescences: 2. The Waves' information and its extraction, with
  and without templates},''
  \href{http://dx.doi.org/10.1103/PhysRevD.57.4566}{{\em Phys. Rev. D}
  {\bfseries 57} (1998) 4566--4587},
  \href{http://arxiv.org/abs/gr-qc/9710129}{{\ttfamily arXiv:gr-qc/9710129}}.

\bibitem{Lindblom:2008cm}
L.~Lindblom, B.~J. Owen, and D.~A. Brown, ``{Model Waveform Accuracy Standards
  for Gravitational Wave Data Analysis},''
  \href{http://dx.doi.org/10.1103/PhysRevD.78.124020}{{\em Phys. Rev. D}
  {\bfseries 78} (2008) 124020},
\href{http://arxiv.org/abs/0809.3844}{{\ttfamily arXiv:0809.3844 [gr-qc]}}.

\bibitem{McWilliams:2010eq}
S.~T. McWilliams, B.~J. Kelly, and J.~G. Baker, ``{Observing mergers of
  non-spinning black-hole binaries},''
  \href{http://dx.doi.org/10.1103/PhysRevD.82.024014}{{\em Phys. Rev. D}
  {\bfseries 82} (2010) 024014},
  \href{http://arxiv.org/abs/1004.0961}{{\ttfamily arXiv:1004.0961 [gr-qc]}}.

\bibitem{Chatziioannou:2017tdw}
K.~Chatziioannou, A.~Klein, N.~Yunes, and N.~Cornish, ``{Constructing
  Gravitational Waves from Generic Spin-Precessing Compact Binary Inspirals},''
  \href{http://dx.doi.org/10.1103/PhysRevD.95.104004}{{\em Phys. Rev. D}
  {\bfseries 95} no.~10, (2017) 104004},
  \href{http://arxiv.org/abs/1703.03967}{{\ttfamily arXiv:1703.03967 [gr-qc]}}.

\bibitem{Pezzella:2024tkf}
L.~Pezzella, K.~Destounis, A.~Maselli, and V.~Cardoso, ``{Quasinormal modes of
  black holes embedded in halos of matter},''
  \href{http://dx.doi.org/10.1103/PhysRevD.111.064026}{{\em Phys. Rev. D}
  {\bfseries 111} no.~6, (2025) 064026},
  \href{http://arxiv.org/abs/2412.18651}{{\ttfamily arXiv:2412.18651 [gr-qc]}}.

\bibitem{Berti:2007fi}
E.~Berti, V.~Cardoso, J.~A. Gonzalez, U.~Sperhake, M.~Hannam, S.~Husa, and
  B.~Bruegmann, ``{Inspiral, merger and ringdown of unequal mass black hole
  binaries: A Multipolar analysis},''
  \href{http://dx.doi.org/10.1103/PhysRevD.76.064034}{{\em Phys. Rev. D}
  {\bfseries 76} (2007) 064034},
  \href{http://arxiv.org/abs/gr-qc/0703053}{{\ttfamily arXiv:gr-qc/0703053}}.

\bibitem{2019A&A...625L..10G}
{\bfseries GRAVITY} Collaboration, R.~Abuter {\em et~al.}, ``{A geometric
  distance measurement to the Galactic center black hole with 0.3\%
  uncertainty},'' \href{http://dx.doi.org/10.1051/0004-6361/201935656}{{\em
  Astron. Astrophys.} {\bfseries 625} (May, 2019) L10},
  \href{http://arxiv.org/abs/1904.05721}{{\ttfamily arXiv:1904.05721
  [astro-ph.GA]}}.

\bibitem{GRAVITY:2021xju}
{\bfseries GRAVITY} Collaboration, R.~Abuter {\em et~al.}, ``{Mass distribution
  in the Galactic Center based on interferometric astrometry of multiple
  stellar orbits},'' \href{http://dx.doi.org/10.1051/0004-6361/202142465}{{\em
  Astron. Astrophys.} {\bfseries 657} (2022) L12},
  \href{http://arxiv.org/abs/2112.07478}{{\ttfamily arXiv:2112.07478
  [astro-ph.GA]}}.

\bibitem{Bhagwat:2021kwv}
S.~Bhagwat, C.~Pacilio, E.~Barausse, and P.~Pani, ``{Landscape of massive
  black-hole spectroscopy with LISA and the Einstein Telescope},''
  \href{http://dx.doi.org/10.1103/PhysRevD.105.124063}{{\em Phys. Rev. D}
  {\bfseries 105} no.~12, (2022) 124063},
  \href{http://arxiv.org/abs/2201.00023}{{\ttfamily arXiv:2201.00023 [gr-qc]}}.

\bibitem{2016A&A...587L...6V}
E.~{Valenti}, M.~{Zoccali}, O.~A. {Gonzalez}, D.~{Minniti},
  J.~{Alonso-Garc{\'\i}a}, E.~{Marchetti}, M.~{Hempel}, A.~{Renzini}, and
  M.~{Rejkuba}, ``{Stellar density profile and mass of the Milky Way bulge from
  VVV data},'' \href{http://dx.doi.org/10.1051/0004-6361/201527500}{{\em
  Astron. Astrophys.} {\bfseries 587} (Mar., 2016) L6},
  \href{http://arxiv.org/abs/1510.07425}{{\ttfamily arXiv:1510.07425
  [astro-ph.GA]}}.

\bibitem{Ansorg:2016ztf}
M.~Ansorg and R.~Panosso~Macedo, ``{Spectral decomposition of black-hole
  perturbations on hyperboloidal slices},''
  \href{http://dx.doi.org/10.1103/PhysRevD.93.124016}{{\em Phys. Rev. D}
  {\bfseries 93} no.~12, (2016) 124016},
  \href{http://arxiv.org/abs/1604.02261}{{\ttfamily arXiv:1604.02261 [gr-qc]}}.

\bibitem{Berti:2016lat}
E.~Berti, A.~Sesana, E.~Barausse, V.~Cardoso, and K.~Belczynski,
  ``{Spectroscopy of Kerr black holes with Earth- and space-based
  interferometers},''
  \href{http://dx.doi.org/10.1103/PhysRevLett.117.101102}{{\em Phys. Rev.
  Lett.} {\bfseries 117} no.~10, (2016) 101102},
\href{http://arxiv.org/abs/1605.09286}{{\ttfamily arXiv:1605.09286 [gr-qc]}}.

\bibitem{Seoane:2021kkk}
P.~A. Seoane {\em et~al.}, ``{The effect of mission duration on LISA science
  objectives},'' \href{http://dx.doi.org/10.1007/s10714-021-02889-x}{{\em Gen.
  Rel. Grav.} {\bfseries 54} no.~1, (2022) 3},
  \href{http://arxiv.org/abs/2107.09665}{{\ttfamily arXiv:2107.09665
  [astro-ph.IM]}}.

\bibitem{PhysRev.185.140}
E.~W. Smith, J.~Cooper, and C.~R. Vidal, ``Unified classical-path treatment of
  stark broadening in plasmas,''
  \href{http://dx.doi.org/10.1103/PhysRev.185.140}{{\em Phys. Rev.} {\bfseries
  185} (Sep, 1969) 140--151}.

\bibitem{PhysRevA.6.1132}
W.~L. Wiese, D.~E. Kelleher, and D.~R. Paquette, ``Detailed study of the stark
  broadening of balmer lines in a high-density plasma,''
  \href{http://dx.doi.org/10.1103/PhysRevA.6.1132}{{\em Phys. Rev. A}
  {\bfseries 6} (Sep, 1972) 1132--1153}.

\bibitem{PhysRevA.11.1854}
W.~L. Wiese, D.~E. Kelleher, and V.~Helbig, ``Variations in balmer-line stark
  profiles with atom-ion reduced mass,''
  \href{http://dx.doi.org/10.1103/PhysRevA.11.1854}{{\em Phys. Rev. A}
  {\bfseries 11} (Jun, 1975) 1854--1864}.

\bibitem{LISA:2022kgy}
{\bfseries LISA} Collaboration, K.~G. Arun {\em et~al.}, ``{New horizons for
  fundamental physics with LISA},''
  \href{http://dx.doi.org/10.1007/s41114-022-00036-9}{{\em Living Rev. Rel.}
  {\bfseries 25} no.~1, (2022) 4},
  \href{http://arxiv.org/abs/2205.01597}{{\ttfamily arXiv:2205.01597 [gr-qc]}}.

\bibitem{Ding:2020bnl}
Q.~Ding, X.~Tong, and Y.~Wang, ``{Gravitational Collider Physics via
  Pulsar-Black Hole Binaries},''
  \href{http://dx.doi.org/10.3847/1538-4357/abd803}{{\em Astrophys. J.}
  {\bfseries 908} no.~1, (2021) 78},
  \href{http://arxiv.org/abs/2009.11106}{{\ttfamily arXiv:2009.11106
  [astro-ph.HE]}}.

\bibitem{Tong:2021whq}
X.~Tong, Y.~Wang, and H.-Y. Zhu, ``{Gravitational Collider Physics via
  Pulsar\textendash{}Black Hole Binaries II: Fine and Hyperfine Structures Are
  Favored},'' \href{http://dx.doi.org/10.3847/1538-4357/ac36db}{{\em Astrophys.
  J.} {\bfseries 924} no.~2, (2022) 99},
  \href{http://arxiv.org/abs/2106.13484}{{\ttfamily arXiv:2106.13484
  [astro-ph.HE]}}.

\bibitem{Du:2022trq}
P.~Du, D.~Egana-Ugrinovic, R.~Essig, G.~Fragione, and R.~Perna, ``{Searching
  for ultra-light bosons and constraining black hole spin distributions with
  stellar tidal disruption events},''
  \href{http://dx.doi.org/10.1038/s41467-022-32301-4}{{\em Nature Commun.}
  {\bfseries 13} no.~1, (2022) 4626},
  \href{http://arxiv.org/abs/2202.01215}{{\ttfamily arXiv:2202.01215
  [hep-ph]}}.

\bibitem{Berti:2019wnn}
E.~Berti, R.~Brito, C.~F.~B. Macedo, G.~Raposo, and J.~L. Rosa, ``{Ultralight
  boson cloud depletion in binary systems},''
  \href{http://dx.doi.org/10.1103/PhysRevD.99.104039}{{\em Phys. Rev. D}
  {\bfseries 99} no.~10, (2019) 104039},
  \href{http://arxiv.org/abs/1904.03131}{{\ttfamily arXiv:1904.03131 [gr-qc]}}.

\bibitem{zener1932non}
C.~Zener, ``{Non-Adiabatic Crossing of Energy Levels},'' {\em Proceedings of
  the Royal Society of London} {\bfseries 137} no.~833, (1932) 696--702.

\bibitem{landau1932theorie}
L.~Landau, ``{Zur Theorie der Energie\"ubertragung},'' {\em Z. Sowjetunion}
  {\bfseries 2} (1932) 46--51.

\bibitem{wigner}
E.~P. Wigner, {\em Group Theory and Its Application to the Quantum Mechanics of
  Atomic Spectra}.
\newblock Academic Press, New York, 1959.

\bibitem{Leaver:1985ax}
E.~Leaver, ``{An Analytic Representation for the Quasi-Normal Modes of Kerr
  Black Holes},''
\href{http://dx.doi.org/10.1098/rspa.1985.0119}{{\em Proc. Roy. Soc. Lond.}
  {\bfseries A402} (1985) 285--298}.

\bibitem{Amaro-Seoane:2012lgq}
P.~Amaro-Seoane, ``{Relativistic dynamics and extreme mass ratio inspirals},''
  \href{http://dx.doi.org/10.1007/s41114-018-0013-8}{{\em Living Rev. Rel.}
  {\bfseries 21} (2018) 4}, \href{http://arxiv.org/abs/1205.5240}{{\ttfamily
  arXiv:1205.5240 [astro-ph.CO]}}.

\bibitem{LISA:2022yao}
{\bfseries LISA} Collaboration, P.~A. Seoane {\em et~al.}, ``{Astrophysics with
  the Laser Interferometer Space Antenna},''
  \href{http://dx.doi.org/10.1007/s41114-022-00041-y}{{\em Living Rev. Rel.}
  {\bfseries 26} no.~1, (2023) 2},
  \href{http://arxiv.org/abs/2203.06016}{{\ttfamily arXiv:2203.06016 [gr-qc]}}.

\bibitem{Mckernan:2017ssq}
B.~Mckernan {\em et~al.}, ``{Constraining Stellar-mass Black Hole Mergers in
  AGN Disks Detectable with LIGO},''
  \href{http://dx.doi.org/10.3847/1538-4357/aadae5}{{\em Astrophys. J.}
  {\bfseries 866} no.~1, (2018) 66},
  \href{http://arxiv.org/abs/1702.07818}{{\ttfamily arXiv:1702.07818
  [astro-ph.HE]}}.

\bibitem{Levin:2006uc}
Y.~Levin, ``{Starbursts near supermassive black holes: young stars in the
  Galactic Center, and gravitational waves in LISA band},''
  \href{http://dx.doi.org/10.1111/j.1365-2966.2006.11155.x}{{\em Mon. Not. Roy.
  Astron. Soc.} {\bfseries 374} (2007) 515--524},
  \href{http://arxiv.org/abs/astro-ph/0603583}{{\ttfamily
  arXiv:astro-ph/0603583}}.

\bibitem{Tong:2022bbl}
X.~Tong, Y.~Wang, and H.-Y. Zhu, ``{Termination of superradiance from a binary
  companion},'' \href{http://dx.doi.org/10.1103/PhysRevD.106.043002}{{\em Phys.
  Rev. D} {\bfseries 106} no.~4, (2022) 043002},
  \href{http://arxiv.org/abs/2205.10527}{{\ttfamily arXiv:2205.10527 [gr-qc]}}.

\bibitem{Fan:2023jjj}
K.~Fan, X.~Tong, Y.~Wang, and H.-Y. Zhu, ``{Modulating binary dynamics via the
  termination of black hole superradiance},''
  \href{http://dx.doi.org/10.1103/PhysRevD.109.024059}{{\em Phys. Rev. D}
  {\bfseries 109} no.~2, (2024) 024059},
  \href{http://arxiv.org/abs/2311.17013}{{\ttfamily arXiv:2311.17013 [gr-qc]}}.

\bibitem{Goodman:2002gv}
J.~Goodman, ``{Selfgravity and QSO disks},''
  \href{http://dx.doi.org/10.1046/j.1365-8711.2003.06241.x}{{\em Mon. Not. Roy.
  Astron. Soc.} {\bfseries 339} (2003) 937},
  \href{http://arxiv.org/abs/astro-ph/0201001}{{\ttfamily
  arXiv:astro-ph/0201001}}.

\bibitem{Goodman:2003sf}
J.~Goodman and J.~C. Tan, ``{Supermassive stars in quasar disks},''
  \href{http://dx.doi.org/10.1086/386360}{{\em Astrophys. J.} {\bfseries 608}
  (2004) 108--118}, \href{http://arxiv.org/abs/astro-ph/0307361}{{\ttfamily
  arXiv:astro-ph/0307361}}.

\bibitem{Amaro-Seoane:2012jcd}
P.~Amaro-Seoane, C.~F. Sopuerta, and M.~D. Freitag, ``{The role of the
  supermassive black hole spin in the estimation of the EMRI event rate},''
  \href{http://dx.doi.org/10.1093/mnras/sts572}{{\em Mon. Not. Roy. Astron.
  Soc.} {\bfseries 429} no.~4, (2013) 3155--3165},
  \href{http://arxiv.org/abs/1205.4713}{{\ttfamily arXiv:1205.4713
  [astro-ph.CO]}}.

\bibitem{Babak:2017tow}
S.~Babak, J.~Gair, A.~Sesana, E.~Barausse, C.~F. Sopuerta, C.~P.~L. Berry,
  E.~Berti, P.~Amaro-Seoane, A.~Petiteau, and A.~Klein, ``{Science with the
  space-based interferometer LISA. V: Extreme mass-ratio inspirals},''
  \href{http://dx.doi.org/10.1103/PhysRevD.95.103012}{{\em Phys. Rev. D}
  {\bfseries 95} no.~10, (2017) 103012},
  \href{http://arxiv.org/abs/1703.09722}{{\ttfamily arXiv:1703.09722 [gr-qc]}}.

\bibitem{Hannuksela:2018izj}
O.~Hannuksela, K.~Wong, R.~Brito, E.~Berti, and T.~Li, ``{Probing the Existence
  of Ultralight Bosons with a Single Gravitational-Wave Measurement},''
  \href{http://dx.doi.org/10.1038/s41550-019-0712-4}{{\em Nat. Astron.}
  {\bfseries 3} (2019) 447--451},
\href{http://arxiv.org/abs/1804.09659}{{\ttfamily arXiv:1804.09659
  [astro-ph.HE]}}.

\bibitem{Cao:2023fyv}
Y.~Cao and Y.~Tang, ``{Signatures of ultralight bosons in compact binary
  inspiral and outspiral},''
  \href{http://dx.doi.org/10.1103/PhysRevD.108.123017}{{\em Phys. Rev. D}
  {\bfseries 108} no.~12, (2023) 123017},
  \href{http://arxiv.org/abs/2307.05181}{{\ttfamily arXiv:2307.05181 [gr-qc]}}.

\bibitem{Mathisson:1937zz}
M.~Mathisson, ``{Neue mechanik materieller systemes},'' {\em Acta Phys. Polon.}
  {\bfseries 6} (1937) 163--200.

\bibitem{Dixon:2015vxa}
W.~G. Dixon, ``{The New Mechanics of Myron Mathisson and Its Subsequent
  Development},'' \href{http://dx.doi.org/10.1007/978-3-319-18335-0_1}{{\em
  Fund. Theor. Phys.} {\bfseries 179} (2015) 1--66}.

\bibitem{Dolan:2023enf}
S.~R. Dolan, L.~Durkan, C.~Kavanagh, and B.~Wardell, ``{Metric perturbations of
  Kerr spacetime in Lorenz gauge: circular equatorial orbits},''
  \href{http://dx.doi.org/10.1088/1361-6382/ad52e3}{{\em Class. Quant. Grav.}
  {\bfseries 41} no.~15, (2024) 155011},
  \href{http://arxiv.org/abs/2306.16459}{{\ttfamily arXiv:2306.16459 [gr-qc]}}.

\bibitem{Dolan:2021ijg}
S.~R. Dolan, C.~Kavanagh, and B.~Wardell, ``{Gravitational Perturbations of
  Rotating Black Holes in Lorenz Gauge},''
  \href{http://dx.doi.org/10.1103/PhysRevLett.128.151101}{{\em Phys. Rev.
  Lett.} {\bfseries 128} no.~15, (2022) 151101},
  \href{http://arxiv.org/abs/2108.06344}{{\ttfamily arXiv:2108.06344 [gr-qc]}}.

\bibitem{1999ApJ...513..252O}
E.~C. {Ostriker}, ``{Dynamical Friction in a Gaseous Medium},''
  \href{http://dx.doi.org/10.1086/306858}{{\em Astrophys. J.} {\bfseries 513}
  no.~1, (Mar., 1999) 252--258},
  \href{http://arxiv.org/abs/astro-ph/9810324}{{\ttfamily
  arXiv:astro-ph/9810324 [astro-ph]}}.

\bibitem{2007MNRAS.382..826B}
E.~{Barausse}, ``{Relativistic dynamical friction in a collisional fluid},''
  \href{http://dx.doi.org/10.1111/j.1365-2966.2007.12408.x}{{\em Mon. Not. Roy.
  Astron. Soc.} {\bfseries 382} no.~2, (Dec., 2007) 826--834},
  \href{http://arxiv.org/abs/0709.0211}{{\ttfamily arXiv:0709.0211
  [astro-ph]}}.

\bibitem{Chua:2020stf}
A.~J.~K. Chua, M.~L. Katz, N.~Warburton, and S.~A. Hughes, ``{Rapid generation
  of fully relativistic extreme-mass-ratio-inspiral waveform templates for LISA
  data analysis},''
  \href{http://dx.doi.org/10.1103/PhysRevLett.126.051102}{{\em Phys. Rev.
  Lett.} {\bfseries 126} no.~5, (2021) 051102},
  \href{http://arxiv.org/abs/2008.06071}{{\ttfamily arXiv:2008.06071 [gr-qc]}}.

\bibitem{Katz:2021yft}
M.~L. Katz, A.~J.~K. Chua, L.~Speri, N.~Warburton, and S.~A. Hughes, ``{Fast
  extreme-mass-ratio-inspiral waveforms: New tools for millihertz
  gravitational-wave data analysis},''
  \href{http://dx.doi.org/10.1103/PhysRevD.104.064047}{{\em Phys. Rev. D}
  {\bfseries 104} no.~6, (2021) 064047},
  \href{http://arxiv.org/abs/2104.04582}{{\ttfamily arXiv:2104.04582 [gr-qc]}}.

\bibitem{Hughes:2021exa}
S.~A. Hughes, N.~Warburton, G.~Khanna, A.~J.~K. Chua, and M.~L. Katz,
  ``{Adiabatic waveforms for extreme mass-ratio inspirals via multivoice
  decomposition in time and frequency},''
  \href{http://dx.doi.org/10.1103/PhysRevD.103.104014}{{\em Phys. Rev. D}
  {\bfseries 103} no.~10, (2021) 104014},
  \href{http://arxiv.org/abs/2102.02713}{{\ttfamily arXiv:2102.02713 [gr-qc]}}.
  [Erratum: Phys.Rev.D 107, 089901 (2023)].

\bibitem{Redondo-Yuste:2023snb}
J.~Redondo-Yuste, V.~Cardoso, C.~F.~B. Macedo, and M.~van~de Meent, ``{Eternal
  binaries},'' \href{http://dx.doi.org/10.1103/PhysRevD.107.124025}{{\em Phys.
  Rev. D} {\bfseries 107} no.~12, (2023) 124025},
  \href{http://arxiv.org/abs/2304.02039}{{\ttfamily arXiv:2304.02039 [gr-qc]}}.

\bibitem{Dosopoulou:2023umg}
F.~Dosopoulou, ``{Dynamical friction in dark matter spikes: Corrections to
  Chandrasekhar\textquoteright{}s formula},''
  \href{http://dx.doi.org/10.1103/PhysRevD.110.083027}{{\em Phys. Rev. D}
  {\bfseries 110} no.~8, (2024) 083027},
  \href{http://arxiv.org/abs/2305.17281}{{\ttfamily arXiv:2305.17281
  [astro-ph.HE]}}.

\bibitem{Cardoso:2019rou}
V.~Cardoso and A.~Maselli, ``{Constraints on the astrophysical environment of
  binaries with gravitational-wave observations},''
  \href{http://dx.doi.org/10.1051/0004-6361/202037654}{{\em Astron. Astrophys.}
  {\bfseries 644} (2020) A147},
  \href{http://arxiv.org/abs/1909.05870}{{\ttfamily arXiv:1909.05870
  [astro-ph.HE]}}.

\bibitem{Miller:2020bft}
J.~Miller and A.~Pound, ``{Two-timescale evolution of extreme-mass-ratio
  inspirals: waveform generation scheme for quasicircular orbits in
  Schwarzschild spacetime},''
  \href{http://dx.doi.org/10.1103/PhysRevD.103.064048}{{\em Phys. Rev. D}
  {\bfseries 103} no.~6, (2021) 064048},
  \href{http://arxiv.org/abs/2006.11263}{{\ttfamily arXiv:2006.11263 [gr-qc]}}.

\bibitem{Gair:2017ynp}
J.~R. Gair, S.~Babak, A.~Sesana, P.~Amaro-Seoane, E.~Barausse, C.~P.~L. Berry,
  E.~Berti, and C.~Sopuerta, ``{Prospects for observing extreme-mass-ratio
  inspirals with LISA},''
  \href{http://dx.doi.org/10.1088/1742-6596/840/1/012021}{{\em J. Phys. Conf.
  Ser.} {\bfseries 840} no.~1, (2017) 012021},
  \href{http://arxiv.org/abs/1704.00009}{{\ttfamily arXiv:1704.00009
  [astro-ph.GA]}}.

\bibitem{Wang:2022obu}
M.~Wang, Y.~Ma, and Q.~Wu, ``{Accretion-modified stellar-mass black hole
  distribution and milli-Hz gravitational wave backgrounds from galaxy
  centre},'' \href{http://dx.doi.org/10.1093/mnras/stad422}{{\em Mon. Not. Roy.
  Astron. Soc.} {\bfseries 520} no.~3, (2023) 4502--4516},
  \href{http://arxiv.org/abs/2212.05724}{{\ttfamily arXiv:2212.05724
  [astro-ph.HE]}}.

\bibitem{Nasim:2022rvl}
S.~S. Nasim, G.~Fabj, F.~Caban, A.~Secunda, K.~E.~S. Ford, B.~McKernan, J.~M.
  Bellovary, N.~W.~C. Leigh, and W.~Lyra, ``{Aligning Retrograde Nuclear
  Cluster Orbits with an Active Galactic Nucleus Accretion Disc},''
  \href{http://dx.doi.org/10.1093/mnras/stad1295}{{\em Mon. Not. Roy. Astron.
  Soc.} {\bfseries 522} no.~4, (2023) 5393--5401},
  \href{http://arxiv.org/abs/2207.09540}{{\ttfamily arXiv:2207.09540
  [astro-ph.GA]}}.

\bibitem{2023MNRAS.522.1763G}
A.~{Generozov} and H.~B. {Perets}, ``{Capture of stars into gaseous discs
  around massive black holes: alignment, circularization, and growth},''
  \href{http://dx.doi.org/10.1093/mnras/stad1016}{{\em Mon. Not. Roy. Astron.
  Soc.} {\bfseries 522} no.~2, (June, 2023) 1763--1778},
  \href{http://arxiv.org/abs/2212.11301}{{\ttfamily arXiv:2212.11301
  [astro-ph.GA]}}.

\bibitem{2024MNRAS.528.4958W}
Y.~{Wang}, Z.~{Zhu}, and D.~N.~C. {Lin}, ``{Stellar/BH population in AGN discs:
  direct binary formation from capture objects in nuclei clusters},''
  \href{http://dx.doi.org/10.1093/mnras/stae321}{{\em Mon. Not. Roy. Astron.
  Soc.} {\bfseries 528} no.~3, (Mar., 2024) 4958--4975},
  \href{http://arxiv.org/abs/2308.09129}{{\ttfamily arXiv:2308.09129
  [astro-ph.GA]}}.

\bibitem{1979ApJ...233..857G}
P.~{Goldreich} and S.~{Tremaine}, ``{The excitation of density waves at the
  Lindblad and corotation resonances by an external potential.},''
  \href{http://dx.doi.org/10.1086/157448}{{\em Astrophys. J.} {\bfseries 233}
  (Nov., 1979) 857--871}.

\bibitem{2004ApJ...602..388T}
H.~{Tanaka} and W.~R. {Ward}, ``{Three-dimensional Interaction between a Planet
  and an Isothermal Gaseous Disk. II. Eccentricity Waves and Bending Waves},''
  \href{http://dx.doi.org/10.1086/380992}{{\em Astrophys. J.} {\bfseries 602}
  no.~1, (Feb., 2004) 388--395}.

\bibitem{Leigh:2017wff}
N.~W.~C. Leigh {\em et~al.}, ``{On the rate of black hole binary mergers in
  galactic nuclei due to dynamical hardening},''
  \href{http://dx.doi.org/10.1093/mnras/stx3134}{{\em Mon. Not. Roy. Astron.
  Soc.} {\bfseries 474} no.~4, (2018) 5672--5683},
  \href{http://arxiv.org/abs/1711.10494}{{\ttfamily arXiv:1711.10494
  [astro-ph.GA]}}.

\bibitem{Bellovary:2015ifg}
J.~M. Bellovary, M.-M. Mac~Low, B.~McKernan, and K.~E.~S. Ford, ``{Migration
  Traps in Disks Around Supermassive Black Holes},''
  \href{http://dx.doi.org/10.3847/2041-8205/819/2/L17}{{\em Astrophys. J.
  Lett.} {\bfseries 819} no.~2, (2016) L17},
  \href{http://arxiv.org/abs/1511.00005}{{\ttfamily arXiv:1511.00005
  [astro-ph.GA]}}.

\bibitem{McKernan_2012}
B.~McKernan, K.~E.~S. Ford, W.~Lyra, and H.~B. Perets, ``Intermediate mass
  black holes in agn discs - i. production and growth: Imbh in agn discs - i,''
  \href{http://dx.doi.org/10.1111/j.1365-2966.2012.21486.x}{{\em Mon. Not. Roy.
  Astron. Soc.} {\bfseries 425} no.~1, (July, 2012) 460–469}.

\bibitem{McKernan_2014}
B.~McKernan, K.~E.~S. Ford, B.~Kocsis, W.~Lyra, and L.~M. Winter,
  ``Intermediate-mass black holes in agn discs – ii. model predictions and
  observational constraints,''
  \href{http://dx.doi.org/10.1093/mnras/stu553}{{\em Mon. Not. Roy. Astron.
  Soc.} {\bfseries 441} no.~1, (May, 2014) 900–909}.

\bibitem{McKernan_2020}
B.~McKernan, K.~E.~S. Ford, R.~O’Shaugnessy, and D.~Wysocki, ``Monte carlo
  simulations of black hole mergers in agn discs: Low $\chi$eff mergers and
  predictions for ligo,'' \href{http://dx.doi.org/10.1093/mnras/staa740}{{\em
  Mon. Not. Roy. Astron. Soc.} {\bfseries 494} no.~1, (Apr., 2020)
  1203–1216}.

\bibitem{Toubiana:2020drf}
A.~Toubiana {\em et~al.}, ``{Detectable environmental effects in GW190521-like
  black-hole binaries with LISA},''
  \href{http://dx.doi.org/10.1103/PhysRevLett.126.101105}{{\em Phys. Rev.
  Lett.} {\bfseries 126} no.~10, (2021) 101105},
  \href{http://arxiv.org/abs/2010.06056}{{\ttfamily arXiv:2010.06056
  [astro-ph.HE]}}.

\bibitem{Sberna:2022qbn}
L.~Sberna {\em et~al.}, ``{Observing GW190521-like binary black holes and their
  environment with LISA},''
  \href{http://dx.doi.org/10.1103/PhysRevD.106.064056}{{\em Phys. Rev. D}
  {\bfseries 106} no.~6, (2022) 064056},
  \href{http://arxiv.org/abs/2205.08550}{{\ttfamily arXiv:2205.08550 [gr-qc]}}.

\bibitem{1998MNRAS.293L...1V}
D.~{Vokrouhlicky} and V.~{Karas}, ``{Stellar capture by an accretion disc},''
  \href{http://dx.doi.org/10.1046/j.1365-8711.1998.01213.x}{{\em Mon. Not. Roy.
  Astron. Soc.} {\bfseries 293} no.~1, (Jan., 1998) L1--L5},
  \href{http://arxiv.org/abs/astro-ph/9710169}{{\ttfamily
  arXiv:astro-ph/9710169 [astro-ph]}}.

\bibitem{1999A&A...352..452S}
L.~{{\v{S}}ubr} and V.~{Karas}, ``{An orbiter crossing an accretion disc},''
  \href{http://dx.doi.org/10.48550/arXiv.astro-ph/9910401}{{\em Astron.
  Astrophys.} {\bfseries 352} (Dec., 1999) 452--458},
  \href{http://arxiv.org/abs/astro-ph/9910401}{{\ttfamily
  arXiv:astro-ph/9910401 [astro-ph]}}.

\bibitem{MacLeod:2019jxd}
M.~{MacLeod} and D.~N.~C. {Lin}, ``{The Effect of Star-Disk Interactions on
  Highly Eccentric Stellar Orbits in Active Galactic Nuclei: A Disk Loss Cone
  and Implications for Stellar Tidal Disruption Events},''
  \href{http://dx.doi.org/10.3847/1538-4357/ab64db}{{\em Astrophys. J.}
  {\bfseries 889} no.~2, (Feb., 2020) 94},
  \href{http://arxiv.org/abs/1909.09645}{{\ttfamily arXiv:1909.09645
  [astro-ph.SR]}}.

\bibitem{Li:2025zgo}
Y.-P. Li, H.~Yang, and Z.~Pan, ``{Extreme mass-ratio inspirals in active
  galactic nucleus disks: The role of circumsingle disks},''
  \href{http://dx.doi.org/10.1103/PhysRevD.111.063074}{{\em Phys. Rev. D}
  {\bfseries 111} no.~6, (2025) 063074},
  \href{http://arxiv.org/abs/2503.04042}{{\ttfamily arXiv:2503.04042
  [astro-ph.HE]}}.

\bibitem{Hopman:2005vr}
C.~Hopman and T.~Alexander, ``{The Orbital statistics of stellar inspiral and
  relaxation near a massive black hole: Characterizing gravitational wave
  sources},'' \href{http://dx.doi.org/10.1086/431475}{{\em Astrophys. J.}
  {\bfseries 629} (2005) 362--372},
  \href{http://arxiv.org/abs/astro-ph/0503672}{{\ttfamily
  arXiv:astro-ph/0503672}}.

\bibitem{1939PCPS...35..405H}
F.~{Hoyle} and R.~A. {Lyttleton}, ``{The effect of interstellar matter on
  climatic variation},''
  \href{http://dx.doi.org/10.1017/S0305004100021150}{{\em Proceedings of the
  Cambridge Philosophical Society} {\bfseries 35} no.~3, (Jan., 1939) 405}.

\bibitem{1944MNRAS.104..273B}
H.~{Bondi} and F.~{Hoyle}, ``{On the mechanism of accretion by stars},''
  \href{http://dx.doi.org/10.1093/mnras/104.5.273}{{\em Mon. Not. Roy. Astron.
  Soc.} {\bfseries 104} (Jan., 1944) 273}.

\bibitem{1987gady.book.....B}
J.~{Binney} and S.~{Tremaine}, {\em {Galactic dynamics}}.
\newblock Princeton University Press, 1987.

\bibitem{2013MNRAS.429.3114C}
D.~{Chapon}, L.~{Mayer}, and R.~{Teyssier}, ``{Hydrodynamics of galaxy mergers
  with supermassive black holes: is there a last parsec problem?},''
  \href{http://dx.doi.org/10.1093/mnras/sts568}{{\em Mon. Not. Roy. Astron.
  Soc} {\bfseries 429} no.~4, (Mar., 2013) 3114--3122},
  \href{http://arxiv.org/abs/1110.6086}{{\ttfamily arXiv:1110.6086
  [astro-ph.GA]}}.

\bibitem{Vicente:2019ilr}
R.~Vicente, V.~Cardoso, and M.~Zilh\~ao, ``{Dynamical friction in slab
  geometries and accretion disks},''
  \href{http://dx.doi.org/10.1093/mnras/stz2526}{{\em Mon. Not. Roy. Astron.
  Soc.} {\bfseries 489} no.~4, (2019) 5424--5435},
  \href{http://arxiv.org/abs/1905.06353}{{\ttfamily arXiv:1905.06353
  [astro-ph.GA]}}.

\bibitem{2021ApJ...916...48D}
A.~J. {Dittmann}, M.~{Cantiello}, and A.~S. {Jermyn}, ``{Accretion onto Stars
  in the Disks of Active Galactic Nuclei},''
  \href{http://dx.doi.org/10.3847/1538-4357/ac042c}{{\em Astrophys. J.}
  {\bfseries 916} no.~1, (July, 2021) 48},
  \href{http://arxiv.org/abs/2102.12484}{{\ttfamily arXiv:2102.12484
  [astro-ph.GA]}}.

\bibitem{Berry:2019wgg}
C.~P.~L. Berry, S.~A. Hughes, C.~F. Sopuerta, A.~J.~K. Chua, A.~Heffernan,
  K.~Holley-Bockelmann, D.~P. Mihaylov, M.~C. Miller, and A.~Sesana, ``{The
  unique potential of extreme mass-ratio inspirals for gravitational-wave
  astronomy},'' {\em Bull. Am. Astron. Soc.} {\bfseries 51} (2019) 42,
  \href{http://arxiv.org/abs/1903.03686}{{\ttfamily arXiv:1903.03686
  [astro-ph.HE]}}.

\bibitem{Zrake:2020zkw}
J.~Zrake, C.~Tiede, A.~MacFadyen, and Z.~Haiman, ``{Equilibrium Eccentricity of
  Accreting Binaries},'' \href{http://dx.doi.org/10.3847/2041-8213/abdd1c}{{\em
  Astrophys. J. Lett.} {\bfseries 909} no.~1, (2021) L13},
  \href{http://arxiv.org/abs/2010.09707}{{\ttfamily arXiv:2010.09707
  [astro-ph.HE]}}.

\bibitem{DOrazio:2021kob}
D.~J. D'Orazio and P.~C. Duffell, ``{Orbital Evolution of Equal-mass Eccentric
  Binaries due to a Gas Disk: Eccentric Inspirals and Circular Outspirals},''
  \href{http://dx.doi.org/10.3847/2041-8213/ac0621}{{\em Astrophys. J. Lett.}
  {\bfseries 914} no.~1, (2021) L21},
  \href{http://arxiv.org/abs/2103.09251}{{\ttfamily arXiv:2103.09251
  [astro-ph.HE]}}.

\bibitem{Siwek:2023rlk}
M.~Siwek, R.~Weinberger, and L.~Hernquist, ``{Orbital evolution of binaries in
  circumbinary discs},'' \href{http://dx.doi.org/10.1093/mnras/stad1131}{{\em
  Mon. Not. Roy. Astron. Soc.} {\bfseries 522} no.~2, (2023) 2707--2717},
  \href{http://arxiv.org/abs/2302.01785}{{\ttfamily arXiv:2302.01785
  [astro-ph.HE]}}.

\bibitem{Tiede:2023dwq}
C.~Tiede and D.~J. D'Orazio, ``{Eccentric binaries in retrograde discs},''
  \href{http://dx.doi.org/10.1093/mnras/stad3551}{{\em Mon. Not. Roy. Astron.
  Soc.} {\bfseries 527} no.~3, (2023) 6021--6037},
  \href{http://arxiv.org/abs/2307.03775}{{\ttfamily arXiv:2307.03775
  [astro-ph.GA]}}.

\bibitem{Klein:2022rbf}
A.~Klein {\em et~al.}, ``{The last three years: multiband gravitational-wave
  observations of stellar-mass binary black holes},''
  \href{http://arxiv.org/abs/2204.03423}{{\ttfamily arXiv:2204.03423
  [astro-ph.HE]}}.

\bibitem{Garg:2023lfg}
M.~Garg, S.~Tiwari, A.~Derdzinski, J.~G. Baker, S.~Marsat, and L.~Mayer, ``{The
  minimum measurable eccentricity from gravitational waves of LISA massive
  black hole binaries},'' \href{http://dx.doi.org/10.1093/mnras/stad3477}{{\em
  Mon. Not. Roy. Astron. Soc.} {\bfseries 528} no.~3, (2024) 4176--4187},
  \href{http://arxiv.org/abs/2307.13367}{{\ttfamily arXiv:2307.13367
  [astro-ph.GA]}}.

\bibitem{Wang:2023tle}
H.~Wang, I.~Harry, A.~Nitz, and Y.-M. Hu, ``{Space-based gravitational wave
  observatories will be able to use eccentricity to unveil stellar-mass binary
  black hole formation},''
  \href{http://dx.doi.org/10.1103/PhysRevD.109.063029}{{\em Phys. Rev. D}
  {\bfseries 109} no.~6, (2024) 063029},
  \href{http://arxiv.org/abs/2304.10340}{{\ttfamily arXiv:2304.10340
  [astro-ph.HE]}}.

\bibitem{Romero-Shaw:2024klf}
I.~M. Romero-Shaw, S.~Goorachurn, M.~Siwek, and C.~J. Moore, ``{Eccentric
  signatures of stellar-mass binary black holes with circumbinary discs in
  LISA},'' \href{http://dx.doi.org/10.1093/mnrasl/slae081}{{\em Mon. Not. Roy.
  Astron. Soc.} {\bfseries 534} no.~1, (2024) L58--L64},
  \href{http://arxiv.org/abs/2407.03869}{{\ttfamily arXiv:2407.03869
  [astro-ph.HE]}}.

\bibitem{Samsing:2024syt}
J.~Samsing, K.~Hendriks, L.~Zwick, D.~J. D'Orazio, and B.~Liu, ``{Gravitational
  Wave Phase Shifts in Eccentric Black Hole Mergers as a Probe of Dynamical
  Formation Environments},'' \href{http://arxiv.org/abs/2403.05625}{{\ttfamily
  arXiv:2403.05625 [astro-ph.HE]}}.

\bibitem{Armitage:2002uu}
P.~J. Armitage and P.~Natarajan, ``{Accretion during the merger of supermassive
  black holes},'' \href{http://dx.doi.org/10.1086/339770}{{\em Astrophys. J.
  Lett.} {\bfseries 567} (2002) L9--L12},
  \href{http://arxiv.org/abs/astro-ph/0201318}{{\ttfamily
  arXiv:astro-ph/0201318}}.

\bibitem{Dittmann:2023dss}
A.~J. Dittmann, G.~Ryan, and M.~C. Miller, ``{The Decoupling of Binaries from
  Their Circumbinary Disks},''
  \href{http://dx.doi.org/10.3847/2041-8213/acd183}{{\em Astrophys. J. Lett.}
  {\bfseries 949} no.~2, (2023) L30},
  \href{http://arxiv.org/abs/2303.16204}{{\ttfamily arXiv:2303.16204
  [astro-ph.HE]}}.

\bibitem{ONeill:2025bjd}
D.~O'Neill, C.~Tiede, D.~J. D'Orazio, Z.~Haiman, and A.~MacFadyen,
  ``{Gravitational Wave Decoupling in Retrograde Circumbinary Disks},''
  \href{http://arxiv.org/abs/2501.11679}{{\ttfamily arXiv:2501.11679
  [astro-ph.HE]}}.

\bibitem{Sarmah:2024nst}
P.~Sarmah, H.~Verma, K.~Cheung, and J.~Silk, ``{Effects of Superradiance in
  Active Galactic Nuclei},''
  \href{http://dx.doi.org/10.1093/mnras/staf326}{{\em Mon. Not. Roy. Astron.
  Soc.} {\bfseries 538} no.~2, (2025) 943--962},
  \href{http://arxiv.org/abs/2404.09955}{{\ttfamily arXiv:2404.09955
  [astro-ph.HE]}}.

\bibitem{1988Natur.331..687H}
J.~G. {Hills}, ``{Hyper-velocity and tidal stars from binaries disrupted by a
  massive Galactic black hole},''
  \href{http://dx.doi.org/10.1038/331687a0}{{\em Nature} {\bfseries 331}
  no.~6158, (Feb., 1988) 687--689}.

\bibitem{Gultekin:2004pm}
K.~Gultekin, M.~C. Miller, and D.~P. Hamilton, ``{Growth of intermediate - mass
  black holes in globular clusters},''
  \href{http://dx.doi.org/10.1086/424809}{{\em Astrophys. J.} {\bfseries 616}
  (2004) 221--230}, \href{http://arxiv.org/abs/astro-ph/0402532}{{\ttfamily
  arXiv:astro-ph/0402532}}.

\bibitem{Samsing:2013kua}
J.~Samsing, M.~MacLeod, and E.~Ramirez-Ruiz, ``{The Formation of Eccentric
  Compact Binary Inspirals and the Role of Gravitational Wave Emission in
  Binary-Single Stellar Encounters},''
  \href{http://dx.doi.org/10.1088/0004-637X/784/1/71}{{\em Astrophys. J.}
  {\bfseries 784} (2014) 71}, \href{http://arxiv.org/abs/1308.2964}{{\ttfamily
  arXiv:1308.2964 [astro-ph.HE]}}.

\bibitem{Ori:2000zn}
A.~Ori and K.~S. Thorne, ``{The Transition from inspiral to plunge for a
  compact body in a circular equatorial orbit around a massive, spinning black
  hole},'' \href{http://dx.doi.org/10.1103/PhysRevD.62.124022}{{\em Phys. Rev.
  D} {\bfseries 62} (2000) 124022},
  \href{http://arxiv.org/abs/gr-qc/0003032}{{\ttfamily arXiv:gr-qc/0003032}}.

\bibitem{Apte:2019txp}
A.~Apte and S.~A. Hughes, ``{Exciting black hole modes via misaligned
  coalescences: I. Inspiral, transition, and plunge trajectories using a
  generalized Ori-Thorne procedure},''
  \href{http://dx.doi.org/10.1103/PhysRevD.100.084031}{{\em Phys. Rev. D}
  {\bfseries 100} no.~8, (2019) 084031},
  \href{http://arxiv.org/abs/1901.05901}{{\ttfamily arXiv:1901.05901 [gr-qc]}}.

\bibitem{Hughes:2019zmt}
S.~A. Hughes, A.~Apte, G.~Khanna, and H.~Lim, ``{Learning about black hole
  binaries from their ringdown spectra},''
  \href{http://dx.doi.org/10.1103/PhysRevLett.123.161101}{{\em Phys. Rev.
  Lett.} {\bfseries 123} no.~16, (2019) 161101},
  \href{http://arxiv.org/abs/1901.05900}{{\ttfamily arXiv:1901.05900 [gr-qc]}}.

\bibitem{Becker:2024xdi}
D.~R. Becker and S.~A. Hughes, ``{Transition from adiabatic inspiral to plunge
  for eccentric binaries},''
  \href{http://dx.doi.org/10.1103/PhysRevD.111.064003}{{\em Phys. Rev. D}
  {\bfseries 111} no.~6, (2025) 064003},
  \href{http://arxiv.org/abs/2410.09160}{{\ttfamily arXiv:2410.09160 [gr-qc]}}.

\bibitem{Kuchler:2024esj}
L.~K\"uchler, G.~Comp\`ere, L.~Durkan, and A.~Pound, ``{Self-force framework
  for transition-to-plunge waveforms},''
  \href{http://dx.doi.org/10.21468/SciPostPhys.17.2.056}{{\em SciPost Phys.}
  {\bfseries 17} no.~2, (2024) 056},
  \href{http://arxiv.org/abs/2405.00170}{{\ttfamily arXiv:2405.00170 [gr-qc]}}.

\bibitem{Ferguson:2020xnm}
D.~Ferguson, K.~Jani, P.~Laguna, and D.~Shoemaker, ``{Assessing the readiness
  of numerical relativity for LISA and 3G detectors},''
  \href{http://dx.doi.org/10.1103/PhysRevD.104.044037}{{\em Phys. Rev. D}
  {\bfseries 104} no.~4, (2021) 044037},
  \href{http://arxiv.org/abs/2006.04272}{{\ttfamily arXiv:2006.04272 [gr-qc]}}.

\bibitem{Yunes:2009ke}
N.~Yunes and F.~Pretorius, ``{Fundamental Theoretical Bias in Gravitational
  Wave Astrophysics and the Parameterized Post-Einsteinian Framework},''
  \href{http://dx.doi.org/10.1103/PhysRevD.80.122003}{{\em Phys. Rev. D}
  {\bfseries 80} (2009) 122003},
  \href{http://arxiv.org/abs/0909.3328}{{\ttfamily arXiv:0909.3328 [gr-qc]}}.

\bibitem{Yunes:2013dva}
N.~Yunes and X.~Siemens, ``{Gravitational-Wave Tests of General Relativity with
  Ground-Based Detectors and Pulsar Timing-Arrays},''
  \href{http://dx.doi.org/10.12942/lrr-2013-9}{{\em Living Rev. Rel.}
  {\bfseries 16} (2013) 9}, \href{http://arxiv.org/abs/1304.3473}{{\ttfamily
  arXiv:1304.3473 [gr-qc]}}.

\bibitem{Okounkova:2017yby}
M.~Okounkova, L.~C. Stein, M.~A. Scheel, and D.~A. Hemberger, ``{Numerical
  binary black hole mergers in dynamical Chern-Simons gravity: Scalar field},''
  \href{http://dx.doi.org/10.1103/PhysRevD.96.044020}{{\em Phys. Rev. D}
  {\bfseries 96} no.~4, (2017) 044020},
  \href{http://arxiv.org/abs/1705.07924}{{\ttfamily arXiv:1705.07924 [gr-qc]}}.

\bibitem{Okounkova:2020rqw}
M.~Okounkova, ``{Numerical relativity simulation of GW150914 in Einstein
  dilaton Gauss-Bonnet gravity},''
  \href{http://dx.doi.org/10.1103/PhysRevD.102.084046}{{\em Phys. Rev. D}
  {\bfseries 102} no.~8, (2020) 084046},
  \href{http://arxiv.org/abs/2001.03571}{{\ttfamily arXiv:2001.03571 [gr-qc]}}.

\bibitem{Yuan:2024duo}
X.~Yuan, J.-d. Zhang, and J.~Mei, ``{Distinguish the environmental effects and
  modified theory of gravity with multiple massive black-hole binaries},''
  \href{http://arxiv.org/abs/2412.00915}{{\ttfamily arXiv:2412.00915 [gr-qc]}}.

\bibitem{Yunes:2010sm}
N.~Yunes, M.~Coleman~Miller, and J.~Thornburg, ``{The Effect of Massive
  Perturbers on Extreme Mass-Ratio Inspiral Waveforms},''
  \href{http://dx.doi.org/10.1103/PhysRevD.83.044030}{{\em Phys. Rev. D}
  {\bfseries 83} (2011) 044030},
  \href{http://arxiv.org/abs/1010.1721}{{\ttfamily arXiv:1010.1721
  [astro-ph.GA]}}.

\bibitem{Meiron:2016ipr}
Y.~Meiron, B.~Kocsis, and A.~Loeb, ``{Detecting triple systems with
  gravitational wave observations},''
  \href{http://dx.doi.org/10.3847/1538-4357/834/2/200}{{\em Astrophys. J.}
  {\bfseries 834} no.~2, (2017) 200},
  \href{http://arxiv.org/abs/1604.02148}{{\ttfamily arXiv:1604.02148
  [astro-ph.HE]}}.

\bibitem{Hendriks:2024zbu}
K.~Hendriks, L.~Zwick, and J.~Samsing, ``{Eccentric features in the
  gravitational wave phase of dynamically formed black hole binaries},''
  \href{http://arxiv.org/abs/2408.04603}{{\ttfamily arXiv:2408.04603 [gr-qc]}}.

\bibitem{DeLuca:2025bph}
V.~De~Luca, L.~Del~Grosso, F.~Iacovelli, A.~Maselli, and E.~Berti, ``{Tainted
  Love: Systematic biases from ignoring environmental tidal effects in
  gravitational wave observations},''
  \href{http://arxiv.org/abs/2503.10746}{{\ttfamily arXiv:2503.10746 [gr-qc]}}.

\bibitem{Robson:2018jly}
T.~Robson and N.~J. Cornish, ``{Detecting Gravitational Wave Bursts with LISA
  in the presence of Instrumental Glitches},''
  \href{http://dx.doi.org/10.1103/PhysRevD.99.024019}{{\em Phys. Rev. D}
  {\bfseries 99} no.~2, (2019) 024019},
  \href{http://arxiv.org/abs/1811.04490}{{\ttfamily arXiv:1811.04490 [gr-qc]}}.

\bibitem{Baghi:2021tfd}
Q.~Baghi, N.~Korsakova, J.~Slutsky, E.~Castelli, N.~Karnesis, and J.-B. Bayle,
  ``{Detection and characterization of instrumental transients in LISA
  Pathfinder and their projection to LISA},''
  \href{http://dx.doi.org/10.1103/PhysRevD.105.042002}{{\em Phys. Rev. D}
  {\bfseries 105} no.~4, (2022) 042002},
  \href{http://arxiv.org/abs/2112.07490}{{\ttfamily arXiv:2112.07490 [gr-qc]}}.

\bibitem{Burke:2025bun}
O.~Burke, S.~Marsat, J.~R. Gair, and M.~L. Katz, ``{Mind the gap: addressing
  data gaps and assessing noise mismodeling in LISA},''
  \href{http://arxiv.org/abs/2502.17426}{{\ttfamily arXiv:2502.17426 [gr-qc]}}.

\bibitem{Strub:2024kbe}
S.~H. Strub, L.~Ferraioli, C.~Schmelzbach, S.~C. St\"ahler, and D.~Giardini,
  ``{Global analysis of LISA data with Galactic binaries and massive black hole
  binaries},'' \href{http://dx.doi.org/10.1103/PhysRevD.110.024005}{{\em Phys.
  Rev. D} {\bfseries 110} no.~2, (2024) 024005},
  \href{http://arxiv.org/abs/2403.15318}{{\ttfamily arXiv:2403.15318 [gr-qc]}}.

\bibitem{Katz:2024oqg}
M.~L. Katz, N.~Karnesis, N.~Korsakova, J.~R. Gair, and N.~Stergioulas,
  ``{Efficient GPU-accelerated multisource global fit pipeline for LISA data
  analysis},'' \href{http://dx.doi.org/10.1103/PhysRevD.111.024060}{{\em Phys.
  Rev. D} {\bfseries 111} no.~2, (2025) 024060},
  \href{http://arxiv.org/abs/2405.04690}{{\ttfamily arXiv:2405.04690 [gr-qc]}}.

\bibitem{Vicente:2025gsg}
R.~Vicente, T.~K. Karydas, and G.~Bertone, ``{A fully relativistic treatment of
  EMRIs in collisionless environments},''
  \href{http://arxiv.org/abs/2505.09715}{{\ttfamily arXiv:2505.09715 [gr-qc]}}.

\bibitem{Witek:2012tr}
H.~Witek, V.~Cardoso, A.~Ishibashi, and U.~Sperhake, ``{Superradiant
  Instabilities in Astrophysical Systems},''
  \href{http://dx.doi.org/10.1103/PhysRevD.87.043513}{{\em Phys. Rev.}
  {\bfseries D87} (2013) 043513},
\href{http://arxiv.org/abs/1212.0551}{{\ttfamily arXiv:1212.0551 [gr-qc]}}.

\bibitem{Alcubierre:2009ij}
M.~Alcubierre, J.~C. Degollado, and M.~Salgado, ``{The Einstein-Maxwell system
  in 3+1 form and initial data for multiple charged black holes},''
  \href{http://dx.doi.org/10.1103/PhysRevD.80.104022}{{\em Phys. Rev. D}
  {\bfseries 80} (2009) 104022},
  \href{http://arxiv.org/abs/0907.1151}{{\ttfamily arXiv:0907.1151 [gr-qc]}}.

\bibitem{MUNZ2000484}
C.-D. Munz, P.~Omnes, R.~Schneider, E.~Sonnendr{\"u}cker, and U.~Voss,
  ``Divergence correction techniques for maxwell solvers based on a hyperbolic
  model,'' {\em Journal of Computational Physics} {\bfseries 161} no.~2, (2000)
  484--511.

\bibitem{Abgrall:2014rfb}
R.~Abgrall and H.~Kumar, ``{Robust Finite Volume Schemes for Two-Fluid Plasma
  Equations},'' \href{http://dx.doi.org/10.1007/s10915-013-9809-6}{{\em Journal
  of Scientific Computing} {\bfseries 60} no.~3, (2014) 584--611}.

\bibitem{Wang:2022hra}
Z.~Wang, T.~Helfer, K.~Clough, and E.~Berti, ``{Superradiance in massive vector
  fields with spatially varying mass},''
  \href{http://dx.doi.org/10.1103/PhysRevD.105.104055}{{\em Phys. Rev. D}
  {\bfseries 105} no.~10, (2022) 104055},
  \href{http://arxiv.org/abs/2201.08305}{{\ttfamily arXiv:2201.08305 [gr-qc]}}.

\bibitem{Hilditch:2013sba}
D.~Hilditch, ``{An Introduction to Well-posedness and Free-evolution},''
  \href{http://dx.doi.org/10.1142/S0217751X13400150}{{\em Int. J. Mod. Phys. A}
  {\bfseries 28} (2013) 1340015},
  \href{http://arxiv.org/abs/1309.2012}{{\ttfamily arXiv:1309.2012 [gr-qc]}}.

\bibitem{Thorne1982}
K.~S. Thorne and D.~MacDonald, ``{Electrodynamics in Curved Spacetime - 3+1
  Formulation},'' \href{http://dx.doi.org/10.1093/mnras/198.2.339}{{\em Mon.
  Not. Roy. Astron. Soc.} {\bfseries 198} (1982) 339}.

\bibitem{Zilhao:2013hia}
M.~Zilh\~ao and F.~L\"offler, ``{An Introduction to the Einstein Toolkit},''
  \href{http://dx.doi.org/10.1142/S0217751X13400149}{{\em Int. J. Mod. Phys. A}
  {\bfseries 28} (2013) 1340014},
  \href{http://arxiv.org/abs/1305.5299}{{\ttfamily arXiv:1305.5299 [gr-qc]}}.

\bibitem{Bernard:2019nkv}
L.~Bernard, V.~Cardoso, T.~Ikeda, and M.~Zilh\~ao, ``{Physics of black hole
  binaries: Geodesics, relaxation modes, and energy extraction},''
  \href{http://dx.doi.org/10.1103/PhysRevD.100.044002}{{\em Phys. Rev. D}
  {\bfseries 100} no.~4, (2019) 044002},
  \href{http://arxiv.org/abs/1905.05204}{{\ttfamily arXiv:1905.05204 [gr-qc]}}.

\bibitem{Cunha:2017wao}
P.~V.~P. Cunha, J.~A. Font, C.~Herdeiro, E.~Radu, N.~Sanchis-Gual, and
  M.~Zilh\~ao, ``{Lensing and dynamics of ultracompact bosonic stars},''
  \href{http://dx.doi.org/10.1103/PhysRevD.96.104040}{{\em Phys. Rev. D}
  {\bfseries 96} no.~10, (2017) 104040},
  \href{http://arxiv.org/abs/1709.06118}{{\ttfamily arXiv:1709.06118 [gr-qc]}}.

\bibitem{Zilhao:2015tya}
M.~Zilh\~ao, H.~Witek, and V.~Cardoso, ``{Nonlinear interactions between black
  holes and Proca fields},''
  \href{http://dx.doi.org/10.1088/0264-9381/32/23/234003}{{\em Class. Quant.
  Grav.} {\bfseries 32} (2015) 234003},
  \href{http://arxiv.org/abs/1505.00797}{{\ttfamily arXiv:1505.00797 [gr-qc]}}.

\bibitem{Sanchis-Gual:2022zsr}
N.~Sanchis-Gual, M.~Zilh\~ao, and V.~Cardoso, ``{Electromagnetic emission from
  axionic boson star collisions},''
  \href{http://dx.doi.org/10.1103/PhysRevD.106.064034}{{\em Phys. Rev. D}
  {\bfseries 106} no.~6, (2022) 064034},
  \href{http://arxiv.org/abs/2207.05494}{{\ttfamily arXiv:2207.05494 [gr-qc]}}.

\bibitem{Pollney:2009yz}
D.~Pollney, C.~Reisswig, E.~Schnetter, N.~Dorband, and P.~Diener, ``{High
  accuracy binary black hole simulations with an extended wave zone},''
  \href{http://dx.doi.org/10.1103/PhysRevD.83.044045}{{\em Phys. Rev. D}
  {\bfseries 83} (2011) 044045},
  \href{http://arxiv.org/abs/0910.3803}{{\ttfamily arXiv:0910.3803 [gr-qc]}}.

\bibitem{Reisswig:2012nc}
C.~Reisswig, R.~Haas, C.~D. Ott, E.~Abdikamalov, P.~M\"osta, D.~Pollney, and
  E.~Schnetter, ``{Three-Dimensional General-Relativistic Hydrodynamic
  Simulations of Binary Neutron Star Coalescence and Stellar Collapse with
  Multipatch Grids},'' \href{http://dx.doi.org/10.1103/PhysRevD.87.064023}{{\em
  Phys. Rev. D} {\bfseries 87} no.~6, (2013) 064023},
  \href{http://arxiv.org/abs/1212.1191}{{\ttfamily arXiv:1212.1191
  [astro-ph.HE]}}.

\bibitem{bender78:AMM}
C.~M. Bender and S.~A. Orszag, {\em {Advanced Mathematical Methods for
  Scientists and Engineers}}.
\newblock McGraw-Hill, 1978.

\bibitem{NIST:DLMF}
{\em {\it NIST Digital Library of Mathematical Functions}}.
\newblock \url{http://dlmf.nist.gov/}.

\bibitem{Sago:2002fe}
N.~Sago, H.~Nakano, and M.~Sasaki, ``{Gauge problem in the gravitational
  selfforce. 1. Harmonic gauge approach in the Schwarzschild background},''
  \href{http://dx.doi.org/10.1103/PhysRevD.67.104017}{{\em Phys. Rev. D}
  {\bfseries 67} (2003) 104017},
  \href{http://arxiv.org/abs/gr-qc/0208060}{{\ttfamily arXiv:gr-qc/0208060}}.

\bibitem{Rosa:2011my}
J.~Rosa and S.~Dolan, ``{Massive Vector Fields on the Schwarzschild Spacetime:
  Quasi-Normal Modes and Bound States},''
  \href{http://dx.doi.org/10.1103/PhysRevD.85.044043}{{\em Phys. Rev.}
  {\bfseries D85} (2012) 044043},
\href{http://arxiv.org/abs/1110.4494}{{\ttfamily arXiv:1110.4494 [hep-th]}}.

\bibitem{Leaver:1986}
E.~W. Leaver, ``{Solutions to a generalized spheroidal wave equation:
  Teukolsky's equation in general relativity, and the two-center problem in
  molecular quantum mechanics},'' {\em Journal of Mathematical Physics}
  {\bfseries 27} (1986) 1238.

\bibitem{Pincherele}
S.~Pincherle, ``{Sur la g\'{e}n\'{e}ration de syst\`{e}mes r\'{e}currents au
  moyen d'une \'{e}quation lin\'{e}aire diff\'{e}rentielle},'' {\em Acta Math.}
  {\bfseries 15} (1892) 341--363.

\bibitem{Gautschi:1967cat}
W.~Gautschi, ``{Computational Aspects of Three-Term Recurrence Relations},''
  {\em SIAM Review} {\bfseries 9} (1967) 24--82.

\bibitem{Nollert:1993zz}
H.-P. Nollert, ``{Quasinormal modes of Schwarzschild black holes: The
  determination of quasinormal frequencies with very large imaginary parts},''
  \href{http://dx.doi.org/10.1103/PhysRevD.47.5253}{{\em Phys. Rev. D}
  {\bfseries 47} (1993) 5253--5258}.

\bibitem{Onozawa:1996ux}
H.~Onozawa, ``{A Detailed study of quasinormal frequencies of the Kerr black
  hole},'' \href{http://dx.doi.org/10.1103/PhysRevD.55.3593}{{\em Phys. Rev. D}
  {\bfseries 55} (1997) 3593--3602},
  \href{http://arxiv.org/abs/gr-qc/9610048}{{\ttfamily arXiv:gr-qc/9610048}}.

\bibitem{Berti:2003jh}
E.~Berti, V.~Cardoso, K.~D. Kokkotas, and H.~Onozawa, ``{Highly damped
  quasinormal modes of Kerr black holes},''
  \href{http://dx.doi.org/10.1103/PhysRevD.68.124018}{{\em Phys. Rev. D}
  {\bfseries 68} (2003) 124018},
  \href{http://arxiv.org/abs/hep-th/0307013}{{\ttfamily arXiv:hep-th/0307013}}.

\bibitem{Berti:2004um}
E.~Berti, V.~Cardoso, and S.~Yoshida, ``{Highly damped quasinormal modes of
  Kerr black holes: A Complete numerical investigation},''
  \href{http://dx.doi.org/10.1103/PhysRevD.69.124018}{{\em Phys. Rev. D}
  {\bfseries 69} (2004) 124018},
  \href{http://arxiv.org/abs/gr-qc/0401052}{{\ttfamily arXiv:gr-qc/0401052}}.

\bibitem{Sundararajan:2007jg}
P.~A. Sundararajan, G.~Khanna, and S.~A. Hughes, ``{Towards adiabatic waveforms
  for inspiral into Kerr black holes. I. A New model of the source for the time
  domain perturbation equation},''
  \href{http://dx.doi.org/10.1103/PhysRevD.76.104005}{{\em Phys. Rev. D}
  {\bfseries 76} (2007) 104005},
  \href{http://arxiv.org/abs/gr-qc/0703028}{{\ttfamily arXiv:gr-qc/0703028}}.

\bibitem{Wardell:2024yoi}
B.~Wardell, C.~Kavanagh, and S.~R. Dolan, ``{Sourced metric perturbations of
  Kerr spacetime in Lorenz gauge},''
  \href{http://arxiv.org/abs/2406.12510}{{\ttfamily arXiv:2406.12510 [gr-qc]}}.

\bibitem{Spiers:2024src}
A.~Spiers, ``{Analytically separating the source of the Teukolsky equation},''
  \href{http://dx.doi.org/10.1103/PhysRevD.109.104059}{{\em Phys. Rev. D}
  {\bfseries 109} no.~10, (2024) 104059},
  \href{http://arxiv.org/abs/2402.00604}{{\ttfamily arXiv:2402.00604 [gr-qc]}}.

\bibitem{Warburton:2021kwk}
N.~Warburton, A.~Pound, B.~Wardell, J.~Miller, and L.~Durkan,
  ``{Gravitational-Wave Energy Flux for Compact Binaries through Second Order
  in the Mass Ratio},''
  \href{http://dx.doi.org/10.1103/PhysRevLett.127.151102}{{\em Phys. Rev.
  Lett.} {\bfseries 127} no.~15, (2021) 151102},
  \href{http://arxiv.org/abs/2107.01298}{{\ttfamily arXiv:2107.01298 [gr-qc]}}.

\end{thebibliography}\endgroup
